\documentclass[openany,graybox,envcountsame,envcountsect]{svmono}

\usepackage[mathcal]{euscript}
\usepackage{mathrsfs}

\usepackage{hyperref}
\usepackage{mathptmx}       
\usepackage{helvet}         
\usepackage{courier}        
\usepackage{makeidx}         
\usepackage{graphicx}        
\usepackage{multicol}        
\usepackage[bottom]{footmisc}

\usepackage[small,nohug]{diagrams}
\diagramstyle[labelstyle=\scriptstyle]
\usepackage{tikz}
\usepackage{tikz-cd}

\usetikzlibrary{decorations.markings}

\makeindex

\usepackage{amscd,amssymb,amsmath,latexsym,bm}
\usepackage{enumerate}
\usepackage{stackengine}
\usepackage{epsfig}
\usepackage{fancybox}
\usepackage{verbatim}
\usepackage{mathtools}
\usepackage{ulem}

\usepackage{paralist}

\newcommand{\CM}{{\mathbb C}}
\newcommand{\NM}{{\mathbb N}}
\newcommand{\RM}{{\mathbb R}}
\newcommand{\SM}{{\mathbb S}}
\newcommand{\TM}{{\mathbb T}}
\newcommand{\ZM}{{\mathbb Z}}
\newcommand{\PM}{{\mathbb P}}

\newcommand{\Aa}{{\cal A}}
\newcommand{\Ee}{{\cal E}}
\newcommand{\Pp}{{\cal P}}
\newcommand{\PP}{{\bf P}}

\newcommand{\BB}{{\bf B}}
\newcommand{\DD}{{\bf D}}

\newcommand{\FF}{{\bf F}}
\newcommand{\GG}{{\bf G}}
\newcommand{\Bb}{{\cal B}}

\newcommand{\Ff}{{\cal F}}

\newcommand{\Uu}{{\cal U}}

\newcommand{\Ss}{{\cal S}}
\newcommand{\Oo}{{\cal O}}
\newcommand{\Tt}{{\cal T}}
\newcommand{\Rr}{{\cal R}}
\newcommand{\Nn}{{\cal N}}
\newcommand{\Mm}{{\cal M}}
\newcommand{\Cc}{{\cal C}}

\newcommand{\Ll}{{\cal L}}
\newcommand{\Qq}{{\cal Q}}
\newcommand{\Kk}{{\cal K}}
\newcommand{\Hh}{{\cal H}}
\newcommand{\HhO}{{\cal H}_0}

\def\esssup{\mathop{\rm ess\,sup}}

\newcommand{\one}{{\bf 1}}

\newcommand{\Tr}{\mbox{\rm Tr}}

\newcommand{\Ch}{{\rm Ch}} 
\newcommand{\Ind}{{\rm Ind}} 
\newcommand{\Ker}{{\rm Ker}} 
\newcommand{\Ran}{{\rm Ran}} 
\newcommand{\sgn}{{\rm sgn}} 
 
\newcommand{\diag}{{\rm diag}}

\newcommand{\ToepProj}{\Pi}
\newcommand{\weight}{w} 
\newcommand{\dualvar}{k} 
\newcommand{\ScaleInd}{j} 
\newcommand{\Weight}{\zeta}
\newcommand{\GenGNS}{X}

\newcommand{\LpDyn}{\alpha^{(p)}}
\newcommand{\LtwoDyn}{\alpha^{(2)}}
\newcommand{\LtwoDynAv}{\alpha}
\newcommand{\Arvesonaction}{\beta} 
\newcommand{\Diffaction}{\beta} 
\newcommand{\Dualityactionop}{\xi} 
\newcommand{\Dualityactiondualop}{\beta} 
\newcommand{\Dualityactionnp}{\theta} 
\newcommand{\Chernaction}{\alpha} 
\newcommand{\Smoothaction}{\alpha}
\newcommand{\Besovaction}{\beta} 
\newcommand{\Indexaction}{\alpha}
\newcommand{\Halfspaceaction}{\xi}
\newcommand{\Halfspaceunitvector}{{v_\xi}}

\newcommand{\Toep}{{\rm T}(\Aa, G, \Dualityactionop)} 
\newcommand{\ToepR}{{\rm T}(\Aa, \bbR, \Dualityactionop)}
\newcommand{\ToepT}{{\rm T}(\Aa, \bbT, \Dualityactionop)}

\newcommand{\bbC}{\mathbb{C}}

\newcommand{\bbN}{\mathbb{N}}
\newcommand{\bbP}{\mathbb{P}}
\newcommand{\bbQ}{\mathbb{Q}}
\newcommand{\bbR}{\mathbb{R}}
\newcommand{\bbT}{\mathbb{T}}
\newcommand{\bbZ}{\mathbb{Z}}

\newcommand{\calB}{\mathcal{B}}
\newcommand{\calC}{\mathcal{C}}
\newcommand{\calD}{\mathcal{D}}

\newcommand{\calF}{\mathcal{F}}
\newcommand{\calK}{\mathcal{K}}

\newcommand{\calN}{\mathcal{N}}
\newcommand{\calP}{\mathcal{P}}

\newcommand{\calT}{\mathcal{T}}

\newcommand{\calV}{\mathcal{V}}

\newcommand{\scrA}{\mathscr{A}}
\newcommand{\scrB}{\mathscr{B}}

\newcommand{\scrF}{\mathscr{F}}

\newcommand{\scrS}{\mathscr{S}}

\newcommand{\Mmc}{\Mm^c_{\Tt,\alpha}}

\newcommand{\conjugate}[1]{ \overline{#1}}
\newcommand{\difd}{\textup{d}}
\newcommand{\idmap}{\textup{id}}
\newcommand{\Exp}{\textup{Exp}}
\providecommand{\abs}[1]{\left \lvert#1 \right \rvert} 
\providecommand{\norm}[1]{\left \lVert#1 \right \rVert}
\DeclareMathOperator*{\slim}{s-lim}

\newarrow {Congruent} 33333

\begin{document}

\title{Harmonic analysis in operator algebras}

\subtitle{and its applications to index theory and topological solid state systems}

\author{Hermann Schulz-Baldes and Tom Stoiber}


\maketitle

\frontmatter


\tableofcontents


\mainmatter

\setcounter{chapter}{-1}

\chapter{Preface and overview}
\label{chap-Preface}

The central theme of classical harmonic analysis  is to characterize regularity properties of functions $f:\RM^n\to \RM$ or $f:\TM^n\to\RM$ such as (weak) differentiability, integrability and mixtures of these two properties. If these functions are seen as (parts of) symbols of (possibly singular) integral operators (with Calderon-Zygmund kernels), it is then of crucial interest to characterize boundedness and traceclass properties of these operators in terms of the regularity of their symbols. This naturally leads to a wealth of function spaces like: H\"older, $L^p$, Sobolev, Besov and BMO spaces. There are numerous modern accounts of the subject, each with different point of view \cite{Ste,Tri,Gra,MS}. Some of the key elements are Littlewood-Paley decompositions, Fourier multipliers and maximal functions. The theory also extends to functions on Riemannian manifolds \cite{Tri}.

\vspace{.2cm}

Of relevance for the present work is the particular case of classical Hankel operators of the form $H_f=Pf(1-P)$ where $P$ is the Hardy projection in $L^2(\TM^1)$ onto functions on the torus $\TM^1=[0,1)$ with positive frequencies and $f$ is viewed as the multiplication operator with a function $f:\TM^1\to\RM$. The following classical results are then well-known and nicely exposed in \cite{Pel,Connes94}. Kronecker proved that $H_f$ is of finite rank if and only if $f$ is rational. On the other hand, results of Nehari \cite{Neh} and Feffermann \cite{Fef} imply that $H_f$ is bounded if and only if $f$ is BMO. Furthermore, Peller showed that $H_f$ is in the Schatten ideal $\Ll^p$ if and only if $f$ is in the Besov space $B^{\frac{1}{p}}_{p,p}$. In particular, $H_f$ is traceclass if $f\in B^1_{1,1}$. For invertible such functions, the associated Toeplitz operator $T_f=PfP$ is a Fredholm operator on $\Ran(P)$ and its index is given by 
\begin{equation}
\label{eq-Noether}
\Ind(T_f)
\;=\;
-\,\Tr\big(f^{-1}[P,f]\big)
\;=\;
-\,\int_{\TM^1}f^{-1}df
\;,
\end{equation}
where latter equality follows from Noether-Gohberg-Krein index theorem for the winding number of a differentiable function $f$.

\vspace{.2cm}

The Peller criterion has been generalized in many directions. Peller himself extended it to vector-valued functions \cite{Peller82} and Janson and Wolff to higher dimensions \cite{JW}. Janson and Peetre proved it for paracommutators \cite{JansonPeetre88} and  Zhu for Hankel operators on Bergman spaces \cite{Zhu}. Necessary and sufficient conditions for the Hankel operators to be of Dixmier trace-class were given by Engli\v{s} and Rochberg \cite{EnglisRochberg09}. This was generalized by Goffeng and Usachev who showed how Besov spaces can be used to characterize Hankel operators in Macaev ideals \cite{GU}.

\vspace{.2cm}

This work further extends the definition of Besov spaces and associated Peller criteria to noncommutative von Neumann algebras $\Mm$ equipped with a weakly continuous $G$-action $\alpha$ and an $\alpha$-invariant semi-finite, normal, faithful (s.n.f.) trace $\Tt$ where $G=\TM^{n_1}\times\RM^{n_2}$ with $n_1+n_2=n$. The main application here is not a trace formula in this context, but rather a generalization of the index formula \eqref{eq-Noether} in the spirit of Connes' non-commutative geometry \cite{Connes94}. In that framework, a natural cyclic cocycle on sufficiently smooth elements $a_0,\ldots,a_n$ of $\Mm$ is defined by
$$
\Ch_{\Tt,\alpha}(a_0,\ldots,a_n) 
\;=\; 
c_n \,
\sum_{\rho \in S_n} (-1)^\rho\, \Tt\big(a_0 \nabla_{\rho(1)} a_1 \cdots \nabla_{\rho(n)} a_n\big)
\;,
$$
where the sum carries over the symmetric group $S_n$ and $(-1)^\sigma$ is the signature of the permutation $\sigma$, the $\nabla_j$ are the derivations associated to the $G$-actions and finally the normalization constants $c_n$ are chosen as in {\rm Definition~\ref{def-ChernCocycleFrechet}} below. It then follows from semi-finite index theory (a history of relevant contributions follows shortly)  that, if $n$ is odd and $u\in\Mm$ a sufficiently smooth unitary, 
\begin{equation}
\label{eq-OddPairingIntro}
\hat{\Tt}_\alpha \mbox{-}\Ind(\PP  \pi(u) \PP  + 1-\PP ) 
\;=\; 
\Ch_{\Tt,\alpha}(u^*-1,u-1,\ldots,u^*-1,u-1)
\;,
\end{equation}
where the index is understood in the Breuer-Fredholm sense w.r.t. the dual trace $\hat{\Tt}_\alpha$ on the von Neumann crossed product $\Mm \rtimes_\alpha G$ and $\PP  = \chi(\DD>0)$ is the Hardy projection of the Dirac operator constructed from the generators $\gamma_1,\ldots,\gamma_n$ of the complex Clifford algebra via
$$
\DD
\;=\; 
\sum_{i=1}^n \gamma_i \otimes D_i 
\;,
$$
with $D_1,\ldots,D_n$ being the self-adjoint generators of the action $\alpha$ in a covariant representation $\pi$. Phrased differently, one considers the Toeplitz operator $\PP \pi(u) \PP$ for non-commutative symbols $u$ contained in a suitable smooth subalgebra $\scrA \subset \Mm$ and finds that its (Breuer)-index is given by a higher-dimensional analogue of the winding number. This is a non-commutative generalization of the Noether-Gohberg-Krein theorem.  
 In a $C^*$-algebraic setting of such a norm continuous action and for $G=\RM$, Connes proved a precursor of \eqref{eq-OddPairingIntro} for smooth unitaries $u$ in the form of a trace formula using the Connes-Thom isomorphism \cite{Connes81}. The interpretation as a semi-finite Breuer-Fredholm index was then refined in \cite{Lesch91} and \cite{PR}. Newer works such as \cite{Carey06,CPRS2,CPRS3,CGRS,Wahl10,Andersson15,PS,BourneSchuba} further extend these results and interpret them in terms of semi-finite spectral triples and spectral flow.  Let us note, however, that all purely $C^*$-algebraic approaches to index theory require some smoothness w.r.t. the operator norm topology, notably in the cases described above, $u$ has to be at least norm-differentiable w.r.t. the action $\alpha$ (and satisfy additional summability conditions).  Also the approach using unbounded spectral triples \cite{CPRS2,CGRS} usually requires at least that the $\nabla_i u$ define bounded operators. This is already too restrictive for some applications (in physical systems that will be described below).


\vspace{.2cm}

This work (see Chapter~\ref{chap-BreuerToep}) proves \eqref{eq-OddPairingIntro} and also its even analogue for differentiable symbols, but the main point is that it provides more general and natural regularity assumptions on the symbols for the validity of this index theorem which do not require the symbols to be either differentiable or continuous ({\it i.e.} lie in an underlying $C^*$-algebra).  Generalizing the classical theory sketched above, one introduces Besov spaces for a quadruple $(\Mm,\alpha,G,\Tt)$ consisting of a $W^*$-dynamical system with an invariant trace.  A lot is known for such $W^*$-dynamical systems, its $W^*$-crossed product $\Mm\rtimes_\alpha G$ and the associated non-commutative $L^p$-spaces $L^p(\Mm)$ and $L^p(\Mm\rtimes_\alpha G)$ \cite{Takesaki2003,Pisier03}, see Section~\ref{sec-CrossedProd} and the appendices. Recalling that $\Mm$ is considered as the space of symbols, the Toeplitz operators $T_a = \PP \pi(a)\PP$ are by construction elements of the $W^*$-crossed product algebra $\Mm\rtimes_\alpha G$ and applicability of the index theorem turns out to be governed by the regularity of the Hankel operator $H_a = \PP \, \pi(a) (\one-\PP) \in \Mm\rtimes_\alpha G$. In particular, the index theorem can be established for $H_a \in L^{n+1}(\Mm\rtimes_\alpha G)$, {\it i.e.} if the commutator $[\PP, \pi(a)]$ is $n$-summable. This justifies the search for a generalization of Peller's criterion.

\vspace{.2cm} 

The approach to harmonic analysis on operator algebras is based on the theory of abelian groups of automorphisms by Arveson \cite{Arveson73} (see also \cite{Pedersen79} and \cite{Takesaki2003}), in which the analogue of a Fourier multiplier $f\in \mathcal{F}(L^1(G))$ acts by convolution of the $G$-action $\alpha$ with the inverse Fourier transform of $f$ by
$$
\widehat{f}  * a
\;=\;
\int_{G} (\Ff^{-1} f)(-t) \,\alpha_t(a)\, \difd{t}
\;, \qquad a \in \Mm
\;.
$$
For some smooth function $\varphi:\RM\to[0,1]$ with support $[2^{-1}, 2]$ and such that $\sum_{\ScaleInd  \in \bbZ} \varphi(2^{-\ScaleInd } x) =1$ for all $x \in \bbR^+$, one introduces the dyadic Littlewood-Paley decomposition $(W_\ScaleInd )_{\ScaleInd \in\bbN}$ by
$$
W_\ScaleInd (t) \;=\; \varphi(2^{-\ScaleInd } |t|) \quad \mbox{for } t\in\RM^n\,,\;\ScaleInd  > 0\;,
\qquad
W_0 \;= \;1 - \sum_{\ScaleInd >0} W_\ScaleInd 
\;,
$$
and then defines, given $p,q\in[1,\infty)$ and $s > 0$, the Besov norm of $a\in \Mm$ in generalization of the classical multiplier definition by
$$
\norm{a}_{B^s_{p,q}}
\;=\;  
\Big(\sum_{\ScaleInd  \geq 0} \,2^{q s \ScaleInd }\, \lVert \widehat{W}_\ScaleInd  * a \rVert_{L^p(\Mm)}^q\Big)^{\frac{1}{q}}
\;.
$$
Chapter~\ref{sec-Besov} shows that elements of $L^p(\Mm)$ with bounded Besov norm form a Banach space $B^s_{p,q}(\Mm)$ and that $\Mm\cap B^s_{p,q}(\Mm)$ is a $*$-algebra. It provides an equivalent norm in terms of finite differences of functions (as in the classical case), uses interpolation theory \cite{BerghLofstrom76,Lunardi2018} to derive properties of the full scale of Besov spaces and establishes connections to the domains of the derivations $\nabla_1,\ldots,\nabla_n$ and the noncommutative Sobolev spaces $W^m_p(\Mm)$. Most importantly,  Theorem~\ref{theo-HankelBesovP} proves a Peller criterion and Theorem~\ref{theorem:hankel_converse} characterizes trace class properties of Hankel operators in terms of Besov space (just as in the classical case \cite{Pel,Zhu}). A particular form of the Peller criterion is part of the following main result of this work (see Chapter~\ref{chap-BreuerToep} and, in particular, Theorem~\ref{theo-Index} for a detailed statement that also covers the case of even pairings):

\begin{theorem}[Sobolev index theorem]
\label{theo-BesovIntro}
Let $n$ be odd and $p>n$. Suppose that $u-1\in \Mm \cap W^1_{p}(\Mm)$. Then the Hankel operator $H_u$ is in $L^{n+1}(\Mm\rtimes_\alpha G)$ and the index formula \eqref{eq-OddPairingIntro} holds.
\end{theorem}

Let us stress that the stated Sobolev property is not necessary for the index formula \eqref{eq-OddPairingIntro} to hold. In fact, for the case of even and integer-valued pairings, it has been shown that the Hankel operators only need to lie in a Macaev ideal that allows supplementary logarithmic divergences \cite{BES,PLB}. This is based on an identity for the Dixmier trace. The recent contribution \cite{MSX} shows a similar result for the rotation algebra. Here a more analytical reasoning closer to Peller's original argument is presented. 

\vspace{.2cm}

Apart from the Sobolev index theorem, this work contains in Chapter~\ref{sec-DualityToep} another result of pure mathematics that is worth mentioning in this introduction. If $\Aa$ is a $C^*$-algebra and $\Dualityactionop$ is norm continuous $\RM$-action, then the Connes-Thom isomorphisms  $\partial_i^\Dualityactionop:K_i(\Aa)\to K_{i+1}(\Aa\rtimes_\Dualityactionop \RM)$ connect the $K$-theory of $\Aa$ to that of the $C^*$-crossed product $\Aa\rtimes_\Dualityactionop \RM$ \cite{Connes81}. If one has another norm-continuous $G$-action $\theta$ commuting with $\Dualityactionop$ and a s.n.f. trace $\Tt$ that is invariant under both $\Dualityactionop$ and $\theta$, one also obtains a $(G\times \RM)$-action $\theta\times\Dualityactionop$. 
There exist natural dense Fr\'echet subalgebras of $\Aa$ and $\Aa\rtimes_\Dualityactionop \RM$ which can be used to define cocycles $\Ch_{\Tt,\theta\times\Dualityactionop}$ and $\Ch_{\hat{\Tt}_\Dualityactionop,\theta}$ that provide well-defined pairings with $K$-theory. Note that if $n$ is even, then $\Ch_{\Tt,\theta\times\Dualityactionop}$ is an odd cocycle which pairs with odd $K$-theory. 

\begin{theorem}[Duality theorem]
\label{theo-DualityIntro}
For $[u]_1\in K_1(\Aa)$ and $n$ even, 
$$
\langle \Ch_{\Tt,\theta \times\Dualityactionop},[u]_1\rangle
\;=\;
-\,
\langle \Ch_{\hat{\Tt}_\Dualityactionop,\theta},\partial_1^\Dualityactionop [u]_1\rangle
\;,
$$
where $\hat{\Tt}_\Dualityactionop$ is the dual trace on  $\Aa\rtimes_\Dualityactionop \RM$.
\end{theorem}

Chapter~\ref{sec-DualityToep} (see, in particular, Theorem~\ref{theo-smooth_duality}) contains a proof of this statement and also a similar one for even $n$. While Theorem~\ref{theo-DualityIntro} essentially follows by combining results from the literature \cite{ENN88,Pim,KRS,KS04} and certainly can also be proved by $KK$-theory (such as in \cite{BR2018,AMZ}), we provide a self-contained proof based on cyclic cohomology. For this we review and slightly generalize the approach of \cite{ENN88} which relates the pairings of $K_i(\Aa)$ with a cyclic cocycle $\varphi$ densely defined on a Fr\'echet subalgebra $\scrA$ of $\Aa$ to those of $K_{1-i}(\Aa \rtimes_\alpha \bbR)$ with a dual cocycle $\#_\alpha \varphi$ densely defined on a subalgebra of $\Aa \rtimes_\alpha \bbR$. An important technical issue is that those dense subalgebras must be spectrally invariant in $\Aa$ respectively $\Aa \rtimes_\alpha \bbR$ such that the K-theoretical index pairings are well-defined. Therefore the arguments of \cite{ENN88} are adapted in such a way that Schweitzer's notion of strong spectral invariance of $\scrA$ in $\Aa$ \cite{Schweitzer} can be applied as a (constructive) sufficient condition. 

\vspace{.2cm}

For applications of this formula, it is often desirable to replace the abstract Connes-Thom isomorphism by the connecting map of a more concrete exact sequence of $C^*$-algebras such as the Wiener-Hopf extension. Likewise, many applications require analogous formulas for crossed products with $\bbZ$-actions based on the connecting maps of the Pimsner-Voiculescu sequence, which have to be proved independently. All of these issues can be addressed in a unified manner by noting that associated to the data $(\Aa, \Dualityactionop, \bbR)$ is an exact sequence of $C^*$-algebras 
\begin{equation*} 
0 \;\to\; \Aa \rtimes_\Dualityactionop \bbR
\;\hookrightarrow \;
\ToepR
\;{\to} \;
\Aa 
\;\to \;0\;.
\end{equation*}
where the middle algebra is the smooth Toeplitz extension introduced by \cite{Ji90}. This was used {\it e.g.} in \cite{Lesch91}  to generalize the index theory of Toeplitz operators to flows in non-commutative symbol algebras. The $K$-theoretical connecting maps of the Toeplitz extensions are known to give isomorphisms related to the Connes-Thom isomorphisms and the connecting maps of Rieffel's Wiener-Hopf extension \cite{Rieffel82}. This is revisited in Section~\ref{sec-ConnectSmoothToep} which also studies an analogous extension for $\bbT$-actions, a generalization of the discrete Toeplitz extension used by Pimsner and Voiculescu \cite{PV}. By applying the interrelations between these $K$-theoretical maps and combining them with Takai duality, the above duality theorem appears as one of several such results for a fairly flexible class of extensions. As an additional result, the analogue of the duality theorem for $\bbZ$-actions also follows directly from Theorem~\ref{theo-DualityIntro} already. Section~\ref{sec-BoundaryCurrents} later on provides an application of these results to the bulk-boundary-correspondence for half-spaces with irrational cutting angles. 

\vspace{.2cm}

The remainder of this introduction describes the application of the above theory and results to the mathematical physics of solid state systems. This makes up Chapter~\ref{sec-Applications}. Let us begin with a short historic account and by documenting some of the enormous recent activity in the field. This work is placed within the operator algebraic framework developed by Bellissard in the early 1980's which allowed to understand quantum Hall systems as examples of Connes' noncommutative geometry \cite{Bel}. More recent accounts thereof are \cite{BES} and \cite{PSbook}. For physical reasons, namely in order to deal with Anderson localized systems without a spectral gap, it was necessary to go beyond the standard $C^*$-algebraic formulation and prove index theorems for certain elements of the enveloping von Neumann algebra, see again \cite{BES}. Another advance was to understand the bulk-boundary correspondence (BBC) in quantum Hall systems as a result of the $K$-theoretic exponential map of Pimsner-Voiculescu exact sequence for the discrete Toeplitz extension and the corresponding duality theory \cite{SKR,KRS}. Being such a robust mathematical concept, the BBC could then be extended to other dimensions and therefore allowed to describe numerous situations of physical interest in the growing field of topological insulators \cite{PSbook,Kub,SaS,BourneProdan,EM,BourneMesland,KP,CS,STo}, some by extending or modifying the initially proposed framework of \cite{Bel}. There are also countless more analytical contributions for these systems and, while it is impossible to cite them all here, let us mention contributions proving the BBC in presence of a mobility gap regime by Elgart, Graf and Schenker \cite{EGS} as well as Graf and Shapiro \cite{ShapiroGraf2018} and for semimetals by Mathai and Thiang \cite{MT} and Carey and Thiang \cite{CarTia}, all of which have not yet (prior to this work) been dealt with within the operator algebraic approach.

\vspace{.2cm}

There are four main novel contributions of this work:

\begin{itemize}

\item The first is to provide sufficient conditions for the existence of the weak bulk Chern numbers. This covers the mobility gap (for Anderson localized systems) and pseudo-gap regime (for certain semimetals). It uses the index theoretical approach with Besov symbols (Theorem~\ref{theo-BesovIntro}), but requires a substantial amount of supplementary analytical preparations.

\item The second contribution, less substantial, is an extension of the range of applicability of the BBC for spectrally gapped insulators to half-spaces with arbitrary orientation w.r.t. the lattice directions. This is based on the duality result Theorem~\ref{theo-DualityIntro}.

\item The third proves that the surface states associated to non-vanishing  weak Chern numbers cannot be localized.

\item The fourth contribution is to establish the BBC for the for weak winding number invariants of chiral Hamiltonians in the absence of a bulk gap (again for the mobility gap and pseudogap regime). This implies the existence of flat bands of surface states in these systems and is also based on the Sobolev index theorem. 

\end{itemize}

\noindent Detailed descriptions of these results are given in Sections~\ref{sec-BulkInv}, \ref{sec-BoundaryCurrents},  \ref{sec-DelocBoundary}  and \ref{sec-flat} respectively. This introduction gives a flavor of these results and the techniques involved, but only states one result (Theorem~\ref{theo-SurfaceIntro}) in some detail.

\vspace{.2cm}

The algebraic set-up for the description of solid state systems is as follows \cite{Bel,PSbook} (see Section~\ref{sec-AlgSetUp}). The set of configurations of the solid is a compact probability space $(\Omega,\PM)$ which is equipped with an invariant and ergodic $\ZM^d$-action. The bulk observables are then elements of the  disordered rotation algebra $\bbT^d_{\BB,\Omega}=C(\Omega)\rtimes_\BB \ZM^d$, a twisted crossed product associated to an anti-symmetric real matrix $\BB$ of constant magnetic fields. Associated to the probability measure $\PM$ on $\Omega$ is a tracial state $\Tt$ on $\bbT^d_{\BB,\Omega}$ whose GNS representation allows to identify an element $a\in \bbT^d_{\BB,\Omega}$ with a covariant family $(a_\omega)_{\omega\in\Omega}$ of bounded operators on $\ell^2(\ZM^d)$. In this faithful representation, $\bbT^d_{\BB,\Omega}$ is generated by the magnetic translations on $\ell^2(\ZM^d)$ and multiplication operators (potentials) obtained by shifting functions $f\in C(\Omega)$.  Furthermore, there is a dual action $\rho$ of the torus $\bbT^d$ whose unbounded generators are the position operators $X_1,\ldots ,X_d$ on the lattice $\bbZ^d$ and which induce derivations $\nabla_1,\ldots,\nabla_d$ on $\bbT^d_{\BB,\Omega}$. For periodic systems the action and derivations are given by the translations and coordinate derivatives on the Brillouin torus, respectively. For the application of the Sobolev index theorem, one further goes over to the von Neumann algebra $\Mm = (\bbT^d_{\BB,\Omega})''=L^\infty(\Mm,\Tt)$ with normal faithful finite trace $\Tt$. Then given any restriction $\theta$ of $\rho$ to an $n$-parameter subgroup of $\TM^d$, there are associated Chern cocycles $\Ch_{\Tt,\theta}$ which lead to semi-finite Breuer-Fredholm indices in the von Neumann algebra $\Mm \rtimes_\theta \bbR^n$ by the Sobolev index theorem. 

\vspace{.2cm}

To explain the physical motivation that leads to consider non-smooth symbols, let us describe how the projections and unitary elements subject to the index pairings arise. Let be given a self-adjoint Hamiltonian $h=h^*$ in a matrix algebra $M_{N}(\bbT^d_{\BB,\Omega})$ which is smooth w.r.t. the dual action $\rho$, namely having a rapid decay of the off-diagonal matrix elements. Assume that $N$ is even and $h$ has a chiral symmetry given by
\begin{equation}
\label{eq-ChiralSym}
J h J 
\;=\; -h\;, 
\qquad 
J \;=\; 
\begin{pmatrix}
    \one_{\frac{N}{2}} &  0\\
    0 &  -\one_{\frac{N}{2}}
  \end{pmatrix}
\;,
\end{equation}
which is equivalent to $h$ and its phase $\sgn(h)=h\,|h|^{-1}$ being of the form
\begin{equation}
\label{eq-FermiUnitary}
h\; =\; 
\begin{pmatrix}
    0 &  a^*\\
    a &  0
  \end{pmatrix}
\;,
\qquad
\sgn(h)\; =\; 
\begin{pmatrix}
    0 &  u_F^*\\
    u_F &  0
  \end{pmatrix}
\;,
\end{equation}
with some operators $a,u_F \in M_{\frac{N}{2}}(\Mm)$. If $h$ is invertible and therefore has a spectral gap around $0$, then $u_F$ is called the Fermi unitary. It is then smooth and defines an element $[u_F]_1 \in K_1(\bbT^d_{\BB,\Omega})$ so that one can define the numerical index pairings $\Ch_{\Tt,\theta}(u_F)=\langle \Ch_{\Tt,\theta}, [u_F]_1 \rangle$ with Chern cocycles for which $\theta$ is generated by an odd number $n$ of generators. Those invariants are called the odd (bulk) Chern numbers of $u_F$. Likewise, the class of the Fermi projection $p_F = \chi(h \leq 0)$ in $K_0(\bbT^d_{\BB,\Omega})$ is used to define pairings with even cocycles which are called even (bulk) Chern numbers. The Chern number with $n=d$ is called the strong Chern number and can be written as an ordinary Fredholm index \cite{Bel,PLB,PSbook}, while those with $n<d$ are called weak Chern numbers and are in general described by semi-finite Breuer-Fredholm indices as above \cite{PS,BourneSchuba}. If the assumption of a spectral gap is dropped, the elements $u_F$ and $p_F$ will in general not be smooth, but in certain cases they satisfy the conditions of the Sobolev index theorem (Theorem~\ref{theo-BesovIntro}) and the Chern numbers continue to be well-defined. In Section~\ref{sec-BulkInv} two sufficient conditions are demonstrated. The first is that all Chern numbers of $h$ can be defined if the Fermi energy $E_F=0$ is in a region of Anderson localization, which describes the regime of strongly disordered topological insulators. The second condition is geared towards topological semimetals, namely if the density of states of $h$ vanishes rapidly enough at the Fermi level, some of the weak Chern numbers continue to be well-defined. Let us, however, emphasize that in the absence of a bulk energy gap the Sobolev index theorem does not imply the invariance of the Chern numbers under perturbations of the Hamiltonian since the dependence of $u_F$ and $p_F$ on $h$ is not norm-continuous, but merely weakly continuous. As the example further below already shows, the Chern numbers can vary continuously with $h$ and take arbitrary real values even for periodic models.

\vspace{.2cm}

The BBC for general half-planes with normal vector $\Halfspaceunitvector \in \SM^{d-1}$ is then based on a smooth Toeplitz extension. The basic idea is to note that the projection $P$ to a half-space can be written as the positive spectral projection $P = \chi(D_\xi \geq 0)$ of the self-adjoint operator $D_\xi = \sum_{i=1}^d (\Halfspaceunitvector)_i X_i$ which is the suitable scalar Dirac operator in this case. Let us for now assume that the components of $v$ are rationally independent, which means that the boundary hypersurface intersects with the lattice $\bbZ^d$ in exactly one point (for the general case see Section~\ref{sec-HalfSpace}). The adjoint action of $\Dualityactionop_t = \mathrm{Ad}_{\exp(\imath D_\xi t)}$ then defines a free $\bbR$-action on $\bbT^d_{\BB,\Omega}$ which is used to form a smooth Toeplitz extension:
$$
0 \;\to\; \bbT^d_{\BB,\Omega} \rtimes_\Dualityactionop \RM
\;\hookrightarrow \;
{\rm T}(\bbT^d_{\BB,\Omega}, \RM, \Dualityactionop)
\;{\to} \;
\bbT^d_{\BB,\Omega}
\;\to \;0\;.
$$
This construction can be interpreted as a bulk-boundary exact sequence, namely Section~\ref{sec-HalfSpace} shows that  the algebra $\bbT^d_{\BB,\Omega}\rtimes_\Dualityactionop \bbR$ can be considered as an algebra of observables which are localized at the hyper-surface with normal vector $v$, while the smooth Toeplitz extension $\mathrm{T}(\bbT^d_{\BB,\Omega}, \RM,\Dualityactionop)$ consists of observables on a half-space that converge to elements of the bulk algebra $\bbT^d_{\BB,\Omega}$ at infinity on one side of the hyper-surface and to $0$ on the other side. Moreover, the dual trace $\hat{\Tt}_\Dualityactionop$ is shown to admit a description in terms of a trace per unit surface area. All of this is essential for the physical interpretation of the results.

\vspace{.2cm}

The $K$-theoretical approach to the BBC presented  in Section~\ref{sec-BoundaryCurrents} is based on this exact sequence. Combined with the duality theorem for smooth Toeplitz extensions, it reproduces and generalizes the BBC of \cite{PSbook} for the Chern numbers of spectrally gapped Hamiltonians. Let us describe the results briefly. When restricting a gapped bulk Hamiltonian $h \in M_{N}(\bbT^d_{\BB,\Omega})$ with non-trivial Chern numbers to a half-space using Dirichlet boundary conditions $\hat{h}=P h P$ perturbed by an arbitrary local operator $\hat{k}\in M_N(\bbT^d_{\BB,\Omega} \rtimes_\Dualityactionop \RM)$ on the boundary, the bulk energy gap is generically filled with additional boundary states to which one can also associate topological invariants, called boundary Chern numbers. The main result (Theorem~\ref{theo-index_map_bb_sm}) is that all boundary Chern numbers can be computed purely in terms of the bulk Chern numbers. In particular, it is shown that the boundary topological invariants have an explicit smooth dependence on the cutting angles of the half-space and are independent of $\hat{k}$. As a technical improvement over existing results \cite{PSbook,BR2018}, it is also proved that the boundary states associated to non-trivial weak Chern numbers are delocalized (see Section~\ref{sec-DelocBoundary}).

\vspace{.2cm}
 
Section~\ref{sec-flat} then considers the special case of the weak odd Chern number with a single generator, namely the winding numbers
\begin{equation}
\label{eq-WindLin}
\Ch_{\Tt,\Dualityactionop}(u)\;=\;-\,\imath \;\Tt\big(u^* \nabla_\xi u\big)\;=\;-\,\imath \,\sum_{i=1}^d (\Halfspaceunitvector)_i \, \Tt\big(u^* \nabla_i u\big)
\;.
\end{equation}
It turns out that the BBC for this invariant is closely related to the index theory studied in the first part of this work, since the auxiliary von Neumann algebra $\Mm \rtimes_\Dualityactionop \bbR$ with dual trace $\hat{\Tt}_\Dualityactionop$ which allows to write $\Ch_{\Tt,\Dualityactionop}$ as a semi-finite index coincides precisely with the von Neumann completion of the edge algebra $\bbT^d_{\BB,\Omega}\rtimes_\Dualityactionop \RM$ w.r.t. the GNS representation of the dual trace $\hat{\Tt}_\Dualityactionop$. Hence one can reformulate the BBC as a problem concerning Breuer-Fredholm operators. Let $\hat{h}= P h P+\hat{k}$ be a chirally symmetric half-space Hamiltonian as above and $\hat{u} \in \Mm\rtimes_\Dualityactionop \bbR$ the off-diagonal part of its polar decomposition as in \eqref{eq-FermiUnitary}. If $h$ is not spectrally gapped, then $\hat{h}$ may have a non-trivial kernel and hence $\hat{u}$ is in general not unitary, but only a partial isometry. Section~\ref{sec-flat} provides conditions under which $\hat{u}$ is Breuer-Fredholm and a compact perturbation of the Toeplitz operator $P u_F P$ with symbol given by the Fermi unitary $u_F$ of the bulk Hamiltonian. If the Sobolev index theorem applies to $u_F$, this allows to strengthen the smooth version of the BBC and make a statement about the kernel of $\hat{h}$.

\begin{theorem}[Flat bands of surface states]
\label{theo-SurfaceIntro}
For a chiral Hamiltonian $h\in M_N(\bbT^d_{\BB,\Omega})$ with Fermi unitary $u_F$ and half-space restriction $\hat{h}= P h P+\hat{k}$ as described above,
\begin{equation}
\label{eq-BBCIntro}
\Ch_{\Tt,\Dualityactionop}(u_F) \;=\; \hat{\Tt}_\Dualityactionop\big(J \;\Ker(\hat{h})\big) 
\;,
\end{equation}
whenever one of the following conditions holds:

\begin{itemize}

\item[{\rm (i)}] The Hamiltonian $h$ has a bulk gap, {\it i.e.}  $0\not\in\sigma(h)$.

\item[{\rm (ii)}] $0$ lies in a {\rm (}Anderson localized{\rm )} mobility gap of $h$.

\item[{\rm (iii)}]
$0$ is a pseudogap of the density of states with sufficiently high order, namely there exist  $\gamma>\frac{3}{2}$ and $C_\gamma<\infty$ such that the spectral projections $\chi(|h|\leq \epsilon)$ satisfy
\begin{equation*}
\Tt\big(\chi(|h|\leq \epsilon)\big)
\;\leq\;
C_\gamma\, \epsilon^{\gamma}
\;.
\end{equation*}


\end{itemize}
\end{theorem}

This shows that a nontrivial bulk winding number $\Ch_{\Tt,\Dualityactionop}(u_F)$ implies the existence of states in the kernel of the half-space Hamiltonian and that the signed surface density of these zero-energy states is determined by the weak bulk winding numbers. These edge states are said to form a flat band. Condition (i) is already covered by \cite{PSbook}. Part (ii) generalizes the one-dimensional result of \cite{ShapiroGraf2018} to higher dimensional systems, thereby adding to the short list of rigorously proven results on the BBC for mobility gapped topological insulators. The sufficient condition (iii) applies {\sl e.g.} chiral Dirac-semimetals in dimensions larger than one. A particularly vivid example is provided by a standard model of (pure) graphene based on the discrete Laplacian on a hexagonal lattice \cite{Wal}. Choosing a parametrization of the hexagonal lattice such that a termination in the directions $\Halfspaceunitvector=e_1$ and $\Halfspaceunitvector=e_2$ gives so-called zigzag edges and $\Halfspaceunitvector=2^{-\frac{1}{2}}(e_2-e_1)$ an armchair zigzag edges, the two weak Chern numbers (winding numbers) can be computed explicitly to be $-\imath\,\Tt(u^*\nabla_iu)=\frac{1}{3}$  (see Section~\ref{sec-Graphene}). Therefore Theorem~\ref{theo-SurfaceIntro} implies
\begin{equation}
\label{eq-GrapheneIntro}
\hat{\Tt}_\Dualityactionop\big(J \;\Ker(\hat{h})\big) \;=\;\frac{1}{3}\,((\Halfspaceunitvector)_1 +(\Halfspaceunitvector)_2)
\;
\end{equation}
and thus armchair edges need not have edge states at all, while the signed density of surface states is maximal for zigzag edges. This was known for a long time by an elementary analysis \cite{Nakada96,RH}, and the formula  \eqref{eq-BBCIntro} for rationally dependent normal vectors $\Halfspaceunitvector$ follows from a non-rigorous Zak-phase argument \cite{DUM}. The existence of edge states (but not the connection with the bulk topology) has also been shown for continuum operators with a hexagon symmetry and rational edges \cite{FLW}. Theorem~\ref{theo-SurfaceIntro} also proves the equality \eqref{eq-BBCIntro} for irrational angles and shows stability, for example, under adding surface disorder $\tilde{k}$ to the half-space Hamiltonian $\hat{h}=P h P + \tilde{k}$, hence showing that the actual structure of the surface is not essential. Let us stress that the numerical range of the Chern numbers is not discrete and the value $\frac{1}{3}$ is in this case essentially given by the projected distance between two band-touching points in momentum space, which changes continuously as the bulk model is varied, {\it e.g.} when other hopping parameters are added to the graphene model, see the model \eqref{eq:honeycombMod} in Section~\ref{sec-Graphene}. Robust is the relation \eqref{eq-BBCIntro}, similar as in an extension of Levinson's theorem to surface states where the density of surface states is equal to a time delay density \cite{Sch}. For higher dimensions one can also write down Hamiltonians which satisfy the conditions of the BBC, for example, by stacking the model above and adding a weak interlayer coupling.  A more interesting possibility in $d=3$ is given by nodal line semimetals where the spectrum at the Fermi level consists of a loop in momentum space . One may again have non-vanishing winding numbers, which in turn lead to flat bands of surface states whose signed density depends linearly on the components of the surface normal vector \cite{MatsuuraEtAl}.

\vspace{.2cm}

This works is organized as follows. Chapter~\ref{sec-CrossedProd} contains preliminaries on $C^*$- and $W^*$-crossed products for abelian group actions and semi-finite tracial states on them. This includes Takai and Takesaki duality, Arveson spectra and basic analysis of smoothness of elements w.r.t. the action. Many of the results are taken from the literature without proofs, but a precise formulation is needed in the arguments later on. The experienced reader can rapidly skim through Chapter~\ref{sec-CrossedProd} as it is not the heart of the matter. Chapter~\ref{sec-Besov} presents the construction of Besov spaces for abelian actions on semi-finite von Neumann algebras. The Peller criteria are proved in Chapter~\ref{chap-BreuerToep} and then combined with known index calculations to prove the Sobolev index theorem. This chapter is the mathematical core of this work.  Chapter~\ref{sec-DualityToep} concerns $C^*$-crossed products with $\RM$-actions and various associated exact sequences. The $K$-theoretic connecting maps are reviewed with care and the proof of the duality theorem is given. This is new, but uses several results from the literature.  The long Chapter~\ref{sec-Applications} then presents the applications to solid state systems. Finally the appendices review and extend results on integration in quasi-Banach spaces and non-commutative $L^p$-spaces, interpolation theory and Breuer-Fredholm semi-finite index theory.


\vspace{.4cm}

\noindent{\bf Acknowledgements:} We are grateful for having received financial support by the DFG through grant SCHU 1358/6-2. Moreover, T.~S. was supported through a scholar\-ship of the {\sl Studienstiftung des Deutschen Volkes}.

\vspace{.2cm}

\vspace{\baselineskip}
\begin{flushright}\noindent
Erlangen, May 2022,\hfill {\it  Hermann Schulz-Baldes and Tom Stoiber}
\\
\end{flushright}

\newpage

\chapter{Preliminaries on crossed products}
\label{sec-CrossedProd}

This chapter reviews several facts about crossed products. Many aspects are covered in the standard reference \cite{Pedersen79}, but some are taken from other works like \cite{Takesaki2003,Blackadar06,Raeburn88,Takesaki73,PR,Lesch91}. Many results hold for general locally compact abelian group actions, but for sake of concreteness we will only restrict to the case of abelian $n$-parameter groups which are relevant for the applications that we have in mind. Hence throughout $G=\bbT^{n_0} \oplus \bbR^{n_1}$ where $n=n_0+n_1$ and we make the identification $\TM=\RM\backslash \ZM=[0,1]/\sim$. Unless there is an explicitly different choice of normalization (which only applies in Chapter~\ref{sec-Applications}), we fix the Haar measures on $\bbT^{d}$ and $\bbR^{d}$ to be the usual Lebesgue measure and on $\bbZ^d$ the counting measure. Combined with the conventions for the Fourier transform as given below, one can then identify the dual groups $\hat{\bbT}=\bbZ$, $\hat{\bbZ}=\bbT$ and $\hat{\bbR}=\bbR$ in such a way that the Plancherel theorem holds without proportionality constants. 

\section{$C^*$-dynamical systems}
\label{sec-CStar}

\begin{definition}
\label{def-C*dyn}
A $C^*$-dynamical system is a triple $(\Aa,G,\alpha)$ consisting of a $C^*$-algebra $\Aa$ and a strongly continuous $G$-action $\alpha: G \to \mbox{\rm Aut}(\Aa)$, namely $t\in G \mapsto \alpha_t(a)$ is norm-continuous for each $a \in \Aa$.  A covariant representation of a $C^*$-dynamical system is a pair  $(\pi, U)$ with $\pi$ a non-degenerate representation of $\Aa$ on a Hilbert space $\HhO$ and $U$ a strongly continuous unitary representation of $G$ on $\HhO$ such that 
$$
\pi(\alpha_t(a)) \;=\; U(t)\, \pi(a)\, U(t)^*\;, 
\qquad \; a \in \Aa\,,\; t \in G\,.
$$ 
\end{definition}

Following Raeburn \cite{Raeburn88}, such a $C^*$-dynamical system determines a unique (up to isomorphism) crossed product $C^*$-algebra $\Aa \rtimes_\alpha G$ that contains $\Aa$ and $G$ as multipliers and for which any covariant representation extends to a $*$-representation.  Given a $C^*$-subalgebra $\Cc$ of the bounded operators $\Bb(\Hh)$ on some Hilbert space $\Hh$, let $M(\Cc)$ denote the set of multipliers and $UM(\Cc)$ the subset of unitary multipliers.

\begin{definition}
\label{def-CrossedProd}
The crossed product $\Aa \rtimes_\alpha G$ for the dynamical system $(\Aa,G,\alpha)$ is a $C^*$-algebra $\Cc\subset \Bb(\Hh)$ with a homomorphism $i_\Aa: \Aa \to M(\Cc)$ and a strictly continuous homomorphism $i_G: G \to UM(\Cc)$ such that

\begin{itemize}

\item[{\rm (i)}] $\;i_\Aa(\alpha_t(a)) = i_G(t) i_\Aa(a) i_G(t)^*$ for all $ a \in \Aa$ and $ t\in G$.

\item[{\rm (ii)}]  If $(\pi,U)$ is a covariant representation on some Hilbert space $\HhO$, there is a unique non-degenerate representation $\pi \times U$ of $\Cc$ {\rm (}and its multipliers{\rm )} on $\HhO$ with $\pi = (\pi \times U) \circ i_\Aa$ and $U = (\pi \times U) \circ i_G$.

\item[{\rm (iii)}]  $i_G$ extends to a representation of the compactly supported continuous functions $C_c(G)$ via 
integration and linear combinations of products of the form $i_\Aa(a) i_G(f)$ with $a \in \Aa$ 
and $ f \in C_c(G)$ are dense in $\Cc$.
\end{itemize}
\end{definition}

Let us collect a few facts from \cite{Raeburn88}. Both $i_\Aa$ and $i_G$ must be injective due to non-degeneracy. Moreover, if  $C_c(G, \Aa)$ is equipped with the operations
\begin{equation}
\label{form:mult}
(fg)(t) \;= \;\int_G f(s) \,\alpha_s(g(t-s))\, \difd{s}\;,
\qquad 
f^*(t) \;= \;\alpha_t(f(-t)^*)\;,
\qquad
f,g \in C_c(G, \Aa)\;,
\end{equation}
with integration w.r.t. a Haar measure on $G$, then $i_\Aa$ and $i_G$ extend to a $*$-homomor\-phism $i_\Aa\times i_G:C_c(G, \Aa)\to \Aa \rtimes_\alpha G$ with dense range. It can be written as 
\begin{equation}
\label{eq-iGdef}
(i_\Aa\times i_G)(f)
\;=\;
\int_G \,i_\Aa(f(t))\,i_G(t)\, \difd{t}\;,
\qquad
f\in C_c(G,\Aa)
\;
\end{equation}
with the integral evaluated in the strong operator topology of an arbitrary faithful non-degenerate representation.
For a complex-valued $f\in C_c(G)$, this formula provides the multiplier $i_G(f)$ of item (iii). For a given covariant representation $(\pi,U)$ on some Hilbert space $\HhO$, one can choose $i_\Aa(a)=\pi(a)$ and $i_G(t)=U(t)$ in Definition~\ref{def-CrossedProd} and then \eqref{eq-iGdef} becomes the integrated representation $\pi \times U$ on $\HhO$, densely defined by
\begin{equation}
\label{eq-IntegratedRep}
(\pi \times U)(f) 
\;=\; 
\int_G \pi(f(t))\, U(t) \,\text{d}t
\;,
\qquad
f\in C_c(G, \Aa)
\;.
\end{equation}

As already stressed above, we will focus on the case where $G=\bbT^{n_0} \oplus \bbR^{n_1}$ is an abelian $n$-parameter group where $n=n_0+n_1$ and the torus is chosen to be $\TM=\RM/\ZM\cong[0,1)$. It is then possible to concretize the description of the crossed product for a given covariant representation $(\pi,U)$, {\it e.g.} \cite{Lesch91}. Let $D=(D_1,\ldots ,D_n)$ be the commuting, selfadjoint, densely defined generators of $t\in G\mapsto U(t)$ chosen such that
\begin{equation}
\label{eq-GenDef}
U(t) \;= \;  e^{2\pi \imath\,  D\cdot t} 
\;,
\qquad
D\cdot t\;=\;\sum_{j=1}^n D_j t_j
\;.
\end{equation}
Choosing $i_\Aa(a)=\pi(a)$ and $i_G(t)=U(t)$ in Definition~\ref{def-CrossedProd}, the multiplier $i_G(f)$, given by \eqref{eq-iGdef} with $f\in C_c(G)$, can be expressed in terms of the Fourier transform
$$
(\calF f)(\dualvar )
\;=\;
\int_G \overline{\langle \dualvar ,t\rangle}\,f(t)\, \difd{t}
\;,
\qquad
f\in L^1(G)
\;.
$$
Here $\dualvar $ is a character of $G$, namely continuous group homomorphism $\dualvar :G\to\SM^1$, and $\langle \dualvar ,t\rangle\in\SM^1$ is the duality paring with $t\in G$. The set of characters equipped with the natural product forms the dual group by $\hat{G}$. For a $n$-parameter group as above, $\hat{G}=\ZM^{n_0}\oplus \RM^{n_1}$ and we choose the dual pairing as $\langle \dualvar ,t\rangle=e^{-2\pi\imath \,\dualvar  \cdot t}$. Then \eqref{eq-iGdef} becomes 
$$
i_G(f)
\;=\;
(\calF f)(D)
\;,
\qquad
f\in C_c(G)
\;,
$$
where the r.h.s. is given in terms of continuous functional calculus of the generators with $\Ff f\in C_0(\hat{G})$. Next let us introduce the inverse Fourier transform
$$
(\calF^{-1} g)(t)
\;=\;
\int_{\hat{G}} \langle \dualvar ,t\rangle\,g(\dualvar )\, \difd{\dualvar }
\;,
\qquad
g\in L^1(\hat{G})
\;
$$
with a Haar measure $\difd{\dualvar }$ on $\hat{G}$ normalized such that the Plancherel formula 
\begin{equation}
\label{eq-Plancharel}
\|\calF f\|_{L^2(\hat{G})}
\;=\;
\|f\|_{L^2(G)}
\end{equation}
holds for $f\in L^1(G)\cap L^2(G)$. Then for $g \in C_c(\hat{G})$
\begin{equation}
\label{eq-FourierD}
g(D)
\;=\;
\big(\Ff(\Ff^{-1}g)\big)(D)
\;=\;
\int_G \,(\Ff^{-1}g)(t)\,e^{2\pi\imath \,D \cdot t}\, \difd{t}
\;.
\end{equation}
Therefore by (iii) and the density of $C_c(\hat{G})$ in $C_0(\hat{G})$
\begin{equation}
\label{eq-CrossedProd}
(\pi \times U)(\Aa \rtimes_\alpha G)
\;=\; 
C^* \left \{\pi(a) g(D)\,: \, a\in \Aa\,,\;\; g \in C_0(\hat{G})\right\}
\;,
\end{equation}
where on the r.h.s. the algebraic closure followed by norm completion in $\calB(\HhO)$ is taken.

\vspace{.2cm}

In the following, a particular covariant representation called the left regular representation will play a crucial role. As essentially all constructions (most importantly of the von Neumann crossed product) and arguments below will be carried out in this regular representation, it will simply be denoted by $(\pi,U)$, even though in other works ({\it e.g.} \cite{Takesaki2003}) the regular representation is often denoted by $(\pi,\lambda)$. It supposes the following set-up \cite{Pedersen79}. The $C^*$-algebra $\Aa\subset\Bb(\Hh)$ is a subalgebra of the bounded operators $\Bb(\Hh)$  on some Hilbert space $\Hh$. This is possibly given after identifying $\Aa$ with its image under some faithful representation (like, later on, the GNS representation of a state). Then the regular representation $(\pi,U)$ on $L^2(G,\Hh)$ is given by
\begin{equation}
\label{eq-RegRepr}
(\pi(a) \psi)(s)\; =\; \alpha^{-1}_{s}(a) \psi(s)  \;,
\qquad
(U(t) \psi)(s) \;=\; \psi(s-t)\;,
\end{equation}
where $\psi\in L^2(G,\Hh)$. Note that the Hilbert space $\HhO$ in Definition~\ref{def-C*dyn} is $L^2(G,\Hh)$. The generators of $U$ in the regular representation are
\begin{equation}
\label{eq-GenDefReg}
(D_j \psi)(s) 
\;=\; 
\frac{\imath}{2\pi}\, (\partial_{s_j} \psi)(s)
\;.
\end{equation}
The only other covariant representation used below will be the GNS representation of an $\alpha$-invariant s.n.f. tracial state on $\Aa$. It will be constructed in Section~\ref{sec:traces} below.  Let us also remark that the $n$ generators of the action defined in an arbitrary integrated representation $(\pi \times U)$ may differ considerably from $D_1,\dots,D_n$ even if the representation is faithful. In particular, the spectrum of $D_j$ in a regular representation is clearly absolutely continuous, whereas the generators of the action in a GNS-representation may have e.g. dense point spectrum.

\section{$W^*$-dynamical systems}
\label{sec-WStar}

\begin{definition}
\label{def-WDynSys}
A $W^*$-dynamical system is a triple $(\Mm,G,\alpha)$ consisting of a von Neumann algebra $\Mm$ and a weakly continuous action $\alpha: G \to \mbox{\rm Aut}(\Mm)$, namely $t\in G \mapsto \alpha_t(a)$ is weakly continuous for any $a\in\Mm$. A covariant representation of a $W^*$-dynamical system is a pair  $(\pi, U)$ with $\pi$ a non-degenerate normal (weakly continuous) representation of $\Mm$ on a Hilbert space $\HhO$ and $U$ a strongly continuous unitary representation of $G$ on $\HhO$, such that 
$$
\pi(\alpha_t(a)) 
\;=\; 
U(t)\, \pi(a)\, U(t)^*
\;, 
\qquad  a \in \Mm\,,\;\;t  \in G\,.
$$ 
\end{definition}

Let us stress that the important difference w.r.t. Definition~\ref{def-C*dyn} is the notion of continuity used. One way to obtain a $W^*$-dynamical system is the following  \cite{Takesaki2003}. Suppose that there is a strongly continuous unitary representation $U$ of $G$ on $\HhO$ and a von Neumann algebra $\Mm\subset \Bb(\HhO)$ that satisfies $U(t)\Mm \,U(t)^* = \Mm$ for all $t \in G$,  then $\alpha_t(a) = U(t) a\, U(t)^*$ defines a weakly continuous action. A general $W^*$-dynamical system $(\Mm,G,\alpha)$ can always be written in this manner by considering the regular representation $(\pi,U)$ on $\HhO=L^2(G,\Hh)$ given by the same formula as in \eqref{eq-RegRepr}. The $W^*$-crossed product is defined in this representation \cite{Takesaki73,Takesaki2003}.

\begin{definition}
\label{def-W*Crossed}
Let $(\pi,U)$ be a regular representation of a $W^*$-dynamical system $(\Mm,G,\alpha)$ on $\HhO=L^2(G,\Hh)$.  The $W^*$-crossed product $\Mm \rtimes_\alpha G$ is the smallest von Neumann algebra in $\calB(\HhO)$ containing $\pi(\Mm)$ and $U(G)$.
\end{definition}

For  $G=\bbT^{n_0} \oplus \bbR^{n_1}$, the $W^*$-crossed product can be written in a similar manner as in \eqref{eq-CrossedProd}:
\begin{equation}
\label{eq-W*CrossedProd}
\Mm\rtimes_\alpha G 
\;=\; 
W^* \left \{\pi(a) g(D)\,: \, a\in \Mm\,,\;\; g \in L^\infty(\hat{G})\right\}
\;,
\end{equation}
where $(\pi,U)$ is the regular representation and $D$ the generator of $U$ as in \eqref{eq-GenDef} and \eqref{eq-GenDefReg}. On the r.h.s. the algebraic closure is taken followed by the weak closure.

\vspace{.2cm}

Next let us comment on the compatibility of $C^*$- and $W^*$-crossed products. Let $(\pi,U)$ be a regular representation on $\HhO=L^2(G,\Hh)$ of a $C^*$-dynamical system $(\Aa,G,\alpha)$. Then the double commutant $\pi(\Aa){''}$ is a von Neumann subalgebra of $\Bb(\HhO)$ on which a weakly continuous $G$-action is defined by
$$
\tilde{\alpha}_t(\pi(a)) \;= \;U(t) \,\pi(a) \,U(t)^*\;,
\qquad 
t\in G
\;.
$$
The $W^*$-dynamical system $(\pi(\Aa){''},G, \tilde{\alpha})$ can hence be considered to be the $W^*$-completion of the $C^*$-dynamical system. Since $\pi(\Aa)$ is $\sigma$-weakly dense in $\pi(\Aa)''$, it is possible to verify that
$$
\pi(\Aa){''} \rtimes_{\tilde{\alpha}}G 
\;=\; 
\big((\pi \times U)(\Aa\rtimes_\alpha G)\big){''}
$$
namely $C^*$- and $W^*$-crossed products are compatible. In the following, $\tilde{\alpha}$ will again simply be denoted by $\alpha$. 

\vspace{.2cm}

\noindent {\bf Remark} In the $W^*$-algebraic case the defining representation is chosen as a regular representation to ensure uniqueness of the crossed product since the universal property from the $C^*$-algebraic case does no longer hold.
In particular, a covariant representation $(\pi,U)$ in the sense of Definition~\ref{def-WDynSys} of $(\Mm, G, \alpha)$ on some Hilbert space $\Hh_0$ does not in general give rise to an integrated representation of $\Mm \rtimes_\alpha G$ on $\Hh_0$. It is also possible to characterize crossed products more abstractly as so-called $G$-products: Given a faithful covariant representation $(\pi,U)$ which also represents $G$ faithfully, one can consider the $W^*$-algebra $\Rr\subset \Bb(\HhO)$ generated by $\pi(\Mm)$ and $U(G)$. That algebra is isomorphic to the crossed product $\Mm \rtimes_\alpha G \simeq \Rr$ if and only if the dual action of $\hat{G}$ (see Section \ref{sec-Duality}) can also be implemented on $\Rr$ as a weakly continuous group of automorphisms \cite[7.10.4]{Pedersen79}.
\hfill $\diamond$

\section{Invariant traces and GNS representation}
\label{sec:traces}

This section recalls basic facts about traces on $C^*$- and von Neumann algebras based on \cite{Dixmier81} and extends traces on $C^*$-algebras to their von Neumann closure and crossed products associated to a given $G$-action using the Hilbert algebra approach of \cite{Dixmier81,Lesch91,PR,Takesaki2003}. In fact, there is a one-to-one correspondence between (full left) Hilbert algebras and faithful semi-finite normal weights on (left) von Neumann algebras \cite[Section VII.2]{Takesaki2003}. 

\vspace{.2cm}

Let us begin with a few standard definitions. For a $C^*$-algebra  $\Aa$  let $\Aa^+$ be its positive cone consisting of non-negative elements. If $\Aa$ is not already unital, then let ${\Aa}^\sim$ be its unitization and the adjoint unit by $\one^{\sim}$. A weight $\phi$ on $\Aa$ is a function $\phi: \Aa^+ \to [0, \infty]$ such that $\phi(\lambda a) = \lambda \phi(a)$ for all $  a\in \Aa^+$ and $ \lambda \in [0,\infty)$, and $\phi( a + b ) =  \phi(a) + \phi(b)$ for all $a,b\in \Aa^+$.  A weight $\phi$ is called a trace if in addition $\phi(u a u^{-1}) = \phi(a)$ for all $a\in \Aa$ and unitaries $u \in \Aa^\sim$. Furthermore, a weight $\phi$ is called faithful if $\phi(a) = 0$ implies $a=0$ for all positive $ a \in \Aa^+$. It is called finite if $\phi(a) < \infty$ for all $a \in \Aa^+$, and lower semi-continuous if $\phi(a) \leq \lim \inf \phi(a_n)$ for all norm-convergent sequences $a_n \to a$ in $\Aa^+$.  A weight $\phi$ on a von Neumann algebra $\Mm $  is called semi-finite if for all $a\in \Mm^+$ 
$$
\phi(a) 
\;=\; 
\sup\big\{ \phi(b) \;:\;b \in \Mm^+\,,\;\; b \leq a \;\mbox{\rm with } \phi(b) < \infty\big\}
\;,
$$ 
and it is called normal if for any increasing net $(a_\lambda)_{\lambda \in \Lambda}$ with supremum in $\Mm$ 
$$
\phi(\sup a_\lambda) 
\;=\; 
\sup \phi(a_\lambda)
\;.
$$

\vspace{.2cm}

Let now $\Aa^+_\phi$ denote the set of all positive elements $a \in \Aa^+$ such that $\phi(a)<\infty$ and call $\phi$ densely defined if $\Aa^+_\phi$ is dense in $\Aa^+$. The linear span of $\Aa^+_\phi$ is denoted $\Aa_\phi$ and forms a $*$-subalgebra of $\Aa$. Any weight $\phi$ extends to a linear functional on $\Aa_\phi$ which will be denoted by the same letter. 
Furthermore, if $\phi$ is a densely defined lower semicontinuous trace on $\Aa$, then $\Aa_\phi$ is a Banach algebra w.r.t. the norm \cite{PR}
\begin{equation}
\label{eq-BanachNorm}
\norm{a}_{\phi} 
\;=\; 
\norm{a} \,+\, \phi(\abs{a})
\;.
\end{equation}

Next define 
$$
\Aa^2_{\phi}
\;=\;
\{a \in \Aa \;:\; \phi(a^* a)<\infty\}
\;,
$$
which is a left ideal in $\Aa$ containing $\Aa_\phi$ and a two-sided ideal if $\phi$ is a trace. In the latter case $\phi(ab) = \phi(ba)$ for $a,b \in \Aa^2_\phi$ since any element of $\Aa$ can be written as a linear combination of four unitary elements in its unitization $\Aa^\sim$.  

\vspace{.2cm}

The main tool for the construction of traces are Hilbert algebras.

\begin{definition}
Let $\Cc$ be a $*$-algebra over $\bbC$ with a scalar product $(x|y)$ which makes it into a pre-Hilbert space whose completion is denoted by $\Hh$. Then $\Cc$ is called a Hilbert algebra, if
\begin{enumerate}
\item[{\rm (i)}] $(x|y) = (y^*|x^*)$ for all $x,y \in \Cc$.
\item[{\rm (ii)}] $(xy|z) = (y|x^*z)$ for all $ x,y,z \in \Cc$.
\item[{\rm (iii)}] For each $x\in \Cc$, the maps $u_x,v_x: \Cc \to \Cc$ given by $y \mapsto xy$ and $y \mapsto yx$ are continuous.
\item[{\rm (iv)}] The set of all products $xy$ is total in $\Cc$, namely the span of such products is dense in $\Cc$.
\end{enumerate}
\end{definition}

The maps $u_x$,$v_x$ can be continued uniquely to bounded operators on $\Hh$, thereby defining two $*$-representations $\pi_l$ and $\pi_r$ of $\Cc$ on $\Hh$. An element $a \in \Hh$ is called (left-)bounded if there is a bounded operator $u_a \in \Bb(\Hh)$ such that $u_a x = v_x a$ for all $x\in \Cc$. The set of all bounded elements of $\Hh$ is again a Hilbert algebra with product $(u_a|u_b) =\langle a|b\rangle_\Hh$, called the full left Hilbert algebra of $\Cc$ and denoted $\hat{\Cc}$ (this is denoted $\Cc''$ in \cite{Takesaki2003}, but we feel this to be a bit misleading). To a Hilbert algebra $\Cc$, one associates the (left) von Neumann algebra generated by its bounded elements $\Uu(\Cc)= \pi_l(\Cc)'' = \pi_l(\hat{\Cc})''$, where the bicommutant is given by the strong completion of $\pi_l(\Cc)$ since $\pi_l$ is a non-degenerate representation. 
The representation $\pi_l: \hat{\Cc} \to \Uu(\hat{C})$ is faithful and can be considered an inclusion, which is compatible with the respective topologies in such a way that $\Cc$ is mutually dense:

\begin{theorem}[\cite{Takesaki2003}, p. 18-19]
\label{theorem-hilbert_algebra_approx}
\begin{enumerate}
\item[{\rm (i)}] The map $\pi_l: \hat{\Cc} \to \Uu(\hat{\Cc})$ is closed w.r.t. the subspace topology of $\hat{\Cc}\subset \Hh$ and the strong operator topology on $\Hh$.
\item[{\rm (ii)}] For $\phi \in \hat{\Cc}$ there is a sequence $(\phi_n)_{n\in \bbN}$ in $\Cc$ such that 
$$
\norm{\phi - \phi_n}_{\Hh}\;\to\; 0\;, 
\qquad 
\norm{\pi_l(\phi_n)} \;\leq\; \norm{\pi_l(\phi)}
\;,
$$ 
and $\pi_l(\phi) = \slim_{n\to \infty} \pi_l(\phi_n)$. 
\end{enumerate}
\end{theorem}

The above algebraic rules can be motivated by the fact that for a $C^*$-algebra with a (densely defined) trace $\Tt$, the image under the (semi-cyclic) GNS representation $\pi_\Tt$ defines a Hilbert algebra with the usual scalar product $(\pi_\Tt(x)|\pi_\Tt(y)) = \Tt(x^*y)$. The main result used here is

\begin{theorem}[Chapter~I.6.2 of \cite{Dixmier81} and p.~58 of \cite{Takesaki2003}]
\label{th:trace}
Let $\Cc$ be a Hilbert algebra with completion $\Hh$ and the  $\Mm=\Uu(\Cc)$ its left von Neumann algebra. Define a weight $\phi$ on $\Mm$ by setting  for $s \in \Mm^+$,
\begin{equation}
\label{form:trace}
\phi(s) 
\;=\;  
\begin{cases} 
(a|a)\;, & \text{if } \sqrt{s} = \pi_l(a) \text{ for some } a \in \hat{\Cc}\;,
\\
\infty\;, & \text{ otherwise .} 
\end{cases}
\end{equation}
Then $\phi$ extends to a semi-finite normal faithful {\rm (}s.n.f.{\rm )} trace on $\Mm$ such that 
$$
\Mm^2_\phi
\;=\; 
\{x \in \Mm \;:\; \phi(x^* x)<\infty\}
\;=\;  
\{x \in \Mm \;:\; x= \pi_l(a) \text{ for some }a \in \hat{\Cc}\}
\;.
$$
Moreover, $\phi(\pi_l(a)^* \pi_l(b)) = (a | b)$ by polarization.
\end{theorem}

This theorem allows to extend traces on $C^*$-algebras to the von Neumann algebras that they generate:

\begin{proposition}
\label{prop:traceext}
Let $(\Aa, G, \alpha)$ be a $C^*$-dynamical system and $\Tt$ be a densely defined, faithful and lower semi-continuous trace on $\Aa$. If $\Tt$ is $\alpha$-invariant, namely 
$$
\Tt(\alpha_t(a)) \;= \;\Tt(a)\;, 
\qquad \forall\;\;a  \in \Aa_\Tt^+\,,\;\; t \in G\;,
$$
then $\Aa$ embeds covariantly into a von Neumann algebra denoted by $L^\infty(\Aa,\Tt)$ and $\alpha$ extends to a $G$-action $\tilde{\alpha}$ on this algebra such that $(L^\infty(\Aa,\Tt),G,\tilde{\alpha})$ is a $W^*$-dynamical system with a s.n.f. trace on $L^\infty(\Aa,\Tt)$ which extends $\Tt$ and is invariant under $\tilde{\alpha}$.
\end{proposition}

\noindent {\bf Proof.} 
Define a scalar product on $\Aa^2_\Tt$ through $( a  | b) = \Tt(a^*b)$. This makes $\Aa_\Tt^2$ into a Hilbert algebra. Clearly, the completion of $\Aa$ coincides with the GNS representation space $\Hh_\Tt$, on which $\Aa$ is represented faithfully since $\Tt$ is faithful. Then set $L^\infty(\Aa,\Tt) = \pi_\Tt(\Aa)''$ which also coincides with the von Neumann algebra generated by the bounded elements of $\Hh_\Tt$. Therefore Theorem~\ref{th:trace} allows to extend $\Tt$ to $L^\infty(\Aa,\Tt)$ in the way described there. If $\Tt$ is finite, the extension of the trace is again finite as $\Tt(x) \leq \lVert x \rVert \,\Tt(1)<\infty$ for all $x \in L^\infty(\Aa,\Tt)$.

\vspace{.1cm}

As $\Tt$ is invariant under $\alpha$, one has
$$
(\alpha_t(a) | \alpha_t(a)) 
\;=\; (a|a)
\;, 
\qquad \forall\; a \in \Aa_\Tt^2
\;,
$$
and hence $\alpha_t$ is an isometry on the pre-Hilbert space $\Aa_\Tt^2$ that  extends to a unitary operator $V(t)$ on $\Hh_\Tt$ for any $t\in G$. As $\alpha$ is strongly continuous on $\Aa$ and $\Tt$ is lower semicontinuous, one gets from the standard trace estimates $|\Tt(ab)|\leq\|a\|\Tt(|b|)$ and $\Tt(|a+b|)\leq \Tt(|a|)+\Tt(|b|)$,
\begin{align*}
\lVert a - \alpha_t(a) \rVert^2_{\Hh_\Tt} 
&
\;=\; 
\Tt\big((a - \alpha_t(a))^*(a - \alpha_t(a))\big)
\\
&
\;\leq \;
\norm{ a - \alpha_t(a)}  \,2\,\Tt(\abs{a})\;, 
\end{align*}
for all $a \in \Aa_\Tt$. Hence by density of $\Aa_\Tt$ the $G$-action $t \mapsto V(t)$ is a strongly continuous action on $\Hh_\Tt$. Therefore the action $\alpha$ can be extended to $L^\infty(\Aa,\Tt)$ by setting $\tilde{\alpha}_t(a) = V(t) a V(t)^*$. This defines a weakly continuous action since it is already defined through a covariant representation. Finally, as $\tilde{\alpha}$ preserves the scalar product of $\Hh_\Tt$, the extension of $\Tt$ is also $\tilde{\alpha}$-invariant.
\hfill $\Box$

\vspace{.2cm}

Whenever in the set-up of Proposition~\ref{prop:traceext}, we will use the notation 
$$
\Mm
\;=\;
L^\infty(\Aa,\Tt)
\;,
$$
for the associated von Neumann algebra with s.n.f. trace $\Tt$, identify $\Aa \subset L^\infty(\Aa,\Tt)$ and denote $\tilde{\alpha}=\alpha$ by the same symbol for simplicity. Note that $\Mm$ is a subalgebra $\pi_\Tt(\Aa)''$ of the bounded operators $\Bb(\Hh_\Tt)$ on the GNS-Hilbert space $\Hh_\Tt$ and that the extension $\tilde{\alpha}$ of $\alpha$ is constructed as in Section~\ref{sec-WStar}.

\vspace{.2cm}

Just as for a general semi-finite von Neumann algebra $\Mm$, it is now possible to construct non-commutative $L^p$-spaces as described in Appendix~\ref{app-Lp}. These spaces are denoted by $L^p(\Mm)$ for $0< p\leq \infty$ and the (quasi-)norm therein by $\|\,.\,\|_p$. For the Hilbert space $L^2(\Aa, \Tt)= L^2(\Mm)$ the construction above can be described concisely as follows:

\begin{proposition}
\label{prop-FullLeftHilbert}
The GNS-Hilbert space $\Hh_\Tt$ is isomorphic to $L^2(\Aa,\Tt)$ in such a way, that the image of the full left Hilbert algebra of $\Aa_\Tt^2$ under $\pi_l$ is equal to $L^\infty(\Aa,\Tt)\cap L^2(\Aa,\Tt)$.
\end{proposition}
%
\noindent {\bf Proof.}
From the definition of the trace, one has $\norm{\pi_l(h)}_2 = \norm{h}_{\Hh_\Tt}$ for every bounded element of $\Hh_\Tt$ and hence $\pi_l$ extends to an isometry $\pi_l: \Hh_\Tt \to L^2(\Aa,\Tt)$. Likewise, one can associate to every $x \in
L^\infty(\Aa,\Tt)_+ \cap L^2(\Aa,\Tt)$ a bounded element of $h \in \Hh_\Tt$ such that $\pi_l(h)=x$ and hence the image of $\pi_l$ is dense in $L^2(\Aa,\Tt)$ by polarization.
\hfill $\Box$

\vspace{.2cm}

In the following let $(\Mm,G,\alpha)$ be a $W^*$-dynamical system with an $\alpha$-invariant s.n.f. trace $\Tt$, which is not necessarily constructed from a $C^*$-dynamical system as above. The $\alpha$-invariance of $\Tt$ implies that the action $\alpha$ is isometric on $L^\infty(\Mm) \cap L^p(\Mm)$ and hence extends to an isometric action on $L^p(\Mm)$ which is denoted by $\LpDyn$.

\begin{proposition}[Lemma 13.4 in \cite{Masuda83}]
\label{prop-LpCont}
The action $\LpDyn: G \times L^p(\Mm) \to L^p(\Mm)$ is strongly continuous w.r.t. the $L^p$-norm for $1 \leq p < \infty$.
\end{proposition}

Let us note that by definition
\begin{equation}
\label{eq-DynEq}
\LpDyn|_{\Mm\cap L^p(\Mm)}
\;=\;
\alpha|_{\Mm\cap L^p(\Mm)}
\;.
\end{equation}
and hence we will often drop the superscript. Note that $\Mm$ is faithfully represented on $L^2(\Mm)$ by left-multiplication in the sense of $\Tt$-measurable operators (see Appendix~\ref{app-Lp}), which also coincides with the GNS-representation under the identification $\Hh_\Tt \simeq L^2(\Mm)$. In the particular case $p=2$, $\LtwoDyn$ is therefore a strongly continuous unitary group on a Hilbert space which we denote by $t\in G\mapsto V(t)$, just  as in the proof of Proposition~\ref{prop:traceext} above.  The strongly continuous unitary group $t\in G\mapsto V(t)$ on $L^2(\Mm)$ has again a set of generators $\GenGNS=(\GenGNS_1,\ldots,\GenGNS_n)$ defined as in  \eqref{eq-GenDef} by
\begin{equation}
\label{eq-GenDefGNS}
V(t) \;= \;  e^{2\pi \imath\,  \GenGNS\cdot t} 
\;,
\qquad
\GenGNS\cdot t\;=\;\sum_{j=1}^n \GenGNS_j t_j
\;.
\end{equation}
%

\begin{proposition}
\label{prop-L2Rep}
The {\rm GNS} representation $\pi_\Tt$ of $\Mm$ on $\Hh_\Tt=L^2(\Mm)$  is a  covariant representation $(\pi_\Tt,V)$ of the $W^*$-dynamical system $(\Mm,G,\alpha)$.
\end{proposition}

\noindent {\bf Proof.}
By definition, any $x\in L^2(\Mm)$ is an $L^2$-limit of a sequence $(b_m)_{m\geq 1}$ in $\Mm_\Tt^2$. Then for any $a\in\Mm$, 
$$
\pi_\Tt(a)x\;=\;L^2\mbox{-}\lim_m ab_m\;,
\qquad
\LtwoDyn_t(x)
\;=\;
L^2\mbox{-}\lim_m \alpha_t(b_m)
\;.
$$
Hence
\begin{align*}
\pi_\Tt(\alpha_t(a))x
&
\;=\;
L^2\mbox{-}\lim_m \alpha_t(a)b_m
\\
&
\;=\;
L^2\mbox{-}\lim_m \alpha_t\big(a\, \alpha_{-t}(b_m)\big)
\\
&
\;=\;
\LtwoDyn_t\big(\pi_\Tt(a)\LtwoDyn_{-t}(x))\big)
\;,
\end{align*}
namely the covariance relation. The strong continuity holds by Proposition~\ref{prop-LpCont}.
\hfill $\Box$

\vspace{.2cm}

Let us note that assuming the conditions of Proposition~\ref{prop:traceext} and $\Mm=L^\infty(\Aa,\Tt)$ the covariant representation $(\pi_\Tt,V)$ leads to a representation formula \eqref{eq-IntegratedRep} of the $C^*$-algebraic crossed product $\Aa \rtimes_\alpha G$ on the GNS-Hilbert space, namely
\begin{equation}
\label{eq-IntegratedRepGNS}
(\pi_\Tt \times V)(f) 
\;=\; 
\int_G \pi_\Tt(f(t))\, e^{2\pi \imath\,  \GenGNS\cdot t} \,\text{d}t
\;,
\qquad
f\in C_c(G, \Aa)
\;.
\end{equation}

Let us also note that if $\Tt$ is a finite trace,  the GNS representation has the cyclic vector $\Omega=\pi_\Tt(\one)$ and
$$
V(t) (\pi_\Tt(a) \Omega)
\;=\; 
\pi_\Tt(\alpha_t(a))\Omega
\;, 
\quad a \in \Mm
\;.
$$
%

\section{Arveson spectrum and spectral decomposition}
\label{sec-SpecDecomp}

In the proceeding sections there appeared various isometric (abelian) $G$-actions on Banach spaces, namely the automorphic action on a $C^*$-algebra $\Aa$, its extension to a von Neumann $\Mm$ and the extensions $\LpDyn$ to non-commutative $L^p$-spaces. This section, basically a review of \cite{Arveson73} and Chapter XI in \cite{Takesaki2003}, discusses the spectral analysis w.r.t. any of these actions. Throughout again $G=\bbT^{n_0} \oplus \bbR^{n_1}$ with $n=n_0+n_1$. 

\vspace{.2cm}

Hence let $E$ be a Banach space with a linear isometric $G$-action $\Arvesonaction$ that is assumed to be strongly continuous for simplicity (as explained in \cite{Takesaki2003}, weaker notions of continuity are also sufficient, in particular $\sigma$-weak continuity if $E$ is a von Neumann algebra). For any $f \in L^1(G)$, let us define a bounded operator $\Arvesonaction_f\in\Bb(E)$ on $E$ as a Riemann integral:
\begin{equation}
\label{eq-LyapDynFunct}
\Arvesonaction_f(x) 
\;=\; 
\int_{G} f(t) \,\Arvesonaction_t(x)\, \difd{t}
\;,
\qquad
x\in E
\;.
\end{equation}
For every fixed $x\in E$,  this gives a representation of the convolution algebra $(L^1(G),\ast)$ on $E$, namely for $f,g\in L^1(G)$ and $\lambda\in\CM$,
\begin{equation}
\label{eq-ConvRep}
\Arvesonaction_{f+\lambda g}(x)
\;=\;
\Arvesonaction_f(x)+\lambda\,\Arvesonaction_g(x)
\;,
\qquad
\Arvesonaction_f(\Arvesonaction_g(x))
\;=\; 
\Arvesonaction_{f\ast g}(x)
\;.
\end{equation}
Next let $FA(\hat{G})= \calF L^1(G) \subset C_0(\hat{G})$ denote the Fourier algebra on $\hat{G}$ with pointwise multiplication. The action of a function $f \in FA(\hat{G})$ is written as
\begin{equation}
\label{eq-FAaction}
\widehat{f} * x 
\;=\; 
\Arvesonaction_{\calF^{-1} f}(x)
\;.
\end{equation}
It defines a bounded linear operator on $E$ with operator norm in $\Bb(E)$ bounded by
\begin{equation}
\label{eq-FAbound}
\big\|\widehat{f} * \cdot\,\big\|
\;\leq \;
\norm{\calF^{-1} f}_1
\;.
\end{equation}
In the classical case where $E$ is a space of functions on $\RM^n$ for which the Fourier transform is densely defined and $\Arvesonaction$ is the shift, the operator $\widehat{f} \,*$ for $f\in FA(\RM^n)$ is called a Fourier multiplier. In the general case, let us note that by \eqref{eq-ConvRep} 
\begin{equation}
\label{eq-ConvRep2}
(f+\lambda g)^\wedge*x
\;=\;
\widehat{f} *x\,+\,\lambda\, \widehat{g}\ast x
\;,
\qquad
\widehat{f} * (\widehat{g}\ast x)
\;=\;
\widehat{fg}\ast x
\;.
\end{equation}
This in turn implies that
$$
I(x) 
\;=\; 
\{f \in FA(\hat{G})\;: \; \widehat{f}*x = 0\}
$$
is for any $x\in E$ an ideal in $FA(\hat{G})$. Arveson's $\Arvesonaction$-spectrum of $x$ is the corresponding zero point set:

\begin{definition}
\label{def-AversonSpec}
For an isometry $G$-action $\Arvesonaction$ on a Banach space $E$, Arveson's $\Arvesonaction$-spectrum $\sigma_\Arvesonaction(x)$ of $x\in E$ is defined by
$$
\sigma_\Arvesonaction(x) 
\;=\; 
\{\lambda \in \hat{G}\;: \; f(\lambda)=0\;\;\forall \;f \in I(x)\}
\;.
$$
\end{definition}

The following basic result is used freely below.

\begin{proposition}[\cite{Arveson73,Takesaki2003}]
For $f\in FA(\hat{G})$ and $x \in E$,
$$
\sigma_\Arvesonaction(\widehat{f}*x)
\;\subset\;\text{\rm supp}(f)
\;.
$$
\end{proposition}

Let us now consider the cases that are relevant for our application, namely of a $W^*$-dynamical system $(\Mm,G,\alpha)$ with an invariant s.n.f. trace $\Tt$ and an isometric action $\LpDyn$ on the associated non-commutative $L^p$-space $L^p(\Mm)$. For an element $a\in \Mm\cap L^p(\Mm)$, there are then two Arveson spectra, which due to \eqref{eq-DynEq} and \eqref{eq-LyapDynFunct} are equal:

\begin{proposition}
\label{prop-AvSpec}
For $a\in \Mm\cap L^p(\Mm)$, one has $\sigma_\alpha(a)=\sigma_{\LpDyn}(a)$.
\end{proposition}

Next let us focus on the case of the Hilbert space $E=L^2(\Mm)$, even though all the formulas and facts below hold for a general Hilbert space. As already noted in \eqref{eq-GenDefGNS}, Stone's theorem then implies $\LtwoDyn_t=e^{2\pi\imath \GenGNS\cdot t}$ for some commuting selfadjoint generators $\GenGNS=(\GenGNS_1,\dots,\GenGNS_n)$. This spectral decomposition of $\GenGNS$ gives a direct integral decomposition
\begin{equation} 
\label{eq:fourierdecomp}
L^2(\Mm) 
\;=\; 
\int_{\sigma(\GenGNS)}^\oplus \mu(\difd \lambda)\;\Hh_\lambda
\;,
\end{equation}
with Hilbert spaces $\Hh_\lambda$ and a Borel-measure $\mu$ (the spectral measure of $\GenGNS$) on the spectrum $\sigma(\GenGNS)\subset \hat{G}$ of $\GenGNS$. Hence any $a\in L^2(\Mm)$  has a "Fourier"-decomposition
\begin{equation} 
\label{eq-aFourierDecomp}
a 
\;=\; 
\int_{\sigma(\GenGNS)}^\oplus \mu(\difd \lambda)\;a_\lambda 
\;,
\end{equation}
with $a_\lambda$ a section of the field of Hilbert spaces $(\Hh_\lambda)_{\lambda \in \sigma(\GenGNS)}$. The decomposition~\eqref{eq-aFourierDecomp} is only defined uniquely up to null sets w.r.t. $\mu$. The domain $\sigma(\GenGNS)$ of the integral in \eqref{eq-aFourierDecomp} can, moreover, be replaced by the $\mu$-a.s. support of $a_\lambda$ which coincides with the Arveson spectrum $\sigma_{\LtwoDyn}(a)$ and thus, due to Proposition~\ref{prop-AvSpec}, also with $\sigma_\alpha(a)$. 
Note that this is a decomposition~\eqref{eq-aFourierDecomp} of $a$ as a vector in the Hilbert space $L^2(\Mm)$ and not as a linear operator, namely the multiplication law cannot in general be written in terms of the components $a_\lambda$. However, one has 
\begin{equation} 
\label{eq-aFourierDecomp2}
\LtwoDyn_t(a) 
\;=\; 
\int_{\sigma_{\LtwoDynAv}(a)}^\oplus \mu(\difd \lambda)\;e^{2\pi\imath\,t\cdot \lambda}\,a_\lambda 
\;,
\qquad
a\in \Mm\cap L^2(\Mm)
\;.
\end{equation}
More generally, the action \eqref{eq-FAaction} of $f\in L^1(\hat{G})$ can be expressed by functional calculus of the generator $X$:
\begin{equation}
\label{eq-FourierMultAct}
\widehat{f} * a
\;=\;
f(\GenGNS)\,a
\;=\;
\int_{\sigma_{\LtwoDynAv}(a)}^\oplus \mu(\difd \lambda)\;f(\lambda)\,a_\lambda 
\;,
\qquad
a\in \Mm\cap L^2(\Mm)
\;.
\end{equation}
Let us note that if $\Tt$ is a finite trace, then in particular $\Mm \subset L^2(\Mm)$ and hence every $a \in \Mm$ has a Fourier decomposition~\eqref{eq-aFourierDecomp}.

\section{Dual traces on crossed products}
\label{sec-DualTraces}

In this section let $(\Mm, G, \alpha)$ be a $W^*$-dynamical system with  an abelian $n$-parameter group $G$ and  a von Neumann algebra $\Mm$ with a s.n.f. trace $\Tt$ that is left invariant by $\alpha$. One can then canonically construct a s.n.f. trace $\hat{\Tt}_\alpha$ on the crossed product $\Mm \rtimes_\alpha G$  which is invariant under both $\alpha$ and the so-called dual action $\hat{\alpha}$.

\begin{definition}
	\label{def-W*DualAct}
Let $(\Mm, G, \alpha)$ be $W^*$-dynamical system with abelian group $G$ and dual group $\hat{G}$.  There is a weakly continuous action $\hat{\alpha}: \hat{G} \to \mbox{\rm Aut}(\Mm \rtimes_\alpha G)$ which acts on the generators by
$$
\hat{\alpha}_\dualvar \big(\pi(a)\big) 
\;=\; 
\pi (a)\,, 
\qquad \forall \;\,a\in \Mm, \dualvar \in \hat{G}
$$
and
$$
\hat{\alpha}_\dualvar \big(U(t)\big) 
\;=\; 
\overline{\langle \dualvar, t \rangle} U(t)
\,, 
\qquad \forall\;\, t\in G, \dualvar \in \hat{G}
\;.
$$
\end{definition}

To define the dual trace one can apply a construction similar to Proposition~\ref{prop:traceext}.

\begin{proposition}
\label{prop-DualTrace}
Let $(\Mm, G, \alpha)$ be a $W^*$-dynamical system with  an $\alpha$-invariant s.n.f. trace $\Tt$. The crossed product $\Mm \rtimes_\alpha G$ has a s.n.f. trace $\hat{\Tt}_\alpha$ that is left invariant by the dual action $\hat{\alpha}$ and is called the dual trace.
\end{proposition}

\noindent {\bf Proof.} 
We only describe the construction and refer to \cite{Takesaki2003} for technical details.
Consider the space $C_c(G,\Mm)$ of $\sigma^*$-strongly continuous functions of compact support which is a $*$-algebra with the algebraic relations as in \eqref{form:mult}
\begin{equation}
(fg)(t) \;= \;\int_G f(s) \,\alpha_s(g(t-s))\, \difd{s}\;,
\qquad 
f^*(t) \;= \;\alpha_t(f(-t)^*)\;,
\qquad
f,g \in \mathcal{C}\;.
\end{equation}
Define the $*$-subalgebra 
\begin{equation}
\label{eq-C_kHilbertAlg}
\mathcal{C}(G,\Mm)
\;=\; 
\text{span} \{\Mm_\Tt \cdot C_c(G, \Mm)\} 
\,\cap\,  
\text{span}\{C_c(G, \Mm) \cdot \Mm_\Tt\}
\;,
\end{equation}
with the algebraic span and $\Mm_\Tt$ acting by pointwise left respectively right multiplication. The elements of $\mathcal{C}(G,\Mm)$ are again in $C_c(G, \Mm)$ and define a Hilbert algebra with the scalar product
\begin{equation}
\label{eq:scalarprod}
(f | g)_G 
\;=\; 
\Tt\big( (f^* g)(0)\big) 
\;= \;
\int_G  \Tt\big( f(s)^*g(s) \big)\,\text{d}s
\;,
\end{equation}
where the integral is w.r.t. a Haar-measure on $G$. The Hilbert space completion of $\mathcal{C}(G,\Mm)$ is then isomorphic to $L^2(G, \Hh_\Tt)$ in such a way that its left von Neumann algebra $\mathcal{U}(\mathcal{C}(G,\Mm))$ coincides with the crossed product $\Mm \rtimes_\alpha G$ in the left regular representation. Theorem~\ref{th:trace} then defines a trace $\hat{\Tt}_\alpha$ on the crossed product. As the dual action $\hat{\alpha}$ acts on $f \in \mathcal{C}(G,\Mm)$ as $(\hat{\alpha}_\dualvar f)(s) =\langle \dualvar,s \rangle\, f(s)$, the scalar product $(f | g)_G$ is $\hat{\alpha}$-invariant on a dense subset of $L^2(G, \Hh_\Tt)$. Hence the dual trace $\hat{\Tt}_\alpha$ is also $\hat{\alpha}$-invariant.
\hfill $\Box$

\vspace{.2cm}

The trace $\hat{\Tt}_\alpha$ for a locally compact group is only determined uniquely up to a choice of normalization of the Haar measure which here is fixed as in the preamble of Chapter~\ref{sec-CrossedProd}. In the following we simply write 
$$
\Nn 
\;=\; 
\Mm \rtimes_\alpha G
$$
for the $W^*$-crossed product and always equip it with the above s.n.f. trace $\hat{\Tt}_\alpha$ whenever $\Mm$ has a trace. It is now possible to construct $L^p$-spaces $L^p(\Nn)$, again as in Appendix~\ref{app-Lp}.  Noting that $\Cc(G,\Mm)$ consists of functions that take values in $L^2(\Mm)\cap \Mm$, it is possible to give a better description of $L^2(\Nn)=L^2(\Mm \rtimes_\alpha G, \hat{\Tt}_\alpha)$:

\begin{lemma}
\label{lem-L2BoundedRep}
One has $L^2(G, L^2(\Mm)) \simeq L^2(\Nn)$ where the isomorphism is densely defined through
\begin{equation}
\label{eq-L2rep_integrated}
 (\pi\times U): f \in \Cc(G,\Mm) 
 \;\mapsto\; 
\int_{G} \pi(f(t))\, U(t)\, \difd{t}
\;,
\end{equation}
with $(\pi,U)$ the regular representation associated to $\pi_\Tt$ and the integral understood in the strong operator topology. In particular, for any $\hat{b}\in L^2(\Nn)\cap\Nn$, there exists $f\in L^2(G, L^2(\Mm))$ and a sequence $(f_n)_{n\in \bbN}$ in $\Cc(G,\Mm)$ such that $f_n \to f$ and $\hat{b}=\slim_{n\to\infty} \int_{G} \pi(f_n(t))\, U(t)\, \difd{t}$.
\end{lemma}
%
\noindent {\bf Proof.} 
The left representation $\pi_l: \Cc(G,\Mm) \to \Uu(\Cc(G,\Mm))=\Mm\rtimes_\alpha G$ of the Hilbert algebra $\Cc(G,\Mm)$ is defined exactly by the same formula as the integrated representation \eqref{eq-L2rep_integrated} (see \cite[Lemma X.1.8]{Takesaki2003}), however, the integral must now be understood in the $\sigma^*$-strong sense. By the same reasoning as in Proposition \ref{prop-FullLeftHilbert}, the map extends to an isomorphism of the $L^2$-spaces such that the image of the full left Hilbert algebra $\hat{\Cc}(G,\Mm)$  coincides with $L^2(\Nn)\cap\Nn$. Hence the last statement is the content of Theorem \ref{theorem-hilbert_algebra_approx}(ii).
\hfill $\Box$

\vspace{.2cm}

Any faithful covariant representation of $\Nn$ also comes with an induced representation \eqref{eq-L2rep_integrated} of the Hilbert algebra $\tilde{\Cc}(G,\Mm) = (\pi \times U)(\Cc(G,\Mm))$ and in order to compute the trace of an element $f \in L^1(\Nn)$, one has to find a factorization $f =  \pi_l(h)^* \pi_l(g)$ with $g,h \in \hat{\Cc}(G,\Mm)$, {\it i.e.}  elements of $\Hh$ corresponding to elements of $\Nn^2_{\hat{\Tt}_\alpha}$, which then implies
$$
\hat{\Tt}_\alpha(f) 
\;=\; 
(g | h)_G 
\;.
$$
If $G$ is a $n$-parameter group this can be done canonically, at least on the generators of the crossed product \cite{Lesch91}:

\begin{proposition}
\label{prop-DualTraceCalc}
Let $G=\bbT^{n_0} \oplus \bbR^{n_1}$, $(\Mm,G,\alpha)$, $\Tt$ as in {\rm Proposition~\ref{prop-DualTrace}}, $\Nn$ as defined above and  $D=(D_1,\ldots ,D_n)$ the generators of $U$ as in \eqref{eq-GenDef}. In terms of the Haar measure $\difd{\dualvar }$ on $\hat{G}$ with a normalization fixed by the Plancherel formula \eqref{eq-Plancharel}, one has
\begin{equation}
\label{eq:traceL2}
\hat{\Tt}_\alpha\big(\conjugate{g}(D) \pi(a^*b) h(D)\big)
\;=\; 
\Tt(a^*b)  \int_{\hat{G}} \conjugate{g(\dualvar )}\,h(\dualvar ) \,\difd{\dualvar }
\;,
\end{equation}
for $g,h \in L^2(\hat{G})\cap L^\infty(\hat{G})$ and $a,b \in \Mm^2_\Tt$.
\end{proposition}

\noindent {\bf Proof.} 
In view of Lemma \ref{lem-L2BoundedRep} and by the density of $\mathcal{F}(C_c(G))$ in $L^2(G)$ it will be enough to prove this for $g,h \in \mathcal{F}(C_c(G))$ and $a,b \in \Mm \cap L^2(\Mm)$. One can then write
$$
\pi(a) g(D) \;=\;  (\pi\times U)(a \cdot (\mathcal{F}^{-1}g)) 
\;=\; 
\int_G \pi(a) \,(\mathcal{F}^{-1}g)(t) \,U(t) \,\difd{t}
$$
and similarly for $\pi(b) h(D)$. Hence one has
\begin{align*}
\hat{\Tt}_\alpha\big(\conjugate{g}(D) \pi(a^*b) h(D)\big) 
&
\;=\; 
(\pi(a)(\mathcal{F}^{-1}g)(\cdot)|\pi(b)(\mathcal{F}^{-1}h)(\cdot))_G
\\
&\; =\; \int_G \Tt\left(a^*b\, \conjugate{(\mathcal{F}^{-1}g)(s)}\,(\mathcal{F}^{-1}h)(s)\right) \,\difd{s} \\
&
\;=\; \Tt(a^*b) \,\int_G  \conjugate{(\mathcal{F}^{-1}g)(s)}\,(\mathcal{F}^{-1}h)(s) \,\difd{s} 
\\
&\;=\;  
\Tt(a^*b) \int_{\hat{G}} \conjugate{g(\dualvar )}\,h(\dualvar ) \,\difd{\dualvar }
\;,
\end{align*}
as the measure on $\hat{G}$ is normalized such that the Plancherel theorem holds without an additional constant. 
\hfill $\Box$

\vspace{.2cm}

Another refinement of the trace formula \eqref{eq:traceL2} is given by:
\begin{corollary}
\label{coro-TraceL1Calc}
For $a \in L^1(\Mm) \cap \Mm$ and $f \in L^1(\hat{G}) \cap L^2(\hat{G})$ such that $\pi(a)f(D) \in L^1(\Nn)$, one has
\begin{equation}
\label{eq:trace}
\hat{\Tt}_\alpha\big( \pi(a) f(D)\big)
\;=\; 
\Tt(a)  \int_{\hat{G}} f(\dualvar ) \,\difd{\dualvar }
\;.
\end{equation}
\end{corollary}

\noindent {\bf Proof.}  
For  a sequence $(g_m)_{m\in \bbN}$ of uniformly bounded Borel functions $g_m \in L^2(\hat{G})$ that converges pointwise to $\one \in L^\infty(\hat{G})$, one has $\slim_{m\to \infty} g_m(D)=\one$ and hence by the normality of the trace
\begin{align*}
\hat{\Tt}_\alpha\big( \pi(a) f(D)\big) \;
&=\; \lim_{m\to \infty}\hat{\Tt}_\alpha\big(g_m(D) \pi(a) f(D)\big) \\
 &= \;\lim_{m\to \infty}\hat{\Tt}_\alpha\big(g_m(D) \pi(u b^*) \pi(b) f(D)\big) \\
 &= \;\lim_{m\to \infty}\Tt(ub^*b) \int_{\hat{G}} g_m(\dualvar) f(\dualvar) \,\difd{\dualvar }\\
&= \;\Tt(a)\,  \int_{\hat{G}} f(\dualvar ) \,\difd{\dualvar }
\end{align*}
where the polar decomposition of $a$ was used to factor $a=u b^*b$ with $b=\sqrt{|a|}$ and $u$ a partial isometry and then Proposition~\ref{prop-DualTraceCalc} was applied.
\hfill $\Box$

\vspace{.2cm}

Let us note that the dual action $\hat{\alpha}$ generates translations on functions of $D$, namely 
$$
\hat{\alpha}_\dualvar (h(D))
\;=\; 
h(D+ \dualvar )
\;,
\qquad
h \in L^2(\hat{G})\cap L^\infty(\hat{G})\;.
$$ 
Replacing in \eqref{eq:traceL2} shows again that the dual trace is invariant under the dual action. The definition of the dual trace gives us a complete description of  
$$
L^2(\Nn)
\;=\;
L^2(\Mm \rtimes_\alpha G, \hat{\Tt}_\alpha)
\;\simeq \;
L^2(G, \Hh_\Tt)
$$ 
and, in particular, the $L^2$-norm. In the following also the other non-commutative $L^p$-spaces $L^p(\Nn)$ will be needed. In these cases it becomes impossible to compute  the norm in terms of the generators of $\Mm\rtimes_\alpha G$ because it involves functional calculus. For products of the generators we obtain a simple estimate:

\begin{proposition}
\label{prop:lp-embedding}
Let $2 \leq p \leq \infty$ and $f \in L^p(\hat{G})$. The map 
$$
(a,f) \,\in\, (\Mm \cap L^p(\Mm)) \,\times\, (L^\infty(\hat{G}) \,\cap\, L^p(\hat{G})) 
\;\mapsto\; 
\pi(a)f(D) \,\in\, \Nn
$$
is $L^p(\Mm) \times L^p(\hat{G}) \to L^p(\Nn)$-bounded with
\begin{equation}
\label{eq:LPforproducts} 
\norm{\pi(a)f(D)}_p 
\;\leq\; 
\norm{a}_p \norm{f}_p
\;.
\end{equation}
For $p=2$, one even has equality
\begin{equation}
\label{eq:L2forproducts} 
\norm{\pi(a)f(D)}_2
\;=\; 
\norm{a}_2 \norm{f}_2.
\end{equation}
\end{proposition}

\noindent {\bf Proof.}  
For $p=\infty$, the estimate \eqref{eq:LPforproducts} is obvious and for $p=2$ it follows from the trace formula \eqref{eq:traceL2}. The claim for all $2 \leq p \leq \infty$ follows from the non-commutative Riesz-Thorin theorem~\ref{theorem:rieszthorin} by noting that the r.h.s. of \eqref{eq:LPforproducts} is the norm of the non-commutative $L^p$-space $(L^p(\Mm) \otimes L^p(\hat{G}), \Tt \otimes \int \difd{\dualvar })$.
\hfill $\Box$

\vspace{.2cm}

The bound \eqref{eq:LPforproducts} allows to interpret some products of unbounded elements of $L^p(\Mm)$ and $L^p(\hat{G})$ as their images under the bounded extension of the map $(a,f)\mapsto \pi(a)f(D)$. Let us note that the classical estimates for the singular values of integral operators of the type $f(X)g(-\imath \nabla)$ are a special case since they arise from the crossed product $L^\infty(\bbR)\rtimes_\lambda \bbR$ with $\lambda$ being right translation. Hence the range of $p$ cannot be extended to $0 < p < 2$ with the same r.h.s. \cite[Chapter 4]{Simon05}, instead one must impose some more stringent conditions. We will return to this issue in Section~\ref{sec-Peller}.

\section{Dual action and duality of crossed products}
\label{sec-Duality}

On a $C^*$-crossed product with an abelian group $G$ one also has a dual action which is again strongly continous \cite{Raeburn88}.
\begin{definition}
\label{def-C*DualAct}
The dual action $\hat{\alpha}: \hat{G} \to \mbox{\rm Aut}(\Aa \rtimes_\alpha G)$ of the dual group $\hat{G}$ on a crossed product $\Aa \rtimes_\alpha G$ is defined by
$$
\hat{\alpha}_\dualvar \big(i_\Aa(a)\,i_G(f)\big) 
\;=\; 
i_\Aa(a) \,i_G(\overline{\langle \dualvar ,\cdot\rangle} f)
\;, 
\qquad \dualvar \in \hat{G}\,,\; \;a \in \Aa\,,\;\; f \in C_c(G)\,.
$$
\end{definition}
One can therefore construct the crossed product $\Aa \rtimes_\alpha G \rtimes_{\hat{\alpha}} \hat{G}$ leading to the periodicity result (see again \cite{Raeburn88} for a short proof):

\begin{theorem}[Takai duality]
\label{eq-TakaiDuality}
The second crossed product  $\Aa \rtimes_\alpha G \rtimes_{\hat{\alpha}} \hat{G}$ is isomorphic to $\Aa \otimes \Kk(L^2(G))$ where $\Kk(L^2(G))$ denotes the compact operators on $L^2(G)$. The isomorphism $i_T:\Aa \rtimes_\alpha G \rtimes_{\hat{\alpha}} \hat{G}\to\Aa \otimes \Kk(L^2(G))$ can be chosen in such a way that the second dual action $\hat{\hat{\alpha}}$  acts as $\alpha \otimes \mbox{\rm Ad}_{\rho_G}$, with $\rho_G$ the regular representation of $G$ on $L^2(G)$, i.e. acting by translation.
\end{theorem}

Next let us consider the dual action $\hat{\alpha}: \hat{G} \to \mbox{\rm Aut}(\Mm \rtimes_\alpha G)$ on the $W^*$-crossed product $\Mm \rtimes_\alpha G$ as defined in Definition~\ref{def-W*DualAct} for $\Mm \rtimes_\alpha G$ defined using a left regular representation on $L^2(G,\Hh)$ denoted by $(\pi_\alpha, U_\alpha)$. The dual action is spatially implemented by the strongly continuous group of unitaries  
$$
(\hat{U}_{\hat{\alpha}}(k) \phi)(t) 
\;=\; 
\langle k, t \rangle\, \phi(t)\;, 
\qquad k \in \hat{G}\;, \;\;
\phi \in L^2(G,\Hh)\;,\;\;
t\in G
\;,
$$
with
$$
\hat{\alpha}_k(\hat{x}) 
\;=\; 
\hat{U}_{\hat{\alpha}}(k) \,\hat{x}\, \hat{U}_{\hat{\alpha}}(k)^*\;,
\qquad \forall\; \hat{x} \in \Mm \rtimes_\alpha G
\;.
$$
The dual action also defines a $W^*$-dynamical system $(\Mm\rtimes_\alpha G, \hat{G}, \hat{\alpha})$ and one has:

\begin{theorem}[Takesaki duality \cite{Takesaki73}]
The second $W^*$-crossed product  $\Mm \rtimes_\alpha G \rtimes_{\hat{\alpha}} \hat{G}$ is isomorphic to $\Mm \otimes \calB(L^2(G))$, with the second dual action $\hat{\hat{\alpha}}$ again acting as $\alpha \otimes \mbox{\rm Ad}_{\rho_G}$.

If $\Tt$ is an $\alpha$-invariant s.n.f. trace on $\Mm$, then the second dual trace $\hat{\hat{\Tt}}= \widehat{(\hat{\Tt}_\alpha)}_{\hat{\alpha}}$  is carried by the isomorphism into $\Tt \otimes \Tr$ with $\Tr$ the usual trace on $\calB(L^2(G))$.
\end{theorem}

For further use, the isomorphism is now described in some detail. Suppose that $\Mm$ is acting on a Hilbert space $\Hh$ and let the crossed product $\Mm \rtimes_\alpha G$ be defined on $L^2(G,\Hh)$ as the $W^*$-span of the covariant representation $(\pi_\alpha, U_\alpha)$. The second crossed product is then defined on $L^2(\hat{G}) \otimes L^2(G, \Hh)$ as the $W^*$-algebra generated by a regular representation $\pi_{\hat{\alpha}}$ of $\Mm \rtimes_\alpha G$ and a representation $\tilde{U}_{\hat{\alpha}}$ of $\hat{G}$, or given in terms of generators $a\in \Mm$, $s,t\in G$, $k,q\in \hat{G}$ acting on $\phi \in L^2(G \times \hat{G}, \Hh)$ by
\begin{align}
\label{eq:secondcrossed_generators}
((\pi_{\hat{\alpha}} \circ \pi_\alpha)(a)\phi)(t,k) &\;=\; \alpha_{-t}(a)\,\phi(t,k)\;, \nonumber \\
(\pi_{\hat{\alpha}}(U_\alpha(s)) \phi)(t,k) &\;=\; \overline{\langle k, s\rangle} \, \phi(t-s,k)\;,\\
(\widetilde{U}_{\hat{\alpha}}(q) \phi)(t,k) &\;=\; \phi(t,k-q)\;. \nonumber
\end{align}
The isomorphism $\Psi: \Mm \rtimes_\alpha G \rtimes_{\hat{\alpha}} \hat{G} \to \Mm \otimes \calB(L^2(G))$ acts by mapping the generators  on $L^2(G \times \hat{G}, \Hh)$ to operators in the regular representation on $L^2(G, \Hh)$
$$
\Psi( (\pi_{\hat{\alpha}} \circ \pi_\alpha)(x)) \;=\; \pi_\alpha(x)
\;,
\quad
\Psi(\pi_{\hat{\alpha}}(U_\alpha(s))) \;=\; U_\alpha(s) 
\;,
\quad
\Psi(\widetilde{U}_{\hat{\alpha}}(q)) \;=\; \hat{U}_{\hat{\alpha}}(q)
\;.
$$
To see that those operators generate $\Mm \otimes \calB(L^2(G))$, one needs to exhibit the commutating representations of $\Mm \otimes \one$ and $\one \otimes \calB(L^2(G))$. An approximation argument shows that the identical representation $\Mm \otimes \one$ is contained in the image $\Psi(\Mm \rtimes_\alpha G \rtimes_{\hat{\alpha}}\hat{G})$ (even though there is apparently no convenient expression in terms of the generators) and since the covariant pair $U_\alpha$, $\hat{U}_{\hat{\alpha}}$ satisfies the Heisenberg-Weyl commutation relations, its $W^*$-span is naturally isomorphic to $\one \otimes \calB(L^2(G))$.

\vspace{.2cm}

Let us now verify the relation for the second dual trace. In view of the Hilbert algebra construction, the second dual trace $\hat{\hat{\Tt}}$  is uniquely defined by the fact that for an element $\hat{\hat{x}}\in \Mm \rtimes_\alpha G \rtimes_{\hat{\alpha}}\hat{G}$ of the form
$$
\hat{\hat{x}} 
\;=\; 
\int_{G\times \hat{G}} \pi_\alpha(f(s,k)) \pi_{\hat{\alpha}}(U_\alpha(s))\, \hat{U}_{\hat{\alpha}}(k) \,\difd{s}\,\difd{k}
$$
with $f \in L^2(G \times \hat{G}, L^2(\Mm))$, 
one has
$$
\hat{\hat{\Tt}}(\hat{\hat{x}}^*\hat{\hat{x}}) 
\;=\; 
\int_{G\times \hat{G}} \Tt(\abs{f(s,k)}^2)\, \difd{s}\,\difd{k}
\;.
$$
Indeed, $\Psi(\hat{\hat{x}})$ is an $\Mm$-valued integral operator on $L^2(G, \Hh)$
\begin{align*}
(\Psi(\hat{\hat{x}})\phi)(t) &
\;=\; \int_{G\times \hat{G}} \alpha_{-t}(f(s,k))\, \langle k, t-s\rangle\, \phi(t-s)\, \difd{s}\,\difd{k} \\
&
\;=\; 
\int_{G} \left(\int_{\hat{G}} \alpha_{-t}(f(t-s,k))\, \langle k, s \rangle\, \difd{k}\right) \phi(s) \,\difd{s}
\end{align*}
with integral kernel
$$
K(t,s)
\;=\;
\int_{\hat{G}} \alpha_{-t}(f(t-s,k)) \langle k, s \rangle\, \difd{k}
$$
One can then compute for $f$ smooth and rapidly decaying
\begin{align*}
(\Tt \otimes & \Tr)(\Psi(\hat{\hat{x}}^*\hat{\hat{x}})) 
\\
&\;=\; \int_{G^2} \Tt(\abs{K(s,t)}^2)\, \difd{t} \,\difd{s}\\ 
&\;=\; \int_{G^2} \Tt\left(\int_{\hat{G}^2} \alpha_{-t}(f(t-s,k_1)^*f(t-s,k_2)) \langle k_1-k_2, s \rangle \,\difd{k_1}\,\difd{k_2}\right) \,\difd{t} \,\difd{s}\\ 
&\;=\; \int_{G^2} \int_{\hat{G}^2}  \Tt\left(f(t-s,k_1)^*f(t-s,k_2)\right) \langle k_1-k_2, s \rangle \,\difd{k_1}\,\difd{k_2}\, \difd{t}\,\difd{s}\\
&\;=\; \int_{G^2} \int_{\hat{G}^2}  \Tt\left(f(t,k_1)^*f(t,k_2)\right) \langle k_1-k_2, s \rangle \,\difd{k_1}\,\difd{k_2}\, \difd{t}\,\difd{s}\\
&\;=\; \int_{G} \int_{\hat{G}} \Tt\left(f(t,k)^*f(t,k)\right) \,\difd{k}\, \difd{t} 
\\
& \;=\; \hat{\hat{\Tt}}(\hat{\hat{x}}^*\hat{\hat{x}}).
\end{align*}

For the Takai duality one can write down the same algebraic relations as above if one considers $\Aa$ to be acting on $\Hh$ and the crossed product in a regular representation on $L^2(G,\Hh)$, with the difference that the generators 
\eqref{eq:secondcrossed_generators} are in general only multipliers in $M(\Aa \rtimes_\alpha G \rtimes_{\hat{\alpha}}\hat{G})$. A norm dense subset of $\Aa \rtimes_\alpha G \rtimes_{\hat{\alpha}}\hat{G}$ is then given by the elements of the form
$$
\hat{\hat{a}} 
\;=\; 
\int_{G\times \hat{G}} \pi_\alpha(f(s,k)) \pi_{\hat{\alpha}}(U_\alpha(s))\, \hat{U}_{\hat{\alpha}}(k) \,\difd{s}\,\difd{k}
\;,
$$
with $f\in C_c(G\times \hat{G}, \Aa)$ and the Takai isomorphism $i_T: \Aa \rtimes_\alpha G \rtimes_{\hat{\alpha}}\hat{G} \to \Aa \otimes \mathcal{K}(L^2(G))$ is densely defined by the same formula as $\Psi$, {\sl i. e.}
$$
i_T\big(\hat{\hat{a}}\big) 
\;=\; 
\int_{G\times \hat{G}} f(s,k) \,U_\alpha(s) \, \widetilde{U}_{\hat{\alpha}}(k) \,\difd{s}\,\difd{k}
\;.
$$
The image of $i_T$ consists of $\Aa$-valued integral operators and can through an approximation argument be identified with $\Aa \otimes \mathcal{K}(L^2(G))$ \cite{Raeburn88}.

\section{Spaces of differentiable elements}
\label{sec-DiffElements}

As a preparation for Chapter~\ref{sec-Besov}, we consider here an isometric $G$-action $\Diffaction$ on a general Banach space $E$. The group is still restricted to be $G = \bbT^{n_0} \oplus \bbR^{n_1}$ and in our applications later on $E$ will be either $\Aa$, or $\Aa_\Tt$ (in Chapter~\ref{sec-DualityToep}) or the non-commutative $L^p$-spaces $L^p(\Mm)$ (for Chapter~\ref{sec-Besov}). For the purposes of this section we consider the $G$-action as a $\bbZ^{n_0}$-periodic $\bbR^{n}$-action.

\vspace{.2cm}

The action $\Diffaction$ is called strongly continuous if the map $t\in G \mapsto \Diffaction_t(x)$ is norm-continuous for every $x \in E$. If $E$ has a pre-dual $E_*$, then the action is called weak$^*$ continuous if $t\in G \mapsto \phi(\Diffaction_t(x))$ is continuous for any $x \in E$ and $\phi \in E_*$. In this work, we only consider weak$^*$ continuous actions in the case of von Neumann algebras, on which the notions of weakly, $\sigma$-weakly and $\sigma$-strongly continuous actions coincide since orbits are uniformly norm-bounded.

\vspace{.2cm}

For the subgroup generated by $v \in \bbR^n$, the directional derivation is defined by
\begin{equation}
\label{eq:nabla}
 \nabla_v x \;=\; \frac{-1}{2\pi}\,\lim_{\epsilon\to 0} \,\frac{\Diffaction_{v\epsilon}(x) - x}{\epsilon}
\end{equation}
with the domain given by all elements for which the limit exists in norm respectively in the weak$^*$ sense. If $E$ is a Banach algebra and $\Diffaction$ an algebra homomorphism then $\nabla_v$ is a derivation. Next let us define spaces of differentiable elements. For $e_1,\ldots ,e_n$ being the standard basis of the Lie algebra of $G$, let $\nabla_1,\ldots ,\nabla_n$ be their associated derivations defined as above. Then the subspaces of the $m$-times differentiable 
$$
C^m(E,\Diffaction) \;=\; 
\big\{
x \in E: \, x \in \textup{Dom }(\nabla_{i_1}\cdots\nabla_{i_k})\; \forall\; 0 \leq k \leq m\,,\; i_1,\ldots ,i_k \in \{1,\ldots ,n\}
\big\}
\;,
$$
and the subspace of smooth elements is
$$
C^\infty(E, \Diffaction)
\;=\; \bigcap_{m\in \bbN} C^m(E,\Diffaction)
\;.
$$
It is well known that the smooth subspaces are norm-dense respectively $\sigma$-weakly dense in $E$ \cite{Bratteli86,BR1976}.
Furthermore $C^m(E,\Diffaction)$ is a Banach space in the norm
$$
\norm{x}_{\Diffaction,m} \;=\; \sum_{0 \leq \abs{j}\leq m} \norm{\nabla^j x}
\;,
$$
with multi-indices $j$ and $\nabla^j = \nabla_{j_1}\cdots\nabla_{j_k}$. The same family of norms makes $C^\infty(E,\Diffaction)$ into a Frech\'et space.
 
\vspace{.2cm}

Let now $E=\Aa$ be a $C^*$-algebra with a strongly continuous isomorphic action $\alpha$ and a densely defined, faithful, lower semi-continuous and $\alpha$-invariant trace $\Tt$. One can associate to $\Aa$ different hierarchies of differentiable elements, for one there are the above differentiable elements $C^m(\Aa)$ in the norm or uniform sense. Moreover, there is the generated von Neumann algebra $\Mm = L^\infty(\Aa, \Tt)$ with weak$^*$ continuous action $\alpha$ and differentiable elements in the weak$^*$ sense. If $\alpha$ is spatially implemented, {\sl i.e} $\alpha_t= \mathrm{Ad}_{\exp(2\pi \imath X\cdot t)}$, then the weak$^*$ differentiable elements can be characterized as those elements $a \in \Mm$ for which $[v\cdot X, a]$ is densely defined and extends to a bounded operator \cite{BR1976} for all $v\in S^n$.
The normalization of \eqref{eq:nabla} was adjusted such that
$\nabla_{v} a = -\imath [v \cdot X, a]$
in that case.

\vspace{.2cm}

Let the $C^*$-algebra $\Aa$ now be equipped with a densely defined lower semicontinuous trace $\Tt$ and a $G$-action $\alpha$ that leaves $\Tt$ invariant. Then $\alpha$ restricts to a strongly continuous isometric action on the Banach algebra $E=\Aa_\Tt$ with norm $\norm{\cdot}_\Tt$ from \eqref{eq-BanachNorm}, see Section~\ref{sec:traces}. Hence one define the Fr\'echet algebra 
$$
\Aa_{\Tt, \alpha} 
\;=\; C^\infty(\Aa_\Tt, \alpha)
$$ 
where differentiability is considered in the norm $\norm{\cdot}_\Tt$. A fairly general argument shows that $\Aa_{\Tt,\alpha}$ is dense in $\Aa$ and invariant under holomorphic functional calculus and hence all elements of $K_j(\Aa)$ can be represented by projections respectively unitaries in matrix algebras over $\Aa_{\Tt,\alpha}$ \cite{Connes81,Connes94}. The Fr\'echet algebra $\Aa_{\Tt, \alpha}$ turns out to be a natural domain for the smooth Chern character, see Section~\ref{sec:smooth_chern} below.


\vspace{.2cm}

There is yet another way to apply the above construction. Proposition~\ref{prop-LpCont} implies that the smooth elements w.r.t. the action $\LpDyn$ are dense in $L^p(\Mm)$.  One can now define the next hierarchy,  namely the non-commutative Sobolev spaces
$$
W^m_{p}(\Mm) 
\;=\; 
C^m(L^p(\Mm),\LpDyn)
\;,
$$
with norms
\begin{equation}
\label{eq-SobolevNorm}
\|x\|_{W_p^m}
\;=\;
\sum_{0 \leq \abs{j}\leq m} \norm{\nabla^j x}_p
\;.
\end{equation}
If $\Tt$ is finite, one has $W^m_{p}(\Mm) \subset W^m_{q}(\Mm)$ for all  $p>q$. Let us stress that in the non-commutative setting there is in general no Sobolev lemma, that is elements from $W^m_{p}(\Mm)$ are not necessarily norm-differentiable or embed into $\Mm$ even if $m$ is large. All these constructions also apply and will be used for differentiable elements of the crossed products $A \rtimes_\alpha G$ and $\Mm \rtimes_\alpha G$ and their associated Sobolev spaces, since the dual trace $\hat{\Tt}_\alpha$ is again $\alpha$-invariant.

\newpage 


\chapter{Besov spaces for isometric $G$-actions}
\label{sec-Besov}

Classical Sobolev, Besov and Triebel-Lizorkin spaces measure smoothness and integrability properties of functions on $\RM^n$ simultaneously. From several points of view, the Besov spaces are particularly well-behaved \cite{Tri} which is why they appear naturally in numerous contexts. Relevant for the present context is Peller's characterization of traceclass properties of Hankel operators by Besov regularity of their symbols \cite{Pel}. Furthermore, it is remarkable that in trace theorems there is no loss of Besov regularity (other than for Sobolev spaces) \cite{Tri}. Several results on classical Besov spaces have been extended to the rotation algebra, also called a quantum torus \cite{CXY,XXY}. For the bulk-boundary correspondence as described in the introduction, the regularity of $\RM^n$-actions on a von Neumann algebra and their associated crossed products algebras has to be analyzed. Many techniques from \cite{XXY} can be transposed to this case. In particular, a finite difference characterization (\`a la Zygmund) of the non-commutative Besov spaces will be proved below, as well as an extension of Peller's results \cite{Pel,Peller82} to the present context.

\section{Motivation, definition and basic properties of Besov spaces}
\label{sec-BesovDef}

While only Besov spaces for $\RM^n$-actions on non-commutative $L^p$-spaces associated to a finite von Neumann algebra will be used later on, it is natural to introduce them in the slightly more general framework of abelian actions on Banach spaces already described in Section~\ref{sec-SpecDecomp}. Hence let $E$ be a Banach space with a linear isometric $\bbR^n$-action $\Besovaction$. We assume that $\beta$ is either strongly continuous, weakly continuous or weak$^*$-continuous (w.r.t. a pre-dual $E_*$), {\it i.e.} assume that the orbits under $\beta$ are continuous in the respective topology. For brevity we call an action continuous if it is continuous in any of those senses. Recall the definition \eqref{eq-FAaction} of the (Fourier multiplier) action $\widehat{f} * x=\Besovaction_{\calF^{-1} f}(x)$ of a function  $f \in FA(\bbR^n)$  on $x\in E$. The bound $\|\widehat{f} * \cdot\,\|\leq\|\calF^{-1} f\|_1$ given in \eqref{eq-FAbound} as well as the linearity \eqref{eq-ConvRep2} in $f$ hold for any continuous action. Also for $(f_\epsilon)_{\epsilon>0}$ an approximate unit for the convolution algebra $L^1(\bbR^n)$ one has
$$
\lim_{\epsilon\to 0} \widehat{\calF f_\epsilon}*x \;= \;x\;, 
\qquad \forall \; x\in E
\;,
$$
with convergence in norm, respectively the weak($^*$)-topology (see {\sl e.g} \cite[Lemma XI.1.3]{Takesaki2003}). The modern definition of classical Besov spaces uses smooth Littlewood-Payley decompositions of Fourier multipliers from $FA(\bbR^n)$ \cite{Tri}. It will now be extended to isometric $\RM^n$-actions on a Banach space. Let $\varphi:\RM\to[0,1]$ be a smooth function with support $[-2, -2^{-1}] \cup [2^{-1}, 2]$ such that for all $x \in \bbR \setminus \{0\}$
$$
\sum_{\ScaleInd  \in \bbZ} \varphi(2^{-\ScaleInd } x) \;=\; 1
\;.
$$
Given a smooth function $\psi:\RM\to[0,1]$ with support $[-2, -2^{-1}] \cup [2^{-1}, 2]$, this normalization condition can always be achieved by setting 
$$
\varphi(x) \;= \;\psi(x)\left(\sum_{\ScaleInd  \in \bbZ} \psi(2^{-\ScaleInd } x)\right)^{-1}
\;.
$$
For any such $\varphi$ one now introduces the so-called dyadic decomposition $(W_\ScaleInd )_{\ScaleInd \in\bbN}$ by
\begin{equation}
\label{eq-WkChoice}
W_\ScaleInd (t) \;=\; \varphi(2^{-\ScaleInd } |t|) \quad \mbox{for } t\in\RM^d\,,\;\ScaleInd  > 0\;,
\qquad
W_0 \;= \;1 - \sum_{\ScaleInd >0} W_\ScaleInd 
\;
\end{equation}
with the abbreviation $\abs{t}=\norm{t}_2$ generally for both $t \in \bbZ^{n_0}\oplus \bbR^{n_1}$ and $t \in \bbT^{n_0}\oplus \bbR^{n_1}$. Note that all $W_\ScaleInd $ are smooth functions and are supported by the annulus 
$$
\{t\in\RM^n\,:\,2^{\ScaleInd -1}\leq |t|\leq 2^{\ScaleInd +1}\}
\;.
$$
In particular, $W_\ScaleInd  \in FA(\bbR^n)$.  As the $W_j$ form a partition of unity one has
\begin{equation}
\label{eq-aRingDecomp}
x 
\;=\; 
\sum_{j=0}^\infty \widehat{W_j}*x, \qquad \forall \;x\in E
\end{equation}
with convergence in norm for a strongly continuous action and in the weak(-$*$)-sense for a weak(-$*$)-continuous action.

\begin{definition}
\label{def-Besov}
Given $q\in[1,\infty)$, $s > 0$ and a dyadic decomposition $(W_\ScaleInd )_{\ScaleInd \in\bbN}$ as above, the Besov norm of $x\in E$ is defined by
$$
\norm{x}_{B^s_q(E, \Besovaction)}
\;=\;  
\Big(\sum_{\ScaleInd  \geq 0} \,2^{q s \ScaleInd }\, \lVert \widehat{W}_\ScaleInd  * x \rVert_E^q\Big)^{\frac{1}{q}}
\;,
\qquad
\norm{x}_{B^s_\infty(E, \Besovaction)}
\;=\;
\sup_{\ScaleInd \geq 0} 2^{s\ScaleInd } \norm{\widehat{W}_\ScaleInd  * x}_E
\;.
$$
The Besov space of scale $s$ and $q$ is then
$$
B^s_{q}(E,\Besovaction)
\;=\; 
\Big\{x \in E\;:\; \norm{x}_{B^s_q(E, \Besovaction)}< \infty \Big\}
\;.
$$
The group action $\Besovaction$ and thus the dimension of the $\bbR^d$-action   will be omitted for notational convenience if it is clear from the context. Furthermore let ${}^0 B^s_q(E)$ denote the subset of $B^s_q(E)$ consisting of those elements for which the sequence $(\widehat{W}_\ScaleInd *x)_{\ScaleInd \in\bbN}$ has only finitely many non-zero elements.
\end{definition}

Intuitively, the parameter $s$ measures the (fractional) order of the smoothness while $q$ fixes the $\ell^q$-norm used to measure the sequence $(2^{s \ScaleInd }\, \lVert \widehat{W}_\ScaleInd  * x \rVert_E)_{\ScaleInd \geq 0}$. Elementary properties are the inclusions
\begin{equation}
\label{eq-BesovIncl1}
B^{s'}_q(E)\;\subset\;B^s_q(E)\;,
\qquad
s\,\leq\,s'\;,
\end{equation}
and
\begin{equation}
\label{eq-BesovIncl2}
B^s_q(E)\;\subset\;B^s_{q'}(E)\;,
\qquad
q\,\leq\,q'\;,
\end{equation}
which are inherited from the corresponding inclusions of the sequential $\ell^q$-spaces.

\vspace{.2cm}

If $E$ is the classical function space $L^p(\RM^n,E)$ with values in a Banach space $E$ equipped with the $L^p$-norm and the shift action $\Besovaction$, then the above definition of $B^s_{q}(E)$ reduces to the classical vector-valued inhomogeneous Besov space $B^s_{p,q}(\RM^n,E)$, however, in a reformulation due to Peetre (see the first chapter of \cite{Tri} for a historic account). Another special case of the above construction are Besov spaces on $L^p(\bbR^n)$ associated to selfadjoint operators  \cite{JensenNakamura94}. The original definition used an equivalent Besov norm based on finite differences, similar to what is described in Section~\ref{sec-EquivBesov} below.  As in the classical theory, one can prove that the set $B^s_{q}(E)$ is independent of the choice of $\varphi$. Let us add that it is also possible to consider the cases $s< 0$ and $q<1$, but then one has to deal with distributions and quasi-Banach spaces.

\vspace{.2cm}

It is apparent from the definition that the Besov norm is tightly linked to sequence space in $E$ which for $s>0$, $q\in [1,\infty]$ are defined by 
\begin{equation}
\label{eq-SeqSpaces}
\ell^q_s(E)
\;=\; 
\big\{x \in E^\bbN\;:\;  \norm{x}_{\ell_s^q(E)}< \infty
\big\}
\;,
\end{equation}
with norms for $q\in [1,\infty)$ and $\infty$ respectively given by
$$ 
\norm{x}_{\ell_s^q(E)}
\;=\;
\Big(\sum_{j\geq 0} 2^{sjq} \norm{ x_j}^q_E\Big)^{\frac{1}{q}}
\;,
\qquad
\norm{x}_{\ell^\infty_s(E)}
\;=\;
\sup_{k\geq 0} 2^{sj} \norm{ x_j}_E
\;.
$$
These are Banach spaces with nice behavior w.r.t. complex interpolation, see \cite[Section 5.6]{BerghLofstrom76} and Appendix~\ref{app-Interpol}. The link is established via the linear map
\begin{equation}
\label{eq-EtaDef}
\eta\;:\; B^s_q(E)\;\to\;\ell_{s}^q(E)
\;,
\qquad
\eta(x)
\;=\;
\Big( \widehat{W}_{\ScaleInd } * a\Big)_{\ScaleInd \geq 0}
\;,
\end{equation}
is well-defined, injective and isometric, that is $\norm{x}_{B^s_q(E, \Besovaction)}
=\norm{\eta(x)}_{\ell^\infty_s(E)}$. This allows us to prove the following property.

\begin{proposition}
$(B^s_{q}(E),\norm{\,\cdot\,}_{B^s_q(E)})$ is a Banach space and $\norm{x}_E\leq C\norm{x}_{B^s_q(E)}$.
\end{proposition}

\noindent {\bf Proof.} 
Applying to \eqref{eq-aRingDecomp} the triangle and H\"older inequality for the sequence spaces,
$$
\norm{x}_E 
\;\leq\; 
\sum_{\ScaleInd  \geq 0} \lVert \widehat{W}_\ScaleInd  * x \rVert_E 
\;\leq\; 
(1-2^{-sp})^{-\frac{1}{p}}
\Big(\sum_{\ScaleInd  \geq 0} 2^{q s \ScaleInd } \lVert \widehat{W}_\ScaleInd  * x \rVert_E^q\Big)^{\frac{1}{q}} 
\;=\;C\, \norm{x}_{B^s_q(E)}
\;
$$ 
for $1 = \frac{1}{p}+\frac{1}{q}$ (note that this requires that indeed  $s>0$). Hence $\norm{x}_{B^s_q(E)}=0$ implies $x=0$. Next let us use the isometric map $\eta$ defined in \eqref{eq-EtaDef}. Hence the triangle inequality for $\norm{\,\cdot\,}_{B^s_q(E)}$ is nothing but the Minkovski inequality in $\ell^q_s( E)$. Completeness is now checked similarly as in \cite[Proposition~3.3]{XXY}. If $(x_n)_{n\geq 1}$ is a Cauchy-sequence in $B^s_{q}(E)$, then $(\eta(x_n))_{n\geq 1}$ is a Cauchy sequence in the Banach space $\ell^q_s(E)$. Hence it is convergent to some $y=(y_\ScaleInd )_{\ScaleInd \geq 0}\in\ell^q_s( E)$. Thus $E\mbox{-}\lim_n \widehat{W}_\ScaleInd * x_n=y_\ScaleInd $ for any $\ScaleInd \geq 0$. Now set
$$
x\;=\;\sum_{\ScaleInd \geq 0} y_\ScaleInd 
\;.
$$
This sum is norm-convergent in $E$ since $(y_\ScaleInd )_{\ScaleInd \geq 0}\in \ell^q_s(E)\subset \ell^q_0(E)$ for $s\geq 0$.
Since the multipliers are norm-bounded $\norm{\widehat{W_j}\,*\,\cdot}\leq 1$ and thus continuous, one also has
\begin{align*}
\widehat{W}_\ScaleInd  * x 
&
\,=\,
\sum_{i \geq 0} \widehat{W}_\ScaleInd  * y_i
\\
&
\,=\,
\lim_{n\to\infty} \sum_{i\geq 0} \widehat{W}_\ScaleInd   *( \widehat{W}_i*x_n )
\\
&
\,=\, 
\lim_{n\to\infty} \sum_{i\geq 0} (W_\ScaleInd W_i)^\wedge\,*x_n 
\\
&
\,=\, 
\lim_{n\to\infty} \widehat{W}_\ScaleInd  * x_n 
\\
&
\,=\, 
y_\ScaleInd \;.
\end{align*}
Therefore $\eta(x)=y$. As $\eta$ is isometric, $x_n \to x$ in $B^s_q(E)$-norm. 
\hfill $\Box$

\vspace{.2cm}

The next property of the Besov spaces is their naturality w.r.t. to complex interpolation. The definition of an interpolation couple $(E_0,E_1)$ of Banach spaces and the corresponding interpolation spaces $(E_0,E_1)_\theta$ is recalled in Appendix~\ref{app-Interpol}.

\begin{proposition}
\label{prop-BesovInterpol}
Let $(E_0,E_1)$ be an interpolation couple of Banach spaces and continuous linear isometric $\bbR^n$-actions $\beta^i: \bbR^n \times E_i \to E_i$. Assume that $\beta^0$ is strongly continuous and $\beta^i_t(x) = \beta^{1-i}_t(x)$ for all $x \in E_0 \cap E_1$. Then $\beta^0$ extends to a strongly continuous action on $E_\theta:=(E_0,E_1)_\theta$ for $\theta\in(0,1)$.
For parameters $s_0,s_1 > 0$ and $q_0,q_1\in[1,\infty]$, one has for the interpolation space of the Besov spaces of order $\theta\in(0,1)$
$$
\big(B^{s_0}_{q_0}(E_0),B^{s_1}_{q_1}(E_1)\big)_\theta
\;=\;
B^{s}_{q}\big(E_\theta\big)
\;,
$$
where $s=(1-\theta)s_0+\theta s_1$ and $\frac{1}{q}=\frac{1-\theta}{q_0} + \frac{\theta}{q_1}$. Moreover, for $x\in B^{s_0}_{q_0}(E_0)\cap B^{s_1}_{q_1}(E_1)$,
$$
\big\|x\big\|_{B^{s}_{q}((E_0,E_1)_\theta)}
\;\leq\;
\big\|x\big\|_{B^{s_0}_{q_0}(E_0)}^{1-\theta}\,
\big\|x\big\|_{B^{s_1}_{q_1}(E_1)}^{\theta}
\;.
$$
\end{proposition}

\noindent {\bf Proof.} 
By interpolation $\beta^i_t=\beta^{1-i}_t$ densely defines a bounded map $\beta^\theta: E_\theta \to E_\theta$ and from the definition of the interpolation norm (see Appendix~\ref{app-Interpol}) on $E_\theta$ it is clear that $\beta^\theta$ is an isometry. Since $(\beta^\theta_t \circ \beta^\theta_s)(x) = \beta^\theta_{s+t}(x)$ holds for all $x\in E_0 \cap E_1$ continuity implies that $\beta^\theta$ is again an $\bbR^n$-action.  Since 
$$
\lim_{t \to 0} \norm{\beta^0_t(x)-x}_{E_\theta} 
\;\leq\; 
\lim_{t \to 0} \norm{\beta^0_t(x)-x}_{E_0}^{1-\theta}\, \norm{\beta^0_t(x)-x}^\theta_{E_1} 
\;=\; 0
$$
holds in the dense subset $E_0 \cap E_1 \subset E_\theta$, the uniform bound $\norm{\beta^0_t -\idmap}_{E_\theta \to E_\theta} \leq 2$ implies that $\beta^\theta$ is strongly continuous. Also note that the multipliers $\widehat{W_j}*x$ for $x \in E_0\cap E_1$ do not depend on the choice of action $\beta^i$ or $\beta^\theta$, so that we will suppress it from the notation entirely.

\vspace{.1cm}

The notations and results of Appendix~\ref{app-Interpol} will be used. Let us consider the constant map $z\in S\mapsto \eta_z$ on the interpolation strip $S$ defined by
$$
\eta_z\;:\;
{}^0B^{s_0}_{q_0}(E_0)\cap {}^0B^{s_1}_{q_1}(E_1)
\;\to\;
\ell^{s_0}_{q_0}(E_0)+\ell^{s_1}_{q_1}(E_1)
\;,
\qquad
\eta_z\,=\,\eta|_{{}^0B^{s_0}_{q_0}(E_0)\,\cap\, {}^0\,B^{s_1}_{q_1}(E_1)}
\;,
$$
with $\eta$ the embedding defined in \eqref{eq-EtaDef} for either of $E_0$ or $E_1$ since they coincide on the intersection. It satisfies all the assumptions of Theorem~\ref{theo-Interpol} and therefore $\eta_\theta$ extends to a bounded operator from the interpolation space $(B^{s_0}_{q_0}(E_0), B^{s_1}_{q_1}(E_1))_\theta$ to $(\ell^{s_0}_{q_0}(E_0),\ell^{s_1}_{q_1}(E_1))_\theta =\ell^{s}_{q}((E_0,E_1)_\theta)$, the latter due to \eqref{eq-InterpollittleLp}. Writing out the norm in this latter space this implies the norm bound and thus that 
$$
(B^{s_0}_{q_0}(E_0), B^{s_1}_{q_1}(E_1))_\theta
\;\subset\;  
B^{s}_{q}((E_0,E_1)_\theta)
\;.
$$ 
It remains to prove the inverse inclusion. For this purpose, let us consider the map
$$
\tilde{\eta}\;:\; B^s_q(E)\;\to\;\ell_{s}^q(E)
\;,
\qquad
\tilde{\eta}(x)
\;=\;
\Big( 
(\delta_{\ScaleInd \not=0}\,\widehat{W}_{\ScaleInd -1}+\widehat{W}_{\ScaleInd }+\widehat{W}_{\ScaleInd +1}) * x\Big)_{\ScaleInd \geq 0}
\;,
$$
which is well-defined, injective and bounded with a norm bounded by $3$. Furthermore, let us introduce
$$
\psi\;:\; \ell_{s}^q(E) \;\to\;B^s_q(E)
\;,
\qquad
\psi\Big((x_\ScaleInd )_{\ScaleInd \geq 0}\Big)\;=\; 
\sum_{\ScaleInd \geq 0} \widehat{W}_\ScaleInd *x_\ScaleInd 
\;,
$$
for $E = E_0,E_1, E_\theta$ which is surjective (but clearly not injective). To prove that its image really lies in $B^s_q(E)$ and that $\psi$ is bounded, let us recall that for any scaled function $\phi_\ScaleInd (t)=\phi( 2^{-\ScaleInd } t)$ one has $\|\Ff^{-1} \phi_\ScaleInd \|_1=\|\Ff^{-1} \phi_1\|_1$ for all $\ScaleInd $. Using \eqref{eq-FAbound} this readily allows to show that $j$ is bounded. Moreover, one has $\psi \circ \tilde{\eta} = \idmap$ on ${}^0B^{s_0}_{q_0}(E_0)\cap {}^0B^{s_1}_{q_1}(E_1)$  because $W_\ScaleInd (\delta_{\ScaleInd \not=0}\,W_{\ScaleInd -1}+W_\ScaleInd +W_{\ScaleInd +1})=W_\ScaleInd $ and due to \eqref{eq-aRingDecomp}. Now construct the interpolation maps $\tilde{\eta}_\theta$ and $\psi_\theta$ as above and extend them to bounded maps from $i_\theta:(B^{s_0}_{q_0}(E_0), B^{s_1}_{q_1}(E_1))_\theta\to\ell_{s}^q((E_0,E_1)_\theta)$ and $\psi_\theta:\ell_{s}^q((E_0,E_1)_\theta)\to(B^{s_0}_{q_0}(E_0), B^{s_1}_{q_1}(E_1))_\theta$. Moreover, $\psi_\theta \circ \tilde{\eta}_\theta = \idmap$. Now let $x\in B^{s}_{q}\big((E_0,E_1)_\theta\big)$ then $\tilde{\eta}(x)\in\ell_{s}^q((E_0,E_1)_\theta)$ so that also $\psi_\theta \circ \tilde{\eta}(x)\in  (B^{s_0}_{q_0}(E_0), B^{s_1}_{q_1}(E_1))_\theta$. But $\psi_\theta \circ \tilde{\eta}(x)=x$ which implies the desired inclusion.
\hfill $\Box$

\vspace{.2cm}

In this work, the Banach space $E$ used in the above constructions will generally be one of the non-commutative $L^p$-spaces $L^p(\Mm)$ associated to a von Neumann algebra $\Mm$ with s.n.f. trace $\Tt$, so the norm is $\|\,.\,\|_E=\|\,.\,\|_p$. The $G$-action $\Besovaction$ on $\Mm$ can then be extended continuously and isometrically to $L^p(\Mm)$, see Proposition~\ref{prop-LpCont}. The Besov spaces associated to these Banach spaces $E=L^p(\Mm)$ are denoted by 
\begin{equation}
\label{eq-LpBesovDef}
B_{p,q}^s(\Mm)\;=\;B^s_q(L^p(\Mm), \LpDyn)
\;,
\end{equation}
and the Besov norm by $\|a\|_{B_{p,q}^s}$. If $\Tt$ is a finite trace, then $\|a\|_{p'}=\|\one \, a\|_{p'}\leq\|a\|_{p}$ for $p'\leq p$, so that, on top of  \eqref{eq-BesovIncl1} and \eqref{eq-BesovIncl2},
\begin{equation}
\label{eq-BesovIncl3}
B_{p,q}^s(\Mm)\;\subset\;B_{p',q}^s(\Mm)\;,
\qquad
p'\leq p
\;.
\end{equation}
Moreover, Proposition~\ref{prop-BesovInterpol} combined with Example~1 in Appendix~\ref{app-Interpol} implies
\begin{equation}
\label{eq-LpBesovInterpol}
\big(B^{s_0}_{p_0,q_0}(\Mm),B^{s_1}_{p_1,q_1}(\Mm)\big)_\theta
\;=\;
B^{s}_{p,q}(\Mm)
\;,
\end{equation}
where $s=(1-\theta)s_0+\theta s_1$, $\frac{1}{p}=\frac{1-\theta}{p_0} + \frac{\theta}{p_1}$ and $\frac{1}{q}=\frac{1-\theta}{q_0} + \frac{\theta}{q_1}$. Furthermore, for $a$ in the intersection of $B^{s_0}_{p_0,q_0}(\Mm)$ and $B^{s_1}_{p_1,q_1}(\Mm)$, 
\begin{equation}
\label{eq-LpBesovInterpolBound}
\big\|a\big\|_{B^{s}_{p,q}}
\;\leq\;
\big\|a\big\|_{B^{s_0}_{p_0,q_0}}^{1-\theta}
\big\|a\big\|_{B^{s_1}_{p_1,q_1}}^\theta
\;.
\end{equation}

\vspace{.2cm}

The classical Besov spaces are closely connected to fractional Sobolev spaces. In particular, setting $W_{p,s}(\bbR^n) = B^s_{p,p}(\bbR^n)$ for $p\notin \bbN$ is one possible definition of these spaces which is well-behaved w.r.t. interpolation \cite{Tri}. In the Hilbert-space $L^2(\Mm)$ where spectral decomposition is possible, one can check directly that this definition makes sense:

\begin{proposition}
\label{prop-FractionalSobolev}
The fractional Sobolev norm 
\begin{equation}
\label{eq-FractionalSobolev}
\|a\|_{2,s}
\;=\;
\left(\int_{\sigma(\GenGNS)} (1+\abs{\lambda}^{2s})\, \norm{a_\lambda}^2 \mu(\difd \lambda)\right)^{\frac{1}{2}}
\;.
\end{equation}
is equivalent to the norm on $B^s_{2,2}(\Mm)$.
\end{proposition}

\noindent {\bf Proof.} 
As $L^2(\Mm)$ is a Hilbert space the action is generated by the self-adjoint commuting operators $\GenGNS=(\GenGNS_1,\ldots ,\GenGNS_n)$ and by functional calculus and \eqref{eq-FourierMultAct}
$$
\widehat{W}_\ScaleInd *a
\;=\; 
W_\ScaleInd (\GenGNS) a
\;=\;
\int^\oplus_{\sigma_\Besovaction(a)}\mu(\difd \lambda)\;W_\ScaleInd (\lambda)\,a_\lambda
\;.
$$
Hence
$$
\sum_{\ScaleInd \geq 0} 2^{2\ScaleInd s} \|\widehat{W}_\ScaleInd *a\|_2^2
\;=\; 
\int_{\sigma(\GenGNS)} \sum_{\ScaleInd \geq 0} 2^{2\ScaleInd s} \|W_\ScaleInd (\lambda) a_\lambda\|^2 \,
\mu(\difd \lambda)
\;.
$$
For all $\ScaleInd \geq 1$ and $\abs{\lambda} \in (2^{\ScaleInd },2^{\ScaleInd +1})$, one has 
$$
W_i(\lambda)\,=\,0\;\;\;\mbox{ for }i\not\in\{\ScaleInd ,\ScaleInd +1\}\;,
\qquad
W_\ScaleInd (\lambda)+W_{\ScaleInd +1}(\lambda)
\,=\,1
\;.
$$
Hence, as $s\geq 0$,
\begin{align*}
\frac{1}{2} 
&
\;\leq\; 
W_\ScaleInd (\lambda)^2+W_{\ScaleInd +1}(\lambda)^2
\\
&
\;\leq\;
W_\ScaleInd (\lambda)^2+ 2^{2s} W_{\ScaleInd +1}(\lambda)^2 
\\
&
\;\leq\;
2^{2s} (W_\ScaleInd (\lambda)^2+ W_{\ScaleInd +1}(\lambda)^2 )
\\
&
\;\leq\;
2^{2s} 
\;.
\end{align*}
From this one can conclude
$$
\frac{1}{2^{2s+1}} \abs{\lambda}^{2s} 
\;\leq \;
\sum^\infty_{\ScaleInd  \geq 1} 2^{2\ScaleInd s} \abs{W_\ScaleInd (\lambda)}^2 
\;\leq\; 
2^{2s} \abs{\lambda}^{2s}
\;,
$$
for all $\lambda \in \bbR$ with $\abs{\lambda} > 1$. The definitions of the norms now allow to conclude.
\hfill $\Box$

\vspace{.2cm}

Let us also discuss the special case of periodic actions, since they will be important in the following. A bit more generally, let $E$ be a Banach space equipped with a continuous $G$-action $\beta$ where $G=\bbT^{n_0} \times \bbR^{n_1}$. Decompose $\beta = \beta^{(0)}\times \beta^{(1)}$ with $\beta^{(0)}: \bbT^{n_0} \times E \to E$ and $\beta^{(1)}: \bbR^{n_1} \times E \to E$ and let $\tilde{\beta}^{(0)}$, $\tilde{\beta}$ be the $\bbR^{n_0}$ respectively  $\bbR^{n_0+n_1}$-action obtained by identifying $\bbT \simeq [0,1)/\sim$. One can consider  Fourier multipliers on $E$ w.r.t. any of these actions, and here we are primarily interested in comparing $\tilde{\beta}$ and $\beta$, {\it i.e.} for $f \in FA(\bbR^n)$ 
$$
\widehat{f}*x
\;=\;
\tilde{\beta}_{\calF^{-1}f}(x) 
\;=\; 
\int_{\bbR^{n_0}\times \bbR^{n_1}} \calF^{-1}(f)(t) \,\tilde{\beta}_{t}(x)\,\difd t
\;,
$$
and for $g \in FA(\hat{G}) = \calF L^1(G)$
$$
\beta_{\calF^{-1}g}(x) 
\;=\; 
\int_{\bbT^{n_0}\times \bbR^{n_2}} \calF^{-1}(g)(t) \,\tilde{\beta}_{t}(x)\, \difd t
$$
with their respective Fourier transforms.

\vspace{.2cm}

Since the dual group $\bbZ^{n_0}$ of $\bbT^{n_0}$ is discrete, one can decompose $E$ into a family of closed subspaces $(E_k)_{k\in \bbZ^{n_0}}$ where $E_k$ is the image of $E$ under the (idempotent) coefficient map defined by
\begin{equation}
\label{eq:fourier_coeff}
x\, \in\, E \; \mapsto \;x_k \;=\; \int_{\bbT^{n_0}} \langle k, t\rangle \, \beta^{(0)}_t(x)\, \difd{t} \, \in\, E
\;.
\end{equation}
These coefficients allow to represent elements as generalized Fourier series.

\begin{lemma}
\label{lem:fourier_series}
The Fourier coefficients $(x_k)_{k\in \bbZ^{n_0}}$ of $x\in E$ have the properties
\begin{enumerate}
\item[{\rm (i)}] $x_k \in E$ and $\beta^{(0)}_{t_0}( x_k) = \overline{\langle k, t_0\rangle}\, x_k$ for all $k\in \bbZ^{n_0}$.

\item[{\rm (ii)}] For $(W^{(0)}_j)_{j\in \bbN}$ a dyadic decomposition of $\bbR^{n_0}$ with the properties \eqref{eq-WkChoice}, one has
$$
x\; =\; 
\sum_{j\in \bbN}\;\sum_{k\in \mathrm{supp}(W_j^{(0)})}\!  W^{(0)}_j(k)\, x_k
\;,
$$
with convergence in norm or in the weak($^*$)-sense depending on the continuity of $\beta^{(0)}$. 
\end{enumerate}
Moreover,  the properties {\rm (i)} and {\rm (ii)} uniquely determine the Fourier coefficients.
\end{lemma}

\noindent{\bf Proof.}
Property (i) clearly holds and for (ii) one computes with the Poisson summation formula
\begin{align*} 
\tilde{\beta}^{(0)}_{(\calF^{-1}W^{(0)}_j)}(x) 
&
\;=\;  \int_{\bbR^{n_0}}\left( \calF^{-1}W^{(0)}_j\right)(t_0) \, \tilde{\beta}^{(0)}_{t_0} (x) \, \difd{t_0} \\
&
\;=\; \sum_{k\in \bbZ^{n_0}}\int_{\bbT^{n_0}}\left( \calF^{-1}W^{(0)}_j\right)(k+t_0)\,  \beta^{(0)}_{t_0} (x) \, \difd{t_0} \\
&
\;=\; \sum_{k\in \bbZ^{n_0}}\int_{\bbT^{n_0}}W^{(0)}_j(k) \, \langle k,t_0\rangle \,  \beta^{(0)}_{t_0} (x) \, \difd{t_0} 
\\
&
\;=\;
\sum_{k\in \mathrm{supp}(W_j^{(0)})} \! W^{(0)}_j(k) \,x_k
\end{align*}
and thus the resummation 
$$
x
\;=\; 
\sum_{j\in \bbN} \left(\tilde{\beta}^{(0)}_{(\calF^{-1}W^{(0)}_j)}(x)\right) 
\;=\; 
\sum_{j\in \bbN}\sum_{k\in \mathrm{supp}(W_j^{(0)})} W^{(0)}_j(k)\, x_k
$$
converges in norm respectively in the weak($^*$)-sense since $W^{(0)}_j$ is an approximate unit of Fourier multipliers. For uniqueness one just notes that when a collection $(\tilde{x}_k)_{k\in \bbZ^{n_0}}$ satisfies the two properties, the Fourier coefficients of $x$ can be computed with \eqref{eq:fourier_coeff} and the orthogonality of the characters $\langle k, \cdot\rangle$ implies $\tilde{x}_k=x_k$ for all $k\in \bbZ^{n_0}$.
\hfill $\Box$

\vspace{.2cm}

One can therefore represent any $x\in E$ uniquely as a Fourier series
$$
x
\;=\; 
\sum_{k \in \bbZ^{n_0}} x_k
\;,
$$
where the sum is to be understood in the sense of Lemma~\ref{lem:fourier_series}, since it does not necessarily converge. Clearly one can also use other regularizations such as Ces\`{a}ro summation, however, the dyadic version above is convenient and sufficient for the present purposes.

\vspace{.2cm}

Since the actions commute, the subspaces $E_k$ are $\beta^{(1)}$-invariant and the action is thus partially diagonalized in the sense that
$$
\beta_{(t_0,t_1)}\left(\sum_{k\in \bbZ^{n_0}} x_k\right) 
\;=\; 
\sum_{k\in \bbZ^{n_0}}\overline{\langle k, t_0\rangle}\, \beta^{(1)}_{t_1}( x_k)
\;.
$$
One can now define Besov spaces for $E$ and $\beta$ in terms of a dyadic decomposition w.r.t. either of $FA(\bbR^n)$ or of $FA(\hat{G})$. That both approaches lead to the same spaces is implied by the following lemma:

\begin{lemma}
\label{lemma:periodic_mult}
Let $f \in \scrS(\bbR^{n_0+n_1})$ be a Schwartz function and denote its restriction to $\hat{G}$ by $f^{r}: \bbZ^{n_0}\times \bbR^{n_1} \to \bbC$. Then 
$$
\tilde{\beta}_{\calF^{-1}f}(x)
\;=\;\beta_{\calF^{-1}f^r}(x)
\;,
$$
and both are given by 
$$
\beta_{\calF^{-1}f^r}(x) 
\;=\; 
\sum_{k_0 \in \bbZ^{n_0}} \int_{\bbR^{n_{1}}}\left( \int_{\bbR^{n_1}} f(k_0, k_1)\, \langle k_1, t_1\rangle\, \difd{k_1}\right) 
\beta^{(1)}_{t_1} (x_{k_0}) \, \difd{t_1}
\;.
$$
In particular, for $n_1 = 0$ the action of the multiplier is given by the simple form
$$
\tilde{\beta}_{\calF^{-1}f}(x) 
\;=\; 
\beta_{\calF^{-1}f^r}\left(\sum_{k_0\in \bbZ^{n_0}} x_{k_0}\right) 
\;=\; 
\sum_{k_0\in \bbZ^{n_0}} f(k_0)\, x_{k_0}
\;.
$$ 
\end{lemma}

\noindent{\bf Proof.}
By definition,
$$
\tilde{\beta}_{\calF^{-1}f}(x) 
\;=\; 
\int_{\bbR^{n_0}\times \bbR^{n_1}} \left(\int_{\bbR^{n_0}\times \bbR^{n_1}} f(k)\, \langle k,t\rangle\, \difd{k}\right) 
\tilde{\beta}_{t}(x)\,\difd t
$$
and
$$
\beta_{\calF^{-1}f^r}(x) 
\;=\; 
\int_{\bbT^{n_0}\times \bbR^{n_1}} \left(\sum_{k_0\in \bbZ^{n_0}} \int_{\bbR^{n_1}} f^r(k_0,k_1) \langle k_0,t_0\rangle\,\langle k_1,t_1\rangle \difd{k_1}\right) 
\beta_{(t_0,t_1)}(x)\,\difd t
\;.
$$
If the Fourier series $(x_{k_0})_{k_0\in \bbZ^{n_0}}$  of $x$ has only finitely many non-vanishing terms, one has
$$
\tilde{\beta}_{(t_0,t_1)}(x)
\;=\;
\beta_{(t_0,t_1)}(x) 
\;=\; 
\sum_{k_0\in \bbZ^{n_0}} \, \overline{\langle k_0, t_0 \rangle} \,
\beta^{(1)}_{t_1}(x_{k_0})
\;,
$$
Replacing the Fourier inversion relation 
$$
\int_{\bbR^{n_0}} \langle k_0, t_0\rangle \overline{\langle p_0, t_0\rangle} \difd{t_0}\;=\; \delta(k_0-p_0)
$$ 
in the formula for $\tilde{\beta}_{\calF^{-1}f}(x) $ and the orthogonality relation 
$$
\int_{\bbT^{n_0}} \sum_{k_0\in \bbZ^{n_0}} \langle k_0, t_0\rangle \overline{\langle p_0, t_0\rangle} \difd{t_0}
\;=\; \delta_{k_0,p_0}
$$  
into the formula for $\beta_{\calF^{-1}f^r}(x)$,  both expressions reduce to the desired form. For arbitrary $a$ one uses the convergent resummation.
\hfill $\Box$

\vspace{.2cm}

Hence the Besov spaces for a Banach space $E$ and $G$-action are unambiguously defined by $B^s_q(E, \tilde{\beta})$. In particular, for $n_1=0$, one obtains
$$
\norm{x}_{B^s_q(E)}
\;=\; 
\Big\|\sum_{k\in \bbZ^n} x_k\Big\|_{B^s_q(E)} 
\;=\; 
\left(\sum_{j\in \bbN} 2^{qsj} \Big\|\sum_{k \in \bbZ^n} W_j(k) x_k \Big\|_E\right)^{\frac{1}{q}}
\;.
$$
For $E= L^p(\bbT^n)$ with the translation action, one recovers the classical periodic Besov spaces.

\section{Finite difference norm for Besov spaces}
\label{sec-EquivBesov}

As already hinted at after Definition~\ref{def-Besov}, there is an equivalent norm based on finite differences that also enter the definition of Zygmund spaces. Here such a characterization of the Besov spaces $B^s_q(E)$ is proved by adapting the arguments in \cite[Theorem 3.16]{XXY} to the present setting. The finite difference operator $\Delta_t: E \to E$ is introduced by 
$$
\Delta_t(x)\;=\; 
\Besovaction_t(x)-x
\;.
$$
Then the $N$-th modulus of smoothness $\omega_E^N: E \times \bbR_{>} \to \bbR_\geq$ is defined by
\begin{equation}
\label{eq:mod_sm}
\omega_E^N(x,t) 
\;=\; 
\sup_{\abs{r} \leq t} \norm{\Delta_r^N(x)}_E
\;.
\end{equation}

\begin{theorem}
\label{theo-EqNorm}
For $q <\infty$ and any integer $N>s>0$, the Besov norm $\norm {\,\cdot\,}_{{B}^s_q(E)}$ is equivalent to the norm
$$
\norm {x}_{\widetilde{B}^s_q(E)}
\;=\; 
\norm{x}_E \,+ \,\left(\int_{[0,1]} t^{-sq}\; \omega_E^N(x,t)^q\; \frac{\difd t}{t}\right)^{\frac{1}{q}}
\;.
$$
For $q=\infty$ and $N > s > 0$ it is equivalent to 
$$
\norm {x}_{\widetilde{B}^s_\infty(E)}
\;=\; 
\norm{x}_E \,+ \, \sup_{t\in [0,1]} \left( t^{-s}\; \omega_E^N(x,t)\right)
\;.
$$
\end{theorem}

Several technical preparations are needed for the proof. The first are merely algebraic properties of the Fourier transform.

\begin{lemma}
\label{lem:beta_multiplier}
For $f \in FA(\bbR^n)$ and $x\in E$, 
$$
\Besovaction_t(\widehat{f}*x) 
\;=\; 
(e^{2 \pi \imath t \cdot} f)^\wedge*x
\;,
\qquad
\Delta_t^n(\widehat{f}*x) 
\;=\; 
\left((e^{2 \pi\imath t \cdot}-1)^n f \right)^\wedge *x
\;.
$$
\end{lemma}

\begin{lemma}
\label{lem-IWbound}
Denote by $I^s:\RM^n\to\RM_\geq $ the function $I^s(\lambda) = \abs{\lambda}^s$.
For $s >0$ there is a constant $K$ such that, uniformly in $\ScaleInd  >0$,
$$
\norm{(I^s W_\ScaleInd )^\wedge * x }_E 
\;\leq \;
K\, 2^{s \ScaleInd }\, \big\|\widehat{W}_\ScaleInd  * x \big\|_E
\;.
$$
\end{lemma}

\noindent {\bf Proof.} 
Since $I^s\varphi$ is smooth and compactly supported, one has $\norm{\calF^{-1}( I^s \varphi )}_1 = C < \infty$. Rescaling thus gives
\begin{equation}
\label{eq-FourierL1Est}
\norm{\calF^{-1}( I^s W_\ScaleInd  )}_1 \;=\; C\, 2^{s \ScaleInd }
\;.
\end{equation}
Moreover, the support condition on $\varphi$ and $W_\ScaleInd $ combined with the partition of unity  property implies for all $\ScaleInd \geq2$ that
$$
I^s W_\ScaleInd  
\;=\;
I^s (W_{\ScaleInd -1}+W_{\ScaleInd }+W_{\ScaleInd +1})W_\ScaleInd 
\;.
$$
Thus using \eqref{eq-ConvRep2} and \eqref{eq-FAbound}
$$
\norm{(I^s W_\ScaleInd )^\wedge * x }_E 
\;\leq\;
\norm{\Ff^{-1}(I^s (W_{\ScaleInd -1} +W_\ScaleInd +W_{\ScaleInd +1}))}_1 \big\|\widehat{W}_\ScaleInd * x \big\|_E 
\;,
$$
so that the bound \eqref{eq-FourierL1Est} concludes the proof.
\hfill $\Box$

\vspace{.2cm}

\noindent {\bf Proof} of Theorem~\ref{theo-EqNorm}. 
First let us assume $q < \infty$ and note that due to the monotonicity of $\omega_E(x,t)$ and $t^{-sq}$ in $t$, there exist two constants $C_{1,2}$ such that
\begin{align}
C_1\, \sum_{\ScaleInd\geq 0} \,2^{\ScaleInd sq}\, 
\omega_E^N(x,2^{-\ScaleInd-1})^q
&
\;  \leq\; 
\int_{[0,1]} t^{-sq} \,\omega_E^N(x,t)^q \,\frac{\difd t}{t} 
\nonumber
\\
&
\;\leq \;
C_2  \, \sum_{\ScaleInd\geq 0} \,2^{\ScaleInd sq} \,\omega_E^N(x,2^{-\ScaleInd})^q
\;,
\label{eq:disc}
\end{align} 
namely one can discretize the integral into dyadic intervals. 

\vspace{.1cm}

Next define $d^N_r(\lambda) = (e^{-\imath r \cdot \lambda}-1)^N$ such that formally 
$$
\Delta^N_t(x) 
\;=\; 
\widehat{d^N_t}* x
\;.
$$
Further choose a Schwartz function $h$ with $h=1$ on the open ball $B_2(0)$ and $h=0$ on $\bbR^n \setminus B_4(0)$. For $\abs{r}\leq 1$, $i\in \bbN$ and $-i \leq \ScaleInd  \leq 0$, one can then write
\begin{align*}
d^N_{2^{-i}r}(\lambda) \,{W}_{i+\ScaleInd }(\lambda)
&
\;=\;
 2^{\ScaleInd N} \,\frac{d^N_{2^{-i}r}\,(\lambda)}{\abs{2^{-i} r \lambda}^N}\, h(2^{-i}\lambda)\, \abs{2^{-i-\ScaleInd }r \lambda}^N\, {W}_{i+\ScaleInd }(\lambda) 
\\
&
\;=\; 
2^{\ScaleInd N}\, \eta_{i,r}(\lambda)\, \rho_{i+\ScaleInd ,r}(\lambda)
\;,
\end{align*}
with 
$$
\eta_{i,r}(\lambda) \;= \;\frac{d^N_{2^{-i}r}(\lambda)}{\abs{2^{-i} r \lambda}^N}\, h(2^{-i}\lambda)\;,
\qquad
\rho_{i,r}(\lambda) 
\;=\;2^{-i}|r|^NI^N(\lambda){W}_{i}(\lambda)
\;.
$$
Since $\eta_{i,r}$ is smooth and has compact support one has
$$
C_3\;=\;\sup_{i\geq 1}\,\sup_{\abs{r} \leq 1} \norm{\calF^{-1} \eta_{i,r}}_1 \; <\; \infty
\;,
$$
and therefore due to \eqref{eq-FAbound}
\begin{align*}
\big\|\Delta^N_{2^{-i}r}( \widehat{W}_{i+\ScaleInd }\,*x)\big\|_E
&
\;=\;
\big\|(d^N_{2^{-i}r} W_{i+\ScaleInd })^\wedge *x\big\|_E 
\\
&
\;=\;
2^{\ScaleInd N}\, \big\|\widehat{\eta}_{i,r}*(\widehat{ \rho}_{i+\ScaleInd ,r}*x)\big\|_E 
\\
&
\;\leq\; 
C_3\, 2^{\ScaleInd N}\, \norm{ \widehat{\rho}_{i+\ScaleInd ,r}*x}_E 
\;.
\end{align*}
Replacing the definition of $\rho_{i+\ScaleInd ,r}$ and applying Lemma~\ref{lem-IWbound}, it follows for $-i \leq \ScaleInd  \leq 0$ that
\begin{align*}
\omega_E(\widehat{W}_{i+\ScaleInd }*x, 2^{-i}) 
&
\;=\; 
\sup_{\abs{r} \leq 2^{-i}} \big\|\Delta_r^N(\widehat{W}_{i+\ScaleInd }*x)\big\|_E
\\
&
\;\leq\; 
C_3\, 2^{\ScaleInd N}\, \sup_{\abs{r} \leq 1} \,
\norm{ \widehat{\rho}_{i+\ScaleInd ,r}*x}_E
\\
& \;\leq\;
C_3\, 2^{-iN}\,\big\|
(I^N W_{i+\ScaleInd })^\wedge \,*x\big\|_E
\\
&
\;\leq\;
C_3\, 
K\, 2^{N\ScaleInd }\, \big\|\widehat{W}_{i+\ScaleInd } * x \big\|_E
\;.
\end{align*}
Independently, by the linearity and isometry of $\Besovaction$ one has for all $i\geq 0$ and $t \in \bbR^d$
$$
\omega_E(\widehat{W}_{i}*x, t) 
\;\leq\; 
2^N \,\big\|\widehat{W}_{i}*x\big\|_E
\;,
$$
because $\Delta^N_\ScaleInd $ has $2^N$ terms.  As $(W_\ScaleInd )_{\ScaleInd \geq 0}$ is a partition of unity,
$$
x\; =\; \sum_{\ScaleInd  \geq 0} \widehat{W}_\ScaleInd *x
\;,
$$
for all $x\in B^s_q(E)$ and with convergence in $E$. Substituting this expression into the r.h.s. of \eqref{eq:disc} and using the triangle inequality for the $\ell^q$-norm gives
\begin{align*}
\Big(\sum_{i=0}^\infty 2^{isq}\, & \omega_E^N(x,2^{-i})^q \Big)^{\frac{1}{q}} 
\\
&
\;\leq\; 
\Big(\sum_{i=0}^\infty  2^{isq}\Big( \sum_{\ScaleInd =-i}^\infty \omega_E^N(\widehat{W}_{i+\ScaleInd }*x,2^{-i})  \Big)^q \Big)^{\frac{1}{q}}
\\
& \leq \;\Big(\sum_{i=0}^\infty 2^{isq}\Big( \sum_{\ScaleInd =-i}^{-1} C_4\, 2^{\ScaleInd N} \big\|\widehat{W}_{i+\ScaleInd }*x\big\|_E +  \sum_{\ScaleInd =0}^\infty 2^{N}\big\|\widehat{W}_{i+\ScaleInd }*x\big\|_E \Big)^q \Big)^{\frac{1}{q}}\\
& \leq \;
  C_4 \sum_{\ScaleInd =-\infty}^{-1}  2^{\ScaleInd N}  \Big(\sum_{i=-\ScaleInd }^\infty 2^{isq} \big\|\widehat{W}_{i+\ScaleInd }*x\big\|^q_E\Big)^{\frac{1}{q}} 
+  2^N\sum_{\ScaleInd =0}^\infty   \Big(\sum_{i=0}^\infty 2^{isq} \big\|\widehat{W}_{i+\ScaleInd }*x\big\|^q_E\Big)^{\frac{1}{q}} \\
&\leq \;
  C_4 \sum_{\ScaleInd =-\infty}^{-1}  2^{\ScaleInd (N-s)}  \Big(\sum_{i=0}^\infty 2^{isq} \big\|\widehat{W}_{i}*x\big\|^q_E\Big)^{\frac{1}{q}} 
+ 2^N\sum_{\ScaleInd =0}^\infty 2^{-\ScaleInd s}  \Big(\sum_{i=0}^\infty 2^{isq} \big\|\widehat{W}_{i}*x\big\|^q_E\Big)^{\frac{1}{q}} \\
&\leq \;
C_5 \,\norm{x}_{B^s_q(E)}
\;,
\end{align*}
with $C_5$ depending on $N$. For the other inequality let us note that one can choose finitely many $r_1,\ldots ,r_L \in B_1(0)$ such that
$$
\sum_{l=1}^L \abs{d_{r_l}^N(\lambda)} \;> \;0\;, \qquad \forall\; \lambda \in B_2(0)
\;.
$$
Writing $B_2(0)= \bigcup_{l=1}^L \Omega_l$ with open intervals $\Omega_l$ such that $d_{r_l}^N >0$ on $\Omega_l$ and using smooth partitions of unity one can decompose
$$
W_1 \;= \;\varphi \;= \;\sum_{l=1}^L \phi_{l}\;,
\qquad 
\text{supp }(\phi_{l})\subset \Omega_l
\;.
$$
In the decomposition
$$
\phi_{l}(2^{-i}\lambda) 
\;=\; 
\left(\frac{\phi_{l}(2^{-i}\lambda)}{d^N_{r_l 2^{-i}}(\lambda)}\, h(2^i \lambda)\right) \,d^N_{r_l 2^{-i}}(\lambda)
$$
the first factor is again $L^1$-bounded uniformly in $i > 0$. Therefore
$$
\big\|\widehat{W}_{i}*x\big\|_E 
\;\leq \;
\sum_{l=1}^L \big\|\widehat{\phi_{l}(2^{-i}\cdot)}*x\big\|_E 
\;\leq \;
C_6 \sum_{l=1}^L \big\|\Delta_{2^{-i}r_l}^N(x)\big\|_E 
\;\leq \;
L \,C_6\, \omega_E^N(x, 2^{-i})$$
with some constant $C_6$ uniform in $i$. Thus
\begin{align*}
\Big(\sum_{i=0}^\infty 2^{qsi} \big\|\widehat{W}_i * x\big\|_E^q\Big)^{\frac{1}{q}} 
& \;\leq\; 
\big\|\widehat{W}_0 * x\big\|_E \;+ \;L \,C_6\, \Big(\sum_{i=1}^\infty 2^{qsi} \, \omega_E^N(x, 2^{-i})^q\Big)^{\frac{1}{q}}
\\
&
\;\leq\; 
C_7 
\Big(\norm{x}_E \;+ \;\Big(\int_{[0,1]} t^{-sq}\, \omega_E^N(x,t)^q\, \frac{\difd t}{t}\Big)^{\frac{1}{q}}\Big)
\end{align*}
with the last inequality again a consequence of the discretization \eqref{eq:disc}. In the case $q=\infty$, the equivalence of norms follows from similar arguments as above applied to the inequality
$$
 \sup_{\ScaleInd\geq 0} \,2^{\ScaleInd s}\, 
\omega_E^N(x,2^{-\ScaleInd-1})
\;  \leq\; 
\sup_{t\in [0,1]} \left(t^{-s} \,\omega_E^N(x,t) \right)
\;\leq \;
 \sup_{\ScaleInd\geq 0} \,2^{\ScaleInd s} \,\omega_E^N(x,2^{-\ScaleInd})
\;,
$$
which can readily be checked.
\hfill $\Box$

\begin{corollary}
\label{coro-EqNorm}
Let $\Mm$ be a von Neumann algebra with a s.n.f. trace $\Tt$ that is invariant under $\Besovaction$.
For $0 < s  < 1$, $1 \leq q \leq \infty$ and $1\leq p < \infty$, the subspaces $B^s_{p,q}(\Mm)\cap \Mm$ form a Banach-$*$-algebra with the norm $\norm{\cdot}_{B^s_{p,q}} +\norm{\cdot}$.
\end{corollary}

\noindent {\bf Proof.} It is clear that $B^s_{p,q}(\Mm)\cap \Mm$ is a Banach space. Furthermore, for $a,b\in\Mm$ 
$$
\Delta_r(ab)
\;=\;
(\Besovaction_r(a)-a)\Besovaction_r(b)\,+\,a((\Besovaction_r(b)-b)
\;=\;
\Delta_r(a)\Besovaction_r(b)\,+\,a\Delta_r(b)
\;,
$$
which implies $\norm{ab}_{\tilde{B}^s_{p,q}} \leq \norm{ab}_{\tilde{B}^s_{p,q}} \, \norm{b} + \norm{a} \norm{b}_{\tilde{B}^s_{p,q}}$. Hence it is also a Banach algebra with continuous multiplication.
\hfill $\Box$

\section{Differentiability and Besov spaces}
\label{sec-BesovSobolev}

This section presents differentiability criteria for being in the Besov space. 

\begin{proposition}
\label{prop:diff_besov0}
Let $l \in \bbN$ be an integer and $x \in C^l(E, \beta)$. For $s<l$ and $q\in[1,\infty]$, there is a constant $C>0$ such that 
$$
\norm{x}_{B^s_q(E, \Besovaction)} 
\;\leq\; 
C\,  \norm{x}_{C^l(E,\beta)}
\;.
$$
\end{proposition}

Let us begin with some preparations for the proof. From Lemma~\ref{lem:beta_multiplier} one immediately has

\begin{lemma}
\label{lem-DiffMult}
For $\varphi \in FA(\bbR^n)$ and a multi-index $m\in\NM^n$, set 
$$
\varphi_{;m}(\lambda) \;=\; (-\imath)^{|m|} \,\lambda^m \varphi(\lambda)
\;.
$$
Then $\varphi_{;m} \in FA(\bbR^n)$ and for $x \in C^{|m|}(E, \beta)$ one has
\begin{equation}
\label{eq:diff_multiplier}
\widehat{\varphi_{;m}} * x 
\;=\; 
\nabla^m (\widehat{\varphi}*x)
\;=\;
\widehat{\varphi}*(\nabla^m x)
\;.
\end{equation}
\end{lemma}

\begin{lemma}
\label{lem:diff_besov}
Let $l\in \bbN$ be an integer and $x \in C^l(E, \beta)$. There is a constant $C>0$ such that for all $j\geq 1$
$$
\big\|\widehat{W_j}*x\big\|_E 
\;\leq\; 
C\, 2^{-jl}\, \norm{x}_{C^l(E,\beta)}
\;.
$$
\end{lemma}

\noindent{\bf Proof.}
Choose non-negative functions $g_1,\ldots,g_K \in C_c^\infty(\bbR^n)$ with $\sum_{i=1}^K g_i(\lambda)=1$ on $\text{supp}(W_1)$ and such that the diameter of each $\text{supp}(g_i)$ is smaller than $\frac{1}{2^{l+1}}$. Then one can choose signs $\sigma^{(i)}_{1},\ldots,\sigma^{(i)}_{n} \in \{-1,1\}$ such that $\sum_{k=1}^n \sigma^{(i)}_{k} \lambda^l_k > 0$ on $\text{supp}(g_i)$. Next let us write 
$$
W_j(\lambda) 
\;=\; 
\frac{1}{\abs{\lambda}^l} \, W_j(\lambda)
\sum_{i=1}^K \frac{\abs{\lambda}^l}{(\sum_{k=1}^n \sigma^{(i)}_{k} \lambda^l_k)}\,g_i(2^{-j} \lambda) \,\big(\sum_{k=1}^n \sigma^{(i)}_{k} \lambda^l_k\big)
\;,
$$
and view $m_{i,l,j}(\lambda)=\frac{\abs{\lambda}^l}{(\sum_{k=1}^n \sigma^{(i)}_{k} \lambda^l_k)}\,g_i(2^{-j} \lambda)$ as a multiplier whose norm is uniformly bounded in $j$.  Since the first factor is $I_{-l}$ from Lemma~\ref{lem-IWbound} and one can use \eqref{eq-ConvRep2} to move factors to the desired order, 
\begin{align*}
\norm{\widehat{W_j}* x}_E
&
\;\leq\; 
C \,2^{-jl} 
\norm{\widehat{W_j}*\big(\sum_{i=1}^Km_{i,l,j}(\lambda) \sum_{k=1}^n \sigma^{(i)}_{k} \lambda^l_k\big)^\wedge*  x}_E
\\
&
\;\leq\; 
C' \,2^{-jl} 
\norm{\big(\sum_{k=1}^n \sigma^{(i)}_{k} \lambda^l_k W_j \big)^\wedge*  x}_E
\\
&
\;\leq\;
C' \,2^{-jl} \sum_{k=1}^n\norm{\widehat{W_j}*\nabla_k^l x}_E
\\
&
\;\leq \;
C'' \,2^{-jl} \norm{x}_{C^l(E,\beta)}
\;,
\end{align*}
where the estimate \eqref{eq-FAbound} was used several times and \eqref{eq:diff_multiplier} in the third step.
\hfill $\Box$

\vspace{.2cm}

\noindent{\bf Proof} of Proposition~\ref{prop:diff_besov0}. Replacing Lemma~\ref{lem:diff_besov} in Definition~\ref{def-Besov} immediately allows to complete the proof. 
\hfill $\Box$

\begin{lemma}
\label{lem:diff_besov2}
For $l \in \bbN$ one has the continuous inclusions 
$$
B^l_1(E,\beta) 
\;\subset \;
C^l(E,\beta) 
\;\subset\; 
B^l_\infty(E,\beta)
\;.
$$
\end{lemma}

\noindent{\bf Proof.}
For $x\in B^s(E,\beta)$ any multi-index $m \in \bbN^n$ with $\abs{m}\leq l$,  Lemma~\ref{lem-DiffMult} implies
$$
\norm{\widehat{W_j}*(\nabla^m  x)}_E 
\;=\; 
\norm{\widehat{W_{j;m}} * x}_E 
\;\leq \;
c\, 2^{\abs{m}j} \norm{\widehat{W_{j}} * x}_E 
\;,
$$
where the estimate follows by a similar scaling argument as in Lemma~\ref{lem-IWbound} since the polynomial $\lambda^m$ is homogeneous of order $\abs{m}$. Hence
$$
\norm{\nabla^m x}_E 
\;\leq\; 
\sum_{j=0}^\infty \norm{\widehat{W_j}*(\nabla^m x)}_E 
\;\leq\; 
c \norm{x}_{B^l_1(E,\beta)}
\;.
$$
The second inclusion follows directly from Lemma~\ref{lem:diff_besov} due to the definition of the norm in $B^l_\infty(E,\beta)$.
\hfill $\Box$

\vspace{.2cm}

One can also compare the Besov- and Sobolev spaces for a semi-finite von Neumann algebra:

\begin{proposition} 
\label{prop-besov-sufficient}
Let $\Mm$ be a von Neumann algebra with a s.n.f. trace $\Tt$ that is invariant under $\Besovaction$ and $1 \leq p \leq \infty$.
\begin{enumerate}
\item[{\rm (i)}] One has $W^l_{p}(\Mm) \subset B^s_{p,q}(\Mm)$ with continuous embedding for all $0 < s < l$ and $1 \leq q \leq \infty$.
\item[{\rm (ii)}] If $a \in \Mm \cap B^s_{p,p}(\Mm)$, then $a \in B^{s\frac{p}{q}}_{q,q}(\Mm)$ for all $q \geq p$ with
$$
\norm{a}_{B^{s\frac{p}{q}}_{q,q}} 
\;\leq\; 
\norm{a}_{B^{s}_{p,p}}^{\frac{p}{q}}\, \norm{a}^{1-\frac{p}{q}}_\Mm
\;.
$$
\item[{\rm (iii)}] 
For $0 < \epsilon \leq 1$ and some constant $C_p$, one has 
$$
\norm{a}_{B^{\frac{p}{p+1}}_{p+1,p+1}} 
\;\leq\; 
C_p
\big( 
\norm{a}\,+\,\norm{a}_{W^1_{p+\epsilon}}
\big)
\;,
$$
so that $W^1_{p+\epsilon}(\Mm)\cap\Mm\subset B^{\frac{p}{p+1}}_{p+1,p+1}(\Mm)\cap\Mm$.
\end{enumerate}
\end{proposition}

\noindent {\bf Proof.}  
(i) Lemma~\ref{lem:diff_besov} applied to this case gives $\|\widehat{W}_\ScaleInd *a\|_p \leq C 2^{-\ScaleInd l } \|a\|_{W^l_p}$ for some $C >0$ independent of $a$ which directly implies the claim.

\vspace{.1cm}

(ii) 
The sequence $(\widehat{W}_\ScaleInd *a)_{\ScaleInd \in \bbN}$ lies in the intersection of $\ell_s^p(L^p(\Mm))$ and $\ell^\infty_0(\Mm)=\ell^\infty_0(L^\infty(\Mm))$ as defined in \eqref{eq-SeqSpaces}. For $1-\theta = \frac{p}{q}$, one therefore gets
$$
\norm{a}_{B^{s\frac{p}{q}}_{q,q}} 
\;\leq\; 
\norm{a}_{B^{s}_{p,p}}^{\frac{p}{q}}\, \norm{a}^{1-\frac{p}{q}}_{\ell^\infty_0(\Mm)}
\;\leq\; 
\norm{a}_{B^{s}_{p,p}}^{\frac{p}{q}}\, \norm{a}^{1-\frac{p}{q}}_\Mm
\;,
$$
implying the claim.

\vspace{.1cm}

(iii) 
Applying (ii) for $s=\frac{p}{p+\epsilon}<1$ and $q=p+1$, one has 
$$
\norm{a}_{B^{\frac{p}{p+1}}_{p+1,p+1}} 
\;=\; 
\norm{a}_{B^{s\frac{p+\epsilon}{p+1}}_{p+1,p+1}} 
\;\leq\; 
\Big(
\norm{a}_{B^{s}_{p+\epsilon,p+\epsilon}} 
\Big)^{\frac{p+\epsilon}{p+1}}
\Big(\norm{a}\Big)^{1-\frac{p+\epsilon}{p+1}}
\;.
$$
Next applying part (i) shows
$$
\norm{a}_{B^{\frac{p}{p+1}}_{p+1,p+1}} 
\;\leq\; 
\Big(C\,
\norm{a}_{W^1_{p+\epsilon}} 
\Big)^{\frac{p+\epsilon}{p+1}}
\Big(\norm{a}\Big)^{1-\frac{p+\epsilon}{p+1}}
\;\leq\;
C_p\big(\norm{a}\,+\,\norm{a}_{W^1_{p+\epsilon}}\big)
\;,
$$
concluding the proof. 
\hfill $\Box$

\vspace{.2cm}

Let us finally introduce another linear space that will be useful in the remainder of the chapter. Note that any Besov space $B^s_{p,q}(\Mm)$ with $0< s <\infty$, $1 \leq p,q \leq \infty$ has a dense subspace given by those elements whose dyadic decomposition terminates
\begin{equation}
\label{eq-besovzero}
{}^0B^s_{p,q}(\Mm)
\;=\;
\{a\in L^p(\Mm)\;: \exists N \in \bbN: \,\widehat{W_j}*a = 0 \text{ for all }j > N\} 
\;
\end{equation}
since the sum $a= \sum_{j=0}^\infty \widehat{W_j}*a$ converges in Besov norm for every $a \in B^s_{p,q}(\Mm)$.

\vspace{.2cm}

It will be useful to consider approximation with bounded elements of $\Mm$ and hence let us define the space of bounded and integrable elements with compact Arveson spectrum 
\begin{equation}
\label{eq-Mmc}
\Mmc
\;=\;
\{a\in \Mm \cap L^1(\Mm)\;: \; \sigma_\alpha(a) \text{ is compact}\}
\;.
\end{equation}
In the special case $\Mm=L^\infty(\bbT)$, this gives the algebra of trigonometric polynomials while for $\Mm=L^\infty(\bbR)$ this is the union of all Paley-Wiener spaces.

\vspace{.2cm}

For any $a \in \Mmc$, the sum $a = \sum_{j=0}^\infty \widehat{W_j}*a$ has only finitely many non-vanishing terms and the identity $\alpha_t(\widehat{W_j}*a)= (W_j e^{2\pi \imath \,t\cdot})^\wedge * a$ therefore implies smoothness in $t$ w.r.t. the $L^p$-norms, thereby justifying the notation similar to the one for the Fr\'echet algebra introduced in Section~\ref{sec-DiffElements}. Let us collect some further elementary properties which will be used frequently in the following:

\begin{lemma}
\begin{enumerate}
\item[{\rm (i)}] $\Mmc$ is a $*$-algebra.
\item[{\rm (ii)}] For any $a\in \Mmc$, the function $t\in G \mapsto \alpha_t(a)$ is smooth w.r.t. the $L^p$-norms for all $1\leq p \leq \infty$. Note that this statement includes the operator norm as case $p=\infty$.
\item[{\rm (iii)}] $\Mmc$ is dense in $L^p(\Mm)$ for all $1\leq p < \infty$ and weakly dense in $\Mm$.
\item[{\rm (iv)}] $\Mmc$ is contained in all Besov spaces $B^s_{p,q}(\Mm)$ for $s>0$ and $1\leq p,q \leq \infty$ and dense for $1 \leq p < \infty$. It is also contained and dense in all Sobolev spaces $W^m_p(\Mm)$ for $1 \leq p < \infty$.
\end{enumerate} 
\end{lemma}

\noindent {\bf Proof.}
{\rm (i)} First of all, $\Mm \cap L^1(\Mm)$ is a $*$-algebra and one has the relations $\sigma_\alpha(a+b)\subset \sigma_\alpha(a) \cup \sigma_{\alpha}(b)$, $\sigma_\alpha(a^*)=-\sigma_\alpha(a)$, $\sigma_{\alpha}(ab)\subset \sigma_\alpha(a)+\sigma_\alpha(b)$ for $a,b \in \Mmc$ \cite[Corollary XI.1.8]{Takesaki2003}, which directly imply the claim.

\vspace{.1cm}

{\rm (ii)} For any $f \in C^\infty_c(\bbR^n)$ with $f(\lambda)=1$ on $\sigma_{\alpha}(a)$ and $\mathrm{supp}(f) \subset B_R(0)$ for some ball with radius $R$, one can write $a = \widehat{f}*a$ and thus have for $v\in \bbR^n$ a series expansion 
$$
\alpha_{v t}(a) 
\;=\; 
\sum_{k=0}^\infty \frac{(2\pi \imath\, t)^k}{k!} (\widehat{f_k} * a)
\;,
$$ 
with $f_k(\lambda) = f(\lambda) (\lambda\cdot v)^k$ which converges w.r.t. all $L^p$-norms, $1 \leq p \leq \infty$, due to the estimate $\norm{\calF^{-1} f_k}_1 \leq \norm{v}_2^k R^{k+1} \norm{f}_\infty$.

\vspace{.1cm}

{\rm (iii)} For $1 \leq p < \infty$ the sum $a = \sum_{j=0}^\infty \widehat{W_j}*a$ converges in $L^p$-norm for $a \in \Mm \cap L^1(\Mm)$. Hence the span of $\Mmc$ contains the dense subspace $\Mm \cap L^1(\Mm)$. For $p=\infty$ the closure of $\Mmc$ contains the weakly dense subspace $\Mm \cap L^1(\Mm)$ by the same reasoning.

\vspace{.1cm}

{\rm (iv)} The elements $b \in L^p(\Mm)$ with $\sigma_\alpha(b)$ compact are contained and dense in $B^s_{p,q}(\Mm)$ and $W^m_p(\Mm)$ since $b = \sum_{j=0}^\infty \widehat{W_j}*b$ converges in the respective norms. For $b = \widehat{f}*b$ with some fixed $f\in C_c^\infty(\bbR^n)$, one can argue as in (ii) to find constants such that $\|\widehat{f}*(a-b)\|_{B^s_{p,q}}\leq C_{s,q,f} \|{a-b}\|_p$ respectively $\|{\widehat{f}*(a-b)}\|_{W^m_{p}}\leq C_{m,f} \|{a-b}\|_p$ for all $a \in L^1(\Mm)\cap \Mm$. By density of $\Mm \cap L^1(\Mm)$ in $L^p(\Mm)$, one can choose $a$ such that $\widehat{f}*a \in \Mmc$ and $\|{\widehat{f}*a-b}\|_{B^s_{p,q}}$ respectively $\|{\widehat{f}*a-b}\|_{W^m_p}$ becomes arbitrarily small.
\hfill $\Box$

\vspace{.2cm}

Note that $\Mmc$ is not dense in the scale of spaces $B^s_{\infty,q}(\Mm)$, a fact that is again known in the classical case.

\chapter{Quantum differentiation and index theorems}
\label{chap-BreuerToep}

As already stressed in the overview in Chapter~\ref{chap-Preface}, this chapter is the mathematical core of this book. In the next Section~\ref{sec-Peller} the Hankel and Toeplitz operators associated to a $W^*$-dynamical system are introduced and then the traceclass Peller criterion is proved. Combining an $L^2$-criterion with interpolation theory, Section~\ref{sec-HigherPeller} then proves Peller criteria for higher Schatten classes. The converse and hence characterization of Schatten class properties of Hankel operators are given in Section~\ref{sec-PellerConverse}. The following Section~\ref{sec-Sobolev} then discusses how Sobolev symbols are linked to weak $L^p$-spaces of Hankel operators. Finally Section~\ref{sec-BreuerToep} introduces Chern cocycles on the Besov spaces and proves the Sobolev index theorem described in the introduction. 

\section{Besov spaces and Hankel operators}
\label{sec-Peller}

From this section on, the set-up is again a $W^*$-dynamical system $(\Mm,G,\alpha)$ with $G=\TM^{n_0}\times\RM^{n_1} $ with a von Neumann algebra $\Mm$ equipped with an $\alpha$-invariant s.n.f. trace $\Tt$. The associated $L^p$-spaces are denoted by $L^p(\Mm)$ and the extended $\RM^n$-action by $\LpDyn$ for $p\in[1,\infty)$.  Finally, extending $\alpha$ and $\LpDyn$ to an $\RM^n$-action by the identification $\bbT \simeq (0,1]/\sim$, there are also associated Besov spaces $B^s_{p,q}(\Mm)= B^s_{q}(L^p(\Mm),\LpDyn)$ as constructed in Chapter~\ref{sec-Besov}.

\vspace{.2cm}

On $\Nn = \Mm \rtimes_\alpha G$ there is the dual trace $\hat{\Tt}_\alpha$ (see Section~\ref{sec-DualTraces}) which is also semi-finite and given by the formula in Proposition~\ref{prop-DualTraceCalc} and Corollary~\ref{coro-TraceL1Calc}. The associated $L^p$-spaces are denoted by $L^p(\Nn)$ and  $\Nn=L^\infty(\Nn)$. By definition of the crossed product in the regular representation $(\pi,U)$, there is a canonical embedding $a \in \Mm \mapsto\pi(a) \in \Nn$ that allows us to consider $\Mm$ as a subalgebra of $\Nn$. The aim here is to study Hankel and Toeplitz operators in $\Nn$ with non-commutative symbols in $\Mm$. Recall that $\alpha_t$ is implemented by $n$ selfadjoint commutating generators $D=(D_1, \ldots , D_n)$ acting on the regular representation space $L^2(G,\Hh)$. An unbounded selfadjoint Dirac operator is now introduced by
\begin{equation}
\label{eq:dirac_def}
\DD
\;=\; 
\sum_{i=1}^n \gamma_i \otimes D_i 
\;=\; 
\gamma \cdot D
\;,
\end{equation}
with $\gamma_1,\ldots ,\gamma_n$ generators of the complex Clifford algebra $Cl_n$ with conventions 
\begin{equation}
\label{eq:CliffGen}
\gamma_i^*\;=\;\gamma_i\;,
\qquad
\gamma_j\gamma_i\;=\;\gamma_i\gamma_j
\;\mbox{ for } i\not=j
\;.
\end{equation}
We will always assume $Cl_n$ to be given by a concrete irreducible representation on $\CM^{2^{\lfloor n/2 \rfloor}}$. To fix this representation up to unitary equivalence it is furthermore imposed that  $\gamma_1 \cdots \gamma_n = \imath^{\frac{n-1}{2}}\one$ for odd $n$. Then $\Mm$ and $\Nn$ are inflated to $\Mm \otimes M_{2^{\lfloor n/2 \rfloor}}(\bbC)$ and $\Nn \otimes M_{2^{\lfloor n/2 \rfloor}}(\bbC)$, with extended s.n.f. traces $\Tt \otimes \Tr$ and $\hat{\Tt}_\alpha \otimes \Tr$, which will be suppressed in the notations.  All of this in turn allows to construct the following:

\begin{definition}
\label{def-HankelToep}
Given a symbol $a \in \Mm$, the associated Hankel and Toeplitz operators are
$$
H_a 
\;=\; 
\PP \, \pi(a) (\one-\PP) \;\in\; L^\infty(\Nn)
\;,
\qquad
T_a 
\;=\; 
\PP \, \pi(a) \PP  \;\in\; L^\infty(\Nn)
\;,
$$
where $\PP  = \chi_{(0,\infty]}(\DD)\in L^\infty(\Nn)$ is the Hardy projection of $\DD$.
\end{definition}

Let us note that $\DD$ is affiliated with the algebra $\Nn$ and generates an $\RM$-action. Thus Definition~\ref{def-HankelToep} is closely related the notion of Toeplitz operators associated with a flow, as studied in \cite{Lesch91,PR,Wahl10}. Particular focus will be on traceclass properties of the Hankel operators $H_a$ w.r.t. $\hat{\Tt}_\alpha$, which in turn will be crucial for index theoretic applications in Section~\ref{sec-BreuerToep}. Similar as in the classical commutative theory developed by Peller in the 1980's (see \cite{Pel}, for the vector-valued case \cite{Peller82}), such traceclass properties hold whenever the symbols are in a Besov space $B_{p,q}^s$ as defined in \eqref{eq-LpBesovDef}, with a suitable choice of the parameters $s$, $p$ and $q$. In this section, we prove the Peller criterion for $n=p=1$ and then in Section~\ref{sec-HigherPeller} for higher $n$. Their converse is stated and proved in Section~\ref{sec-PellerConverse}.

\vspace{.2cm}

As a preparation let us first show some general statements that are also needed for the case larger $n\geq 1$. For the combined spectral projections of the commuting generators, we will always use the notation 
\begin{equation}
\label{eqref:spec_generator}
P_I \; = \; \chi\big((D_1,\ldots,D_n) \, \in \, I\big)
\end{equation}
for $I \subset \bbR^n$ a Borel set.

\vspace{.2cm}

Let us note that $\Mm\cap L^1(\Mm)=L^\infty(\Mm)\cap L^1(\Mm)\subset L^2(\Mm)$ by the H\"older inequality. Hence any $a\in \Mm\cap L^1(\Mm)$ has a decomposition \eqref{eq-aFourierDecomp} into spectral subspaces of the selfadjoint operator $\GenGNS$ on $L^2(\Mm)$ which is the generator of the continuous extension $\LtwoDyn_t=V(t)$ of $\alpha$ to $L^2(\Mm)$, as given in \eqref{eq-GenDefGNS}. In this spectral decomposition, the dynamics is given by \eqref{eq-aFourierDecomp2}. Also recall that the Arveson  spectra of $\sigma_\alpha(a)$ and $\sigma_{\LpDyn}(a)$ coincide for $a\in\Mm\cap L^p(\Mm)$ by Proposition~\ref{prop-AvSpec}.

\begin{lemma}
\label{lemma:help}
Let $a\in \Mm\cap L^1(\Mm)$  and $I,J\subset \RM^n$ be closed sets.  Then
\begin{equation}
\label{eq-help}
P_I\,\pi(a)
\;=\;
P_I\,\pi(a)\,P_{I-\sigma_{\LtwoDynAv}(a)}
\;,
\qquad
\pi(a)\,P_J
\;=\;
P_{J+\sigma_{\LtwoDynAv}(a)}\,\pi(a)\,P_J
\;.
\end{equation}
If $I,J$ or $\sigma_{\LtwoDynAv}(a)\cap(I-J)$ is bounded, one has 
$$
P_I\, \pi(a)\, P_J 
\;=\; 
P_I \,
\pi\left(\int_{\sigma_{\LtwoDynAv}(a) \cap (I - J)}^\oplus \mu(\difd \lambda)\; a_\lambda \right) 
P_J
\;.
$$
\end{lemma}

\noindent {\bf Proof.}  Let us expand the element $h(D) \pi(a) g(D)$ with $h,g \in C_c^\infty(\hat{G})$ in terms of the multiplication law (\ref{form:mult})
$$
h(D)\, \pi(a) \,g(D) 
\;=\;
\int_G f(t)\, e^{2\pi\imath\, D\cdot t}\,\difd{t}
\;,
$$ 
with the convolution kernel $f: G \to \Mm$ which, using \eqref{eq-aFourierDecomp2}, is given by
\begin{align*}
\label{eq:multinfcoeff}
f(s) 
&\;=\;\int_{G} (\calF^{-1} h)(t)\,(\calF^{-1} g)(s-t)\, \pi(\LtwoDyn_t(a))\, \difd{t} \\
&\;=\; \int_{G} (\calF^{-1} h)(t)\, (\calF^{-1} g)(s-t) \int_{\sigma(\GenGNS)}^{\oplus} \mu(\difd \lambda)\; e^{ 2\pi\imath\,  \lambda\cdot t} \,a_\lambda 
\,\difd{t}\\
&\;= \;\int_{\hat{G}} \left(\int_{\sigma(\GenGNS)}^{\oplus} \mu(\difd \lambda)\;h(\dualvar + \lambda )\,g(\dualvar )\, a_\lambda  \right) e^{-2 \pi\imath \,\dualvar\cdot s}\,\difd{\dualvar }
\;.
\end{align*}
Here $\sigma(\GenGNS)$ can be replaced by the Arveson spectrum $\sigma_{\LtwoDynAv}(a)$. Let us first suppose that $I$ is bounded and choose formally $h=\chi_{I}$ and $g=\chi_{[-m,m]^n}$ the characteristic function of a cube $[-m,m]^n$ for $m > 0$. Non-vanishing contributions to the inner integral only appear for $\lambda\in\sigma_{\LtwoDynAv}(a)$ and $k + \lambda\in I$, which requires $k\in I - \sigma_{\LtwoDynAv}(a)$ independently of $m$. Approximating $h$ and $g$ by smooth functions it therefore follows that 
$$ 
P_{I} \,\pi(a) \,P_{[-m,m]^n} \;=\; P_{I} \,\pi(a) \,P_{I-\sigma_{\LtwoDynAv}(a)}
\;,
$$
for all $m$ sufficiently large. As $P_{[-m,m]^n}$ converges to $\one$ strongly by functional calculus and is uniformly bounded, the first claimed identity follows for bounded $I$. For unbounded $I$, one can approximate it by the bounded intervals $I\cap [-m,m]^n$. The second equality follows in a similar manner, as does the last claim.
\hfill $\Box$

\vspace{.2cm}

The first application provides $L^p$-norm estimates for products of the generators $\pi(a)f(D)$ for the case $1 \leq p < 2$ which is not covered by Proposition~\ref{prop:lp-embedding}. Those are important for the applications in Chapter~\ref{sec-DualityToep} and \ref{sec-Applications}, but are also of independent interest. 

\vspace{.2cm}

Let $Q_{y} = y + [-\frac{1}{2},\frac{1}{2})^n$ be the unit cube with center $y$. Then $1 = \sum_{y\in \bbZ^n} \chi_{Q_y}(\lambda)$ for all $\lambda$. Now $\ell^p(L^2(\hat{G}))$ is defined as the closed subspace of $L^2(\hat{G})$ with
$$
\norm{f}_{\ell^p(L^2)}
\;=\;
\Big(\sum_{y \in \bbZ^n} \norm{\chi_{Q_y} f}^p_2\Big)^{\frac{1}{p}} 
\;<\; 
\infty
\;.
$$

\begin{proposition}
\label{prop:l1-embedding}
For $a \in \Mm \cap B^{\frac{n}{2}}_{1,1}(\Mm)$ and $f \in L^\infty(\hat{G}) \cap \ell^1(L^2(\hat{G}))$, there is a constant $C>0$ independent of $a$ and $f$ such that
\begin{equation}
\label{eq:L1forproducts} 
\norm{\pi(a)\,f(D)}_1
\;\leq\; C
\norm{a}_{B^{\frac{n}{2}}_{1,1} } \, \norm{f}_{\ell^1(L^2)}
\;.
\end{equation}
\end{proposition}

\noindent {\bf Proof.}  
Let us use the notations $a_j = \widehat{W_j}*a$ and $f_y = \chi_y f$. Then Lemma~\ref{lemma:help} implies 
$$
\pi(a_j)\, f_y(D) 
\;=\; 
\chi_{Q_y + \mathrm{supp}(W_j)}(D) \,\pi(a_j)\, f_y(D)
$$
for all $k,m\in \bbZ^n$. Using polar decomposition, one has $a_j = u_j b^*_j b_j$ with $u_j \in \Mm$ unitary and $b_j=|a_j|^{\frac{1}{2}} \in L^2(\Mm)$. Applying the H\"older inequality and \eqref{eq:L2forproducts} from Proposition~\ref{prop:lp-embedding} thus leads to the bound
\begin{align*}
\norm{\pi(a_j)\, f_y(D)}_1 
&
\;\leq\; 
\norm{\chi_{Q_y + \mathrm{supp}(W_j)}(D) \,\pi(u b_j^*)}_2 \, \norm{\pi(b_j)\, f_y(D)}_2 \\
&
\;\leq\; 
\norm{\chi_{Q_y + \mathrm{supp}(W_j)}}_2\, \norm{b_j}^2_2 \, \norm{f_y}_2\\
&
\;\leq\; 
2^{\frac{n}{2}(j+1)} \norm{a_j}_1 \, \norm{f_y}_2
\end{align*}
Performing the sums over $j$ and $y$ yields the desired bound.
\hfill $\Box$

\vspace{.2cm}

This bound is not sharp, since it does not coincide with the sharp $\ell^1(L^2)$-condition known for the integral operators of type $f(X)g(-\imath \nabla)$ \cite[Chapter 4]{Simon05}. To also obtain a similar bound for $1 < p < 2$ one can use interpolation:
\begin{proposition}
	\label{prop:lpsmall-embedding}
	For $a \in \Mm \cap B^{\frac{n}{2}(\frac{2}{p}-1)}_{p,p}(\Mm)$ and $f \in L^\infty(\hat{G}) \cap \ell^p(L^2(\hat{G}))$, there is a constant $C>0$ independent of $a$ and $f$ such that
	\begin{equation}
		\label{eq:Lpforproducts} 
		\norm{\pi(a)\,f(D)}_p
		\;\leq\; C_p
		\norm{a}_{B^{\frac{n}{2}(\frac{2}{p}-1)}_{p,p} } \, \norm{f}_{\ell^p(L^2)}
		\;.
	\end{equation}
\end{proposition}
\noindent {\bf Proof.}  
As in the proof of Proposition~\ref{prop-BesovInterpol} define the map
$$\psi: \ell^p_s(L^p(\Mm)) \to L^p(\Mm), \quad \psi(x) = \sum_{j=0}^\infty \widehat{W_j}*x_j$$
which is bounded in particular for $s=\frac{n}{2}(\frac{2}{p}-1)$ and $1 \leq p \leq 2$.

Setting $a_j=\widehat{W}_j*x_j$ in the proof of Proposition~\ref{prop:l1-embedding} one deduces the lower endpoint estimate
$$\norm{\pi(\psi(x))f(D)}_1 \leq C_1 \norm{x}_{\ell^1_{s_1}(L^1(\Mm))}\, \norm{f}_{\ell^1(L^2)}$$
with $s_1=\frac{n}{2}$ and since $\norm{\psi(x)}_{L^2(\Mm)} \leq 3 \norm{x}_{\ell^2(L^2(\Mm))}$ one also has from Proposition~\ref{prop:lp-embedding} the upper endpoint 
$$\norm{\pi(\psi(x))f(D)}_2 \leq 3 \norm{x}_{\ell^2_{s_2}(L^2(\Mm))}\, \norm{f}_{\ell^2(L^2)}$$
with $s_2=0$. By interpolation of the sequence spaces (cf. Example 2 of Appendix~\ref{app-Interpol}) there is a bounded extension $$\norm{\pi(\psi(x))f(D)}_p \leq C_p \norm{x}_{\ell^p_{s_p}(L^2(\Mm))}\, \norm{f}_{\ell^p(L^2)}$$
with $s_p=s_1 \frac{2}{p}-1) =\frac{n}{2}(\frac{2}{p}-1)$
and composing it with the map $\tilde{\eta}: B^{s_p}_{p,p}(\Mm)\to \ell^p_s(L^p(\Mm))$ from the proof of Proposition~\ref{prop-BesovInterpol} gives the result. 
\hfill $\Box$

\vspace{.2cm}

In the remainder of this section, we consider the Hankel operators for $p=1$, $n=1$ since that case requires a slightly different approach than the general case treated in the following Section~\ref{sec-HigherPeller}.

\begin{theorem}
\label{theo-Hankel}
For $a\in \Mm \cap B^{1}_{1,1}(\Mm)$, the Hankel operator satisfies $H_a \in L^1(\Nn)$.
\end{theorem}

The proof is a variation of the one in \cite{Pel} for the classical vector-valued case and is based on a simple appeal to homogeneity. 
Since $\alpha$ only has a single generator, one has $\DD=D$ and $\PP= P_{(0,\infty]}$. The norm bounds follow from the simple observation:

\begin{lemma}
\label{lemma:help2}
Let $a \in \Mm\cap L^1(\Mm) $. Then, with norms in $L^1(\Nn)$ and $L^1(\Mm)$ respectively,
$$
\norm{P_{(0,m]} \pi(a) P_{[-m,0]}}_1
\;\leq \; \,(m+1)\,\norm{a}_1
\;.
$$
\end{lemma}
\noindent{\bf Proof.}
Writing again $a = (u \abs{a}^{\frac{1}{2}})\abs{a}^{\frac{1}{2}}$, then by the H\"older inequality and Proposition~\ref{prop:lp-embedding} 
\begin{align}
\norm{ P_{(0,m]} \pi(a) P_{[-m,0]}}_1 
&
\;\leq\;
\big\|P_{(0,m]} \pi(u |a|^{\frac{1}{2}}) \big\|_2 \, \big\|\pi(|a|^{\frac{1}{2}}) P_{[-m,0]} \big\|_2
\nonumber
\\
&
\;\leq\;
\big\|\chi_{[0,m]}\big\|_2^2 \,\big\||a|^{\frac{1}{2}}\big\|^2_2 
\label{eq-PaPestimate}
\\
&
\;= \; \,(m+1)\,\norm{a}_1
\nonumber
\;,
\end{align}
as claimed.
\hfill $\Box$

\vspace{.1cm}

\noindent {\bf Proof} of Theorem~\ref{theo-Hankel}. 
One has $a \;=\; \sum_{\ScaleInd  \geq 0} \widehat{W}_\ScaleInd  * a$ with convergence in the weak operator topology. Hence
$$
H_a 
\;=\;
\sum_{\ScaleInd  \geq 0} H_{\widehat{W}_\ScaleInd  * a}
$$
also converges weakly because $\pi$ is a normal representation. As the support of $W_\ScaleInd $ is contained in $[-2^{\ScaleInd +1},2^{\ScaleInd +1}]$ the same holds for $\sigma_{\LtwoDynAv}(\widehat{W}_\ScaleInd  * a)\subset  [-2^{\ScaleInd +1},2^{\ScaleInd +1}]$. Thus by Lemma~\ref{lemma:help}
\begin{equation}
\label{eq-HankelProj}
H_{\widehat{W}_\ScaleInd  * a}
\;=\;
P_{(0,2^{\ScaleInd +1}]}\,\pi(\widehat{W}_\ScaleInd  * a)\,P_{[-2^{\ScaleInd +1},0]}
\;.
\end{equation}
Hence Lemma~\ref{lemma:help2} implies
$$
\lVert H_{\widehat{W}_\ScaleInd  * a} \rVert_1
\;\leq \; C\,2^{\ScaleInd }\,\lVert \widehat{W}_\ScaleInd  *a \rVert_1
\;
$$
and therefore
$$ 
\lVert H_a \rVert_1
\;\leq\;
\sum_{\ScaleInd  \geq 0} \lVert H_{\widehat{W}_\ScaleInd  *a} \lVert_1
\;\leq \;
C \sum_{\ScaleInd  \geq 0} 2^{\ScaleInd } \,\lVert \widehat{W}_\ScaleInd  * a \lVert_1
\;\leq \;
C \, \lVert a \lVert_{B^{1}_{1,1}(\Mm)}
\;,
$$
which shows the claim.
The sum $H_{\widehat{W}_0*a} + \sum_{\ScaleInd  \in \bbN} H_{\widehat{W}_\ScaleInd  * a}$ converges in $L^1(\Mm \rtimes_\alpha G)$ and its limit must coincide with its weak limit $H_a$ by Lemma \ref{lemma:convergence}.
\hfill $\Box$

\vspace{.2cm}

Using the identities
$$
[P, \pi(a)] \;=\; 
H_a - (H_{a^*})^*
\;, \qquad [\sgn(\DD), \pi(a)] = 2[P,\pi(a)] + [\chi(D=0),\pi(a)]
$$ 
one deduces from Theorem~\ref{theo-Hankel} and Proposition~\ref{prop:l1-embedding}
\begin{corollary}
\label{cor:besov}
For an operator $a\in \Mm \cap B^{1}_{1,1}(\Mm)$,  one has $[P , \pi(a)] \in L^1(\Nn)$ and $[\sgn(\DD), \pi(a)] \in L^1(\Nn)$.
\end{corollary}

In the following section, a second estimate combined with interpolation theory is used to generalize Theorem~\ref{theo-Hankel} to establish criteria for Hankel operators to lie in higher Schatten classes.

\section{Peller criterion for higher Schatten classes}
\label{sec-HigherPeller}

The set-up in this section is the same as in Section~\ref{sec-Peller}. The main result generalizes Theorem~\ref{theo-Hankel}.

\begin{theorem}
\label{theo-HankelBesovP}
For $p > n$ and symbol $a\in \Mm \cap B^{\frac{n}{p}}_{p,p}(\Mm)$, the associated Hankel operator satisfies $H_a \in L^p(\Nn)$. 
\end{theorem}

The proof given in the remainder of this section is based on \cite{Peller82,JansonPeetre88} and uses complex interpolation theory \cite{Kur,BerghLofstrom76,Lunardi2018} as outlined in Appendix~\ref{app-Interpol}. A key difficulty is that it is {\it not} true that $a\in B^{\frac{n}{2}}_{2,2}(\Mm)$ implies $H_a\in L^2(\Nn)$ for $n\geq 2$. Hence it is necessary to consider Hankel operators which are weighted with decaying functions of $\DD$. This allows to use the Hilbert space $B^{\frac{n}{2}+\Weight}_{2,2}(\Mm)$ as a lower endpoint of the interpolation. Thus let us introduce for a symbol $a \in \Mm $ and a given $\Weight \in \bbR$ the weighted Hankel operator %
$$
H_a^{(\Weight)}
\;=\; 
\weight (D)^{\frac{\Weight}{2}}\, H_a \,\weight(D)^{\frac{\Weight}{2}}
\;,
$$
with  weight function 
$$
\weight (\dualvar )
\;=\; 
(1+\abs{\dualvar })
\;.
$$ 
The case $\Weight=0$ coincides with the unperturbed Hankel operator. The upper endpoint of the interpolation will be the embedding $B^{\Weight}_{\infty,\infty}(\Mm)\mapsto H^{(\Weight)}_a \in \Nn$
which fails to hold for $\Weight \leq 0$. Hence it is also necessary to consider  weighted Hankel operators with positive $\Weight$. In that case the weights are unbounded and hence the product is only defined for sufficiently regular symbols.

\vspace{.2cm}

The first aim is to show that the map $a \in \Mmc \mapsto H^{(\zeta)}_a \in \Nn$ is bounded as an operator from $B^{\frac{n}{2}+\Weight}_{2,2}$ to $L^2(\Nn)$. This relies on the following relation between the Fourier decomposition in $L^2(\Mm)$ and the formula for the dual trace \eqref{form:trace}. 

\begin{lemma}
\label{lemma:l2norm}
For bounded Borel functions $f,g,h \in B(\hat{G})$ with $f,g,h \in L^2(\hat{G})$ and $a\in L^2(\Mm)$ with Fourier resolution $(a_\lambda)_{\lambda \in \hat{G}}$, one has
\begin{equation} 
\label{form:comm}
\norm{f(D)\pi(a) g(D)}^2_{L^2(\Nn)} 
\;=\; 
\int_{\hat{G}} \int_{\sigma_{\LtwoDynAv}(a)}  \abs{f(\dualvar + \lambda)g(\dualvar )}^2 \norm{a_\lambda}^2_{\Hh_\lambda}
\;\mu(\difd \lambda)\;\difd{\dualvar }
\;
\end{equation}
and 
\begin{align} 
&
\norm{f(D)[h(D),\pi(a)] g(D)}^2_{L^2(\Nn)} 
\nonumber
\\
&
\;\;
\;=\; 
\int_{\hat{G}}  \int_{\sigma_{\LtwoDynAv}(a)}  \abs{f(\dualvar + \lambda)(h(\dualvar+\lambda)-h(\dualvar))g(\dualvar )}^2 \norm{a_\lambda}^2_{\Hh_\lambda}\;\mu(\difd \lambda)\;\difd{\dualvar }
\;.
\label{form:comm2}
\end{align}
\end{lemma}


\noindent {\bf Proof.} 
By density we can assume $f,g \in C^\infty_c(\bbR^d)$ and $a\in \Mmc$ such that due to \eqref{eq-FourierD}
\begin{align*}
f(D) & \pi(a) g(D) 
\\
&
\;=\; 
\left(\int_G (\mathcal{F}^{-1}f)(t) e^{2\pi \imath  \,D\cdot t}\difd{t}\right) \pi(a) \left(\int_G (\mathcal{F}^{-1}g)(s)e^{2\pi \imath\,  D \cdot s}\difd{s}\right)
\\
&
\;=\; 
\int_{G\times G} \pi(\alpha_t(a)) (\mathcal{F}^{-1}f)(t)(\mathcal{F}^{-1}g)(s-t) e^{2\pi \imath\,  D \cdot s}\,\difd{t}\,\difd{s}\\
&
\;=\; 
\int_{G\times G} \pi\left(\int_{\sigma_{\LtwoDynAv}(a)} \mu(\difd{\lambda})\,e^{2\pi \imath\,  t \cdot\lambda}\, a_\lambda\right) (\mathcal{F}^{-1}f)(t)(\mathcal{F}^{-1}g)(s-t)\, e^{2\pi \imath\,  D \cdot s}\difd{t}\,\difd{s}
\;.
\end{align*}
The inner $t$ integral is the Fourier transform of a product and the convolution theorem gives 
\begin{align*}
\int_G e^{2\pi \imath\,  t\cdot \lambda} (\Ff^{-1}f)(t)(\Ff^{-1} g)({s-t})\, \difd{t} 
& 
\;=\; 
\int_{\hat{G}} f(\dualvar + \lambda ) g(\dualvar ) e^{-2\pi \imath \, \dualvar \cdot s}\, \difd{\dualvar }
\;,
\end{align*}
and hence
$$
f(D) \pi(a) g(D) 
\;=\; 
\int_G  \pi\left(\int_{\hat{G}} \int_{\sigma_{\LtwoDynAv}(a)} 
\,f(\dualvar + \lambda) g(\dualvar )\, a_\lambda\,e^{-2\pi \imath\,  \dualvar  \cdot s}
\;\mu(\difd \lambda)\;\difd{\dualvar }
\right)\,e^{2\pi \imath\,  D \cdot s}\;\difd{s}
\;.
$$
This can also be rewritten in terms of Fourier multipliers as
$$
f(D) \pi(a) g(D) 
\;=\; 
\int_G 
\pi\left(\int_{\hat{G}}(\widehat{\psi}_{\dualvar}*a) \,e^{-2\pi \imath\,  {\dualvar} \cdot s}\;\difd{\dualvar } \right)
\,e^{2\pi \imath\,  D \cdot s}\;\difd{s}
\;,
$$
with 
\begin{equation}
\label{eq:l2kernel1}
\psi_{\dualvar}(\lambda)
\;=\; f({\dualvar}+\lambda) g({\dualvar})
\;.
\end{equation}
The $\hat{G}$-integral is an inverse Fourier transform $\calF^{-1}: L^2(\hat{G}, L^2(\Mm)) \to L^2(G, L^2(\Mm))$ and using the definition of the dual trace and the Plancherel identity \eqref{eq-Plancharel} gives
\begin{align*} 
\norm{f(D)\pi(a) g(D)}^2_{L^2(\Nn)} 
&
\;=\; 
\int_G  \norm{\int_{\hat{G}} (\widehat{\psi}_{\dualvar}*a)\, e^{-2\pi \imath \, {\dualvar}\cdot s} \;\difd{{\dualvar}}}_{L^2(\Mm)}^2 \;\difd{s}\\
&
\;=\; 
\int_{\hat{G}}  \norm{\widehat{\psi}_{\dualvar}*a}_{L^2(\Mm)}^2 \,\difd{{\dualvar}}
\\
&
\;=\; 
\int_{\hat{G}}  \int_{\sigma_{\LtwoDynAv}(a)}  |\psi_{\dualvar}(\lambda)|^2 \norm{a_\lambda}^2_{\Hh_\lambda}  \;\mu(\difd{\lambda})\;\difd{{\dualvar}}
\;,
\end{align*}
concluding the proof of the first equality. The second follows from the same computation with the obvious modification to the kernel function in  \eqref{eq:l2kernel1}.
\hfill $\Box$

\begin{corollary}
\label{coro-l2scalprod}
For bounded Borel functions $f,g,h,\tilde{f},\tilde{g} \in B(\hat{G})$ with $f,g,h \in L^2(\hat{G})$ and $a,b\in L^2(\Mm)$ with Fourier resolutions $(a_\lambda)_{\lambda \in \hat{G}}$ and $(b_\lambda)_{\lambda \in \hat{G}}$, one has
%
%
\begin{align*}
\langle f(D)&[h(D), \pi(a)] g(D),f(D)[h(D),\pi(b)] g(D)\rangle_{L^2(\Nn)}
\\
&
\;=\; 
\int_{\hat{G}}\int_{\hat{G}} 
\abs{\psi_k(\lambda)}^2\, 
\langle a_\lambda,b_\lambda\rangle_{\Hh_\lambda}
\mu(\difd{\lambda})\,\difd{{\dualvar}}
\;
\end{align*}
for $$\psi_k(\lambda)=f(\dualvar + \lambda)(h(\dualvar+\lambda) - h(\dualvar))g(\dualvar).$$
\end{corollary}
%
\noindent {\bf Proof.} 
This follows from \eqref{form:comm} using the polarisation identity.
\hfill $\Box$

\begin{proposition}
\label{prop:interp-lower}
For $0 < \frac{n}{2} + \Weight < 1$ and $a \in \Mm \cap B^{\frac{n}{2}+\Weight}_{2,2}(\Mm)$ one has for some uniform constant
$$
\norm{H^{(\Weight)}_a}_2
\;\leq\; 
C \,\norm{a}_{B^{\frac{n}{2}+\Weight}_{2,2}(\Mm)}
$$
\end{proposition}

\noindent {\bf Proof.} 
Applying Lemma \ref{lemma:l2norm} componentwise to truncations of $\sgn$ and $\weight$ yields
\begin{equation}
\label{eq-NormForumula}
\norm{\weight ^{\frac{\Weight}{2}}(D) [\sgn(\DD), \pi(a)] \weight ^{\frac{\Weight}{2}}(D)}_2^2 
\;= \;  
\int_{\hat{G}}  \int_{\sigma_{\alpha}(a)}   |\psi_{\dualvar}(\lambda)|^2_2 \,\norm{a_\lambda}^2_{\Hh_\lambda}\mu(\difd{\lambda})\;\difd{{\dualvar}}
\;,
\end{equation}
with $|\,.\,|_2$ denoting the Hilbert-Schmidt matrix norm and the matrix-valued multiplier 
$$
\psi_{\dualvar}(\lambda) 
\;=\; 
(1+\abs{{\dualvar}+\lambda})^{\frac{\Weight}{2}} 
\big(\sgn(\gamma\cdot(\dualvar + \lambda))-\sgn(\gamma\cdot {\dualvar})\big) 
(1+\abs{{\dualvar}})^{\frac{\Weight}{2}}
\;.
$$
As in \cite[p.~149]{PSbook}, let us now use the asymptotic bound
\begin{equation}
\label{eq-SgnSgn}
|\sgn(\gamma\cdot({\dualvar}+\lambda))-\sgn(\gamma\cdot {\dualvar}) |_2
\;\leq\; 
C_1\,\frac{\abs{\lambda}}{\abs{{\dualvar}}}
\quad
\mbox{ for large } {\dualvar}
\;.
\end{equation}
This implies that the ${\dualvar}$-integral in \eqref{eq-NormForumula} is finite for fixed $\lambda$ as long as $2\Weight -2 < -n $ which is equivalent to $ \Weight < 1- \frac{n}{2}$. A quantitative upper bound is obtained by further elementary estimates
\begin{align*}
\int_{\hat{G}} 
(1+\abs{{\dualvar}+\lambda})^{\Weight}& (1+\abs{{\dualvar}})^{\Weight} |\sgn(\gamma\cdot({\dualvar}+\lambda))-\sgn(\gamma\cdot {\dualvar})|_2^2  \,\difd{{\dualvar}}
\\
\leq\; &\int_{ \{\abs{{\dualvar}}<\frac{\abs{\lambda}}{2}\}\cup \{|{\dualvar}+\lambda|<\frac{|\lambda|}{2}\} }(1+\abs{{\dualvar}+\lambda})^{\Weight} (1+\abs{{\dualvar}})^{\Weight} \,4\,\difd{{\dualvar}}
\\
& \,+ \,
\int_{ \{\abs{{\dualvar}}>\frac{\abs{\lambda}}{2}\}\cap \{|{\dualvar}+\lambda|>\frac{|\lambda|}{2}\} }
(1+\abs{{\dualvar}+\lambda})^{\Weight} (1+\abs{{\dualvar}})^{\Weight} \,C_1^2 \,\frac{\abs{\lambda}^2}{\abs{{\dualvar}}^2} \,\difd{{\dualvar}}
\\
\leq \; &C_2(1+ \abs{\lambda}^{n+2\Weight})
\;,
\end{align*}
with the last estimate resulting from the volume form $\sim \abs{{\dualvar}}^{n-1}$ giving all integrals  asymptotics of $\abs{\lambda}^n$ for small $\lambda$ and $\abs{\lambda}^{n+2\Weight}$ for large $\lambda$. Due to Proposition~\ref{prop-FractionalSobolev}, this leads to 
$$
\norm{\weight ^{\frac{\Weight}{2}}(D) [\sgn(\DD), \pi(a)] \weight^{\frac{\Weight}{2}}(D)}_2 
\;\leq\; 
C_3 
\norm{a}_{B^{\frac{n}{2}+\Weight}_{2,2}}\;,
$$ 
and hence the same bound for $H^{(\Weight)}_a$ follows from 
$$
H_a \;=\; 
-\frac{1}{2}P[\sgn(\DD)-\chi(D=0),\pi(a)]
$$ 
and Proposition~\ref{prop:lp-embedding}.
\hfill $\Box$

\vspace{.2cm}
For $n=1$, one readily obtains a converse result which is also worth while stating explicitly:

\begin{lemma}
\label{lemma:besov2norm}
For $n=1$, an equivalent norm on $B^{\frac{1}{2}}_{2,2}(\Mm)$ is given by $\norm{a}_2 + \norm{[\sgn(\DD),a]}_2$.
\end{lemma} 

\noindent{\bf Proof.}
Substituting $\psi_k(\lambda)=\sgn({\dualvar}+\lambda) - \sgn(\dualvar)$ into \eqref{eq-NormForumula} one evaluates
$$
\norm{[\sgn(\DD), \pi(a)]}_2^2 
\;\simeq \;  4 \int_{\sigma_{\LtwoDynAv}(a)}   |\lambda| \,\norm{a_\lambda}^2_{\Hh_\lambda}\,\mu(\difd{\lambda})
\;,
$$
and comparing with Proposition~\ref{prop-FractionalSobolev} gives the result.
\hfill $\Box$

\vspace{.2cm}

The next step is the embedding for the upper endpoint, {\it i.e.}  that $H^{(\Weight)}_a$ is a bounded operator for $\Weight\in(0,1)$ and a symbol $a \in {}^0 B^{\Weight}_{\infty,\infty}(\Mm)$ (see Proposition~\ref{prop:interp-upper}). This requires to prove decay estimates to control the increasing weight factors. The argument is split into several technical lemmata and follows the general strategy of \cite{JansonPeetre88} with the necessary adaptations. The choice of constants in the conditions on $y_1$ and $y_2$ in the first lemma is tailored for the application in the proof of Lemma~\ref{lemma:comm_diagonal} below.

\begin{lemma}
\label{lemma:comm_decay}
Let $P_{L,y}= \chi_{Q_{L,y}}(D)$ be the projection to the half-open cube $Q_{L,y}$ with center $y \in \bbR^n$ and sides $L$. Then there is a constant $C$ such that
\begin{equation} 
\label{eq:norm_mult} 
\norm{P_{L, y_1} [\sgn(\DD), \pi(a)] P_{L,y_2}}
\;\leq\; 
C\, \frac{L}{\abs{y_1}+\abs{y_2}}\, \norm{a}
\;,
\end{equation} 
uniformly for all $a\in \Mm$, $L>0$ and $y_1,y_2 \in \bbR^n$ satisfying $\abs{y_1},\abs{y_2} \geq 4L$ and $\abs{y_1-y_2} \leq 2^4 L$.
\end{lemma}

\noindent {\bf Proof.} We consider $a\in \Mmc$ by weak density.
For $f,g,h \in C^\infty_c(\bbR^n)$ one has as in the proof of Lemma~\ref{lemma:l2norm} that
$$
f(D) [g(D),\pi(a)] h(D) 
\;=\; 
\int_G \pi\left(
\int_{\hat{G}} (\widehat{\psi}_{\dualvar}*a)\, e^{-2\pi \imath\,  {\dualvar}\cdot s} \difd{{\dualvar}}\right)
e^{2\pi \imath\,  D\cdot s}\;\,\difd{s}
\;,
$$
with 
$$
\psi(\dualvar,\lambda)\;=\; f({\dualvar}+\lambda) (g({\dualvar}+\lambda)-g({\dualvar})) h({\dualvar})
\;.
$$
Independently of $G$ we may also write this as
$$
f(D) [g(D),\pi(a)] h(D) 
\;=\; 
\int_{\bbR^n} \pi\left(
\int_{\bbR^n} (\widehat{\psi}_{\dualvar}*a)\, e^{-2\pi \imath\,  {\dualvar}\cdot s}\difd{{\dualvar}}\right)
e^{2\pi \imath\,  D\cdot s}\,\difd{s}
.
$$
An easy way to see this is to consider the isometric action $\beta: \hat{G}^2 \times \Nn  \to \Nn$ defined by
$$\beta_{t,s}(b) = \alpha_t(b)e^{2\pi \imath D \cdot s}$$
and note that in the notation of Lemma~\ref{lemma:periodic_mult}
$$f(D) [g(D),\pi(a)] h(D) = \beta_{\calF^{-1}\psi^r} (\pi(a)) = \tilde{\beta}_{\calF^{-1}\psi} (\pi(a)).$$
Due to \eqref{eq-FAbound} one obtains the operator norm bound
\begin{align*}
&
\norm{f(D) [g(D),\pi(a)] h(D)} 
\\
&
\;\;\;\;
\;\leq\; 
\int_{\bbR^n}  \norm{\int_{\bbR^n} \pi (\widehat{\psi}_{\dualvar}*a) e^{-2\pi \imath \, {\dualvar}\cdot s}\;\difd{{\dualvar}}}\;\difd{s}
\\ 
&
\;\;\;\;
\;\leq\; 
\int_{\bbR^n}  \int_{\bbR^n}  \abs{\int_{\bbR^n}  \int_{\bbR^n} 
\psi(\dualvar,\lambda)\, e^{-2\pi \imath  \,{\dualvar}\cdot s}\,e^{2\pi \imath  \,\lambda \cdot t}\,\difd{{\dualvar}}\,\difd{\lambda}
} \norm{a}\;\difd{t}\;\difd{s}
\;.
\end{align*}
On the r.h.s. appears the $L^1$-norm of the Fourier transform of $\psi(\dualvar,\lambda)$ in both variables. This is a standard object for estimates on a Fourier multiplier $\psi \in FA(\bbR^{2n})$ and hence we write it as $\norm{\psi({\dualvar},\lambda)}_{1,\calF}$. Let now $\widehat{\chi}_{L, y}({\dualvar})$ be a smooth function which is equal to $1$ in $Q_{L,y}$ and has compact support contained in $Q_{\frac{3}{2} L,y}$. Then one can bound and rescale as follows:
\begin{align*}
&
\norm{P_{L, y_1} [g(D), \pi(a)] P_{L,y_2}} 
\\
&
\;\;\;
\;\leq\;
\norm{(g({\dualvar}+\lambda)-g({\dualvar})) \widehat{\chi}_{L, y_1}({\dualvar}+\lambda)\widehat{\chi}_{L, y_2}({\dualvar})}_{1,\calF}\, \norm{a}
\\
&
\;\;\;
\;=\; 
\norm{(g(L ({\dualvar}+\lambda)+y_1)-g(L{\dualvar} + y_2)) \widehat{\chi}_{1, 0}({\dualvar}+\lambda) \widehat{\chi}_{1, 0}({\dualvar})}_{1,\calF}\, \norm{a}
\\
&
\;\;\;
\;=\; 
\norm{(g(L ({\dualvar}+\lambda+\tilde{y}_1))-g(L({\dualvar} + \tilde{y}_2))) \widehat{\chi}_{1, 0}({\dualvar}+\lambda) \widehat{\chi}_{1, 0}({\dualvar})}_{1,\calF}\, \norm{a}
\;,
\end{align*}
where $\tilde{y}_j=\frac{y_j}{L}$. It is now possible to substitute $g({\dualvar})=\sgn(\gamma \cdot {\dualvar})$ since the restriction to the support of $\widehat{\chi}_{L, y_1}({\dualvar}+\lambda)\widehat{\chi}_{L, y_2}({\dualvar})$ coincides with a smooth function considering that  $\abs{y_1},\abs{y_2} \geq 4L$. Therefore, using the scale invariance of $\sgn$, the above norm $\|\,.\,\|_{1,\calF}$ can be bounded by
\begin{align*}
\frac{L}{\abs{y_1}+\abs{y_2}}
\,& \sup\;
(\abs{\tilde{y}_1}+\abs{\tilde{y}_2} ) \cdot
\\
&
\norm{(\sgn(({\dualvar}+\lambda +\tilde{y}_1)\cdot \gamma)-\sgn(({\dualvar} +\tilde{y}_2)\cdot \gamma)) \, \widehat{\chi}_{1, 0}({\dualvar}+\lambda) \widehat{\chi}_{1, 0}({\dualvar})}_{1,\calF}
\;,
\end{align*}
with a supremum carrying over all $\tilde{y}_1,\tilde{y}_2 \in \bbR^d$ satisfying $\abs{\tilde{y}_1},\abs{\tilde{y}_2}\geq 4$ and $\abs{\tilde{y_1}-\tilde{y}_2} \leq 2^4$. On the other hand, using the asymptotics \eqref{eq-SgnSgn} allows to show that the supremum of the appearing multiplier can be bound by
$$
\sup
(\abs{\tilde{y}_1}+\abs{\tilde{y}_2} )
\norm{(\sgn(({\dualvar}+\lambda +\tilde{y}_1)\cdot \gamma)-\sgn(({\dualvar} +\tilde{y}_2)\cdot \gamma)) \, \widehat{\chi}_{1, 0}({\dualvar}+\lambda) \widehat{\chi}_{1, 0}({\dualvar})} 
\;<\;
\infty
\;,
$$
with a supremum carrying over all $\tilde{y}_1,\tilde{y}_2 \in \bbR^n$ as above and, moreover, over all ${\dualvar},\lambda \in \bbR^n$. Furthermore, all partial derivatives of the multiplier w.r.t. ${\dualvar}$ and $\lambda$ can also be bounded uniformly on the same set. Since the multiplier has a compact support this also implies the same bounds for the $L^1$-norms of its derivatives, uniformly in $y_1$,$y_2$,$L$ subject to the constraints given. We can therefore conclude \eqref{eq:norm_mult} by estimating the $L^1$-norm of the Fourier transform in the standard manner in terms of derivatives up to order $n+1$. 
\hfill $\Box$

\vspace{.2cm}

On the l.h.s. of \eqref{eq:norm_mult}, the operator $a$ will be replaced by its Littlewood-Payley decomposition \eqref{eq-aRingDecomp}. Then only finitely many terms contribute as shows the following lemma which is a higher-dimensional extension of Lemma~\ref{lemma:help}.

\begin{lemma}
\label{lemma:supp0}
Let $\ScaleInd \geq 2$ and $a\in\Mm$. For boxes on scale $i=\ScaleInd -2$ centered at $2^iy_1$ and $2^iy_2$, the operator
\begin{equation} 
\label{eq:supp0} 
P_{2^i, 2^iy_1} \pi(\widehat{W}_\ScaleInd *a) P_{2^i,2^iy_2}
\end{equation} 
vanishes unless $2^{\ScaleInd -2}(2-\sqrt{2})\leq 2^i|y_1-y_2|\leq 2^{\ScaleInd +2}$.
\end{lemma}

\noindent {\bf Proof.} 
Assuming that $a \in L^2(\Mm) \cap \Mm$ it follows from \eqref{form:comm} that
$$
\norm{P_{2^i, 2^iy_1} \pi(\widehat{W}_\ScaleInd *a) P_{2^i,2^iy_2}}^2_2 
\;=\; 
\int_{\hat{G}}  \int_{\sigma_{\LtwoDynAv}(a)}  |\psi_{\dualvar}(\lambda)|^2 \,\norm{a_\lambda}^2_{\Hh_\lambda}
\,\mu(\difd{\lambda})\;\difd{{\dualvar}}\;,
$$
with the kernel 
$$
\psi_{\dualvar}(\lambda)
\;=\;
\chi_{Q_{2^i,2^iy_1}}({\dualvar}+\lambda)\,\chi_{Q_{2^i,2^iy_2}}({\dualvar}) 
\,\chi_{[2^{\ScaleInd -1},2^{\ScaleInd +1}]}(|\lambda|)\, W_\ScaleInd (\lambda)
\;,
$$
because $W_\ScaleInd $ is supported on the annulus $\{\lambda\in\RM^n:2^{\ScaleInd -1}\leq |\lambda|\leq 2^{\ScaleInd +1}\}$. By a geometric argument, this function $\psi_{\dualvar}$ vanishes unless $2^{\ScaleInd -1}-\sqrt{2}\, 2^i\leq 2^i|y_1-y_2|\leq 2^{\ScaleInd +1}+\sqrt{2} \,2^i$. This implies the claim by weak density of $L^2(\Mm)\cap \Mm$ in $\Mm$.
\hfill $\Box$

\vspace{.2cm}

The second preparation for the proof of Lemma~\ref{lemma:comm_diagonal} (and that of Proposition~\ref{prop:interp-upper} as well) is of general nature:

\begin{lemma}
\label{lemma-FiniteRangeBound}
Let $(Q_j)_{j\in \ZM^n}$ be partition of unity in a von Neumann algebra $\Mm$, namely $Q_j$ are pairwise orthogonal projections with $\sum_{j}Q_j=\one$ in the weak {\rm (}or strong{\rm )} operator topology. 

\begin{enumerate}

\item[{\rm (i)}]
Suppose that $a\in\Mm$ is of finite range $N\in \bbR^+$ w.r.t. this partition, that is,
$$
a
\;=\;
\sum_{|i|\leq N}\sum_j\,Q_{j+i}\,a\,Q_{j}
\;.
$$
Then 
$$
\|a\|
\;\leq\;N^d\,
\sup_{i,j}\; \|Q_j\, a\,Q_i\|
\;.
$$

\item[{\rm (ii)}] Suppose that $a\in\Mm$ has off-diagonal decay
$$
\|Q_{j+i}\,a\,Q_{j}\|\;\leq\;C\,e^{-\kappa|i|}
\;,
$$
with $\kappa>0$ and $C$ uniform in $j$, then there is a constant $c_n$ depending on $n$ such that
$$
\|a\|
\;\leq\;c_n\,C\,\kappa^{-n}
\;.
$$
\end{enumerate}
\end{lemma}

\noindent {\bf Proof.} Inserting the partition of unity twice, using the triangle inequality and $C^*$-equation one has
\begin{align*}
\|a\|
&
\;=\;
\big\|\sum_{|i|\leq N}\sum_{j}Q_{j+i}\,a\,Q_{j}\big\|
\\
&
\;\leq\;
\sum_{|i|\leq N}
\big\|\sum_{j}Q_{j+i}\,a\,Q_{j}\big\|
\\
&
\;=\;
\sum_{|i|\leq N}
\big\|\sum_{j}Q_j\,a^*\,Q_{j+i}\,a\,Q_j\big\|^{\frac{1}{2}}
\;,
\end{align*}
the latter because of the pairwise orthogonality. Now $\sum_{j}Q_ja^*Q_{j+i}aQ_j$ is a block diagonal operator. Thus
\begin{align*}
\|a\|
&
\;\leq\;
\sum_{|i|\leq N}
\big(\sup_j \|Q_j\,a^*\,Q_{j+i}\,a\,Q_j\|\big)^{\frac{1}{2}}
\\
&
\;=\;
\sum_{|i|\leq N}
\,\sup_j \|Q_{j+i}\,a\,Q_j\|
\\
&
\;\leq\;
\sum_{|i|\leq N}
\sup_{i',j} \|Q_{i'}\,a\,Q_{j}\|
\;,
\end{align*}
completing the proof of (i). For the proof of (ii), let the sum over $i$ run over all $\ZM^n$ and use the hypothesis in the last estimate. Note that $c_n\approx \Gamma(n)\mbox{Vol}(\SM^{n-1})$.
\hfill $\Box$

\vspace{.2cm}

Let us now introduce the supports corresponding to the dyadic decomposition  \eqref{eq-WkChoice}
\begin{equation}
\label{eq-DyadicDecomp}
\Lambda_0 \;= \;[-2,2]
\;,
\qquad
\Lambda_m \;= \;[-2^{m+1},-2^m) \cup (2^m,2^{m+1}] \;\mbox{ for }\; m>0
\;,
\end{equation}
and $P_{\Lambda_m}= \chi_{\Lambda_m}(\DD)=\chi_{(2^m,2^{m+1}]}(|\DD|)$.

\begin{lemma}
\label{lemma:comm_diagonal}
For each $0< \Weight < 1$ there is constant $C$ uniform in $a \in {}^0B^{\Weight}_{\infty,\infty}(\Mm)$, $m \in \bbN$ and $l \in \{-1,0,1\}$ such that
\begin{equation} 
\norm{\weight(D)^{\frac{\Weight}{2}} P_{\Lambda_{m}} [\sgn(\DD),\pi(a)] P_{\Lambda_{m+l}}\weight(D)^{\frac{\Weight}{2}}}
\;\leq\; 
C \norm{a}_{B^{\Weight}_{\infty,\infty}}
\;.
\end{equation}
\end{lemma}

\noindent {\bf Proof.} 
For $m=0,1,2$ and thus $m+l\leq 3$, using $a = \sum_{j=0}^\infty \widehat{W_j}*a$ one can simply bound 
\begin{align*}
\norm{\weight(D)^{\frac{\Weight}{2}} P_{\Lambda_{m}} [\sgn(\DD),\pi(a)] P_{\Lambda_{m+l}}\weight(D)^{\frac{\Weight}{2}}}
&
\;\leq\; 
C_1\,\sum_{\ScaleInd \geq 0}\|\widehat{W}_\ScaleInd *a\|
\\
&
\;\leq\;
C_1 \sum_{\ScaleInd \geq 0}e^{-\Weight \ScaleInd }
\norm{a}_{B^{\Weight}_{\infty,\infty}}
\;,
\end{align*}
due to Definition~\ref{def-Besov}. Hence let now $m\geq 2$ and assume, by symmetry,  that $l \geq 0$. Let us first bound the contributions of the terms $\widehat{W}_\ScaleInd *a$  
\begin{align}
\big\|\weight(D)^{\frac{\Weight}{2}} & P_{\Lambda_{m}} [\sgn(\DD),\pi(a)] P_{\Lambda_{m+l}}\weight(D)^{\frac{\Weight}{2}}\big\| 
\nonumber
\\
&
\;\leq\;  
\sum_{\ScaleInd \geq 0} \norm{\weight(D)^{\frac{\Weight}{2}} P_{\Lambda_{m}} [\sgn(\DD),\pi(\widehat{W}_\ScaleInd *a)] P_{\Lambda_{m+l}} \weight(D)^{\frac{\Weight}{2}} }  
\nonumber
\\
&
\;\leq\;  
2^{(m+2)\Weight}\,\sum_{\ScaleInd \geq 0} \norm{ P_{\Lambda_{m}} [\sgn(\DD),\pi(\widehat{W}_\ScaleInd *a)] P_{\Lambda_{m+l}}  }  
\;,
\label{eq-intermed}
\end{align}
(Lemma~\ref{lemma:supp} shows that the sum over $\ScaleInd $ can be restricted to $\ScaleInd \leq m+l+3$, but this is irrelevant for the present argument.) In order to bound the matrix elements on the r.h.s. of \eqref{eq-intermed}, let us consider the partition of unity $(P_{2^i,2^i y})_{y \in \bbZ^n}$ where $i=\ScaleInd -2$. Then
$$
P_{\Lambda_{m}} [\sgn(\DD),\pi(\widehat{W}_\ScaleInd *a)] P_{\Lambda_{m+l}}
\;=
\!\sum_{y_1,y_2\in\ZM^n}\!
 P_{2^i, 2^i y_1} P_{\Lambda_{m}}[\sgn(\DD),\pi(\widehat{W}_\ScaleInd *a)]P_{\Lambda_{m+l}} P_{2^i, 2^i y_2} 
\;.
$$
By Lemma~\ref{lemma:supp0}, the appearing matrix elements vanish unless $2^{\ScaleInd -2}(2-\sqrt{2}) \leq 2^i\abs{y_1-y_2} \leq 2^{\ScaleInd +2}$, and clearly also unless  $2^{m-1} \leq 2^i\abs{y_1} \leq 2^{m+2}$ and $2^{m+l-1} \leq 2^i\abs{y_2} \leq 2^{m+l+2}$.  This restricts the number of summands to be finite (but increasing with $m$). More importantly, the operator is also of finite range w.r.t. the partition of unity $(P_{2^i,2^i y})_{y \in \bbZ^n}$ because non-vanishing entries appear only for $\abs{y_1-y_2}\leq 2^4$. Therefore by Lemma~\ref{lemma-FiniteRangeBound}(i)
$$
\norm{ P_{\Lambda_{m}} [\sgn(\DD),\pi(\widehat{W}_\ScaleInd *a)] P_{\Lambda_{m+l}}}
\;\leq\;
2^{4n}
\sup_{y_1,y_2\in\ZM^n}\,
p^{(\ScaleInd )}_{y_1,y_2}
\;,
$$
with the notation
$$
p^{(\ScaleInd )}_{y_1,y_2}
\;=\;
\norm{P_{\Lambda_{m}} P_{2^i, 2^i y_1} [\sgn(\DD),\pi(\widehat{W}_\ScaleInd *a)] P_{2^i, 2^i y_2} P_{\Lambda_{m+l}}}
\;.
$$
Let us next bound this matrix element using Lemma~\ref{lemma:comm_decay} for the points $2^iy_1$ and $2^iy_2$ with $L=2^i$. The conditions are satisfied as long as $\ScaleInd \leq m-1$ (which implies $2^i|y_1|\geq 2^{m-1}$), and the other constraints above hold. Then  with a constant $C_2$ uniformly for all permissible $y_1$,$y_2$,$m$,$l$ and $\ScaleInd $,
\begin{align*}
p^{(\ScaleInd )}_{y_1,y_2} 
&
\;\leq\; 
\norm{P_{2^i, 2^i y_1} [\sgn(\DD),\pi(\widehat{W}_\ScaleInd *a)] P_{2^i, 2^i y_2} }
\\
&
\;\leq\;
C_2\, \frac{2^i}{2^i(\abs{y_1}+\abs{y_2})} \,\norm{\widehat{W}_\ScaleInd *a} 
\\
&
\;\leq\; 
C_2\, 2^{i-m+1}\,2^{-\ScaleInd  \Weight}\,  \norm{a}_{B^{\Weight}_{\infty,\infty}}
\\
&
\;= \;
\frac{1}{2}\,C_2\, 2^{\ScaleInd (1-\Weight)-m} \norm{a}_{B^{\Weight}_{\infty,\infty}}
\;.
\end{align*}
Replaced in the above yields, still for $\ScaleInd \leq m-1$,
$$
\norm{P_{\Lambda_{m}} [\sgn(\DD),\pi(\widehat{W}_\ScaleInd *a)] P_{\Lambda_{m+l}}} 
\;\leq\; 
2^{4n}\,\frac{1}{2}\,C_2\,2^{\ScaleInd (1-\Weight)-m} \norm{a}_{B^{\Weight}_{\infty,\infty}}
\;.
$$
For $\ScaleInd \geq m$, it is sufficient to bound as follows:
$$
\norm{P_{\Lambda_{m}} [\sgn(\DD),\pi(\widehat{W}_\ScaleInd *a)] P_{\Lambda_{m+l}}} 
\;\leq\; 
2\,\norm{\pi(\widehat{W}_\ScaleInd *a)} 
\;\leq\; 
2\,2^{-\Weight \ScaleInd } \norm{a}_{B^{\Weight}_{\infty,\infty}}
\;.
$$
Replacing these last two bounds in \eqref{eq-intermed} yields
\begin{align*}
\big\|\weight(D)^{\frac{\Weight}{2}} & P_{\Lambda_{m}} [\sgn(\DD),\pi(a)] P_{\Lambda_{m+l}}\weight(D)^{\frac{\Weight}{2}}\big\| 
\\
&
\;\leq\;  
2^{(m+2)\Weight}
\left(\sum_{\ScaleInd =0}^{m-1}2^{4n}\,\frac{1}{2}\,C_2\,2^{\ScaleInd (1-\Weight)-m} \;+\;\sum_{\ScaleInd \geq m}2\,2^{-\Weight \ScaleInd }\right)
\norm{a}_{B^{\Weight}_{\infty,\infty}}
\;,
\end{align*}
so that summing the series and using $\Weight\in(0,1)$ completes the proof.
\hfill $\Box$

\vspace{.2cm}

The final preparation is another extension of Lemma~\ref{lemma:help}.

\begin{lemma}
\label{lemma:supp}
For $a \in \Mm$ and $m,l\in \bbN$, the operator 
$$
P_{\Lambda_m} [\sgn(\DD),\pi(\widehat{W}_\ScaleInd *a)] P_{\Lambda_l}
$$
vanishes unless 
$$
\chi(|m-l|>1)\,\big(\max\{m,l\}\,-2\big)\;\leq\;\ScaleInd  \;\leq\; \max\{m,l\}\,+\,3
\;.
$$
\end{lemma}

\noindent {\bf Proof.} As $\sgn(\DD)=2P-\one$, only the operators $P_{\Lambda_m} (\one-P) \pi(\widehat{W}_\ScaleInd  *a) P  P_{\Lambda_l}$ that are off-diagonal in the grading of $P$ have to be considered. As in Lemma~\ref{lemma:supp0}, assuming that $a \in L^2(\Mm) \cap L^\infty(\Mm)$ it follows from \eqref{form:comm} that
$$
\norm{P_{\Lambda_m} (\one-P) \pi(\widehat{W}_\ScaleInd  *a) P  P_{\Lambda_l}}^2_2 
\;=\; 
\int_{\hat{G}}  \int_{\sigma_{\LtwoDynAv}(a)} \norm{\psi_{\dualvar}(\lambda)}^2_2 \norm{a_\lambda}^2_{\Hh_\lambda}\;\mu(\difd{\lambda})
\;\difd{{\dualvar}}
\;,
$$
with the matrix-valued kernel 
\begin{align*}
\psi_{\dualvar}(\lambda)
&
\;=\;
\chi_{\Lambda_m}(\gamma\cdot({\dualvar}+\lambda))\,\chi_{\Lambda_l}(\gamma\cdot {\dualvar}) 
\,\chi_{\bbR_-}(\gamma\cdot({\dualvar}+\lambda))\,\chi_{\bbR_+}(\gamma\cdot {\dualvar})\, W_\ScaleInd (\lambda)
\\
&
\;=\;
\chi_{(2^m,2^{m+1}]}(|{\dualvar}+\lambda|)\,\chi_{(2^l,2^{l+1}]}(|{\dualvar}|)\, \chi_{[2^{\ScaleInd -1},2^{\ScaleInd +1}]}(|\lambda|)
\cdot
\\
&
\;\;\;\;\;\;\;
\,\cdot \chi_{\bbR_-}(\gamma\cdot({\dualvar}+\lambda))\,\chi_{\bbR_+}(\gamma\cdot {\dualvar})\, W_\ScaleInd (\lambda)
\;,
\end{align*}
Now one has a non-vanishing contribution only if
$$
2^{\ScaleInd -1}\;\leq\;|\lambda|\;\leq\;|{\dualvar}+\lambda|+|{\dualvar}|\;\leq\;2^{m+1}+2^{l+1}\;\leq\;2^{\max\{m,l\}+2}
\;,
$$
and
$$
2^{\ScaleInd +1}\;\geq\;|\lambda|\;\geq\;\big||{\dualvar}+\lambda|-|{\dualvar}|\big|\;\geq\;
\max\{
2^{m}-2^{l+1},2^{l}-2^{m+1},0\}
\;,
$$
namely the stated bound. The lemma follows by weak density of $L^2(\Mm) \cap L^\infty(\Mm)$ in $\Mm$.
\hfill $\Box$

\begin{proposition}
\label{prop:interp-upper}
For $0 < \Weight < 1$ and $a\in {}^0 B^{\Weight}_{\infty,\infty}(\Mm)$ one has for some uniform constant
$$
\norm{H^{(\Weight)}_a} 
\;\leq\; 
C \,\norm{a}_{B^{\Weight}_{\infty,\infty}}
\;.
$$
\end{proposition}

\noindent {\bf Proof.} 
The weighted Hankel operator is given by the formal sum
$$
H_a^{(\Weight)}
\;= \;
\sum_{\ScaleInd \geq 0} \weight(D)^{\frac{\Weight}{2}} (\one-P)\, \pi(\widehat{W}_\ScaleInd  *a)\, P\,  \weight(D)^{\frac{\Weight}{2}}
\;,
$$
where it is not even evident that the individual terms are elements of $\Nn$. Hence let us define the regularization
\begin{equation}
\label{eq-HankelInf}
H^{(\Weight)}_a 
\;=\; 
\slim_{N \to \infty} 
\,
\sum_{m,l,\ScaleInd =0}^N P_{\Lambda_m} \weight(D)^{\frac{\Weight}{2}} (\one-P) \,\pi(\widehat{W}_\ScaleInd  *a) \,P\,  \weight(D)^{\frac{\Weight}{2}} P_{\Lambda_l}
\;.
\end{equation}
Lemma \ref{lemma:supp} implies that, assuming $\abs{m-l} > 1$, the operator vanishes unless $\max\{m,l\}-2 \leq \ScaleInd  \leq \max\{m,l\}+3$. Let us therefore assume $l < m -1$ and focus on the matrix elements 
\begin{align*}
P_{\Lambda_m}H^{(\Weight)}_a P_{\Lambda_l} 
&
\;=\; 
\sum_{\ScaleInd  \geq 0} P_{\Lambda_m}   \weight(D)^{\frac{\Weight}{2}} (\one-P)\, \pi(\widehat{W}_\ScaleInd  *a) \,P\,  P_{\Lambda_l} \weight(D)^{\frac{\Weight}{2}}
\\
&
\;= \;  
\sum_{\ScaleInd  = m-2}^{m+3} P_{\Lambda_m}  \weight(D)^{\frac{\Weight}{2}} (\one-P) \,\pi(\widehat{W}_\ScaleInd  *a)\, P\,  P_{\Lambda_l} \weight(D)^{\frac{\Weight}{2}}
\;.
\end{align*}
(The upper bound $\ScaleInd \leq m+3$ holds, but is not relevant for the following.) Hence bounding $\weight$ implies
\begin{align*}
\big\|P_{\Lambda_m}H_a^{(\Weight)} P_{\Lambda_l}\big\| 
&
\;\leq\;
2^{(m+1)\frac{\Weight}{2}}\,2^{(l+1)\frac{\Weight}{2}} \,\sum_{\ScaleInd =m-2}^{m+3}
\big\|\widehat{W}_{\ScaleInd } *a\big\|
\\
&
\;\leq\; 
2^{(m+1)\frac{\Weight}{2}}\,2^{(l+1)\frac{\Weight}{2}} 2^{-(m-2)\Weight} \big(\sum_{j=0}^5 2^{-j\Weight}\big)
\norm{a}_{B^\Weight_{\infty,\infty}} \;,
\end{align*}
due to the definition of the norm in $B^\Weight_{\infty,\infty}(\Mm)$. Swapping $m$ and $l$, respectively using Lemma \ref{lemma:comm_diagonal} for $\abs{m-l}\leq 1$ one finds that for all $m,l\in \bbN$ there is a constant $C_\Weight$ uniform in $a$ such that
$$
\norm{P_{\Lambda_m}H^{(\Weight)}_a P_{\Lambda_l}} 
\;\leq\; 
C \norm{a}_{B^\Weight_{\infty,\infty}}\, 2^{-\frac{\Weight}{2}\abs{m-l}}
\;.
$$
For any $\Weight < \tilde{\Weight} < 1$ one can further take a factor $\weight(D)^{(\tilde{\Weight}-\Weight)}$ out of the sum to bound
$$\norm{P_{\Lambda_m}H^{(\Weight)}_a P_{\Lambda_l}} 
\;\leq\; \tilde{C} \norm{a}_{B^{\tilde{\Weight}}_{\infty,\infty}}\, 2^{-\frac{\tilde{\Weight}}{2}\abs{m-l}} 2^{-(m+l)(\tilde{\Weight}-\Weight)}$$
and hence the sum \eqref{eq-HankelInf} converges absolutely. Lemma~\ref{lemma-FiniteRangeBound}(ii) then completes the proof because $(P_{\Lambda_m})_{m\geq 0}$ is an orthogonal partition of unity. 
\hfill $\Box$

\vspace{.2cm}

Now all is prepared for the application of complex interpolation, as outlined in  Appendix~\ref{app-Interpol}. The reader is referred to this appendix for the definitions of the interpolation strip $S$ as well as interpolation couples and spaces, and also the interpolation theorem.

\vspace{.2cm}

\noindent {\bf Proof} of Theorem~\ref{theo-HankelBesovP}. (The following argument is essentially the same as in \cite{Pel}.)
Let  $p>n$ and set
$$
\Weight_0 \,=\, - \frac{n}{2} \,\frac{p-2}{p}
\;, 
\qquad 
\Weight_1 \,=\, \frac{n}{p}\;,
\qquad
s_0\,=\,\frac{n}{2} + \Weight_0
\;, 
\qquad 
s_1\,=\,\Weight_1 
\;,
$$
as well as, for $z\in S$,
$$
\Weight_z \,= \,(1-z)\Weight_0 + z \Weight_1
\;,
\qquad
s_z \,=\, (1-z)s_0 + z s_1
\;.
$$

Now let us describe in detail the set-up for the application of Theorem~\ref{theo-Interpol}, by choosing two interpolation couples $(E_0,E_1)$ and $(F_0,F_1)$ and a densely defined linear operator $T_z$ from $E_0\cap E_1$ to $F_0+F_1$ with suitable continuity, analyticity and boundedness properties. The interpolation couples are
$$
(E_0,E_1)
\;=\;
\big(
B^{s_0}_{2,2}(\Mm) , B^{s_1}_{\infty,\infty}(\Mm)
\big)
\;,
\qquad
(F_0,F_1)
\;=\;
\big(L^2(\Nn ),L^\infty(\Nn)\big)
\;.
$$
The corresponding interpolation spaces $(E_0,E_1)_\theta$ and $(F_0,F_1)_\theta$ are given by \eqref{eq-LpBesovInterpol} and  \eqref{eq-InterpolLp} respectively. Then for $z \in S$, the interpolation operators are chosen to be
$$
T_z\;:\; 
\mathcal{D} \subset B^{s_0}_{2,2}(\Mm) \cap B^{s_1}_{\infty,\infty}(\Mm)
\; \to\; L^2(\Nn ) + L^\infty(\Nn)
\;,
\qquad
T_z(a)\;=\; H^{(\Weight_z)}_{a}
\;,
$$ 
where
$$
H^{(\Weight_z)}_{a} 
\;=\; 
\sum_{\ScaleInd  \geq 0} \weight(D)^{\frac{\Weight_z}{2}} \,(\one-P)\, \pi(\widehat{W}_\ScaleInd  * a) \,P \,  \weight(D)^{\frac{\Weight_z}{2}}
\;
$$
is defined on the subspace $\mathcal{D}={}^0 B^{s_0}_{2,2}(\Mm) \cap {}^0B^{s_1}_{\infty,\infty}(\Mm)$ which is dense w.r.t. the combined norm. To see that for fixed $a\in \mathcal{D}$ the map $z \mapsto T_z(a) \in L^2(\Nn) + L^\infty(\Nn)$ is analytic for $z \in S^\circ$ and continuous on the boundary choose $s_1 <\tilde{\zeta}<1$ and note that $H^{(\tilde{\zeta})}_{a}$ is norm-bounded with $
H^{(\Weight_z)}_{a}= \weight(D)^{\frac{\Weight_z-\tilde{\zeta}}{2}} H^{(\tilde{\zeta})}_{a} \weight(D)^{\frac{\Weight_z-\tilde{\zeta}}{2}}$. Since $T_z(a)$ depends on $z$ only through a bounded and analytic family of operators it is even analytic w.r.t. the operator norm, which is more restrictive than the norm of $L^2(\Nn) + L^\infty(\Nn)$.

\vspace{.1cm}

Next let us analyze the continuity properties of the operator $T_z$ on the boundary of $S$ where $\Re e( z)=0$ and $\Re e( z) = 1$. For this purpose, it is crucial that
$$
H^{(\Weight_z)}_{a} 
\;=\; 
U_z H^{(\Re e( \Weight_z))}_{a} U_z^*
$$ 
for some unitary operator $U_z \in \Nn $ and where $H^{(\Re e( \Weight_z))}_{a}$ is the weighted Hankel operator as defined above. For $\Re e( z)=0$ this operator is bounded from $E_0=B^{s_0}_{2,2}(\Mm) $ to $F_0=L^2(\Nn )$ by Proposition~\ref{prop:interp-lower} because $\Weight_0 < 1 - \frac{n}{2} \iff n < p$. For $\Re e( z) = 1$ the operator $T_z$ is bounded due to Proposition~\ref{prop:interp-upper} because $\Weight_1 < 1 \iff n < p$. In conclusion, all the hypothesis of Theorem~\ref{theo-Interpol} in Appendix~\ref{app-Interpol}) are satisfied and therefore $T_{\theta}$ extends to a bounded operator from $(E_0,E_1)_\theta$ to $(F_0,F_1)_\theta$. The suitable choice is $\theta = \frac{p-2}{p}\in (0,1)$ which restrict $p$ to be in $(2,\infty)$. Then $\Weight_\theta=0$, $s_\theta=\frac{n}{p}$ and the interpolation spaces are $(E_0,E_1)_\theta=B^{\frac{n}{p}}_{p,p}(\Mm)$ to $(F_0,F_1)_\theta=L^p(\Nn)$ so that $T_{\theta}$ extends to a bounded operator
$$
T_{\theta}\;:\; B^{\frac{n}{p}}_{p,p}(\Mm) \; \to\; L^p(\Nn)
\;,
\qquad
T_{\theta}(a)\;=\;H_{a}
\;, 
\qquad \forall\; a\in \mathcal{D}
\;.
$$

Finally we note that for general $a \in \Mm \cap B^{\frac{n}{p}}_{p,p}(\Mm)$ we can write $a= \sum_{j=0}^\infty \widehat{W_j}*a$ and $H_a = \sum_{j=0}^\infty H_{\widehat{W_j}*a}$ with convergence in the weak operator topology. Since $T_\theta(\sum_{j=0}^N \widehat{W_j}*a) = H_{\sum_{j=0}^N \widehat{W_j}*a}=\sum_{j=0}^N  H_{\widehat{W_j}*a}$ holds due to $\sum_{j=0}^N \widehat{W_j}*a \in {}^0 B^{s_1}_{\infty,\infty}(\Mm)$ this is a Cauchy-sequence in $L^p(\Nn)$-norm for $N\to \infty$ and hence we have $T_\theta a = H_a$ by Lemma~\ref{lemma:convergence}(ii).

\vspace{.1cm}

It remains to show the claim for $n=1$ and $1 \leq p \leq 2$. This follows again by interpolation, albeit a more elementary version simply using the boundedness of the two maps $a\in B_{1,1}^1(\Mm)\mapsto H_a\in L^1(\Nn)$  (see  Theorem \ref{theo-Hankel})  and $a\in B_{2,2}^{\frac{1}{2}}(\Mm)\mapsto H_a\in L^2(\Nn)$ (see Proposition~\ref{prop:interp-lower}). Indeed the corresponding interpolation spaces for $\theta=\frac{2(p-1)}{p}$ are $(B_{1,1}^1(\Mm),B_{2,2}^{\frac{1}{2}}(\Mm))_\theta=B_{p,p}^{\frac{1}{p}}(\Mm)$ and $(L^1(\Nn),L^2(\Nn))_\theta=L^p(\Nn)$ so that again Theorem~\ref{theo-Interpol} allows to conclude.
\hfill $\Box$

\section{Converse of the Peller criterion}
\label{sec-PellerConverse}

This section proves the following converse of the Peller criterion (Theorem~\ref{theo-HankelBesovP}) which includes a statement about the commutator version of the Hankel operator
\begin{equation}
\label{eq:Hankel_commutator}
\hat{H}_a
\;=\;  [\sgn(\DD ),\pi(a)]
\;.
\end{equation}

\begin{theorem}
\label{theorem:hankel_converse}
For $n < p < \infty$ or $n=p=1$ and $a \in \Mm \cap L^p(\Mm)$ the following statements are equivalent
\begin{enumerate}
\item[{\rm (i)}] $a \in B^{\frac{n}{p}}_{p,p}(\Mm)$
\item[{\rm (ii)}] $\hat{H}_a \in L^p(\Nn)$
\item[{\rm (iii)}] $H_a, H_{a^*} \in L^p(\Nn)$
\end{enumerate}
Moreover, the linear map $a\in\Mm \cap B^{\frac{n}{p}}_{p,p}(\Mm) \mapsto (a,\hat{H}_a)\in \Mm\oplus \Nn$ extends to a bounded and invertible map from $B^{\frac{n}{p}}_{p,p}(\Mm)$ to $L^p(\Mm) \oplus L^p(\Nn)$. In particular, $\norm{a}_p +\|\hat{H}_a\|_p$  is an equivalent norm for $B^{\frac{n}{p}}_{p,p}(\Mm)$.
\end{theorem}

At least for $n=1$, it is clear that (iii) cannot be weakened to $H_a \in L^p(\Nn)$ since there are symbols with $a\neq 0$, but $H_a=0$. In the version by Peller \cite{Peller82}, this is circumvented by projecting to the part with $\sigma_\alpha(a)\subset (-\infty, 0]$. For higher dimensional generalizations, similar differences between one sided-problems involving $H_a$ and two-sided problems involving the commutator arise \cite{Xia2008,FX2013}. The proof of Theorem~\ref{theorem:hankel_converse} implements a duality argument introduced in \cite{Peller82} and refined in \cite{JansonPeetre88}. It requires $L^p$-estimates, $1<p<2$, for a commutator version of the weighted Hankel operators 
$$
\hat{H}^{(\zeta)}_{b,r,\delta} \;=\; \hat{\weight}_{r,\delta}(D)^{\frac{\zeta}{2}}  [\sgn(\DD+ \gamma\cdot r),\pi(b)]\hat{\weight}_{r,\delta}(D)^{\frac{\zeta}{2}}
\;
$$
with a shift $r \in [0,1]^n$ and a regulator $\delta > 0$ entering into the weight function
$$
\hat{\weight}_{r,\delta}(\dualvar)
\;=\; 
\chi(|\dualvar+r|^2 \geq \delta^2)\, \abs{\gamma\cdot (\dualvar +r)}
\;.
$$ 
Note that for $G=\bbR^n$, the offset $r$ is inessential due to $\hat{H}^{(\zeta)}_{b,r,\delta} = \hat{\alpha}_r(\hat{H}^{(\zeta)}_{b,0,\delta})$, but it cannot be eliminated for more general $G$. The lower endpoint for the interpolation is obtained by once again adapting an argument of \cite{JansonPeetre88} that only needs to be supplemented with numerical estimates appropriate for the present setting.

\begin{lemma}
\label{lem:hankelbesovl1}
For any $\zeta > 1-n$, there is a constant $C>0$ that is uniform in $R>0$,  $r\in [0,1]^n$ and $\delta > 0$,  such that for all $a\in L^1(\Mm)$ with $\sigma_\alpha(a)\subset B_R$, one has 
$$
\big\|\hat{H}_{a,r,\delta}^{(\zeta)}\big\|_1 
\;\leq\; 
C\, R^{n+\zeta}\, \norm{a}_1
\;.
$$
\end{lemma}

\noindent {\bf Proof.}
For $y\in \bbR^n$,  let $Q_{L,y}$  be the cube with center $y$ and sides $L$, {\it cf.} Lemma~\ref{lemma:comm_decay}. Set $L=2R$ such that $\pi(a) P_{Q_{L,y}} = P_{Q_{3L,y}} \pi(a) P_{Q_{L,y}}$ due to Lemma~\ref{lemma:help}. We abbreviate $\DD_r = \DD + \gamma \cdot r$ and again use a smooth indicator function $\hat{\chi}_{Q,y}$ and a smooth approximation $g$ of the sign function exactly as in the proof of Lemma~\ref{lemma:comm_decay}. For $\abs{y_1+r},\abs{y_2+r} > 4L$ one can thus write
\begin{align*}
P_{Q_{L,y_1}} &  [\sgn(\DD_r), \pi(a)] P_{Q_{L,y_2}} 
\\
&
\;=\; 
P_{Q_{L,y_1}} \hat{\chi}_{L,{y_1}}(D) [g(\DD_r), \pi(a)] \hat{\chi}_{L,{y_2}}(D) P_{Q_{L,y_2}}\\
&
\;=\; 
P_{Q_{L,y_1}} \int_{\bbR^n} \pi\left(
\int_{\bbR^n} (\widehat{\psi}_{\dualvar,r}*a)\, e^{-2\pi \imath\,  {\dualvar}\cdot s}\;\difd{{\dualvar}}\right)
e^{2\pi \imath\,  D\cdot s} \,P_{Q_{L,y_2}}\;\difd{s}
\;,
\end{align*}
with the smooth multiplier 
$$
\psi_{\dualvar,r}(\lambda) 
\;=\; 
\hat{\chi}_{L,{y_1}}(k+\lambda)(g(k+r+\lambda)-g(k+r))\hat{\chi}_{L,{y_2}}(k)
\;.
$$ 
Hence proceeding as in \eqref{eq-PaPestimate}
\begin{align*}
&
\norm{P_{Q_{L,y_1}} [\sgn(\DD_r), \pi(a)] P_{Q_{L,y_2}}}_1 
\\
\;\;\;\;
&\;\leq\;  \int_{\bbR^n}  \norm{P_{Q_{L,y_1}}\;\pi\left(
	\int_{\bbR^n}(\widehat{\psi}_{\dualvar,r}*a)\, e^{-2\pi \imath\,  {\dualvar}\cdot s}\,\difd{{\dualvar}} \right)
	e^{2\pi \imath\,  D\cdot s}P_{Q_{L,y_2}}}_1\difd{s}\\
\;\;\;\;
&\;\leq\; \int_{\bbR^n} \norm{P_{Q_{L,y_1}}}_2\;
\norm{\int_{\bbR^n}(\widehat{\psi}_{\dualvar,r}*a)\, e^{-2\pi \imath\,  {\dualvar}\cdot s}\, \difd{{\dualvar}} }_1 \norm{P_{Q_{L,y_2}}}_2\difd{s}\\
\;\;\;\;
&\;\leq\; (2 R)^n \int_{\bbR^n}  \int_{\bbR^n}  \abs{\int_{\bbR^n}  \int_{\bbR^n}
	\psi_{\dualvar,r}(\lambda)\, e^{-2\pi \imath  \,{\dualvar}\cdot s}\,e^{2\pi \imath  \,\lambda \cdot t} \difd{{\dualvar}} \,\difd{\lambda}}
	\norm{a}_1\difd{t}\,\difd{s} ,
\end{align*}
where in the last inequality  the norm of the multiplier was estimated in the same manner as in Lemma~\ref{lemma:comm_decay}  in terms of the $L^1$-norm of its Fourier transform.  Note that the integral is up to the shift by $r$ identical to the one in Lemma~\ref{lemma:comm_decay}. Since the estimate of the multiplier norm there is invariant under translation, one obtains by the same reasoning a bound
$$
\norm{P_{Q_{L,y_1+r}} [\sgn(\DD_r), \pi(a)] P_{Q_{L,y_2+r}}}_1 
\;\leq\; 
c_1 R^n \frac{L}{\abs{y_1}+\abs{y_2}} \norm{a}_1
\;,
$$
with a constant that is uniform in $r \in [0,1]^n$ and which holds for all $\abs{y_1}>4L, \abs{y_2}>4L$ with $\abs{y_1-y_2}\leq 2^4 L$. Applying Lemma~\ref{lemma:help} and then writing $Q_{3L,y+r}$ as a union of a fixed number of cubes $Q_{L,y_2+r}$, one thus has
\begin{align*}
\norm{[\sgn(\DD_r), \pi(a)] P_{Q_{L,y+r}}}_1 
&
\;=\; 
\norm{P_{Q_{3L,y+r}}[\sgn(\DD_r), \pi(a)] P_{Q_{L,y+r}}}_1 
\\
&
\;\leq\; 
c_2 \,R^n\, \frac{L}{\abs{y}}\, \norm{a}_1
\;,
\end{align*}
with a uniform constant $c_2$ for $\abs{y} > 8 L$.
This bound is sufficient for large $y$. The remaining cubes for a partition of unity are regrouped to $E_r = \bigcup_{\abs{y}\leq 8L}Q_{L,y+r}$. With $\tilde{E}_r=\bigcup_{\abs{y}\leq 8L} Q_{3L,y+r}$, $P_{E_r}=\chi_{E_r}(D)$ and $P_{\tilde{E}_r}=\chi_{\tilde{E}_r}(D)$ let us use Proposition~\ref{prop:lp-embedding} to estimate
\begin{align*}
&
\norm{\hat{\weight}_{r,\delta}^{\frac{\zeta}{2}}(D)[\sgn(\DD_r ),\pi(a)] \hat{\weight}_{r,\delta}^{\frac{\zeta}{2}}(D) P_{E_r}}_1  
\\
&
\;\;\;\;
\; = \;\norm{P_{\tilde{E}_r} \hat{\weight}_{r,\delta}^{\frac{\zeta}{2}}(D) [\sgn(\DD_r ),\pi(a)] \hat{\weight}_{r,\delta}^{\frac{\zeta}{2}}(D)  P_{E_r}}_1 \\
& 
\;\;\;\;
\;\leq \;
2 \norm{P_{\tilde{E}_r} \hat{\weight}_{r,\delta}^{\frac{\zeta}{2}}(D) \pi(a) \hat{\weight}_{r,\delta}^{\frac{\zeta}{2}}(D)  P_{E_r}}_1 \\
&
\;\;\;\;
\;\leq\; 
2\norm{P_{\tilde{E}_r} \hat{\weight}_{r,\delta}^{\frac{\zeta}{2}}(D)  \pi(\abs{a}^{\frac{1}{2}})}_2\, \norm{\pi(\abs{a}^{\frac{1}{2}}) \hat{\weight}_{r,\delta}^{\frac{\zeta}{2}}(D)  P_{E_r}}_2\\
&
\;\;\;\;
\;\leq\; c_3 \,R^{n+\zeta}\, \norm{a}_1
\;,
\end{align*}
uniformly in $\delta > 0$ and $r\in [0,1]^n$ since the weight function stays integrable for $\delta=0$. Combining the estimates gives
\begin{align*}
&
\norm{\hat{\weight}_{r,\delta}^{\frac{\zeta}{2}}(D)[\sgn(\DD_r),\pi(a)] \hat{\weight}_{r,\delta}^{\frac{\zeta}{2}}(D)}_1 
\\
&
\;\;\;\;\;\;
\;=\;  
\norm{\sum_{y\in \bbZ^n} \hat{\weight}_{r,\delta}^{\frac{\zeta}{2}}(D)[\sgn(\DD_r),\pi(a)] \hat{\weight}_{r,\delta}^{\frac{\zeta}{2}}(D) P_{Q_{L,Ly+r}}}_1
\\
&
\;\;\;\;\;\;
\;\leq\; 
\sum_{y\in \bbZ^n,\, \abs{y}>8} \norm{P_{Q_{3L,Ly+r}}\hat{\weight}_{r,\delta}^{\frac{\zeta}{2}}(D)[\sgn(\DD_r ),\pi(a)] \hat{\weight}_{r,\delta}^{\frac{\zeta}{2}}(D) P_{Q_{L,Ly+r}}}_1 
\\ 
&
\;\;\;\;\;\;
\;\;\;\;\;\;\;\;\;\;\;\;\;\;
+\; \norm{P_{\tilde{E}_r}\hat{\weight}_{r,\delta}^{\frac{\zeta}{2}}(D)[\sgn(\DD_r),\pi(a)] \hat{\weight}_{r,\delta}^{\frac{\zeta}{2}}(D) P_{E_r}}_1
\\
&
\;\;\;\;\;\;
\;\leq\; 
\sum_{y\in \bbZ^n ,\, \abs{y}>8} c_4 \,R^{n+\zeta} \abs{y}^{\zeta-1} \norm{a}_1  \;+\; c_3\, R^{n+\zeta} \norm{a}_1 
\\
&
\;\;\;\;\;\;
\;\leq \;c_5\, R^{n+\zeta} \norm{a}_1
\;.
\end{align*}
This concludes the proof.
\hfill $\Box$

\vspace{.2cm}

The same argument as in the proof of Theorem~\ref{theo-Hankel} (with Lemma~\ref{lem:hankelbesovl1} replacing Lemma~\ref{lemma:help2})  then readily allows to show the following

\begin{proposition}
\label{prop:hankelbesovl1}
For $-n < \zeta < 1-n$ and $a \in {}^0B_{1,1}^{n+\zeta}(\Mm)$, there is a uniform constant such that
$$
\big\|\hat{H}_{a,r,\delta}^{(\zeta)}\big\|_1 
\;\leq\; 
C \norm{a}_{B_{1,1}^{n+\zeta}}
\;.
$$
\end{proposition}

Furthermore, the proof of Proposition~\ref{prop:interp-lower} can be adapted to show

\begin{proposition}
\label{prop:hankelbesovl2}
For $0 < \frac{n}{2} + \Weight < 1$ and $a \in \Mmc$ there is a constant that is uniform in $a$, $r \in [0,1]^n$ and $\delta > 0$ such that
$$
\norm{\hat{H}^{(\Weight)}_{a,r, \delta}}_2
\;\leq\; 
C \,\norm{a}_{B^{\frac{n}{2}+\Weight}_{2,2}(\Mm)}
$$
\end{proposition}

Now one can interpolate between those endpoints:

\begin{lemma}
\label{lemma:hankelbesovl1tol2}
For $1<p<2$ and $0 <\zeta + n < 1$ the map $a \mapsto \hat{H}_{a,r,\delta}^{(\frac{\zeta}{p})}$ is bounded from $B^{\frac{n+\zeta}{p}}_{p,p}(\Mm)$ to $L^p(\Nn)$ with norm that is uniformly bounded in $r \in [0,1]^n$ and $\delta > 0$.
\end{lemma}

\noindent{\bf Proof.}
Let us consider the two maps $T_i: B^{s_i}_{i+1,i+1} \to L^{i+1}(\Nn)$, $i=0,1$, given by $a \mapsto \hat{H}_{a,r,\delta}^{(\frac{\zeta}{i+1})}$ and where
$$
s_0\;=\;n \,+\, \zeta\;, 
\qquad 
s_1
\;=\;\frac{n}{2} \,+\, \frac{\zeta}{2}
\;.
$$
Due to the choice of $\zeta$, they are bounded by the Propositions~\ref{prop:hankelbesovl1} and \ref{prop:hankelbesovl2} respectively.
For the interpolation between these operators, let us choose 
$$
\frac{1}{p} 
\;=\; 
\frac{1-\theta}{1}\, +\, \frac{\theta}{2} 
\quad \iff\quad 
\theta\;=\;2\,-\,\frac{2}{p}
\;.
$$ 
Then
$$
s_\theta 
\;=\; \Big(\frac{2}{p}-1\Big)s_0 \,+\, \Big(2-\frac{2}{p}\Big)s_1 
\;=\; 
\frac{n}{p} \,+\, \frac{\zeta}{p}
\;,
$$
and hence the argument can be completed as in the proof of Theorem~\ref{theo-HankelBesovP}.
\hfill $\Box$

\vspace{.2cm}

For the duality argument, a characterization of the dual spaces of Besov spaces are needed. In the classical case those are Besov spaces with negative smoothness which embed into the space of tempered distributions. Since there is no such convenient ambient topology in the present situation, we do not pursue this approach further in this work.

\begin{proposition}
Let $0< s < \infty, 1\leq q<\infty$ and $1 =\frac{1}{q}+\frac{1}{\tilde{q}}$.
Let $E$ be a Banach space with strongly continuous $\bbR^n$-action $\beta$ and let $\beta^*: E' \times \bbR^n \to E'$ be the weak$^*$-continuous action $\beta^*_t(\phi)=\phi \circ \beta_t$. 
For any continuous linear functional $\phi \in B^{s}_{q}(E)'$, the sequence $(\widehat{W_j}*\phi)_{j\in \bbN}$ lies in $\ell_{-s}^{\tilde{q}}(E')$ and $\phi = \sum_{j\in \bbN} \widehat{W_j}*\phi$ converges in the weak$^*$-topology of $(B^{s}_{q}(E))'$. 
\end{proposition} 

\noindent{\bf Proof.}
First note that $(\widehat{W_j}*\phi)(a)=\beta^*_{\calF^{-1}W_j}(a)=\phi(\widehat{W_j}*a)$ so that $\widehat{W_j}*\phi$ extends continuously to an element of $E'$ since $a \in E \mapsto \widehat{W_j}*a\in B^s_q(E)$ is a bounded map.
Convergence of $\phi = \sum_{j\in \bbN} \widehat{W_j}*\phi$in the weak$^*$-topology follows from the weak$^*$-continuity and the fact that $(W_j)_{j\in \bbN}$ forms a partition of unity. To see the summability condition we adapt an argument due to Peetre \cite{Peetre72}, which is based on the fact that $(\ell_s^q(E))'= \ell_{-s}^{\tilde{q}}(E')$ for any Banach space $E$ and $1 \leq q <\infty$. By the Hahn-Banach theorem, there is a bounded functional $T \in  \ell_s^q(E)'$ which extends
$$
T((f_j)_{j\in \bbN}) 
\;=\; 
\phi(b)
$$
from the subspace spanned by sequences of the form $(f_j)_{j\in\bbN} = (\widehat{W_j}*b)_{j\in \bbN}\in\ell^q_s(E)$ for some $b \in B^s_{q}(E)$. Hence there is a sequence $(T_j)_{j\in \bbN} \in \ell_{-s}^{\tilde{q}}(E')$ such that
$$
T((f_j)_{j\in \bbN})
\;=\; 
\sum_{j=0}^\infty T_j(f_j)
$$
and thus one can write for all $b\in B^s_{q}(E)$
$$
\sum_{j=0}^\infty (\widehat{W_j}*\phi)(b)=\phi(b)
\;=\;  
\sum_{j=0}^\infty T_j(\widehat{W_j}*b)
\;=\;
\sum_{j=0}^\infty (\widehat{W_j}*T_j)(b)
\;.
$$
The identity $\widehat{W_{j}} = \widehat{W_{j}}*(\widehat{W_{j-1}}+\widehat{W_{j}}+\widehat{W_{j+1}})$ implies
$$
(\widehat{W}_j*\phi)(b)
\;=\; 
\phi(\widehat{W}_j*b) 
\;=\; 
(\widehat{W}_j*(\widehat{W}_{j-1}*T_{j-1}+\widehat{W}_{j}*T_j+\widehat{W}_{j+1}*T_{j+1}))(b)
\;,
$$ 
which leads to $\big\|(\widehat{W_j}*\phi)_{j\in \bbN}\big\|_{\ell_{-s}^{\tilde{q}}(E')} \leq c \norm{(T_j)_{j\in \bbN}}_{\ell_{-s}^{\tilde{q}}(E')} = c \norm{\phi}_{B^s_q(E)'}$. 
\hfill $\Box$

\begin{corollary}
\label{cor:besovdual}
Let $0< s < \infty, 1 \leq p,q<\infty$ and $\frac{1}{p}+\frac{1}{\tilde{p}}= 1 =\frac{1}{q}+\frac{1}{\tilde{q}}$. Under the isomorphism $L^p(\Mm)' \simeq L^{\tilde{p}}(\Mm)$, every element $\phi \in B^s_{p,q}(\Mm)'$ corresponds to a sequence $(\phi_j)_{j\in \bbN}$ in $L^{\tilde{p}}(\Mm)$ where $\phi_j=\widehat{W}_{j}*\phi$ and  with
$$
\Big(\sum_{j=0}^\infty 2^{-\tilde{q}sj} \norm{\phi_j}_{\tilde{p}}^{\tilde{q}}\Big)^{\frac{1}{\tilde{q}}} \,< \;c \norm{\phi}_{B^s_{p,q}(\Mm)'}
\; <\; \infty
$$
and 
$$
\phi(b)\;=\; \sum_{j=0}^\infty \Tt(\phi_j^*b)
$$
for every $b \in B^s_{p,q}(\Mm)$.
\end{corollary}

We can now proceed with the proof of the main claim of Theorem~\ref{theorem:hankel_converse}. The strategy is to employ the sufficient direction of the Peller criterion to construct from a Hankel operator $\hat{H}_a$ a functional on a certain Besov space and then use the characterization of Corollary \ref{cor:besovdual} to relate the Besov norm of $a$ to the norm of that functional.

\begin{proposition}
\label{propf:pellernecessary1}
Let $n < p < \infty$. If $a\in L^p(\Mm) \cap \Mm$ has a $p$-summable commutator $\hat{H}_a \in L^p(\Nn)$, then $a \in B^{\frac{n}{p}}_{p,p}(\Mm)$ and there is a uniform constant $C$ such that
$$
\norm{a}_{B^{\frac{n}{p}}_{p,p}} 
\;\leq\; 
C (\norm{a}_p + \big\|\hat{H}_a\big\|_p)
\;.
$$
\end{proposition}

\noindent{\bf Proof.} Let us first treat the case $n\geq 2$. Since a uniform bound will be proved, one can assume $a\in \Mmc$ by density and note that $a=\sum_{j=0}^\infty \widehat{W}_j*a$ and $H_a =\sum_{j=0}^\infty H_{\widehat{W}_j*a}$ with convergence in their respective $L^p$-norms.

\vspace{.1cm}

Let us now choose any $-n < \zeta < 1-n$ and set $\xi=\frac{\zeta}{q}$   for $1 = \frac{1}{p} + \frac{1}{q}$. Note that then $1<q<2$ so that for $b\in B^{\frac{n}{q}+\xi}_{q,q}(\Mm)$ one has $\hat{H}_{b,r,\delta}^{(\xi)}\in L^{q}(\Nn)$ by Lemma~\ref{lemma:hankelbesovl1tol2}. Furthermore, $a\in B^{\frac{n}{p}}_{p,p}(\Mm)$ implies that $\hat{H}_a \in L^p(\Nn)$ by Theorem~\ref{theo-HankelBesovP}. Hence by Corollary~\ref{cor:besovdual}, the map $\phi_{a,\delta}:  B^{\frac{n}{q}+\xi}_{q,q}(\Mm) \to \bbC$ given by 
$$
\phi_{a,\delta}(b)\;=\;\hat{\Tt}_\alpha\big(\hat{H}_a^* \hat{H}_{b,0,\delta}^{(\xi)}\big)
$$ 
is a bounded linear functional for any $\delta>0$. Let us also shift the commutator by defining
$$
\hat{H}_{a,r} \;=\; \hat{H}_{a,r,0}^{(0)} = [\sgn(\DD+\gamma\cdot r),\pi(a)] 
\;.
$$
Since the function $g_r(k)= \sgn(\gamma\cdot (k+r))-\sgn(\gamma\cdot k)$ decays like $\frac{1}{\abs{k}}$, see \eqref{eq-SgnSgn}, it defines an element of $L^p(\hat{G})$ and thus Proposition~\ref{prop:lp-embedding} shows 
$$
\big\|\hat{H}_{a,r}\big\|_p 
\;\leq\; \big\|\hat{H}_a\big\|_p + 2 \norm{g_r}_p \, \norm{a}_p 
\;\leq\; 
\big\|\hat{H}_{a}\big\|_p \,+\, c_1 \norm{a}_p
$$
with a constant $c_1$ that is uniform for $r \in [0,1]^n$. By Lemma~\ref{lemma:hankelbesovl1tol2} there is also a bound
$$
\big\|\hat{H}_{b,r,\delta}^{(\xi)}\big\|_{q} 
\;\leq \;
c_2 \norm{b}_{B^{\frac{n}{q}+\xi}_{q,q}}
\;,
$$
uniformly in $r \in [0,1]^n$ and $\delta > 0$. Thus
\begin{equation}
\label{eq:hankelbesovfunc}
\tilde{\phi}_{a,\delta}\;:\; B^{\frac{n}{q}+\xi}_{q,q}(\Mm)\; \mapsto\; \int_{[0,1]^n}  \hat{\Tt}_\alpha\big(\hat{H}_{a,r}^* \hat{H}_{b,r,\delta}^{(\xi)}\big)\difd{r}
\end{equation}
 is a bounded linear functional with  
 $$
 \norm{\tilde{\phi}_{a,\delta}}_{B^{\frac{n}{q}+\xi}_{q,q}(\Mm)'} 
 \;\leq \;
 c_3 \big(\norm{a}_p \,+\, \big\|\hat{H}_a\big\|_p\big)
 \;.
 $$ 
 For $b \in \Mmc$, one can write out \eqref{eq:hankelbesovfunc}  using Lemma~\ref{lemma:l2norm} and  Corollary~\ref{coro-l2scalprod} as a scalar product to obtain
\begin{align*}
\tilde{\phi}_{a,\delta}(b)
\;=\; 
\int_{\hat{G}}\Psi^{(\xi)}_\delta(\lambda) \,\langle a_\lambda, b_\lambda\rangle_{\Hh_\lambda}\,\mu(\difd \lambda)
\;=\; 
\Tt\Big(\big(\widehat{\Psi_\delta^{(\xi)}}*a\big)^* b\Big)
\;,
\end{align*}
with the unbounded multiplier
\begin{align*}
\Psi^{(\xi)}_\delta(\lambda) 
&
\;=\; \int_{[0,1]^n} \int_{\hat{G}}  \abs{k+r+\lambda}^{\frac{\xi}{2}}\,  \chi((k+r+\lambda)^2 \geq \delta^2)\, \chi((k+r)^2 \geq \delta^2)
\\ 
&
\hspace{2.6cm}
\abs{\sgn(\gamma\cdot(k+r+\lambda))-\sgn(\gamma\cdot (k+r))}_2^2 \abs{k+r}^{\frac{\xi}{2}}\difd k \, \difd{r} \\
&
\;\;
\;=\;
\int_{\bbR^n}  \abs{k+\lambda}^{\frac{\xi}{2}}\abs{\sgn(\gamma\cdot(k+\lambda))-\sgn(\gamma\cdot k)}_2^2 \abs{k}^{\frac{\xi}{2}} 
\\
&
\hspace{2.6cm}
\chi((k+\lambda)^2 \geq \delta^2)\, \chi(k^2 \geq \delta^2)\,\difd k 
\;,
\end{align*}
which is implicitly restricted to a bounded multiplier by applying a large enough cut-off. 
Note that the integral converges for $\xi < 2-n$ which is always satisfied due to $q < \frac{n}{n-1}$. The singularities also stay integrable for $\delta \to 0$ and $\lambda \not= 0$. It becomes apparent that the average in \eqref{eq:hankelbesovfunc} was introduced so that one may always integrate over $\bbR^n$ for all different cases of $\hat{G}$. Note that $\Psi_\delta^{(\xi)}$ is a radial function $\Psi_\delta^{(\xi)}(\lambda)=\Psi_\delta^{(\xi)}(\abs{\lambda}e_1)$ and for $\delta=0$ it is also homogeneous of degree $\xi+n$. Hence it takes the simple form $\Psi_0^{(\xi)}(\lambda) =  \abs{\lambda}^{n+\xi} \Psi_0^{(\xi)}(e_1)$. This is the multiplier corresponding to the Riesz potential $I^{n+\xi}$ which also appeared in Lemma~\ref{lem-IWbound}. By dominated convergence the limit
$$
\tilde{\phi}_{a,0}(b) 
\;=\; 
\lim_{\delta\to 0} \tilde{\phi}_{a,\delta}(b)
\; =\; 
\Tt((\widehat{\Psi^{(\xi)}_0} *a)^*b)
$$
exists for all $b \in \Mmc$ and therefore densely defines a bounded functional with 
$$
\norm{\tilde{\phi}_{a,0}}_{B^{\frac{n}{q}+\xi}_{q,q}(\Mm)'}  
\;\leq\; 
\liminf_{\delta\to 0}\norm{\tilde{\phi}_{a,\delta}}_{B^{\frac{n}{q}+\xi}_{q,q}(\Mm)'} 
\;\leq\; 
c_3 \big(\norm{a}_p + \norm{\hat{H}_a}_p\big)
\;.
$$
By Corollary \ref{cor:besovdual} there exists  $(\tilde{a}_j)_{j\in \bbN} = (\widehat{W_j}*\tilde{\phi}_{a,0})_{j\in \bbN}\in \ell_{-n/q-\xi}^{p}(L^p(\Mm))$ representing $\tilde{\phi}_{a,0}$ such that
$$
\tilde{\phi}_{a,0}(b)
\;=\; 
\Tt((\widehat{\Psi_0^{(\xi)}}*a)^* b)
\;=\;
\sum_{j=0}^\infty\Tt(\tilde{a}_j^* b)
\;.
$$
Substituting the Ansatz $b=\widehat{W_j}*\tilde{b}$ and comparing coefficients yields 
$$
\big(\widehat{\Psi_0^{(\xi)}} * \widehat{W_j}\big)* a
\;=\;
\widehat{W_j}*(\tilde{a}_{j-1} + \tilde{a}_j +\tilde{a}_{j+1})
\;,
$$ 
and, since $\Psi_0^{(\xi)}$ is strictly positive on the support of $W_j$ for $j>0$, the relation can be inverted to 
$$
\widehat{W_j} * a 
\;=\; 
\left(\frac{W_j}{\Psi_0^{(\xi)}}\right)^\wedge * (\tilde{a}_{j-1} + \tilde{a}_j +\tilde{a}_{j+1}) 
\;=\; 
\left(\frac{W_{j-1}+W_j+W_{j+1}}{\Psi_0^{(\xi)}}\right)*\tilde{a}_j
\;,
$$ 
for all $j>1$. To estimate the norm of the multiplier on the right let us use the homogeneity to bound the norm of its Fourier transform by
$$
\norm{\calF^{-1} \frac{W_j}{\Psi_0^{(\xi)}}}_{1} 
\;=\; 
\norm{\calF^{-1} \frac{W_1} {\Psi_0^{(\xi)}(2^{j-1} \cdot)}}_{1} 
\;\leq\; 
c_4\, 2^{-j(n+\zeta)}
$$
with a constant that is uniform for $j>0$. Hence one has
$$
\norm{\widehat{W_j} * a}_p 
\;=\; 
\norm{\left(\frac{W_{j-1}+W_j+W_{j+1}}{\Psi_0^{(\xi)}}\right)^\wedge * \tilde{a}_j}_p 
\;\leq\; 
c_5\, 2^{-j(n+\xi)} \norm{\tilde{a}_j}_p
$$
uniformly for $j>1$. This implies 
$$
\norm{a}_{B^{\frac{p}{n}}_{p,p}} 
\;\leq\; 
c_6\, \norm{\tilde{a}}_{\ell_{-n/q- \xi}^{p}(L^p(\Mm))} 
\;+\; c_7\, \norm{a}_p 
\;\leq\; 
c_8\big( \norm{a}_p + \norm{\hat{H}_a}_p\big)
\;.
$$
Let us now comment on $n=1$. The same strategy as above applies for the range $2 < p < \infty$ and for $1 < p < 2$ one uses analogous reasoning for the functionals $b \in B^{1/q}_{q,q}(\Mm) \mapsto \hat{\Tt}_\alpha(\hat{H}_a^* \hat{H}_b)$ which are bounded by Theorem \ref{theo-HankelBesovP} since $2 < q < \infty$. The corresponding multiplier is then computed explicitly to be $\Psi_0^{(0)}(\lambda) = 4 \abs{\lambda}$, so that one can complete  the proof as above.  
\hfill $\Box$

\vspace{.2cm}

Trying to apply the argument to the case $p=1=n$ directly fails, since the dual space of $B^{s}_{\infty,\infty}(\Mm)$ is too large to be useful here. It is preferable to use preduals which requires to strengthen the $L^\infty$- endpoint estimate:

\begin{lemma}
\label{lemma:hankellinfty_refined}
For $n=1$ and $s>0$, there is a bounded map $\ell^\infty_s(\Mm) \to \Nn$, densely defined on finite sequences by
$$
(a_j)_{j\in \bbN} 
\;\mapsto\; 
H^{(s)}_{\eta(a)}
$$ 
with 
$$
\eta\,:\,\ell_s^\infty(\Mm)\to \Mm, \quad (a_j)_{j\in \bbN} 
\;\mapsto\; 
\sum_{j=0}^\infty \widehat{W_j}*a_j
\;.
$$
\end{lemma} 

\noindent {\bf Proof.} 
Since $n=1$, the spectral conditions simplify the Hankel operator greatly to
$$
H_{\eta(a)}^{(s)}
\;= \;
\sum_{j\geq 0} \weight(D)^{\frac{s}{2}}\, P_{[-2^{j+2},0)}\, \pi(\widehat{W_j} *a_j) \,P_{[0,2^{j+2})}\, \weight(D)^{\frac{s}{2}}
\;,
$$
with a sum consisting of only finitely many terms non-zero terms. To prove a bound on this, let us use a dyadic decomposition $\one = \sum_{l=0}^\infty P_{\Lambda_l}$ with supports $\Lambda_l$ as in \eqref{eq-DyadicDecomp} and then focus on the matrix elements $P_{\Lambda_m}H_{\eta(a)}^{(s)} P_{\Lambda_l}$, first for the case $0 < m \leq l$: 
\begin{align*}
P_{\Lambda_m}\,H^{(s)}_{\eta(a)}\, P_{\Lambda_l} 
&
\;=\; 
\sum_{j \in \bbN} P_{\Lambda_m}  \, \weight(D)^{\frac{s}{2}}\, P_{[-2^{j+2},0)}\, \pi(\widehat{W_j} *a_j)\, P_{[0,2^{j+2})}\, P_{\Lambda_l} \weight(D)^{\frac{s}{2}}
\\
&
\;= \; 
\sum_{j \geq l-1} P_{\Lambda_m} \, \weight(D)^{\frac{s}{2}} \,P_{[-2^{j+2},0)} \,\pi(\widehat{W_j} *a_j) \,P_{\Lambda_l}\, \weight(D)^{\frac{s}{2}}\\
&
\;=\; 
\sum_{j \geq l-1} P_{\Lambda_m} \, \weight(D)^{\frac{s}{2}} \,P_{[-2^{j+2},0)} \,P_{\Lambda_l + \text{supp}(W_j)} \,\pi(\widehat{W_j} *a_j) \, P_{\Lambda_l} \,\weight(D)^{\frac{s}{2}}
\;,
\end{align*}
where in the last step \eqref{eq-help} was used. An examination of the constraints shows that all terms except those with $j \in [l-1,l+1]$ must vanish. Hence bounding $\weight$ implies
\begin{align*}
\big\|P_{\Lambda_m}
\,H^{(s)}_{\eta(a)}\, P_{\Lambda_l}\big\| 
&
\;\leq\;
2^{(m+1)\frac{s}{2}}\, 2^{(l+1)\frac{s}{2}} \,\left(
\big\|\widehat{W}_{l-1} *a_{l-1}\big\|\,+\,\big\|\widehat{W}_{l}*a_{l}\big\|\,+\,\big\|\widehat{W}_{l+1} *a_{l+1}\big\|
\right)
\\
&
\;\leq\; 
2^{(m+1)\frac{s}{2}}\,2^{(l+1)\frac{s}{2}} \,\norm{a}_{\ell_s^\infty(\Mm)}2^{-(l-1)s} (1\, +\, 2^{-s}+2^{-2s})
\;,
\end{align*}
due to the definition of the norm in $\ell_s^\infty(\Mm)$. Treating the other cases analogously, one finds that for all $m,l\in \bbN$ there is a constant $C$ uniform in $a$ such that
$$
\norm{P_{\Lambda_m}H^{(s)}_{\eta(a)} P_{\Lambda_l}} 
\;\leq\; 
C \norm{a}_{\ell_s^\infty(\Mm)}\, 2^{-\frac{s}{2}\abs{l-m}}
\;.
$$
As the $P_{\Lambda_m}$ are an orthogonal partition of unity, it follows that
\begin{align*}
\norm{H^{(s)}_{\eta(a)}} 
&
\;\leq\; 
\sum_{m\in\bbZ}
\big\|
\sum_{l\in\bbZ} P_{\Lambda_{m+l}}H^{(s)}_{\eta(a)} P_{\Lambda_l}
\big\|
\\
&
\;\leq\;  
\sum_{m\in \bbN}\sup_{l\in\bbN}
\big\|P_{\Lambda_{m+l}}H^{(s)}_{\eta(a)} P_{\Lambda_l}\big\|
\\ 
&
\;\leq\; C \,
\norm{a}_{\ell_s^\infty(\Mm)}  
\,\sum_{m \in \bbZ} 2^{-\frac{\beta}{2}\abs{m}} 
\;,
\end{align*}
which carrying out the sum shows the claim.
\hfill $\Box$

\begin{proposition}
Let $n=p=1$. Suppose that $a\in L^1(\Mm) \cap \Mm$ has a trace-class commutator $\hat{H}_a \in L^1(\Nn)$. Then $a \in B^{1}_{1,1}(\Mm)$ and there is a uniform constant such that
$$
\norm{a}_{B^{1}_{1,1}} 
\;\leq\; C \big(\norm{a}_1 + \big\|\hat{H}_a\big\|_1\big)
\;.
$$
\end{proposition}

\noindent{\bf Proof.}
Choose $0 < \zeta < 1$ and assume $a\in {}^0B^{1}_{1,1}(\Mm)\cap \Mm$. Using similar notations as in the proof of Proposition \ref{propf:pellernecessary1}, let us define the functional
$$
\tilde{\phi}\,:\, b \in \ell_\zeta^\infty(\Mm) 
\;\mapsto \;
\hat{\Tt}_\alpha\big(\hat{H}_{a} \hat{H}^{(\zeta)}_{\eta(b)}\big)
\;,
$$
which is bounded by Lemma~\ref{lemma:hankellinfty_refined}. A priori it is an element of the dual space $\ell_\zeta^\infty(\Mm)'$, however, the same reasoning as in Proposition~\ref{propf:pellernecessary1} shows that is represented by $\tilde{\psi}(b) =\sum_{j=0}^\infty \Tt(\tilde{a}_j^*b_j)$ with %
$$
\tilde{a}_j
\;=\; 
\Psi^{(\zeta)}*(\widehat{W_{j-1}}+\widehat{W_{j}}+\widehat{W_{j+1}})*a
$$
where
\begin{align*}
\Psi^{(\xi)}(\lambda)
\;=\; 
&\int_{\hat{G}} \difd k \; \weight(k+\lambda)^{\frac{\xi}{2}} \abs{\sgn(k+\lambda)-\sgn(k)}^2 \weight(k)^{\frac{\xi}{2}}
\end{align*}
again behaves like $\abs{\lambda}^{1+\xi}$ and thus $(\tilde{a}_j)_{j\in \bbN} \in \ell_{-\zeta}^1(L^1(\Mm)$. Hence $\tilde{\phi}$ is actually an element of the predual $ \ell_\zeta^\infty(\Mm)_*=\ell_{-\zeta}^1(L^1(\Mm))$. Since the predual embeds isometrically into the dual, one has
$$
\norm{\tilde{a}}_{\ell_{-\zeta}^1(L^1(\Mm))} 
\;\leq\; 
c_1 \,
\|\tilde{\phi}\|_{\ell_\zeta^\infty(\Mm)'} 
\;\leq\; 
c_2\big(\norm{a}_1 + \|\hat{H}_a\|_1\big)
\;.
$$
With this inequality the proof can be completed essentially as above.
\hfill $\Box$

\section{Breuer index of Toeplitz operators}
\label{sec-BreuerToep}

For clarity we restate the general assumptions of this section, namely $(\Mm,\Indexaction,G)$ is a $W^*$-dynamical system with a group action of $G=\bbT^{n_0} \oplus \RM^{n_1}$, and on $\Mm$ is given an $\Indexaction$-invariant s.n.f. trace $\Tt$. Let $(\pi, U)$ be a regular representation of this dynamical system and the von Neumann crossed product $\Nn_\Indexaction = M_{N}(\bbC) \otimes (\Mm \rtimes_\Indexaction G)$ and $\DD$ the canonical Dirac operator \eqref{eq:dirac_def} constructed from the generators of $\Indexaction$ in that representation, and $N=2^{\lfloor n/2 \rfloor}$. The dual trace induced on $\Nn_\Indexaction$ by $\Tt \otimes \Tr_N$ will be denoted $\hat{\Tt}_\Indexaction$ (so we notationally suppress the matrix fibers and recall that the normalization of the dual trace is uniquely fixed by the conventions of Chapter~\ref{sec-CrossedProd}). 

\vspace{.2cm}

Let $\PP=\chi(\DD>0)\in \Nn$ be the Hardy projection and $\FF = \frac{1}{2}(\PP-\one)$. It will be useful to introduce an offset $x_0 \in \bbR^{n} \setminus \bbZ^n$ and set $\FF_{x_0}=\sgn(\DD+\gamma\cdot x_0)$, especially since $\DD$ has discrete spectrum and a non-trivial kernel for pure torus actions $G=\bbT^n$.

\vspace{.2cm}

Further let us introduce the space of $(n+1)$-summable symbols
$$
\Cc_n(\Mm)
\;=\;
\{a \in \Mm: a \in L^{n+1}(\Mm)\, , \, [\PP,\pi(a)] \in L^{n+1}(\Nn_\Indexaction)\}
$$
which is a $*$-subalgebra of $\Mm$ and noting that $x \mapsto \sgn(\gamma\cdot x+x_0)-\sgn(\gamma\cdot x)$ is a matrix-valued function with entries in $L^{n+1}(\bbR^n)$, the estimate \eqref{eq:LPforproducts} shows that also $[F_{x_0}, \pi(a)] \in L^{n+1}(\Nn_\Indexaction)$ for any $x_0 \in \bbR^n$. Hence $(\Cc_n(\Mm),\Nn,\FF_{x_0})$ defines an $(n+1)$-summable semi-finite Fredholm module (in the sense of \cite{CGRS}). This allows to adapt the standard formulation of pairings in non-commutative geometry \cite{Connes94} with unitaries and projections in (matrix algebras over) the unitization $\Cc_n(\Mm)^\sim = \bbC \one + \Cc_n(\Mm) \subset \Mm$. 

\vspace{.2cm}

The non-commutative Peller criterion (Theorem~\ref{theo-HankelBesovP}) gives us a characterization of $(n+1)$-summability purely in terms of the regularity of the symbols. The particular Besov space that is distinguished by the Peller criterion appears repeatedly and will be abbreviated
\begin{equation}
\label{eq-BesovAbbrev}
B_{n}(\Mm)
\;=\;
B^{\frac{n}{n+1}}_{n+1,n+1}(\Mm, \Indexaction)
\;.
\end{equation}
Hence 
$$
\Cc_n(\Mm) \;=\; B_n(\Mm) \cap \Mm
\;.
$$
A practical criterion to lie in the algebra $\Cc_n(\Mm)$ can be directly deduced from Proposition~\ref{prop-besov-sufficient}(iii): 

\begin{proposition} 
\label{prop-besov-sufficient2}
For any $p\in(n,n+1]$, one has $W^1_{p}(\Mm) \cap \Mm \subset B_n(\Mm) \cap \Mm$ with a continuous embedding of Banach algebras, namely there is a constant $C$ such that
%
$$
\norm{a}+\norm{a}_{B_n}
\;\leq\;
C\big(\norm{a}+\norm{a}_{W^1_{p}}\big)
\;,
\qquad
a\in W^1_{p}(\Mm) \cap \Mm
\;.
$$
\end{proposition}

For the classical function spaces over $\bbR^n$ there are also embeddings $W^1_n\subset B_n$ that do no require the intersection with $L^\infty$ as special cases of the so-called Franke-Jawerth embeddings \cite{Franke,Jawerth}, which have recently been extended to the non-commutative torus \cite{XXY}.

\vspace{.2cm}

The Chern cocycle for the action $\Indexaction$ can be defined in two different ways, one that directly extends the usual definition for smooth elements and one that stems from the associated index theorem \cite{PSbook}, but here also has an extended domain of definition. It will be shown below that these Chern cocycle coincide for elements that are smooth and summable w.r.t. $\Indexaction$.

\begin{definition}
\label{def-ChernCocycle}
Let $p\in[n,n+1]$ and   $\nabla_1,\dots,\nabla_n$ be the derivations \eqref{eq:nabla} on $W^1_p(\Mm,\Indexaction)$ w.r.t. the unit directions of $G$.
The Chern cocycle for the action $\Indexaction$ is a cyclic cocycle on $\Mm \cap W_{{p}}^1(\Mm,\Indexaction)$ defined by 
$$
\Ch_{\Tt,\Indexaction}(a_0,\ldots ,a_n) 
\;=\; 
c_n \,
\sum_{\rho \in S_n} (-1)^\rho\, \Tt\big(a_0 \nabla_{\rho(1)} a_1 \ldots  \nabla_{\rho(n)} a_n\big)
\;,
$$
where the normalization constants are given by
%
$$
c_n 
\;=\; 
\begin{cases}
\frac{(2\pi \imath\,)^k}{k!}\;, \quad &\text{for }n=2k\;,
\\
\frac{\imath\,(\pi\imath)^k}{(2k+1)!!}\;, \quad &\text{for }n=2k+1
\;.
\end{cases}
$$
\end{definition}

Let us note that a multiple H\"older inequality followed by an interpolation bound implies that $\Ch_{\Tt,\Indexaction}$ is well-defined and continuous on its domain:
\begin{align*}
\Tt\big(a_0 \nabla_{\rho(1)} a_1 \ldots  \nabla_{\rho(n)} a_n\big)
&
\;\leq\;
\|a_0\|_{\frac{p}{p-n}}
\|\nabla_{\rho(1)} a_1\|_{p} \cdots  \|\nabla_{\rho(n)}a_n\|_{p}
\\
&
\;\leq\;
\|a_0\|_{\frac{p}{p-n}}
\|a_1\|_{W^1_p} \cdots  \|a_n\|_{W^1_p}
\\
&
\;\leq\;
\|a_0\|_{p}^{p-n}\|a_0\|_{\infty}^{n+1-p}
\|a_1\|_{W^1_p} \cdots  \|a_n\|_{W^1_p}
\;.
\end{align*}
Let us also point out that neither of the endpoints $p=n$ and $p=n+1$ in the Sobolev scale is always natural. For example, already for functions in $M_2(L^\infty(\RM^d))$ the $L^p$-regularity of derivatives may fail precisely at certain integer values due to either slow inverse polynomial decay at infinity or finite-order poles.

\vspace{.2cm}

For use in the following let us also note that the Chern cocycles depends only on the action $\Indexaction$ up to orientation:

\begin{proposition}
	\label{prop-ChernRotInv}
	The Chern cocycle $\Ch_{\Tt,\Indexaction}$ is invariant under orientation conserving rotations $O\in \mbox{\rm SO}(n)$, namely
	\begin{equation}
	\label{eq-ChernRotInv}
	\Ch_{\Tt,\Indexaction}(a_0,\ldots ,a_n) 
	\;=\; 
	c_n \,
	\sum_{\rho \in S_n} (-1)^\rho\, \Tt\big(a_0 \nabla_{Oe_{\rho(1)}} a_1 \cdots  \nabla_{O e_{\rho(n)}} a_n\big)
	\;,
	\end{equation}
	where $\nabla_v$ is the derivation \eqref{eq:nabla} in direction $v\in\SM^n$.
\end{proposition}

\noindent {\bf Proof.} Let $\Ch^O_{\Tt,\Indexaction}(a_0,\ldots ,a_n) $ denote the r.h.s. of \eqref{eq-ChernRotInv}. If $O=(O_{j,k})_{j,k=1,\ldots,n}$ denotes the matrix elements, then 
$$
\Ch^O_{\Tt,\Indexaction}(a_0,\ldots ,a_n) 
\,=\,
c_n \!
\sum_{j_1,\ldots,j_n=1}^n \;
\sum_{\rho \in S_n} (-1)^\rho\, O_{j_1,\rho(1)}\cdots  O_{j_n,\rho(n)}\,\Tt\big(a_0 \nabla_{j_1} a_1 \cdots  \nabla_{j_n} a_n\big)
.
$$
Now for any function $b_{\rho(3),\ldots,\rho(n)}$, one has
$$
\sum_{\rho \in S_n} (-1)^\rho\, O_{j,\rho(1)} O_{j,\rho(2)} \,b_{\rho(3),\ldots,\rho(n)}
\;=\;0
\;,
$$
and similarly for other equal pairs $j_i=j_{i'}$. Therefore the sum in $\Ch^O_{\Tt,\Indexaction}(a_0,\ldots ,a_n)$ only contains non-vanishing terms for pairwise different $j_1,\ldots,j_n$ which thus correspond to a permutation $\sigma\in S_n$:
\begin{align*}
&
\Ch^O_{\Tt,\Indexaction}(a_0,\ldots ,a_n) 
\\
&
\;\;\;
\;=\; 
c_n \,
\sum_{\sigma\in S_n} \;
\sum_{\rho \in S_n} (-1)^\rho\, O_{\sigma(1),\rho(1)}\cdots  O_{\sigma(n),\rho(n)}\,\Tt\big(a_0 \nabla_{\sigma(1)} a_1 \cdots  \nabla_{\sigma(n)} a_n\big)
\;.
\end{align*}
Finally
\begin{align*}
\sum_{\rho \in S_n} 
&
 (-1)^\rho\, O_{\sigma(1),\rho(1)}\cdots  O_{\sigma(n),\rho(n)}
\\
& 
\;=\;
(-1)^\sigma\sum_{\rho \in S_n} (-1)^{\rho\circ\sigma^{-1}}\, O_{\sigma(1),\rho\circ\sigma^{-1}(\sigma(1))}\cdots  O_{\sigma(n),\rho\circ\sigma^{-1}(\sigma(n))}
\\
& \;=\;
(-1)^\sigma\;\det(O)
\\
&
\;=\;(-1)^\sigma
\;.
\end{align*}
Replacing completes the proof.
\hfill $\Box$

\vspace{.2cm}

Hence $\Ch_{\Tt,\Indexaction}$ is a natural $n$-cocycle that is constructed from the data $\Tt$ and $\Indexaction$. Another way to obtain a cocycle is in the following way:

\begin{definition}
The Chern cocycle for the bounded Fredholm modules given by $\FF_{x_0}$ is a cyclic cocycle on the algebra $\Cc_n(\Mm)$ defined for any $x_0 \in \bbR^n\setminus \bbZ^n$ by 
\begin{equation}
	\label{eq-chtilde}
\begin{split}
\widetilde{\Ch}_{\Tt,\Indexaction,x_0}(a_0,\ldots ,a_n)
 &
 \;=\;  \tilde{c}_n\,\hat{\Tt}'_{\Indexaction,x_0}\big(\pi(a_0) [\FF_{x_0},\pi(a_1)]\cdots [\FF_{x_0},\pi(a_n)]\big) 
 \\
&
\;=\; 
\frac{\tilde{c}_n}{2}\;\hat{\Tt}_\Indexaction \big(\gamma_0 \FF_{x_0}  [\FF_{x_0},\pi(a_0)] [\FF_{x_0},\pi(a_1)]\cdots [\FF_{x_0},\pi(a_n)]\big)
\end{split}
\end{equation}
with normalization constants
$$
\tilde{c}_n 
\;=\; 
\begin{cases}
(-1)^k\;, \quad &\text{for }n=2k\;,
\\
(-1)^{k+1} \,2^{-2k-1}\;, \quad &\text{for }n=2k+1
\;,
\end{cases}
$$
and with the supertrace 
$$
\hat{\Tt}'_{\Indexaction,x_0}(a) 
\;=\; 
\frac{1}{2} \;
\hat{\Tt}_\Indexaction \big(\FF_{x_0} \,d_{x_0} (\gamma_0 a)\big)
\;,
$$
the graded commutator $d_{x_0} a = \FF_{x_0} a - (-1)^n a \FF_{x_0}$ and $\gamma_0=(-\imath)^{n/2}\gamma_1\cdots\gamma_n$ for $n$ even respectively $\gamma_0=(-\imath)^{(n-1)/2}\gamma_1\cdots\gamma_n=\one$ for $n$ odd such that $\gamma_0 \FF_{x_0}+(-1)^n \FF_{x_0}\gamma_0 =0$. Furthermore, the integrated cocycle is
\begin{equation}
\label{eq-IntegrateCocycle}
\widetilde{\Ch}_{\Tt,\Indexaction}(a_0,\ldots ,a_n) 
\;=\; \,\int_{[0,1]^n}   \widetilde{\Ch}_{\Tt,\Indexaction,x_0}(a_0,\ldots ,a_n)\;\difd^n{x_0}
\;.
\end{equation}
\end{definition}

By using the invariance of $\hat{\Tt}_\Indexaction$ under the dual action of $\hat{G}=\bbZ^{n_0}\oplus \bbR^{n_1}$, one can see that for $n_0=0$ the cocycles $\widetilde{\Ch}_{\Tt,\Indexaction,x_0}=\widetilde{\Ch}_{\Tt,\Indexaction}$ all coincide. For $n_0>0$ they are not numerically identical, but still induce the same index pairings. Both $\Ch_{\Tt,\Indexaction}$ and $\widetilde{\Ch}_{\Tt,\Indexaction}$ are densely defined on $(n+1)$-summable elements of $\Mm$ with Fourier spectrum in a compact interval. To show that they coincide, let us introduce the discretizations that align with the grid for the torus components. More precisely, for $x \in \bbZ^{n_0}$, $y \in \bbR^{n_1}$ define the cubes
$$
Q_{x,y}^{(L)}
\;=\; 
\{x\} \times (y + [0,L)^{n_1}) 
\;\subset\; 
\bbR^n
\;,
$$
along with the projections $P^{(L)}_{x,y} = \chi\big((D_1,\ldots ,D_n) \in Q_{x,y}^{(L)}\big)$, as well as the discretizations
$$
\FF_{x_0}^{(L)} 
\;=\; 
\sum_{x \in \bbZ^m} \sum_{y \in \bbZ^{n_1}} 
P^{(L)}_{x, L y} \, \sgn\big(\gamma\cdot(x + L y + x_0)\big)
\;.
$$
For $L\to 0$, they converge strongly to $\FF_{x_0}=\sgn\big(\DD+\gamma\cdot x_0\big)$. To approximate products of commutators the following technical estimate is needed.


\begin{lemma}
\label{lemma:commutator_discretisation}
Let $a_0,\ldots,a_n \in \Mm \cap L^{n+1}(\Mm)$ be elements with Arveson spectra contained in some compact set, {\sl. i.e.} $\sigma_\alpha(a_i)\subset [-r,r]^n$ for some $r > 0$. Then there is a constant $C>0$ depending on $a_0,\ldots, a_n$ but uniform in $L$ and $z \in \bbR^n$ such that
\begin{align*}
&
\Big|\hat{\Tt}_\Indexaction
\big(\gamma_0 P^{(L)}_{0,0}  \FF_z\,\pi(a_0) [\FF_{z},\pi(a_1)]
\!\cdot\!\cdot\!\cdot\![\FF_{z},\pi(a_n)]\big)
\\
&
\;\;\;\;-\,
\hat{\Tt}_\Indexaction\big(\gamma_0 P^{(L)}_{0,0} \FF^{(L)}_z \,\pi(a_0) [\FF^{(L)}_{z},\pi(a_1)]
\!\cdot\!\cdot\!\cdot\![\FF^{(L)}_{z},\pi(a_n)]\big)\Big|
\\
&
\;\;\;\;\;\;\;\;
\leq \; 
\frac{C\, L^{n_1+1}}{(1+\abs{z})^{n+1}}
\;.
\end{align*}
\end{lemma}

\noindent {\bf Proof.}
Note first that with $\ell > 0$ and the floor function applied componentwise,
\begin{align*}
&
\norm{P^{(\ell r)}_{0,0} [\FF_z-\FF_z^{(L)},\pi(a_j)]P^{(\ell r)}_{0,0}}
\\
&
\;\;\;\;\;\;
\leq\; 
2\, \norm{a} \sup_{|(x,L y)| \leq \ell} 
\abs{\sgn(\gamma\cdot (x+Ly + z)) - \sgn(\gamma\cdot (x+L\lfloor y\rfloor + z))}
\\
&
\;\;\;\;\;\;
\leq \;
2\, \norm{a}\, C\, \sup_{|(x,L y)| \leq \ell} \frac{L \abs{y - \lfloor y \rfloor}}{\abs{x+z}} 
\\
&
\;\;\;\;\;\;
\leq\; C_2 \;\frac{L}{\abs{z}}
\;.
\end{align*}
Due to the condition on the Arveson spectra one may insert $P^{(\ell)}_{0,0}$ for $\ell > (n+1)r + L$:
\begin{align*}
&
\hat{\Tt}_\Indexaction\big(\gamma_0 P^{(L)}_{0,0} \,\pi(a_0) [\FF_{z},\pi(a_1)]
\!\cdot\!\cdot\!\cdot\!
[\FF_{z},\pi(a_n)]\big)
\\
&
\;\;\;\;\;\;\;\;\;\;\;
\;-\;
\hat{\Tt}_\Indexaction\big(\gamma_0 P^{(L)}_{0,0} \,\pi(a_0) [\FF^{(L)}_{z},\pi(a_1)]
\!\cdot\!\cdot\!\cdot\!
[\FF^{(L)}_{z},\pi(a_n)]\big)
\\
&
\;\;\;\;\;\;=\;
\hat{\Tt}_\Indexaction\big(\gamma_0 P^{(L)}_{0,0} \,\pi(a_0) P^{(\ell)}_{0,0}[\FF_{z},\pi(a_1)]
\!\cdot\!\cdot\!\cdot\!
P^{(\ell)}_{0,0}[\FF_{z},\pi(a_n)]P^{(\ell)}_{0,0}\big)
\\
&
\;\;\;\;\;\;\;\;\;\;\;\;
-\;\hat{\Tt}_\Indexaction\big(\gamma_0 P^{(L)}_{0,0} \,\pi(a_0) P^{(\ell)}_{0,0}[\FF^{(L)}_{z},\pi(a_1)]
\!\cdot\!\cdot\!\cdot\!
P^{(\ell)}_{0,0}[\FF^{(L)}_{z},\pi(a_n)]P^{(\ell)}_{0,0}\big)
\;.
\end{align*}
By successively inserting $\FF_z = (\FF_z-\FF^{(L)}_z) +\FF^{(L)}_z$ and bounding terms like
$$
\abs{\hat{\Tt}_\Indexaction\big(\gamma_0 P^{(L)}_{0,0} b P^{(\ell)}_{0,0} [\FF_{z}-\FF^{(L)}_z,\pi(a_1)]P^{(\ell)}_{0,0}c\big)}
\;\leq \;
\big\|P^{(L)}_{0,0}\big\|_1\; \norm{b} \; C_2\; \frac{L}{\abs{z}}\; \norm{c}
$$
together with $\big\|P^{(L)}_{0,0}\big\|_1 = C_3 L^{n_1}$, the result follows.
\hfill $\Box$

\vspace{.2cm}

For the proof of the following geometric fact the reader is referred to 6.4.1 in \cite{PSbook}.

\begin{lemma}
\label{lemma:magic_formula}
Let $x_0,\ldots,x_n \in \bbR^n$ with $x_n=0$ and $\gamma_1,\ldots,\gamma_n$ an irreducible representation of the complex Clifford algebra $Cl_n$ on $\bbC^{2^{\lfloor n/2\rfloor}}$ and $\gamma_0 = (-\imath)^{\lfloor n/2\rfloor}\gamma_1\dots \gamma_n$. Then
$$
\int_{\bbR^n}  \Tr\left(\gamma_0 \prod_{i=1}^n (\sgn(\gamma \cdot (x_i+x))-\sgn(\gamma \cdot (x_{i+1}+x))\right)\,\difd^n{x}
\;=\; 
\hat{c}_n\, \sum_{\rho \in S_n} (-1)^\rho \prod_{i=1}^n x_{i, \rho(i)}
$$
with 
$$
\hat{c}_n
\;=\; 
\begin{cases}
\frac{(2\pi \imath\, )^k}{k!}\;, \quad &\text{for }n=2k\;,\\
\frac{2^{2k+1}(\pi \imath)^k}{(2k+1)!!}\;, \quad &\text{for }n=2k+1\;.
\end{cases}
$$
\end{lemma}

\begin{proposition}
\label{prop-chandchtilde}
Let $p\in[n,n+1]$. For $a_0,\ldots,a_n \in  W^1_{p}(\Mm) \cap \Cc_n(\Mm)$,  one has
$$
\Ch_{\Tt,\Indexaction}(a_0,\ldots,a_n) 
\;=\; 
(-1)^{n-1}\;
\widetilde{\Ch}_{\Tt,\Indexaction}(a_0,\ldots,a_n)
\;.
$$
\end{proposition}

\noindent {\bf Proof.}
One can always write $a_i= \sum_{\ScaleInd=1}^\infty \widehat{W}_\ScaleInd*a_i$ which is uniformly bounded in operator norm and converges w.r.t. the weak$^*$-topology as well as the norms in $W^1_{p}(\Mm)$. Likewise, since $\PP$ is $\Indexaction$-invariant and $\Indexaction$ defines a strongly continuous action on $L^{n+1}(\Nn_\Indexaction)$,
$$
[\FF_{x_0}, a_i] 
\;= \;
\sum_{\ScaleInd=1}^\infty [\FF_{x_0}, \widehat{W}_\ScaleInd*a_i]
\;=\;  
[\FF_{x_0}, \sum_{\ScaleInd=1}^\infty \widehat{W}_\ScaleInd*a_i]
$$ 
with convergence in $L^{n+1}(\Nn_\Indexaction)$.  Using the $\sigma$-weak continuity of the s.n.f. traces it is then possible to check that both cocycles can be approximated simultaneously by truncating the infinite sums for large enough $j$. 

\vspace{.1cm}

Hence it is sufficient to show that the equality holds if $a_0,\ldots,a_n$ have Arveson spectra contained in a compact set. For a given $L > 0$, let us set $Q_L=[0,1]^{n_0} \times [0,L)^{n_1}$ and  manipulate the cocycle as follows:
\begin{align*}
& 
\!\!\!
\widetilde{\Ch}_{\Tt,\Indexaction}(a_0,\ldots,a_n)
\\
&
\;=\; 
\tilde{c}_n\int_{[0,1]^n}  \hat{\Tt}'_{\Indexaction,x_0}\big(\pi(a_0) [\FF_{x_0},\pi(a_1)]\cdots [\FF_{x_0},\pi(a_n)]\big)\,\difd^n{x_0}
 \\
&
\;\;
= \; \frac{\tilde{c}_n}{L^{n_1}}\int_{Q_L} \hat{\Tt}'_{\Indexaction,x_0}\big(\pi(a_0) [\FF_{x_0},\pi(a_1)]\cdots [\FF_{x_0},\pi(a_n)]\big)\,\difd^n{x_0}
\\
&
\;\;
= \;
\frac{\tilde{c}_n}{L^{n_1}}\int_{Q_L}\sum_{(x,y) \in \bbZ^n} \hat{\Tt}_\Indexaction\big(\gamma_0 \FF_{x_0} P^{(L)}_{x,L y} \,\pi(a_0) [\FF_{x_0},\pi(a_1)]\cdots [\FF_{x_0},\pi(a_n)]\big)\,\difd^n{x_0}
\\
&
\;\;
=\; 
\frac{\tilde{c}_n}{L^{n_1}}\int_{Q_L}\sum_{(x,y) \in \bbZ^n} \hat{\Tt}_\Indexaction\big(\gamma_0 P^{(L)}_{0,0} \FF_{x_0-(x,Ly)}\,\pi(a_0) [\FF_{x_0-(x,Ly)},\pi(a_1)]\cdots
\\
&
\hspace{4cm}
\cdot [\FF_{x_0-(x,Ly)},\pi(a_n)]\big)\,\difd^n{x_0}
 \\
&
\;\;
=\;
\frac{\tilde{c}_n}{L^{n_1}}\int_{\bbR^n} \hat{\Tt}_\Indexaction\big(\gamma_0 P^{(L)}_{0,0}\FF_{z} \,\pi(a_0) [\FF_{z},\pi(a_1)]\cdots [\FF_{z},\pi(a_n)]\big)
\, \difd^n{z}
\;.
\end{align*}
Here the first step used invariance under the dual $\bbR^{n_1}$-action which implies that the integral is actually constant in the last $n_1$ components of $x_0$ so that $L$ can be rescaled. Then a  partition of unity was inserted which allowed to rewrite the expression in terms of $\hat{\Tt}_\Indexaction$. From the third to fourth line the invariance under the dual $\bbZ^{n_0} \times \bbR^{n_1}$-action to shift to the origin was used to redefine the variables. Lemma~\ref{lemma:commutator_discretisation} now shows that  $\FF_z$ can be replaced with its discretization in the limit $L \to 0$:
\begin{align*}
&
\widetilde{\Ch}_{\Tt,\Indexaction}(a_0, \ldots,a_n)
\\
&
\;\;\;
\;=\; 
\lim_{L\to 0} \frac{\tilde{c}_n}{L^{n_1}}\int_{\bbR^n} 
\hat{\Tt}_\Indexaction\big(\gamma_0 P^{(L)}_{0,0} \FF^{(L)}_{z}\,\pi(a_0) [\FF^{(L)}_{z},\pi(a_1)]\cdots [\FF^{(L)}_{z},\pi(a_n)]\big)
\; \difd^n{z}\\
&
\;\;\;
\;=\;
\lim_{L\to 0} \frac{\tilde{c}_n}{L^{n_1}} \int_{\bbR^n} \difd^n{z}\,  \sum_{(x_1,y_1),\ldots,(x_n,y_n) \in \bbZ^n} 
\hat{\Tt}_\Indexaction\big(\gamma_0 P^{(L)}_{0,0} \,\pi(a_0)P^{(L)}_{x_1,L y_1} \pi(a_1)\cdots P^{(L)}_{x_n,L y_n} \pi(a_n)\big) 
\\
& 
\;\;\;\;\;\;\;\;\;\;
\Tr\Big(\gamma_0 \prod_{j=1}^n \left(\sgn(\gamma\cdot(x_j + L y_j + z)) - \sgn(\gamma\cdot(x_{j+1} + L y_{j+1} + z)) \right) \Big)
\;,
\end{align*}
where the convention that $(x_{n+1},y_{n+1})=0$ was used and the second equality resulted after inserting further partitions of unity (which are actually finite sums due the spectral conditions). In this form Lemma~\ref{lemma:magic_formula} applies to deduce that $\widetilde{\Ch}_{\Tt,\Indexaction}(a_0, \ldots,a_n)$ is equal to
\begin{align*}
\lim_{L\to 0} &\frac{\tilde{c}_n \hat{c}_n}{L^{n_1}} 
\,\sum \,
\hat{\Tt}_\Indexaction\big(P^{(L)}_{0,0} \,\pi(a_0)P^{(L)}_{x_1,L y_1} \pi(a_1)\cdots P^{(L)}_{x_n,L y_n} \pi(a_n)\big) 
\cdot \\
&
\cdot \sum_{\rho \in S_n} (-1)^\rho \prod_{j=1}^{n_0} x_{j,\rho(j)}  \prod_{i=1}^{n_1} (L y_{i,\rho(n_0+i)})
\;,
\end{align*}
with a sum carrying over $(x_1,y_1),\ldots ,(x_n,y_n) \in \bbZ^n$. Next introduce discretizations of the generators $D_1,\ldots,D_n$ as the multiplication operators 
$$
D^{(L)}_j
\;=\;
\begin{cases} 
D_j\,, \qquad & j=1,\ldots,n_0\;, \\ 
\sum_{z\in \bbZ^d} (L z_j) P_{x,Ly}\,, \qquad & j=n_0+1,\ldots,n \;,
\end{cases}
$$ 
and note that due to the antisymmetrisation one can write $\widetilde{\Ch}_{\Tt,\Indexaction}(a_0, \ldots,a_n)$ as 
$$
\lim_{L\to 0} \frac{\tilde{c}_n \hat{c}_n}{L^{n_1}} 
\,\sum \, \sum_{\rho \in S_n} 
\hat{\Tt}_\Indexaction\big(P^{(L)}_{0,0} \,\pi(a_0)P^{(L)}_{x_1,L y_1} [D^{(L)}_{\rho(1)},\pi(a_1)]\cdots P^{(L)}_{x_n,L y_n} [D^{(L)}_{\rho(n)},\pi(a_n)]\big) 
\;.
$$ 
Replacing the definition of the derivations $\nabla_j = -\imath[D_j, \cdot]$ it follows that
\begin{align*}
&\!\! \widetilde{\Ch}_{\Tt,\Indexaction}  (a_0, \ldots,a_n)
\\
&
\;=\;
\lim_{L\to 0} \frac{\tilde{c}_n\hat{c}_n}{L^{n_1}} \sum_{\rho \in S_n} (-1)^\rho  \, \frac{1}{(-\imath)^n}\, 
\hat{\Tt}_\Indexaction\big(P^{(L)}_{0,0} \, a_0 \pi(\nabla_{\rho(1)} a_1)\cdots \pi(\nabla_{\rho(n)} a_n)\big)\;+\; \mathcal{O}(L)
\\
&
\;=\; 
\frac{\tilde{c}_n\hat{c}_n}{c_n (-\imath)^n}\; \hat{\Tt}_\Indexaction\big(P_{0,0}^{(1)}\big)\; \Ch_{\Tt,\Indexaction}(a_0,\ldots ,a_n)
\\
&
\;=\;
(-1)^{n-1} \Ch_{\Tt,\Indexaction}(a_0,\ldots ,a_n)
\;.
\end{align*}
The $\mathcal{O}(L)$-bounds of the discretization error result from estimates of the form 
\begin{align*}
\abs{\hat{\Tt}_\Indexaction\big(P^{(L)}_{0,0}\, a [D_j - D_j^{(L)}, b] \,c\big)} 
&
\;\leq\; 
\norm{P^{(L)}_{0,0}}_1\, \norm{D_j - D_j^{(L)}}\, \norm{a} \, \norm{b}\, \norm{c} 
\\
&
\;\leq \;
C L^{n_1} \cdot L  \norm{a} \, \norm{b} \,\norm{c}
\;
\end{align*} 
and noting that the error term can be expanded into a finite number of such expressions.
\hfill $\Box$

\vspace{.2cm}

Now one can define and evaluate the usual index pairings with projections and unitaries and since $\widetilde{\Ch}_{x_0}$ is the standard cocycle associated to a Fredholm module those pairings can be written as $\hat{\Tt}_\Indexaction$-indices:

\begin{proposition}
\label{prop-chtildeandindex} 
If $n$ is odd and $u\in M_{N'}(\Cc_n(\Mm)^\sim)$ a unitary with scalar part $s(u)$ and hence $u-s(u) \in M_{N'}(\Cc_n(\Mm))$, then
$$
\hat{\Tt}_\Indexaction \mbox{-}\Ind(\PP_{x_0}  \pi(u) \PP_{x_0}  + 1-\PP_{x_0} ) 
\;=\; 
\widetilde{\Ch}_{\Tt,\Indexaction,x_0}(u^*,u,\ldots ,u^*,u)
\;,
$$
with $\PP_{x_0}  = \frac{1+\FF_{x_0}}{2}$ for any $x_0 \in \bbR^n$. If $n$ is even and $e\in M_N(\Cc_n(\Mm))^\sim$ a projection with $e - s(e) \in M_N(\Cc_n(\Mm))$, then
$$
\hat{\Tt}_\Indexaction\mbox{-}\Ind(\pi(e) \GG_{x_0} \pi(e) + 1-\pi(e)) 
\;=\; 
\widetilde{\Ch}_{\Tt,\Indexaction,x_0}(e,\ldots ,e)
\;,
$$
with $\GG_{x_0}$ defined by $\FF_{x_0} = \begin{pmatrix}
0 & \GG_{x_0}^* \\ \GG_{x_0} & 0
\end{pmatrix}$ in a basis such that $\gamma_0= \begin{pmatrix}
1 & 0 \\ 0 & -1
\end{pmatrix}$ and any $x_0 \in (0,1]^n$.
\end{proposition}

\noindent {\bf Proof.}
Let us first point out that matrix fibers are again suppressed  ({\it i.e.} properly one has tensor products $\PP_{x_0}\otimes \one_{N'}$, $\hat{\Tt}_\Indexaction\otimes\mathrm{Tr}_{N'}$ and so on). Also the cocycles extend naturally to unitizations, since their algebraic properties are consistent with simply dropping the scalar parts, as can be seen in the second formula of \eqref{eq-chtilde}.

\vspace{.1cm}

For readability we also drop the subscript $x_0$ and the representation $a=\pi(a)$. Both cases are based on Proposition~\ref{prop:fedosov}, noting its conditions are satisfied for any $m > \frac{n+1}{2}$. From this point the manipulations are standard (see e.g. \cite{Connes94}), but we repeat them in order to track factors and signs. In the even case one gets, for $m= \frac{n+2}{2}$,
\begin{align*}
(-1)^m\; & \hat{\Tt}_\Indexaction\mbox{-}\Ind(e \GG e + 1-e) 
\\
&
\;=\;(-1)^m \left( 
\hat{\Tt}_\Indexaction\left((e-e \GG^* e \GG e)^m\right)-\hat{\Tt}_\Indexaction\left((e-e \GG e \GG^* e)^m\right)\right) \\
&
\;=\;
\hat{\Tt}_\Indexaction\left(e([\GG^*,e][\GG,e])^m\right)-\hat{\Tt}_\Indexaction\left(e([\GG,e][\GG^*,e])^m\right)\\
&
\;=\;
\hat{\Tt}_\Indexaction(\gamma_0 e [\FF,e]^{2m})
\;.
\end{align*} 
From the other side, one has using the identities $[\FF,e]=e[\FF,e]+[\FF,e]e$, $\FF[\FF,e]=-[\FF,e]\FF$, $\{\gamma_0, \FF\}=0$ and $e[\FF,e]^2=[\FF,e]^2e$:
\begin{align*}
\frac{2}{\tilde{c}_n}\;& \widetilde{\Ch}_{\Tt,\Indexaction}(e-s(e),\ldots ,e-s(e)) 
\\
&
\;=\; 
\hat{\Tt}_\Indexaction(\gamma_0 \FF[\FF,e]^{n+1})= \hat{\Tt}_\Indexaction(\gamma_0 \FF (e[\FF,e]^{n+1}+[\FF,e]^{n+1}e))\\
&
\;=\; 
\hat{\Tt}_\Indexaction(\gamma_0 (\FF e+e\FF) [\FF,e]^{n+1}) 
\\
&
\;=\; 
-\,2\,\hat{\Tt}_\Indexaction(\gamma_0 e[\FF,e] [\FF,e]^{n+1})\,+\, 2\,\hat{\Tt}_\Indexaction(\gamma_0 e \FF e [\FF,e]^{n+1})\\
&
\;=\; 
-\,2\,\hat{\Tt}_\Indexaction(\gamma_0 e[\FF,e]^{n+2})\,+\, 2\,\hat{\Tt}_\Indexaction(\gamma_0 \FF [\FF,e]^ne[\FF,e]e)\\
&
\;=\; 
-\,2\,\hat{\Tt}_\Indexaction(\gamma_0 e[\FF,e]^{n+2})
\;.
\end{align*} 
Therefore 
$$
\hat{\Tt}_\Indexaction\mbox{-}\Ind(e \GG e + 1-e) 
\;=\; 
 \frac{(-1)^{m+1}}{ \tilde{c}_n} 
\;
\widetilde{\Ch}_{\Tt,\Indexaction}(e,\ldots ,e)
\;=\; 
\widetilde{\Ch}_{\Tt,\Indexaction}(e,\ldots ,e)
\;.
$$
In the odd case with $m= \frac{n+1}{2}$ one notes that $\PP$ and $\FF$ commute with $[u,\FF][u^*,\FF]$ and thus
\begin{align*}
(-1)^m & \;\hat{\Tt}_\Indexaction  \mbox{-}\Ind(\PP  u \PP  + 1-\PP ) 
\\
&
\;=\; 
(-1)^m \left(\hat{\Tt}_\Indexaction\left((\PP -\PP  u^* \PP  u \PP )^m\right)-\hat{\Tt}_\Indexaction\left((\PP -\PP  u \PP  u^* \PP )^m\right)\right) \\
&
\;=\;
2^{-2m}\,\left(\hat{\Tt}_\Indexaction\left(\PP ([u^*,\FF][u,\FF])^m\right)-\hat{\Tt}_\Indexaction\left(\PP ([u,\FF][u^*,\FF])^m\right)\right)\\
&
\;=\;
2^{-2m}\,\left(\hat{\Tt}_\Indexaction\left(\FF ([u^*,\FF][u,\FF])^m\right)-\hat{\Tt}_\Indexaction\left(\FF ([u,\FF][u^*,\FF])^m\right)\right)\\
&
\;=\;
2^{-2m}\,\left(\hat{\Tt}_\Indexaction\left(\FF ([u^*,\FF][u,\FF])^m\right)-\hat{\Tt}_\Indexaction\left([u^*,\FF] \FF  [u,\FF] ([u^*,\FF][u,\FF])^{m-1}\right)\right)\\
&
\;=\;
2^{-2m}\,\hat{\Tt}_\Indexaction\left(\FF([u^*,\FF][u,\FF])^{m}\right)
\\
&
\;=\; 
\frac{2^{-2m+1}}{\tilde{c}_n} \;
\widetilde{\Ch}_{\Tt,\Indexaction}(u^*,u,\ldots,u^*,u)
\;,
\end{align*} 
which due to the definition of $\tilde{c}_n$ completes the proof.
\hfill $\Box$

\vspace{.2cm}

Combining the above statements now leads to the main statement of this chapter.

\begin{theorem}[Sobolev index theorem]
\label{theo-Index}
Let $p\in (n,n+1]$.
If $n$ is odd and a unitary $u \in M_{N'}((\Mm\cap W^1_p(\Mm))^\sim)$, then 
$$
\Ch_{\Tt,\Indexaction}(u^*,u,\ldots ,u^*,u)
\;=\; 
-\,\hat{\Tt}_\Indexaction\mbox{-}\Ind(\PP_{x_0}  \pi(u) \PP_{x_0}  + 1-\PP_{x_0} ) 
\;,
$$
with $\PP_{x_0}  = \frac{1+\FF_{x_0}}{2}$ for any $x_0 \in (0,1)^n$. 
If $n$ is even and a projection $e\in M_N( (\Mm\cap W^1_p(\Mm))^\sim)$, then
$$
\Ch_{\Tt,\Indexaction}(e,\ldots ,e)
\;=\; 
\hat{\Tt}_\Indexaction\mbox{-}\Ind(\pi(e) \GG_{x_0} \pi(e) + 1-\pi(e)) 
\;,
$$
for any $x_0 \in (0,1)^n$ with $\GG_{x_0}$ being the off-diagonal part of $\FF_{x_0}$ in the basis of $\gamma_0$:
$$
\FF_{x_0} 
\;=\; \begin{pmatrix}
0 & \GG_{x_0}^* \\ \GG_{x_0} & 0
\end{pmatrix}
\;,
\qquad
\gamma_0
\;=\; 
\begin{pmatrix}
1 & 0 \\ 0 & -1
\end{pmatrix}
\;.
$$
\end{theorem}

\noindent{\bf Proof.}
By Proposition~\ref{prop-besov-sufficient2}, $\Mm\cap W^1_p(\Mm)$ is a common domain for ${\Ch}_{\Tt,\Indexaction}$ and $\widetilde{\Ch}_{\Tt,\Indexaction}$. In view of the Propositions~\ref{prop-chandchtilde} and \ref{prop-chtildeandindex}, one only needs to show that the r.h.s.'s are independent of $x_0\in (0,1)^n$ so that the average over the unit cube in the definition~\eqref{eq-IntegrateCocycle} of $\widetilde{\Ch}_{\Tt,\alpha}$ can be dropped. For $n_0=0$ this is clear, since the dual action is unitarily implemented. For $n_0>0$ note that $\abs{\DD_{x_0}} \geq \mathrm{dist}(x_0,\bbZ^{n_0}\times \bbR^{n_1})$ since $D_1,\ldots,D_{n_0}$ each has spectrum $\bbZ$ and so $\DD_{x_0}$ is, in particular, boundedly invertible.  For another $y_0 \in (0,1)^n$, the commutativity of $\abs{\DD_{x_0}}$ with all $\DD_{y_0}$ leads to an elementary geometric estimate 
\begin{align*}
\norm{\FF_{x_0}-\FF_{y_0}}
&
\;\leq\; 
\norm{\frac{\DD_{x_0}}{\abs{\DD_{x_0}}}\;-\;\frac{\DD_{y_0}}{\abs{\DD_{y_0}}}} 
\\
&
\;=\; 
\norm{\frac{\abs{\DD_{x_0}}-\abs{\DD_{y_0}}}{\,\abs{\DD_{y_0}}}\, \frac{\DD_{x_0}}{\abs{\DD_{x_0}}}
\;-\;\frac{\DD_{x_0}-\DD_{y_0}}{\abs{\DD_{y_0}}}}  
\\
&
\;\leq\; \frac{2 \abs{x_0-y_0}}{\mathrm{dist}(y_0,\bbZ^{n_0}\times \bbR^{n_1})}
\;.
\end{align*}
This shows that $\FF_{x_0}$ depends norm-continuously on $x_0$ and hence the Breuer-Fredholm indices appearing in Proposition~\ref{prop-chtildeandindex} are invariant (due to Proposition~\ref{prop-fredholminv}(i)).
\hfill $\Box$

\vspace{.2cm}

Theorem~\ref{theo-Index} does not cover the lower endpoint $p=n$ in the Sobolev scale which for $n\geq 2$ can be accessed by the alternative technique presented in the next Section~\ref{sec-Sobolev}, see Corollary~\ref{coro-SobolevLower}. However, for all the applications in Chapter~\ref{sec-Applications}
the condition $p>n$ is sufficient.

\section{Sobolev criterion}
\label{sec-Sobolev}

Besides the Schatten classes there are other summability classes which are often used in noncommutative geometry, especially in the context of noncommutative integration using singular traces \cite{Connes94,LSZ}, namely the weak $L^p$ spaces. It turns out that symbols lying in a suitable Sobolev space lead to Hankel operators in the weak $L^p$-spaces. In the remainder of the book, only the criterion of Theorem~\ref{theo-Index} will be used. Therefore the proofs in this section are not spelled out at the same technical level as the previous sections and refer to several results from the literature. 

\vspace{.2cm}

Let us first recall the definition and basic properties of the weak $L^p$-space $L^{(p,\infty)}(\Nn, \hat{\Tt})$ \cite{LSZ}. It is the space of all $\hat{\Tt}$-measurable operators over $\Nn$ for which the quasi-norm 
$$
\norm{a}_{p,\infty} 
\;=\; 
\sup_{t>0} \,t^{\frac{1}{p}}\, \mu_t(a)
$$
is finite and where $\mu_t$ is the singular value function
$$
\mu_t(a) = \inf\{\norm{ae}: \; e \text{ is a projection in }\Nn \text{ and }\hat{\Tt}(1-e)\leq t\}
\;.
$$ 
For $1 < p < \infty$ those spaces are real interpolation spaces between $L^{p_1}(\Nn)$ and $L^{p_2}(\Nn)$ for any $1\leq p_1 < p <p_2 \leq \infty$. In particular, there are equivalent norms that make them into Banach spaces and there is a H\"older inequality
$$
\norm{ab}_{r,\infty} 
\;\leq\; 
c_{p,q} \norm{a}_{p,\infty} \, \norm{b}_{q,\infty} 
$$
for dual exponents $\frac{1}{r}=\frac{1}{q}+\frac{1}{p}$.

\begin{proposition}
\label{prop:lpw-embedding}
Let $2 < p \leq \infty$ and $f \in L^{(p,\infty)}(\hat{G})$. The map
$$
(a,f) \,\in \,(\Mm \cap L^p(\Mm)) \,\times\, (L^\infty(\hat{G})\, \cap \,L^{(p,\infty)}(\hat{G}))\; \mapsto\; \pi(a)f(D) \,\in\, \Nn
$$
is $L^p(\Mm) \times L^{(p,\infty)}(\hat{G}) \to L^{(p,\infty)}(\Nn)$-bounded with
\begin{equation}
		\label{eq:LPwforproducts} 
		\norm{\pi(a)f(D)}_p 
		\;\leq\; C_p
		\norm{a}_p \norm{f}_{p, \infty}
		\;.
\end{equation}
For $1 < p < \infty$, one also has 
	\begin{equation}
		\label{eq:LPwforproducts2} 
		\norm{f(D)\pi(a)g(D)}_p 
		\;\leq\; C_p
		\norm{a}_p \norm{f}_{\frac{p}{2}, \infty}\, \norm{g}_{\frac{p}{2}, \infty}
		\;.
	\end{equation}
\end{proposition}

\noindent{\bf Proof.}
The inequality \eqref{eq:LPwforproducts} follows from the abstract Cwikel estimate \cite[Corollary 3.6]{LSZ20}, essentially since $L^{(p,\infty)}$ is also an interpolation space between $L^2$ and $L^\infty$ for which one has the endpoint estimates from Proposition~\ref{prop:lp-embedding}. The second estimate \eqref{eq:LPwforproducts2} follows from the first one by applying the H\"older inequality to $f(D)\pi(u \abs{a}^{\frac{1}{2}}) \pi(\abs{a}^{\frac{1}{2}})g(D)$ where $a=u|a|$ is the polar decomposition.
\hfill $\Box$

\vspace{.2cm}

Next a result analogous the that of \cite{LMSZ,MSX,MSX2} is presented, and the proof is sketched along the same lines, mainly to show that it extends to the present setting without any issues. It is based on the notion of a double-operator integrals on which we will not give an extensive amount of technical details, but only cite necessary results. Let $A,B$ be densely defined self-adjoint operators affiliated to a von Neumann algebra $\Nn$ and let $(E_A(\lambda))_{\lambda \in \bbR}$ and $(E_B(\mu))_{\mu\in \bbR}$ be their spectral resolutions. A double-operator integral is formally given by the integral
$$
T_\varphi^{A,B}(a) 
\;=\; 
\int \varphi(\lambda, \mu) \;E_A(\difd \lambda) \;a\; E_B(\difd \mu)\;, 
\qquad a \in \Nn
\;,
$$
for suitable functions $\varphi$ and with the integral understood in an appropriate sense. For example, if $\varphi$ is continuous with compact support, then the integral can be understood as an operator-valued Riemann-Stieltjes-integral. Another important special case is if $\varphi(\lambda,\mu)=f(\lambda-\mu)$ for a Schwartz function $f \in \Ss(\bbR)$, in which case $T_\varphi^{A,A}(a)= \int_{\RM} e^{\imath A t} a e^{-\imath A t} \hat{f}(t)\, \difd{t}$ where the Fourier transform is given by $\hat{f}(t) = (2\pi)^{-1}\int_{\RM} f(\lambda)e^{-\imath \lambda t} \difd{t}$. In particular, the Fourier multipliers introduced in Section~\ref{sec-SpecDecomp} can also be considered as double-operator integrals if the group action is inner. An important fact about double-operator integrals is that functions which depend only on either of the variables factor out of them, {\sl i.e.}
$$
T_{\varphi}^{A,B}(x)
\;=\; 
\varphi_1(A)\, T_{\varphi_2}^{A,B}(x) \, \varphi_3(B)=\, T_{\varphi_2}^{A,B}(\varphi_1(A) x \varphi_3(B))
$$
for $\varphi(\lambda,\mu)=\varphi_1(\lambda)\varphi_2(\lambda,\mu)\varphi_3(\mu)$.

\begin{theorem}
\label{theo-SobolevEmb}
Assume that $2 \leq n < \infty$. There is a bounded map from $W^1_n(\Mm,\Indexaction) \to L^{(n,\infty)}(\Nn)$ extending the map $a \in \Mm^c_{\Tt,\theta} \mapsto [\sgn(\DD), \pi(a)] \in \Nn$. 
\end{theorem}

\noindent{\bf Proof.}
First let us note that due to Proposition~\ref{prop:lp-embedding} (respectively Proposition~\ref{prop:lpsmall-embedding} in the case $n=2$) both terms on the r.h.s. of 
\begin{align*}
\Big[\sgn(\DD) & -\frac{\DD}{\sqrt{1+\DD^2}}, \pi(a)\Big]
\\
&
\;=\;\Big(\sgn(\DD)-\frac{\DD}{\sqrt{1+\DD^2}}\Big)\pi(a)\;-\;\pi(a)\Big(\sgn(\DD)-\frac{\DD}{\sqrt{1+\DD^2}}\Big) 
\end{align*}
lie in $L^p(\Nn)\cap \Nn \subset L^{(n,\infty)}(\Nn)$ for $1 < p < n$ since the real-valued function $\lambda \in \RM^d \mapsto \sgn(\gamma\cdot\lambda )-\gamma\cdot\lambda|1+\lambda^2|^{-\frac{1}{2}}$ behaves asymptotically like $\Oo(\abs{\lambda}^{-2})$. Therefore it is sufficient to replace $\sgn(\DD)$ by $\DD|1+\DD^2|^{-\frac{1}{2}}$.

\vspace{.1cm}

The remainder of the proof relies crucially on a decomposition of the commutator which is derived and proved in \cite[Lemma 7 and 8]{LMSZ}.
The starting point is to write
$$
\left[\frac{\DD}{\sqrt{1+\DD^2}}, \pi(a)\right] 
\;=\; 
T_g^{\DD,\DD}(\pi(a)) 
\;=\; 
T_{\tilde{g}}^{\DD,\DD}\left(\frac{1}{(1+\DD^2)^{\frac{1}{4}}} [\DD, \pi(a)] \frac{1}{(1+\DD^2)^{\frac{1}{4}}}\right)
\;,
$$
with $g(\lambda, \mu) = \frac{\lambda}{\sqrt{1+\lambda^2}}-\frac{\mu}{\sqrt{1+\mu^2}}$ and $\tilde{g}(\lambda,\mu) = \frac{g(\lambda,\mu)}{\lambda-\mu} (1+\lambda^2)^{\frac{1}{4}}(1+\mu)^{\frac{1}{4}}$. Factoring out of $\tilde{g}$ parts that do not depend on either of the two variables, this gives
\begin{equation} 
\label{eq-CommutatorDouble}
\left[\frac{\DD}{\sqrt{1+\DD^2}}, \pi(a)\right] 
\;=\; 
\sum_{i=1}^3 f_i(\DD)\; T_{\varphi}^{\DD,\DD}\,\left(\frac{1}{(1+\DD^2)^{\frac{1}{4}}} [\DD, \pi(a)] \frac{1}{(1+\DD^2)^{\frac{1}{4}}}\right)f_i(\DD)
\;,
\end{equation}
with the three functions $f_1(\lambda)=1$, $f_2(\lambda)=|1+\lambda^2|^{-\frac{1}{2}}$, $f_3(\lambda)=\lambda|1+\lambda^2|^{-\frac{1}{2}}$ and appropriate function $\varphi$. While $T_{\varphi}^{\DD,\DD}$ is originally defined as a double-operator integral, it can by a variable transformation be rewritten in the form 
$$
T_{\varphi}^{\DD,\DD}(x)
\;=\; 
\int_{\RM} e^{\imath A t} \,x\, e^{-\imath A t} \,\frac{1}{4 \cosh(\frac{\pi}{2}t)} \,\difd{t}
\;,
$$
in terms of the self-adjoint operator $A=\frac{1}{4}\log(1+D^2)$. Since  $\frac{1}{\cosh}$ is an $L^1$-function, the expression obviously extends to a bounded map $T_{\varphi}^{\DD,\DD}:L^{p}(\Nn)\to L^{p}(\Nn)$. By real interpolation \cite[Theorem 3.2]{DDP}, this implies that also  $T_{\varphi}^{\DD,\DD}:L^{(n,\infty)}(\Nn)\to L^{(n,\infty)}(\Nn)$ in bounded. 

\vspace{.1cm}

To complete the proof, it only remains to show that the argument of $T_{\varphi}^{\DD,\DD}$ in \eqref{eq-CommutatorDouble} is in $L^{(n,\infty)}(\Nn)$ which is precisely the result of \eqref{eq:LPwforproducts2} in Proposition~\ref{prop:lpw-embedding}.
\hfill $\Box$

\vspace{.2cm}

Combined with the results of Section~\ref{sec-BreuerToep} one now deduces the following:

\begin{corollary}
\label{coro-SobolevLower}
Let $n\geq 2$. Then the statement of the Sobolev index theorem {\rm (}Theorem~\ref{theo-Index}{\rm )} also hold for a unitary $u\in M_{N'}((\Mm\cap W^1_n(\Mm))^\sim) $ in the odd case, and a projection $e \in M_N( (\Mm\cap W^1_n(\Mm))^\sim)$ in the even case.
\end{corollary}

It does not seem to be known if Theorem~\ref{theo-SobolevEmb} is also true for $n=1$ in general. However, for classical Hankel operators that is indeed the case \cite[Chapter 6.4]{Pel}.

\newpage

\chapter{Duality for Toeplitz extensions}
\label{sec-DualityToep}

Let $(\Aa, G, \Dualityactionop)$ be a $C^*$-dynamical system, where $\Aa$ is a  separable $C^*$-algebra and $G=\bbR$ or $G=\bbT$. This section will relate $\Aa$ and $\Aa\rtimes_\Dualityactionop G$ in an exact sequence of $C^*$-algebras and then show a duality for pairings of cyclic cocycles with $K_j(\Aa)$ respectively $K_{j+1}(\Aa \rtimes_\Dualityactionop G)$. This is a generalization of well-known relations for the Wiener-Hopf and Toeplitz extensions. In the application in Section~\ref{sec-BoundaryCurrents}, the algebra $\Aa$ will take the form $\Aa= \Bb \rtimes \bbZ^d$ and describes the bulk algebra of an infinitely extended homogeneous quantum system  and the $G$-action will be given by the dual action of a subgroup of the dual group $\TM^d$ on the crossed product $\Aa$. Elements in the crossed product $\Aa \rtimes_\Dualityactionop G$ will then have a natural interpretation as observables supported on the edge of a semi-infinite system, obtained by cutting $\ZM^d$ into two pieces along a hyperplane with possibly irrational angles. 

\section{The smooth Toeplitz extension}
\label{sec-toep}

Let us assume that $\Aa$ acts on a Hilbert space $\Hh_0$ and define $\Aa \rtimes_\Dualityactionop G$ in the regular representation $\pi \times U$ on $\Hh = L^2(G,\Hh_0)$, where the action is implemented by exponentiation of the self-adjoint generator $D$ as above in \eqref{eq-GenDef}.
Recall that $\Aa \rtimes_\Dualityactionop G$ is then the $C^*$-algebraic span of
$$
\Aa \rtimes_\Dualityactionop G  
\;= \;
C^*\left\{\pi(a) f(D)\;:\; \, a\in \Aa\,,\;\; f \in C_{0}(\hat{G}) \right\} \;\subset \;\calB(\Hh)
\;.
$$
Let $C_{0,*}(\hat{G})$ denote the continuous functions which vanish in $-\infty$ and admit a limit in $+\infty$. The smooth Toeplitz extension is then obtained by supplementing the crossed product with $C_{0,*}(\hat{G})$-functions of the generator:

\begin{definition}[\cite{Lesch91,Ji90}]
\label{def-ToepExt}
The smooth Toeplitz extension $\Toep  $ associated to $(\Aa,G,\Dualityactionop)$ is defined as
\begin{equation}
\label{eq:sm_toeplitz}
\Toep  
\;= \;
C^*\left\{\pi(a) f(D)\;:\; \, a\in \Aa\,,\;\; f \in C_{0,*}(\hat{G}) \right\} \;\subset \;\calB(\Hh)
\;.
\end{equation}
\end{definition}
Let $\chi_s\in C_{0,*}(\hat{G})$ be some fixed smooth, non-decreasing function that for some $\epsilon > 0$ satisfies
\begin{equation}
\label{eq-SmoothCutOff}
\chi_s(t) \;=\; 
\begin{cases} 
      0\;, & \text{if } t \leq 0 \;,\\
      1\;, & \text{if } t > \epsilon\;. 
\end{cases}
\end{equation}
The operator $\mathcal{P} = \chi_s(D)$ is a smooth approximation of $P=\chi(D > 0)$ and one can take $\mathcal{P}=P$ if $G=\bbT$ since the spectrum of $D$ is then discrete. In the case of an $\bbR$-action, the Toeplitz operators $P\pi(a)P$ are not elements of $\ToepR$ such that one can only consider their smooth approximations $\mathcal{P}\pi(a)\mathcal{P}$, hence the extension carries the adjective "smooth". The following lemma is well-known in the case of $G=\bbR$ ({\it e.g.} \cite{Lesch91,Ji90}).

\begin{lemma}
\label{lemma:smtoep_help}
Let $a\in \Aa$. Then $[\mathcal{P}, \pi(a)] \in \Aa \rtimes_\Dualityactionop G$ and hence $\Aa \rtimes_\Dualityactionop G$ is a two-sided ideal in $\Toep $. An element of the form $\pi(a) \mathcal{P}$ is in $\Aa \rtimes_\Dualityactionop G$ if and only if $a=0$. In fact, for all $a\in \Aa$ and $e\in \Aa\rtimes_\Dualityactionop G$ one has
\begin{equation}
\label{eq:toeplitzineq}
\norm{\pi(a)\Pp + e} \;\geq \;
\norm{a}
\;.
\end{equation}
\end{lemma}

\noindent {\bf Proof.}
For the first part it is enough to show that $[\mathcal{P}, \pi(a)]\in \Aa \rtimes_\Dualityactionop G$ for $a \in C^\infty(\Aa)$, the dense subset of smooth elements w.r.t. $\Dualityactionop$. Let $h_n$ be a sequence of rapidly decaying smooth functions converging to $\chi_s$. One computes using the multiplication law
$$
[h_n(D), \pi(a)] 
\;=\; 
\int_G f(t) \,e^{2\pi\imath \,D t}\, \difd{t}
$$
with $f(t) = \pi(\Dualityactionop_t(a)-a) (\calF^{-1} h_n)(t)$.  For $G=\bbR$ one has 
$$
(\calF^{-1} h_n)(t) \;=\; \frac{(\calF^{-1} h_n')(t)}{2\pi\imath \,t}
\;=\; 
\frac{(\calF^{-1} \Delta h)(t)}{e^{-2\pi \imath \,t}-1}
\;,
$$
and thus 
$$
[h_n(D), \pi(a)]
\;=\;
\int_G \frac{\pi(\Dualityactionop_t(a)-a)}{2\pi\imath \,t}\, (\calF^{-1} h_n')(t)\,e^{ 2\pi \imath \,D t} \difd{t} \in \Aa \rtimes_\Dualityactionop \bbR
$$
while for $G = \bbT$ one can write
$$
(\calF^{-1} h_n)(t) \;=\; \frac{(\calF^{-1} \Delta h_n)(t)}{e^{-2\pi \imath\, t}-1}
\;,
$$
with the forward difference $\Delta h_n(p)=h_n(p+1)-h_n(p)$ and finds that 
$$
[h_n(D), \pi(a)]
\;=\;
\int_G \frac{\pi(\Dualityactionop_t(a)-a)}{e^{-2\pi \imath \,t}-1}\, (\calF^{-1} \Delta h_n)(t)\,e^{2\pi \imath \,D t} \difd{t} 
\;\in\; 
\Aa \rtimes_\Dualityactionop \bbT
\;.
$$
In both cases one can take the norm limit $n \to \infty$ since the integral kernels converge to smooth functions and therefore indeed specify elements in $\Aa \rtimes_\Dualityactionop G$ as the integral kernels are smooth and decay at infinity. To show that $\Aa \rtimes_\Dualityactionop G$ is an ideal in $\Toep $ one notes that, {\it e.g.},
$$
\chi_s(D) (\pi(a) f(D)) 
\;=\; 
[\chi_s(D),\pi(a)]f(D) + \pi(a) (\chi_sf)(D) \;\in\; \Aa \rtimes_\Dualityactionop G
\;,
$$
for all $a \in \Aa$, $f \in C_0(\hat{G})$ and $\chi_s\in C_{0,*}(\hat{G})$, because $\chi_sf\in C_0(\hat{G})$. 

\vspace{.1cm}

For the last claim, let us note that by density it is sufficient to prove \eqref{eq:toeplitzineq} for $e$ in the linear span of the generators $e= \sum_{i=1}^N \pi(a_i) f_i(D)$ with $a_i\in \Aa$ and $f_i\in C_c(\RM)$.  Also by density one can find for any $\epsilon>0$ a large enough $R > 0$ and a unit vector of the form $\phi=\chi(\abs{D}< R)\phi \in \Hh$ such that 
$$
\norm{\pi(a) \phi}_\Hh 
\;\geq \;
(1-\epsilon) \norm{a}
\;.
$$
The dual action $\hat{\Dualityactionop}$ acts on generators by translation $\hat{\Dualityactionop}_r(\pi(a_i) f(D))=\pi(a)f(D+r)$, thus
\begin{align*}
\norm{\hat{\xi}_{-r}\left(\pi(a)\Pp + e\right)\phi}_\Hh 
&
\,=\,
\Big\| \left(\pi(a)\chi_s(D-r) + \sum_{i=1}^N \pi(a_i) f_i(D-r)\right) \chi(\abs{D}<R)\phi\Big\|_\Hh \\
&
\,=\, 
\norm{\pi(a) \phi}_\Hh 
\\
&
\,\geq\, 
(1-\epsilon) \norm{a}
\end{align*}
for $r$ large enough. This shows $\|\hat{\xi}_{-r}\left(\pi(a)\Pp + e\right)\| \geq (1-\epsilon) \norm{a}$ and since $\hat{\Dualityactionop}$ is an isometry one concludes \eqref{eq:toeplitzineq}.
\hfill $\Box$

\vspace{.2cm}

The statements from Lemma~\ref{lemma:smtoep_help} taken together imply that
$$
\Toep 
\;=\; 
\pi(\Aa) \Pp + \Aa\rtimes_\Dualityactionop G
$$
without any further algebraic or topological completion and that each element of $\Toep$ can be written uniquely (for any fixed choice of $\Pp$)  in the form $\pi(a) \mathcal{P} + c$ with $a\in \Aa$ and $c \in \Aa \rtimes_\Dualityactionop G$. Thus:

\begin{proposition}
The map $\ToepProj: \Toep  \to \Aa$ defined through the equality 
$$
\ToepProj\big(\pi(a) \mathcal{P} + \Aa \rtimes_\Dualityactionop G\big)\; = \;a
$$
is a surjective homomorphism and hence there is the exact sequence
\begin{equation} 
0 \;\to\; \Aa \rtimes_\Dualityactionop G 
\;\hookrightarrow \;
\Toep  
\;\stackrel{\ToepProj}{\to} \;
\Aa 
\;\to \;0\;.
\label{seq:1ptoep} 
\end{equation}
\end{proposition}

As will be seen in the applications in Section~\ref{sec-BoundaryCurrents}, the smooth Toeplitz extension can be applied to generalize many arguments in which the less flexible Wiener-Hopf-extension or the discrete Toeplitz extension are commonly used.

\vspace{.2cm}

Note that one can also deduce from Lemma~\ref{lemma:smtoep_help} that $C_{0,*}(\hat{G})$-functions of the generator are contained in the multiplier algebra $M(\Aa \rtimes_\Dualityactionop G)$ and therefore the definition of $\Toep$ does not depend on the representation up to natural isomorphisms.  In particular, if $\hat{\pi}: \Aa \rtimes_\Dualityactionop G \to \Bb(\Hh')$ is a faithful non-degenerate representation of the crossed product then it extends uniquely to a representation of $\Toep$ that acts on the generators by $\hat{\pi}( \pi(a) f(D)) = \hat{\pi}(a) f(X)$ where $X$ is the generator of the unitary group $t \mapsto \hat{\pi}(e^{2\pi \imath D t})$ on $\Hh'$ and the first factor is given by the induced representation of $\Aa \subset M(\Aa \rtimes_\Dualityactionop G)$. To simplify the notation in the following, we will therefore assume that $\Aa \rtimes_\Dualityactionop G$ and $\Toep$ act on some Hilbert space $\Hh$ and identify $\Aa$ with its image in $M(\Aa \rtimes_\Dualityactionop G)$ such that the auxiliary representation $\pi$ can be omitted.

\section{Connecting maps of the smooth Toeplitz extension}
\label{sec-ConnectSmoothToep}

Let us now turn the attention towards the connecting maps in $K$-theory induced by the exact sequence 
\eqref{seq:1ptoep}  which we denote as following
$$
\Ind_G^\Dualityactionop \;:\; K_1(\Aa)\; \to\; K_0(\Aa \rtimes_\Dualityactionop G)\;,
\qquad
\Exp_G^\Dualityactionop \;:\; K_0(\Aa) \;\to\; K_1(\Aa \rtimes_\Dualityactionop G)
\;.
$$ 
In the case $G=\bbR$, it is well-known \cite{Connes81,FackSkandalis} that $K_{j}(\Aa) \simeq K_{1-j}(\Aa \rtimes_\Dualityactionop \bbR)$ with isomorphisms given by the Connes-Thom maps 
$$
\partial^\Dualityactionop_{1}\;:\; K_{1}(\Aa)\; \to\; K_{0}(\Aa \rtimes_\Dualityactionop \bbR)
\;,
\qquad
\partial^\Dualityactionop_{0}\;:\; K_{0}(\Aa)\; \to\; K_{1}(\Aa \rtimes_\Dualityactionop \bbR)
\;.
$$  
There is another related exact sequence, namely the Wiener-Hopf extension
\begin{equation}
\label{eq-WienerHopf}
0 
\;\to\; 
C_0(\bbR, \Aa)\rtimes_{\lambda \otimes \Dualityactionop} \bbR 
\;\to\; 
C_{0,*}(\bbR, \Aa)\rtimes_{\lambda \otimes \Dualityactionop} \bbR 
\;\to\; 
\Aa \rtimes_\Dualityactionop \bbR 
\;\to\; 0
\end{equation}
with $\lambda$ the left-translation on $C_{0,*}(\bbR)$. By Takai duality as stated in Theorem~\ref{eq-TakaiDuality}, one has $C_0(\bbR, \Aa)\rtimes_{\lambda \otimes \Dualityactionop} \bbR \cong \Aa\otimes\Kk(L^2(\RM))$, the connecting maps of the Wiener-Hopf extension can be seen as maps
$$
\Exp^\RM_\Dualityactionop \;:\; K_0(\Aa\rtimes_\Dualityactionop \bbR) \;\to\; K_1(\Aa )
\;,
\qquad
\Ind^\RM_\Dualityactionop \;:\; K_1(\Aa\rtimes_\Dualityactionop \bbR)\; \to\; K_0(\Aa )
\;.
$$ 
Let us stress that the indices on the connecting maps are chosen to be in exchanged order, since they are maps in the opposite direction compared to $\Exp_\bbR^\Dualityactionop$, $\Ind_\bbR^\Dualityactionop$. Rieffel \cite{Rieffel82} proved that those maps also are isomorphisms that are related to the Connes-Thom isomorphisms. This will be discussed further below. Ji has shown that the smooth Toeplitz extension can be considered the $KK$-inverse of the Wiener-Hopf extension \cite{Ji90} (which requires the usual separability conditions on $\Aa$ to apply the isomorphism between the $KK$-groups and extensions). As a consequence, one has

\begin{theorem}
\label{th:whext}
For separable $\Aa$ the connecting maps $\Ind^\Dualityactionop_\RM$ and $\Exp^\Dualityactionop_\RM$ of the smooth Toeplitz extension associated to $(\Aa,\Dualityactionop,\bbR)$ are isomorphisms and inverse to the connecting maps of the  Wiener-Hopf extension
$$
\Ind^\Dualityactionop_\RM
\;=\;
(\Exp^\RM_\Dualityactionop)^{-1}
\;,
\qquad
\Exp^\Dualityactionop_\RM
\;=\;
(\Ind^\RM_\Dualityactionop)^{-1}
\;.
$$
\end{theorem}

In the following, an elementary derivation of those facts is presented, which also makes the connection to the Connes-Thom isomorphisms apparent.  Let us recall the main result of Connes \cite{Connes81}, which gives an axiomatic characterization of his isomorphisms:

\begin{theorem}
\label{theo-Connes_axioms}
The Connes-Thom isomorphisms $(\partial_\Aa)_j^\Dualityactionop: K_j(\Aa) \to K_{1-j}(\Aa \rtimes_\Dualityactionop \bbR)$ are a functorial assignment of homomorphisms to any $C^*$-algebra $\Aa$ with an $\bbR$-action $\Dualityactionop$ which is uniquely determined by the properties
\begin{itemize}
\item[{\rm (i)}] {\bf(Normalization)} For the $\Aa = \bbC$ with trivial action $\Dualityactionop=\idmap$, i.e. $\Aa\rtimes_\Dualityactionop \bbR \simeq C_0(\bbR)$, the generator $[1]_0$ of $K_0(\bbC)$ is mapped by $(\partial_\bbC)^\idmap_0$ to the positive generator of $K_1(C_0(\bbR))$.
\item[{\rm (ii)}] {\bf(Naturality)} If $\Aa,\Bb$ are $C^*$-algebras with $\RM$-actions $\Dualityactionop, \beta$ and $\rho:\Aa \to \Bb$ is a covariant homomorphism, then 
$$\rho_* \circ (\partial_\Aa)_j^\Dualityactionop \;=\; (\partial_\Bb)_j^\beta \circ \rho_*\;.
$$
\item[{\rm (iii)}] {\bf(Suspensions)} For  the suspension map $s^1_\Aa: K_1(\Aa) \to K_0(S\Aa)$ and  the Bott map $s^0_\Aa: K_0(\Aa) \to K_1(S\Aa)$, one has 
$$
s^j_{\Aa\rtimes_{\Dualityactionop} \bbR}\circ (\partial_{\Aa})_{1-j}^\Dualityactionop 
\;=\; 
(\partial_{S\Aa})_j^{\Dualityactionop} \circ s^{1-j}_{\Aa}
\;,
$$
for $j=0,1$ where the action induced by $\mbox{\rm id}\times \Dualityactionop$ on $S\Aa$ is denoted by the same symbol $\Dualityactionop$. 
\end{itemize}
\end{theorem}

For concreteness and later use, let us also state the precise definitions of the suspension maps under the identification $S\Aa = C_0(\bbR) \otimes \Aa$. The Bott map $s_0: K_0(\Aa) \to K_1(S\Aa)$ takes a class $[e]_0 -[\one_K^\sim]_0 \in K_0(\Aa)$ with representative $e\in M_N(\Aa^\sim)$ having a scalar part $s(e)=\one_K^\sim$ to the loop 
\begin{equation}
\label{eq-s0map}
s_0([e]_0-[\one^\sim_K]_0) 
\;=\; 
[(\one^{\sim}_N-e + e \otimes f )(\one^{\sim}_N-\one_K^\sim + \one_K^\sim \ \otimes \overline{f})\big]_1
\end{equation}
for any function $f\in 1+ C_0(\bbR)$ with winding number one, such as $f(x)=(x-\imath)(x+\imath)^{-1}$.
The suspension map $s_1: K_1(\Aa) \to K_0(S\Aa)$ is the connecting map from the exact sequence
$$
0 
\;\to\; 
C_0(\bbR, \Aa) 
\;\to\; 
C_{0,*}(\bbR, \Aa) 
\;\to\; 
\Aa 
\;\to\; 
0
\;,
$$
with $C_{0,*}(\bbR, \Aa)$ being the continuous functions that admit a limit in $+\infty$ and vanish at $-\infty$. More explicitly, for $[u]_1 \in K_1(\Aa)$ represented by a unitary $u\in M_N(\Aa^\sim)$ one chooses a unitary element $v\in C_{0,*}(\bbR, M_{2N}(\Aa^\sim))^\sim$ such that $v(-\infty) = \one_2$ and $v(\infty)= \mathrm{diag}(u, u^*)$, and then sets
\begin{equation}
\label{eq-s1map}
s_1[u] 
\;=\; 
\left[
v \begin{pmatrix}
1&0\\0&0
\end{pmatrix} v^*
\right]_0 
\,-\, 
\left[\begin{pmatrix}
1&0\\0&0
\end{pmatrix}\right]_0
\;.
\end{equation}

The idea is now to verify the axioms of the Connes-Thom isomorphisms for the connecting maps of the smooth Toeplitz extension. 

\begin{proposition}
\label{prop:connecting_maps_axiomatic}
The connecting maps of the smooth Toeplitz extension $\ToepR$, here written as $(\Exp_\Aa)_\bbR^\Dualityactionop$ and $(\Ind_\Aa)_\bbR^\Dualityactionop$ for clarity, are related to the Connes-Thom isomorphisms by
$$
(\Exp_{\Aa})^\Dualityactionop_\RM 
\;=\; 
-\,
(\partial_\Aa)^\Dualityactionop_0
\;, 
\qquad 
(\Ind_{\Aa})_\RM^\Dualityactionop \;=\; -\,(\partial_\Aa)^\Dualityactionop_1
\;.
$$ 
\end{proposition}

\noindent{\bf Proof.}
It is sufficient to show that $-(\Exp_{\Aa})^\Dualityactionop_\RM$ and $-(\Ind_{\Aa})_\RM^\Dualityactionop$ satisfy the properties (i)-(iii) of Theorem~\ref{theo-Connes_axioms}:

\vspace{.1cm}

\noindent {\rm (i)} For  the generator $D$ of the regular representation of $\bbC \rtimes_\idmap \bbR$, the element $1 \in \bbC$ is lifted to the self-adjoint $a = f(D)$ for any real function $f\in C_{0,*}(\bbR)$ with $\lim_{t\to\infty}f(t)=1$. With $f(t)=\frac{1}{\pi} \arctan(t) + \frac{1}{2}$, one has 
$$
(\Exp_{\bbC})_\bbR^\idmap([1]_0)
\;=\;
[\exp(-2\pi \imath\, a)]_1 
\;=\; 
\left[\frac{D + \imath \,1^\sim}{D - \imath \,1^\sim}\right]_1
\;=\;
-\,[(\partial_\bbC)_0^\idmap]_1
\;,
$$
since the function $t \mapsto \frac{t+\imath}{t-\imath}$ has  winding number $-1$, {\it i.e.} represents the negative of the positive generator of $K_1(C_0(\bbR))$.

\vspace{.1cm}

\noindent {\rm (ii)}  The covariant homomorphism $\rho$ extends to the multiplier algebras and thereby induces a homomorphism $\rho: M(\Aa\rtimes_\Dualityactionop \bbR) \to M(\Bb\rtimes_\beta \bbR)$ in the obvious manner which acts on the generators of the crossed product by $\rho(a \,e^{2\pi \imath \,D_\Dualityactionop t})= \rho(a)\, e^{2 \pi \imath \,D_\beta t}$ and thus $\rho(a \,f(D_\Dualityactionop))=\rho(a)\,f(D_\beta)$ with the obvious notations. Hence one obtains the commutative diagram
$$
\begin{tikzcd}
0 \arrow[r] &  \Aa \rtimes_\Dualityactionop \bbR \arrow[d, "\rho"] \arrow[r] & {{\rm T}(\Aa, \bbR, \Dualityactionop)} \arrow[d, "\rho"] \arrow[r,"\ToepProj"] & \Aa \arrow[d, "\rho"] \arrow[r] & 0 \\
0 \arrow[r] &  \Bb \rtimes_\beta \bbR \arrow[r] & {{\rm T}(\Bb, \bbR, \beta)} \arrow[r,"\ToepProj"] & \Bb \arrow[r] & 0
\end{tikzcd}
$$
and the naturality then follows from the naturalness of the connecting maps.

\vspace{.1cm}

\noindent {\rm (iii)}  Since crossed products commute with suspensions, the definition of the exponential map using Bott periodicity reads 
$$
(\Exp_\Aa)_\bbR^\Dualityactionop 
\;=\; 
(s^1_{\Aa\rtimes_\Dualityactionop \bbR})^{-1} \circ (\Ind_{S\Aa})_\bbR^\Dualityactionop \circ s_\Aa^0
\;,
$$
which is precisely the suspension property (iii) for the case $j=1$. A second application of Bott periodicity then further implies that the case $j=0$ also holds.
\hfill $\Box$

\vspace{.2cm}

As emphasized by \cite{Ji90}, the Wiener-Hopf-extension \eqref{eq-WienerHopf} is naturally isomorphic to a smooth Toeplitz extension with the dual action
$$
\begin{tikzcd}
0 \arrow[r] &  \Aa \otimes \Kk(L^2(\RM))  \arrow[d, ""] \arrow[r] & 
C_{0,*}(\bbR, \Aa)\rtimes_{\lambda \otimes \Dualityactionop} \bbR  \arrow[d, ""] \arrow[r,""] & 
\Aa \rtimes_\Dualityactionop \bbR \arrow[d, ""] \arrow[r] & 0 \\
0 \arrow[r] &  \Aa \rtimes_\Dualityactionop \bbR \rtimes_{\hat{\Dualityactionop}} \bbR \arrow[r] & {\rm T}(\Aa\rtimes_\Dualityactionop \bbR, \bbR, \hat{\Dualityactionop}) \arrow[r,"\ToepProj"] & \Aa\rtimes_\Dualityactionop \bbR \arrow[r] & 0
\end{tikzcd}
$$
and therefore one can express the connecting maps in terms of the Connes-Thom isomorphisms 
\begin{align*}
\Ind^\bbR_\Dualityactionop 
&\;=\; (i_T)_* \circ \Ind_\bbR^{\hat{\Dualityactionop}} \;=\; -\,(i_T)_* \circ \partial^{\hat{\Dualityactionop}}_1\;, \\
 \Exp^\bbR_\Dualityactionop &
 \;=\; (i_T)_* \circ \Exp_\bbR^{\hat{\Dualityactionop}} \;=\; -\,(i_T)_* \circ \partial^{\hat{\Dualityactionop}}_0
 \;,
\end{align*} 
with the Takai isomorphism $i_T:\Aa \rtimes_\Dualityactionop \bbR \rtimes_{\hat{\Dualityactionop}} \bbR\to \Aa\otimes \Kk(L^2(\RM))$, see Theorem~\ref{eq-TakaiDuality}. Connes also proved in \cite{Connes81} that
$$
(i_T)_* \circ \partial^{\hat{\Dualityactionop}}_{1-j} \circ \partial^{\Dualityactionop}_{j} 
\;=\; 
\idmap_{K_j(\Aa)}
\;,
\qquad
j=0,1\;,
$$
under the identification of $K_j(\Aa)\simeq K_j(\Aa \otimes \Kk(L^2(\RM)))$. Combining those relations with Proposition~\ref{prop:connecting_maps_axiomatic} in various ways, one directly deduces further interrelations between the connecting maps of both smooth Toeplitz and Wiener-Hopf-extensions. This also concludes the proof of Theorem~\ref{th:whext}.

\begin{proposition}
\label{prop-ConnectingTakai}
The connecting maps of the smooth Toeplitz extensions and the  Wiener-Hopf extensions associated to $(\Aa,\Dualityactionop,\bbR)$ and $(\Aa\rtimes_\Dualityactionop\RM,\hat{\Dualityactionop},\bbR)$ satisfy
$$
\Exp^\Dualityactionop_\bbR 
\;=\; 
\Exp_{\hat{\Dualityactionop}}^\RM \circ (i_T^{-1})_*
\;,
\qquad
\Ind^\Dualityactionop_\bbR 
\;=\; 
\Ind_{\hat{\Dualityactionop}}^\RM \circ (i_T^{-1})_*
\;,
$$
and
\begin{align*}
\Exp_\Dualityactionop^\bbR 
\;&=\; 
(i_T)_*\circ\Exp^{\hat{\Dualityactionop}}_\RM 
\;=\; 
(\Ind_\bbR^\Dualityactionop)^{-1} \;\;=\; -\,(\partial_1^\Dualityactionop)^{-1}
\;,
\\
\Ind_\Dualityactionop^\bbR 
\;&=\; 
(i_T)_* \circ\Ind^{\hat{\Dualityactionop}}_\RM 
\;\;=\; 
(\Exp_\bbR^\Dualityactionop)^{-1} 
\;=\; -\,(\partial_0^\Dualityactionop)^{-1}
\;.
\end{align*}
\end{proposition}

Let us remark that Rieffel \cite{Rieffel82} apparently shows that the Connes-Thom isomorphisms are inverse to the connecting maps of the Wiener-Hopf extension instead of their negatives. Translating to the present notations, to find the pre-image of a projection $e \in M_n(\Aa)$  Rieffel constructs an isometry $V_e \in M_N({\rm T}(\Aa\rtimes_\Dualityactionop \bbR, \bbR, \hat{\Dualityactionop})^\sim)$ whose corange projection $E_e$ defines the same class in $K_0(A)$ as $e$, and then finds that $\ToepProj(V_e) \in M_N(\Aa\rtimes_\Dualityactionop \bbR)^\sim$ is a unitary that defines the same class as the Connes-Thom isomorphism $[\ToepProj(V_e)]_1=\partial_0^\Dualityactionop([e]_0)$. Rieffel then claims that $[\ToepProj(V_e)]_1$ is the preimage of $[e]_0$ under the index map. However, when lifting to partial isometries, one has
$$
\Ind_\Dualityactionop^\bbR([\Pi(V_e)]_1)
\;=\; [\Ker( V_e)]_0 \;-\; [\Ker( V_e^*)]_0 \;=\; -\,[E_e]_0\;=\; -\,[e]_0
\;,
$$
which is consistent with Proposition~\ref{prop-ConnectingTakai}.

\vspace{.2cm}

For the computations leading to main duality result (Theorem~\ref{theo-smooth_duality} below), it will be essential to write out  the Connes-Thom isomorphism $\partial^{\Dualityactionop}_0$ in a particularly simple and explicit form. This is proved in Proposition~\ref{prop-index_map_preimage} below, which is essentially due to Connes \cite{Connes81}. It will later on be applied to the dual action $\hat{\Dualityactionop}$ on the crossed product $\Aa \rtimes_\Dualityactionop \bbR$ instead of $\Dualityactionop$ on $\Aa$ and since that algebra is not unital  we always adjoin a unit. The unit of the unitization $\Aa^{\sim}$ will be denoted by $\one^{\sim}$ and its tensor product by $\one^{\sim}_N = \one^{\sim} \oplus \ldots  \oplus \one^{\sim} \in M_N(\Aa^{\sim})$. Unless indicated otherwise, the standard picture of $K$-theory will be used \cite{RLL} where elements in $K_0(\Aa)$ are represented by formal differences $[e]_0 - [\one^{\sim}_K]_0$ of projections $e \in M_N(\Aa^{\sim})$ with scalar part $s(e)=\one_K^{\sim}$ and elements of $K_1(\Aa)$ by equivalence classes $[v]_1$ of unitaries $v\in M_N(\Aa^{\sim})$ with scalar part $s(v)=\one_N^{\sim}$.

\begin{proposition}
\label{prop-index_map_preimage}
Let $(\Aa, \bbR, \Dualityactionop)$ be a $C^*$-dynamical system and let $D$ be the unbounded generator of $\Dualityactionop$ in some faithful representation of $\Aa \rtimes_\Dualityactionop \bbR$. If $[e]_0 - [\one^{\sim}_K]_0 \in K_0(\Aa)$ is represented by a differentiable projection $e \in M_N(\Aa^{\sim})$, {\it i.e.} the commutator $[D,e]\in\Aa$ is well-defined, then its image under the Connes-Thom isomorphism is given by
$$
\partial_0^\Dualityactionop
\big([e]_0-[\one^{\sim}_K]_0\big) 
\;=\; 
\big[((\one^{\sim}_N-e) + e \,u_e) ((\one^{\sim}_N-\one_K^\sim) + \one_K^\sim \,u_e^*)\big]_1
$$
with the unitary $u_e = f(D_e)$ constructed from $f(x)=\frac{\sinh(x)-\imath}{\sinh(x)+\imath}$ and  
$$
D_e 
\;=\; 
e\,D\,e \;+\; (\one^{\sim}_N -e)\,D\,(\one^{\sim}_N -e) 
\;=\; 
D \;-\; [[D,e],e]
\;.
$$ 
\end{proposition}

\noindent{\bf Proof.}
Let us describe the construction of this isomomorphism by Connes \cite{Connes81}. If $e$ is invariant under $\Dualityactionop$, then consider the subalgebra $\Aa_0$ of fixed points under $\Dualityactionop$ such that $[e]_0-[\one_K^\sim]_0 \in K_0(\Aa_0)$. The inclusion $\imath_0: \Aa_0 \to \Aa$ is equivariant and hence gives rise to a canonical homomorphism $\hat{\imath}_0: \Aa_0 \rtimes_\Dualityactionop \bbR \to \Aa \rtimes_\Dualityactionop \bbR$. Since the action is trivial action on $\Aa_0$, one can identify $\Aa_0\rtimes_\Dualityactionop \bbR = S\Aa$ and pulling back the Connes-Thom isomorphism to the Bott map by naturality implies
\begin{align*}
\partial_0^\Dualityactionop([e]_0-[\one_K^\sim]_0) &= (\partial_0^\Dualityactionop \circ (\imath_0)_*)([e]_0-[\one_K^\sim]_0) \\
&
\;=\; 
((\hat{\imath}_0)_* \circ s_0)([e]_0-[\one_K^\sim]_0) \\
&
\;=\; 
(\hat{\imath}_0)_*\big[(\one^{\sim}_N-e + e \otimes f ) (\one^{\sim}_N-\one_K^\sim + \one_K^\sim \ \otimes \overline{f})\big]_1  \\
&
\;=\; 
\big[(\one^{\sim}_N-e + e \,f(D)  ) (\one^{\sim}_N-\one_K^\sim + \one_K^\sim \,\overline{f}(D)  )\big]_1\,
\end{align*}
for an arbitrary unitary function $f \in 1+ C_0(\bbR)$ with winding number $1$.

\vspace{.1cm}

To compute $\tilde{\partial}^{\Dualityactionop}_0$ for an arbitrary differentiable projection $e \in M_N(\Aa^{\sim})$ representing a class in $K_0(\Aa)$, one considers a new $\bbR$-action $\Dualityactionop'$ that leaves $e$ invariant and is outer equivalent to $\Dualityactionop$, {\it i.e.} there is a continuous family $(w_k)_{k\in \bbR}$ of unitaries $w_k\in M_N(\Aa^\sim)$ which satisfies the cocycle condition
$$
\Dualityactionop_{k}'(a) 
\;=\; 
w_k \,\Dualityactionop_k(a)\,w_k^*
\;,
\qquad
\forall\,k\in\RM, \;a\in M_N(\Aa^\sim)
\;.
$$ 
One particular choice that fixes $e$ is given by Connes' unitary cocycle 
$$
w_k 
\;=\; 
e^{2\pi \imath D_e k}\,e^{-2\pi \imath\,Dk}
$$ 
for $D_e$ the unbounded multiplier defined above in some faithful representation of the crossed product $\Aa^\sim \rtimes_\Dualityactionop \bbR$. By definition $\Dualityactionop_{k}'(e)=e$ and to see that $k\mapsto w_k \in M_N(\Aa^\sim)$ is continuous and independent of the representation one notes that it is equivalently defined by the perturbative expansion 
\begin{equation}
\label{eq:connes_cocycle_def}
w_k 
\;=\;
\one\;+\;
\sum_{m=1}^\infty  (2\pi \imath)^m\int_{0 < s_1 < \ldots  < s_m < k} \Dualityactionop_{s_1} ([e,[D,e]])\cdots\Dualityactionop_{s_m}([e,[D,e]])\; \difd s_1 \cdots \difd s_m
\end{equation}
which converges absolutely in $M_N(\Aa^\sim)$.  Then introduce the action 
$$
\Dualityactionop'\;:\; M_{2N}(\Aa) \times \bbR \;\to\; M_{2N}(\Aa)
\;, 
\qquad \Dualityactionop_k' \begin{pmatrix}
a & b \\ c & d
\end{pmatrix}
\;=\;  
\begin{pmatrix}
\Dualityactionop_k(a) & \Dualityactionop_k(b) w_k^* \\
w_k \Dualityactionop_k(c) & w_k \Dualityactionop_k(d) w^*_k
\end{pmatrix}
\;.
$$
By construction, $\Dualityactionop_k'(\mathrm{diag}(e,0))= \mathrm{diag}(\Dualityactionop_k(e),0)$ and $\Dualityactionop_k'(\mathrm{diag}(0,e))= \mathrm{diag}(0,e)$. On the upper left corner the action reduces to $\Dualityactionop$, hence the inclusion $\imath_N: M_N(\Aa) \to M_{2N}(\Aa)$ is equivariant and gives rise to a canonical inclusion $\hat{\imath}_N: M_N(\Aa)\rtimes_\Dualityactionop \bbR \to M_{2N}(\Aa)\rtimes_{\Dualityactionop'}\bbR$. Let further $\Aa_e \subset M_{2N}(\Aa)$ be the fixed point subalgebra under $\Dualityactionop'$. The inclusion $\imath_e: \Aa_e \to M_{2N}(\Aa)$ is again equivariant and induces a map $\hat{\imath}_e: S\Aa_e \simeq \Aa_e\rtimes_{\Dualityactionop'}\bbR \to M_{2N}(\Aa)\rtimes_{\Dualityactionop'}\bbR$.
Writing $g=f-1$ for brevity and using simple homotopies that interchange the upper left and lower right corners this implies for the Connes-Thom isomorphisms
\begin{align*}
\partial^{\Dualityactionop'}_0 &
\left(\left[\begin{pmatrix} e & 0 \\ 0 & 0_N \end{pmatrix}\right]_0-\left[\begin{pmatrix}  \one_K^\sim & 0 \\ 0 & 0_N\end{pmatrix}\right]_0\right) 
\\
&
\;=\;
\partial^{\Dualityactionop'}_0\left(\left[\begin{pmatrix} 0_N & 0 \\ 0 & e \end{pmatrix}\right]_0-\left[\begin{pmatrix}  0_N & 0 \\ 0 & \one_K^\sim\end{pmatrix}\right]_0\right)  \\
&
\;=\; 
\left[\begin{pmatrix}\one_N^\sim & 0  \\ 0 & (\one^{\sim}_N + e g(D_e) ) (\one^{\sim}_N + \one_K^\sim \overline{g}(D_e))\end{pmatrix}\right]_1 \\
&
\;=\; 
\left[\begin{pmatrix}(\one^{\sim}_N + e g(D_e) ) (\one^{\sim}_N + \one_K^\sim\ \overline{g}(D_e))  & 0 \\ 0 & \one_N^\sim \end{pmatrix}\right]_1 
\,,
\end{align*}
since $0_N \oplus e$ is an invariant projection and $D\oplus D_e$ the generator of $\Dualityactionop'$. This computation shows
$$
((\hat{\imath}_N)_* \circ \partial^{\Dualityactionop}_0)
([e]_0-[\one_K^\sim]_0 )  
\;=\; 
(\hat{\imath}_N)_*\big[(\one^{\sim}_N+ \,g(D_e)  ) (\one^{\sim}_N + \one_K^\sim \,\overline{g}(D_e)  )  \, \big]_1\,
$$ 
and since $(\imath_N)_*$ is an isomorphism the same is true for $(\hat{\imath}_N)_* = (\partial^{\Dualityactionop'}_0)^{-1} \circ (\imath_N)_* \circ \partial^{\Dualityactionop}_0 $ by naturality. Hence this equality determines $\partial^\alpha_0([e]_0-[\one_K^\sim]_0)$ uniquely.
\hfill $\Box$

\vspace{.2cm}

Next let us examine the connecting maps in the case $G=\bbT$. A complete characterization is given by

\begin{proposition}
\label{prop:sm_connecting_maps}
Consider a given $\bbT$-action $\Dualityactionop$ on $\Aa$ as a $\bbR$-action by setting $\Dualityactionop_{t+1}=\Dualityactionop_t$. Then the connecting maps of $\ToepT$ are given by
\begin{equation}
 \label{eq-sm_toep_factor}
\Ind^\Dualityactionop_\bbT 
\;=\; 
q_* \circ \Ind^\Dualityactionop_\bbR 
\;,
\qquad 
\Exp^\Dualityactionop_\bbT 
\;=\; 
q_* \circ \Exp^\Dualityactionop_\bbR 
\;,
\end{equation}
with $q: \Aa \rtimes_\Dualityactionop \bbR \to \Aa \rtimes_\Dualityactionop \bbT$ the natural surjection defined by
$$
q\left(\int_\bbR f(x) e^{2\pi\imath\, D_\bbR x}\, \difd{x}\right)
\;=\;
\int_\bbT \left(\sum_{k \in \bbZ} f(x + k)\right) \,e^{2\pi\imath\, D_\bbT x} \, \difd{x}
$$
for all $f\in C_c(\bbR, \Aa)$ with $D_\bbR$ and $D_\bbT$ being the generators of the respective actions in a faithful covariant representation. 
\end{proposition}

\noindent {\bf Proof.}
Since $q$ is surjective, it extends uniquely to the multiplier algebras ({\it e.g.} Proposition~2.2.16 in \cite{WO})  and thus to $\ToepR$. Clearly $q(a g(D_\bbR)) = a g(D_\bbT)$ for all $a\in \Aa$ and $g\in C_0(\bbR)$. Hence the extension actually gives to a surjection $q: \ToepR \to \ToepT$. Since the diagram
$$
\begin{tikzcd}
0 \arrow[r] &  \Aa \rtimes_\Dualityactionop \bbR \arrow[d, "q"] \arrow[r] & \ToepR \arrow[d, "q"] \arrow[r,"\ToepProj"] & \Aa \arrow[d, equal] \arrow[r] & 0 \\
0 \arrow[r] &  \Aa \rtimes_\Dualityactionop \bbT \arrow[r] & \ToepT \arrow[r,"\ToepProj"] & \Aa \arrow[r] & 0
\end{tikzcd}
$$
commutes, \eqref{eq-sm_toep_factor} follows immediately from the naturality of the connecting maps.
 \hfill $\Box$
 
\vspace{.2cm}

One can also relate the connecting maps of the smooth $\bbT$-Toeplitz extension with the six-term exact sequence 
 for crossed products with $\bbT$ \cite[10.6]{Bla}
\begin{equation*}
\begin{tikzcd}
K_0(\Aa \rtimes_\Dualityactionop \bbT) \arrow[r, "\idmap-\hat{\Dualityactionop}_*"] & K_0(\Aa \rtimes_\Dualityactionop \bbT) \arrow[r] & K_0(\Aa)\arrow[d, "\Exp^\ZM_{\hat{\Dualityactionop}}"] \\
K_1(\Aa) \arrow[u, "\Ind^\ZM_{\hat{\Dualityactionop}}"]  & K_1(\Aa \rtimes_\Dualityactionop \bbT) \arrow[l] & K_1(\Aa \rtimes_\Dualityactionop \bbT) \arrow[l,"\idmap-\hat{\Dualityactionop}_*"]
\end{tikzcd}
\end{equation*}
which is obtained by applying the Pimsner-Voiculescu sequence
\begin{equation*}
\begin{tikzcd}
K_0(\Bb) \arrow[r, "\idmap-\beta_*"] & K_0(\Bb) \arrow[r] & K_0(\Bb\rtimes_{\beta} \bbZ)\arrow[d, "\Exp^\ZM_{\beta}"] \\
K_1(\Bb \rtimes_{\beta} \bbZ) \arrow[u, "\Ind^\ZM_{\beta}"]  & K_1(\Bb) \arrow[l] & K_1(\Bb) \arrow[l,"\idmap-\beta_*"]
\end{tikzcd}
\end{equation*}
to the special case $\Bb= \Aa \rtimes_\Dualityactionop \bbT$ and $\beta = \hat{\Dualityactionop}$ and using Takai duality $\Bb \rtimes_\beta \bbZ \cong\Aa\otimes \Kk(L^2(\TM))$ to identify $K_i(\Bb\rtimes_\beta \bbZ) \simeq K_i(\Aa)$ and  $K_i(\Bb) = K_i(\Aa \rtimes_\Dualityactionop \bbT)$.
The connecting maps here are also given by $\Exp^\bbZ_{\hat{\Dualityactionop}} = q_* \circ \partial_0^\Dualityactionop$ and $\Ind^\bbZ_{\hat{\Dualityactionop}} = q_* \circ \partial_1^\Dualityactionop$, again with $q: \Aa \rtimes_\Dualityactionop \bbR \to \Aa \rtimes_\Dualityactionop \bbT$ and the Connes-Thom isomorphisms. Comparing with Proposition \ref{prop:connecting_maps_axiomatic} and Proposition \ref{prop:sm_connecting_maps}, this recovers the connecting maps of the smooth $\bbT$-Toeplitz extension up to a sign
$$
\Exp_\bbT^\Dualityactionop
\;=\;
-\,\Exp_{\hat{\Dualityactionop}}^\ZM
\;,
\qquad
\Ind_\bbT^\Dualityactionop
\;=\;
-\,\Ind_{\hat{\Dualityactionop}}^\ZM
\;.
$$ 
Note that due to the supplementary Takai duality involved, the statement does not involve the inverses as in Theorem~\ref{th:whext}. 

\vspace{.2cm}

Since the Pimsner-Voiculescu sequence can also be obtained from the six-term-sequence of the discrete Toeplitz extension  \cite[10.2]{Bla}
\begin{equation} 
\label{eq:discreteToeplitz_ext}
0 
\;\to \;
C_0(\bbZ, \Bb)\rtimes_{\lambda \otimes \beta} \bbZ 
\;\hookrightarrow\; 
C_{0,*}(\bbZ, \Bb)\rtimes_{\lambda \otimes  \beta} \bbZ 
\;\to\; 
\Bb \rtimes_\beta \bbZ 
\;\to\;
 0
\end{equation}
with $\lambda$ left translation, this gives an alternative way to compute the connecting maps of the smooth $\bbT$-Toeplitz extension. A similar relation also holds for the other direction:

\begin{lemma}
\label{lemma:discrete_toeplitz}
For a $C^*$-algebra $\Bb$ with $\bbZ$-action $\beta$, the exact sequence \eqref{eq:discreteToeplitz_ext} is naturally isomorphic to the smooth $\bbT$-Toeplitz extension 
$$
0 \;\to\; (\Bb \rtimes_\beta \bbZ) \rtimes_{\hat{\beta}} \bbT 
\;\hookrightarrow \;
\rm T(\Bb \rtimes_\beta \bbZ, \hat{\beta}, \bbT)
\;\stackrel{\ToepProj}{\to} \;
\Bb \rtimes_\beta \bbZ
\;\to \;0\;.
$$
\end{lemma}

\noindent{\bf Proof.}
Let us describe an explicit representation of  the algebra $C_{0,*}(\bbZ, \Bb)\rtimes_{\lambda \otimes  \beta} \bbZ$ which we denote by ${\rm T}_\bbZ $. Assume that $\Bb$ acts on a Hilbert space $\Hh_0$  in which $\beta$ is implemented by a unitary $u$ on $\Hh_0$, namely $\beta_k(b)= u^k b (u^*)^k$. Using the operators $P_j = |j\rangle \langle j|$ and $P_+ = \sum_{j=0}^\infty P_j$ on $\ell^2(\bbZ)$, let us represent a function $f \in C_{0,*}(\bbZ, \Bb)$ as the multiplication operator 
$$
\sum_{j \in \bbZ} f(j) \otimes P_j
$$ 
on $\Hh \otimes \ell^2(\bbZ)$. The action $(\lambda \otimes \beta)_k(f)(n)= \beta_k(f(n-k))$ is then implemented by conjugation with the unitary $U=u \otimes S$ where $S$ is the shift $S|j\rangle = |j-1\rangle$. This is a faithful covariant representation of $(C_{0,*}(\bbZ, \Bb), \lambda \otimes \beta, \bbZ)$ on $\Hh \otimes \ell^2(\bbZ)$ and hence ${\rm T}_\bbZ$ can be identified with the $C^*$-algebra generated by elements of the form $(b \otimes P_j)U^k$ and $(b \otimes P_+)U^k$. The surjection $\Pi:{\rm T}_\bbZ \to \Bb \rtimes_\beta \bbZ$ is then densely defined on the generators by $(b\otimes P_+) U^k + \Bb \otimes \Kk(\ell^2(\ZM)) \mapsto b u^k$.

\vspace{.1cm}

In this explicit representation the unbounded operator $X = \sum_{j\in \bbZ} j (\one \otimes P_j)$ commutes with $C_{0,*}(\bbZ,\Bb)$ and satisfies 
$$
e^{2\pi \imath\, Xt}\, U^k \, e^{-2\pi \imath\, Xt} 
\;=\; 
\overline{\langle k, t\rangle} \;U^k
\;.
$$ 
Hence it generates the $\bbT$-action $\hat{\beta}$. It is then straightforward to check that ${\rm T}_\bbZ$ coincides with the $C^*$-algebra generated by all elements of the form $(b\otimes \one) U^k f(X)$ for $b \in \Bb$ and $f\in C_{0,*}(\bbZ)$, namely $\rm T(\Bb \rtimes_\beta \bbZ, \beta, \bbT) \cong {\rm T}_\bbZ$ in such a way that $C_0(\bbZ,\Bb) \rtimes_{\lambda \otimes \beta} \bbZ \cong \Bb \otimes \Kk(\ell^2(\ZM)) \cong \Bb \rtimes_{\beta} \bbZ \rtimes_{\hat{\beta}} \bbT$, with the last isomorphism being given by Takai duality.
\hfill $\Box$

\vspace{.2cm}

Hence the connecting maps of any discrete Toeplitz extension can be related to those of a smooth $\bbT$-Toeplitz extension and vice versa. In this sense, the smooth $\bbT$-Toeplitz extension is dual to the discrete Toeplitz extension in a similar way that the smooth $\bbR$-Toeplitz extension is dual to the Wiener-Hopf extension (compare \cite{Ji90}).

\section{Cyclic cohomology and smooth subalgebras}
\label{sec:smooth_chern}

Section~\ref{sec-BreuerToep} already introduced the Chern cocycles and their pairings with projections and unitaries in a special case. In that setting the coycles were defined on large (non-separable) and fairly intractable algebras such that analytical aspects were dominant while other properties of the index pairings such as their homotopy invariance only played a minor role. This is different for cocycles that are densely defined on (separable) $C^*$-algebras and hence let us introduce further notions of cyclic cohomology \cite{Connes94}.

\begin{definition}
A cyclic $n$-cocycle on an algebra $A$ is an $n+1$-linear functional $\varphi: A^{n+1}\to \bbC$ which is cyclic
$$
\varphi(a_0,\ldots ,a_n)\;=\;(-1)^n \varphi(a_1,\ldots ,a_{n},a_0)
$$
and a cocycle w.r.t. the Hochschild boundary operator $b$ defined by
\begin{align*}
(b\varphi)(a_0,\ldots ,a_{n+1})\;=\; & \sum_{j=0}^n (-1)^j \varphi(a_0, \ldots , a_j a_{j+1},\ldots , a_{n+1})
\\
& \;+\; (-1)^{n+1} \varphi(a_{n+1} a_0, a_1, \ldots , a_{n})\;,
\end{align*}
that is, $b\varphi=0$.
\end{definition}

Natural constructions are more easily carried out using the equivalent formalism of $n$-cycles:

\begin{definition}
	An $n$-cycle $(\Omega, d, \varphi)$ over an algebra $A$ consists of the following:
\begin{itemize}

\item[{\rm (i)}] A graded algebra $\Omega = \bigoplus_{j\neq 0} \Omega_j$ where $\Omega_{j_1} \Omega_{j_2} \subset \Omega_{j_1+j_2}$ together with a homomorphism $\rho: A \to \Omega_0$. The elements of $\Omega_j$ are called homogeneous with degree $j$. 

\item[{\rm (ii)}]  A graded differential $d: \Omega \to \Omega$, {\it i.e.} a linear map with  $d\Omega_j \subset \Omega_{j+1}$, $d^2=0$ and $d(ab)= (da)b + (-1)^{\mathrm{deg}(a)} a (db)$ for elements of homogeneous degree.

\item[{\rm (iii)}] A closed graded trace $\varphi: \Omega \to \bbC$ of top degree $n$, which means $\varphi(ab)=(-1)^{\mathrm{deg}(a) \mathrm{deg}(b)}\varphi(ba)$, $\varphi(da)=0$ and finally $\varphi(a)=0$ for $\mathrm{deg}(a) > n$.

\end{itemize}
\end{definition}

Every $n$-cycle defines an $(n+1)$-linear cyclic cocycle over $A$ via its character
$$
\varphi(a_0, \ldots ,a_n) \;=\; \varphi\big(\rho(a_0)d\rho(a_1)\cdots d\rho(a_n)\big)
\;,
$$
and vice versa any cyclic cocycle defines an $n$-cycle by lifting it to the universal differential graded algebra
$$
\Omega(A) \;=\; \bigoplus_{j\geq 0}\, A^\sim \otimes A^{\otimes j}
$$
which has a natural graded differential for which $(a_0+\lambda\one^\sim)da_1\cdots da_n = (a_0+\lambda \one^\sim) \otimes a_1 \otimes \dots \otimes a_n$. Since the homomorphism $A \to \Omega(A)_0$ is the identity map it is often omitted from the notation. One can also use this lift to define an extension $\varphi \# \Tr_N$ of any cyclic cocycle to the matrix algebras $M_N(A^\sim)$ in the obvious manner.

\vspace{.2cm}

We assume without loss of generality that $A$ is unital and let $\varphi$ be a cyclic $n$-cocycle over $A$, respectively its corresponding $n$-cycle denoted by the same letter. For $n$ even it pairs with the algebraic $K$-group of $K^{\mathrm{alg}}_0(A)$ via
\begin{equation}
\label{eq:evenpairing}
\langle [\varphi], [e]_0\rangle \;=\; \varphi(e, e, \ldots , e) 
\;=\; 
\varphi\big(e (de)^n\big)
\;,
\end{equation}
with $[e]_0 \in K^{\mathrm{alg}}_0(A)$ represented by an idempotent $e\in M_N(A)$.
For odd $n$ the pairing is instead 
\begin{equation}
\label{eq:oddpairing}
\langle [\varphi], [u]_1\rangle \;=\; \varphi(u^{-1}-1, u-1, u^{-1}-1, \ldots , u-1) \;=\; \varphi\big(u^{-1}-1) (du du^{-1})^{\frac{n-1}{2}}\big)
\;,
\end{equation}
with $u \in \mathrm{GL}_N(A)$ representing a class $[u]_1 \in K^{\mathrm{alg}}_1(A)$, which is the quotient of $\mathrm{GL}_\infty(A)=\lim_{n\to\infty}\mathrm{GL}_n(A)$ by its commutator subgroup $\mathrm{GL}_\infty(A)_0$. In either case the pairing is a homomorphism w.r.t. the group operation of $K_j^{\mathrm{alg}}(A)$ and depends only on the cohomology class of the cycle $\varphi$ \cite{Connes94}.

\vspace{.2cm}

Let us now consider the case of topological algebras. Since continuous $n$-cocycles that are defined on a $C^*$-algebra $\Aa$ always result in trivial pairings with $K_j(\Aa)$, one must pass to dense subalgebras with a finer topology. 

\vspace{.2cm}

A Fr\'echet algebra is an algebra $\scrA$ that is a complete metrizable locally convex space with a jointly continuous multiplication. For an increasing family of seminorms $(\norm{\cdot}_j)_{j\in \bbN}$ generating the topology of $\scrA$, this holds if and only if there is for each $j \in \bbN$ some $j'\in \bbN$ and a constant $C_j$ such that
$$
\norm{ab}_j 
\;\leq\; 
C_j \norm{a}_{j'} \, \norm{b}_{j'}
$$
for all $a,b \in \scrA$ \cite{Phillips91,PhillipsSchweitzer94,Schweitzer}.
If one can further choose the family of seminorms to be submultiplicative for each fixed $j$, one says that $\scrA$ is $m$-convex (note that in some works like \cite{Ren} a Fr\'echet algebra is always required to be $m$-convex). An $m$-convex Fr\'echet algebra can always be written (non-uniquely) as an inverse limit of Banach algebras $\scrA=\varprojlim \scrA_j$ (with $\scrA_j$ the completion in the norm $\norm{\cdot}_j$), and it therefore has a holomorphic functional calculus \cite{Phillips91}.

\begin{definition}
Let $\scrA$ be a Fr\'echet $*$-algebra which is a dense subalgebra of a $C^*$-algebra $\Aa$. Then $\scrA$ is called smooth in $\Aa$ if the inclusion $\imath: \scrA \hookrightarrow \Aa$ is continuous and $\imath(\scrA)$ is closed under the holomorphic functional calculus of $\Aa$.
\end{definition}

If $\scrA$ is smooth in $\Aa$, then $M_N(\scrA^\sim)$ is also smooth in $M_N(\Aa^\sim)$ and therefore every class in $K_j(\Aa)$ can be represented by elements of $M_N(\scrA^\sim)$. Every continuous cyclic $n$-cocycle over $\scrA$ then defines a pairing with the topological $K$-group $K_j(\Aa)$ via
$$\langle [\varphi], [e]_0-[\one_K]_0\rangle \;=\; \varphi(e, e, \ldots , e)$$
for $n$ even and $[e]_0-[\one_K]_0$ represented by any $e\in M_N(\scrA^\sim)$,
respectively
$$\langle [\varphi], [u]_1\rangle \;=\; \varphi(u^{-1}-1, u-1, u^{-1}-1, \ldots , u-1)$$
for $n$ odd and $[u]_1$ represented by any invertible $u \in M_N(\scrA^\sim)$ \cite{Connes86}.

\vspace{.2cm}

Let $\Aa$ be a $C^*$-algebra equipped with a faithful densely defined lower semicontinuous trace $\Tt$ and an $\bbR^n$-action $\Chernaction$ that leaves $\Tt$ invariant. From this data, one can define  the Fr\'echet algebra $\Aa_{\Tt, \Chernaction}$, see Section~\ref{sec-DiffElements}. It is possible to show that $\Aa_{\Tt, \Chernaction}$ is smooth in $\Aa$, and this will also follow from the criterion of Theorem~\ref{theo-Schweiterzer} below.

\begin{definition}
	\label{def-ChernCocycleFrechet}
	The Chern cocycle for the action $\Chernaction$ is a cyclic $n$-cocycle on $(\Aa_{\Tt, \Chernaction} )^{n+1}$ defined by
	$$
	\Ch_{\Tt,\Chernaction}(a_0,\ldots ,a_n) 
	\;=\; 
	c_n \,
	\sum_{\rho \in S_n} (-1)^\rho\, \Tt\big(a_0 \nabla_{\rho(1)} a_1 \cdots  \nabla_{\rho(n)} a_n\big)
	\;,
	$$
	where $\nabla_1,\ldots ,\nabla_n$ are the derivations  \eqref{eq:nabla} on $\Aa_{\Tt,\Chernaction}$ w.r.t. to an orthonormal basis $e_1,\ldots ,e_n$ of the Lie algebra $\RM^n$ of $G$,  $S_n$ is the symmetric group and $(-1)^\sigma$ the signum of a permutation $\sigma\in S_n$ and the normalization constants are as in {\rm Definition~\ref{def-ChernCocycle}} given by
	\begin{equation}
	\label{eq-chern_normalizationFrechet}
	c_n 
	\;=\; 
	\begin{cases}
	\frac{(2\pi \imath\,)^k}{k!}\;, \quad &\text{for }n=2k\;,
	\\
	\frac{\imath\,(\pi\imath)^k}{(2k+1)!!}\;, \quad &\text{for }n=2k+1
	\;.
	\end{cases}
	\end{equation}
\end{definition}

For even $n$, the cocycles pair with projections $e \in M_N(\Aa_{\Tt,\Chernaction}^{\sim})$ and for odd $n$ with unitaries $v \in M_N(\Aa_{\Tt,\Chernaction}^{\sim})$ using the natural extensions to matrix algebras
\begin{equation}
\label{eq-EvenPairing}
\langle \Ch_{\Tt,\Chernaction}, [e]_0\rangle 
\;=\; 
\Ch_{\Tt \otimes \mbox{\rm\tiny Tr}_N,\Chernaction}(e - s(e),  \ldots, e - s(e))
\end{equation}
and
\begin{equation}
\label{eq-OddPairing}
\langle \Ch_{\Tt,\Chernaction}, [v]_1\rangle 
\;=\; 
\Ch_{\Tt \otimes \mbox{\rm\tiny Tr}_N,\Chernaction}(v^* - s(v^*),v - s(v),v^* - s(v^*),  \ldots, v - s(v))
\;,
\end{equation}
with $s: M_N(\Aa^{\sim}) \to M_N(\bbC)$ the homomorphism extracting the scalar part. This gives precisely the pairings in Section~\ref{sec-BreuerToep} which can be expressed as semifinite Breuer-Fredholm indices.

\vspace{.2cm}

Here we are interested in relating the pairing of $K_j(\Aa)$ with Chern cocycles $\Ch_{\Tt, \theta}$ on a $C^*$-algebra $\Aa$ furnished with an additional $\bbR$-action $\xi$ to the pairing of $K_{j+1}(\Aa \rtimes_\xi \bbR)$ with dual Chern cocycles $\Ch_{\hat{\Tt},\theta \times \hat{\xi}}$, by using  the Connes-Thom isomorphisms or other boundary maps in $K$-theory. It is helpful to consider the problem more broadly and construct such a dual $(n+1)$-cocycle for more general $n$-cycles densely defined on $\Aa$ since this allows to apply the equivariance property of the Connes-Thom isomorphisms more effectively. As a preparation, let us  first construct  the natural domain for the dual cocycle, which will be the so-called smooth crossed product \cite{ENN88}.

\begin{definition}
	Let $\scrA$ be a Fr\'echet algebra with increasing family of seminorms $(\norm{\cdot}_m)_{m\in \bbN}$. An action $\Smoothaction: \bbR \times \scrA  \to \scrA$ is called smooth if the orbits $t \mapsto \Smoothaction_t(a)$ are infinitely often differentiable and for all $m_1,k \in \bbN$ there exist $C_{m_1,k} > 0$ and $m_2, j\in \bbN$ such that
	$$\norm{\frac{\difd^{k}}{\difd t^{k}} \Smoothaction_t(a)\big|_{t=0}}_{m_1} \leq C_{m_1,k} (1+t^2)^{\frac{j}{2}} \norm{a}_{m_2}.$$	
\end{definition}

For  a Fr\'echet algebra $\scrA$ with smooth action $\Smoothaction$, the smooth crossed product $\scrA \rtimes_\Smoothaction \bbR = \scrS^*(\bbR, \scrA, \Smoothaction)$ is defined as the set of smooth and rapidly decaying functions with the twisted convolution multiplication
$$
(f_1*f_2)(t) 
\;=\; 
\int_\bbR f_1(s) \Smoothaction_{s}(f_2(t-s))\difd{s}
\;.
$$
This is again a Fr\'echet algebra with a natural family of seminorms  $(\norm{f}_{j,m})_{j,m\in \bbN}$ such as
$$
\norm{f}_{j,m} \;=\; \sum_{\beta=0}^m\sup_{t\in \bbR}  \norm{(1+\abs{t})^{m} (\nabla^\beta f)(t)}_j 
\;.
$$ 
We write $\scrS^*(\bbR, \Aa, \Smoothaction)$ for the twisted convolution algebra and use $\scrS(\bbR, \Aa)$ for the algebra with pointwise multiplication. Both of these algebras are equal as linear spaces to the projective tensor product $\scrA \otimes \scrS(\bbR)$ since the second factor is nuclear ({\it e.g.} \cite{Tre}). To avoid confusion it is convenient to write elements of $\scrA \rtimes_\Smoothaction \bbR$ as formal integrals
$$
f \in \scrS^*(\bbR, \scrA, \Smoothaction) 
\; \longleftrightarrow \;
\int_{\bbR} f(t) u_t \, \difd{t}
$$
with symbols $(u_t)_{t\in \bbR}$ such that the algebraic structure is defined through the commutation relations
$$
u_t u_s 
\;=\; 
u_{s+t}
\;, 
\qquad 
u_t a 
\;=\; 
\Smoothaction_t(a)u_t
\;, 
\qquad \forall\; a\in \scrA, \; t,s\in \bbR
\;.
$$ 
If there is a continuous, dense and equivariant inclusion $\scrA \hookrightarrow \Aa$ for a $C^*$-dynamical system $(\Aa, \Smoothaction, \bbR)$, then this notation is consistent with the natural dense inclusion $\scrA \rtimes_\Smoothaction \bbR \hookrightarrow \Aa \rtimes_\Smoothaction \bbR$ given by 
$$
\int_{\bbR} f(t) u_t \, \difd{t} 
\;\mapsto\;  
\int_{\bbR} \pi(f(t)) U(t) \, \difd{t}
$$ 
for any regular representaton $(\pi, U)$.

\vspace{.2cm}

An important special case occurs for a trivial action. We will denote by $\scrS\scrA = \scrS(\bbR, \scrA)$ with pointwise multiplication the smooth suspension of a Fr\'echet algebra $\scrA$ and note that $\scrS \scrA \simeq \scrA \rtimes \bbR$ via the isomorphism $f\in \scrS \scrA \leftrightarrow \int_\bbR (\calF^{-1}f)(t) u_t \difd{t}$.

\vspace{.2cm}

If $\scrA$ is smooth in a $C^*$-algebra $\Aa$, then the unitization $\scrA^\sim$ is smooth in $\Aa^\sim$, however, it is not clear that this property is stable under other natural operations such as (smooth) suspensions or crossed products with smooth actions. The following notion is better behaved in this respect:

\begin{definition}[\cite{Schweitzer}]
Let $\scrA$ be a Fr\'echet $*$-algebra with a continuous dense inclusion $\scrA \hookrightarrow \Aa$ into a Banach algebra $\Aa$. 
\begin{enumerate} 
\item[{\rm (i)}] The inclusion is called spectral invariant if $\sigma_{\scrA}(a)=\sigma_{\Aa}(\imath(a))$ holds for every $a\in \scrA$.
	
\item[{\rm (ii)}] The inclusion is called strongly spectral invariant if the topology of $\scrA$ is generated by a family of seminorms  $(\norm{\cdot}_j)_{j\in \bbN}$ arranged such that $\norm{\cdot}_0=\norm{\cdot}_\Aa$ is the norm of $\Aa$ and there is a constant $C>0$ such that for every $m\in \bbN$ there is some $D_m > 0$ and $p_m \in \bbN$ such that for all $a_1,\ldots ,a_n \in \scrA$, one has
\begin{equation}
\label{eq:spectral_invariance}
\norm{a_1 \cdots a_n}_j \leq D_j\, C^{\,n} \sum_{j_1+\ldots +j_n \leq p_j} \norm{a_1}_{j_1}\,\cdots\norm{a_n}_{j_n}
\;,
\end{equation}
independently of $n\in \bbN$.
\end{enumerate}
\end{definition}

If \eqref{eq:spectral_invariance} holds for some family of seminorms with $\norm{\cdot}_0=\norm{\cdot}_\Aa$, then it also holds for all other such families and it can be shown that $\scrA$ is $m$-convex, meaning that the seminorms can be chosen to be submultiplicative. An example for a strongly spectral invariant inclusion is $\Aa_{\Tt, \Chernaction} \hookrightarrow \Aa$ for action $\Chernaction$ and trace $\Tt$ as above, which can be verified for the natural family of increasing semi-norms using the Leibniz rule.

\vspace{.2cm}

Strong spectral invariance implies spectral invariance \cite{Schweitzer}. Indeed, since the inclusion is an homomorphism, one always has $\sigma_{\scrA}(a)\supset \sigma_{\Aa}(\imath(a))$. For the reverse inclusion, one must show that for all $a\in \scrA$ the invertibility of $i(a)$ implies that $i(a)^{-1} \in \Aa$ is an element of $i(\scrA)$ (respectively quasi-inverses in the non-unital case). By constructing inverses as power series, it is sufficient to show for this that the series $\sum_{n\in \bbN} a^n$ converges in $\scrA$ for each $a\in \scrA$ with $\norm{i(a)}_\Aa$ small enough and, indeed, this follows from strong spectral invariance \eqref{eq:spectral_invariance}.

\vspace{.2cm}

The interest in strong spectral invariance comes from its compatibility with crossed products:

\begin{theorem}[\cite{Schweitzer}]
\label{theo-Schweiterzer}
Let $\scrA$ be a Fr\'echet algebra with continuous dense inclusion $\scrA\hookrightarrow\Aa$ into a $C^*$-algebra $\Aa$. Suppose that $\scrA\hookrightarrow\Aa$ is strongly spectral invariant. Then:
\begin{enumerate}
\item[{\rm (i)}] $\scrA$ is smooth in $\Aa$.

\item[{\rm (ii)}] If $\Smoothaction$ is a smooth action on $\scrA$ which is the restriction of a strongly continuous action on $\Aa$, then the inclusion $\scrA\rtimes_\Smoothaction \bbR \hookrightarrow \Aa \rtimes_\Smoothaction \bbR$ is also smooth.
	
\end{enumerate} 
\end{theorem}

Actually \cite{Schweitzer} shows several more general results, but since the proofs are quite involved even when restricting to the case of interest here let us indicate briefly how the conditions of Theorem~\ref{theo-Schweiterzer} are used. 
For part $(i)$ just note that since $\scrA$ is $m$-convex the existence of a holomorphic functional calculus on both $\scrA$ and $\Aa$ implies that an inclusion is smooth if and only if it is spectral invariant. For the assertion $(ii)$ about crossed products one shows that $\scrA\rtimes_\Smoothaction \bbR$ with natural Fr\'echet seminorms is strongly spectral invariant in the convolution  algebra $L^1(\bbR, \Aa, \Smoothaction) \subset \Aa \rtimes_\Smoothaction \bbR$ using the polynomial growth condition on $\Smoothaction$ (a smooth action in the above  sense is also called a tempered action). Finally one proves that the inclusion $L^1(\bbR, \Aa, \Smoothaction) \subset \Aa \rtimes_\Smoothaction \bbR$ is spectral invariant, which holds without condition on $\Smoothaction$ since $\bbR$ is a compactly generated polynomially growing group. It is apparently not known if the inclusion $\scrA\rtimes_\Smoothaction \bbR \hookrightarrow \Aa \rtimes_\Smoothaction \bbR$ is also strongly spectral invariant.

\vspace{.2cm}

Strong spectral invariance is also preserved under some natural operations:

\begin{proposition}
	Let $\scrA$ be a Fr\'echet algebra that is strongly spectral invariant in a $C^*$-algebra $\Aa$ with $(\norm{\cdot}_j)_{j\in\bbN}$ a family of submultiplicative seminorms for $\scrA$ such that $\norm{\cdot}_0=\norm{\cdot}_\Aa$. Then 
\begin{enumerate}

\item[{\rm (i)}] The unitization $\scrA^\sim$ is a Fr\'echet algebra with the family of submultiplicative seminorms defined by the $C^*$-norm $\norm{a +\lambda \one^\sim}_0 = \norm{a +\lambda \one^\sim}$ and $\norm{a+\lambda \one^\sim}_j = \norm{a}_j + \abs{\lambda}$. It is strongly spectral invariant in $\Aa^\sim$.

\item[{\rm (ii)}] The smooth suspension $\scrS\scrA$ is a Fr\'echet algebra with the family of seminorms
		$$\norm{f}_{j,m} \;=\; \sum_{\beta=0}^m\sup_{t\in \bbR}  \norm{(1+\abs{t})^{m} (\nabla^\beta f)(t)}_j$$
		and it is strongly spectral invariant in $S\Aa$.
	\end{enumerate}
\end{proposition}

\noindent{\bf Proof.}
(i) Recall that $\norm{a + \lambda \one^\sim}_0'= \norm{a}_0 + \abs{\lambda}$ and $\norm{\cdot}_0$ are equivalent norms on the $C^*$-algebra $\Aa^\sim$. The unitization $\scrA^\sim$ is an $m$-convex Fr\'echet algebra in the direct sum topology of $\scrA\oplus \bbC$ as a topological vector space and the given family of seminorms generates that topology. Due to equivalence of seminorms, it is enough to show strong spectral invariance when $\norm{\cdot}_0$ is replaced with $\norm{\cdot}_0'$ in \eqref{eq:spectral_invariance} which then follows by expanding the products and the estimates in \cite[Lemma 1.18]{Schweitzer}.

\vspace{.1cm}

(ii) The family of seminorms is known to be one of the equivalent choices for the tensor product of $\scrA$ with the nuclear Fr\'echet algebra $\scrS(\bbR)$ \cite{Phillips91,PhillipsSchweitzer94}. 
	The norm $\norm{\cdot}_{0,0}$ is the $C^*$-norm of $S\Aa$ and applying the Leibniz rule \eqref{eq:spectral_invariance} pointwise we get
\begin{align*}
&\!\!\!
\norm{f_1 \cdots f_n}_{j,m} 
\\
&\;\leq \;\sum_{\beta_1 + \ldots +\beta_n \leq m} \sup_{t\in \bbR}  \norm{(1+\abs{t})^{m} (\nabla^{\beta_1} f_1)(t)\cdots (\nabla^{\beta_n} f_n)(t)}_j \\
	&\;\leq\; \sum_{\beta_1 + \ldots +\beta_n \leq m} \sup_{t\in \bbR}  (1+\abs{t})^{m} D_j C^{\,n} \sum_{j_1+\ldots +j_n\leq p_j}\norm{\nabla^{\beta_1} f_1)(t)}_{j_1}\cdots \norm{\nabla^{\beta_n} f_n)(t)}_{j_n} \\
	&\;\leq\;  \sum_{\beta_1 + \ldots +\beta_n \leq m}  D_j C^{\,n} \sum_{j_1+\ldots +j_n\leq p_j}\norm{f_1}_{j_1,\beta_1}\cdots \norm{ f_n}_{j_n,\beta_n}.
	\end{align*}
For any bijection $\sigma: \bbN \times \bbN \to \bbN$ of the seminorms with $\sigma(0,0)=(0)$ one can now choose $p_{\sigma(j,m)} \in \bbN$ as the maximum of $\sigma(j_1,\beta_1)+\ldots +\sigma(j_n,\beta_n)$ among all tuples with  $\beta_1 + \ldots +\beta_n \leq m$ and $j_1+\ldots +j_n\leq p_j$ which is finite and independent of $n$. Hence
$$
\norm{f_1 \cdots f_n}_{\sigma(j,m)} 
\;\leq\; 
D_j C^{\,n} \sum_{i_1 + \ldots  + i_n \leq p_{\sigma(j,m)}}\norm{f_1}_{\sigma^{-1}(i_1)}\cdots \norm{f_n}_{\sigma^{-1}(i_n)}
\;,
$$
concluding the proof. 
\hfill $\Box$

\vspace{.2cm}

As an alternative to smooth subalgebras it may be advantageous to work with the (representable) $K$-theory of Fr\'echet algebras directly \cite{Phillips91}. Many results have already been obtained in that setting, including generalizations of the Connes-Thom isomorphisms and the Pimsner-Voiculescu sequence \cite{PhillipsSchweitzer94}. However, for applications to $C^*$-algebras spectral invariance conditions must be imposed at some stage or another, hence not much generality is lost when working with the more familiar notions of $K$-theory on $C^*$-algebras.

\section{Duality for smooth crossed products}

This section reviews the construction of the dual cocycle on crossed products with $\bbR$ and show that this construction is dual to the Connes-Thom isomorphism w.r.t. the pairing of $K$-theory and cyclic cohomology. While following the original construction and arguments of \cite{ENN88} fairly closely, the treatment is self-contained and takes a more direct path to the duality result. There are two main motivations for this section: One is to justify chosen numerical and sign factors since we are interested in applying the duality to physical models, the other is that the result in \cite{ENN88} is not stated in full generality, but only for smooth subalgebras of a certain form. While it may appear clear to an expert that the proof generalizes, the required conditions on spectral invariance and associated constructions are  not obvious. For example, since crossed products can be problematic w.r.t. spectral invariance, we replace the use of Takesaki-Takai duality with a suspension argument.

\vspace{.2cm}

Let $\scrA$ be a Fr\'echet algebra with smooth action $\Smoothaction$ and $\varphi$ a continuous $n$-cycle over $\scrA$ defined on the universal differential algebra of top degree $n$ given by
$$
\Omega_n(\scrA)
\;=\; 
\bigoplus_{j=0}^{n} \scrA^\sim \otimes \scrA^{\otimes j}
\;,
$$
where the projective tensor product is used. For clarity, the universal differential on that algebra is denoted by $d_\scrA$ such that its homogeneous elements take the form $(\lambda \one^\sim + a_0)d_\scrA a_1\cdots d_\scrA a_j = (\lambda \one^\sim + a_0) \otimes  a_1 \otimes\cdots  \otimes a_j$ with $j\leq n$. Extending $\Smoothaction$ to an action on the Fr\'echet algebra $\Omega_n(\scrA)$ by letting it commute with $d_\scrA$ again yields a smooth action.

\vspace{.2cm}

From these data one can construct a boundary $(n+1)$-cycle over $\scrA \rtimes_\Smoothaction \bbR$. For the graded differential algebra one uses the linear space
\begin{align*}
\Omega^\Smoothaction_{n+1}(\scrA)
\;=\;\Omega_n(\scrA) \otimes \Omega_1(\scrS(\bbR)) \;=\; \Omega_n(\scrA) \otimes (\scrS(\bbR)^\sim \oplus \scrS(\bbR)^\sim \otimes \scrS(\bbR))
\;,
\end{align*}
a dense subset of which is given by the linear span of elements of the form $\omega \otimes (\lambda \one + f)$ and $\omega \otimes (\lambda \one + f) dg = \omega \otimes (\lambda \one + f) \otimes g$ with $\omega \in \Omega_n(\scrA)$ and $f,g \in \scrS(\bbR)$. In the following, let $\omega, f, g$ with subscripts be homogeneous elements of $\Omega_n(\scrA)$, $\scrS(\bbR)^\sim$ and $\scrS(\bbR)$ respectively.

\vspace{.2cm}

The left multiplication with elements of degree $(m,0)$ extends the ordinary multiplication on $\scrS^*(\bbR, \Omega_n(\scrA),\Smoothaction)$, namely one defines
\begin{align*}
((\omega_1 \otimes g_1)(\omega_2\otimes \one^\sim))(t) 
&
\;=\; \omega_1 \Smoothaction_t(\omega_2) \otimes g_1(t)
\;,
\\
((\omega_1 \otimes \one^\sim)(\omega_2\otimes g_2))(t) 
&
\;=\; 
\omega_1 \Smoothaction_t(\omega_2) \otimes g_2(t)
\;,
\\
((\omega_1 \otimes g_1)(\omega_2 \otimes g_2))(t) 
&
\;=\; 
\int_{\bbR} \omega_1 \Smoothaction_s(\omega_2) \otimes g_1(s)g_2(t-s) \,\difd{s}
\;,
\end{align*}
as functions in $\scrS(\bbR, \scrA)$ and the pointwise (not convolution) product  under the integral. The left multiplication on elements of higher degree is then determined by associativity in $\Omega_1(\scrS(\RM))$
$$ 
(\omega_1 \otimes f_1)(\omega_2 \otimes f_2 dg_2) 
\;=\; 
((\omega_1 \otimes f_1)(\omega_2 \otimes f_2)) dg_2
\;.
$$
The differential $d^\Smoothaction$ acts by 
$$
d^\Smoothaction(\omega \otimes f) 
\;=\; 
(d_\scrA \omega) \otimes f + (-1)^{\mathrm{deg}(\omega)} \omega \otimes df
\;, 
\qquad 
d^\Smoothaction(\omega \otimes f dg) 
\;=\;
(d_\scrA \omega) \otimes f dg
\;,
$$
and left multiplication with elements of higher degree is thereby determined through
$$
(\omega \otimes f dg) x 
\;=\; 
((\omega_1\otimes f) d^\Smoothaction(\one\otimes g))x 
\;=\; 
(\omega\otimes f) (d^\Smoothaction((\one\otimes g)x)- (\one\otimes g)d^\Smoothaction x)
$$
for all $x\in \Omega^\Smoothaction_{n+1}(\scrA)$.

\vspace{0.2cm}

Given a continuous $n$-cycle $(\varphi, d_\scrA, \Omega_n(\scrA))$ on the algebra $\scrA$ let us now introduce an $(n+1)$-cycle $(\#_\Smoothaction \varphi, d^\Smoothaction, \Omega^\Smoothaction_{n+1}(\scrA))$ by setting
$$
(\#_\Smoothaction \varphi)(\omega \otimes f dg) 
\;=\; 
-\,\imath\int_\bbR \difd{t}\int_{0}^{-t}\difd{s} \, \varphi(\Smoothaction_s(\omega)) f(t)g(-t)
$$
for all $\omega \in \Omega_n(\scrA)$, $f,g \in \scrS(\bbR)$ and $(\#_\Smoothaction \varphi)(x)=0$ for all homogeneous elements that are not of this form. 
The algebraic rules and graded trace for more general elements are defined by linearity and continuity and the canonical homomorphism $\scrA \rtimes_\Smoothaction \bbR \to \Omega_n^\Smoothaction(\scrA)_0$ is given by the obvious map which identifies $\scrA \rtimes_\Smoothaction \bbR = \scrS^*(\bbR, \scrA, \Smoothaction)$ with the linear subspace $\scrA \otimes \scrS(\bbR)$.

\begin{lemma}
\label{lemma:dualcocycleconstruction}
In the setting above, $(\Omega^\Smoothaction_{n+1}(\scrA), d^\Smoothaction, \#_\Smoothaction \varphi)$ is an $(n+1)$-cycle over $\scrA \rtimes_\Smoothaction \bbR$. It is natural under equivariant morphisms, {\it i.e.} if $\rho: \scrB \to \scrA$ is an equivariant morphism of Fr\'echet algebras with smooth actions satisfying $\rho \circ \beta = \Smoothaction \circ \rho$ then 
$$
\#_\beta (\rho^*\varphi) 
\;=\; 
\hat{\rho}^*(\#_\Smoothaction \varphi)
$$
with $\rho^*$, $\hat{\rho}^*$ being the pull-backs under the induced morphisms $\rho: \Omega_n(\scrB) \to \Omega_n(\scrA)$ and  $\hat{\rho}: \Omega^\beta_n(\scrB) \to \Omega^\Smoothaction_{n+1}(\scrA)$ of differential graded algebras.
\end{lemma}

\noindent{\bf Proof.}
The algebraic properties are in principle not difficult to check, but challenging to write down since the product is defined in terms of the tensor products $\omega \otimes f$ but does not preserve this form.

\vspace{.1cm}

Let us therefore propose a more efficient notation by writing again elements $f \in \scrS(\bbR)$ as formal convolution operators $f = \int_\bbR f(t) u_t \difd{t}$ and let all differentials commute with integrals and complex numbers to write
$dg = d( \int_\bbR g(t) u_t \difd{t}) =  \int_\bbR f(t) du_t \difd{t}$. In terms of formal operators $(u_t)_{t\in \bbR}$, $(du_s)_{s\in \bbR}$ the algebraic structure on $\Omega^\Smoothaction_{n+1}(\scrA)$ is determined completely from the product of $\Omega_n(\Aa)$ and the fairly intuitive relations
\begin{align}
& u_s u_t 
\;=\; 
u_{s+t}
\;, 
\qquad 
(du_t)u_s 
\;=\; du_{t+s} - u_t du_s\;, 
\qquad 
(du_s)(du_t)
\;=\;0\;, 
\nonumber 
\\ 
\qquad
& u_s \omega  \;=\; \Smoothaction_s(\omega)u_s
\;,
\qquad
(du_t)\omega
\;=\; 
(-1)^{\mathrm{deg}(\omega)}\Smoothaction(\omega)du_t  
\;,
\label{eq:relations_diff}
\\
& 
d^\Smoothaction(u_s)
\;=\; 
du_s
\;, 
\qquad 
d^\Smoothaction(du_t)
\;=\;0\;, 
\qquad 
d^\Smoothaction\omega 
\;=\; 
d_\scrA \omega
\;,
\nonumber
\end{align}
for all $s,t\in \bbR$ and homogeneous $\omega \in \Omega_n(\scrA)$, where it is understood that one can freely perform linear changes of variables under the formal integrals. The trace is in this notation given by
$$
(\#_\Smoothaction \varphi)
\Big(\int_{\bbR}\int_{\bbR} f(t,s) u_t du_s \difd{t}\difd{s}\Big) 
\;=\;
-\,\imath\int_\bbR \difd{t}\int_{0}^{-t}\difd{s} \, \varphi\big(\Smoothaction_s(f(t,-t))\big)
$$
for any $f \in \scrS(\bbR^2, \Omega_n(\scrA))$ and vanishing on all elements that are not of this form.

\vspace{.1cm}

Next let us check that $\#_\Smoothaction\varphi$ is a graded trace. Since most products do not contribute it is enough to expand $\#_\Smoothaction\varphi(\hat{a}\hat{b})$ for all homogeneous elements of the form 
$$
\hat{a}
\;=\;
\int_\RM
f(t_1)\,u_{t_1}
\difd{t_1}
\;,
\qquad
\hat{b}
\;=\;
\int_{\bbR}\int_{\bbR} g(t_2,s_2)\, u_{t_2} du_{s_2} \,\difd{t_2}\difd{s_2}
\;.
$$
Then, using the above relations,
$$
\hat{a}\,\hat{b}
\;=\;
\int_{\bbR}\int_{\bbR}\Big(\int_{\bbR} f(t_1)\,
\alpha_{t_1}\big(g(t_2-t_1,s_2)\big)\,\difd{t_1}\Big) u_{t_2} du_{s_2} \,
\difd{t_2}\difd{s_2}
\;,
$$
so that
\begin{equation}
\label{eq-Phiab}
(\#_\Smoothaction \varphi)
(\hat{a}\,\hat{b})
\;=\;
-\imath
\int_{\bbR}\difd{t_2}\int_0^{-t_2}\difd{s_2}\Big(\int_{\bbR} 
\difd{t_1}\,\varphi\circ\alpha_{s_2}\Big(
f(t_1)\,
\alpha_{t_1}\big(g(t_2-t_1,-t_2)\big)\Big)\Big)
\;.
\end{equation}
Proceeding similarly for $\hat{b}\hat{a}$, one finds after some algebraic manipulations using  the rules in \eqref{eq:relations_diff} that 
%
\begin{align*}
\hat{b}\,\hat{a}
\;=\;
(-1)^{\deg(\hat{a})}\!\!
\int_{\bbR}\int_{\bbR}\Big(\int_{\bbR} 
\Big[&
g(t_2,t_1-s_2)\alpha_{t_1+t_2-s_2}(f(s_2))
\\
&
\,-\,
g(t_2-s_2,s_2)
\alpha_{t_2}(f(t_1))
\Big]
\difd{t_1}\Big) 
u_{t_2} du_{s_2} \,
\difd{t_2}\difd{s_2}
\;.
\end{align*}
Replacing this into $\#_\Smoothaction\varphi$ and using the cyclicity of $\varphi$ together with $\deg(g)=\deg(\hat{b})-1$ as well as $\deg(f)=\deg(\hat{a})$ leads o
\begin{align*}
&
(\#_\Smoothaction \varphi)
(\hat{b}\,\hat{a})
\\
&
\;=\;
-\imath(-1)^{\deg(\hat{a})\deg(\hat{b})}
\int_{\bbR}\difd{t_2} \int_0^{-t_2}\difd{s_2}  \Big(\int_{\bbR} 
\difd{t_1}\Big[
\varphi\circ\alpha_{s_2-t_1}
\Big(
f(t_1)\,
\alpha_{t_1}\big(g(t_2,-t_2-t_1)\big)\Big)\,
\\
&
\hspace{5.3cm}
\,-\,
\varphi\circ\alpha_{s_2+t_2}
\Big(
f(-t_2)\,
\alpha_{-t_2}\big(g(t_2-t_1,t_1)\big)\Big)\,
\Big]\Big)
\;.
\end{align*}
Shifting integration variables in such a way that the integrand coincides with that of \eqref{eq-Phiab} shows that the two summands combine to the integral in \eqref{eq-Phiab}.

\vspace{.1cm}

To see that the trace is closed, note that one must check
\begin{align*}
(\#_\Smoothaction\varphi)\big(d^\Smoothaction(\omega \otimes (\lambda \one + f)\big) 
&
\;=\;
 (\#_\Smoothaction\varphi)\big(d_\scrA\omega \otimes (\lambda \one + f) + (-1)^{\mathrm{deg} \omega} \omega df\big) 
\\ 
&
\;=\; 
-\,\imath \int_\bbR \difd{t}\int_{0}^{-t}\difd{s} \, \varphi\big(d_\scrA\Smoothaction_s(\omega)\big) f(-t) 
\;=\; 0
\end{align*}
and 
\begin{align*}
(\#_\Smoothaction\varphi)\big(d^\Smoothaction(\omega \otimes (\lambda \one + f)dg\big) 
&
\;=\;
(\#_\Smoothaction\varphi)\big(d_\scrA\omega \otimes (\lambda \one + f)dg\big) 
\\ 
&
\;=\; 
-\imath \int_\bbR \difd{t}\int_{0}^{-t}\difd{s} \, \varphi(d_\scrA\Smoothaction_s(\omega)) f(t)g(-t) 
\\
&
\;=\; 0
\;,
\end{align*}
both of which hold since by definition $d_\Aa\Smoothaction_s(\omega)=\Smoothaction_s(d_\Aa\omega)$ and $\varphi$ is closed.

\vspace{.1cm}

Finally, the equivariance is obvious since the homomorphisms $\rho$, $\hat{\rho}$ are well-defined as commuting with $d$ and intertwining  $\Smoothaction$, $d_\scrA$ with $\beta$, $d_\scrB$, hence also $d^\Smoothaction$ with $d^\beta$.
\hfill $\Box$

\vspace{.2cm}

If the action $\Smoothaction$ is trivial then $\Omega^\Smoothaction_{n+1}(\scrA) \simeq \Omega_n(\scrA) \otimes \Omega_1(\scrS(\bbR))$ also holds as an isomorphism of algebras and $\#_\Smoothaction \varphi$ reduces to the cup product $\varphi \# \eta$ with the winding number $1$-cycle $\eta$ over $\scrS(\bbR)$ which after Fourier transform is given by
$$
\eta(f dg) 
\;=\;
 -\imath\int_{\bbR}  t f(t)g(-t) \, \difd{t}
 \;.
 $$ 
 In general, however, the cup product does not define a graded trace since the multiplication is twisted by $\Smoothaction$, hence the construction must be modified as above.

\vspace{.2cm}

Let us next write out the character of the $(n+1)$-cycle. For $\hat{a}_i = \int_\bbR f_i(t) u_t \difd{t} \in \scrA\rtimes_\Smoothaction \bbR$ write $d_\scrA \hat{a}_i = \int_\bbR \big(d_\scrA f_i(t)\big) u_t \difd{t}$ and $\tilde{d}\hat{a}_i = d^\Smoothaction( \hat{a}_i)-d_\scrA \hat{a}_i $ such that 
\begin{align*}
(\#_\Smoothaction\varphi) & (\hat{a}_0\, d^\Smoothaction \hat{a}_1 \cdots d^\Smoothaction\hat{a}_{n+1})
\\
&
\;=\;
\sum_{j=1}^{n+1} (\#_\Smoothaction\varphi)\left((\hat{a}_0 d_\scrA\hat{a}_1 \cdots d_\scrA\hat{a}_{j-1}) (\tilde{d}\hat{a}_j) (d_\scrA\hat{a}_{j+1} \cdots d_\scrA\hat{a}_{n+1})\right)
\end{align*}
since all terms with more than one $\tilde{d}$ or more than $n$ factors of $d$ drop out. Writing out the convolutions and evaluating the dual cycle, one can express the character in the closed form 
\begin{align*}
&
(\#_\Smoothaction\varphi)(\hat{a}_0\, d^\Smoothaction \hat{a}_1 \cdots d^\Smoothaction\hat{a}_{n+1}) 
\\
&
\; =\;-\imath\int_{\substack{t\in \bbR^{n+2}\\{t_0+\ldots +t_{n+1}=0}}}\difd{t} \sum_{j=1}^{n+1} (-1)^{n+1-j} \int_{-t^{(j)}}^{-t^{(j-1)}}\difd{s}\,(\varphi\circ \Smoothaction_s)\\
&
\;\;\;\;\;\;\;\;
\left(f_{0}(t_0)\,
\left(\prod_{m=1}^{j-1} \Smoothaction_{t^{(m-1)}}d_\scrA f_{m}(t_{m})\right)
 \Smoothaction_{t^{(j)}}f_j(t_j)
 \, \left(\prod_{l=j+1}^{n+1} \Smoothaction_{t^{(l-1)}}d_\scrA f_{l}(t_{l})\right)
 \right) \;,
\end{align*}
where $t^{(m)}=t_0+\ldots +t_m$ and all remaining products and differentials act in $\Omega_n(\scrA)$ (the probably easiest way do this computation is to use elements $\hat{a}_j = a_j \otimes g_j = a_j \int_\bbR g_j(t_j)u_{t_j}\difd{t_j}$ in product form, cycle the term $\tilde{d}\hat{a}_j= a_j \otimes dg_j = a_j \int_\bbR g_j(t_j)du_{t_j}\difd{t_j}$ to the right, evaluate $\#_\Smoothaction \varphi$ by inserting a formal $\delta$-function, cycle the terms back to the original order and then adjust the integration variables).

\vspace{.2cm}

If $\varphi$ is $\Smoothaction$-invariant the expression simplifies since the dependence on $s$ becomes trivial and the integral over $s$ results in a mere multiplication by the number $t_j$. Absorbing the factor into the derivation $\hat{\nabla} \hat{a}_j = -\imath \int_{\bbR} t_j f_j(t_j)u_{t_j} \difd{t_j}$ w.r.t. the dual action $\hat{\Smoothaction}$, the remaining integrals are the (twisted convolution) product in $\scrA \rtimes_\Smoothaction \bbR$ evaluated in $0$, {\it i.e.} with $\mathrm{ev}_0\big(\int f(t) u_t \difd{t}\big) = f(0)$
\begin{align}
(\#_\Smoothaction\varphi) & (\hat{a}_0\, d^\Smoothaction \hat{a}_1 \cdots d^\Smoothaction\hat{a}_{n+1}) 
\nonumber
\\
\label{eq:dualcocycle_specialcase}
&\;=\;\sum_{j=1}^{n+1} (-1)^{n+1-j} (\varphi \circ \mathrm{ev}_0)
\big(\hat{a}_0 d_\scrA\hat{a}_1 \, \cdots d_\scrA\hat{a}_{j-1}\, (\hat{\nabla}\hat{a}_j) d_\scrA\hat{a}_{j+1}\cdots d_\scrA\hat{a}_{n+1}
\big)
.
\end{align}


For $(\Omega_n(\scrA), d_\scrA, \varphi)$ an $n$-trace over  a dense Frech\'et subalgebra $\scrA \subset \Aa$ one can simplify the construction to define the suspended $(n+1)$-cycle $(\Omega_{n+1}^s(\scrS\scrA), d^s, \varphi^s)$ by using the graded tensor product
$$
\Omega^s_{n+1}(\scrS\scrA) 
\;=\; 
\scrS(\bbR)^\sim \otimes \Omega_n(\scrA)\, \hat{\otimes}\, \Lambda(\bbC)
$$
with $\Lambda(\bbC)=\bbC \oplus \bbC$ the Grassmann algebra with the two generators $1$, $dy$. The graded differential is determined by
$$
d^s(f ) 
\;=\; 
d_\scrA f \,+\, (-1)^{\mathrm{deg}(f)} (\nabla f) dy
\;, 
\qquad d^s(f dy) \;=\; 0
\;,
$$ 
with $d_\scrA$ acting pointwise and  the usual derivative with the normalization as above
$$
\nabla f 
\;=\; 
-\,\frac{1}{2\pi}\, f'\;.
$$ 
The suspension cocycle is then defined (as in \cite{Pim}) by
$$
\varphi^s
\big((\lambda \one + f) dy\big) 
\;=\; 
\int_{\bbR} \varphi(f(y)) \difd{y}
\;,
$$
and vanishing on elements of lower degree.

\begin{lemma}
\label{lemma:dualcyclesuspension}
Let $\scrA$ be a Fr\'echet algebra and $\varphi$ an $n$-trace on $\Omega_n(\scrA)$.  If $\Smoothaction$ is the trivial action, then the characters of $\varphi^s$ and $\#_\Smoothaction\varphi$ over $\scrS\scrA \simeq \scrA \rtimes_\alpha \bbR$ coincide as $(n+2)$-linear functionals, {\it i.e.} the $(n+1)$-cycles only differ by their domains.
\end{lemma}

\noindent{\bf Proof.}
For $f_0,\ldots ,f_{n+1} \in \scrS(\bbR, \scrA)$ set $\hat{f}_j=\calF^{-1} f_j$ such that the image of $f_j \in \scrS\scrA$ in $\Aa \rtimes_\Smoothaction \bbR$ under the isomorphism $\scrS(\bbR, \scrA) \simeq \scrA \rtimes_\Smoothaction \bbR$ is $\hat{a}_j = \int_{\bbR}\hat{f}_j(t)u_t \difd{t}$. Hence the expression \eqref{eq:dualcocycle_specialcase} for the dual cocycle can be written
\begin{align*}
&(\#_\Smoothaction\varphi)(\hat{a}_0\, d^\Smoothaction \hat{a}_1 \dots d^\Smoothaction\hat{a}_{n+1}) 
\\ 
&=\;
\sum_{j=1}^{n+1} (-1)^{n+1-j} (\varphi \circ \mathrm{ev}_0)
\Big(\hat{f}_0 * (d_\scrA \hat{f}_1)*  \cdots* 
\\
& 
\hspace{4cm}
*(d_\scrA \hat{f}_{j-1})* (\hat{\nabla} \hat{f}_j) * (d_\scrA \hat{f}_{j+1})*\cdots *(d_\scrA \hat{f}_{n+1})
\Big)
\,.
\end{align*}
By the convolution theorem and Fourier inversion the evaluation at $0$ is given by
\begin{align}
\sum_{j=1}^{n+1} (-1)^{n+1-j} \int_{\bbR} 
\varphi\Big(
f_0(y) d_\scrA f_1(y)\cdots & d_\scrA f_{j-1}(y)(\nabla f_j)(y) d_\scrA f_{j+1}(y))\cdot
\nonumber
\\
&
\cdots d_\scrA f_{n+1}(y)\Big)\difd{y}
\;,
\label{eq:suspension_cocycle}
\end{align}
and one readily checks that this is precisely $\varphi^s(f_0 d^s f_1\cdots  d^sf_{n+1})$. 
\hfill $\Box$

\vspace{.2cm}

This also shows that the suspension of cocycles is natural w.r.t. all homomorphisms  of differential graded algebras, namely $\rho^*(\varphi)^s= (\rho^*\varphi)^s$. 

\vspace{.2cm}

The main goal is to show that the dual cocycle is dual to the Connes-Thom isomorphisms of $K$-theory and the strategy is to use the naturalness and equivariance of both constructions to pull the problem back to the case of a trivial action, {\it i.e.} a suspension. Let us therefore first show that the suspension is compatible with Bott periodicity: iterating the suspension gives the suspension map of periodic cyclic cohomology (in the notation of Connes \cite{Connes94}, $(\varphi^s)^s=S\varphi$ up to normalization, but this notation is not used in the following).

\vspace{.2cm}

To make sure that our signs are correct, we compute the image of $s_1 \circ s_0: K_0(\bbC) \to K_0(SS \bbC)$, {\it i.e.} the class of the Bott projection (there are two projections which may plausibly be called the Bott projection, which differ by their orientation and it is surprisingly difficult to pin down the correct one since most sources only vaguely identify $\bbR^2\sim \bbC$). The Bott periodicity isomorphism $K_i(\Aa) \to K_{1-i}(SS\Aa)$ is then the external tensor product with this class.

\vspace{.2cm}

To fix the orientation, elements of $S\bbC=C_0(\bbR)$ and $SS\bbC=C_0(\bbR^2)$ will be written as functions of one variable $y_1$ respectively two variables $y_1,y_2$. Recall from \eqref{eq-s0map} that $s_0([1]_0)=[f]_1$ for any function with winding number $1$ such as $f(y_1)=\frac{y_1-\imath}{y_1+\imath}$. To compute $s_1$, one chooses a continuous family of unitaries $y_2\in\RM \mapsto v(y_2)\in \mbox{\rm U}(2)$ such that $v(-\infty) = \one_2$ and $v(\infty)= \mathrm{diag}(f, f^*)$, and by \eqref{eq-s1map}
$$
s_1[f]_1 
\;=\; 
\Big[v \begin{pmatrix}
1&0\\0&0
\end{pmatrix} v^*\Big]_0 
\,-\, 
\Big[\begin{pmatrix}
1&0\\0&0
\end{pmatrix}
\Big]_0
\;.
$$
A standard choice is 
$$
v(y_2) 
\;=\; 
\begin{pmatrix}
\cos(g(y_2))&\sin(g(y_2))\\-\sin(g(y_2))&\cos(g(y_2))
\end{pmatrix} \begin{pmatrix}
f^*&0\\0&1
\end{pmatrix} \begin{pmatrix}
\cos(g(y_2))&-\sin(g(y_2))\\\sin(g(y_2))&\cos(g(y_2))
\end{pmatrix}\begin{pmatrix}
f&0\\0&1
\end{pmatrix}
\;,
$$
for some function $g$ with $g(-\infty)=0$, $g(\infty)=\frac{\pi}{2}$, such as $g(y_2)= \frac{1}{2}\arctan(y_2) + \frac{\pi}{4}$. The trigonometric functions simplify and after a tedious computation one obtains
$$
v(y_1,y_2) \begin{pmatrix}
1&0\\0&0
\end{pmatrix} v^*(y_1,y_2)
\;=\; 
\frac{1}{1+\tilde{y}_1^2+y_2^2} \begin{pmatrix}
\tilde{y}_1^2 +y_2^2 & \imath \tilde{y}_1 + y_2\\-\imath \tilde{y}_1 + y_2&1
\end{pmatrix}
\;,
$$
where $v(y_1,y_2)=v(y_2)(y_1)$ and with the abbreviation $\tilde{y}_1= \frac{y_1}{\sqrt{1+y_2^2}}$. Hence replacing $\tilde{y}_1\sim y_1$ by homotopy, one retrieves the usual form of the Bott projection.

\begin{proposition}
\label{prop:doublesuspension}
Let the Fr\'echet algebra $\scrA$ be smooth in the $C^*$-algebra $\Aa$ and $\scrS \scrS \scrA$ smooth in $SS\Aa$ and let $\varphi$ be an $n$-cycle over $\scrA$ with odd $n$. Then for the Bott periodicity isomorphism $\Phi=s_{0}\circ s_1: K_1(\Aa) \to K_1(SS\Aa)$ one has
$$
\langle (\varphi^s)^s,\Phi[x]_1 \rangle 
\; =\; 
\frac{c_n}{c_{n+2}}\, \langle  \varphi, [x]_1 \rangle
\;.
$$
\end{proposition}

\noindent{\bf Proof.}
Let us first consider the case $\Aa= \scrA=\bbC$ with trivial actions and the $0$-cycle $\varphi=\mathrm{ev}$ given by  $\mathrm{ev}(\lambda)=\lambda$. Then $\Phi([1]_0) = [e_\Phi]_0-[s(e_\Phi)]_0$ with the Bott projection as above represented by the function
$$
e_\Phi(y_1,y_2) 
\;=\; 
\frac{1}{1+y_1^2+y_2^2} \begin{pmatrix}
y_1^2 +y_2^2 & \imath y_1 + y_2\\-\imath y_1 + y_2&1
\end{pmatrix} \in M_2(SS\bbC)
$$ 
with scalar part $s(e_\Phi) = \mathrm{diag}(1,0)$. 
By composing $y_1,y_2$ with a rapidly increasing function such as $\sinh$, one can also construct representatives $\tilde{e}_\Phi \in M_2(\scrS(\bbR^2)^\sim)$, but this will be suppressed. Now
$$
\langle \mathrm{ev}^{ss}, [{e}_\Phi]_0-[s({e}_\Phi)]_0\rangle 
\;=\; 
\int_{\bbR^2} \Tr({e}_\Phi
\big(\nabla_{y_1} {e}_\Phi \nabla_{y_2} {e}_\Phi -\nabla_{y_2} {e}_\Phi \nabla_{y_1} {e}_\Phi\big)\, \difd{y_1}\difd{y_2} 
\;.
$$
Now one computes $\Tr({e}_\Phi(\nabla_{y_1} {e}_\Phi \nabla_{y_2} {e}_\Phi -\nabla_{y_2} {e}_\Phi \nabla_{y_1} {e}_\Phi))=-\frac{1}{4\pi^2}\,\frac{2\imath}{(1+y_1^2+y_2^2)^2}$ and therefore
$$
\langle \mathrm{ev}^{ss}, [{e}_\Phi]_0-[s({e}_\Phi)]_0\rangle  
\;=\; 
(2\pi\imath)^{-1}
\;=\;
\frac{c_0}{c_2}
\;=\;
\frac{c_0}{c_2}
\,
\langle \mathrm{ev}, [1]_0\rangle  
\;.
$$

Let us now consider the general case and assume that $n$ is odd (since this is the only case that will be applied in the following and anyway the other case is similar). The Bott isomorphism $\Phi=s_0\circ s_1$ is given by 
$$
\Phi([u]_1)
\;=\; 
\big[(\one^\sim_N\otimes \one^\sim_2 + (u-\one^\sim_N) \otimes e_\Phi)(\one^\sim_N\otimes \one^\sim_2 + (u^*-\one^\sim_N) \otimes s(e_\Phi))
\big]_1
$$
for $u\in \one_N^\sim + M_N(\scrA)$. Note that the representative is indeed in $\one^\sim_{2N} + M_{2N}(\scrA \otimes \scrS \scrS \bbC)$. One can consider $\varphi^{ss}$ in the natural way as a cocycle over $\scrA \otimes \scrS(\bbR^2)^\sim$  and therefore use the homorphism property of the pairing to drop the second factor. Let us abbreviate $d' = (d^s)^s$, $\one=\one^\sim$ and $d= d_\scrA$ for sake of readability. With $v=\one_{2N} + (u-\one_N) \otimes \tilde{e}_\Phi$ one has
$$
\langle \varphi^{ss}, [v]_1\rangle 
\;=\; 
\varphi^{ss}\big((v^*-\one_N)(d'v d'v^*)^{\frac{n+1}{2}}d'v\big)
$$
and can expand the product using 
\begin{align*}
d'v 
& \;=\; 
dv + (\nabla_1 v) \, dy_1 + (\nabla_2 v) \, dy_2 
\\
& 
\;=\; 
du \otimes {e}_\Phi + (u-\one_N) \otimes (\nabla_1{e}_\Phi) dy_1 + (u-\one_N) \otimes (\nabla_2{e}_\Phi) dy_2
\;.
\end{align*}
The only contributing terms in the product have $n$ factors of $dv$, $dv^*$ and one of $dy_1$ and $dy_2$ each. Being derivatives of projections, $(\one_N-{e}_\Phi)\nabla_j {e}_\Phi {e}_\Phi=0$ and hence any term in which the factors of $dy_1$ and $dy_2$ are not consecutive must also vanish.
With $n=2k+1$ one obtains
\begin{align*}
\varphi^{ss}(v^*& (d'v d'v^*)^{k+1}d'v) 
\\
&
\;=\; 
\sum_{m=0}^k  \varphi^{ss}
\Big((u^*-\one_N)(du du^*)^m \\
&
\;\;\;\;\;\;\;\;\;\;
\big(du (u^*-\one_N)(u-\one_N) + (u-\one_N)(u^*-\one_N) du) (du^*du)^{k-m}\big) \\
&
\;\;\;\;\;\;\;\;\;\;
\otimes ({e}_\Phi (\nabla_1 \tilde{e}_\Phi \nabla_2 {e}_\Phi - \nabla_2 {e}_\Phi \nabla_1 {e}_\Phi)dy_1 dy_2\Big)
\;.
\end{align*} 
The integral over $y_1$, $y_2$ evaluates the pairing $\langle \mathrm{ev}^{ss}, [{e}_\Phi]_0-[s({e}_\Phi)]_0\rangle = (2\pi\imath)^{-1}$  and with the identity $(u-\one_N)(u^*-\one_N)=2 \cdot \one_N - u - u^*$ it follows that
\begin{align*}
&
\varphi^{ss}(v^*(d'v  d'v^*)^{k+1}d'v) 
\\
&
\;=\;
(\pi \imath)^{-1} (2k+2)\langle \varphi, [u]_1\rangle \, 
\\
&
\;\;\;\;\;\;\;\;
-\frac{1}{2\pi \imath} \sum_{m=0}^k 
\varphi\Big((u^*-\one_N)(du du^*)^m \big(du (u+u^*) + (u+u^*) du\big) (du^*du)^{k-m}\Big)
\;.
\end{align*}
Commuting $u$ and $u^*$ to the left or right using $du u^* = - u du^*$, $du^* u = - u^* du$, using cyclicity and telescoping one can show (see the appendix of \cite{KS04} for a more detailed computation)
$$\sum_{m=0}^k \varphi
\Big(u^*(du du^*)^m (du (u+u^*) + (u+u^*) du) (du^*du)^{k-m}\Big)
\;=\; -\, 2 \,\langle \varphi, [u]_1\rangle
\;,
$$
so that $\langle \varphi^{ss}, [v]_1\rangle =(\pi\imath)^{-1}(2k+3) \langle \varphi, [u]_1\rangle$ which shows the claim.
\hfill $\Box$

\vspace{.2cm}

An analogous statement also holds for even $n$, which follows readily by a direct computation if $\scrA$ and $\Aa$ are unital or from a combination of the results further below if one poses suitable conditions on spectral invariance. 

\vspace{.2cm}

The following suspension argument in its basic form goes back to Pimsner \cite{Pim}, see also \cite{Kel}.

\begin{proposition}
\label{prop:dualitysuspension}
Let $n=2k$ be even and let $\scrA$ be a Fr\'echet $*$-algebra with an $n$-cycle $\varphi$ over $\Omega_n(\scrA)$. For any projection $e\in M_N(\scrA^\sim)$ with scalar part $\one^\sim_K$, one has
$$
\varphi^s\big((v^*-\one_N^\sim) d^s v (d^sv^* d^s v)^n\big) 
\;=\; -\,\imath\; \frac{2^{k}(2k+1)!!}{k!}\; 
\varphi\big((e-\one_K^\sim) (de)^n\big)
$$
with the unitary
$$
v 
\;=\; 
(\one_N - e+ e\otimes f) (\one^{\sim}_N-\one_K^\sim + \one_K^\sim \otimes \overline{f}) \, \in M_N(\scrS \scrA)^\sim$$
where $f \in 1+\scrS(\bbR)$ is the unitary function with winding number $1$ from {\rm Proposition \ref{prop-index_map_preimage}}.

\vspace{.1cm}

If $\scrA$ is smooth in a $C^*$-algebra $\Aa$ in such a way that the smooth suspension $\scrS\scrA$ is also smooth in $S\Aa$ under the induced embedding then this shows
$$
\langle \varphi^s, s_0([e]_0-[\one^\sim_K]_0)\rangle 
\;=\;  
\frac{c_n}{c_{n+1}} \;\langle \varphi, [e]_0-[\one^\sim_K]_0\rangle
$$
for every $[e]_0-[\one^\sim_K]_0 \in K_0(\Aa)$ and with the suspension map $s_0: K_0(\Aa) \to K_1(S\Aa).$
\end{proposition}

\noindent{\bf Proof.}
Note that $\varphi^s$ is the restriction of an $(n+1)$-cycle over $\scrS\scrA^\sim$ and that $u=(\one^\sim_N-e)+e\otimes f = \one^\sim_N + e\otimes g \in  M_N(\scrS\scrA^\sim)^\sim$ and $V_{K,N}= \one^\sim_N + \one_K^\sim \otimes \overline{g} \in M_N(\scrS\scrA^\sim)^\sim$ are separately unitary, hence the homomorphism property of the pairing gives
\begin{align*}
\langle \varphi^s, [v]_0]\rangle 
\;=\; 
&
\varphi^s\big((u^*-\one^\sim_N)d^su (d^su^*d^su)^n\big) 
\\
&\;\,-\, 
\varphi^s\big((V_{N,K}^*-\one^\sim_N)d^sV_{N,K} (d^sV_{N,K}^*d^sV_{N,K})^n\big)
\;.
\end{align*}
The second term can be dropped since its argument vanishes algebraically for $n>0$ respectively due to $\varphi(\one_N^\sim)=0$ by definition in the case $n=0$. To evaluate the first term let us note
$$
d_\scrA u 
\;=\; 
d_\scrA e \otimes g
\;, 
\qquad 
\nabla u 
\;=\; 
-\,\frac{1}{2\pi} \,e \otimes g'
\;,
$$
and 
$$
\big(e (d_\scrA e)^{j-1}\big)\big(e (d_\scrA e)^{n+1-j}\big)
\;=\; 
\begin{cases} e (d_\scrA e)^{n}\;,\quad &j \text{ odd}\;, \\
0\;, \quad &j \text{ even}\;.
\end{cases}
$$
Hence one can drop in \eqref{eq:suspension_cocycle} all terms with $j$ odd. Commuting all functions to the right,
\begin{align*}
\langle \varphi^s & , s_0([e]_0-[\one^\sim_K]_0)\rangle 
\\
&
\;=\; \varphi^s\big((u^*-\one_N)d^su (d^su^*d^su)^n\big) 
\\
&
\;=\;\frac{-1}{2\pi} \,(k+1)\, \varphi\big(e (d_\scrA e)^n\big) \left(\int_{\bbR} g'(y) (g(y))^k \overline{(g(y))}^{k+1}\difd{y}\right)\\
&
\;=\; 
\frac{-1}{2\pi}\,(k+1)\, \langle \varphi, [e]_0 \rangle \int_{\bbR}  \left(\frac{\difd}{\difd t}\,\frac{t-\imath}{t+\imath}\right) \left(\frac{t-\imath}{t+\imath}-1\right)^k \left( \frac{t+\imath}{t-\imath} - 1\right)^{k+1} \difd{t}\\
&
\;=\; 
-\,\imath \,
\;\frac{2^{k}(2k+1)!!}{k!} \,\langle \varphi, [e]_0-[\one^\sim_K]_0 \rangle
\end{align*}
where we applied the substitution $y=\sinh(t)$ and computed the integral by the residue theorem (alternatively, a further substitution $y \mapsto -\cot(\pi x)$ leads up to a sign to exactly the same term as in \cite{KRS,PSbook} which can be evaluated combinatorially). The reformulation in the case of smooth inclusions is then clear from Proposition~\ref{prop-index_map_preimage} since $[v]_1 = s_0([e]_0-[\one_K^\sim]_0)$ and every class in $K_0(\Aa)$ can be represented in this way.
\hfill $\Box$

\vspace{.2cm}

Using naturality of both the dual cocycle and the Connes-Thom isomorphisms under equivariance we can therefore show:

\begin{lemma}
\label{lemma:equvariance_pairing}
Let $\scrA$, $\scrB$ be Fr\'echet algebras with smooth actions $\Smoothaction, \beta$ respectively and let $\rho: \scrB \to \scrA$ be a continuous equivariant homomorphism. Assume that $\scrA$, $\scrB$ are smooth in $C^*$-algebras $\Aa$, $\Bb$ in such a way that $\scrA\rtimes_\Smoothaction \bbR$, $\scrB \rtimes_\beta \bbR$ are smooth in $\Aa \rtimes_\Smoothaction \bbR$, $\Bb \rtimes_\beta \bbR$ respectively. If $\varphi$ is a continuous $n$-cycle over $\scrA$, then the Connes-Thom ismorphisms relate the index pairings
$$
\langle \#_\Smoothaction \varphi, (\partial_j^\Smoothaction \circ \rho_*)[x]_j\rangle 
\;=\; 
\langle \#_\beta (\rho^* \varphi), \partial_j^\beta[x]_j\rangle 
$$
for any $[x]_j \in K_j(\Aa)$ and $j=n\,\mbox{\rm mod}\,2$. 
\end{lemma}

\noindent{\bf Proof.}
Due to $\partial_j^\Smoothaction \circ \rho_* = \hat{\rho}_*\circ  \partial_j^\beta$, it is enough to show
$$
\langle \#_\Smoothaction \varphi, \hat{\rho}_*[\hat{x}]_{j+1}\rangle 
\;=\; 
\langle \#_\beta (\rho^*\varphi), [\hat{x}]_{j+1}\rangle
$$
for all $[\hat{x}]_{j+1} \in K_{j+1}(\Bb \rtimes_\Smoothaction \bbR)$. In the even case, any element of $[\hat{x}]_0 \in K_0(\Bb\rtimes_\Smoothaction \bbR)$ is represented by a projection $\hat{e} \in M_N((\scrB\rtimes_\Smoothaction \bbR)^\sim)$. Denoting the extension to matrix algebras by the same letter, one has
\begin{align*}
\langle \#_\Smoothaction \varphi, \hat{\rho}_*([\hat{e}]_0-[\one^\sim_K]_0)\rangle 
&
\;=\;  
(\#_\Smoothaction \varphi)\big(\hat{\rho}(\hat{e})(d^\Smoothaction \hat{\rho}(\hat{e}))^n\big) 
\;=\; 
(\hat{\rho}^*(\#_\Smoothaction\varphi))\big(\hat{e}(d^\beta\hat{e})^n\big) \\
&
\;=\; 
(\#_\beta(\rho^*\varphi))\big(\hat{e}(d^\beta\hat{e})^n\big) 
\;=\; 
\langle \#_\beta (\rho^*\varphi), [\hat{e}]_0-[\one^\sim_K]_0\rangle
\;,
\end{align*}
where the first equality of the second line is the equivariance property $\#_\beta (\rho^*\varphi) = \hat{\rho}^*(\#_\Smoothaction \varphi)$ from Lemma~\ref{lemma:dualcocycleconstruction}. The odd case works with the obvious modifications.
\hfill $\Box$

\vspace{.2cm}

Now all is set up for the proof of the even version of the main result of this section:

\begin{theorem}
\label{theo:duality_even}
Let $\scrA$ be a Fr\'echet algebra with smooth action $\Smoothaction$ and let $\varphi$ be an $n$-cycle over $\Omega_n(\scrA)$ for even $n$. Further assume that there is a dense equivariant inclusion $\scrA \hookrightarrow \Aa$ in such a way that $\scrA$ is smooth in $\Aa$ and $\scrA \rtimes_\Smoothaction \bbR$ is smooth in $\Aa \rtimes_\Smoothaction \bbR$ under the induced embedding.
Then 
$$
\langle \#_\Smoothaction\varphi, \partial^\Smoothaction_0([e]_0-[\one^\sim_K]_0)\rangle 
\;=\; 
\frac{c_n}{c_{n+1}} \,\langle \varphi, [e]_0-[\one^\sim_K]_0\rangle
$$
holds for all $[e]_0-[\one^\sim_K]_0 \in K_0(\Aa)$.
\end{theorem}

\noindent{\bf Proof.}
The even case for a trivial action is already proved by Proposition~\ref{prop:dualitysuspension} and now Connes' unitary cocycle is used to accomplish the same for non-trivial actions.

\vspace{.1cm}

Let $[e]_0 - [\one^\sim_K]_0 \in K_0(\Aa)$ be represented by a smooth projection $e\in M_N(\scrA^\sim)$ and let $(w_k)_{k\in \bbR}$ be the smooth family of unitaries $w_k=e^{2\pi \imath D_e k} e^{-2\pi \imath Dk}$ defined as in the proof of Proposition~\ref{prop-index_map_preimage} such that $\Smoothaction_k(e) = w_k^* e w_k$. Then introduce the action 
$$
\Smoothaction'\;:\; M_{2N}(\Bb) \times \bbR \;\to\; M_{2N}(\Bb)
\;, 
\qquad \Smoothaction_k' \begin{pmatrix}
a & b \\ c & d
\end{pmatrix}
\;=\;  
\begin{pmatrix}
\Smoothaction_k(a) & \Smoothaction_k(b) w_k^* \\
w_k \Smoothaction_k(c) & w_k \Smoothaction_k(d) w^*_k
\end{pmatrix}
\;.
$$
Note that $\Smoothaction'$ is also a smooth action. By construction, one has $\Smoothaction_k'(\mathrm{diag}(e,0))= \mathrm{diag}(\Smoothaction_k(e),0)$ and $\Smoothaction_k'(\mathrm{diag}(0,e))= \mathrm{diag}(0,e)$. On the upper left corner the action reduces to $\Smoothaction$, hence the inclusion $\imath_N: M_N(\Aa) \to M_{2N}(\Aa)$ is equivariant and this gives rise to a canonical inclusion $\hat{\imath}_N: M_N(\Aa)\rtimes_\Smoothaction\bbR \to M_{2N}(\Aa)\rtimes_{\Smoothaction'}\bbR$. The same is true for the smooth crossed products $\hat{\imath}_N(M_N(\scrA)\rtimes_\Smoothaction \bbR) \subset M_{2N}(\scrA)\rtimes_{\Smoothaction'} \bbR$. Note further that $M_{2N}(\scrA) \rtimes_{\Smoothaction'}\bbR$ is smooth in $M_{2N}(\Aa) \rtimes_{\Smoothaction'}\bbR$ since both crossed products are naturally isomorphic to crossed products with $\Smoothaction$ by outer equivariance.

\vspace{.1cm}

Let $\scrA_e$, $\Aa_e$ be the fixed point algebras of $M_{2N}(\scrA)$ respectively $M_{2N}(\Aa)$ under $\Smoothaction'$. Since $\Smoothaction'$ is trivial on $\Aa_e$ with equivariant inclusion $\imath_e: \Aa_e \to M_{2N}(\Aa)$, there is again a canonical inclusion $\hat{\imath}_e: S\Aa_e \simeq \Aa_e\rtimes_{\Smoothaction'}\bbR \to M_{2N}(\Aa)\rtimes_{\Smoothaction'}\bbR$. Also note that the dense inclusions $\scrA_e \hookrightarrow \Aa_e$ and $\scrS \scrA_e \simeq \scrA_e\rtimes_{\Smoothaction'}\bbR \hookrightarrow \Aa_e\rtimes_{\Smoothaction'}\bbR \simeq S\Aa_e$ are smooth since the holomorphic functional calculus preserves the invariant subalgebras.

\vspace{.1cm}

Using a homotopy that exchanges the upper left with the lower right corner one obtains
\begin{align*}
(\partial^{\Smoothaction'}_0 \circ (\imath_N)_*)
([e]_0 \,-\, [\one^{\sim}_K]_0)   
\;&=\; \partial^{\Smoothaction'}_0([e \oplus 0_N]_0 - [\one_K^\sim \oplus 0_{2N-K}]_0)  \\
\;&=\; 
\partial^{\Smoothaction'}_0([0_N \oplus e]_0 \,-\, [0_N \oplus \one^{\sim}_K]_0)\\
&=\; (\partial^{\Smoothaction'}_0 \circ (\imath_e)_*)([0_N \oplus e]_0-[0_N \oplus \one_K^\sim]_0)
\;.
\end{align*}
On the smooth crossed product $M_{2N}(\scrA) \rtimes_{\Smoothaction'}\bbR$ one has the dual $(n+1)$-cycle $\#_{\Smoothaction'}(\Tr_{2N} \# \varphi)$. Due to the equivariance of inclusions
\begin{align*}
(\hat{\imath}_N)^*\#_{\Smoothaction'}(\Tr_{2N} \# \varphi) 
&
\;=\; 
\#_{\Smoothaction}(\imath_N)^*(\Tr_{2N} \# \varphi) 
\;=\;\#_{\Smoothaction}(\Tr_{N} \# \varphi)  \;,
\\
(\hat{\imath}_e)^*\#_{\Smoothaction'}(\Tr_{2N} \# \varphi) 
&
\;=\; 
\#_{\Smoothaction'}(\imath_e)^*(\Tr_{2N} \# \varphi) 
\;=\; 
(\Tr_{2N} \# \varphi)^s
\;,
\end{align*}
{\it i.e.} the $(n+1)$-cycle reduces to $\#_\Smoothaction\varphi$ when evaluated on the upper left corner and to the suspension $\varphi^s$ when evaluated on $\Aa_e$.  Then
\begin{align*}
\langle \#_\Smoothaction\varphi, \partial^\Smoothaction_0( [e]_0-[\one^\sim_K]_0) \rangle 
&
\;=\; 
\langle \#_{\Smoothaction'}(\Tr_{2N} \# \varphi), (\partial^{\Smoothaction'}_0 \circ (\imath_N)_*)
([e]_0 \,-\, [\one^{\sim}_K]_0)  \rangle  \\
&
\;=\;
\langle \#_{\Smoothaction'}(\Tr_{2N} \# \varphi),(\partial^{\Smoothaction'}_0 \circ (\imath_e)_*)([0_N \oplus e]_0-[0_N \oplus \one_K^\sim]_0)\rangle\\
&
\;=\; 
\langle 
\#_{\Smoothaction'} (\imath_e)^*(\Tr_{2N} \# \varphi),\partial^{\Smoothaction'}_0 ([0_N \oplus e]_0-[0_N \oplus \one_K^\sim]_0)\rangle\\
&
\;=\; 
\langle 
(\Tr_{2N} \# \varphi)^s,\partial^{\mbox{\rm\tiny id}}_0 ([0_N \oplus e]_0-[0_N \oplus \one_K^\sim]_0)\rangle\\
&
\;=\; 
\langle (\Tr_{2N} \# \varphi)^s, s_0 ([0_N \oplus e]_0-[0_N \oplus \one^\sim_K ]_0) \rangle 
\;,
\end{align*}
where in the third equality Lemma~\ref{lemma:equvariance_pairing} for $\rho=\imath_e$ was used to pull back the index pairings to the smaller algebras. On the last expression, Proposition~\ref{prop:dualitysuspension} can be applied to get
$$
\langle \#_\Smoothaction\varphi, \partial^\Smoothaction_0( [e]_0-[\one^\sim_K]_0) \rangle 
\;=\; 
\frac{c_n}{c_{n+1}} \;\langle \varphi, [e]_0-[\one^\sim_K]_0 \rangle
\;,
$$
which concludes the proof.
\hfill $\Box$

\vspace{.2cm}

As emphasized above, strong spectral invariance of the inclusion $\scrA \hookrightarrow \Aa$ is a sufficient condition for Theorem~\ref{theo:duality_even}. For the odd case we argue by suspension and Bott periodicity and hence assume this stronger condition for simplicity instead of asking for spectral invariance of inclusions for all intermediate algebras:

\begin{theorem}
\label{theo:duality_odd}
Let $\scrA$ be a Fr\'echet algebra with smooth action $\Smoothaction$ and an $n$-cycle $\varphi$ over $\Omega_n(\scrA)$ for odd $n$. Further assume that there is a dense equivariant inclusion $\scrA \hookrightarrow \Aa$ such that $\scrA$ is strongly spectral invariant in $\Aa$.
Then 
$$
\langle \#_\Smoothaction\varphi, \partial^\Smoothaction_1[v]_1\rangle
\;=\;
-\,\frac{c_n}{c_{n+1}} \;\langle \varphi, [v]_1 \rangle
$$
holds for all $[v]_1\in K_1(\Aa)$.
\end{theorem}

\noindent{\bf Proof.}
The argument will use the commutative diagram
$$
\begin{tikzcd}
& K_1(\Aa) \arrow[r, "\partial_1^\Smoothaction"] \arrow[d, "s_1"] & K_0(\Aa\rtimes_\Smoothaction\bbR) \arrow[d, "s_0"]  \\
K_1(SS\Aa) \arrow[<-, ru, "\Phi"] &  K_0(S\Aa) \arrow[l, "s_0"]  \arrow[r,"\partial_0^{S\Smoothaction}"] & K_1(S\Aa\rtimes_\Smoothaction\bbR) 
\end{tikzcd}
$$
Let us note that by strong spectral invariance $\scrA \rtimes_\Smoothaction \bbR$ is smooth in $\scrA \rtimes_\Smoothaction \bbR$, $\scrS\scrA$ in $S\Aa$ and also $(\scrS \scrA)\rtimes_\Smoothaction \bbR \simeq \scrS(\scrA\rtimes_\Smoothaction \bbR)$ in $(S \Aa)\rtimes_\Smoothaction \bbR \simeq S(\Aa\rtimes_\Smoothaction \bbR)$. Hence Theorem~\ref{theo:duality_even} applies to the duality of the index pairings for all maps that originate from $K_0$-groups.  It is not difficult to see that $(\#_\Smoothaction \varphi)^s = - \#_\Smoothaction (\varphi^s)$ with natural isomorphisms of the graded differential algebras (note that in the former case an expression must end with $(\omega \otimes f dg) dy$ to evaluate the cocycle while it is $(\omega dy) \otimes f dg  = -(\omega \otimes f dg)dy$ in the latter). Using this identity, duality for even cycles and finally Proposition~\ref{prop:doublesuspension} one obtains
\begin{align*}
\langle \#_\Smoothaction \varphi, \partial^\Smoothaction_1 [v]_1\rangle 
&
\;=\; 
\frac{c_{n+2}}{c_{n+1}} 
\;\langle (\#_\Smoothaction \varphi)^s, (s_0\circ \partial^\Smoothaction_1) [v]_1\rangle 
\\
&
\;=\; 
-\,\frac{c_{n+2}}{c_{n+1}}\;\langle \#_\Smoothaction (\varphi^s), (s_0\circ \partial^\Smoothaction_1) [v]_1\rangle\\
&
\;=\; 
-\,\frac{c_{n+2}}{c_{n+1}}
\;
\langle \#_\Smoothaction (\varphi^s), (\partial^{S\Smoothaction}_0\circ s_1) [v]_1\rangle
\\
&
\;=\; -\,\langle \varphi^s, s_1 [v]_1\rangle
\\
&
\;=\;
-\,\frac{c_{n+2}}{c_{n+1}} \; \langle (\varphi^s)^s, (s_0 \circ s_1) [v]_1\rangle 
\\
&
\;=\;  
-\,\frac{c_{n+2}}{c_{n+1}} \;\frac{c_{n}}{c_{n+2}}\langle \varphi,  [v]_1\rangle
\;,
\end{align*}
concluding the proof.
\hfill $\Box$

\vspace{.2cm}

With the same type of suspension argument one can also conclude the odd version of Proposition~\ref{prop:dualitysuspension}, {\it i.e.}
$$
\langle \varphi^s, s_1[v]_1\rangle
\;=\;
\frac{c_n}{c_{n+1}} \;
\langle \varphi, [v]_1 \rangle
$$
for any odd cycle $\varphi$ (which has no minus sign and this is consistent since the Connes-Thom isomorphism for a trivial action actually reduces to $\partial_1^\idmap=-s_1$).

\section{Duality of Chern cocycles}

Now the results of the previous section are applied to Chern cocycles and the connecting maps of the smooth Toeplitz extension. Let $\Aa$ be a $C^*$-algebra with a strongly continuous $G$-action $\Dualityactionop$, where $G$ is a one-parameter group.  If there is additional strongly continuous $\bbR^n$-action $\Dualityactionnp$ that commutes with $\Dualityactionop$, {\it i.e.} $\Dualityactionop \circ \Dualityactionnp = \Dualityactionnp \circ \Dualityactionop$, one can unambiguously form the combined $\bbR^n\times G$-action $\Dualityactionnp \times \Dualityactionop$, which extends uniquely to $\Aa \rtimes_\Dualityactionop G$ by acting trivially on the generator of $\Dualityactionop$. 

\vspace{.2cm}

The dual cocycle then takes a simple form: 

\begin{lemma}
For a $C^*$-algebra $\Aa$ with commuting actions $\theta$ and $\Smoothaction$ that leave a densely defined faithful lower semicontinuous trace $\Tt$ invariant, one has
$$
\#_\Smoothaction \Ch_{\Tt, \theta} 
\;=\; 
\frac{c_{n}}{c_{n+1}} 
\;
\Ch_{\hat{\Tt}_\Smoothaction, \theta \times \hat{\Smoothaction}}
$$
on $\Aa_{\Tt,\theta} \rtimes_\Smoothaction \bbR \subset (\Aa \rtimes_\Smoothaction \bbR)_{\hat{\Tt}_\Smoothaction,\theta\times \hat{\Smoothaction}}$.
\end{lemma}

\noindent {\bf Proof.}
Let us note the algebraic identity 
\begin{align*}
[\hat{b}_0 & d\hat{b}_1 \cdots d\hat{b}_{j-1}]\, [\hat{b}_j d\hat{b}_{j+1}\cdots d\hat{b}_{n+1}]
\\
&
\;=\;
\sum_{k=1}^{j-1} (-1)^{j-1-k} \hat{b}_0 d\hat{b}_1\cdots d\hat{b}_{k-1}d(\hat{b}_k \hat{b}_{k+1})d\hat{b}_{k+1}\cdots d\hat{b}_{n+1} 
\\
&
\;\;\;\;\;
\;+\; 
(-1)^{j-1} (\hat{b}_0 \hat{b}_1) \,d\hat{b}_2\cdots d\hat{b}_{n+1}
\;.
\end{align*}
Furthermore $\hat{\Tt}_\Smoothaction = \Tt \circ \mathrm{ev}_0$. Let $\nabla_1,\ldots ,\nabla_n$ be the derivations for the unit directions of $\theta$ and $\nabla_{n+1}=\hat{\nabla}$ the derivation corresponding to the dual action. Substituting into $\Ch_{\Tt,\theta}$ and applying the Leibniz rule, almost all terms in the expansion of the product cancel (the only remaining term comes from $k=j-1$ with $\nabla_{\sigma(j-1)}$ acting on $\hat{a}_{j-1}$):
\begin{align*}
&\#_\Smoothaction \Ch_{\Tt, \theta}(\hat{a}_0,\ldots ,\hat{a}_{n+1}) \\
&
\;\;=\; 
c_n \sum_{j=1}^{n+1} (-1)^{n+1-j} \sum_{\sigma \in S_n} (-1)^\sigma 
\\
&
\hspace{1.5cm}
\hat{\Tt}_\Smoothaction(\hat{a}_0 \nabla_{\sigma(1)}\hat{a}_1 \cdots\nabla_{\sigma(j-1)}\hat{a}_{j-1} (\nabla_{n+1}\hat{a}_j) \,\nabla_{\sigma(j)}\hat{a}_{j+1} \cdots\nabla_{\sigma(n)}\hat{a}_{n+1})
\\
&
\;\;=\; 
c_n \sum_{\sigma \in S_{n+1}} (-1)^\sigma 
\hat{\Tt}_\Smoothaction(\hat{a}_0 \nabla_{\sigma(1)}\hat{a}_1 \cdots\nabla_{\sigma(n+1)}\hat{a}_{n+1})
\;,
\end{align*}
which up to a constant is precisely $\Ch_{\hat{\Tt}_\Smoothaction, \theta \times \hat{\Smoothaction}}(\hat{a}_0,\ldots ,\hat{a}_{n+1})$.
\hfill $\Box$

\vspace{.2cm}

It is convenient to first state a duality result for the connecting maps $\Exp_{\Dualityactiondualop}^\RM$ and $ \Ind_{\Dualityactiondualop}^\RM$ of the Wiener-Hopf extension associated to $(\Bb,\RM,\Dualityactiondualop)$. Relating the connecting maps with the Connes-Thom isomorphisms using Proposition \ref{prop-ConnectingTakai}, Theorems~\ref{theo:duality_even} and \ref{theo:duality_odd} then show:

\begin{theorem}
\label{theo-cocycle_duality}
Let $\Bb$ be a $C^*$-algebra with a $\bbR$-action $\Dualityactiondualop$, an $\RM^n$-action $\Dualityactionnp$ commuting with $\Dualityactiondualop$ and an $(\Dualityactionnp\times\Dualityactiondualop)$-invariant densely defined lower semicontinuous trace $\Tt$. Then for $n$ odd and $[e]_0 - [s(e)]_0 \in K_0(\Bb \rtimes_{\Dualityactiondualop} \bbR)$, one has 
\begin{equation}
\label{eq:cocycle_duality_even}
\langle \Ch_{\hat{\Tt}_\Dualityactiondualop,\Dualityactionnp\times \hat{\Dualityactiondualop}}, [e]_0-[s(e)]_0\rangle 
\;=\;  \langle \Ch_{\Tt,\Dualityactionnp}, \Exp_{\Dualityactiondualop}^\RM ([e]_0-[s(e)]_0)\rangle
\;,
\end{equation}
while for $n$ even and $[v]_1 \in K_1(\Bb \rtimes_{\Dualityactiondualop} \bbR)$,
\begin{equation}
\label{eq:cocycle_duality_odd}
\langle \Ch_{\hat{\Tt}_\Dualityactiondualop,\Dualityactionnp\times \hat{\Dualityactiondualop}} , [v]_1 \rangle 
\;=\; -\,
\langle \Ch_{\Tt,\Dualityactionnp}, \Ind_{\Dualityactiondualop}^\RM ([v]_1) \rangle
\;.
\end{equation}
\end{theorem}

The result can in principle also be concluded from \cite{ENN88,KS04}, however, this would have required to at least address several technical issues with regards to domains and spectral invariance. Since these questions can be quite subtle, it was more attractive to present a self-contained proof. In the special case where $\Aa=\Bb \rtimes_\Dualityactionnp \bbR^n$ for suitable algebra $\Bb$, the result was also shown in \cite{BR2018} through an application of $KK$-theory to the Wiener-Hopf extension.

\vspace{.2cm}

The rest of this section is devoted to the proof of the following duality result for the smooth Toeplitz extension, which is essential for the applications in Chapter~\ref{sec-Applications}: 

\begin{theorem}
\label{theo-smooth_duality}
Let $\Aa$ be a $C^*$-algebra with two commuting actions, namely a strongly continuous $\bbR^n$-action $\Dualityactionnp$ and a strongly continuous $G$-action $\Dualityactionop$, where $G$ is a one-parameter group. Let $\Tt$ be a densely defined faithful lower semicontinuous trace on $\Aa$ that is invariant under $\Dualityactionnp \times \Dualityactionop$. With  the connecting maps $\Ind^\Dualityactionop_G$ and $\Exp^\Dualityactionop_G$ of the smooth Toeplitz extension, one has
$$
\langle \Ch_{\Tt,\Dualityactionnp\times \Dualityactionop} ,[e]_0-[s(e)]_0 \rangle 
\;=\; 
\langle \Ch_{\hat{\Tt}_\Dualityactionop,\Dualityactionnp},\Exp^\Dualityactionop_G ([e]_0-[s(e)]_0)\rangle
$$
for $n$ odd and $[e]_0 - [s(e)]_0 \in K_0(\Aa)$, respectively
$$
\langle \Ch_{\Tt,\Dualityactionnp\times \Dualityactionop}, [v]_1 \rangle 
\;=\; -\,\langle \Ch_{\hat{\Tt}_\Dualityactionop,\Dualityactionnp}, \Ind^\Dualityactionop_G [v]_1\rangle
$$
for $n$ even and $[v]_1 \in K_1(\Aa)$.  
\end{theorem}

Let us note two special cases. The first is when $\Aa$ is itself a crossed product with $\bbR$ and $\Dualityactionop$ the dual action such that the smooth Toeplitz extension is isomorphic to the Wiener-Hopf extension. In that case the result is immediately implied by Theorem~\ref{theo-cocycle_duality} and the relations of the connecting maps Proposition~\ref{prop-ConnectingTakai}. The second special case to consider is $G=\bbT$, $\Aa = \Bb \rtimes_\alpha \bbZ$ and $\Dualityactionop=\hat{\alpha}$. In that case the smooth Toeplitz extension is a discrete Toeplitz extension (see Lemma~\ref{lemma:discrete_toeplitz}) and the result can be compared with duality results for the Pimsner-Voiculescu sequence \cite{Nest88,KRS,PSbook}.

\vspace{.2cm}

We first prove the case $G=\bbR$ by combining Theorem~\ref{theo-cocycle_duality} with Takai duality and then deduce the case $G=\bbT$ from that.

\vspace{.2cm}

\noindent {\bf Proof} of Theorem \ref{theo-smooth_duality} for the case $G=\bbR$. As already hinted above, we apply Theorem~\ref{theo-cocycle_duality} to $\Bb=\Aa \rtimes_\Dualityactionop \bbR$ with trace $\Tt$ given by $\hat{\Tt}_\Dualityactionop$ and the $\bbR$-action $\beta=\hat{\Dualityactionop}$. Let $n$ be odd and $e \in M_N(\Aa^{\sim}_{\Tt,\Dualityactionnp\times \Dualityactionop})$ such that $[e]_0 - [\one^{\sim}_K]_0 \in K_0(\Aa)$. Further let $f \in \Kk(L^2(\bbR))$ a rank-one projection onto some smooth and rapidly decaying function. By Takai duality, one has $i^{-1}_T(e \otimes f),i^{-1}_T(\one^{\sim}_K \otimes f) \in M_N(\Aa^{\sim} \rtimes_\Dualityactionop \bbR \rtimes_{\hat{\Dualityactionop}} \bbR)$ and $i^{-1}_T((e-\one^{\sim}_K) \otimes f) \in M_N(\Aa \rtimes_\Dualityactionop \bbR \rtimes_{\hat{\Dualityactionop}} \bbR) = M_N(\Bb \rtimes_{\hat{\Dualityactionop}} \bbR)$ where $i_T:\Aa \rtimes_\Dualityactionop \bbR \rtimes_{\hat{\Dualityactionop}} \bbR\to \Aa\otimes \Kk(L^2(\RM))$ is the Takai isomorphism from Theorem~\ref{eq-TakaiDuality}. Hence 
$$
(i_T^{-1})_*([e]_0 - [\one^{\sim}_K]) 
\;=\; 
[\tilde{e}]_0 - [\one^{\sim}_K]_0 
\;\in\; 
K_0(\Bb \rtimes_{\hat{\Dualityactionop}} \bbR)
$$
with $\tilde{e}=\one^{\sim}_K + i^{-1}_T((e-\one^{\sim}_K) \otimes f)$. Now using the identity  $\Exp^\Dualityactionop_\bbR =\Exp_{\hat{\Dualityactionop}}^\RM \circ (i_T^{-1})_*$ from Proposition~\ref{prop-ConnectingTakai}, it follows that
$$
\Exp_{\hat{\Dualityactionop}}^\RM\big([\tilde{e}]_0 - [\one_K^{\sim}]_0\big) 
\;=\; 
\Exp^\Dualityactionop_\bbR\big([e]_0 - [\one_K^{\sim}]_0\big) 
\;\in\; K_1(\Aa \rtimes_\Dualityactionop \bbR)
\;.
$$
From this and \eqref{eq:cocycle_duality_even} in Theorem~\ref{theo-cocycle_duality} one deduces
\begin{align*}
\langle \Ch_{\hat{\Tt}_\Dualityactionop,\Dualityactionnp},\Exp^\Dualityactionop_\RM ([e]_0-[s(e)]_0)\rangle
&
\;=\;
\langle \Ch_{\hat{\Tt}_\Dualityactionop,\Dualityactionnp}, \Exp_{\hat{\Dualityactionop}}^\RM\big([\tilde{e}]_0 - [\one_K^{\sim}]_0\big) 
\rangle
\\
&
\;=\;
\langle \Ch_{\hat{\hat{\Tt}},\Dualityactionnp\times \hat{\hat{\Dualityactionop}}},[\tilde{e}]_0 - [\one_K^{\sim}]_0 \rangle
\;.
\end{align*} Moreover, the last term is equal to $\langle \Ch_{\hat{\hat{\Tt}},\Dualityactionnp\times \hat{\hat{\Dualityactionop}}}, [\tilde{e}]_0\rangle$. To compute this, let us recall from Theorem~\ref{eq-TakaiDuality} that the Takai isomorphism $i_T$ satisfies 
$$
i_T\circ \Dualityactionnp\;=\;\Dualityactionnp\circ i_T
\;,
\qquad
i_T\circ \hat{\hat{\Dualityactionop}}\;=\;(\Dualityactionop \otimes \lambda_\bbR)\circ i_T
\;,
\qquad
\hat{\hat{\Tt}}\;=\;(\Tt \otimes \Tr_{L^2(\bbR)})\circ i_T
\;.
$$
Furthermore, $\Tr(f \nabla_{\hat{\hat{\Dualityactionop}}} f )=0$ since the derivation of a projection is off-diagonal. Thus
\begin{align*}
\langle & \Ch_{\hat{\hat{\Tt}} ,\Dualityactionnp\times \hat{\hat{\Dualityactionop}}}, [\tilde{e}]_0\rangle 
\\
&
\;=\; 
c_{n+1}\sum_{\sigma \in S_{n+1}} (-1)^\sigma (\hat{\hat{\Tt}}\otimes \Tr_N)\big((\tilde{e}-\one^{\sim}_K)\nabla_{\sigma(1)}\tilde{e}\cdots \nabla_{\sigma(n+1)}\tilde{e}\big)\\
&
\;=\; 
c_{n+1}\sum_{\sigma \in S_{n+1}} (-1)^\sigma (\hat{\hat{\Tt}}\otimes \Tr_N)\big(i^{-1}_T((e-\one^{\sim}_K)\nabla_{\sigma(1)}e\cdots \nabla_{\sigma(n+1)}e \otimes f)\big)\\
&
\;=\; 
c_{n+1}\sum_{\sigma \in S_{n+1}} (-1)^\sigma (\Tt\otimes \Tr_N)\big((e-\one^{\sim}_K)\nabla_{\sigma(1)}e\cdots \nabla_{\sigma(n+1)}e\big)
\\
&
\;=\;
\langle \Ch_{\Tt,\Dualityactionnp\times \Dualityactionop}, [e]_0 \rangle 
\;,
\end{align*}
which concludes the proof for odd $n$. For $n$ even, one uses an analogous computation with a class $[u]_1 \in K_1(\Aa)$ being mapped into $K_1(\Bb \rtimes_{\hat{\Dualityactionop}} \bbR)$ using $u \mapsto \one^{\sim}_K + i_T^{-1}((u-\one^{\sim}_K)\otimes f)$.
\hfill $\Box$

\vspace{.2cm}

To deal with the case $G=\bbT$, it was already pointed out in Proposition~\ref{prop:sm_connecting_maps} that it is useful to view the $\TM$-action as a periodic $\RM$-action by setting $\Dualityactionop_{t+1}=\Dualityactionop_t$. This allows to write out the connecting maps of the smooth Toeplitz extension as in \eqref{eq-sm_toep_factor}. This alone does not suffice, however, since $q$ does not preserve the dual trace. We therefore use an averaging argument:

\begin{lemma}
Let $\Aa$ be as in Theorem \ref{theo-smooth_duality} and, considering $\Dualityactionop$ as a periodic $\bbR$-action, construct $\Aa \rtimes_\Dualityactionop \bbR$ and $\Aa \rtimes_\Dualityactionop \bbT$ with their respective dual traces $\hat{\Tt}_\bbR$ and $\hat{\Tt}_\bbT$. For $0 \leq t \leq 1$, let $q_t: \Aa \rtimes_\Dualityactionop \bbR \to \Aa \rtimes_\Dualityactionop \bbT$ be the surjective homomorphism densely defined through
$$
q_t\left(\int_\bbR f(x) e^{2\pi\imath\, D_\bbR x} \difd{x}\right)
\;=\;
\int_\bbT \left(\sum_{k \in \bbZ} f(x + k)e^{-2\pi \imath (x+k)t}\right) e^{2\pi\imath\, D_\bbT x}  \difd{x}
\;,
$$
for all $f\in C_c(\bbR, \Aa)$, namely acting on the generators as $q_t(a g(D_\bbR + t))=a g(D_\bbT)$. Then the path $t \in [0,1] \mapsto q_t(\hat{a})$ is norm-continuous for every $\hat{a} \in \Aa \rtimes_\Dualityactionop \bbR$ and 
$$
\hat{\Tt}_\bbR(\hat{a}) 
\;=\; 
\int_{\bbT} \hat{\Tt}_\bbT (q_t(\hat{a}))\, \difd{t}
\;
$$
for all $\hat{a} \in (\Aa \rtimes_\Dualityactionop \bbR)_{\hat{\Tt}_\bbR}$.
\end{lemma}
\noindent{\bf Proof.}
The map $\hat{a} \mapsto \{q_t(\hat{a})\}_{t\in [0,1]} \in C([0,1],\Aa \rtimes_\Dualityactionop \bbT)$ for $\hat{a}=\int_\bbR f(x) e^{2\pi\imath\, D_\bbR x} \difd{x}$ and $f\in C_c(\bbR,\Aa)$ densely defines a homomorphism of $C^*$-algebras and therefore $t\mapsto q_t(\hat{a})$ is norm-continuous for each $\hat{a}\in \Aa\rtimes_{\Dualityactionop} \bbR$. By 
Lemma~\ref{lem-L2BoundedRep} we can represent positive $\hat{a} \in (\Aa \rtimes_\Dualityactionop \bbR)_{\hat{\Tt}_\bbR}$ as $\hat{a} =\hat{b}^*\hat{b}$ with $\hat{b} = \int_{\bbR} g(x) e^{2\pi \imath D_\bbR x}\difd{x}$ for $g\in L^2(\bbR, L^2(\Aa,\Tt))$ such that
\begin{align*}
\hat{\Tt}_\bbR(\hat{b}^*\hat{b}) &= \int_{\bbR} \Tt(g(x)^*g(x)) \difd{x} = \int_{\bbT} \sum_{k\in \bbZ} \Tt(g(k + t)^*g(k+t)) \difd{t} =\int_{\bbT} \hat{\Tt}_\bbT(q_t(\hat{b}^*\hat{b}))  \difd{t}
\end{align*}
and hence the trace formula for arbitrary elements follows by linearity and polarization.
\hfill $\Box$

\vspace{.2cm}

\noindent{\bf Proof} of Theorem~\ref{theo-smooth_duality} for $G=\bbT$. We only write out the case of odd $n$.  With $[e]_0 - [s(e)]_0 = \Ind^\Dualityactionop_\bbR ([v]_1) \in K_0(\Aa \rtimes_\Dualityactionop \bbR)$ represented by a smooth projection, the case $G=\RM$ of Theorem~\ref{theo-smooth_duality} and the trace identity give
\begin{align*}
\langle & \Ch_{\Tt,\Dualityactionnp\times \Dualityactionop} , [v]_1 \rangle
\\
&
\;=\;  \langle \Ch_{\hat{\Tt}_\bbR,\Dualityactionnp}, \Ind^\Dualityactionop_\bbR ([v]_1)\rangle\\
&
\;=\; 
c_{n}\sum_{\sigma \in S_{n}} (-1)^\sigma (\hat{\Tt}_\bbR \otimes \Tr_N)\big((e-s(e))\nabla_{\sigma(1)}e \cdots \nabla_{\sigma(n+1)} e\big)\\
&
\;=\; c_{n}\sum_{\sigma \in S_{n}} (-1)^\sigma \int_\bbT (\hat{\Tt}_\bbT \otimes \Tr_N)\big((q_t(e)-s(e))\nabla_{\sigma(1)}q_t(e) \cdots \nabla_{\sigma(n+1)} q_t(e) \big)\,\difd{t}\\
&
\;=\; \, \int_\bbT \langle \Ch_{\hat{\Tt}_\bbT,\Dualityactionnp},((q_t)_* \circ \Ind^\Dualityactionop_\bbR) ([v]_1)\rangle \,\difd{t}\\
&
\;=\; \; \langle \Ch_{\hat{\Tt}_\bbT,\Dualityactionnp},\Ind^\Dualityactionop_\bbT ([v]_1)\rangle 
\;,
\end{align*}
where the last line follows from
$$
\Ind^\Dualityactionop_\bbT 
\;=\; 
q_* \circ \Ind^\Dualityactionop_\bbR 
\;=\;
(q_t)_* \circ \Ind^\Dualityactionop_\bbR
\;,
$$
which holds by \eqref{eq-sm_toep_factor} and homotopy invariance. 
\hfill $\Box$

\newpage

\chapter{Applications to solid state systems}
\label{sec-Applications}

This chapter is about the applications of the results in the earlier chapters to solid state systems. There are numerous new results, and furthermore new or at least improved arguments of statements that can be already be found in the literature. The overview in Chapter~\ref{chap-Preface} essentially only mentions two new results (the bulk-boundary correspondence for irrational edges and the existence of flat bands of edge states for chiral Hamiltonians with a pseudogap and non-vanishing weak invariants). As this chapter is rather long and contains much more than that, let us give a brief more detailed outline of its contents.  

\begin{itemize}

\item[Section~\ref{sec-AlgSetUp}] shows how covariant families of random operators on $\ZM^d$ describing solid state systems can appear as representations of a crossed product algebra that may be twisted by the magnetic field. Then the analysis tools on this so-called bulk algebra are described, namely the non-commutative differentiation given by taking commutators with the position operator and the trace per unit volume. This places these physical systems into the mathematical framework of the above chapters. The whole section is merely a review of parts of the monograph \cite{PSbook} to which we also refer for further physical motivations.

\item[Section~\ref{sec-HalfSpace}] constructs the edge and half-space algebra for the above systems as the algebras in the smooth Toeplitz extension for an $\RM$-action on the bulk algebra given by a subgroup of the dual action. This allows to describe also half-spaces with possibly irrational angles. Furthermore, the representation theory of the edge and half-space algebras is carefully addressed. The (perpendicular) shift of the edge then appears as a one-parameter subgroup of the dual group on which there is a natural invariant measure. It is shown that the trace per surface area along the edge is almost surely constant also w.r.t. this real shift parameter and can be used to extract surface densities of suitable operator families. The proofs are different for rational and irrational edges and are admittedly fairly technical. However, this analysis is necessary in order to place the study of irrational edges on solid ground and later on allows the novel extension of the bulk-boundary correspondence to such systems. 

\item[Section~\ref{sec-BulkInv}] provides several criteria under which the strong and weak bulk topological invariants are well-defined as pairings of the Fermi projection with the cyclic cocycles constructed from the dual action. This requires the Fermi projection to lie in a suitable Besov space. It is a by now classical result that this is the case when the Fermi energy lies in a gap of the bulk Hamiltonian (a case which is included for sake of completeness) and at least in a mobility gap. A novel criterium assures that also the Fermi projection of bulk systems with a pseudogap at the Fermi level (vanishing of the density of states) lie in a Besov space and therefore have well-defined bulk invariants. This case applies to several classes of semimetals. By the index theorem,  these invariants are constant under norm-continuous deformations of the Fermi projection. Unless one has a bulk gap, the Fermi projection is, however, only strongly continuous under norm-continuous deformations of the Hamiltonian. In this situation, it is shown that the bulk invariants nevertheless vary continuously.

\item[Section~\ref{sec-BoundaryCurrents}] uses the smooth Toeplitz extension to prove the bulk-boundary correspondence. The novelty here is that also irrational edges are dealt with.

\item[Section~\ref{sec-DelocBoundary}] shows that a gapped bulk system with non-trivial bulk invariants cannot have edge spectrum that satisfies Aizenman-Molcanov localization bounds for all energies in the bulk gap. This extends the claims of \cite{PSbook} where the same claim was proved under the condition that the strong bulk invariants are non-trivial. 

\item[Section~\ref{sec-flat}] applies the Sobolev index theorem to chiral Hamiltonians for which the Fermi projection lies in Besov space. In dimension $d=1$, this allows to recover results of Graf and Shapiro \cite{ShapiroGraf2018}, while for $d>1$ it can be applied to chiral semimetals with a pseudogap at the Fermi level and proves the equality of the weak winding numbers and the signed surface state density, namely a weak bulk-boundary correspondence.

\item[Section~\ref{sec-Graphene}] works out the example of graphene in detail to illustrate the results of Section~\ref{sec-flat}.

\end{itemize}

\section{Algebraic set-up for solid state systems}
\label{sec-AlgSetUp}

The section briefly reviews the formalism developed by Bellissard \cite{Bel} for the description of one-particle models for solid state physics using covariant operator families on tight-binding Hilbert spaces. This allows to deal naturally with periodic, almost periodic and random systems in possibly irrational magnetic fields. All of this can be described by a disordered non-commutative torus that will be presented in a form that is similar to that given in the monographs \cite{PSbook,Pro}. 

\vspace{.2cm}

In this chapter let $\BB = (B_{i,j})_{1\leq i,j\leq d} \in \bbR^{d\times d}$ be an anti-symmetric real matrix, which is in the applications built out of the components of the magnetic field. Denote its lower triangular part by $\BB_+$ such that $\BB = \BB_+ - \BB_+^T$.

\begin{definition}
Let $\Cc$ be a separable $C^*$-algebra with a $\bbZ^d$-action $\gamma: \Cc \times \bbZ^d\to \Cc$.
The $\Cc$-valued non-commutative torus $\Cc\rtimes_{\gamma,\BB} \bbZ^d$ is the universal $C^*$-algebra generated by $d$ unitary generators $u_1,\ldots ,u_d$ and a representation of $\Cc$ together with the commutation relations
	$$
	u_i u_j \;=\; e^{\imath B_{i,j}} u_j u_i
	\;,
	\qquad
	f u_j \;=\; u_j(\gamma_{e_j}(f))
	\;,
	\qquad f \in \Cc\,, \;\;i,j=1,\ldots ,d
	\;.
	$$
\end{definition}

By universal $C^*$-algebra we mean that any pair $(\pi, u)$ consisting of a non-degenerate representation $\pi:\Cc\to\Bb(\Hh)$ and $d$ unitaries $u_1,\ldots ,u_d \in \Bb(\Hh)$ on some Hilbert space $\Hh$ subject to these commutation relations gives rise to a unique surjective homomorphism $\Cc\rtimes_{\gamma,\BB} \bbZ^d \to C^*(\pi(\Cc), u_1,\ldots ,u_d)$ that maps generators to generators. As already expressed by the notation, $\Cc\rtimes_{\gamma,\BB} \bbZ^d$ can also be described as a twisted (or iterated) crossed product with the unitary $2$-cocycle $\varphi(x,y)=e^{\imath\langle x|\BB_+|y\rangle}$ over $\bbZ^d$ (compare \cite{PSbook}). Viewed in that way, a representation of the commutation relations above corresponds precisely to a covariant representation  of a twisted dynamical system $(\Cc, \gamma, \bbZ^d, \varphi)$. The existence and universal property of twisted crossed products \cite{PackerRaeburn} then implies that the $\Cc$-valued non-commutative torus exists and is well-defined.

\vspace{.2cm}

The algebra $\Cc\rtimes_{\gamma,\BB} \bbZ^d$ is the completion in a universal $C^*$-norm of the algebra spanned by the Fourier series in the monomials $u^x =u_1^{x_1} \cdots u_d^{x_d}$, $x=(x_1,\ldots,x_d)\in\ZM^d$, of the form
$$
a 
\;=\; 
\sum_{x\in \bbZ^d} f_x u^x
\;,
$$
where $f_x \in \Cc$ is non-vanishing for only finitely many $x$. From the invariance of the commutation relations under phase changes and the universal property one obtains the strongly continuous $\bbT^d$-action $\rho$ (corresponding to the action dual to $\gamma$ on the twisted crossed product) acting on the generators of $\Cc\rtimes_{\gamma,\BB} \bbZ^d$ by
\begin{equation}
\label{eq-RhoAct0}
\rho_k (f) \; = \; f, \qquad \rho_k (u^x) 
\;=\; 
 \overline{\langle x, k\rangle} u^x
\;=\; 
 e^{2\pi\imath \, k \cdot x} u^x
\;
\end{equation}
for all $f\in \Cc$, $x\in \bbZ^d$ and $k\in \bbT^d$. We conclude by  Lemma~\ref{lem:fourier_series} that in fact any element $a\in \Cc\rtimes_{\gamma,\BB} \bbZ^d$ has a unique Fourier series 
$$a = \sum_{x\in \bbZ^d} \psi_x(a) u^x$$
with convergence the sense described there and the Fourier coefficients 
\begin{equation}
\label{eq-RhoAct}
\psi_x(a) 
\;=\; 
\int_{\bbT^d} \rho_k(a (u^{x})^*) \,\difd{k} \, \in \Cc
\;,
\end{equation}
where the integral is over the normalized Haar-measure on the torus. Concrete representations are most easily obtained if the action $\rho$ is also implemented spatially:

\begin{lemma}
\label{lem:cov_reps}
A regular covariant representation $(\pi, U, V)$ of the $\Cc$-valued non-commutative torus on a Hilbert space $\Hh$ shall be a triple consisting of a non-degenerate representation $\pi:\Cc \to \Bb(\Hh)$, $d$ unitaries $U=(U_1,\ldots ,U_d)$ in $\Bb(\Hh)$ and a strongly continuous unitary representation $V:\bbT^d\to \Bb(\Hh)$ which satisfy the commutation relations
$$
U_i U_j \;=\; e^{\imath B_{i,j}} U_j U_i
\;,
\qquad
\pi(f) U_j \;=\; U_j\pi(\gamma_{e_j}(f))
\;,
$$
and
$$V(k) \pi(f) V(k)^*= \pi(f), \qquad V(k) U_j V(k)^* = \overline{\langle e_j, k\rangle} U_j$$
for all $f \in \Cc, i,j=1,\ldots ,d$ and $k\in \bbT^d$.

\vspace{.1cm}

Any regular covariant representation induces a representation $\pi: \Cc\rtimes_{\gamma,\BB} \bbZ^d \to \Bb(\Hh)$, which is faithful if and only if $\pi\big|_{\Cc}$ is faithful.
\end{lemma}

\noindent{\bf Proof.}
Since a regular covariant representation in particular implements the universal commutation relations we only need to discuss the faithfulness. Let $a \in \Cc\rtimes_{\gamma,\BB} \bbZ^d$ with Fourier series $a= \sum_{x\in \bbZ^d}\psi_x(a) u^x$ then $\pi(a)=\sum_{x\in \bbZ^d}\pi(\psi_x(a)) U^x$. Conjugation with $V(k)$ defines a strongly continuous action $\bbT^d$-action on the image $\pi(\Cc\rtimes_{\gamma,\BB} \bbZ^d)$ and thus every $\pi(a) \in \pi(\Cc\rtimes_{\gamma,\BB} \bbZ^d)$ also has a unique Fourier series $\pi(a)=\sum_{x\in \bbZ^d} \pi(a)_x$ w.r.t. the spectral subspaces of the action by $V$ and notation as in \eqref{eq:fourier_coeff}. Covariance and continuity of $\rho$ imply that these two notions of Fourier series coincide, $\pi(a)_x = \pi(\psi_x(a))U^x$ for all $x\in \bbZ^d$. Since the Fourier series is unique $\pi(a)=0$ requires $\pi(\psi_x(a))=0$ for all $x\in \bbZ^d$, implying the claim.
\hfill $\Box$

\vspace{.2cm}

We now introduce a space $\Omega$ to model the possibly disordered configurations of a solid. 
Let $(\Omega,T,\ZM^d,\PM)$ consist of a probability space $(\Omega,\PM)$ with $\Omega$ a compact metrizable Hausdorff space and $\PM$ a regular Borel measure with full topological support. Let $\Omega$ be equipped with a continuous action $T: \bbZ^d \times \Omega \to \Omega$ under which $\PM$ is invariant. We further assume that the action $T$ is ergodic, {\it i.e.} any measurable set $A \subset \Omega$ that is $\bbZ^d$-invariant up to sets of measure zero must have $\bbP(A)=1$ or $\bbP(A)=0$. The action is denoted by $T_x(\omega)$ for $x \in \bbZ^d$ and induces an action $T^*$ on $C(\Omega)$ by $f \mapsto f \circ T_x$. Hence $C(\Omega)$ is a separable $C^*$-algebra on which integration w.r.t. $\PM$ defines a continuous finite faithful trace that is invariant under $T^*$.

\begin{definition}
The disordered non-commutative torus $\bbT^d_{\BB,\Omega} = C(\Omega)\rtimes_{T,\BB}\bbZ^d$ is the universal $C^*$-algebra generated by $d$ unitary generators $u_1,\ldots ,u_d$ and a representation of the continuous functions $C(\Omega)$ together with the commutation relations
$$
u_i u_j \;=\; e^{\imath B_{i,j}} u_j u_i
\;,
\qquad
 f u_j \;=\; u_j (f \circ T_{e_j})
 \;,
\qquad f \in C(\Omega)
\;.
$$
\end{definition}
A finite continuous faithful trace $\Tt:\bbT^d_{\BB,\Omega} \to \bbC$ is given by
$$
\Tt(a) 
\;=\; 
\int_\Omega  \psi_0(a,\omega)\;\bbP(\difd \omega)
\;
$$
and as $\psi_0$ is invariant under $\rho$, the trace is also $\rho$-invariant.

\vspace{.2cm}

We introduce representations of $\bbT^d_{\BB,\Omega}$ on the physical Hilbert space $\ell^2(\bbZ^d)$ of tight-binding wave functions, following \cite{Bel,PSbook}. The standard basis of $\ell^2(\bbZ^d)$ is denoted by $|x \rangle$, $x\in \bbZ^d$. Further the shifts $S^y$ by $y\in\ZM^d$ and unbounded position operators $X = (X_1,\ldots,X_d)$ are 
$$
S^y |x \rangle \;=\; |x + y\rangle
\;,
\qquad
X_j |x\rangle \;=\; x_j |x\rangle
\;.
$$

\begin{proposition}
\label{prop-CovRep}
A family $(\pi_\omega)_{\omega \in \Omega}$ of $*$-representations of $\bbT^d_{\BB,\Omega}$ on $\ell^2(\bbZ^d)$ is on the generators given by
$$
u_j = e^{\imath\langle e_j | \BB_+ | X \rangle} S^j
\;,
\qquad
\pi_\omega(f)
\; =\; 
\sum_{x\in\bbZ^d} f(T_x\omega) \,|x\rangle \langle x|
\;,
$$
for all $f\in C(\Omega)$. With $V(k)=e^{2\pi \imath X\cdot k}$ they define regular covariant representations of $\bbT^d_{\Omega,\BB}$ and the induced representations $\pi_\omega: \bbT^d_{\Omega,\BB} \to \Bb(\ell^2(\bbZ^d))$ are non-degenerate and faithful for $\bbP$-almost all $\omega \in \Omega$.
\end{proposition}

\noindent{\bf Proof.}
The commutation relations can be checked as in \cite{PSbook}; thus $(\pi_\omega, u, V)$ is a covariant representation and induces a non-degenerate representation of $\bbT^d_{\Omega,\BB}$. By Lemma~\ref{lem:cov_reps} it is enough to show that $C(\Omega)$ is almost surely represented faithfully. For any fixed $0\neq f\in C(\Omega)$ one has $\bbP$-almost surely
$$
\norm{\pi_\omega(f)}
\;\geq\;
\sup_{x\in \bbZ^d} \abs{\langle x| \pi_\omega(f)|x\rangle}
\;=\;
\sup_{x\in \bbZ^d} \abs{f(T_x\omega)} \;\stackrel{\mathrm{a.s.}}{=} \;
\norm{f}_\infty
\;,
$$
since $T$ is ergodic and the expression is positive and $T$-invariant. As $C(\Omega)$ has a countable dense subset this implies that all of $C(\Omega)$ is almost surely represented faithfully.
\hfill $\Box$

\vspace{.2cm}

The representation $\pi_\omega$ describes a single realization of a random solid state system and the translation action $T$ is supposed to shift the underlying disorder configuration while keeping the lattice fixed. Its ergodicity implies that the solid is in a vague sense homogeneous at large scales.
The matrix elements in these representations satisfy the covariance relation
\begin{equation}
\label{eq:GNSrep_covariant}
\langle x| \pi_\omega(a)| y \rangle 
\;=\; 
e^{\imath \langle x-y|\BB_+| x)} \langle 0| \pi_{T_x \omega}(a)| y-x\rangle
\;,
\end{equation}
such that in particular the $\bbP$-averages of their absolute values are indeed translation-invariant. Using the ergodicity of $\PM$, the trace $\Tt$ can be written as the trace per unit volume
\begin{align}
\Tt(a) 
&
\;=\; 
\int_{\Omega}  \langle 0 | \pi_\omega(a) | 0 \rangle\;\bbP(\difd{\omega})
\nonumber
\\
&
\label{eq-TV}
\;=\;
\lim_{L\to \infty} \frac{1}{(2L)^d} \sum_{\substack{x\in \bbZ^d\,,\, \lVert x \rVert_\infty < L}} \langle x | \pi_\omega(a) | x \rangle
\;,
\end{align}
where the second equality holds for almost every $\omega \in \Omega$ due to Birkhoff's ergodic theorem. Let us relate those representations with the GNS representation for $\Tt$.  The GNS-Hilbert space $L^2(\bbT^d_{\BB,\Omega},\Tt)$ is the completion of $\bbT^d_{\BB,\Omega}$ under the $L^2$-norm induced by
\begin{align*}
\Tt(a^* b) 
&
\;=\;
\sum_{x \in \bbZ^d}\int_\Omega  \overline{\psi_x(a,\omega)}\, \psi_x(b,\omega)\;\bbP(\difd{\omega})
\\
&
\;=\; 
\sum_{x \in \bbZ^d}\int_\Omega 
\langle 0|\pi_\omega(a^*)|x\rangle\,\langle x|\pi_\omega(b)|0\rangle\;\bbP(\difd{\omega})
\;.
\end{align*}
Hence there is an isometric isomorphism between $L^2(\bbT^d_{\BB,\Omega},\Tt)$ and $L^2(\Omega \times \bbZ^d)=L^2(\Omega,\PM)\otimes\ell^2(\ZM^d)$ that maps the cyclic vector $\one \in L^2(\bbT^d_{\BB,\Omega},\Tt)$ of the GNS-represen\-tation to $|0\rangle = 1_\Omega \otimes |0\rangle$
\begin{equation}
\label{eq-embedding}
a\,\one \in \bbT^d_{\BB,\Omega} \subset L^2(\bbT^d_{\BB,\Omega},\Tt)
\;\mapsto\; 
\int^\oplus_\Omega\bbP(\difd{\omega})\;\pi_\omega(a)|0\rangle \;\in\; 
L^2(\Omega,\PM)\otimes\ell^2(\ZM^d)
\;.
\end{equation}
and thereby implements a unitary equivalence between the GNS representation and a fibered representation given by the direct integral
\begin{equation}
\label{eq-GNSfiber}
\pi_\Tt \;=\; \int^\oplus_\Omega\bbP(\difd{\omega})\, \pi_\omega.
\;
\end{equation}
Now the von Neumann algebra $L^\infty(\bbT^d_{\BB,\Omega},\Tt)$ is defined just as in Proposition \ref{prop:traceext} as the double commutant in this representation:
$$
L^\infty(\bbT^d_{\BB,\Omega},\Tt) 
\;=\; 
\big(\pi_\Tt(\bbT^d_{\BB,\Omega})\big)''
\;.
$$
As the operators that are decomposable w.r.t. to a direct integral such as \eqref{eq-GNSfiber} form a von Neumann algebra \cite[Theorem IV.8.18]{Takesaki2001} also the operators in $L^\infty(\bbT^d_{\BB,\Omega},\Tt)$ are decomposable. Furthermore, the $\TM^d$-action $\rho$ extends to a weakly continuous action on $L^\infty(\bbT^d_{\BB,\Omega},\Tt)$ and also $\Tt$ extends to a $\rho$-invariant finite normal faithful trace on $L^\infty(\bbT^d_{\BB,\Omega},\Tt)$ which is still given by \eqref{eq-TV}. In the following we tacitly identify $\bbT^d_{\BB,\Omega}$ and $L^\infty(\bbT^d_{\BB,\Omega},\Tt)$ with their respective images under the faithful representation $\pi_\Tt$. Summing up, the constructions of Chapter~\ref{sec-CrossedProd} will be applied to the $C^*$-algebra $\Aa$ and finite von Neumann algebra $\Mm$ given by
$$
\Aa \;=\;\bbT^d_{\BB,\Omega}
\;,
\qquad
\Mm 
\;=\;
L^\infty(\bbT^d_{\BB,\Omega},\Tt)
\;.
$$ 
%

\vspace{.2cm}

Let us now discuss some analytic aspects of the torus action $\rho$. 
To the action $\rho$ and the standard basis we associate $d$ commuting derivations denoted by $\nabla=(\nabla_1,\ldots,\nabla_d)$. Due to \eqref{eq:nabla} they are explicitly given
$$
\nabla_j (a) = \nabla_j (\sum_{x\in \bbZ^d} \psi_x(a) u^x)
\;=\;
-\imath \,
\sum_{x\in \bbZ^d} x_j \psi_x(a) u^x
\;,
\qquad
j=1,\ldots,d
\;,
$$
for a differentiable element $a \in C^1(\bbT^d_{\Omega,\BB},\rho)$. 
For a unit vector $v \in\SM^{d-1}$ let us also set $\nabla_v=v \cdot\nabla$.  
The action $\rho$ extends to an isometric action on $L^p(\Mm, \Tt)$, which implies that the coefficient maps $\psi_x$ extend continuously and take values in $L^p(\Omega, \bbP)$. This further shows existence of Fourier series representations for all elements $a\in L^p(\Mm, \Tt)$ and therefore the derivatives on the Sobolev spaces $W^s_p(\Mm, \rho)$ act formally by the same expression as above and will still be denoted $\nabla_1, \ldots ,\nabla_d$.

\vspace{.2cm}

We also need the Besov spaces associated to the $\bbT^d$-action $\rho$ on the algebra $\Mm$. The necessary conditions of the index theorem are formulated in terms of the spaces $B^{\frac{n}{n+1}}_{n+1,n+1}(\Mm , \theta)$ where $\theta$ will be a restriction of $\rho$ to an $n$-parameter subgroup of $\TM^d$. By the characterization of the Besov norm by differences given in Section~\ref{sec-EquivBesov}, one deduces the inclusion $B^s_{q,p}(\Mm ,\rho)\subset B^s_{q,p}(\Mm ,\theta)$. Since the regularity of physical observables usually does not depend on the spatial direction, we only work with the smaller Besov space $B^s_{q,p}(\Mm ,\rho)$ for simplicity. Consequently, we will also drop the argument $\rho$ and use the abbreviation $B_n(\Mm)=B^{\frac{n}{n+1}}_{n+1,n+1}(\Mm)$ as in \eqref{eq-BesovAbbrev}.

\vspace{.2cm}

By Lemma~\ref{lemma:periodic_mult} a Fourier multiplier $f \scrS(\bbR^d)$ acts on $a\in L^p(Mm)$ by
\begin{equation}
\label{eq:multiplier_noncomtorus}
\hat{f} * a
\;=\; 
\sum_{x \in \bbZ^d} f(x) \,\psi_x(a) u^x
\end{equation}
and according to Definition~\ref{def-AversonSpec}, the Arveson spectrum of $a$ w.r.t. $\rho$ is hence 
$$
\sigma_\rho(a)
\;=\;
\{x \in \bbZ^d\,:\,\psi_x(a)\not=0\}
\;.
$$
Moreover, the Besov norm $B^s_{p,q}(\Mm )=B^s_{q}(L^p(\Mm ))$ of Definition~\ref{def-Besov} becomes explicitly  
$$ 
\lVert a \rVert_{B^s_{p,q}}
\;=\;
\Big(\sum_{\ScaleInd \geq 0} 2^{qs\ScaleInd } \Big\| \sum_{x \in \bbZ^d}  W_{\ScaleInd} (x) \,\psi_x(a) u^x \Big\|_p^q\Big)^{\frac{1}{q}}
\;,
$$
An element of $a \in \Mm $ whose Fourier coefficients are absolutely summable in the sense that
$\sum_{x\in \bbZ^d} \abs{x} \norm{\psi_x(a)}_{L^1(\Omega)}$
is readily shown to be in any Besov space $B^s_{p,q}(\Mm )$  with $0 < s \leq 1$. This is, however, only a rough sufficient condition. In general, even the classical Sobolev- and Besov-spaces are not easily characterized in terms of the decay of Fourier coefficients.

\section{Half-space and boundary algebras}
\label{sec-HalfSpace}

The physical space $\ZM^d$ (and also the enveloping space $\RM^d$ in which $\ZM^d$ is embedded) will be split into two half-spaces by a hypersurface $\{x\in\RM^d\,:\,x\cdot \Halfspaceunitvector=0\}$ associated to a  unit vector $\Halfspaceunitvector \in \SM^{d-1}$. Here the dot denotes the Euclidian scalar product and we consider $\Halfspaceunitvector$ to be in bijection with the one-parameter group actions $\Halfspaceaction$ on $\Aa=\bbT^d_{\BB,\Omega}$ that are obtained from the action \eqref{eq-RhoAct} by (quasi-)periodic linear flows on the torus
\begin{equation}
\label{eq-ActionHalfSpace}
\Halfspaceaction_t(a) \;=\; \rho_{t \Halfspaceunitvector}(a)
\;,
\qquad t\in\RM\;
\end{equation}
where $\rho$ is continued to a periodic $\bbR^d$-action.
Associated to $\Halfspaceaction$ is also the additive subgroup $\Gamma_\Halfspaceaction = \Halfspaceunitvector \cdot \bbZ^d$ of $\bbR$. We will call $\Halfspaceunitvector$ or $\Halfspaceaction$ rationally dependent if $\Halfspaceunitvector$ is a scalar multiple of a vector in $\bbQ^d$. In that case, there is a smallest positive element $\Lambda_\Halfspaceaction$ of $\Gamma_\Halfspaceaction$ such that $\Gamma_\Halfspaceaction=\Lambda_\Halfspaceaction \ZM$ and hence the action $\Halfspaceaction$ is a periodic $G=\Lambda_\Halfspaceaction^{-1} \bbT$-action. Otherwise, $\Gamma_\Halfspaceaction$ is dense in $\RM$. In the following, $\Halfspaceunitvector$ will be arbitrary and we always fix $G= \Lambda_\Halfspaceaction^{-1}\bbT$ if $\Halfspaceunitvector$ is rationally dependent and $G=\bbR$ otherwise.

\vspace{.2cm}

In the irrational case $G=\bbR$ we use as Haar measures $\mu$ and $\hat{\mu}$ on $\hat{G}$ the normalized Lebesgue measure. In the rationally dependent case $G=\Lambda_\Halfspaceaction^{-1} \bbT$ is isomorphic to $\bbT$, but we will not rescale the action to be $1$-periodic as in earlier chapters since $\Lambda_\Halfspaceaction$ has a physical meaning. The Haar measure on $G$ shall then be the Lebesgue-measure $\mu$ on $[0,\Lambda_\Halfspaceaction)$ giving the volume $\mu(G)=\Lambda_\Halfspaceaction^{-1}$. The dual group will be presented as $\hat{G}=\Lambda_\Halfspaceaction \bbZ$ and has as Haar measure $\Lambda_\Halfspaceaction \hat{\mu}$ with $\hat{\mu}$ the counting measure such that the Parseval theorem holds without constant factors.
The generator of $G$ in a regular representation $(\pi,U)$ on $L^2(G, \Hh)$ is still related to $\Halfspaceaction$ through
$$
\pi(\Halfspaceaction_t(a)) 
\;=\; 
U(t)\pi(a)U(t)\;, 
\quad 
U(t)\;=\;\exp(2\pi \imath D_\Halfspaceaction t)\;, 
\qquad \forall\; t\in G
\;,
$$
independent of $G$, which determines $D_\Halfspaceaction = \partial_t$ as the usual derivative on $[0, \Lambda_\Halfspaceaction^{-1})$ respectively $\bbR$. Note that this choice is consistent with the identification $\sigma(D_\Halfspaceaction) = \hat{G}$.

\vspace{.2cm}

An associated exact sequence of $C^*$-algebras is given by the smooth Toeplitz extension of the $C^*$-dynamical system $(\bbT^d_{\BB,\Omega},\Halfspaceaction,G)$ with $\Halfspaceaction$ as in \eqref{eq-ActionHalfSpace}, constructed as in Section~\ref{sec-toep}:
\begin{equation} 
0 \;\to\; \bbT^d_{\BB,\Omega} \rtimes_\Halfspaceaction G 
\;\hookrightarrow \;
{\rm T}(\bbT^d_{\BB,\Omega}, G, \Halfspaceaction)
\;\stackrel{q}{\to} \;
\bbT^d_{\BB,\Omega}
\;\to \;0\;.
\label{eq-Toep} 
\end{equation}
As we will see in concrete representations, the edge algebra $\Ee =\bbT^d_{\BB,\Omega} \rtimes_\Halfspaceaction G$ is generated by restrictions of bulk operators to $(d\mathrm{-1})$-dimensional strips and is therefore physically interpreted as observables located at the boundary. The half-space algebra $\hat{\Aa}={\rm T}(\bbT^d_{\BB,\Omega}, G, \Halfspaceaction)$ contains in addition the restrictions of the bulk operators to the half-space $\{\Halfspaceunitvector\cdot x > 0\}$ subject to continuous boundary conditions. Note that both $\Ee$ and $\hat{\Aa}$ implicitly depend on $\Halfspaceaction$. With these notations, the exact sequence~\eqref{eq-Toep} takes the form
\begin{equation} 
0 \;\to\; \Ee
\;\hookrightarrow \;
\hat{\Aa}
\;\to \;
\Aa
\;\to \;0\;.
\label{eq-ToepBis} 
\end{equation}
This is the exact sequence behind the smooth bulk-boundary correspondence (BBC) that will be discussed in Section~\ref{sec-BoundaryCurrents}. In order to extract the physical content of this algebraic bulk-boundary-correspondence the remainder of this sections will study representations of those algebras and their associated von Neumann algebras that are more closely aligned with the conventional theory of random Schr\"odinger operators on lattices.

\vspace{.2cm}

Let $V_{\Halfspaceaction}(t)=V(t\Halfspaceunitvector)$ be the periodic extension of $V$ as given in Proposition~\ref{prop-CovRep} to $\bbR^n$. Clearly $(\pi_\Tt, V_\Halfspaceaction)$ is a covariant representation of $(\Aa, \Halfspaceaction, G)$ on the GNS-Hilbert space $\Hh_\Tt=L^2(\Omega,\PM)\otimes\ell^2(\ZM^d)$ (see Proposition~\ref{prop-L2Rep}) and its integrated form, {\it cf.} \eqref{eq-IntegratedRep}, will be denoted by 
$$
\hat{\pi}_\Tt
\;=\;
\pi_\Tt \times V_{\Halfspaceaction}
\;:\;\Aa  \rtimes_\Halfspaceaction G\;\to\;\Bb(\Hh_\Tt)
\;.
$$
According to \eqref{eq-GNSfiber}, the GNS representation of $\Aa$ is fibered into covariant representations $\pi_\omega$ and thus also
\begin{equation}
\label{eq-PiTauHatFiber}
\hat{\pi}_\Tt
\;=\;
\int_\Omega^\oplus \bbP(\difd \omega) \;(\pi_\omega \times V_{\Halfspaceaction})
\;,
\end{equation}
where the integrated representation $\pi_\omega \times V_{\Halfspaceaction}$ again as defined  in \eqref{eq-IntegratedRep}. As is made clear by the subscript, this representation still acts on the GNS-Hilbert space $\Hh_\Tt = L^2(\Omega,\PM)\otimes\ell^2(\ZM^d)$ and, as we will show below, it is faithful and thus we can discuss the physical interpretation of the exact sequence in terms of it.

\vspace{.2cm}

The generator $X_\Halfspaceaction$ of the action $\Halfspaceaction$ extended to $\Hh_\Tt$ is calculated explicitly using the definition \eqref{eq-RhoAct0} of $\rho$ and the embedding \eqref{eq-embedding}. One finds
$$
X_\Halfspaceaction\;=\;\Halfspaceunitvector \cdot X 
\;,
$$
where $X=(X_1,\ldots,X_d)$ is the (unbounded selfadjoint) operator on $\ell^2(\ZM^d)$. More explicitly,  for a dense set of differentiable elements $a=\sum_{x \in \bbZ^d}  a_x u^x\in \bbT^d_{\BB,\Omega}\subset \Hh_\Tt$,
$$
X_\Halfspaceaction a 
\;=\; 
\sum_{x \in \bbZ^d} (\Halfspaceunitvector \cdot x)\, 
a_x u^x
\;.
$$
Since $X_\Halfspaceaction$ diagonal in the standard basis its spectrum is the closure of $\Gamma_\Halfspaceaction$, that is $\Lambda_\Halfspaceaction \ZM$ in the rationally dependent case $G=\Lambda_\Halfspaceaction^{-1}\TM$ and $\RM$ otherwise. Note that in both cases the spectrum of $X_\Halfspaceaction$ is given by the dual group $\hat{G}=\Lambda_\Halfspaceaction \bbZ$ of $G$ and the point spectrum labels the orthogonal distances of lattice points from the hyperplane $\Halfspaceunitvector\cdot \bbR^d = 0$. Furthermore, recall from Section~\ref{sec-CStar} that the crossed product $\hat{\pi}_\Tt(\Aa  \rtimes_\Halfspaceaction G)$ is generated by products of the form $\pi_\Tt(a) f(X_\Halfspaceaction)$ where $a \in\Aa $ and $f \in C_0(\hat{G})$. 

\vspace{.2cm}

Let us note that $f(X_\Halfspaceaction)$ is a multiplication operator on $\Hh_\Tt$ that only depends on the displacement in the direction $\Halfspaceunitvector$ relative to some arbitrary reference point. In particular, $P= \chi(X_\Halfspaceaction>0)$ is the restriction to the half-space of all $x\in \bbZ^d$ with $\Halfspaceunitvector \cdot x > 0$. In the rational case, it is a multiplier of $\Aa \rtimes_\Halfspaceaction \bbT$ which can hence be used to describe restrictions of elements of $\Aa$ to half-spaces, {\it i.e.} physical systems with Dirichlet boundary conditions. In the irrational case we must use a smooth switch function $f(X_\Halfspaceaction)$ instead to identify half-space operators with elements of the multiplier algebra $M(\Aa \rtimes_\Halfspaceaction \bbR)$ since the sharp spectral projections are not contained in the crossed product.
Just as in \cite{KRS,PSbook}, generic elements from half-space algebras such as $\Aa \rtimes_\Halfspaceaction G$ will carry a hat like $\hat{a}$. With the particular choice $\Halfspaceunitvector=e_d$ of a basis vector of the lattice, one recovers the setup of \cite{KRS,PSbook}.

\vspace{.2cm}

\begin{proposition}
\label{prop:halfspace_GNSrep}
The integrated form $\hat{\pi}_\Tt = \pi_\Tt \times V_\Halfspaceaction$ of the GNS-representation is a faithful representation of $\Aa \rtimes_\Halfspaceaction G$ on $L^2(\Omega \times \bbZ^d)$ for $G=\Lambda_\Halfspaceaction^{-1}\bbT$ if $\Halfspaceunitvector$ is rational and for $G=\bbR$ if $\Halfspaceunitvector$ is irrational.
\end{proposition}

\noindent{\bf Proof.}
Let us first exhibit a convenient alternative description of $\Aa \rtimes_\Halfspaceaction G$. We now identify $\Aa \subset \Bb(\Hh_\Tt)$ and define $\Aa \rtimes_\Halfspaceaction G$ in the regular representation $(\pi,U)$ on $L^2(G,\Hh_\Tt)$. Note that with $V(k) = \one_{L^2(G)}\otimes V(k)$ from Proposition~\ref{prop-CovRep} one still has the covariance relation $\pi(\rho(a))=V(k)\pi(a)V(k)^*$ and thus the action $\rho$ is spatially implemented and thereby also extends to $\Aa \rtimes_\Halfspaceaction G$.

\vspace{.1cm}

A dense subalgebra of $\Aa \rtimes_\Halfspaceaction G$ is given by functions $f\in C_c(G, \Aa)$ determining the elements $(\pi \times U)(f)=\int_{G} \pi(f(t)) U(t) \mu(\difd{t})$ and a simple approximation argument shows that $C_c(G, \Aa_c)$ is also dense, where $\Aa_c \subset \Aa$ shall denote the elements whose Fourier series has only finitely many non-vanishing terms. Hence we can write $f\in C_c(G, \Aa_c)$ as $f(t) = \sum_{x\in \bbZ^d} f_x(t) u^x$ with $f_x \in C_c(G,C(\Omega))$ vanishing for $\abs{x}$ large enough. The covariance relation $U(t) \pi(u^x) U(-t)= e^{2\pi \imath(x\cdot \Halfspaceunitvector) t} \pi(u^x)$ then gives
\begin{align*}
(\pi \times U)(f)
&
\;=\;
\int_{G} \sum_{x\in \bbZ^d} \pi(f_x(t) u^x) U(t) \mu(\difd{t}) 
\\
&
\;=\; 
\sum_{x\in \bbZ^d} \left(\int_{G} \pi(f_x(t)) U(t) e^{2\pi \imath(x\cdot \Halfspaceunitvector)t} \mu(\difd{t})\right) \pi(u^x)
\;,
\end{align*}
and the term in brackets can be considered an element of the crossed product $\Cc_\Halfspaceaction := \pi((C(\Omega))\rtimes_\Halfspaceaction G$ with $\Halfspaceaction$ acting trivially on $C(\Omega)$. Thus we have an isomorphism $\Cc_\Halfspaceaction \simeq C_0(\hat{G},C(\Omega))$ which is for $\hat{g} \in C_c(\hat{G},C(\Omega))$ given by the Fourier transform 
$$
\tilde{\pi}(\hat{g}) 
\;=\; 
(\pi\times U)(\calF^{-1}\hat{g})= \int_{G} \pi(\calF^{-1}\hat{g})(t) U(t) \mu(\difd{t})
$$
and one verifies the commutation relation
$$
\pi(u^x) \tilde{\pi}(\hat{g}) \pi(u^{-x}) 
\;=\; 
\tilde{\pi}(T_x^{(\Halfspaceaction)}\hat{g} )
$$
with the action 
$$
T_x^{(\Halfspaceaction)}\,:\, \bbZ^d\times C_0(\hat{G},C(\Omega)) \to C_0(\hat{G},C(\Omega))
\;, 
\qquad  (T_x^{(\Halfspaceaction)}\hat{g})(\omega, r) 
\;=\; \hat{g}(T_x \omega, r + x\cdot \Halfspaceunitvector)
\;.
$$ 
Recognizing these commutation relations as those of a $\Cc_\Halfspaceaction$-valued non-commutative torus these manipulations yield a natural surjective homomorphism from $\Cc_\Halfspaceaction \rtimes_{T^{(\Halfspaceaction)}, \BB} \bbZ^d$ to $\Aa \rtimes_\Halfspaceaction G$ and Lemma~\ref{lem:cov_reps} implies further $\Cc_\Halfspaceaction \rtimes_{T^{(\Halfspaceaction)}, \BB} \bbZ^d \simeq \Aa \rtimes_\Halfspaceaction G$ since $(\tilde{\pi}, \pi(u), V)$ is a regular covariant representation and $\tilde{\pi}$ is faithful.

\vspace{.1cm}

Since $\pi_\Tt$ is a covariant representation of the non-commutative torus $\bbT^d_{\Omega,\BB}$ the integrated form $\hat{\pi}_\Tt= \pi_\Tt \times V_\Halfspaceaction$ is also a regular covariant representation of the non-commutative torus $\Cc_\Halfspaceaction \rtimes_{T^{(\Halfspaceaction)},\BB} \bbZ^d$ in the sense of Lemma~\ref{lem:cov_reps} and thus it is enough to check if $\Cc_\Halfspaceaction$ is represented faithfully on $L^2(\Omega \times \bbZ^d)$. By density it is enough to show that $\hat{\pi}_\Tt$ is isometric for all $\hat{a} \in \Cc_\Halfspaceaction$ given in the form $\hat{a}= \int_G \pi((\calF^{-1} \hat{g})(t)) U(t)\mu(\difd{t})$ for some $\hat{g} \in C_c(\hat{G}, C(\Omega))$. Making the ansatz $\Psi_i = \psi_i \otimes |x\rangle \in L^2(\Omega)\otimes \ell^2(\bbZ^d)$ the matrix elements for $\hat{\pi}_\Tt$ are
\begin{align*}
\langle \Psi_1, \hat{\pi}_\Tt(\hat{a})\Psi_2\rangle_{L^2(\Omega\times\bbZ^d)} 
&
\;=\; 
\langle \Psi_1, \left(\int_{G} \pi_\Tt((\calF^{-1}\hat{g})(t))e^{2\pi\imath X_\Halfspaceaction t} \mu(\difd{t})\right) \Psi_2\rangle_{L^2(\Omega\times\bbZ^d)} \\
&
\;=\; 
\int_{G} e^{2\pi\imath (x\cdot\Halfspaceunitvector) t} \langle \Psi_1,  \pi_\Tt((\calF^{-1}\hat{g})(t)) \Psi_2\rangle_{L^2(\Omega\times \bbZ^d)} \mu(\difd{t}) \\
&
\;=\;
\int_{G} e^{2\pi\imath (x\cdot\Halfspaceunitvector) t} \langle \psi_1, ((\calF^{-1}\hat{g})(t) \circ T_x)  \psi_2\rangle_{L^2(\Omega)}\mu(\difd{t}) \\
&
\;=\; 
\langle \psi_1, (\hat{g}(x\cdot \Halfspaceunitvector)\circ T_x) \psi_2\rangle_{L^2(\Omega)}
\end{align*}
and since $T_x$ induces a unitary map on $L^2(\Omega)$ the density of $\Gamma_\Halfspaceaction$ in $\hat{G}$ and continuity of $\hat{g}$ give 
\begin{align*}
\norm{\hat{\pi}_\Tt(\hat{a})}
&
\;\geq\; 
\sup_{\substack{\Psi_1,\Psi_2 \in L^2(\Omega \times \bbZ^d)\\ \norm{\Psi_1}=1=\norm{\Psi_2}}} 
\abs{\langle \Psi_1, \hat{\pi}_\Tt(\hat{a})\Psi_2\rangle}
\\
&
\;\geq\; 
\sup_{\substack{\psi_1,\psi_2 \in L^2(\Omega)\\ \norm{\psi_1}=1=\norm{\psi_2}
\\
r \in \Gamma_\Halfspaceaction}} \abs{\langle \psi_1, \hat{g}(r)\psi_2\rangle} 
\\
&
\;=\; \norm{\hat{g}}_\infty \;=\; \norm{\hat{a}}
\end{align*}
implying that the representation is isometric.
\hfill $\Box$

\vspace{.2cm}

To obtain almost surely faithful representations on the physical Hilbert space $\ell^2(\bbZ^d)$ one must pose additional conditions on the measurable dynamical system $(\Omega, \bbP, T)$. In the case $d=1$ this is easy to see since one has $\Halfspaceunitvector=\pm e_1$ and hence the element $\pi_\Tt(f)\chi(X_\Halfspaceaction=0)$ is mapped under $\pi_\omega \times V_\Halfspaceaction$ to the rank-$1$-operator $f(\omega)|0\rangle\langle 0|$ for fixed $\omega$. Therefore, the fiber representation $\hat{\pi}_\omega$ is almost surely faithful if and only if $\Omega$ is a singleton set. For higher dimension one can give a convenient sufficient condition.

\begin{definition}
A dynamical system $(\Omega, T, \bbP)$ is said to be strong mixing if for all measurable sets $A,B \subset \Omega$ one has
$$
\label{eq-StrongMixing}
\lim_{x \to \infty} \, \mathbb{P}(A \cap (T_{x}B)) \; = \;  \mathbb{P}(A)\mathbb{P}(B).
$$
\end{definition}

Recall that the general assumption is only that $T$ is ergodic as a $\bbZ^d$-action, which is much weaker considering that it even allows some subgroups of $\bbZ^d$ to act trivially. In many applications strong mixing holds naturally, {\sl e.g.} if the probability space is constructed in the form $\Omega=(\Omega_0)^{\bbZ^d}$ with $\bbP=(\bbP_0)^{\otimes \bbZ^d}$ a product measure and $T$ acting by translation of the index.

\vspace{.2cm}

The following lemma will be essential for combining the lattice translations with irrational flows.

\begin{lemma}
\label{lem-ergodic}
Assume that $(\Omega, T, \bbP)$ is as above and let $x_1,x_2 \in \bbZ^d$ be linearly independent, further $M, \Delta \in \bbR_{>0}$ with $\frac{\Delta}{M} \in \bbR \setminus \bbQ$. For each $r \in[0,M)$ let $n(r)$ be the unique integer such that 
$$
M n(r) \leq r + \Delta \;< \;M (n(r)+1)
$$ 
and define a measurable transformation $S: \Omega \times [0, M) \to \Omega \times [0,M)$ by
$$
S(\omega,r) 
\;=\; 
(T_{x_2- n(r) x_1}\omega, \, r + \Delta \mod M)
\;.
$$
If $(\Omega, T, \bbP)$ is strong mixing then $S$ is ergodic on $\Omega \times [0, M)$ and hence every $S$-invariant measurable function is almost surely constant w.r.t. the product measure of $\bbP$ and the Lebesgue measure.
\end{lemma}

\noindent{\bf Proof.}
The action $S$ is a so-called skew product of the dynamical systems $(M \, \bbT,\tau,\lambda)$, $(\Omega,T,\bbP)$ with $\tau$ an irrational rotation on the torus and $\lambda$ the Lebesgue measure, {\it i.e.} there is a twisting function $\theta: [0,M) \to \bbZ^{d}$, such that the action takes the form
$$
S(\omega,r) 
\;=\; 
( T_{\theta(r)}(\omega),\tau(r))
\;.
$$
A useful criterion for the ergodicity of this type of action is proved in \cite{Brettschneider2007}:
Denote for $r$ the orbit under $\tau$ as $r_n = \tau^n(r)$. If $\tau$ is ergodic and for $\lambda$-almost every $r$, the action $\theta$ is weakly mixing along the orbit, {\it i.e.} if
\begin{equation}
\label{eq-OrbitMixing}
\lim_{N\to \infty} \,\frac{1}{N}\,\sum_{n=0}^{N-1} \,\big| \mathbb{P}(A \cap T^{-1}_{\theta(r_n)}B)- \mathbb{P}(A)\mathbb{P}(B)\big|
\;=\; 
0
\end{equation}
holds for all measurable $A,B \subset \Omega$, then the skew product is ergodic w.r.t. the product measure $\lambda \times \mathbb{P}$. As the compositions act by
$$
T^{-1}_{\theta(r_n)}(\omega) 
\;=\; 
T_{-n x_2 + m(r,n)x_1}(\omega)
$$
for some integers $m(r,n)$ increasing monotonously along each orbit the sequence $(\theta(r_n))_{n\in\bbN}$ diverges to infinity and hence \eqref{eq-OrbitMixing} clearly holds under the assumption of strong mixing.
\hfill $\Box$

\vspace{.2cm}

\begin{proposition}
Let $d \geq 2$ and again let $G=\Lambda_\Halfspaceaction^{-1} \bbT$ in the rationally dependent case and $G=\bbR$ otherwise. If $(\Omega, T, \bbP)$ is strong mixing, then $\pi_\omega \times V_\Halfspaceaction$ is $\bbP$-almost surely a faithful representation of $\Aa \rtimes_\Halfspaceaction G$.
\end{proposition}

\noindent{\bf Proof.}
As in the proof of Proposition~\ref{prop:halfspace_GNSrep} it is again enough to show that $\Cc_\Halfspaceaction$ as defined there is represented faithfully and we consider $\hat{a} \in \Cc_\Halfspaceaction$ in a dense subset of elements which are given in the form $\hat{a} = \int_G \pi(\calF^{-1}\hat{g}(t)) U(t)$ for some $\hat{g} \in C_c(\hat{G}, C(\Omega))$. Since that set contains a countable dense subset of $\Aa \rtimes_\Halfspaceaction G$ it is enough to show that each individual such $\hat{a}$ is almost surely represented faithfully. Computing matrix elements one obtains
$$
\langle x| (\pi_\omega\times V_\Halfspaceaction)(\hat{a})|x\rangle
\;=\;
\langle x| \int_{G} \pi_\omega((\calF^{-1}\hat{g})(t))e^{2\pi (x\cdot\Halfspaceunitvector) t} \mu(\difd{t})|x\rangle 
\;=\; 
\hat{g}(T_x\omega, x\cdot \Halfspaceunitvector)
\;.
$$
Due to $\norm{(\pi_\omega\times V_\Halfspaceaction)(\hat{a})}\geq\sup_{x\in \bbZ^d} \abs{\hat{g}(T_x\omega, x\cdot \Halfspaceunitvector)}$ it is enough to show that this supremum is almost surely equal to $\norm{\hat{g}}_\infty$.
Replacing $\hat{g}$ with $\tilde{g}(\omega, r)=\hat{g}(T_y \omega, r + y\cdot \Halfspaceunitvector)$ does not change the set of diagonal elements and thus we may shift the compact support of $\hat{g}$ to assume $\mathrm{supp}(\hat{g})\subset [0, x_1 \cdot \Halfspaceunitvector)$ for some $x_1 \in \bbZ^d$. 

\vspace{.1cm}

If $\Halfspaceaction$ is rationally dependent choose $x_2\in \bbZ^d \setminus \{0\}$ with $x_2\cdot \Halfspaceunitvector=0$ and define the measurable function 
$$
f(\omega)
\;=\;
\sup_{x\in \bbZ^d} \abs{\hat{g}(T_x\omega, x\cdot \Halfspaceunitvector)}
\;.
$$ 
Note that $f \circ T_{x_2} = f$ and since the strong mixing in particular implies that $T_{x_2}$ is ergodic, $f$ is almost surely constant. Hence 
$$
\norm{(\pi_\omega\times V_\Halfspaceaction)(\hat{a})} 
\;\geq\; 
\sup_{x\in \bbZ^d} \abs{\langle x| (\pi_\omega\times V_\Halfspaceaction)(\hat{a})|x\rangle} 
\;=\; 
f(\omega)\,\stackrel{\mathrm{a.s.}}{=}\, \norm{f}_\infty 
\;=\; 
\norm{\hat{g}}_\infty 
\;=\; 
\norm{\hat{a}}\;,
$$
implying that $\hat{a}$ is almost surely represented isometrically.

\vspace{.1cm}

For $\Halfspaceaction$ not rationally dependent choose any $x_2 \in \bbZ^d$ such that $0 < x_2\cdot \Halfspaceunitvector < x_1 \cdot \Halfspaceunitvector$ and $\frac{x_2 \cdot \Halfspaceunitvector}{x_1\cdot \Halfspaceunitvector} \in \bbR \setminus \bbQ$. Define on $\tilde{\Omega}:=\Omega \times [0, x_1 \cdot \Halfspaceunitvector)$ the measurable function
$$
f(\omega, r) \;=\; 
\sup_{x\in \bbZ^d} \abs{\hat{g}(T_x\omega, x\cdot \Halfspaceunitvector+r)}
$$
and note that it is invariant $f \circ S = f$ under the measurable transformation $S$ from Lemma~\ref{lem-ergodic} with the parameters $M=x_1\cdot \Halfspaceunitvector$, $\Delta=x_2\cdot \Halfspaceunitvector$ and $x_1,x_2$ as above. Since $S$ is ergodic $f$ must be almost surely constant in the product measure and thus there is a set $\Omega_0$ of probability one such that $f(\omega, \cdot)$ is Lebesgue-almost surely constant for fixed $\omega \in \Omega_0$. Since $\hat{g}$ has compact support and is therefore uniformly continuous, $f(\omega, \cdot)$ is also continuous and hence 
$$
f(\omega,0)
\;=\;
\norm{f(\omega, \cdot)}_\infty \,\stackrel{\mathrm{a.s.}}{=}\,
\norm{\hat{g}}_\infty
$$ 
such that $\norm{(\pi_\omega\times V_\Halfspaceaction)(\hat{a})}\geq f(\omega,0)$ completes the proof.
\hfill $\Box$

\vspace{.2cm}

Let us finally note that the faithful representations on $L^2(\Omega \times \bbZ^d)$ respectively $\ell^2(\bbZ^d)$ also extend uniquely to faithful representations of the Toeplitz algebra $\hat{\Aa}=\Toep$.
%

\vspace{.2cm}

In order to be able to use Borel functional calculus and also to treat physical problems involving sharp boundaries in the irrational case, it is necessary to pass to von Neumann algebras. 
As in Section~\ref{sec-WStar} the action \eqref{eq-ActionHalfSpace} extends to a weakly continuous group of automorphisms of $\Mm =L^\infty(\bbT^d_{\BB,\Omega},\Tt)$.  As the trace $\Tt$ is invariant under $\rho$, it is also invariant under $\Halfspaceaction$. Consequently one can construct as in Chapter~\ref{sec-CrossedProd} the $C^*$-crossed product $\Aa \rtimes_\Halfspaceaction G$ and the $W^*$-crossed product $\Mm  \rtimes_\Halfspaceaction G$ which are both equipped with the dual trace $\hat{\Tt}_\Halfspaceaction$. Again we use the notations and identifications
$$
\Nn_{\Halfspaceaction}\;=\;\Mm  \rtimes_\Halfspaceaction G
\;=\;L^\infty(\bbT^d_{\BB,\Omega},\Tt)\rtimes_\Halfspaceaction G
\;=\;
L^\infty(\Nn_{\Halfspaceaction},\hat{\Tt})
\;,
$$
and denote the $L^p$-spaces by $L^p(\Nn_{\Halfspaceaction})$. The trace $\hat{\Tt}_\Halfspaceaction$ is by construction invariant under both $\hat{\rho}$ and the dual action $\hat{\Halfspaceaction}$.

\vspace{.2cm}

Some technical complications arise now since the natural von Neumann algebra for our purposes is not necessarily $\hat{\pi}_\Tt(\Aa \rtimes_\Halfspaceaction G)''$ defined on the GNS-Hilbert space $\Hh_\Tt$, but rather the von Neumann crossed product $\Nn_{\Halfspaceaction}=L^\infty(\Aa ,\Tt)\rtimes_\Halfspaceaction G$ in the regular representation on $L^2(G, \Hh_\Tt)$.
Hence it is a priori not clear how elements of the $W^*$-crossed product relate to operators in the physical representation, which eventually should act on the space $\ell^2(\bbZ^d)$ of physical wavefunctions. 

\vspace{.2cm}

For sake of completeness, let us demonstrate that both constructions produce incompatible von Neumann-algebras in the case of irrational $\Halfspaceunitvector$ and hence that $\Nn_\xi$ has no natural representation on the GNS-Hilbert space. This may seem surprising, since $\hat{\pi}_\Tt$ is a faithful representation of $\Aa\rtimes_\Halfspaceaction \bbR$ on $L^2(\bbZ^d\times \Omega)$ which also extends to multipliers. This correspondence is, however, not compatible with the weaker operator topologies:

\begin{proposition}
\label{prop-NoWeakExtend}
Let $\Halfspaceunitvector$ be irrational and let $\widetilde{\Nn}_{\Halfspaceaction} \subset \mathcal{B}(L^2(\bbZ^d\times \Omega))$ be the von Neumann algebra generated by $\pi_\Tt(\Mm )$ and $ V(\bbR)$. Neither the isomorphism 
$$
\hat{\pi}_\Tt
\;=\;
(\pi_\Tt \times V)
\;:\; \Aa\rtimes_\Halfspaceaction \bbR \;\to\; (\pi_\Tt \times V)(\Aa\rtimes_\Halfspaceaction \bbR)\; \subset\; \mathcal{B}(L^2(\bbZ^d\times \Omega))
$$
nor its inverse can be extended to a normal homomorphism $\Nn_{\Halfspaceaction}\to \widetilde{\Nn}_{\Halfspaceaction}$ respectively $\widetilde{\Nn}_{\Halfspaceaction}\to \Nn_{\Halfspaceaction}$ of von Neumann algebras.
\end{proposition}

\noindent {\bf Proof.}
Let $D_\Halfspaceaction $ be the generator of $\Halfspaceaction$ in the regular representation of $\Nn_{\Halfspaceaction}$. Since $D_\Halfspaceaction $ is the generator of translations on $L^2(\bbR)$ it has purely continuous spectrum and in $\Nn_{\Halfspaceaction}$ one has the strong limit 
$$
\mathrm{s}\mbox{-}\!\lim_{\delta \to 0}\; \chi(D_\Halfspaceaction  \in (-\delta, \delta)) 
\;=\; 0
\;,
$$ 
whereas in $\widetilde{\Nn}_{\Halfspaceaction}$  
$$
\mathrm{s}\mbox{-}\!\lim_{\delta \to 0} \;\chi(X_\Halfspaceaction \in (-\delta, \delta))
\;=\;
\chi(X_\Halfspaceaction=0) 
\;\neq\; 0
\;
$$
considering that $|0\rangle$ is a proper eigenvector of $X_\Halfspaceaction$.
Since $\hat{\pi}_\Tt(f(D_\Halfspaceaction )) = f(X_\Halfspaceaction)$ for every continuous function $f$, the strong limits would have to coincide if the representation is normal which is hence impossible.  There could still be a normal homomorphism $\tilde{\pi}: \widetilde{\Nn}_{\Halfspaceaction} \to \Nn_{\Halfspaceaction}$ extending the inverse of $\hat{\pi}_\Tt$. However, this implies that for any $\lambda \in \bbR$
$$
\chi(\tilde{\pi}(X_\Halfspaceaction) = \lambda) 
\;=\;  
\chi(D_\Halfspaceaction   = \lambda) 
\;=\; 
0
\;,
$$
and thus one has again a contradiction since for any $f\in C_0(\bbR)_+$
$$
f(D_\Halfspaceaction ) 
\;= \;
\tilde{\pi}\left(\sum_{\lambda \in \sigma(X_\Halfspaceaction)} f(\lambda)\, \chi(X_\Halfspaceaction=\lambda)\right)
\;=\; 
\sum_{\lambda \in \sigma(X_\Halfspaceaction)} f(\lambda) \,\chi(D_\Halfspaceaction =\lambda) 
\;=\; 
0
\;,
$$
where the sums are positive and increasing and hence interchange with the normal homomorphism $\tilde{\pi}$. 
\hfill $\Box$

\vspace{.2cm}

The resolution of this problem consists in the choice of a different representation of the crossed product that is closer to the applications than the regular representation. As a preparation, consider the Hilbert space $L^2(\hat{G} \times \Hh_\Tt) = \int_{\hat{G}}^\oplus \hat{\mu}(\difd{r})\, \Hh_\Tt$ and introduce on it the fibered position operator  $\hat{X}_\Halfspaceaction$ by setting
$$
\hat{X}_\Halfspaceaction 
\;=\; 
\int_{\hat{G}}^\oplus \hat{\mu}(\difd{r})\;(X_\Halfspaceaction + r\one)\otimes \one_{\Omega} 
\;,
$$ 
where the direct integral is w.r.t. the Lebesgue integral on $\bbR$ respectively the counting measure on $\Lambda_\Halfspaceaction \bbZ$. The $C^*$-algebraic crossed product $\Aa \rtimes_\Halfspaceaction G$ is represented faithfully on $L^2(\hat{G} \times \bbZ^d \times \Omega)$ using an integrated representation $\hat{\pi}_{\Tt,G}$, which is densely defined on the generators by mapping $a f(D_\Halfspaceaction ) \mapsto (\one \otimes \pi_\Tt(a)) f(\hat{X}_\Halfspaceaction)$ for all $a\in \Aa $, $f\in C_0(\hat{G})$. It is precisely this representation which extends to a normal representation of $\Nn_{\Halfspaceaction}$ as the following holds: 

\begin{proposition}
\label{prop:rational_rep} 
The integrated representation $\hat{\pi}_{\Tt,G}$ of $\Aa  \rtimes_\Halfspaceaction G$ on $L^2(\hat{G} \times \bbZ^d \times \Omega)$ defined on the generators by
\begin{equation}
\label{eq:rep_irrational1}
\hat{\pi}_{\Tt,G}\big(a f(D_\Halfspaceaction )\big)
\;=\; \pi_\Tt(a)  f(\hat{X}_\Halfspaceaction) \;=\;
\pi_\Tt(a) \int_{\hat{G}}^\oplus \hat{\mu}(\difd{r})\,f(X_\Halfspaceaction+ r)
\end{equation}
can be extended to a faithful normal representation of the von Neumann crossed product $\Nn_{\Halfspaceaction}$. The operators in the representation are fibered
$$
\hat{\pi}_{\Tt,G}\big(\hat{b})
\;=\;
\int^\oplus_{\hat{G}} \hat{\mu}(\difd{r}) \int^\oplus_\Omega \PM(\difd\omega)\, \hat{\pi}_{\Tt,G}(\hat{b})_{\omega,r}
\;,
\qquad
\hat{b}\in\Nn_{\Halfspaceaction} \;,
$$
with fibers $\hat{\pi}_{\Tt,G}(\hat{b})_{\omega,r}$ acting on $\ell^2(\ZM^d)$. If $\hat{h} \in \Nn_{\Halfspaceaction}$ is self-adjoint and $f$ a bounded Borel function on $\bbR$, then almost surely w.r.t. the product measure of $\Omega$ and the Haar measure on $\hat{G}$
$$
\hat{\pi}_{\Tt,G}\big(f(\hat{h})\big)_{\omega,r}
\;=\;
f(\hat{\pi}_{\Tt,G}\big(\hat{h})_{\omega,r}\big)
\;.
$$ 
\end{proposition}

\noindent {\bf Proof.}
Recall from Section~\ref{sec-WStar} that the crossed product is defined in the regular representation $(\pi,U)$ on the Hilbert space $L^2(G, \Hh_\Tt)$, namely
$$
(\pi(a) \phi)(t)\; =\; \pi_\Tt(\Halfspaceaction_{-t}(a)) \phi(t)\;,  
\qquad
(U(s) \phi)(t) \;=\; \phi(t-s)
\;,
$$
where $\phi\in L^2(G, \Hh_\Tt)$, $a\in \Mm $, $s,t\in G$. Now let us introduce another representation $(\pi',U')$ on $L^2(G,\Hh_\Tt)$ by setting
$$
(\pi'(a)\phi)(t) \;=\; \pi_\Tt(a) \phi(t)
\;,
\qquad
(U'(s)\phi)(t) \;=\; V(s)\, \phi(s+t)
\;
$$
with $V(s)=e^{2\pi\imath\, X_\Halfspaceaction s}$ the unitary representation of $G$ on $\Hh_\Tt$.
The von Neumann algebra generated by these operators is denoted by $\mathfrak{R}(\Mm ,G)$ and it is spatially isomorphic to $\Mm  \rtimes_\Halfspaceaction G$ since the generators are mapped bijectively to the usual generators \eqref{eq-RegRepr} of $\Mm  \rtimes_\Halfspaceaction G$:
$$
W\pi'(a)W \;=\; \pi(a)
\;,
\qquad
W\,U'(s)\,W \;=\; U(s)
\;,
$$
where $W$ is the involutive unitary on $L^2(G,\Hh_\Tt)$ given by
$$
(W\phi)(t) \;=\; V(-t)\,\phi(-t)
\;.
$$

Let now further $\Ff:L^2(G, \Hh_\Tt)\to L^2(\hat{G}, \Hh_\Tt)$ be the unitary Fourier transform and consider the algebra $\calF\,\mathfrak{R}(\Mm ,G)\,\calF^*$ acting on $L^2(\hat{G}, \Hh)$. On this Hilbert space, the generators act on $\hat{\psi}\in L^2(\hat{G},\Hh)$ as
\begin{align*}
(\calF\, \pi'(a)\,\calF^*)(\hat{\psi}) 
&
\;=\; \pi'(a)\hat{\psi}
\;=\; 
\pi_\Tt(a)\hat{\psi}
\;, 
\\
(\calF \,U'(s)\,\calF^*)(\hat{\psi})(r) 
&
\;=\; 
e^{2\pi\imath\,r s} V(s)\, \psi(r)
\;, 
\end{align*}
where $r \in \hat{G}$. This shows that the generators   decomposable into directs integrals with
$$ 
\calF\, \pi'(a)\,\calF^* 
\;=\; 
\int_{\hat{G}}^\oplus \hat{\mu}(\difd{r})\, \pi_\Tt(a)
\;, 
\qquad 
(\calF \,U'(s)\,\calF^*)
\;=\; 
\int_{\hat{G}}^\oplus \hat{\mu}(\difd{r}) \,e^{2\pi\imath r s} V(s)
\;.
$$
Noting that the fibers of $V(s)$ are precisely $e^{2\pi\imath\, (X_\Halfspaceaction+r) s}$, one can extend the representation of the $C^*$-crossed product simply by setting $\hat{\pi}_{\Tt,G}= \calF\, W( \pi\times U) W\,\calF^*$. Since it is unitarily equivalent to the regular representation, $\hat{\pi}_{\Tt,G}$ is obviously faithful.

\vspace{.1cm}

Next let us come to the decomposition of $\hat{\pi}_{\Tt,G}(\hat{b})$. Every generator of the crossed product has a direct integral decomposition w.r.t. $(\omega,r)$ and since the decomposable operators form a von Neumann algebra \cite[Theorem IV.8.18]{Takesaki2001} weak limits of decomposable operators are also decomposable. Therefore all operators in $\mathfrak{R}(\Mm ,G)$ are decomposable w.r.t. $(\omega,r)$.

\vspace{.1cm}

Finally for all $\psi = \int^\oplus_{\Omega}   \bbP(\difd\omega)  \,\psi_{\omega}\in \ell^2(\bbZ^d) \otimes L^2(\Omega \times \hat{G})$ one has
\begin{align*}
\langle \psi| \hat{\pi}_{\Tt,G}(f(\hat{h})) \psi\rangle 
&
\;=\; 
\langle \psi| f(\hat{\pi}_{\Tt,G}(\hat{h})) \psi\rangle 
\\
&
\;=\;
\int_{\Omega\times \hat{G}} 
\langle \psi_{\omega,r}| f(\hat{\pi}_{\Tt,G}(\hat{h})_{\omega,r}) \psi_{\omega,r} \rangle
\; \bbP(\difd\omega)\,\hat{\mu}(\difd{r})\;,
\end{align*}
where the first equality uses that $\hat{\pi}_{\Tt,G}$ is a normal representation and the second follows from dominated convergence applied to a sequence of polynomials converging pointwise to $f$. Hence the fibers  of $\hat{\pi}_{\Tt,G}(f(\hat{h}))$ are almost surely equal to $f(\hat{\pi}_{\Tt,G}(\hat{h})_{\omega})$.
\hfill $\Box$

\vspace{.2cm}

The representation $\hat{\pi}_{\Tt,G}$ is the main representation of $\Nn_{\Halfspaceaction}$ that will be used, so let us discuss the physical representations recovered from it. For example, let $a \in \Aa $ be a bulk observable which we want to restrict to the half-plane $x\cdot \Halfspaceunitvector > 0$ using Dirichlet boundary conditions. Under the above representation, the restriction is described by a measurable family $(\hat{a}_{\omega,r})_{(\omega,r)\in \Omega\times \hat{G}}$ of operators, which themselves act on the physical Hilbert space $\ell^2(\bbZ^d)$. The fibers are given by
\begin{equation}
\label{eq-DirchletRestriction}
\hat{a}_{r,\omega}
\;=\; 
\chi(X_\Halfspaceaction+ r >0) \,\pi_\omega(a) \,\chi(X_\Halfspaceaction + r > 0)
\;,
\end{equation}
and interpreting $r$ as the orthogonal distance of the boundary w.r.t. the lattice point $0 \in \bbZ^d$, the element $\hat{a}$ takes all possible ways to cut a boundary hypersurface with normal vector $\Halfspaceunitvector$ into account. Passing to more general elements $\hat{a}\in \Nn_{\Halfspaceaction}$, one still has a direct integral decomposition $\hat{a} = \int^\oplus_{\hat{G}\times\Omega } \hat{\mu}(\difd{r})\, \bbP(\difd{\omega})\,\hat{a}_{\omega,r}$ into physical operators, however, the lack of continuity means that the evaluation of a single fiber $a_{\omega,r}$ is not sensible in general. At best, one can use ergodicity to make statements that hold almost surely in $(\omega,r)$. This is completely analogous to the way disorder is incorporated in the bulk.

\vspace{.2cm}

In the rationally dependent case the dual group is discrete and hence it is possible to work directly with a representation on $\Hh_\Tt$ instead:

\begin{corollary}
\label{cor:rational_case}
Let $\Halfspaceunitvector$ be rationally dependent, $\Lambda_\Halfspaceaction$ the smallest positive value of $\bbZ^d\cdot \Halfspaceaction$ and further $G=\Lambda_\Halfspaceaction^{-1}\bbT$ and $\hat{G}=\Lambda_\Halfspaceaction \bbZ$. Then the representation $\hat{\pi}_{\Tt,G}$ decomposes into a direct sum $\hat{\pi}_{\Tt,G}= \bigoplus_{r \in \Lambda_\Halfspaceaction \bbZ} \hat{\pi}_r$ and the summand $\hat{\pi}_0$ is a faithful normal representation of $\Nn_{\Halfspaceaction}$ that extends the representation $\hat{\pi}_\Tt$ of $\Aa  \rtimes_\Halfspaceaction G$.
\end{corollary}

\noindent{\bf Proof. }
Since $\hat{G}$ is discrete the direct integral decomposition of the generators of $\mathfrak{R}(\Mm ,G)$ w.r.t. the counting measure of $\hat{G}$ is a direct sum and hence one can project to the summands using bounded projections $P_r$. Thus $\hat{\pi}_{\Tt,G} = \bigoplus_{r \in \Lambda_\Halfspaceaction \bbZ} P_r \hat{\pi}_{\Tt,G} P_r$ decomposes into a direct sum of normal representations. Set $\hat{\pi}_r=P_r \hat{\pi}_{\Tt,G} P_r$. It is straightforward to see that $\hat{\pi}_0$ is precisely the extension of $\hat{\pi}_\Tt$. It remains to prove that $\hat{\pi}_0$ is faithful. Choose some $x\in \bbZ^d$ with $x\cdot \Halfspaceunitvector =\Lambda_\Halfspaceaction$ and note that the translation action $T$ on $\Omega$ is implemented unitarily on $L^2(\Omega)$ by 
\begin{equation}
\label{eq:unitaryOmega}
(U_x \psi)(\omega) 
\;=\; 
\psi(T_x\omega)
\;, 
\qquad \psi \in L^2(\Omega)
\;.
\end{equation}
With the specific $x$ chosen above and the magnetic translation $u^{x} = u_1^{x_1}\cdots u_d^{x_d}$,
one then has the covariance relation
$$
u^x \pi(a) (u^x)^* 
\;=\; 
U_x \,\pi(a)\, U_x^*
\;,
\qquad
a\in \Mm \;,
$$
and the identity
\begin{align*}
u^x \,V_r(s) \,(u^x)^* 
&
\;=\; 
u^x e^{ 2\pi\imath(X_\Halfspaceaction+r)s} (u^x)^* 
\\
&
\;=\; 
e^{2\pi\imath(X_\Halfspaceaction+r-\Lambda_\Halfspaceaction)s} 
\\
&
\;=\;
V_{r-\Lambda_\Halfspaceaction} (s)
\\
&
\;=\; 
U_x\,V_{r-\Lambda_\Halfspaceaction} (s)\, U_x^*
\;,
\end{align*}
which follows from the relations $u^y (\Halfspaceunitvector\cdot X) (u^{y})^* = \Halfspaceunitvector\cdot X-(\Halfspaceunitvector\cdot y)\one$ and $U_x X = X U_x$ on $\Hh_\Tt$.
Hence the representations $\hat{\pi}_0$ and $\hat{\pi}_{n\Lambda_\Halfspaceaction}$ are spatially isomorphic through conjugation with the unitary $(v_{x}^*u^{x})^n$ for any $n\in\ZM$. Therefore the representation $\hat{\pi}_{\Tt,G}$ is unitarily equivalent to countably many copies of $\hat{\pi}_0$ and hence the latter must be faithful as well. 
\hfill $\Box$

\vspace{.2cm}

We denote the extension of $\hat{\pi}_\Tt$ by the same letter, however, for notational convenience most results will still be formulated  in terms of $\hat{\pi}_{\Tt,G}$ such that no condition on $G$ is needed. One just recalls that a property that holds almost surely w.r.t. the Haar measure of $\hat{G}=\Lambda_\Halfspaceaction \bbZ$ is one that holds deterministically. 

\vspace{.2cm}

Next let us further investigate the dual trace $\hat{\Tt}_\Halfspaceaction$ induced on $\Nn_{\Halfspaceaction}$. It is constructed in Section~\ref{sec-DualTraces} and can be calculated on generators  $\pi(a) f(D_\Halfspaceaction )$ as stated in Proposition~\ref{prop-DualTraceCalc}. The normalizations for $\hat{G}=\bbR$ and $\hat{G}=\Lambda_\Halfspaceaction\ZM$ follow from those of the Haar measures and are such that, respectively,
\begin{equation}
\label{eq-ThatForm}
 \hat{\Tt}_\Halfspaceaction(f(D_\Halfspaceaction )) \;=\; \int_{\bbR} f(x)\, \difd{x}
\;, 
\qquad
\hat{\Tt}_\Halfspaceaction(f(D_\Halfspaceaction )) \;=\; \Lambda_\Halfspaceaction \sum_{k \in \bbZ} f(k{\Lambda_\Halfspaceaction})
\;,
\end{equation}
for $f\in C_c(\bbR)$. Note that the two expressions coincide for $\Lambda_\Halfspaceaction^{-1}\to 0$, which is appropriate considering that physical quantities computed from the dual trace should eventually be as continuous as possible w.r.t. $\Halfspaceunitvector$. In the following we will usually omit the subscript $\Halfspaceaction$ on $\hat{\Tt}_\Halfspaceaction$, since it is the only trace on $\Nn_{\Halfspaceaction}$ that will be used. Next let us show that the dual trace can be calculated almost surely in a single fiber of $\hat{\pi}_{\Tt,G}$, namely as a trace per unit volume along the boundary $\Halfspaceunitvector\cdot\RM^d$ combined with the usual trace in the direction perpendicular to the boundary. This is similar as in \cite{KRS,PSbook} and is the key to converting algebraic statements about elements of $\Nn_{\Halfspaceaction}$ into almost sure statements about physical observables. Let us first make this explicit for rational $\Halfspaceunitvector$:

\begin{proposition}
\label{prop-rationaltrace}
Let $d\geq 2$ and assume that $\Halfspaceunitvector$ has rationally dependent components so that $G=\Lambda_\Halfspaceaction^{-1}\bbT$ and $\hat{G}=\Lambda_\Halfspaceaction \bbZ$.  Define the cubic strips
$$
C_N
\;=\; 
\{x \in \bbZ^d\;:\; \lVert x - (\Halfspaceunitvector\cdot x) \Halfspaceunitvector \rVert_\infty \leq N\}
\;
$$ 
and for $\hat{a}\in \Nn_{\Halfspaceaction} \cap L^1(\Nn_{\Halfspaceaction})$ the trace per surface area by 
\begin{equation}
\label{eq:tr_surfarea_rational}
\hat{\Tt}_{\omega,r}(\hat{a}) 
\;=\; 
\lim_{N \to \infty} \frac{1}{(2N)^{d-1}} \sum_{x \in C_N} \langle x | \hat{\pi}_{\Tt,G}(\hat{a})_{\omega,r} | x \rangle
\;.
\end{equation}
If $(\Omega, T, \bbP)$ is strong mixing then $\hat{\Tt}_{\omega,r}(\hat{a}) =\hat{\calT}_\Halfspaceaction(\hat{a})$ holds for $\PM$-almost all $\omega \in \Omega$ and all $r\in \hat{G}$. 
\end{proposition}

\noindent {\bf Proof.}
By polarisation it is enough to prove the statement for elements of the form $\hat{a}= \hat{b}^*\hat{b}$ with $\hat{b}\in L^2(\Nn_{\Halfspaceaction})\cap\Nn$. By Lemma \ref{lem-L2BoundedRep} and Lemma \ref{eq-L2rep_integrated} there exists a function $f\in L^2(G, L^2(\Mm ))$  such that
\begin{equation}
\label{eq-bDecompHelp}
\hat{b} 
\;=\; 
\int_{G} \pi(f(t))\, U(t)\,\mu(\difd{t})
\;,
\end{equation}
with $(\pi,U)$ the regular covariant representation induced by $\pi_\Tt$, $\Halfspaceaction$ and the integral convergences in the sense that is described in Lemma \ref{eq-L2rep_integrated}. Noting our conventions for the Haar measures and the construction of $\hat{\Tt}_\Halfspaceaction$ in Proposition~\ref{prop-DualTrace} the trace is then given by 
\begin{align*}
\hat{\Tt}_\Halfspaceaction(\hat{a}) 
\;&=\; 
\int_{0}^{\Lambda_\Halfspaceaction^{-1}} \Tt(f(t)^*f(t)) \,\mu(\difd{t}) 
\\
\;&=\; 
\Lambda_\Halfspaceaction \sum_{k\in \bbZ} \Tt\big(\hat{f}(k \Lambda_\Halfspaceaction)^*\hat{f}(k \Lambda_\Halfspaceaction)\big)
\\
\;&=\;\Lambda_\Halfspaceaction \int_{\Omega} \sum_{k\in \bbZ} \langle 0| \pi_\Tt(|\hat{f}(k \Lambda_\Halfspaceaction)|^2)_{\omega})| 0\rangle \,\bbP(\difd\omega)
\;,
\end{align*}
where the Parseval identity with $\hat{f} = \mathcal{F} f \in L^2(\bbZ, L^2(\Mm ))$ was used. We thus have to show that \eqref{eq:tr_surfarea_rational} computes the integral of the $L^1(\Omega \times \hat{G})$-function $F(\omega, r)= \langle 0| \pi_\Tt(|\hat{f}(r)|^2)_{\omega})| 0\rangle$. In the representation $\hat{\pi}_{\Tt,G}(\hat{b})$ acting on $\ell^2(\bbZ^d)$ the fibers are given by
$$
\hat{\pi}_{\Tt,G}(\hat{b})_{\omega,r}
\;=\; 
\int_{G} \pi_\Tt(f(t))_\omega \, e^{ 2\pi\imath (X_\Halfspaceaction+r)  t}\, \mu(\difd{t})
\;.
$$
and hence $\bbP$-almost surely
\begin{align*}
\langle x| \hat{\pi}_{\Tt,G}(\hat{a})_{\omega,r} | x\rangle 
& 
\;=\; 
\int_{G^2} \; e^{\imath(\Halfspaceunitvector \cdot x + r)(t_1-t_2)} \langle x| \pi_\Tt(f(t_1)^*f(t_2))_{\omega} | x\rangle\;\mu(\difd{t_1})\mu(\difd{t_2})
\\
&
\;=\; 
\langle x| \pi_\Tt(|\hat{f}(\Halfspaceunitvector \cdot x + r)|^2)_{\omega}\, | x\rangle
\;=\; 
F(T_x\omega, \Halfspaceunitvector \cdot x + r)
\;.
\end{align*}
with the last line using the definition of $\pi_\Tt=\int_\Omega^\oplus \bbP(\difd \omega) \pi_\omega$ of the representation. 

\vspace{.1cm}

Due to its rational dependence there is some positive $k_0\in \bbN$ such that $k_0 \Lambda_\Halfspaceaction \Halfspaceunitvector \in \bbZ^d$. Thus define $V=\{x\in \bbZ^d: \, 0 \leq x\cdot \Halfspaceunitvector < k_0 \Lambda_\Halfspaceaction\}$ and set
$$
G(\omega, r) 
\;=\; 
\sum_{m \in \bbZ} F(T_{m k_0\Lambda_\Halfspaceaction \Halfspaceunitvector}(\omega), m k_0 \Lambda_\Halfspaceaction+r)
$$
for $r \in \hat{G}$. The average defining the trace per unit volume can thus be written
$$
\hat{\calT}_{\omega,r}(\hat{a})
\;=\;  \lim_{N \to \infty} \frac{1}{(2N)^{d-1}} \sum_{x \in C_N \cap V} G(T_x\omega, \Halfspaceunitvector\cdot x + r).
$$
The number of lattice points in integer dilations of a convex lattice polytope can be estimated using different methods such as Ehrhart polynomials and is essentially equal to its (continuous) volume up to lower order terms in its size (see \cite{Beck}). We omit all further details since they lead to the expected result 
$$
\#\abs{C_N \cap V} 
\;=\; 
(2N)^{d-1} (k_0 \Lambda_\Halfspaceaction)\, +\, \Oo(N^{d-2})
\;.
$$
Let us further decompose $V$ into the $k_0$ disjoint slices $S_k = \{x \in \bbZ^d\,: \, \Halfspaceunitvector \cdot x = k{\Lambda_\Halfspaceaction} \}$ with cardinality $\#\abs{C_N \cap V \cap S_k}\sim (2N)^{d-1}\Lambda_\Halfspaceaction$ each such that
\begin{align*}
\lim_{N \to \infty} &\frac{1}{(2N)^{d-1}} 
\sum_{x \in C_N \cap V} G(T_x\omega, \Halfspaceunitvector\cdot x + r) 
\\
&
\;=\; 
\sum_{k=0}^{k_0-1} \lim_{N \to \infty} \frac{\Lambda_\Halfspaceaction}{\#\abs{C_N \cap V \cap S_k}}  \sum_{x \in C_N \cap V \cap S_k} G(T_x\omega, k\Lambda_\Halfspaceaction + r)
\;.
\end{align*}
Noting that $S_k = k x_0 + S_0$ for any $x_0 \in \bbZ^d$ with $x_0\cdot \Halfspaceunitvector=\Lambda_\Halfspaceaction$ the limit for fixed $k$ can be written as a Birkhoff sum w.r.t. $T$ restricted to the subgroup $S_0 \subset \bbZ^d$. The strong mixing implies that the restriction is still ergodic and therefore $\bbP$-almost surely
\begin{align*}
\hat{\calT}_{\omega,r}(\hat{a})
&
\;=\; \Lambda_\Halfspaceaction\sum_{k=0}^{k_0-1} \int_{\Omega} G(\omega, k\Lambda_\Halfspaceaction + r) \, \bbP(\difd\omega) 
\\
&
\;=\; 
\Lambda_\Halfspaceaction\sum_{k=0}^{\infty} \int_{\Omega} F(\omega, k\Lambda_\Halfspaceaction + r) \, \bbP(\difd\omega) 
\\
&
\;=\; 
\hat{\Tt}_\Halfspaceaction(\hat{a})
\;,
\end{align*}
completing the proof.
\hfill $\Box$

\vspace{.2cm}

Next the dual trace shall be computed almost surely by averaging on $\ell^2(\bbZ^d)$ also in the irrational case. Defining the trace per surface area as an average over boundary-aligned cubes as above is possible, but not very convenient, in particular, since even estimating the number of lattice points in such a cube precisely is already cumbersome. 
Let us rather perform the sum along an arbitrary non-parallel lattice direction and then average over the remaining directions. Hence let $a \in L^1(\Nn_{\Halfspaceaction})$ and assume that $\Halfspaceaction_d = e_d\cdot \Halfspaceunitvector \neq 0$, otherwise relabel the coordinate directions. We first perform the sum over the column $\bbZ e_d$ with the other coordinates held fixed:
$$
Z_{\underline{n}}(\hat{a}, \omega,r) 
\;=\; 
\sum_{n_d \in \bbZ} \,
\langle n_d e_d + \underline{n}|\, \hat{\pi}_{\Tt,\RM}(\hat{a})_{\omega,r}\, | n_d e_d + \underline{n}\rangle
\;,
\qquad 
\underline{n} \in \bbZ^{d-1}\times\{0\}\subset\ZM^d
\;,
$$
and then define the trace per unit surface area as the average over all column sums
$$
\hat{\Tt}_{\omega,r}(\hat{a}) 
\;=\; 
\lim_{L\to \infty} \frac{\abs{\Halfspaceaction_d}}{(2L+1)^{d-1}} \sum_{\norm{\underline{n}}_\infty \leq L} Z_{\underline{n}}(\hat{a}, \omega,r)
\;.
$$ 
The factor $\abs{\Halfspaceaction_d}$ will ensure that the overall normalisation does not depend on the choice of lattice directions.
 
\begin{proposition}
\label{prop-irrationaltrace}
Let $\Halfspaceunitvector$ be rationally independent with $\abs{\Halfspaceaction_d} \neq 0$ and suppose that $(\Omega, T, \PM)$ is strong mixing. If $\hat{a} \in \Nn_{\Halfspaceaction} \cap L^1(\Nn_{\Halfspaceaction})$, 
$$
\hat{\Tt}_{\omega,r}(\hat{a}) 
\;=\; 
\hat{\Tt}_\Halfspaceaction(\hat{a})
\;,
$$
for $\mathbb{P}$-almost all $\omega \in \Omega$ and Lebesgue-almost every $r \in \bbR$.
\end{proposition}

\noindent {\bf Proof.} As in the proof of Proposition~\ref{prop-rationaltrace} it is sufficient to consider elements of the form $\hat{a}= \hat{b}^*\hat{b}$ with $\hat{b}\in L^2(\Nn_{\Halfspaceaction})\cap\Nn_{\Halfspaceaction}$ of the form \eqref{eq-bDecompHelp}. Again $\hat{b}$ is of the form \eqref{eq-bDecompHelp} and the Plancherel identity gives 
$$
\hat{\Tt}_\Halfspaceaction(\hat{a}) 
\;=\; 
\int_{\bbR} \Tt(f(t)^*f(t))\;\difd{t}  
\;=\; 
\int_{\bbR} \Tt(\hat{f}(k)^*\hat{f}(k)) \;\difd{k}
\;,
$$
with $\hat{f} = \mathcal{F} f \in L^2(\bbR, L^2(\Mm ))$. Furthermore, 
\begin{align*}
\langle x| \hat{\pi}_{\Tt,G}(\hat{a})_{\omega,r} | x\rangle 
&
\;=\; 
\int_{\bbR^2}  e^{\imath(\Halfspaceunitvector \cdot x + r)(t_1-t_2)} \langle x| \pi_\Tt(f(t_1)^*f(t_2))_{\omega} | x\rangle\;\difd{t_1}\difd{t_2}
\\
&
\;=\; 
\langle x| \pi_\Tt(|\hat{f}(\Halfspaceunitvector \cdot x + r)|^2)_{\omega}\, | x\rangle
\;.
\end{align*}
For convenience we write $g(\omega,k)= \pi_\Tt(|\hat{f}(k)|^2)_\omega$ and the column sum takes the form
$$
Z_{\underline{n}}(\hat{a}, \omega,r) 
\;=\; 
\sum_{n_d \in \bbZ} \langle n_d e_d + \underline{n}|\, g(\omega,n_d \Halfspaceaction_d + \underline{n}\cdot \Halfspaceunitvector +r) | n_d e_d + \underline{n}\rangle\;.
$$
Using the definition of $\pi_\omega$ the column can be shifted to the origin
$$
Z_{\underline{n}}(\hat{a}, \omega,r) 
\;=\; 
\sum_{n_d \in \bbZ} \langle n_d e_d|\, g(T_{\underline{n}}\omega,n_d \Halfspaceaction_d + \underline{n}\cdot \Halfspaceunitvector  +r) | n_d e_d\rangle
\;=\;
Z_{\underline{0}}(\hat{a},T_{\underline{n}}\omega, r+\underline{n}\cdot \Halfspaceunitvector )
\;,
$$
and acting by $T_{e_d}$ shifts the summation index so that
$$
Z_{\underline{n}}(\hat{a}, T_{e_d}\omega,r)  
\;=\; 
Z_{\underline{n}}(\hat{a}, \omega,r - \Halfspaceaction_d)
\;.
$$
This shows that it is enough to prove $\hat{\Tt}_{\omega,r}(\hat{a}) 
=
\hat{\Tt}_\Halfspaceaction(\hat{a})$ for almost every $(\omega, r) \in \tilde{\Omega}:=\Omega\times[0,\abs{\Halfspaceaction_d})$ since one can otherwise redefine $(\omega, r)$ to restrict $r$ to that range.

\vspace{.2cm}

For simplicity let us assume $\Halfspaceaction_d >0$ with trivial modifications in the negative case. We define a $\bbZ^{d-1}\cong\ZM^{d-1}\times\{0\}$ action on as follows: Let $m(\underline{n},r)$ be the unique integer such that
$$
r + \underline{n}\cdot \Halfspaceunitvector - m(\underline{n},r) \;\in\; [0,\Halfspaceaction_d)
$$
holds, and define $S_{\underline{n}}: \tilde{\Omega} \to \tilde{\Omega}$ by
$$
S_{\underline{n}}(\omega,r) 
\;=\; 
(T_{\underline{n}+ m(\underline{n},r)e_d}\omega, r + \underline{n}\cdot {\Halfspaceunitvector} \mod \Halfspaceaction_d)
\;.
$$
The definition is chosen such that 
$$
Z_{\underline{n}}(\hat{a},\omega,r) 
\;=\; 
Z_{\underline{0}}(\hat{a},S_{\underline{n}}(\omega,r))
\;,
$$
for all $\underline{n} \in \bbZ^{d-1}\times\{0\}$ and  $(\omega,r)\in \tilde{\Omega}$. Hence the trace per unit volume $\hat{\Tt}_{\omega,r}(\hat{a})$ is the limit of a Birkhoff sum of $Z_{\underline{0}}$ for this $\ZM^{d-1}$-action. Assuming ergodicity the pointwise convergence to the average for almost all $(\omega,r)$ follows from the $\bbZ^d$-version of Birkhoff's theorem:
\begin{align*}
\hat{\Tt}_{\omega,r}(\hat{a})
&
\;= \;
\int_0^{\Halfspaceaction_d}  \int_\Omega   Z_{\underline{0}}(\hat{a}, \omega',r')\;\mathbb{P}(\difd{\omega'})\; \difd{r'}
\\
&
\;=\;
\int_0^{\Halfspaceaction_d} \int_\Omega   \sum_{n_d \in \bbZ} \langle 0|\, g(\omega',n_d \Halfspaceaction_d + r') | 0\rangle\;\mathbb{P}(\difd{\omega'})\;\difd{r'}
\\
&
\;=\; 
\int_\bbR  \Tt(|\hat{f}(r')|^2)\;\difd{r'} 
\\
&
\;=\; 
\hat{\Tt}_\Halfspaceaction(\hat{a})
\;,
\end{align*}
where the normalization factor $\Halfspaceaction_d$ cancels against the volume of $\tilde{\Omega}$ and translation invariance of the averages is used.

\vspace{.2cm}

Let us therefore check ergodicity and note for this that $S_{\underline{n}}$ coincides with the measurable transformation $S$ from Lemma~\ref{lem-ergodic} for $M=\Halfspaceaction_d$, $\Delta=\underline{n}\cdot \Halfspaceunitvector$, $x_1=\underline{n}$, $x_2=e_d$ in the notation there. It is sufficient for our argument that a single one $S_{\underline{n}}$ is ergodic, and indeed if the sufficient condition $\frac{\Delta}{M}\in \bbR\setminus \bbQ$ were to fail for all $\underline{n}\in \bbZ^{d-1}$ then $\Halfspaceunitvector$ would necessarily be rationally dependent, a contradiction. 
\hfill $\Box$

\vspace{.2cm}

\begin{corollary}
\label{coro-almostsure}
Let $d>1$ and $\hat{h} \in \Nn_{\Halfspaceaction}$ be self-adjoint. Denote by $\hat{e}$ the projection on the kernel $\Ker(\hat{h})$. If $\hat{e}\neq 0$, then the fibers $\hat{h}_{\omega,r}=\hat{\pi}_{\Tt,G}(\hat{h})_{\omega,r}$ have almost surely {\rm (}w.r.t. the product measure of $\PM$ and the Haar measure on $\hat{G}${\rm )} an infinitely degenerate eigenvalue at $0$.
\end{corollary}

\noindent {\bf Proof.}
By the Propositions~\ref{prop-rationaltrace} and \ref{prop-irrationaltrace} one has almost surely
$$
\hat{\Tt}_{\omega,r}(\hat{e}) 
\;=\; 
\hat{\Tt}_\Halfspaceaction(\hat{e}) 
\;>\; 0
\;,
$$
because $\hat{\Tt}_\Halfspaceaction$ is faithful. Hence the almost sure eigenvalue projections $\chi_{\{0\}}(\hat{h}_{\omega,r})= \hat{\pi}_{\Tt,\RM}(\hat{h})_{\omega,r}$ cannot vanish. The eigenvalue must further be infinitely degenerate, since otherwise
$$
\sum_{x\in \bbZ^d} 
\langle x | \hat{\pi}_{\Tt,G}(\hat{e})_{\omega,r}|x\rangle 
\;<\; \infty
\;,
$$
which would imply that the average vanishes.
\hfill $\Box$

\vspace{.2cm}

Starting from Proposition~\ref{prop-rationaltrace} and up to Corollary~\ref{coro-almostsure}, the one-dimensional case was excluded since the dual trace is not self-averaging. It is, however, possible to make pointwise statements that will be helpful in connection with the invariance properties of the index.

\begin{proposition}
\label{prop-dim1fredholm}
Let $d=1$ and thus $\Halfspaceunitvector = \pm e_1$ with $\Nn_{\Halfspaceaction} = L^\infty(\Aa)\rtimes_\Halfspaceaction \bbT$.
For $\hat{a} \in \Nn$, let us denote $\hat{\pi}_{\Tt}(\hat{a})= \int_\Omega^\oplus \bbP(\difd\omega)\,\hat{a}_\omega $, that is, $\hat{\pi}_{\Tt}(\hat{a})_\omega=\hat{a}_\omega$. 
\begin{enumerate}
\item[{\rm (i)}] If $\hat{a}\in L^p(\Nn_{\Halfspaceaction})$, then $\hat{a}_\omega$ is $\PM$-almost surely a $p$-Schatten operator on $\ell^2(\bbZ)$ with $0 < p < \infty$. In particular, for $\hat{a}\in L^1(\Nn_{\Halfspaceaction})$,
\begin{equation}
\label{eq-dim1trace}
\hat{\Tt}_\Halfspaceaction(\hat{a})
\;=\; 
\int_\Omega \Tr(\hat{a}_\omega)\;\bbP(\difd\omega)
\;.
\end{equation}

\item[{\rm (ii)}] If $\hat{a}$ is $\hat{\Tt}_\Halfspaceaction$-compact, then $\hat{a}_\omega$ is $\PM$-almost surely compact.
\item[{\rm (iii)}] If $\hat{a}$ is $\hat{\Tt}_\Halfspaceaction$-Fredholm, then $\hat{a}_\omega$ is $\PM$-almost surely a Fredholm operator on $\ell^2(\bbZ)$ with 
\begin{equation}
\label{eq-dim1fredholm}
\hat{\Tt}_\Halfspaceaction\mbox{-}\Ind(\hat{a}) 
\;=\; 
\int_\Omega  \Ind(\hat{a}_\omega)\;\bbP(\difd\omega)
\;.
\end{equation}
\item[{\rm (iv)}] If $\hat{a}=\hat{u}\abs{\hat{a}} $ is the polar decomposition of $\hat{a}$, then is $\hat{u}$ also a direct integral $\hat{\pi}_\Tt(\hat{u})=\int_\Omega^\oplus \bbP(\difd\omega)\,\hat{u}_\omega $ and $\hat{a}_\omega= \hat{u}_\omega|\hat{a}_\omega|$ is $\PM$-almost surely the polar decomposition of $\hat{a}_\omega$.
\end{enumerate}
\end{proposition}

\noindent {\bf Proof.}
(i) For $p=2$, the same manipulations as in the proof of Proposition \ref{prop-rationaltrace} lead to 
\begin{align*}
\norm{\hat{a}}^2_2 
&
\;=\; 
\hat{\Tt}_\Halfspaceaction(\hat{a}^*\hat{a}) 
\\
&
\;=\; 
\int_\Omega \sum_{n\in \bbZ} \langle n| \hat{a}_\omega^*\hat{a}_\omega |n \rangle\,\bbP(\difd\omega) 
\\
&
\;=\; 
\int_\Omega  \Tr(\hat{a}_\omega^*\hat{a}_\omega) \,\bbP(\difd\omega)
\\
&
\;=\;
\int_\Omega \norm{\hat{a}_\omega}^2_2
\,\bbP(\difd\omega)
\;,
\end{align*}
which shows that $\hat{a}_\omega$ is $\PM$-almost surely a Hilbert-Schmidt operator. Applied to $|\hat{a}|^{\frac{2}{p}} \in L^2(\Nn_{\Halfspaceaction})$ this shows the claim for general $p$ and the formula \eqref{eq-dim1trace} the trace follows again from the existence of factorisations and polarisation.

\vspace{.1cm}

(ii) By density, $\hat{a}= \lim_{n\to \infty} \hat{a}_n$ as a norm limit with $\hat{a}_n \in \Nn_{\Halfspaceaction} \cap L^1(\Nn_{\Halfspaceaction})$. Because the operator norm under direct integrals takes the form
$$
\norm{\hat{a} - \hat{a}_n} 
\;=\; 
\PM\mbox{-}\!\esssup_{\omega \in \Omega} \|\hat{a}_\omega - (\hat{a}_n)_\omega\|
\;,
$$
there is a set of full measure such that $(\hat{a}_n)_\omega$ converges to $\hat{a}_\omega$ in norm. Item (i) now implies that almost all $\hat{a}_\omega$ are norm limits of trace-class operators and thus compact.

\vspace{.1cm}

(iii) This follows from (ii) since $\hat{a}$ has a parametrix $\hat{b}\in \Nn_{\Halfspaceaction}$ such that $\one - \hat{a} \hat{b}$  and $\one - \hat{b}\hat{a} $ are $\hat{\Tt}_\Halfspaceaction$-compact and thus $\one - \hat{a}_\omega \hat{b}_\omega$  and $\one - \hat{b}_\omega\hat{a}_\omega $ are almost surely compact. Identifying $\Ker(\hat{a})$ with the projection onto this subspace, the equation \eqref{eq-dim1fredholm} then follows from 
$$
\hat{\Tt}_\Halfspaceaction\big(\Ker(\hat{a})-\Ker(\hat{a}^*)\big) 
\;=\; 
\int_\Omega \Tr\big(\Ker(\hat{a}_\omega)-\Ker(\hat{a}^*_\omega)\big)\;\bbP(\difd\omega)
\;.
$$

\vspace{.1cm}

(iv) The polar decomposition is given by the strong limit $\hat{u} = \slim_{\epsilon\downarrow 0} \hat{a}(\hat{a}^*\hat{a}+\epsilon)^{-\frac{1}{2}}$ and is therefore also decomposable w.r.t. the direct integral, since the decomposable operators form a von Neumann-algebra. Hence the fibers are almost surely given by $\hat{u}_\omega = \slim_{\epsilon\downarrow 0} \hat{a}_\omega (\hat{a}_\omega^*\hat{a}_\omega+\epsilon)^{-\frac{1}{2}}$, that is, the polar decompositions of $\hat{a}_\omega$.
\hfill $\Box$

\section{Bulk topological invariants}
\label{sec-BulkInv}

The bulk observable algebra of a spatially homogeneous random lattice system with $N$ on-site degrees of freedom consists of weakly  measurable families $(a_\omega)_{\omega\in \Omega}$ of operators $a_\omega$ on $\ell^2(\bbZ^d) \otimes M_N(\bbC)$ described by an element $a \in M_N(\Mm )$. Let us introduce a few conventions to keep the notations as simple as possible. The representation $\pi_\Tt$ extends to matrices and we always identify elements $a \in M_N(\Mm )$ with their representations $a=\int^\oplus_\Omega \PM(\difd{\omega})\,a_\omega$ on $L^2(\bbC^N \times \Omega \times \bbZ^d)\cong\int^\oplus_\Omega \PM(\difd{\omega})\,\ell^2(\ZM^d)\otimes\CM^N$, namely we will throughout work in the GNS representation of $M_N(\Mm )$ w.r.t. the natural trace  $\Tt_N = \Tt \otimes \Tr$. We usually omit the subscript $N$ if it is obvious from the context. We will write $a \in L^p(\Mm )$ as an abbreviation for $a \in L^p(M_N(\Mm ), \Tt_N)$ and likewise for other derived spaces such as the Sobolev or Besov spaces. Summing up, the size and dependence on the matrix degree of freedom will often be suppressed.

\vspace{.2cm}

The time evolution of a concrete fermionic quantum system is fixed by a bulk one-particle Hamiltonian $h = h^* \in M_N(\Mm )$. The ground state of the system is described by the Fermi projection 
$$
p_F \;=\; \chi(h \leq E_F) \;\in\;  M_N(\Mm )
\;,
$$
with the Fermi level $E_F\in \bbR$ being a given real number. While $E_F$ may lie in the spectrum $\sigma(h)$, it will often be necessary to assume that $E_F$ is not an eigenvalue of $h$. This is equivalent to saying that $E_F$ is almost surely not an eigenvalue of $h_\omega$. Some of the results below are obtained in the setting where the Hamiltonian has a spectral gap:

\begin{definition}[{\bf Bulk gap hypothesis (BGH)}]
\label{def-BGH}
The {\rm BGH} is satisfied for a Hamiltonian $h$ if the Fermi level $E_F$ is contained in a spectral gap of $h$ , i.e. there is a compact interval $\Delta$ with $E_F \in \Delta$ and $\Delta \cap \sigma(h)=\emptyset$.
\end{definition}

The topological invariants associated to $h$ are read off the Fermi projection and the passage from $h$ to $p_F$ is often called spectral flattening. The pairing of the Fermi projection with an even Chern character can always be defined, provided $p_F$ is contained in the respective domain, but for an odd Chern character one needs to construct  a unitary operator from $h$. For this, one assumes that $h$ anti-commutes with a self-adjoint unitary $J$. Physically, this corresponds to a so-called chiral or sublattice symmetry of $h$. For chiral systems we will always assume $N$ to be even and use the distinguished self-adjoint unitary 
\begin{equation}
\label{eq-JDef}
J 
\;=\; 
\begin{pmatrix} \one_{\frac{N}{2}} & 0 \\ 0 &  -\one_{\frac{N}{2}}\end{pmatrix}\; \in\; M_N(\bbC)\,\subset\;M_N(\Mm)
\;.
\end{equation}

\begin{definition}[{\bf Chiral hypothesis (CH)}] 
\label{def-CH}
The {\rm CH} holds for $h=h^*$  if  $JhJ=-h$.
\end{definition}

The CH implies that the spectrum satisfies $\sigma(h)=-\sigma(h)$. It is hence natural (and physically reasonable) to fix $E_F=0$ in a system with CH. Since $h$ is then off-diagonal w.r.t. the grading $J$ and if furthermore $0$ is not an eigenvalue of $h$, the Fermi projection takes the form 
$$
p_F 
\;=\; 
\frac{1}{2}(\one_N - \sgn(h)) 
\;=\; 
\frac{1}{2}\begin{pmatrix} \one_{\frac{N}{2}} & -u_F^* \\ -u_F & \one_{\frac{N}{2}}  \end{pmatrix}
\;,
$$ 
where the off-diagonal term is a unitary element $u_F \in M_{\frac{N}{2}}(\Aa )$, called the Fermi unitary operator. 

\vspace{.2cm}

The focus will now be on the index pairings between $p_F$ or $u_F$ and $\Ch_{\Tt,\theta}$ where $\theta$ is a $\bbR^n$-action obtained from the $\bbT^d$-action $\rho$ through $n$ generators given by
\begin{equation}
\label{eq-ThetaAct}
\theta_{t}(a) 
\;=\; 
\rho_{\hat{e}\cdot t}(a)
\;,
\qquad
t\,=\,(t_1,\ldots ,t_n)\,\in\,\RM^n
\;,
\end{equation}
with $\hat{e}=(\hat{e}_1,\ldots ,\hat{e}_n)$ being unit vectors in $\bbR^d$ that are taken to be linearly independent (otherwise the Chern character vanishes trivially). Without restriction, we will assume that $\hat{e}_1,\ldots ,\hat{e}_n$ are orthonormal. If $h$ satisfies the BGH and is smooth in the sense that it is an element of the smooth subalgebra $M_N(\Aa_{\rho, \Tt})$ (as defined in Section~\ref{sec:smooth_chern}), the Fermi projection can then be obtained from $h$ by continuous instead of Borel functional calculus, namely
$$
p_F 
\;=\; 
\chi(h \leq E_F) 
\;=\; 
g(h) 
\;\in \;M_N(\Aa )
\;,
$$ 
where $g\in C^\infty_0(\RM)$ is a suitable smooth approximation of the indicator function below $E_F$. Therefore $p_F$ also lies in $M_N(\Aa)$, determines a class $[p_F]_0 \in K_0(\Aa )$ and a set of even Chern numbers 
$$
\Ch_{\Tt,\theta}(p_F)
\;=\;
\langle\Ch_{\Tt,\theta},[p_F]_0\rangle
\;,
\qquad
n\;\mbox{even}\;.
$$ 
If $n=d$ is even, then the Chern number  is called the strong invariant, while for $n<d$ (with $n$ even, but $d$ either even or odd) the invariant is said to be weak. If now $h$ satisfies not only the BGH, but also the CH, then the Fermi unitary $u_F$ lies in $M_{\frac{N}{2}}(\Aa)$, fixes a class $[u_F]_1 \in K_1(\Aa )$ and leads to a set of odd Chern numbers 
$$
\Ch_{\Tt,\theta}(u_F)
\;=\;
\langle\Ch_{\Tt,\theta},[u_F]_1\rangle
\;,
\qquad
n\;\mbox{odd}
\;.
$$ 
Again if $n=d$ is odd, this invariant is called strong, otherwise weak. The following result showing that $p_F$ and $u_F$ are smooth is by now standard:

\begin{proposition}
\label{prop:spectralgap} 
For a smooth Hamiltonian $h$ satisfying the {\rm BGH}, one has $p_F \in M_N(\Aa_{\Tt,\rho})$ and the even bulk Chern numbers $\Ch_{\Tt,\theta}(p_F)$ are well-defined. If $h$, moreover, satisfies the {\rm CH} also $u_F \in M_{\frac{N}{2}}(\Aa_{\Tt,\rho})$ and the odd bulk Chern numbers $\Ch_{\Tt,\theta}(u_F)$ are well-defined.
\end{proposition}

\noindent{\bf Proof.}
From \cite[Proposition 3.3.4]{PSbook} follows that a self-adjoint $h$ is smooth if and only if  the matrix elements of a smooth $h$ decay faster than any inverse polynomial
$$
\sup_{\omega \in \Omega} \norm{\langle 0| h_\omega| x\rangle} 
\;\leq\; 
C_j \,\frac{1}{1 + |x^j|}
\;,
$$
with $j \in \NM^d$ any multi-index and any norm on $M_N(\bbC)$. A Combes-Thomas-estimate ({\it e.g.} in the form of \cite{Aizenman94}) implies that the resolvent then also decays faster than any inverse polynomial
$$
\sup_{\omega \in \Omega} \norm{\langle 0| \frac{1}{h_\omega - z}| x\rangle} 
\;\leq\; 
\tilde{C}_j(\delta)\, \frac{1}{1 + |x^j|}
\;, 
\qquad 
\forall\; x \in \bbZ^d
$$
for some constants $\tilde{C}_j(\delta)$ and all $z \in \bbC$ with $\mathrm{dist}(z, \sigma(h)) > \delta > 0$. Using the Riesz projection formula for $p_F$, it is obvious that $p_F$ and $u_F$ also have rapidly decaying matrix elements and are thus smooth in $\Aa$.
\hfill $\Box$

\vspace{.2cm}

As investigated in Section~\ref{sec-BreuerToep} and is well-known \cite{BES,PLB,PS,PSbook}, the existence of the Chern numbers does not require norm differentiability w.r.t. the action nor the BGH. Apparently, viewing the Chern numbers as pairings with $K$-groups does not provide much benefit in the absence of a bulk gap, since there are no natural separable $C^*$-algebras associated to those elements. However, if $p_F$ or $u_F$ lie in  a Sobolev space $W_p^1(\Mm)$ for some $p \in (n,n+1]$, then the Sobolev index theorem (Theorem~\ref{theo-Index}) shows that the Chern numbers are well-defined and satisfy
\begin{align}
\label{eq-ChernNumberEven}
\Ch_{\Tt,\theta}(p_F) 
& \;=\;
\;\hat{\Tt}_\theta \mbox{-}\Ind\big(\pi_\theta(p_F) \GG_{x_0} \pi_\theta(p_F) + \one-\pi_\theta(p_F)\big) 
\;,
\qquad n\;\mbox{even}\;,
\\
\Ch_{\Tt,\theta}(u_F)
&
\;=\;
\;-\,\hat{\Tt}_\theta \mbox{-}\Ind\big(\PP  \pi_\theta(u) \PP  + \one-\PP \big) 
\;,
\qquad \qquad \qquad \quad \;\;n\;\mbox{odd}\;,
\label{eq-ChernNumberOdd}
\end{align}
with $\hat{\Tt}_\theta$ the dual trace on $\Mm \rtimes_\theta \bbR^n$ and $\pi_\theta: \Mm \to \Mm \rtimes_\theta \bbR^n$ any regular representation. It is, however, more challenging to verify the required regularity of $p_F$ and $u_F$ if the BGH is dropped and the Fermi level $E_F$ is embedded into the bulk spectrum $\sigma(h)$. The remainder of this section considers sufficient conditions that are appropriate for non-smooth Fermi projections. The main results are summarized in Theorem~\ref{theo-BesovPseudogap} below.

\vspace{.2cm}

The first type of systems are Anderson insulators with a mobility gap that is characterized by the Aizenman-Molcanov estimate \cite{AM}:

\begin{definition}[{\bf Mobility gap regime (MGR)}]
\label{def-MGR}
A self-adjoint Hamiltonian $h \in M_N(\Mm )$ has a mobility gap in the open spectral interval $\Delta$, if for some $s \in (0,1)$ and every $\delta > 0$ there are $A_s(\delta)$ and $ \beta_s(\delta) >0$ such that
\begin{equation}
\label{eq:exploc}
\int_{\Omega} 
\,\norm{ \langle 0| \frac{1}{h_\omega-z}| x \rangle}^s \;\bbP(\difd\omega)
\;\leq\; 
A_s(\delta)\, e^{-\beta_s(\delta) \lvert x \rvert}
\end{equation}
holds uniformly for all $x\in \bbZ^d$ and $z \in \bbC \setminus \bbR$ with $\mathrm{dist}(z, \sigma(h)\setminus \Delta) > \delta > 0$. If the Fermi energy $E_F$ lies in a mobility gap $\Delta$, the system is said to be in the {\rm MGR}.
\end{definition}

A standard result which goes back to \cite{AizenmanGraf} shows that if the Fermi level lies in a mobility gap then the Fermi projection also has exponentially decaying matrix elements:

\begin{proposition}
\label{prop-MBG}
If $h$ has a mobility gap in $\Delta$, then no $E \in \Delta$ is an eigenvalue of $h$ {\rm (}namely of $h_\omega$ for a set of $\omega\in\Omega$ of positive $\PM$-measure{\rm )} and for $E_F\in \Delta$ the Fermi projection is exponentially localized in the sense 
\begin{equation}
\label{eq:exploc2}
\int_{\Omega} 
\,\norm{ \langle 0|(p_F)_\omega| x \rangle} \;\bbP(\difd\omega)
\;\leq\; 
\tilde{A}\, e^{-\beta \lvert x \rvert}
\end{equation}
for all $x\in \bbZ^d$. In particular, $p_F$ is an element of any Besov space $B^{\tilde{s}}_{q,p}(\Mm)$ and any Sobolev space $W^m_p(\Mm)$ for all $0< \tilde{s}<\infty$, $1 \leq p,q < \infty$ and $m\geq 0$.  If $h$ satisfies the {\rm CH}, then the decay \eqref{eq:exploc2} holds for the Fermi unitary $u_F$ which then also lies in the same Besov and Sobolev spaces.
\end{proposition}

For the proof and for further use in the following let us introduce approximations by analytic functions, which are related to Stone's formula for spectral projections:

\begin{lemma}
\label{lemma:contours}
Let $h$ be a bounded self-adjoint operator and $\Cc^{\pm}_\epsilon$ be the piecewise-linear contour in $\bbC$ connecting $(\imath \epsilon, \imath,  \imath  \pm(\norm{h}+1), -\imath  \pm (\norm{h}+1), -\imath, -\imath \epsilon)$ which lies to the left respectively to the right of the spectrum $\sigma(h)$. Setting 
$$
\chi_\epsilon(h)
\;=\;
\frac{1}{2}\,-\,\frac{1}{\pi} \arctan\left(\frac{1}{\epsilon} \,h\right) 
\;,
\qquad
\sgn_\epsilon(h) 
\;=\; 
\frac{2}{\pi} \arctan\left(\frac{1}{\epsilon} \,h\right)
\;,
$$
one has
$$
\chi_\epsilon(h)
\;=\; 
\frac{-1}{2\pi \imath}\int_{\Cc^-_\epsilon} \frac{1}{h-z} \,\difd{z}
\;,
\qquad
\sgn_\epsilon(h) 
\;=\; 
\frac{1}{2\pi \imath}\sum_{\sigma \in\{-,+\}} \int_{\Cc^\sigma_\epsilon} \frac{1}{h-z} \,\difd{z}
\;.
$$
Hence 
$$
\slim_{\epsilon\downarrow 0}\; \chi_\epsilon(h)
\;=\; 
\chi(h < 0) \,+\, \frac{1}{2}\,\chi(h=0)
\;, 
\qquad 
\slim_{\epsilon\downarrow 0} \;\sgn_\epsilon(h)
\;=\;
\sgn(h)
\;.
$$
\end{lemma}

\noindent{\bf Proof.}
Applying the spectral representation of $h$ it is sufficient to show 
\begin{align}
\label{eq-IntegralId}
\int_{\Cc^-_\epsilon} \frac{1}{E-z} \,\difd{z} 
&
\;=\; 
-\,\pi \imath \,+\,
2\imath \arctan\left(\frac{E}{\epsilon} \right), \qquad \forall\; E \in \RM
\;.
\end{align}
For $E\leq 0$ one can deform the contour to a circular arc with center in $E$ and radius $R$ such that $\pm \imath \epsilon = E+R e^{\pm\imath \theta}$, namely $\tan(\theta)=\frac{\epsilon}{\abs{E}}$. This path is explicitly given by $t\in[\theta,2\pi-\theta]\mapsto E+Re^{\imath t}$ so that
\begin{align*}
\int_{\Cc^-_\epsilon} \frac{1}{E-z}\, \difd{z} 
&
\;=\;
\int_{\theta}^{2\pi -\theta}(-\imath) \,\difd{t} 
\\
&
\;=\; 
-\,2\pi \imath \,+\, 2\imath \arctan\left(\frac{\epsilon}{\abs{E}} \right)
\\
&
\;=\; 
-\,\pi \imath \,-\, 2\imath \arctan\left(\frac{\abs{E}}{\epsilon} \right)
\;,
\end{align*}
where the last step follows from the trigonometric identity 
$$
\arctan\left(\frac{1}{y}\right)
\;=\; 
\frac{\pi}{2}\,\sgn(y) \,-\,  \arctan(y)
\;.
$$
This implies \eqref{eq-IntegralId} for $E\leq 0$. For $E > 0$ the contour can be taken as the straight line $t\in[-\epsilon,\epsilon]\mapsto -\imath t$ and hence
$$
\int_{\Cc^-_\epsilon} \frac{1}{E-z}\, \difd{z} 
\;=\;
\int_{-\epsilon}^{\epsilon} \frac{1}{E+\imath t}\,(-\imath)\, \difd{t} 
\;=\;
-2\imath \int_{0}^{\epsilon} \frac{E}{E^2+t^2} \,\difd{t} 
\;=\; 
-2\imath\, \arctan\left(\frac{\epsilon}{E} \right)
\;,
$$
which by the above trigonometric identity again implies \eqref{eq-IntegralId}.
\hfill $\Box$

\vspace{.2cm}

\noindent{\bf Proof (of Proposition~\ref{prop-MBG}). }
Choose $\delta>0$ with $\mathrm{dist}(E_F,\sigma(h)\setminus \Delta) > \delta$ such that \eqref{eq:exploc} holds for some $s\in (0,1)$. Combined with $\norm{\frac{1}{h_\omega - z}}\leq \frac{1}{\abs{\Im m(z)}}$ one estimates
$$
\int_{\Omega} 
\,\norm{ \langle 0| \frac{1}{h_\omega-z}| x \rangle} \;\bbP(\difd\omega)
\;\leq\; \frac{1}{\abs{\Im m(z)}^{1-s}}
\,A_s(\delta)\, e^{-\beta_s(\delta) \lvert x \rvert}
\;
$$
for $\mathrm{dist}(z,\sigma(h)\setminus \Delta) > 0$. To check that $E \in \Delta$ is not an eigenvalue, it is enough to show that $\mathbb{P}$-almost surely $E$ is not an eigenvalue of $h_\omega$. Choosing $\delta>0$ small enough, the general formula $\chi(h_\omega=E)=\slim_{\epsilon\downarrow 0} \frac{\imath \epsilon}{h_\omega-E + \imath \epsilon}$ implies 
\begin{align*}
\int_{\Omega}  
\,\norm{ \langle x| \chi(h_\omega=E) | y \rangle} \;\bbP(\difd\omega)
&
\;=\;
\int_{\Omega}  
\,\norm{ \langle 0| \chi(h_\omega=E) | y-x \rangle} \;\bbP(\difd\omega)
\\
&
\;\leq\; \lim_{\epsilon\downarrow 0} \epsilon^s 
A_s(\delta)\, e^{-\beta_s(\delta) \lvert y-x \rvert} 
\;=\; 0
\;,
\end{align*}
for all $x,y \in \bbZ^d$ and therefore $\chi(h_\omega=E)=0$ almost surely. Since $E_F$, in particular, is not an eigenvalue one can use the approximations from Lemma~\ref{lemma:contours} to write 
$$
p_F
\;=\;
\slim_{\epsilon\downarrow 0}\, \chi_\epsilon(h-E_F)
\;=\;-\,
\frac{1}{2\pi \imath }\; 
\slim_{\epsilon\downarrow 0}\int_{\Cc^-_\epsilon} \frac{1}{h-E_F - z}\difd{z}.
$$
The contour is chosen such that the spectrum of $h$ is approached only in a neighborhood of $E_F$. For $\delta$ small enough one has by the Lemma of Fatou
\begin{align*}
\int_{\Omega} 
\,\norm{ \langle 0|(p_F)_\omega| x \rangle} \;\bbP(\difd\omega)
\;&\leq\;
\liminf_{\epsilon\downarrow 0} \,\frac{1}{2\pi }\, 
\int_{\Cc^-_\epsilon} 
\int_{\Omega} \norm{ \langle 0|\frac{1}{h-E_F-z}|x \rangle}\;\bbP(\difd\omega) \;\difd z
\;\\
&\leq \;
A_s(\delta)\, e^{-\beta_s(\delta) \lvert x \rvert} \left(
\liminf_{\epsilon\downarrow 0} \,\frac{1}{2\pi }\,
\int_{\Cc^-_\epsilon}  \frac{1}{\abs{\Im m(z)}^{1-s}} \;\difd z\right)
\;\\
&= \;\tilde{A}_s\, e^{-\beta_s \lvert x \rvert}
\;,
\end{align*}
since the limit in brackets exists and is finite due to the integrability of the singularity.

\vspace{.1cm}

For the statement about the Besov spaces note that the triangle inequality implies for $a \in M_N(\Mm)$ with $a = \sum_{x\in \bbZ^d} \sum_{i,j=1}^N a_{x,i,j} (e_{i,j}\otimes u^x)$ with $a_{x,i,j}\in L^\infty(\Omega)$, one has
$$
\norm{a}_p
\;\leq \;
\sum_{x\in \bbZ^d} \sum_{i,j=1}^N \norm{a_{x,i,j} u^x}_p
\;=\; 
\sum_{x\in \bbZ^d} \sum_{i,j=1}^N \left(\int_{\Omega} 
\,\abs{a_{x,i,j}(\omega)}^p \mathbb{P}(\difd \omega)\right)^{\frac{1}{p}}
\;,
$$
and hence the definition of the representation $\pi_\omega$ and formula \eqref{eq:multiplier_noncomtorus} for Fourier multipliers allow to bound
\begin{align*}
\norm{\widehat{W_j}*p_F}_p
\;&\leq\; 
N^2 \sum_{x\in \bbZ^d} \abs{W_j(x)} \left(\int_{\Omega} 
\,\norm{\langle 0|( p_F)_\omega| x\rangle}^p \mathbb{P}(\difd \omega)\right)^{\frac{1}{p}}\\
&\leq N^2 \sum_{x\in \bbZ^d} \abs{W_j(x)} \left(\int_{\Omega} 
\,\norm{\langle 0|( p_F)_\omega| x\rangle} \mathbb{P}(\difd \omega)\right)^{\frac{1}{p}}\;.
\end{align*}
where  $\norm{\langle 0|( p_F)_\omega| x\rangle} \leq \norm{p_F}\leq 1$ was applied for the second inequality.
Recalling $\abs{W_j(x)} \leq 1$ and $W_j(x)=0$ for $\abs{x} \notin [2^{j-1},2^{j+1}]$, this gives
\begin{align*}
\norm{p_F}_{B_{p,1}^{\tilde{s}}} 
&
\;\leq \;
\norm{\widehat{W_0}*p_F}_p \;+\; N^2 \sum_{x\in \bbZ^d} \sum_{j=1}^\infty 2^{j\tilde{s}} \abs{W_j(x)} \left(\int_\Omega \norm{\langle 0|( p_F)_\omega| x\rangle} \mathbb{P}(\difd \omega)\right)^{\frac{1}{p}}
\\ 
&
\;\leq \;
\norm{\widehat{W_0}*p_F}_p \;+ \;N^2 \sum_{x\in \bbZ^d} 3 \abs{4 x}^{\tilde{s}} \left(\int_\Omega \norm{\langle 0|( p_F)_\omega| x\rangle} \mathbb{P}(\difd \omega)\right)^{\frac{1}{p}}
\\ 
&
\;\leq \; 
\norm{\widehat{W_0}*p_F}_p\; +\; N^2 \sum_{x\in \bbZ^d} 3 \abs{4 x}^{\tilde{s}} \tilde{A_s} e^{-\frac{\beta_s}{p}\abs{x}}
\;,
\end{align*} 
which clearly is finite for all $1\leq p < \infty$ and arbitrarily large $\tilde{s}>0$. Proposition~\ref{prop-besov-sufficient}(ii) and the basic inclusions between the Besov spaces then imply $p_F \in B^{\tilde{s}}_{p,q}(\Mm)$ for all $0< \tilde{s}<\infty$, $1 \leq p,q < \infty$. The Sobolev regularity then follows from Lemma~\ref{lem:diff_besov2} which implies the inclusion $B^m_{p,1}(\Mm)\subset W^m_p(\Mm)$ because $\Tt$ is a finite trace.

\vspace{.1cm}

If $h$ satisfies the CH then applying the reasoning above to $\sgn(h)$ shows the analogous statements for $u_F$.
\hfill $\Box$

\vspace{.2cm}

 For a periodic Hamiltonian, {\it i.e.} non-random with vanishing magnetic field, the necessary Besov or Sobolev regularity generically does not hold if the Fermi level is in the interior of the spectrum. Exceptions to this rule can only occur at isolated points in the spectrum at which the density of states vanishes polynomially. Examples of this type in dimension two and three are Weyl and Dirac semimetals or nodal-line semimetals which also give rise to gapless topologically protected phases (see for example \cite{MT,ArmitageEtAl,MatsuuraEtAl}).

\newpage
 
\begin{definition}[{\bf Density of states (DOS), its regularity and pseudogap}]
\label{def-DOS}
For a selfadjoint $h \in \Mm $, the {\rm DOS} measure $\nu_h$ is defined by
$$
\nu_h(I) 
\;=\; 
\Tt(\chi_I(h))
\;,
\qquad
I\subset\RM\;\;\mbox{\rm Borel}\,.
$$
The {\rm DOS} of $h$ is $\gamma$-H\"older continuous at $E_0$ if there is an open interval $I$, $E_0 \in I$ and a constant $C$ such that for all $\epsilon>0$ with $[E_0-\epsilon,E_0+\epsilon]\subset I$
\begin{equation}
\label{eq-dosbound}
\nu_h([E_0-\epsilon, E_0+\epsilon]) \;\leq\; C\,\epsilon^{\gamma}\;.
\end{equation}
If $\gamma > 1$ and $E_0$ is in the interior of the spectrum of $h$, one says that $h$ has a pseudogap at $E_0$ of order $\gamma$.
\end{definition}

Let us note that the normalization is such that $\nu_h(\bbR)=\Tt(\one_N)=N$. Furthermore, \eqref{eq-dosbound} is in particular fulfilled if the DOS measure of $h$ is absolutely continuous with a Radon-Nykodym derivative satisfying 
$$
\abs{\frac{\difd \nu_h}{\difd E}(E)}
\;\leq \;
C\, |E-E_0|^{\gamma-1}
\;.
$$

\begin{proposition}
\label{prop:resolvent_bound} 
Assume that the {\rm DOS} of $h=h^* \in \Mm$ is $\gamma$-H\"older continuous at $E_0$ with the bound \eqref{eq-dosbound} satisfied in the interval $I=[-M,M]$. For $\kappa \in (0,1)$ introduce the set
$$
D_{M,\kappa}
\;=\;
\big\{R e^{\imath\theta} \in \bbC\;:\; \abs{R \cos( \theta)} \leq \tfrac{M}{2}\;\mbox{ and }\; \abs{\cos (\theta)} \leq \kappa
\big\}
\;.
$$

\begin{enumerate}
\item[{\rm (i)}] The inverse $(h-E_0+z)^{-1}$ exists as an element of $L^p(\Mm)$ for all $p\in(0,\gamma)$ and its $L^p$-norm is bounded uniformly for all $z\in D_{M,\kappa}$.
 
\item[{\rm (ii)}] For all $p\in(0,\gamma)$ and $r\in (0,\gamma-p)$, one has
\begin{equation}
\label{eq:resolvent_continuity}
\Big\| \frac{1}{h-E_0} \;-\; \frac{1}{h-E_0 + z} \Big\|_p
\;\leq\; 
K_\kappa \abs{\Im m(z)}^{{s}}
\;
\end{equation}
with ${s}=\min\{\frac{r}{p}, 1\}$ and a constant $K_\kappa$ uniformly for all $z\in D_{M,\kappa}$.
\end{enumerate}
\end{proposition}

\noindent {\bf Proof.}  
The $\gamma$-H\"older continuity implies $\Tt(\chi_{\{E_0\}}(h)) \leq \lim_{\epsilon\downarrow 0}\nu_h([E_0-\epsilon, E_0 + \epsilon]) = 0$ and since $\Tt$ is a faithful trace $E_0$ is therefore not an eigenvalue of $h$. Also let us take $E_0=0$.

\vspace{.1cm}

(i) Since $h$ is injective it has a closed and densely defined inverse, affiliated to $M_N(\Mm )$. The same is true for $\abs{h}$ and $h^{-1} = u\abs{h}^{-1} $ with $u$ the unitary from the polar decomposition of $h$. Hence it is enough to demonstrate $ \abs{h}^{-1}\in L^p(\Mm )$. Its spectral projections for $0 < a < b$ are related to those of $h$ by
$$
\chi_{[a,b]}(\abs{h}^{-1}) 
\;=\; 
\chi_{[\frac{1}{b},\frac{1}{a}]}(\abs{h}) 
\;=\;  
\chi_{[-\frac{1}{a},-\frac{1}{b}]}(h) 
\;+\; \chi_{[\frac{1}{b},\frac{1}{a}]}(h)
\;,
$$
and thus it is, in particular, $\Tt$-measurable (see Appendix~\ref{app-Lp} for the definition). Bounding $\abs{\frac{1}{h}}^p$ in terms of its spectral projections one estimates
\begin{align*}
\Tt\left(\abs{\frac{1}{h}}^p\right)
&\;=\; 
\int_{-\norm{h}}^{\norm{h}} \frac{1}{\abs{\lambda}^p}\,\nu_h(\difd \lambda) 
\\
&\;\leq\; \sum_{k =0}^\infty 2^{kp} \nu_h([-2^{-k}M,-2^{-k-1} M) \cup (2^{-k-1} M, 2^{-k}M]) ) 
\\
&
\;\;\;\;\;\;\;
\;+\; \frac{1}{M^p}\,\nu_h(\bbR \setminus [-M,M])\\
&
\;\leq\; C\;\sum_{k =0}^L 2^{k(p-\gamma)} M^\gamma \;+\; \frac{N}{M^p} \;<\; \infty\;,
\end{align*}
where also $\nu_h(\bbR)=N$ was used. This shows $\frac{1}{h} \in L^p(\Mm)$ and since the trace of an unbounded positive operator is defined by its spectral resolution, one gets for arbitrary $z\in D_{M,\kappa}$
\begin{align*}
\Tt\left(\abs{\frac{1}{h+z}}^p\right)
\;&=\; 
\int_{-\norm{h}}^{\norm{h}} \frac{1}{\abs{\lambda+z}^p}\,\nu_h(\difd \lambda) 
\\
\;&
\leq\; 
\int_{-\norm{h}}^{\norm{h}} \left(1+\frac{\abs{z}}{\abs{\Im m( z)}}\right)^p \frac{1}{\abs{\lambda}^p}\,\nu_h(\difd \lambda) \\
\;&\leq\; 
\int_{-\norm{h}}^{\norm{h}} \left(1+(1-\kappa^2)^{-\frac{1}{2}}\right)^p\, \frac{1}{\abs{\lambda}^p}\,\nu_h(\difd \lambda) \\
\;&\leq\; 
 \left(1+(1-\kappa^2)^{-\frac{1}{2}}\right)^p\,
\Tt\left(\abs{\frac{1}{h}}^p\right)
\;
\end{align*}
where we used $\frac{1}{\lambda+z}=\left(1-\frac{z}{\lambda+z}\right) \frac{1}{\lambda}$ and $\frac{\abs{z}}{\abs{\Im m (z)}}< (1-\kappa^2)^{-\frac{1}{2}}.$ 

\vspace{.1cm}

(ii) In the case $r \geq p$, one has $s=1$ and $2p<\gamma$. Hence by the H\"older inequality
$$
\norm{\frac{1}{h} - \frac{1}{h + z}}_p 
\;=\; 
\abs{z}\,\norm{\frac{1}{h + z}\,\frac{1}{h}}_p
\;\leq\; 
\abs{z}\,\norm{\frac{1}{h + z}}_{2p}\,\norm{\frac{1}{h}}_{2p}
\;,
$$
so that (i) provides the claimed estimate. Now let $r<p$. One then applies to $a\in L^{p+r}(\Mm) \cap \Mm$ and $b\in L^p(\Mm) \cap L^{p+r}(\Mm)$ first the H\"older inequality and then $\log$-convexity \eqref{eq-LogConvex} of the $p$-norms 
\begin{equation}
\label{eq:lp_convexity}	
\norm{ab}_p 
\;\leq\; 
\norm{a}_{p(1+\frac{p}{r})}\, \norm{b}_{p+r}
\;\leq\; 
\norm{a}_{p+r}^{\frac{r}{p}} \,\norm{a}_\infty^{1-\frac{r}{p}}\,  \norm{b}_{p+r}
\;.
\end{equation}
Hence
\begin{align*}
\norm{\frac{1}{h} - \frac{1}{h + z}}_p 
& 
\;=\; 
\abs{z}\,\norm{\frac{1}{h + z}\,\frac{1}{h}}_p
\\
&
\;\leq\; 
\abs{z}\, \norm{\frac{1}{h + z}}^{1-\frac{r}{p}}_{\infty}\norm{\frac{1}{h + z}}^{\frac{r}{p}}_{p+r}\,\norm{\frac{1}{h}}_{p+r} 
\\ 
&
\;\leq\; 
\frac{\abs{z}}{\abs{\Im m (z)}^{1-\frac{r}{p}}}\ \, \norm{\frac{1}{h + z}}^{\frac{r}{p}}_{p+r}\,\norm{\frac{1}{h}}_{p+r}.
\end{align*}
As $\frac{\abs{z}}{\abs{\Im m (z)}}< (1-\kappa^2)^{-\frac{1}{2}}$ for $z\in D_{M,\kappa}$ and $p+r<\gamma$ item (i) completes the proof.
\hfill $\Box$

\vspace{.2cm}

There is a partial converse to Proposition~\ref{prop:resolvent_bound}(i) which shows that the H\"older continuity of $\nu_h$ is captured well by the summability of the resolvent:

\begin{proposition}
\label{prop:res_dos_inverse}
If $\frac{1}{h-E_0} \in L^p(\Mm )$ for a self-adjoint $h \in M_N(\Mm )$ and some $0 < p < \infty$, then $h$ is $p$-H\"older continuous at $E_0$.
\end{proposition}

\noindent {\bf Proof.} 
Again let us set $E_0=0$. One has for any $L>0$
\begin{align*}
&
\left(\sum_{k\in \bbZ} 2^{k p} \,L^{-p} \,\norm{\chi\left(\abs{h} \in (2^{-k-1}L, 2^{-k}L] \right)}_1\right)^{\frac{1}{p}} 
\\ 
& \;\;\;\;
\;=\;\left(\sum_{k\in \bbZ} 2^{k p} \,L^{-p} \,\norm{\chi\left(\abs{\frac{1}{h}} \in [2^{k+1} L^{-1},2^{k} L^{-1}) \right)}_1\right)^{\frac{1}{p}}  
\\ 
& \;\;\;\;
\;\leq \;2  \norm{\frac{1}{h}}_p 
\;,
\end{align*}
and hence
$$ 
\norm{\chi\left(\abs{h} \in (2^{-k-1}L, 2^{-k}L] \right)}_1 
\;\leq\; 
2^{-kp}\, L^p\, 2^p\,  \norm{\frac{1}{h}}^p_p
$$
which implies
$$
\norm{\chi\left(\abs{h} \in [0, L) \right)}_1 
\;\leq\; 
\sum_{k=0}^{\infty} 2^{-kp}\, L^p \,2^p \, \norm{\frac{1}{h}}^p_p 
\;=\; 
\frac{4^p}{2^p-1}  \norm{\frac{1}{h}}^p_p L^p
\;,
$$
completing the proof.
\hfill $\Box$

\vspace{.2cm}

The bound of Proposition~\ref{prop:resolvent_bound} allows to control the resolvent $\frac{1}{h+z}$ at zero well enough to estimate the functional calculus with functions that are not continuous at $E_0$:

\begin{proposition}
\label{prop:pseudogap_sufficient}
If $h$ is norm-smooth and has a pseudogap at $E_F$ of order $\gamma=mp+\delta$ with  $p \geq 1$, $m \in \bbN_+$ and $\delta>0$, then 
$$
p_F\; \in\; W^m_{p}(\Mm).
$$
\end{proposition}

\noindent {\bf Proof.}  
Let us again first note that $E_F$ is not an eigenvalue because the DOS is H\"older continuous. Therefore  the same functional calculus as in Lemma~\ref{lemma:contours}  can be used as in the proof of Proposition~\ref{prop-MBG}. Since the operator norm of the net is uniformly bounded in $\epsilon$ and $\Mm$ has a finite trace, the SOT convergence already implies that the net $\epsilon \mapsto \chi_\epsilon(h)$ convergences to $p_F$ in $L^p$-norm. To show $p_F\in W^m_p(\Mm)$ it is therefore is enough to show that the net $\epsilon \mapsto \chi_\epsilon(h)$ is Cauchy in the norm of $W^m_p(\Mm)$ since its limit must then be $p_F$.  
For $\epsilon > 0$ one has norm-differentiability on any of the curves $\calC_\epsilon^\sigma$ and, setting $E_F=0$ for  simplicity, one can write
\begin{equation}
\label{eq:pseudogap_fncalc}
\nabla^j \chi_\epsilon(h) 
\;=\; 
\frac{1}{2\pi \imath }\,
\int_{\Cc^-_\epsilon} \nabla^j \frac{1}{h-z} \;\difd z
\end{equation}
for any multi-index $j \in \bbN^d$. Using the Leibniz rule and the identity 
$$
\nabla_i \frac{1}{h-z} 
\;=\; 
- \,\frac{1}{h-z}\, (\nabla_i h)\,\frac{1}{h-z}
\;,
\qquad
i=1,\ldots, d
\;,
$$
one can expand $\nabla^j \frac{1}{h-z}$ into a linear combination of products that involve only derivatives of $h$ and at most $\abs{j}+1$ resolvents. Since  $\norm{\nabla^j h}_\infty$ is finite for every multi-index $j$ there are by the H\"older inequality constants $C_i$ depending only on $h$ such that
\begin{align}
\norm{\nabla^j \frac{1}{h-z}}_p 
&
\;\leq\; 
\sum_{i=1}^{\abs{j}+1} C_i \norm{\frac{1}{h-z}}^i_{ip} 
\nonumber
\\
&
\;\leq\; 
\sum_{i=1}^{\abs{j}+1} C_i \norm{\frac{1}{h-z}}^{\frac{p-r}{p}}_{\infty} \norm{\frac{1}{h-z}}_{(i-1)p+r}^{\frac{(i-1)p+r}{p}} 
\nonumber
\\
&
\;\leq\; 
\sum_{i=1}^{\abs{j}+1} C_i \abs{\Im m( z)}^{-\frac{p-r}{p}}  \norm{\frac{1}{h-z}}_{(i-1)p+r}^{\frac{(i-1)p+r}{p}}
\;, 
\label{eq:pseudogap_resolvent_exp2}
\end{align}
where the log-convexity \eqref{eq-LogConvex} of the $p$-norms was used for some $0< r <p$, as well as the standard resolvent estimate in the last step. For $\abs{j}\leq m$ and $0 < r < \min\{\delta,p\}$ the norms on the r.h.s.  of \eqref{eq:pseudogap_resolvent_exp2} are bounded uniformly in $\epsilon$ on any curve $\calC_\epsilon^\sigma$ by Proposition \ref{prop:resolvent_bound}. Hence there is a constant $c$ depending only on $0 < r < \delta$, the norms $\norm{\nabla^j h}_\infty$ and the parameters of the pseudo-gap such that for all $0 < \tilde{\epsilon} <\epsilon  < 1$
\begin{align*}
\norm{\nabla^j(\chi_{\tilde{\epsilon}} - \chi_\epsilon)}_p 
&
\;=\; 
\norm{
\frac{1}{2\pi \imath }\,
\int_{\Cc^-_{\tilde{\epsilon}}\setminus\Cc^-_{\epsilon}} \nabla^j \frac{1}{h-z} \;\difd z}_p 
\\
&
\;\leq\; 
2 \int^\epsilon_{\tilde{\epsilon}} \norm{\nabla^j \frac{1}{h-\imath z}}_p \difd{z} 
\\
&
\;\leq\; 
c (\epsilon^\frac{r}{p}-\tilde{\epsilon}^\frac{r}{p})
\;,
\end{align*}
which allows to conclude $\norm{p_F - \chi_\epsilon(h)}_{W^m_p} \leq c' \epsilon^\frac{r}{p}$.
\hfill $\Box$

\vspace{.2cm}

Using Proposition~\ref{prop:pseudogap_sufficient} as input for Proposition~\ref{prop-besov-sufficient2} also gives the Sobolev regularity required for the index theorem:

\begin{corollary}
\label{coro-BesovPseudogap}
If $h$ has a pseudogap at $E_F$ of order $\gamma = n+\delta$ with $n\geq 1$ and $\delta >0$, then $p_F, u_F \in W^1_{n+\frac{1}{2}\delta}(\Mm) $. 
\end{corollary}

Let us now present a rich class of examples of periodic Hamiltonians which generically have pseudogaps (they were also studied by \cite{MT}). The magnetic field is supposed to vanish $\BB=0$ and we also assume that there is no disorder, {\it i.e.} $\Omega$ has only a single element. Then $\Aa_d \cong \bbC \rtimes \bbZ^d = C(\bbT^d)$ with the isomorphism given by the Fourier transform that maps $u^x$ to the function $k \mapsto e^{2\pi \imath k\cdot x}$. Let $h \in M_{N}(\bbC)\otimes C(\bbT^d)$ be given in the form
\begin{equation}
\label{eq:dirac_weyl}
h(k) 
\;=\; 
\sum_{j=1}^d \gamma_j h_j(k)
\;,
\end{equation}
with $\gamma_1,\dots,\gamma_d$ being a representation of the Clifford matrices as in \eqref{eq:CliffGen} and $h_1,\dots, h_d \in C(\bbT^d)$ real-valued trigonometric polynomials. We note that $h(k) = \sum_{j=1}^d h_j(k)^2$ and assume that $h^2(k)$ has at most isolated simple zeroes (counting degrees of freedom suggests that this is in some sense almost surely the case). Labeling the zeroes $k^{(1)},\dots, k^{(m)}$, one then finds 
$$
h(k^{(l)}+k) 
\;\sim \;
\sum_{j=1}^d \gamma_j a^{(l)}_j k_j 
\;+\; \Oo(k^2)
$$
for some coefficients $a^{(l)}_j\in\CM$. Such band-touching points in momentum space with a linear dispersion are called Dirac or Weyl points for $d$ even or $d$ odd respectively. If the spectrum around $0$ is given purely by Dirac- or Weyl points the integrated density of states scales as 
\begin{equation}
\label{eq-DWScaling}
\nu_h([-\epsilon,\epsilon]) \;\sim\; \epsilon^{d}
\;,
\end{equation}
and hence by Corollary~\ref{coro-BesovPseudogap}
all Chern numbers with $n<d$ are well-defined. An explicit example for a two-dimensional chiral Dirac-semimetal with non-trivial weak Chern numbers will be given in Section~\ref{sec-Graphene}.

\vspace{.2cm}

In higher dimensions there is also the possibility  that energy bands meet not just in isolated points, but {\it e.g.} on some $(d-2)$-dimensional manifold. For $d=3$ such a manifold is often called a nodal line and generically occurs in the so-called non-centrosymmetric nodal superconductors \cite{Schnyder2012}. In this case, one again has a density of states that vanishes rapidly enough as long as the dispersion around the Fermi energy is linear in the directions perpendicular to the nodal line and may also have non-vanishing weak invariants.

\vspace{.2cm}

Let us again sum up the sufficient criteria given by the Propositions~\ref{prop:spectralgap}, ~\ref{prop-MBG} and \ref{prop:pseudogap_sufficient}:

\begin{theorem}
\label{theo-BesovPseudogap}
The even weak Chern numbers $\Ch_{\Tt,\theta}(p_F) $ associated to the $\RM^n$-action $\theta$ given in \eqref{eq-ThetaAct} are well-defined and given by \eqref{eq-ChernNumberEven} if either of the following conditions hold:
\begin{enumerate}
\item[{\rm (i)}] $E_F$ lies in a spectral gap of $h$.
\item[{\rm (ii)}] $E_F$ lies in a mobility gap of $h$.
\item[{\rm (iii)}] $h$ has a pseudogap at $E_F$ of order $\gamma = n+\delta$ with $n\geq 1$ and $\delta >0$.
\end{enumerate}
If $h$ satisfies, moreover, the {\rm CH}, then the odd Chern numbers $\Ch_{\Tt,\theta}(u_F) $ are well-defined and given by \eqref{eq-ChernNumberOdd}.
\end{theorem}

It is important to note that adding disorder to a periodic model like \eqref{eq:dirac_weyl} will usually eliminate the pseudogap (at least unless the disorder respects a chiral symmetry). For example, the lower Wegner estimate (similar as in \cite{HM2008}) implies that for sufficiently regular diagonal disorder the density of states has a strictly positive density everywhere inside the spectrum. Therefore  methods for establishing well-definedness of the Chern numbers that are based on the density of states likely do not apply in the disordered case. Moreover, it is a wide-open question to prove or disprove localization for the states that fill the pseudogap. For three-dimensional Weyl semimetals on a lattice (thus with two Weyl points) there are indications for delocalization \cite{AB,ArmitageEtAl}. Furthermore another general open question is how much regularity and decay the Fermi projections of disordered lattice models possess outside the regime of exponential localization. 

\vspace{.2cm}

The index theorem implies that the Chern numbers are constant on any norm continuous path of projections and unitaries. Note, however, that in the absence of a bulk gap, small perturbations of the Hamiltonian do not result in norm continuous perturbations of the Fermi projection, but only in strongly continuous perturbations. Nevertheless, it is possible to obtain a fairly general continuity result that still applies in this case:

\begin{proposition}
\label{prop-WeakInvCont}
Let $t \in [0,1] \to h_t \in  M_N(\Mm )$ be a {\rm SOT}-continuous path of Hamiltonians such that the Fermi projections $p_{F,t}$ are in $W^1_{n+\epsilon}(\Mm ,\theta)$ for some $ \epsilon > 0$ with $\norm{p_{F,t}}_{W^1_{n+\epsilon}} < C$ uniformly bounded for all $t \in [0,1]$. If $E_F$ is not an eigenvalue of any $h_t$ and the {\rm CH} holds pointwise in the odd case, then $t\mapsto \Ch_{\Tt,\theta}(p_{F,t})$ respectively $t\mapsto \Ch_{\Tt,\theta}(u_{F,t})$ are continuous functions for $n$ even or odd respectively.
\end{proposition}

\noindent{\bf Proof.}
Since $E_F$ is not an eigenvalue, the spectral projection $p_{F,t}$ changes continuously in the SOT \cite[Theorem VIII.24]{ReedSimon1}. Hence $t \mapsto \|\widehat{W_j}*p_{F,t}\|_{n+1}$ is also continuous. By Proposition \ref{prop-besov-sufficient}, it follows that $p_{F,t} \in B^{\frac{n}{n+1} + \tilde{\epsilon}}_{n+1,n+1}(\Mm )$ for any $0 < \tilde{\epsilon} < \frac{1}{n+1}$ with a norm that is uniformly bounded in $t$ by some constant $\tilde{C}$. Writing out the Besov norm, this gives
$$
\big\|\widehat{W_j}*p_{F,t}\big\|_{n+1} 
\;\leq\; 
2^{-j (\frac{n}{n+1}+\tilde{\epsilon})}\norm{p_{F,t}}_{B^{\frac{n}{n+1}+\tilde{\epsilon}}_{n+1,n+1}} 
\;\leq\; 
\tilde{C} \,2^{-j (\frac{n}{n+1}+\tilde{\epsilon})}
\;.
$$
Hence continuity of $t \mapsto p_{F,t}$ w.r.t. the norm in $B_n(\Mm)=B^{\frac{n}{n+1}}_{n+1,n+1}(\Mm)$ follows from dominated convergence applied to 
$$
\norm{p_{F,t}-p_{F,s}}^{n+1}_{B_n} 
\;=\; 
\sum_{j=0}^\infty 2^{nj} \big\|\widehat{W_j}*(p_{F,t}-p_{F,s})\big\|_{n+1}^{n+1}
\;.
$$
Due to the definition \eqref{eq-ChernNumberEven}, the index theorem (Theorem~\ref{theo-Index}) and the continuity of the cocycle  $\widetilde{\Ch}_{\Tt,\Halfspaceaction,x_0}$ w.r.t. the $B_n(\Mm)$-norm the claim follows. The odd case follows by implementing the chiral symmetry.
\hfill $\Box$

\vspace{.2cm}

If there is a uniform spectral gap throughout the perturbation then the Chern numbers must moreover be constant due to the index theorems since $t\mapsto p_{F;t}$ respectively $t \mapsto u_{F,t}$ are then norm-continuous paths. In the absence of a spectral gap there are more possibilities.
Using index theorems and ergodicity one can show that the strong invariants, namely those with $n = d$, only take discrete values (see {\it e.g} \cite{PSbook}). Hence continuity in that case implies that the Chern numbers are still invariant under perturbations. It is important to emphasize that this is not necessarily the case for the weak topological invariants with $n < d$ and the Chern numbers may change continuously. An explicit example for this phenomenon will be given in Section~\ref{sec-Graphene}. Some numerical investigations on the stability of weak topological invariants can be found in {\it e.g.} \cite{ClaesHughes2019}.

\section{Smooth bulk-boundary correspondence}
\label{sec-BoundaryCurrents}

This section spells out the bulk-boundary correspondence (BBC) for insulators satisfying the BGH. This allows to use the $C^*$-algebraic exact sequence given by \eqref{eq-ToepBis} and the associated $K$-theory. For particular half-space Hamiltonians with a smooth boundary condition, the BBC will then essentially follow from the results of Chapter~\ref{sec-DualityToep}. Using the index theory of Chapter~\ref{chap-BreuerToep} they can then be extended to more general situations.

\vspace{.2cm}

Let us first recapitulate the set-up from Section~\ref{sec-HalfSpace} and then define the boundary invariants. Given a normal vector $\Halfspaceunitvector \in \SM^d$, one considers the associated $G$-action $\Halfspaceaction$ defined in \eqref{eq-ActionHalfSpace} where $G$ is either $\RM$ or $\TM$ for $\Halfspaceunitvector$ rationally independent or dependent respectively. One then has the smooth Toeplitz extension \eqref{eq-ToepBis}. Throughout, we will identify these algebras with their faithful GNS representation $\hat{\pi}_{\Tt,G}$ given in Proposition~\ref{prop:rational_rep}. In particular, $\Ee$ and $\hat{\Aa}$ are both subsets of  $\Nn_{\Halfspaceaction}$. All operators in these algebras are of the form $\int^\oplus_{\Omega \times \hat{G}} \PM(\difd{\omega}) \hat{\mu}(\difd{r}) \,\hat{a}_{\omega,r}$ with fiber operators $\hat{a}_{\omega,r}$ acting on $\ell^2(\ZM^d)$. We also identify $\Mm $ with its embedding into $\Nn_{\Halfspaceaction}$ since the inclusion map is just $ \pi_{\Tt} \otimes \one_{L^2(\hat{G})}$ already $\Mm  = \pi_{\Tt}(\Mm )$. Finally the units in $\Ee^\sim$ and $\Nn_\Halfspaceaction$ are also identified. Recall that the generator $D_\Halfspaceaction$ is represented by $\hat{X}_\Halfspaceaction = \int^\oplus_{\hat{G}}\hat{\mu}(\difd{r})\,(X_\Halfspaceaction+r)$ and hence the smooth projection $\Pp$ from Section~\ref{sec-toep} is in this representation is given by $\Pp=\int^\oplus_{\hat{G}}\hat{\mu}(\difd{r})\,\chi_s(X_\Halfspaceaction+r)$ with $\chi_s$ as in \eqref{eq-SmoothCutOff}. It can be used to construct a lift of an operator $a\in\Aa$ to $\hat{\Aa}$ as $\hat{a}=\Pp a\Pp+\tilde{k}$ with $\tilde{k}\in\Ee$. In this manner, one can construct half-space Hamiltonians given by a smooth restriction. The following proposition now shows how $K$-theoretic information transposes from the bulk to the boundary. It is proved exactly as in \cite{PSbook}:

\begin{proposition}
\label{prop-index_map_bb}
Let $h \in M_N(\Aa )$ be self-adjoint satisfying the {\rm BGH} with $E_F$ lying in the spectral gap $\Delta$. Further let $\hat{h} \in M_N(\hat{\Aa })$ be a self-adjoint lift of $h$ in the exact sequence \eqref{eq-ToepBis}, e.g. $\hat{h} = \calP h \calP + \tilde{k}$ with a boundary term $\tilde{k} \in M_N(\Ee)$.
\begin{enumerate}
\item [{\rm (i)}] The exponential map $\Exp: K_0(\Aa ) \to K_1(\Ee)$ maps $[p_F]_0$ to $[\hat{u}_\Delta]_1$ where 
$$
\hat{u}_\Delta 
\;=\; 
\exp(2\pi \imath \,f_\Exp(\hat{h})) \;\in\; M_N(\Ee^{\sim}) \,\subset\, M_N(\Nn_{\Halfspaceaction})
\;,
$$
with $f_\Exp \in C^\infty(\bbR)$ a smooth function with $f_\Exp(t) =0$ below $\Delta$ and $f_\Exp(t)=1$ above $\Delta$.

\item [{\rm (ii)}] If the {\rm CH} holds for both $h$ and $\hat{h}$, the index map $\Ind: K_1(\Aa ) \to K_0(\Ee)$ maps $[u_F]_1$ to $[\hat{p}_\Delta]_0 - [0_{\frac{N}{2}} \oplus \one_{\frac{N}{2}}]_0$ where 
$$
\hat{p}_\Delta 
\;=\; 
e^{-\imath \,\frac{\pi}{2} \,f_\Ind(\hat{h})} 
\begin{pmatrix} \one_{\frac{N}{2}} & 0 \\ 0 & 0_{\frac{N}{2}} \end{pmatrix} 
e^{\imath \,\frac{\pi}{2}\, f_\Ind(\hat{h})}\; \in\; M_N(\Ee^{\sim}) \,\subset \,M_N(\Nn_{\Halfspaceaction})
\;,
$$
with $f_\Ind \in C^\infty(\bbR)$ an odd symmetric smooth function with $f_\Ind(t) =-1$ below $\Delta$ and $f_\Ind(t)=1$ above $\Delta$.
\end{enumerate}
\end{proposition}
Here we are using the modified standard picture of $K$-theory for the unital algebra $\Aa$, where classes in $K_i(\Aa)$ are represented by projections and unitaries over $\Aa$ instead of $\Aa^\sim$. A minor difference compared to \cite{PSbook} is that $\hat{\Aa}$ is now non-unital and hence one must construct a lift of $u_F-\one_{N/2} + \one_{N/2}^\sim$ for the computation of the index map, which still leads to exactly the same expressions. In the end the representatives are included into $\Nn_\Halfspaceaction$ by identifying the unit of $\Ee^\sim$ with that of $\Nn_\Halfspaceaction$.

\vspace{.2cm}

Proposition~\ref{prop-index_map_bb} provides natural $K$-theoretic boundary invariants. The next aim is to extract numerical invariants by pairing these $K$-group elements with suitable Chern characters that will be constructed next. First recall that the action $\rho$ also extends to a strongly continuous $\RM^{d}$-action on the crossed product $\Ee$ by acting trivially on functions $f(\hat{X}_\Halfspaceaction)$ of the generator. An $n$-dimensional subaction $\theta$ is then generated by a choice of an orthonormal set $\hat{e}_1,\ldots,\hat{e}_n \in \bbR^d$ with $\hat{e}_1,\ldots,\hat{e}_n \perp \Halfspaceunitvector$. This action $\theta$ leaves the dual trace $\hat{\Tt}_\Halfspaceaction$ invariant so that one can introduce smooth Chern characters $\Ch_{\hat{\Tt}_\Halfspaceaction,\theta}$ on $\Ee$ as in Section~\ref{sec:smooth_chern}. For smoothly summable projections $\hat{p} \in \Ee^{\sim}_{\hat{\Tt}_\Halfspaceaction, \theta}$ or unitaries $\hat{u} \in  \Ee^{\sim}_{\hat{\Tt}_\Halfspaceaction, \theta}$, one can now consider the index pairings $\langle \Ch_{\hat{\Tt}_\Halfspaceaction,\theta}, [\hat{p}]_0\rangle$ and $\langle \Ch_{\hat{\Tt}_\Halfspaceaction,\theta}, [\hat{u}]_1\rangle$ respectively. In order to actually evaluate these invariants, it needs to be checked that the images under the exponential and index map are smooth and summable. The idea for the following proof goes back to \cite{EG}.

\begin{proposition}
\label{prop-index_map_bb_sm}
Let $h$, $\hat{h}$, $\Delta$, $\hat{u}_\Delta$ and $\hat{p}_{\Delta}$ be as in {\rm Proposition~\ref{prop-index_map_bb}}, with $\hat{p}_{\Delta}$ requiring $\hat{h}$ to satisfy the {\rm CH}. Then $\hat{u}_{\Delta}-\one_{N} \in \Ee_{\hat{\Tt}_\Halfspaceaction,\theta}$ and $\hat{p}_{\Delta} - 0_{\frac{N}{2}} \oplus \one_{\frac{N}{2}} \in \Ee_{\hat{\Tt}_\Halfspaceaction,\theta}$.
\end{proposition}

\noindent{\bf Proof.}
Noting that $\hat{p}_{\Delta} = 0_{\frac{N}{2}} \oplus \one_{\frac{N}{2}} + \frac{1}{2} J \left(e^{-\imath \pi f_{\Ind,\epsilon}} + 1\right)(\hat{h})$ one sees that both $\hat{u}_\Delta - \one_N$ and $\hat{p}_\Delta - 0_{\frac{N}{2}} \oplus \one_{\frac{N}{2}}$ are of the form $g(\hat{h})$ with $g$ a smooth function whose support is contained in the bulk gap $\Delta$. In particular, $g(h)=0$. Hence it is sufficient to consider $g(\hat{h})$ for such functions. The functional calculus will be written out using the Dynkin-Helffer-Sj\"ostrand formula 
\begin{equation}
\label{eq-DHS}
g(\hat{h}) 
\;=\; 
\frac{1}{2 \pi}\int_D (\partial_{\overline{z}}\tilde{g}_K)(z) \,\frac{1}{\hat{h}-z} \, \difd{z}
\;,
\end{equation}
with $\tilde{g}_K$ a pseudo-analytic continuation of $g$ satisfying $\abs{\tilde{g}_K(z)} \leq C_K \abs{\Im m(z)}^K$ where $K \in \bbN$ can be chosen arbitrarily large and $\sigma(\hat{h}) \subset D \subset \bbC$ is an open set. 

\vspace{.1cm}

Let $P = \chi(\hat{X}_\Halfspaceaction > 0)$ be the sharp half-space projection and note that $g(0)=0$ and $\hat{h}=P \hat{h} P$ imply
\begin{align*}
g(\hat{h}) 
&
\;=\; 
\frac{1}{2 \pi}\int_D (\partial_{\overline{z}}\tilde{g}_K)(z) \,\left(-\frac{1}{z}(\one-P) + \frac{P}{\hat{h}-z P}\right) \, \difd{z}
\\
&
\;=\; 
\frac{1}{2 \pi}\int_D (\partial_{\overline{z}}\tilde{g}_K)(z) \,\frac{P}{\hat{h}-z P} \, \difd{z}
\end{align*}
with $\frac{P}{\hat{h}-zP} = P \frac{1}{\hat{h}+\one-P-z}P$ denoting the inverse in the algebra $P\Nn_{\Halfspaceaction} P$. Using the geometric resolvent identity
\begin{equation}
\label{eq-GeomResol}
\frac{P}{\hat{h}-zP} 
\;=\; 
P \,\frac{1}{h-z } \,P\; +\; \frac{P}{\hat{h}-zP}\, (Ph\,-\,\hat{h}\,)\,\frac{1}{h-z}\,P
\;,
\end{equation}
one deduces
\begin{equation}
\label{eq-GeomResolConcl}
g(\hat{h}) 
\;=\; 
g(\hat{h})- P g(h) P 
\;=\; 
\frac{1}{2 \pi}\int_D (\partial_{\overline{z}}\tilde{g}_K)(z) \,\frac{P}{\hat{h}-z P}\, V\, \frac{1}{h-z} \,P \;\difd{z}
\;,
\end{equation}
with 
$$
V 
\;=\; 
(P-\calP) h \calP \,+\, P h (P-\calP)  \,+\,P h (\one-P) \, -\, \tilde{k}
\;.
$$
Note that $P h (\one-P)$ is the Hankel operator of Section~\ref{sec-Peller} and since smoothness is readily seen to be a sufficient condition for Theorem~\ref{theo-Hankel} one deduces  $\norm{V}_1 < \infty$ so that $V \in  L^1(\Nn_{\Halfspaceaction})$ and hence also $g(\hat{h}) \in  L^1(\Nn_{\Halfspaceaction})$. It remains to show that $g(\hat{h})$ is smooth w.r.t. $\theta$. For this purpose, let us first note that by hypothesis $h$, $\tilde{k}$, $\calP$ and thus also $\hat{h}$ are infinitely often differentiable w.r.t. $\theta$ and the operator norm  (note in particular that $\calP$ is invariant under $\theta$). By iteration of the resolvent identity 
\begin{equation}
\label{eq:res_diff}
\nabla_{\hat{e}_j} \frac{1}{a-z} 
\;=\; \frac{1}{a-z} \, (\nabla_{\hat{e}_j} a)\, \frac{1}{a-z}
\end{equation} 
one derives bounds for multi-indices $j \in \bbN^{n}$
$$
\Big\|\nabla^j \frac{1}{h-z}\Big\|
\;\leq\; 
C_1(\abs{j}) \,\frac{1}{\abs{\Im m(z)}^{\abs{j}+1}}
\;, 
\qquad 
\Big\|\nabla^j \frac{1}{\hat{h}-z P}\Big\| 
\;\leq \;
C_2(\abs{j}) \,
\frac{1}{\abs{\Im m(z)}^{\abs{j}+1}}
\;,
$$
with constants that are uniform in $\Im m(z)$. Since $V$ is norm-smooth w.r.t. $\theta$, it follows from the Leibniz rule applied to \eqref{eq-GeomResolConcl} with $K=\abs{j}+2$ that $g(\hat{h})$ is infinitely often differentiable w.r.t. $\theta$ and the norm $\norm{\cdot}_\infty + \norm{\cdot}_1$. By definition, $g(\hat{h}) \in \Ee_{\hat{\Tt}_\Halfspaceaction,\theta}$.
\hfill $\Box$

\vspace{.2cm}

After these analytical preparations, one can now apply Theorem~\ref{theo-smooth_duality} to extract the numerical information about the bulk and boundary Chern numbers from the $K$-theoretic statement of Proposition~\ref{prop-index_map_bb}. 


\begin{theorem}[Smooth BBC]
\label{theo-index_map_bb_sm}
Let $\hat{h}$, $h$, $\hat{p}_\Delta$ and $\hat{u}_\Delta$ be as in {\rm Proposition~\ref{prop-index_map_bb_sm}}. 
\begin{enumerate}
\item [{\rm (i)}] For $n$ odd, 
$$
\langle \Ch_{\Tt,\theta \times \Halfspaceaction}, [p_F]_0\rangle 
\;=\; 
\langle \Ch_{\hat{\Tt}_\Halfspaceaction,\theta}, [\hat{u}_{\Delta}]_1\rangle
\;.
$$

\item [{\rm (ii)}] For $n$ even and $h, \hat{h}$ satisfying the {\rm CH}, 
$$
\langle \Ch_{\Tt,\theta \times \Halfspaceaction}, [u_F]_1\rangle 
\;=\; 
-\,\langle \,\Ch_{\hat{\Tt}_\Halfspaceaction,\theta}, [\hat{p}_{\Delta}]_0\rangle
\;.
$$
\end{enumerate}
\end{theorem}

By Proposition~\ref{prop-ChernRotInv} the Chern character $\Ch_{\Tt, \theta \times \Halfspaceaction}$ up to a sign only depends on the hyperplane spanned by $\hat{e}_1,\dots,\hat{e}_n, \Halfspaceunitvector$. If one considers the case $n=d-1$, then the l.h.s. in both (i) and (ii) are the strong invariants which are independent of the choice of  $\Halfspaceunitvector$ up to orientation. 
Hence also the boundary invariants on the r.h.s. are independent of $\Halfspaceunitvector$.
 For $d=2$ and $n=1$, the boundary invariant is equal to the boundary current \cite{KRS,PSbook} and hence these boundary currents are constant and independent of the choice of the cutting angles $\Halfspaceunitvector$. This was recently also proved by other means \cite{LT}. For further physical implications of Theorem~\ref{theo-index_map_bb_sm} and, in particular, links to response coefficients the reader is referred to \cite{PSbook}.

\vspace{.2cm}

The smooth BBC can be extended to significantly less regularity using the index theorems. On the bulk side, one can perturb the bulk Hamiltonian with a non-smooth perturbation that is small enough such as not to close the spectral gap. This allows to deal with discontinuously distributed disorder or long-range potentials that lead to a Hamiltonian which is not an element of $\bbT^d_{\BB,\Omega}$.  On the boundary side, it may be of interest to consider a much larger class of boundary conditions, such as Dirichlet boundary conditions. As will be shown below, the only restriction is that $\hat{u}_\Delta$ and $\hat{p}_\Delta$ must still satisfy the Besov space conditions of the index theorem (Theorem~\ref{theo-Index}). To be more precise, we will deal with following reasonably large class of Hamiltonians:

\begin{proposition}
\label{prop-index_map_bb_nsm}
Let $h \in M_N(\Mm )$ satisfy the {\rm BGH} and let $\hat{h} = P h P + \tilde{k}$ for some $\tilde{k} \in M_N(\Nn_{\Halfspaceaction})$. Assume the Sobolev differentiability $h \in W^2_{n+1+\delta}(\Mm ,\rho)$ and $\tilde{k} \in W^1_{n+\delta}(\Nn_{\Halfspaceaction},\theta)$ for some $\delta>0$ and further that $h$ and $\tilde{k}$ are weakly differentiable w.r.t. $\theta$, namely the difference quotients converge in the weak-$*$ sense. Then: 
\begin{enumerate}
\item [{\rm (i)}]
$p_{F} \in B_{n+1}(\Mm , \rho)$ and $\hat{u}_\Delta-\one_N \in B_n(\Nn_{\Halfspaceaction},\theta)$. 
\item [{\rm (ii)}]
If $h$ and $\hat{h}$ satisfy the {\rm CH}, also $u_{F} \in  B_{n+1}(\Mm , \rho)$ and $\hat{p}_\Delta - 0_{\frac{N}{2}} \oplus \one_{\frac{N}{2}} \in B_n(\Nn_{\Halfspaceaction},\theta)$. 
\end{enumerate}
\end{proposition}

\noindent{\bf Proof.} 
As to the bulk, one just notes that the resolvents $\frac{1}{h-z}$ are also in $W^1_{n+1+\delta}(\Mm,\rho)$ due to \eqref{eq:res_diff}  and therefore the claims follow directly from the Riesz projection formula and Proposition~\ref{prop-besov-sufficient}(i). For the boundary systems, one proceeds as in Proposition~\ref{prop-index_map_bb_sm} to obtain \eqref{eq-GeomResolConcl}, but because of a different half-space Hamiltonian $\hat{h}$ one now has
$$
V \;=\; P h (\one-P) -  \tilde{k}
\;.
$$
By the hypothesis the difference quotient $t^{-1}(\theta_{t\hat{e}_j}(h)-h)$ converges w.r.t. the $W^1_{n+1+\delta}(\Mm ,\rho)$ norm and, $\Tt$ being a finite trace, also w.r.t. $W^1_{n+\delta}(\Mm ,\rho)$. Proposition~\ref{prop-besov-sufficient}(i) then shows that it also converges in $B^{\frac{1}{n+\delta}}_{n+\delta,n+\delta}(\Mm ,\rho)$. Using Theorem~\ref{theo-HankelBesovP} (Peller's criterion), one obtains 
$$
\Big\| P \,\frac{\theta_{t\hat{e}_j}(h)-h}{t}\,(1-P) \,- \,P\, \nabla_{\hat{e}_j} h\, (1-P)\Big\|_{n+\delta} 
\;\leq\; 
\Big\|
\frac{\theta_{t\hat{e}_j}(h)-h}{t} \,-\, \nabla_{\hat{e}_j} h
\Big\|_{B^{\frac{1}{n+\delta}}_{n+\delta,n+\delta}(\Mm ,\Halfspaceaction)}
\;,
$$ 
and hence, due to the hypothesis $h \in W^2_{n+1+\delta}(\Mm ,\rho)$, one has $V \in W^1_{n+\delta}(\Nn_{\Halfspaceaction}, \theta)$. Next let us note that resolvents of weakly differentiable elements are again weakly differentiable with the derivative of the resolvent given by \eqref{eq:res_diff}. Furthermore, it is a general fact that, if $a,c$ are weakly differentiable and $b\in W^1_p(\Nn_{\Halfspaceaction},\theta)$, then $abc \in W^1_p(\Nn_{\Halfspaceaction},\theta)$ and the Leibniz rule holds. Using the default estimate for the resolvent, this implies
\begin{align*}
\Big\|\nabla_{\hat{e}_i} \frac{P}{\hat{h}-zP} \,V\, \frac{1}{h-z}\, P\Big\|_{n+\delta} 
&
\;\leq \;
\frac{\lVert \nabla_{\hat{e}_i} \hat{h}\rVert}{\abs{\Im m(z)}^3}  \norm{V}_{n+\delta} 
\,+\,  
\frac{1}{\abs{\Im m(z)}^2}  \norm{V}_{W^1_{n+\delta}(\Nn_{\Halfspaceaction},\theta)}
\\
&
\;\;\;\;\;\;
\,+\, 
\frac{\norm{\nabla_{\hat{e}_i} h}}{\abs{\Im m(z)}^3}  \norm{V}_{n+\delta}
,
\end{align*}
and hence \eqref{eq-GeomResolConcl} converges absolutely in the $L^{n+\delta}(\Nn_\Halfspaceaction)$ norm, provided that $K\geq 3$. Then one can estimate the $n$-dimensional Fourier multipliers $W^\theta_j$ w.r.t. $\theta$ by
\begin{align*}
\Big\|\widehat{W}^\theta_j* g(\hat{h})\Big\|_{n+\delta} 
&
\;\leq\; 
\int_D \abs{\partial_{\overline{z}}\tilde{g}_K(z)} \,\Big\|\widehat{W^\theta_j}* \frac{P}{\hat{h}-zP} \,V\, \frac{1}{h-z}\, P\Big\|_{n+\delta}\, \difd{z} 
\\
&
\;\leq\; 
C\, 2^{-j} 
\int_D \abs{\partial_{\overline{z}}\tilde{g}_K(z)}\,\Big\|\frac{P}{\hat{h}-zP} \,V\, \frac{1}{h-z} \,P\Big\|_{W^1_{n+\delta}(\Nn_{\Halfspaceaction},\theta)} 
\,\difd{z} 
\\
&
\;\leq\; C'\,2^{-j}
\;.
\end{align*}
This shows $g(\hat{h})\in B^1_{n+\delta, \infty}(\Nn_{\Halfspaceaction}, \theta)$ and Proposition~\ref{prop-besov-sufficient}(ii) can be applied to deduce $g(\hat{h})\in B_n(\Nn_{\Halfspaceaction},\theta)$.
\hfill $\Box$
  
\begin{proposition}
Let $t\in[0,1] \mapsto h(t) \in M_N(\Mm )$ and $t\in[0,1] \mapsto \hat{h}(t)$ be norm-continuous paths that satisfy the hypothesis of {\rm Proposition~\ref{prop-index_map_bb_nsm}} pointwise. Let $n$ be odd. Then the Chern numbers $\langle \Ch_{\Tt,\theta \times \Halfspaceaction}, [p_F(t)]_0\rangle$ and $\langle \Ch_{\hat{\Tt}_\Halfspaceaction,\theta}, [\hat{u}_{\Delta}(t)]_1\rangle$ are constant along the path. If, moreover, $h(0) \in M_N(\Aa )$, they satisfy 
$$
\langle \Ch_{\Tt,\theta \times \Halfspaceaction}, [p_F(t)]_0\rangle 
\;=\; \,
\langle \Ch_{\hat{\Tt}_\Halfspaceaction,\theta}, [\hat{u}_{\Delta}(t)]_1\rangle
\;.
$$
For $n$ even, suppose that, moreover, $h_t$ and $\hat{h}_t$ both satisfy the {\rm CH}. Then both pairings $\langle \Ch_{\Tt,\theta \times \Halfspaceaction}, [u_F(t)]_1\rangle$ and $\langle \Ch_{\hat{\Tt}_\Halfspaceaction,\theta}, [\hat{p}_{\Delta}(t)]_0\rangle$ are constant along the path and satisfy
$$
\langle \Ch_{\Tt,\theta \times \Halfspaceaction}, [u_F(t)]_1\rangle 
\;=\; -\,
\langle \Ch_{\hat{\Tt}_\Halfspaceaction,\theta}, [\hat{p}_{\Delta}(t)]_0\rangle
\;,
$$
again provided that $h(0) \in M_N(\Aa )$.
\end{proposition}

\noindent{\bf Proof.}
Let us focus on the case $n$ odd. The first statement is clear since continuous functional calculus maps norm-continuous paths to norm-continuous paths and hence the index theorem shows that the Chern numbers are constant.

\vspace{.1cm}

Assume now in addition $h(0)=h_0 \in M_N(\Aa )$ and let $\hat{h}_0= \calP h_0 \calP$ be a reference  Hamiltonian with smooth boundary condition. Since the smooth BBC works on the level of $K$-theory, the equality of the index pairings also holds for the not necessarily differentiable Hamiltonians $(h_0,\hat{h}_0)$ since they are still in the continuous subalgebras $M_N(\Aa )$ respectively $M_N(\hat{\Aa})$. Next let us note that the straight-line path $\lambda \in [0,1] \to \lambda \hat{h}_0 + (1-\lambda) \hat{h}(0)$ between $\hat{h}_0$ and $\hat{h}(0)$ also satisfies the conditions of Proposition~\ref{prop-index_map_bb_nsm} pointwise. Hence with obvious choice of notations, one then has
\begin{align*}
\langle \Ch_{\Tt,\theta \times \Halfspaceaction}, [p_{F}(t)]_0\rangle
&
\;=\; 
\langle \Ch_{ \Tt,\theta \times \Halfspaceaction}, [p_{F}(0)]_0\rangle 
\\
&
\;=\;
\langle \Ch_{\Tt,\theta \times \Halfspaceaction}, [p_{F,0}]_0\rangle  
\\
&
\;=\;  \langle \Ch_{\hat{\Tt}_\Halfspaceaction,\theta}, [\hat{u}_{\Delta,0}]_1\rangle 
\\
&
\;=\; 
\, \langle \Ch_{\hat{\Tt}_\Halfspaceaction,\theta}, [\hat{u}_{\Delta}(0)]_1\rangle 
\\
&
\;=\; 
\,\langle \Ch_{\hat{\Tt}_\Halfspaceaction,\theta}, [\hat{u}_{\Delta}(t)]_1\rangle 
\;,
\end{align*}
concluding the proof in the case of odd $n$.
\hfill $\Box$

\section{Delocalization of boundary states}
\label{sec-DelocBoundary}

In \cite{PSbook} it is shown that, if the bulk Hamiltonian $h$ has a spectral gap $\Delta$, the half-space Hamiltonian $\hat{h}$ cannot satisfy the Aizenman-Molchanov bound in $\Delta$ unless its strong Chern number (the one with $n=d$) vanishes. This suggests that the boundary states are resistant to dynamical localization. An important ingredient in the proof was the quantization of the strong Chern numbers which implies invariance under perturbations that are continuous in a Sobolev topology. This section will demonstrate that the quantization of the strong invariants is actually not essential and give a general argument that also the boundary states corresponding to the weak invariants are protected from localization in the same sense.

\vspace{.2cm}

For sake of simplicity, we will assume that $h \in M_N(\Aa )$ is smooth and let $\hat{h} \in M_N(\hat{\Aa})$ with $\hat{h} = \Pp h\Pp + \tilde{k}$ and $\tilde{k} \in M_N(\Ee_{\hat{\Tt}_\Halfspaceaction,\rho})$.  What will be proved is not delocalization in the actual sense, but rather that certain decay properties of the Hamiltonian which are expected to hold in the localized regime constrain the boundary invariants to vanish.

\vspace{.2cm}

For a spectral interval $I$, let us introduce an eigenfunction correlator for the half-space operator, following \cite{Aizenman94,AizenmanWarzel}
$$
\mathcal{Q}(\hat{h}_{\omega,r}, x,y, I) 
\;=\; 
\sup_{f \in \Bb(I)\,,\, \norm{f}_\infty \leq 1}\; 
\big\|\langle x| f(\hat{h}_{\omega,r})|y \rangle\big\|_2
\;,
$$
with the supremum taken over all bounded Borel functions and  the Hilbert-Schmidt-norm on $M_N(\bbC)$. As usual, we consider the disorder average of the eigenfunction correlator and in addition it will also be useful to average over a compact subset $[0,R)\cap\hat{G}$ with an  arbitrary, but fixed positive $R\in \Gamma_\Halfspaceaction=\Halfspaceunitvector \cdot \bbZ^d$:
$$
\overline{\mathcal{Q}}_2(\hat{h}, x,y, I) 
\;=\;  
\int_{[0,R)\cap \hat{G}}  \int_{\Omega}  \mathcal{Q}(\hat{h}_{\omega,r}, x,y, I)^2\;\bbP(\difd\omega)\,\hat{\mu}(\difd{r})
\;.
$$ 
Since the average extends only over a finite interval, it can characterize the decay w.r.t. the hypersurface uniformly. The interval itself is arbitrary since one can always shift the boundary using a combination of magnetic translations and translations in $\Omega$.  

\vspace{.2cm}

A standard localization condition is rapid off-diagonal decay of the averaged eigenfuncion correlator:

\begin{definition}
\label{def-BoundaryLoc}
A Hamiltonian $\hat{h}$ is said to satisfy strong dynamical localization in a spectral interval $I$ if
\begin{equation}
\label{eq:correlator_bound_natural}
	\sup_{y\in \bbZ^d}\; \sum_{x\in \bbZ^d} \;\, (1+\abs{x-y})^{k} \, \overline{\mathcal{Q}}_2(\hat{h},x,y,I) 
	\; < \infty
\end{equation}
holds for each $k\in \bbN$. 
\end{definition}

If \eqref{eq:correlator_bound_natural} holds for an interval $I\subset \Delta$, it is possible to show (under mild assumption on the degeneracy of eigenvalues) that $\hat{h}_{\omega,r}$ almost surely has only pure-point spectrum in $I$ with eigenfunctions that decay polynomially in space ({\sl e.g.} \cite[Chapter 7]{AizenmanWarzel}). The bulk gap automatically improves such off-diagonal decay to rapid decay w.r.t. the boundary hypersurface:

\begin{lemma}
\label{eq:lemma_correlatordecay}
Let $\hat{h}=\Pp h \Pp + \tilde{k}$ be a smooth Hamiltonian as above with bulk gap $\Delta$ and rapidly decaying boundary term in the sense that 
$$
\sup_{\omega\in \Omega, r\in \hat{G}} (1+ \abs{x-y})^j(1+\abs{\Halfspaceunitvector\cdot x-r})^j (1+\abs{\Halfspaceunitvector\cdot y-r})^j \norm{\langle x| \tilde{k}_{\omega, r}|y\rangle} 
\;<\; \infty
$$
for each $j\in \bbN$. If $\hat{h}$ satisfies strong dynamical localization in an interval $I \subset \Delta$, then it also satisfies the condition
\begin{equation}
	\label{eq:correlator_bound}
	\sup_{y\in \bbZ^d}\; \sum_{x\in \bbZ^d} \;(1+\abs{y \cdot \Halfspaceunitvector}^{k+1}) \, (1+\abs{x-y})^{k} \, \overline{\mathcal{Q}}_2(\hat{h},x,y,I) 
	\;<\; 
	\infty
	\;,
\end{equation}
for any $k\in \bbN$.
\end{lemma}

\noindent{\bf Proof.}
Choose an interval $\tilde{I}$ with $I \subset \tilde{I} \subset \Delta$ and let $g\in C_0^\infty(\Delta)$ be a function with $g\rvert_I = 1\rvert_I$ and vanishing outside $\tilde{I}$. We note the standard estimate
$$
\sup_{\omega\in \Omega, r\in \hat{G}} \abs{\Im m (z)}^{j+1} (1+ \abs{x-y})^j \norm{\langle x| \frac{1}{\hat{h}_{\omega, r}-z}|y\rangle} 
\;<\; \infty
\;,
$$ 
which follows from the norm-estimate $\norm{\nabla^j  \frac{1}{\hat{h}_{\omega, r}-z}}< c \abs{\Im m(z)}^{-\abs{j}-1}$ and the relation 
$$
\langle x|\nabla_{\hat{e}_i}\hat{a}_{\omega,r}| y\rangle 
\;=\; 
-\imath(y_i-x_j)\langle x|\hat{a}_{\omega,r}| y\rangle
\;.
$$ 
For the bulk resolvent we also have the Combes-Thomas estimate from the proof of Proposition~\ref{prop:spectralgap}. Substituting those estimates for matrix elements into the Dynkin-Helffer-Sj\"ostrand representation for $g(\hat{h})$ (see the proof of Proposition~\ref{prop-index_map_bb_sm}) it is then possible to show that there is for any $j\in \bbN$ a constant $c_j$ independent of $r \in \hat{G}$ such that
$$
\sup_{\omega\in \Omega} \norm{\langle x| g(\hat{h}_{\omega, r})|y\rangle} 
\;\leq\; 
\frac{c_j}{(1+ \abs{x-y}^j)(1+\abs{\Halfspaceunitvector\cdot x-r}^j) (1+\abs{\Halfspaceunitvector\cdot y-r}^j)}
\;.
$$
Consider the $L^1$-averaged eigenfunction correlator
$$
\overline{\mathcal{Q}}_1(\hat{h}, x,y, I) 
\;=\;  
\int_{[0,R)}  \int_{\Omega} \mathcal{Q}(\hat{h}_{\omega,r}, x,y, I)
\;\bbP(\difd\omega)\,\hat{\mu}(\difd{r})
\;,
$$
which is readily seen to satisfy $\overline{\mathcal{Q}}_2 \leq  \overline{\mathcal{Q}}_1 \leq c_{R} \sqrt{\overline{\mathcal{Q}}_2}$ due to $\Qq \leq 1$ and since $L^1 \subset L^2$ holds for finite measure spaces.
Hence $\overline{\mathcal{Q}}_1$ also decays faster than any polynomial in $\abs{x-y}$ which means that for any $j,\ell \in \bbN$ there are constants such that
\begin{align*}
&\!\!\overline{\mathcal{Q}}_1(\hat{h}, x,y, I) 
\\
\;&=\;  
\int_{[0,R)\cap\hat{G}}  \int_{\Omega} \sup_{f \in \Bb(I)\,,\, \norm{f}_\infty \leq 1}\; 
\big\| \sum_{z\in \bbZ^d} \langle x| f(\hat{h}_{\omega,r})| z\rangle \langle z| g(\hat{h}_{\omega,r})|y \rangle\big\|_1\;\bbP(\difd\omega)\,\hat{\mu}(\difd{r})\\
\;&\leq\;   \sum_{z\in \bbZ^d} \int_{[0,R)\cap \hat{G}}  \int_{\Omega} \norm{\langle z|g(\hat{h}_{\omega,r}) |y\rangle} \mathcal{Q}(\hat{h}_{\omega,r}, x,z, I) \;\bbP(\difd\omega)\,\hat{\mu}(\difd{r})\\
\;&\leq\;  \sum_{z\in \bbZ^d} \left(\sup_{\omega\in \Omega, r\in [0,R) \cap \hat{G}} \norm{\langle z|g(\hat{h}_{\omega,r}) |y\rangle}\right) \overline{\mathcal{Q}}_1(\hat{h}, x,z, I)\\
\;&\leq\;   \sum_{z\in \bbZ^d} \frac{c_{j,R}}{(1+ \abs{x-y}^j)(1+\abs{\Halfspaceunitvector\cdot z}^j) (1+\abs{\Halfspaceunitvector\cdot y}^j)} \;\frac{\tilde{c}_\ell}{1+\abs{x-z}^\ell}
\;.
\end{align*}
As $j$ and $\ell$ are arbitrary, $\overline{\mathcal{Q}}_1(\hat{h}, x,y, I)$ and thus $\overline{\mathcal{Q}}_2(\hat{h}, x,y, I)$ can be seen to decay faster than any polynomial in $\abs{x-y}$ and $\abs{\Halfspaceunitvector\cdot y}$. Substituting this upper bound into \eqref{eq:correlator_bound} completes the proof. \hfill $\Box$ 

\begin{lemma}
	\label{lem-l2matrixelements}
	Let $\hat{a}\in \Nn_{\Halfspaceaction} \cap L^2(\Nn_{\Halfspaceaction})$ then 
	$$
	\big\|\hat{a}\big\|_2^2 
	\;=\; 
	\int_{\hat{G}} \int_{\Omega} \sum_{x\in \bbZ^d} \norm{\langle x| \hat{a}_{\omega,r}| 0\rangle}_2^2 \;\PM(\difd \omega)\, \hat{\mu}(\difd{r}).
	$$
	For any $x_0 \in \bbZ^d$ with $R:= x_0 \cdot \Halfspaceunitvector > 0$ one further has
	$$
	\big\|\hat{a}\big\|_2^2 
	\;=\; 
	\int_{\hat{G}\cap [0,R)} \sum_{m\in \bbZ} \int_{\Omega} \sum_{x\in \bbZ^d} \norm{\langle x + m x_0| \hat{a}_{\omega,(r-mR)}| m x_0\rangle}_2^2 \;\PM(\difd \omega)\, \hat{\mu}(\difd{r}).
	$$
\end{lemma}
\noindent{\bf Proof.}
As in the proofs of Propositions~\ref{prop-rationaltrace} and ~\ref{prop-irrationaltrace} there is a function $\hat{g}\in L^2(\hat{G}, L^2(\Mm,\Tt))$ such that $\hat{a}= \int_{G} \pi(\calF^{-1}g)(t) U(t)\mu(\difd{t})$, 
$$\big\|\hat{a}\big\|_2^2 = \int_{\hat{G}} \Tt\left((\hat{g}(r))^*\hat{g}(r)\right)\hat{\mu}(\difd{r})$$
and by the isometric embedding of $L^2(\Mm)$ into $\Hh_\Tt$ one further has
$$\Tt\left((\hat{g}(r))^*\hat{g}(r)\right) =  \int_{\Omega} \sum_{x\in \bbZ^d} \norm{\langle x| \hat{g}(r)_\omega| 0\rangle}_2^2 \;\PM(\difd \omega).$$
The relation between $\hat{g}$ and the matrix elements of $\hat{a}$ is
\begin{align*}
\langle x| \hat{a}_{\omega,r}| y\rangle 
&
\;=\; 
\int_G  \langle x| (\calF^{-1}g)(t)_\omega e^{2\pi \imath (X_\Halfspaceaction+r)t}| y\rangle \;\mu(\difd{t})
\\
&
\;=\; 
\int_G  \langle x| (\calF^{-1}g)(t)_\omega e^{2\pi \imath (\Halfspaceunitvector \cdot y + r)t}| y\rangle 
\;\mu(\difd{t})
\\
&
\;=\; 
\langle x| \hat{g}(\Halfspaceunitvector \cdot y + r)_{\omega}| y\rangle
\;
\end{align*}
which implies both of the desired formulas.
\hfill $\Box$

\begin{lemma}
\label{lem-CorrelatorConcl}
If \eqref{eq:correlator_bound} holds, then 
$$
\sup_{f \in \Bb(I)\,,\, \norm{f}_\infty \leq 1}\; 
\big\|f(\hat{h})\big\|_{B^{\frac{k}{2}}_{2,2}} 
\;<\; \infty
\;.
$$
Moreover, if $(f_l)_{l\in \bbN}$ is a uniformly bounded sequence in the Borel functions $\Bb(I)$ with $f=\lim_{l\to \infty} f_l$, then $f_l(\hat{h})$ converges to $f(\hat{h})$ in $B^{\frac{k}{2}}_{2,2}(\Nn_{\Halfspaceaction})$.
\end{lemma}

\noindent{\bf Proof.}
For $f\in C^\infty_0(I)$ with $\norm{f}_\infty \leq 1$ one applies Lemma~\ref{lem-l2matrixelements} to $\hat{a}:= f(\hat{h}) \in \Ee_{\hat{\Tt}_\Halfspaceaction,\rho}$ to obtain
\begin{align*}
\big\|f(\hat{h})\big\|_2^2 
&
\;=\; 
\int_{[0,R)\cap \hat{G}}\int_{\Omega}  \sum_{m\in \bbZ}  \sum_{x\in \bbZ^d} \big\|\langle x + m x_0| f(\hat{h})_{\omega,(r- m R)} | m x_0\rangle\big\|_2^2 \;\PM(\difd \omega)\, \hat{\mu}(\difd{r})  \\
&
\;\leq \;
\sum_{m \in \bbZ} \sum_{x\in \bbZ^d} \overline{\mathcal{Q}}_2(\hat{h}, x + m x_0, m x_0, I)
\;.
\end{align*}
Similarly 
\begin{equation}
\label{eq:multiplier_correlator}
\big\|\widehat{W_j}* f(\hat{h})\big\|_2^2 
\;\leq \;
\sum_{m\in \bbZ} \sum_{x\in \bbZ^d} \abs{W_j(x)}^2 \,  \overline{\mathcal{Q}}_2(\hat{h}, x + m x_0, m x_0, I)
\;.
\end{equation}
Thus by \eqref{eq:correlator_bound} one obtains a bound that is uniform for all smooth $f$. An arbitrary $f \in \Bb(I)$ can be written as a pointwise limit $f_l \to f$ of smooth functions, hence $f_l(\hat{h})\to f(\hat{h})$ in the SOT and therefore in the weak $L^2$-topology by Lemma~\ref{lemma:convergence}(iii). Since multipliers are continuous, this implies $\widehat{W_j}*f_l(\hat{h}) \to \widehat{W_j}*f(\hat{h})$ also in the weak $L^2$-topology and hence
$$
\big\|\widehat{W_j}*f(\hat{h})\big\|_2^2 
\;\leq\; 
\liminf_{l\to \infty}
\big\|\widehat{W_j}*f_l(\hat{h})\big\|_2^2
\;.
$$ 
This shows that \eqref{eq:multiplier_correlator} actually holds for all $f \in \Bb(I)$ with $\norm{f}_\infty \leq 1$.  Using the same manipulations as in the proof of Proposition~\ref{prop-FractionalSobolev}, one can bound the equivalent norm of $B^{\frac{k}{2}}_{2,2}$ 
\begin{align*}
&\!\!
\big\|f(\hat{h})\big\|^2_{B^{\frac{k}{2}}_{2,2}} 
\\
&
\;\leq \;
\sum_{m\in \bbZ} \sum_{x\in \bbZ^d} (1+\abs{x})^k \, \overline{\mathcal{Q}}_2(\hat{h}, x + m x_0, m x_0, I) \\
&
\;=\; 
\sum_{m\in \bbZ} \frac{1}{1+\abs{m R}^{k+1}}\sum_{x\in \bbZ^d} (1+\abs{m x_0 \cdot \Halfspaceunitvector}^{k+1})\, (1+\abs{x}^k) \, \overline{\mathcal{Q}}_2(\hat{h}, x + m x_0, m x_0, I) 
\,.
\end{align*}
Together with \eqref{eq:correlator_bound} this shows that $f(\hat{h})\in B^{\frac{k}{2}}_{2,2}(\Nn_{\Halfspaceaction})$ with a bound on the norm that is uniform for $f \in \Bb(I)$ with $\norm{f}_\infty \leq 1$.

\vspace{.1cm}

Having established that all spectral projections of $\hat{h}$ in the interval $I$ have finite trace, a dominated convergence argument shows that for a pointwise convergent sequence $f_l \to f$, one also has convergence in the norm $L^2$-sense $\|f(\hat{h}) - f_l(\hat{h})\|_2 \to 0$. As multipliers are continuous, this gives 
$$
\big\|\widehat{W_j}*f(\hat{h}) - \widehat{W_j}*f_l(\hat{h})\big\|_2 
\;\to\; 0
\;,
$$ 
and dominated convergence using \eqref{eq:multiplier_correlator} implies convergence in $B^{\frac{k}{2}}_{2,2}(\Nn_{\Halfspaceaction})$.
\hfill $\Box$

\vspace{.2cm}

\begin{proposition}
Let $n\leq d$ and $E_F$ in a bulk gap $\Delta$.
If $\Ch_{\Tt,\theta \times \Halfspaceaction}(p_F) \neq 0$, then the bound \eqref{eq:correlator_bound} cannot hold  for any $k\geq n$ and in any open interval $I \subset \Delta$. If $h$ and $\hat{h}$ satisfy the {\rm CH} and $\Ch_{\Tt,\theta \times \Halfspaceaction}(u_F) \neq 0$, then the bound \eqref{eq:correlator_bound} with $k\geq n$ cannot  hold in an interval $I \subset \Delta$ containing $0$ unless $0$ is an eigenvalue of $\hat{h}$.
\end{proposition}

\noindent{\bf Proof.}
Let $\chi_I = \chi(\hat{h} \in I)$ and set
$$
\hat{u}_{\Delta,\epsilon} 
\;=\; 
\exp\big(2\pi \imath \,f_{\Exp,\epsilon}(\hat{h})\big)
\;,
$$
for any $\epsilon>0$ and $f_{\Exp,\epsilon} \in C^\infty(\RM)$ a choice of smooth functions which take the value $0$ below $I$, $1$ above $I$ and converge to  $\chi(E> E_0)$ as $\epsilon\to0$ for an arbitrary, but fixed $E_0 \in I$. By Lemma~\ref{lem-CorrelatorConcl}, one has $B^{\frac{k}{2}}_{2,2}$-convergence 
$$
B^{\frac{k}{2}}_{2,2}\mbox{-}\lim_{\epsilon\downarrow 0} 
\exp\big(2\pi\imath\, f_{\Exp,\epsilon}(\hat{h})\big)\, \chi_I 
\;=\; 
\exp\big(2\pi\imath\, \chi(\hat{h} > E_0)\big) \,\chi_I 
\;=\; 
\chi_I
\;.
$$ 
Note that $\exp(2\pi \imath \,f_{\Exp,\epsilon}(\hat{h})) = \exp(2\pi \imath f_{\Exp,\epsilon}(\hat{h}))\chi_I + (\one_N - \chi_I)$ and hence also
$$
B^{\frac{k}{2}}_{2,2}\mbox{-}
\lim_{\epsilon\downarrow 0} \hat{u}_{\Delta,\epsilon} \, -\, \chi_I 
\;=\; 
B^{\frac{k}{2}}_{2,2}\mbox{-}\lim_{\epsilon\downarrow 0}
\exp\big(2\pi \imath \,f_{\Exp,\epsilon}(\hat{h})\big)
\,\chi_I \;-\; \one_N 
\;=\; 
0\;.
$$ 
Appealing to Proposition \ref{prop-besov-sufficient}(ii), one can bound 
$$
\big\|\hat{a}\big\|_{B_n} 
\;\leq \;
C\, 
\lVert \hat{a}\rVert^{\frac{2}{n+1}}_{B^{\frac{n}{2}}_{2,2}} \, \norm{\hat{a}}_\infty^{1-\frac{2}{n+1}}
\;
$$
for all $\hat{a}\in \Nn_\Halfspaceaction \cap B^{\frac{n}{2}}_{2,2}(\Nn_\Halfspaceaction, \rho)$, 
which shows that the boundary cocycle is continuous w.r.t. the $B^{\frac{n}{2}}_{2,2}$-norm on uniformly operator-norm bounded sets.
Hence for $k \geq n$
$$
0
\;=\;
\langle \Ch_{\hat{\Tt}_\Halfspaceaction,\theta},[\one_N]_1\rangle 
\;=\; 
\lim_{\epsilon \to 0}\; 
\langle \Ch_{\hat{\Tt}_\Halfspaceaction,\theta},[\hat{u}_{\Delta,\epsilon}]_1\rangle 
\;=\; 
\langle \Ch_{\Tt,\theta \times \Halfspaceaction},[p_F]_0\rangle
\;,
$$
in contradiction to the assumption. In the odd case one applies an analogous convergence argument to 
$$
p_{\Delta,\epsilon} 
\;=\; 
e^{-\imath \frac{\pi}{2} f_{\Ind,\epsilon}(\hat{h})} \begin{pmatrix} \one_N & 0 \\ 0 & 0_N \end{pmatrix} e^{\imath \frac{\pi}{2} f_{\Ind,\epsilon}(\hat{h})}
\;,
$$
with $f_{\Ind,\epsilon}$ anti-symmetric functions that converge to $\sgn$ as $\epsilon\to 0$. Then 
$$
B^{\frac{k}{2}}_{2,2}\mbox{-}\lim_{\epsilon\downarrow 0} p_{\Delta,\epsilon}\;-\; 0_{\frac{N}{2}} \oplus \one_{\frac{N}{2}} 
\;=\; 
\frac{1}{2} \,J \;B^{\frac{k}{2}}_{2,2}\mbox{-}\lim_{\epsilon\downarrow 0} \,
\left(e^{-\imath \pi f_{\Ind,\epsilon}} + 1\right)(\hat{h}) 
\;=\; 
J \,\chi(\hat{h}=0)
\;.
$$
Hence
\begin{align*}
\langle \Ch_{\hat{\Tt}_\Halfspaceaction,\theta},[0_N \oplus \one_{\frac{N}{2}} + J \chi(\hat{h}=0)]_0 - [\one_{\frac{N}{2}}]_0 \rangle 
&
\;=\; 
\lim_{\epsilon \to 0} \;\langle \Ch_{\hat{\Tt}_\Halfspaceaction,\theta},[p_{\Delta,\epsilon}]_0  - [\one_{\frac{N}{2}}]_0 \rangle 
\\
&
\;=\; 
\langle \Ch_{\Tt,\theta \times \Halfspaceaction},[u_F]_1\rangle
\;,
\end{align*}
allowing to complete the proof as above.
\hfill $\Box$

\vspace{.2cm}

In particular, strong dynamical localization cannot hold for the boundary states corresponding to even $n$ and the same is true for odd $n$ unless the Fermi level is an eigenvalue with positive probability. 

\section{Flat bands of edge states}
\label{sec-flat}

This section considers bulk Hamiltonians with chiral symmetry that are not necessarily gapped, but have a pseudogap or are in the MBGR at the Fermi energy $E_F=0$. The main result (already stated in Theorem~\ref{theo-SurfaceIntro} in the introduction) states that, under suitable conditions, one can nevertheless apply the Sobolev index theorem (Theorem~\ref{theo-Index}) and the associated Breuer-Fredholm index can then be interpreted as the signed density of a flat band of edge states. The relevant bulk invariant is the odd Chern number $\Ch_{\Tt,\Halfspaceaction}(u_F)$, {\it i.e.} the non-commutative winding number associated to the $G$-action $\Halfspaceaction$ given by \eqref{eq-ActionHalfSpace} and generated by the direction $\Halfspaceunitvector$ perpendicular to the boundary. In this situation, the constructions for the half-space and the semi-finite index pairing both take place in the algebra $\Nn_{\Halfspaceaction}= \Mm  \rtimes_{\Halfspaceaction} G$ which already suggests a closer interrelation than merely the smooth BBC.

\vspace{.2cm}

Let us describe the set-up building, in particular, on Sections~\ref{sec-AlgSetUp} and \ref{sec-HalfSpace}. The bulk Hamiltonian $h \in M_{N}(\Mm )$ satisfies the CH of Definition~\ref{def-CH}, namely $N$ is even and $JhJ=-h$. Throughout it will be assumed that $h_\omega$ has almost surely no eigenvalue at $E_F=0$. Stronger regularity conditions will be imposed later on, but this is not yet necessary at this stage. In this section it is algebraically convenient to work with the sharp half-space projection $P = \chi(\hat{X}_\Halfspaceaction > 0)$, though smooth boundary conditions are also possible.  The half-space Hamiltonian $\hat{h}$ will be a matrix with coefficients in $P\Nn_{\Halfspaceaction}P$ which takes the form
\begin{equation}
\label{eq-HamHalfSpace}
\hat{h} \;=\;P h P + \tilde{k}
\;,
\end{equation}
with $\tilde{k} \in P\Nn_{\Halfspaceaction}P \cap \calK_{\hat{\Tt}_\Halfspaceaction}$ also satisfying the CH, {\it i.e.} $\tilde{k}$ is a chiral $\hat{\Tt}_\Halfspaceaction$-compact boundary term. As also $P$ commutes with the chiral symmetry operator $J$ (which only acts on the matrix degrees of freedom), the half-space Hamiltonian has the chiral symmetry 
\begin{equation}
\label{eq-HamHalfSpaceChiral}
J\, \hat{h} \,J 
\;=\; 
-\,\hat{h}
\;.
\end{equation}
The physical representations $\hat{h}_{r,\omega}$ of $\hat{h}$ are then given by a sum of a term of the form \eqref{eq-DirchletRestriction} and a boundary term stemming from $\tilde{k}$, see \eqref{eq-HhatRep} below. Let us stress that $PJ=JP$ implies that the matrix degrees of freedom over a given site of the lattice are not split.  For the special case of  the graphene Hamiltonian presented in Section~\ref{sec-BulkInv}, this means that the unit cells are not broken up, see the discussion towards the end of this section.

\vspace{.2cm}

The chiral symmetries of $h$ and $\hat{h}$ imply that angular parts of the polar decompositions of $h$ and $\hat{h}$ are of the form
$$
\text{sgn}(h)
\; =\; 
\begin{pmatrix}
    0 &  u_F\\
    u_F^* &  0
  \end{pmatrix}
\; , \qquad 
\text{sgn}(\hat{h}) 
\;=\; 
\begin{pmatrix}
    0 &  \hat{u}\\
    \hat{u}^* &  0
  \end{pmatrix}
\;,
$$
with the matrices being in the grading of $J$ given in \eqref{eq-JDef}. The partial isometry $\hat{u} \in P \Nn_{\Halfspaceaction}P$ is connected to the projection $\Pi_{\Ker(\hat{h})} $ onto the kernel of $\hat{h}$ by
$$
\hat{u}^*\hat{u} \oplus \hat{u} \hat{u}^* 
\;=\; 
P -\Pi_{\Ker(\hat{h})} 
\;=\; 
P\,-\,
\begin{pmatrix}
    \hat{p}_{0,+} &  0\\
    0 &  \hat{p}_{0,-}
\end{pmatrix}
\;.
$$
Otherwise stated, 
\begin{equation}
\label{eq:polardecomp}
\hat{p}_{0,+}
\;=\; 
P - \hat{u}\hat{u}^* 
\;,
\qquad
\hat{p}_{0,-} 
\;=\; 
P - \hat{u}^*\hat{u}
\;.
\end{equation}
The following result now establishes a link between the bulk winding number $\Ch_{\Tt,\Halfspaceaction}(u_F)$ and the kernel of $\hat{h}$. There are two non-trivial assumptions that will be discussed in the following. 

\begin{proposition}
\label{prop:flat_band_suff}
Let as above $h\in M_{N}(\Mm )$ and $\hat{h} \in M_{N}(P \Nn_{\Halfspaceaction} P)$ be self-adjoint Hamiltonians satisfying the {\rm CH}. Suppose that $\Ker(h)=\{0\}$ and:
\begin{enumerate}
\item[{\rm (i)}]
The Fermi unitary $u_F$ is an element of $W^1_p(\Mm)$ for some $1<p \leq 2$.
\item[{\rm (ii)}]
The off-diagonal part $\hat{u}\in M_N(P \Nn_{\Halfspaceaction} P)$ from the polar decomposition of $\hat{h}$ is a compact perturbation of $P u_F P$, namely
\begin{equation}
\label{eq:hatu_compact_perturbation}
\hat{u} \,-\, P u_F P  \;\in \;\calK_{\hat{\Tt}_\Halfspaceaction}
\;.
\end{equation}
\end{enumerate}
Then
$$
\hat{\Tt}_\Halfspaceaction\big(J\, \Ker(\hat{h})\big) 
\;=\; \Ch_{\Tt,\Halfspaceaction}(u_F)
\;.
$$
\end{proposition}

\noindent {\bf Proof.}
Note that $P$ is (up to the isomorphism $\hat{\pi}_{\Tt,G}$) the positive spectral projection $\PP$ of the Dirac operator $\DD\sim {X}_\Halfspaceaction$ as it appears in the Fredholm module that localizes $\Ch_{\Tt,\Halfspaceaction}$ in $\Mm  \rtimes_\Halfspaceaction G$. Hence the Sobolev index theorem (Theorem~\ref{theo-Index})  states 
$$
\Ch_{\Tt,\Halfspaceaction}(u_F) 
\;=\; -\,
\hat{\Tt}_\Halfspaceaction\mbox{-}\Ind(P u_F P\, +\, \one-P) 
\;.
$$
Due to condition (ii) and the stability of the $\hat{\Tt}_\Halfspaceaction$-Breuer index under $\hat{\Tt}_\Halfspaceaction$-compact perturbation, this index is given by
$$
\Ch_{\Tt,\Halfspaceaction}(u_F) 
\;=\; -\,
\hat{\Tt}_\Halfspaceaction\mbox{-}\Ind(\hat{u} + \one-P) 
\;.
$$
The r.h.s. can also be viewed as the index $\hat{\Tt}_\Halfspaceaction\mbox{-}\Ind(P\hat{u}P)$ in the algebra $P \Nn_{\Halfspaceaction} P$ which has $P$ as the identity. Then \eqref{eq:polardecomp} shows that $\hat{u}$ is  partial isometry which is  a compact perturbation of the identity. Hence $\hat{p}_{0,\pm}$ are trace-class and the Calderon-Fedosov formula (Theorem~\ref{prop:fedosov}) implies
$$
\Ch_{\Tt,\Halfspaceaction}(u_F) 
\;=\; 
-(
\hat{\Tt}_\Halfspaceaction(P - \hat{u}^*\hat{u}) \;-\; \hat{\Tt}_\Halfspaceaction(P - \hat{u}\hat{u}^*))
\;=\; 
\hat{\Tt}_\Halfspaceaction(\hat{p}_{0,+}) \;-\; \hat{\Tt}_\Halfspaceaction(\hat{p}_{0,-})
\;.
$$
The r.h.s. is just a rewriting of $\hat{\Tt}_\Halfspaceaction(J\, \Ker(\hat{h}))$. 
\hfill $\Box$

\vspace{.2cm}

Before discussing the assumptions of Proposition~\ref{prop:flat_band_suff}, let us show that the signed surface density of states can be computed as an almost sure quantity. As the bulk Chern number is also an almost sure quantity, the equality in Proposition~\ref{prop:flat_band_suff} actually holds almost surely pointwise. The case $d=1$ of the following result was already analyzed in \cite{ShapiroGraf2018}, though in a slightly different setting. 

\begin{proposition}
\label{prop-AlmostSureSurface}
Assume the situation of {\rm Proposition~\ref{prop:flat_band_suff}}.
\begin{enumerate}
\item[{\rm (i)}] If $d>1$ and $\hat{p}_{0,\pm}= \int_{\Omega\times \hat{G}}^\oplus  \bbP(\difd{\omega})\,\hat{\mu}(\difd{r})\;(\hat{p}_{0,\pm})_{\omega,r}$, one has
$$
\hat{\Tt}_\Halfspaceaction\big(J\, \Ker(\hat{h})\big) 
\;=\; \hat{\Tt}_{\omega,r}\big((\hat{p}_{0,+})_{\omega,r}-(\hat{p}_{0,-})_{\omega,r}\big) 
\;,
$$
almost surely w.r.t. $\bbP$ and the Haar measure of $\hat{G}$. In particular, if the weak Chern number does not vanish, then the physical Hamiltonians $\hat{\pi}_{\Tt,G}(\hat{h})_{\omega,r}$ almost surely have a non-trivial infinitely degenerate kernel.

\item[{\rm (ii)}] 
If $d=1$ and $\hat{p}_{0,\pm}= \bigoplus_{r\in \bbZ} \int_{\Omega}^\oplus\bbP(\difd{\omega})\, (\hat{p}_{0,\pm})_{\omega,r} $, then $\bbP$-almost surely
$$
\hat{\Tt}_\Halfspaceaction\big(J\, \Ker(\hat{h})\big) 
\;=\; 
\Tr\big(J\, \Ker(\hat{h}_{\omega,r})\big) 
\;.
$$
Hence $\hat{h}_{\omega,r}$ almost surely has a non-trivial finite-dimensional kernel if the index does not vanish. 
\end{enumerate}
\end{proposition}

\noindent {\bf Proof.} The claim for $d>1$ is a direct consequence of the Propositions~\ref{prop-rationaltrace} and \ref{prop-irrationaltrace} combined with Corollary~\ref{coro-almostsure}. In the one-dimensional case, the fibers for different $r$ are unitarily equivalent and hence one can assume $r=0$ and drop the subscript (which corresponds to using the representation $\hat{\pi}_\Tt$ from above). By Proposition~\ref{prop-dim1fredholm},  $\hat{u}_\omega$ is almost surely the polar decomposition of $\hat{h}_\omega$ and hence
$$
\Tr\big(J\, \Ker(\hat{h})_\omega\big) 
\;=\; 
\Ind(\hat{u}_\omega \,+\, \one - P)
\;.
$$
Proposition~\ref{prop-dim1fredholm} also shows that $\hat{u}_\omega + \one -P$ and $P u_\omega P + \one - P$ are almost surely Fredholm and that $\hat{u}_\omega - P (u_F)_{\omega} P$ is almost surely a compact operator.  Hence almost surely 
$$
\Ind(\hat{u}_\omega + \one - P) 
\;=\;  
\Ind\big(P (u_F)_\omega P + \one - P\big)
$$
and 
$$
\hat{\Tt}_\Halfspaceaction\mbox{-}\Ind(T_u) 
\;=\; 
\int_{\Omega} \, \Ind\big(P (u_F)_{\omega} P + \one - P\big)
\;\bbP(\difd\omega)\;.
$$
It is enough to show that the index on the r.h.s. is almost surely constant. Since $\bbP$ is ergodic under translations, one only needs to check that it is invariant under the transformation $T_{e_1}$. Due to covariance, 
$$
P (u_F)_{T_{e_1}(\omega)} P 
\;=\; 
S P (u_F)_{T_{e_1}(\omega)} P S^* \;+\; |0\rangle\langle 0| (u_F)_{T_{e_1}(\omega)}| 0 \rangle \langle 0|
\;,
$$
with $S$ the right shift on the lattice. By the invariance of the Fredholm index under unitary conjugation and compact perturbations, one thus has
$$
\Ind\big(P (u_F)_{T_{e_1}(\omega)} P + \one - P\big)
\;=\;
\Ind\big(P (u_F)_{\omega} P + \one - P\big)
\;.
$$
This concludes the proof.
\hfill $\Box$

\vspace{.2cm}

Now let us discuss the two conditions in Proposition~\ref{prop:flat_band_suff}.
The first condition (i) requires the bulk Fermi unitary $u_F$ to have sufficient Sobolev regularity for the Sobolev index theorem for a winding number. This is clearly satisfied in presence of a bulk spectral gap.  Sufficient to assure this Besov regularity is also a MBGR (see Proposition~\ref{prop-MBG}) or a pseudogap of the DOS at $E_F=0$  (see Corollary~\ref{coro-BesovPseudogap}). 
Summing up, condition (i) is already dealt with in a somewhat satisfactory manner.

\vspace{.2cm}

The main object of the remainder of this section is hence to establish condition (ii) of Proposition~\ref{prop:flat_band_suff}. The condition states that the Fermi projections of $h$ and $\hat{h}$ differ only by a compact term, {\it i.e.} one that is essentially localized at the boundary. For differences of smooth functions of the Hamiltonian this follows from similar arguments as proof of Proposition~\ref{prop-index_map_bb_sm}, however, the functional calculus now involves the sign function which is not smooth. Before starting with analytical arguments, let us provide a natural interpretation of the condition~\eqref{eq:hatu_compact_perturbation} in terms of an alternative exact sequence that can be viewed as another bulk-boundary exact sequence and will be construct next. In the classical theory of Toeplitz operators, one can study Toeplitz operators with so-called quasi-continuous symbols \cite{Connes94}. They are characterized by having a compact commutator with the Szego projection. For the present setting, this corresponds to the $C^*$-algebra 
$$
Q 
\;=\; 
\big\{a \in \Mm \;: \; [P,a] \,\in\, \calK_{\hat{\Tt}_\Halfspaceaction}\big\}
\;.
$$
One can then introduce a larger version of the Toeplitz extension by taking the $C^*$-algebra $T \subset \Nn_{\Halfspaceaction}$ spanned by the restrictions of $Q$, namely $T= P Q P + P \calK_{\Tt}P$. It can be shown that no element of the form $P aP$ with $a\in \Mm $ can be $\hat{\Tt}_\Halfspaceaction$-compact and hence there is a well-defined symbol map $q: PaP + P\calK_{\hat{\Tt}_\Halfspaceaction}P \mapsto a$ and an exact sequence
\begin{equation}
\label{eq-ClassToepl}
0 
\;\to\; 
P\calK_{\hat{\Tt}_\Halfspaceaction}P 
\;\to\; 
T 
\;\stackrel{q}{\to}\;  
Q 
\;\to \;
0
\;.
\end{equation}
Hence \eqref{eq:hatu_compact_perturbation} states that $\hat{u}$ is a lift of $u_F$ and therefore 
$$
\Ind[u_F] 
\;=\;
[\hat{p}_{0,+}]_0 \,-\, [\hat{p}_{0,-}]_0 
\;\in \;
K_0(P \calK_{\hat{\Tt}_\Halfspaceaction}P)\,\simeq\, K_0(\calK_{\hat{\Tt}_\Halfspaceaction})
$$ 
for the index map associated to the exact sequence \eqref{eq-ClassToepl}.

\vspace{.2cm}

The first sufficient condition for (ii) of Proposition~\ref{prop:flat_band_suff} is the presence of a bulk gap:

\begin{proposition}
\label{prop-hInvertible}
Let $h \in M_N(\Mm )$ and $\hat{h}\in M_N(P \Nn_{\Halfspaceaction} P)$ be Hamiltonians for which the {\rm CH} holds such that $h \in W^1_{1+\delta}(\Mm ,\rho)$ for some $\delta > 0$ and that $\hat{h}- PhP \in \calK_{\hat{\Tt}_\Halfspaceaction}$. If $0 \notin \sigma(h)$, then the conditions {\rm (i)} and {\rm (ii)} of {\rm Proposition~\ref{prop:flat_band_suff}} are satisfied.
\end{proposition}

\noindent{ \bf Proof.}
The Sobolev condition on $h$ together the bulk gap implies the Besov condition on $u_F$, namely condition (i) is satisfied.
As $h$ is invertible, one also has $h \in W^1_{1+\delta}(\Mm ,\rho)\subset B^1_{1,1}(\Mm,\rho)$ and thus $[P, h] \in L^1(\Nn_{\Halfspaceaction})$ by the Peller criterion (Theorem~\ref{theo-HankelBesovP}). This  implies $[P, f(h)] \in \calK_{\hat{\Tt}_\Halfspaceaction}$ for every function $f$ that is continuous on $\sigma(h)$. If one considers the polar decomposition $\sgn(\hat{h})$ with functional calculus in $P \Nn_{\Halfspaceaction} P$, then in the Calkin algebra
\begin{align*}
P \abs{h} \sgn(h) P + \calK_{\hat{\Tt}_\Halfspaceaction} 
&
\;=\; P h P \;+\; \calK_{\hat{\Tt}_\Halfspaceaction} 
\\
&
\;=\; \hat{h} \;+ \calK_{\hat{\Tt}_\Halfspaceaction} 
\\
&
\;=\;
 \lvert \hat{h}\rvert\, \sgn(\hat{h}) \;+\; \calK_{\hat{\Tt}_\Halfspaceaction}\\
&
\;=\; 
\lvert P h P\rvert \, \sgn(\hat{h})\;+\; \calK_{\hat{\Tt}_\Halfspaceaction} 
\\
&
\;=\;  P\lvert h\rvert P \, \sgn(\hat{h})\;+\; \calK_{\hat{\Tt}_\Halfspaceaction}
\;.
\end{align*}
Multiplying from the left with $P\abs{h}^{-1}P$ then gives
$$
P\, \sgn(h)\,P \,+\,  \calK_{\hat{\Tt}_\Halfspaceaction} 
\;=\; 
\sgn(\hat{h}) \,+\, \calK_{\hat{\Tt}_\Halfspaceaction}
\;,
$$
which implies the claim.
\hfill $\Box$

\vspace{.2cm}

Let us now investigate the condition (ii) in Proposition~\ref{prop:flat_band_suff} in the absence of a bulk gap. 
As already stressed above, this is more delicate than the argument in  Proposition~\ref{prop-index_map_bb_sm} because measurable functional calculus is needed. There is no purely algebraic argument that shows that the difference $\sgn(\hat{h})-P\, \sgn(h)P$ is $\hat{\Tt}_\Halfspaceaction$-compact. In fact, a counterexample due to Krein shows that even a rank-one perturbation of a self-adjoint operator can result in non-compact differences of spectral projections \cite{Kostrykin}.

\vspace{.2cm}

The functional calculus will again be carried out using the analytic approximations from Lemma~\ref{lemma:contours}. This allows to use the geometric resolvent identity to write the approximate sign functions $\sgn_\epsilon(\hat{h})$ as compact perturbations of $P\sgn_\epsilon(\hat{h})P$ since the resolvent differences are $\hat{\Tt}$-compact. However, since the operator norm of the resolvents always diverges at the same rate when approaching the spectrum, the strong limit of $\sgn_\epsilon(\hat{h})-P\, \sgn_\epsilon(h)P$ is not approximated uniformly in operator norm and can therefore fail to be compact in general. Recalling $L^p(\Nn_\Halfspaceaction)\subset \Kk_{\hat{\Tt}_\Halfspaceaction}$ we therefore shift to estimating the $L^p$-(quasi-)-norms which tend to be better behaved close to the spectrum. We will be lead to conditions on the $L^p$-smoothness of the bulk Hamiltonian and its resolvent which can be checked for physically relevant examples.

\vspace{.2cm}

In the remainder of the section we impose the following finite-range condition on the Hamiltonian and the boundary term for technical simplicity (though the conditions could be relaxed to polynomial decay as in Lemma~\ref{eq:lemma_correlatordecay} without much difficulty):

\begin{definition} 
\label{def-HalfSpaceHamConst}
Let $h \in M_N(\Mm )$ and $\hat{h} \in M_N(P \Nn_{\Halfspaceaction} P)$ be self-adjoint and related by
$$
\hat{h} 
\;=\;
P h P \,+\, \tilde{k}
\;,
$$
with $\tilde{k} \in M_N(P \Nn_{\Halfspaceaction} P)$. Then $h$ and $\hat{h}$ are said to satisfy the finite hopping range conditions if there is some $m>0$ such that
$$
P h (\one -P) \;+\; \tilde{k} 
\;=\; 
(P h (\one -P) \;+\; \tilde{k})\,P_{[-m,m]}
\;,
$$
where $P_{[-m,m]}=\chi(\hat{X}_\Halfspaceaction \in [-m,m])$, namely $h$ has finite hopping range and $\tilde{k}$ is supported by a finite strip. 
\end{definition}

The first sufficient condition for \eqref{eq:hatu_compact_perturbation} from Proposition~\ref{prop:flat_band_suff} that also applies without a spectral gap  is a sufficiently steep pseudogap of $h$. Then Proposition~\ref{prop:resolvent_bound} allows to control the $L^2$-norms of the resolvent $\frac{1}{h- z}$ for $z \to 0$:

\begin{proposition}
\label{prop:dos_splitting}
Let $h \in M_N(\Mm )$ and $\hat{h} = 
P h P \,+\, \tilde{k} \in M_N(P \Nn_{\Halfspaceaction} P)$ satisfy the {\rm CH} and the finite hopping range conditions. If $h$ has a pseudogap at $0$ of order $\gamma>2$, then for $p>\gamma$ 
$$
\sgn(\hat{h})\,-\,P\, \sgn(h)\, P 
\;\in\; L^p(P\Nn_{\Halfspaceaction} P)
\;.
$$
Thus condition {\rm (ii)} of {\rm Proposition~\ref{prop:flat_band_suff}} holds.
\end{proposition}

\noindent {\bf Proof.} 
Let us decompose $\gamma = p + s$ with $s >0$ and $p\geq 2$.  For the functional calculus one again uses Lemma~\ref{lemma:contours} to write 
\begin{equation}
	\label{eq-SgnRep}
	\sgn(\hat{h})
	\;=\; 
	\slim_{\epsilon\downarrow 0}\; \sgn_{\epsilon}(\hat{h}) 
	\;=\; 
	\frac{1}{2\pi \imath} \slim_{\epsilon\downarrow 0} \sum_{\sigma\in\{-,+\}}  \int_{\Cc^\sigma_\epsilon} \frac{1}{\hat{h}-z} \;\difd{z}\;,
\end{equation}
and notes that this holds without hypothesis on the kernel of $\hat{h}$.  The same representation \eqref{eq-SgnRep} holds also for $\sgn(h)$. Using the geometric resolvent identity \eqref{eq-GeomResol}, one can thus write the operator difference $\Delta_\epsilon = \sgn_\epsilon(\hat{h}) - P\, \sgn_\epsilon(h)P$ as 
\begin{equation}
\label{eq-cRep}
\Delta_\epsilon  
\;=\;
\frac{1}{2 \pi \imath}\sum_{\sigma\in\{-,+\}}
\int_{\calC^\sigma_\epsilon}  \frac{P}{\hat{h}-zP} \,(P h(\one-P)  - k)\,\frac{1}{h-z}\, P\; \difd{z}
\;.
\end{equation}
Let us set $V=P h(\one-P)  - \tilde{k}$. Then by the short range hypothesis, there is some $m \in \bbN$ with $V P_{[-m,m]}= V$. Inserting $P_{[-m,m]}$ leads to a $\hat{\Tt}_\Halfspaceaction$-traceclass element on the r.h.s. because by Propositions~\ref{prop:lp-embedding} and \ref{prop:resolvent_bound}
\begin{equation}
\label{eq:dos_splitting_tech1}
\Big\| P_{[-m,m]}\, \frac{1}{h-z}  \Big\|_p
\;\leq\; 
C_p \norm{ P_{[-m,m]}}_p\,\Big\|\frac{1}{h-z} \Big\|_p
\;, 
\qquad \forall \;2 \leq p < \gamma
\;.
\end{equation}
The integrand in \eqref{eq-cRep} is analytic in the Banach space $L^p(\Mm)$ at any point $z \in \calC_\epsilon^\sigma$ and hence the integral exists as a convergent Riemann integral in the $L^p$-norm. Let us now set $g(z) = \frac{1}{h-z} - \frac{1}{h}$ and note that 
\begin{equation*}
\begin{split}
 \Delta_\epsilon 
 &
 \;=\; \frac{1}{2 \pi \imath}\sum_{\sigma\in\{-,+\}}
 \int_{\calC^\sigma_\epsilon}   \,\frac{P}{\hat{h}-zP} \,V \,P_{[-m,m]} \,\left(g(z)\,+\,\frac{1}{h}\right) P \;\difd{z}\\
 &
\;=\; 
\sgn_\epsilon(\hat{h}) \,V\, P_{[-m,m]}\,\frac{1}{h}\,P 
\;+\; \frac{1}{2 \pi \imath}\sum_{\sigma\in\{-,+\}}
\int_{\calC^\sigma_\epsilon}  \,\frac{P}{\hat{h}-zP} \,V\, P_{[-m,m]}\,g(z) \,P\;\difd{z}
\;,
\end{split}
\end{equation*}
where the $L^p$-Riemann integral can be evaluated since the $z$-dependent part converges in $L^\infty$-norm. The first term converges in the $L^p(\Nn_{\Halfspaceaction})$-quasi-norm (see Lemma \ref{lemma:convergence}(v)) and the same has to be shown for the second term. Most of the straight line parts converge simply since the Riemann integrals exist due to analyticity of the resolvent. The only critical parts are the four line segments that approach $z=0$. Each of them is bounded in the same way, so let us only consider
$$
\int_\epsilon^1\, f(z)\; \frac{\difd{z} }{2 \pi} 
\qquad
\mbox{ with }
\;\;
f(z)\;=\;
\frac{P}{\hat{h}-\imath\, zP} \,V\, P_{[-m,m]}\, \left( \frac{1}{h-\imath \,z}\;-\;\frac{1}{h} \right)\, P
\;.
$$
By Proposition~\ref{prop:resolvent_bound} we may choose some $0 < r <s$ such that there is a constant $K$ with
\begin{equation}
\label{eq:dos_splitting_tech2}
\norm{f(z)}_p 
\;\leq\;
\frac{1}{\abs{z}}\, \norm{V}\, \norm{P_{[-m,m]}}_p \, K \abs{z}^{\tilde{s}}
\;,
\qquad
z>0
\;,
\end{equation}
for $\tilde{s}=\min\{\frac{r}{p}, 1\}$. Hence the integral converges absolutely in $L^p$-norm for $\epsilon \to 0$.

\vspace{.1cm}

In consequence, the Riemann integral defining $\Delta_\epsilon$ converges in the $L^p$-norm and since it is also bounded uniformly in operator norm by $\norm{\Delta_\epsilon} \leq 2$, its limit coincides with its strong limit $\sgn(h)-\sgn(\hat{h})$ by Lemma~\ref{lemma:convergence}(iii). Hence $\sgn(\hat{h})-P \,\sgn(h)P \in L^p(\Nn_{\Halfspaceaction})\cap \Nn_{\Halfspaceaction}$. The last claim $\sgn(\hat{h})-P \,\sgn(h)P \in \calK_{\hat{\Tt}_\Halfspaceaction}$ follows from $\calK_{\hat{\Tt}_\Halfspaceaction} = \conjugate {L^\infty(\Nn) \cap L^q(\Nn)}$.
\hfill $\Box$

\vspace{.2cm}

The sufficient condition on the DOS is satisfied for Dirac- or Weyl-semimetals in dimension $d\geq 3$, while it fails marginally in $d=2$, due to the DOS scaling $\nu_h([-\epsilon,\epsilon]) \sim \epsilon^{d}$, see \eqref{eq-DWScaling}. The main obstruction to the extension of the proof above to less restrictive conditions on the DOS (possibly only its H\"older continuity at $0$) stems from the fact that Proposition~\ref{prop:lp-embedding} fails for $p<2$ such that no norm bound on terms of the form $P_{[-N,N]} \frac{1}{h-z}$ can be obtained. This problem can be circumvented by imposing additional smoothness conditions on the resolvent:

\begin{proposition}
\label{prop:dos_splitting2}
Let $h \in M_N(\Mm )$ and $\hat{h} = 
P h P \,+\, \tilde{k} \in M_N(P \Nn_{\Halfspaceaction} P)$ satisfy the {\rm CH} and the finite hopping range conditions.
If $h$ has a pseudogap and the inverse of $h$ in the sense of {\rm Proposition~\ref{prop:resolvent_bound}} satisfies $\frac{1}{h} \in B^{\frac{1}{2}}_{1,1}(\Mm)$ with a resolvent estimate
\begin{equation}
\label{eq:resolvent_besov}
\norm{\frac{1}{h}\,-\,\frac{1}{h-\imath \epsilon}}_{B^{\frac{1}{2}}_{1,1}} \;\leq \;K \,\epsilon^s\;, \qquad \forall \;\epsilon \in (0,1)\;,
\end{equation}
for constants $K,s>0$, then $$\sgn(\hat{h})\,-\,P\, \sgn(h)\, P 
\;\in\; L^p(P\Nn_{\Halfspaceaction} P)\;, 
\qquad \forall \;1 \leq p \leq \infty 
\;
$$ 
and thus condition {\rm (ii)} of {\rm Proposition~\ref{prop:flat_band_suff}} holds.
\end{proposition}

\noindent {\bf Proof.}
This is a direct modification of the proof of Proposition~\ref{prop:dos_splitting} where one now takes $p=1$. Note that the resolvent $(h-z)^{-1}$ for $z \notin \sigma(h)$ is smooth and hence in the Besov space $B^{\frac{1}{2}}_{1,1}(\Mm)$ with a bound on the norm that depends only on the inverse distance from the spectrum due the Combes-Thomas estimate (see Proposition~\ref{prop:spectralgap}). For $z$ close to the spectrum, Proposition~\ref{prop:l1-embedding} and \eqref{eq:resolvent_besov} provide the necessary estimates to obtain replacements for \eqref{eq:dos_splitting_tech1} and \eqref{eq:dos_splitting_tech2}.
\hfill $\Box$

\vspace{.2cm}

Fortunately, the technical additional condition \eqref{eq:resolvent_besov} on the regularity of the resolvent can also be derived as a consequence of the pseudogap:

\begin{proposition}
	\label{prop:dirac2d}
	Assume that $h$ is smooth and has a pseudogap of order $\gamma>1$ at $E=0$. Then:
	\begin{itemize}
		\item[{\rm (i)}] $\frac{1}{h} \in B^{s}_{1,1}(\Mm)$ for each $0 < s < \gamma-1$.
		\item[{\rm (ii)}] For any $0 < s < \gamma-1$, there exists some $0 < \tilde{s} < 1$ for which there is a constant $C>0$ such that
		\begin{equation}
			\label{eq:besovresolventbound}
			\norm{\frac{1}{h} \,-\, \frac{1}{h-\imath \epsilon}}_{B^{s}_{1,1}} \;\leq\; C \abs{\epsilon}^{\tilde{s}} 
		\end{equation}
		for $\epsilon$ small enough.
	\end{itemize} 
\end{proposition} 

The following result is needed as a preparation for the proof.

\begin{lemma}
\label{lem-SmoothHSmoothF} Let $h$ be smooth and $g:\RM\to\CM$ a smooth function. Then for all $p,q\geq 1$ and $0 < s < l$ with $l\in \bbN$ there is a constant $C_{s,l}$ independent of $g$ such that
$$
\|g(h)\|_{B^s_{p,q}}\;< c_{s,l} \sum_{m=1}^{l+1} \norm{g^{(m)}}_\infty 
\;.
$$ 
\end{lemma} 

\noindent {\bf Proof.} The functional calculus is written out using the Dynkin-Helffer-Sj\"ostrand formula 
\begin{equation}
\label{eq-DHS2}
g(h) 
\;=\; 
\frac{1}{2 \pi}\int_D (\partial_{\overline{z}}\tilde{g}_K)(z) \,\frac{1}{h-z} \, \difd{z}
\;,
\end{equation}
with $\tilde{g}_K$ a pseudo-analytic continuation of $g$ satisfying $\abs{\tilde{g}_K(z)} \leq C_K \abs{\Im m(z)}^K$ where $K \in \bbN$ can be chosen arbitrarily large and $D$ is an open set such that $\sigma({h}) \subset D \subset \bbC$. By iteration of the identity 
$$
\nabla_i \frac{1}{h-z} 
\;=\; 
-\,\frac{1}{h-z} (\nabla_i h) \frac{1}{h-z} 
\;,
$$ 
one finds that 
$$
\norm{\nabla^j \frac{1}{h-z}} 
\;\leq\; 
\sum_{m=1}^{\abs{j}+1} \frac{1}{\abs{\Im m(z)}^{m+1}} \left(1+\norm{h}_{W_\infty^{\abs{j}-m}}\right)^{\abs{j}}
$$
for any multi-index $j$ and the norms
$$
\|x\|_{W_\infty^l} 
\;=\;
\sum_{0 \leq \abs{j}\leq l} \norm{\nabla^j x}
\;.
$$
Choosing $K \geq \abs{j}+1$ one therefore has with a norm-convergent integral
$$
\nabla^j g(h) 
\;=\; 
\frac{1}{2 \pi}\int_D (\partial_{\overline{z}}\tilde{g}_K)(z) \,\nabla^j \frac{1}{h-z} \, \difd{z}
\;,
$$
since one can construct the continuation to satisfy \cite{Davies95}
$$
\abs{\partial_{\overline{z}}\tilde{g}_K)(z)} 
\;\leq\; 
\sum_{m=1}^{K} c_m \|g^{(m)}\|_\infty \, \abs{\Im m(z)}^m \chi_U(z) + c_{K+1} \|g^{(K+1)}\|_\infty \, \abs{\Im m(z)}^{K} \chi_V(z)
$$
for universal constants and bounded compactly supported functions $\chi_U$, $\chi_V$ of which $\chi_U$ vanishes on a neighborhood of the real line. Hence
$$
\norm{\nabla^j g(h)} 
\;\leq\; 
C_j \left( \sum_{m=1}^{l+1} \norm{g^{(m)}}_\infty \right) \left(\sum_{m=1}^{\abs{j}+1} \frac{1}{\abs{\Im m(z)}^{m+1}} \left(1+\norm{h}_{W_\infty^{\abs{j}-m}}\right)^{\abs{j}}\right)
$$
and since the trace is finite the same estimate applies to $\norm{\nabla^j g(h)}_p$ for any $0 < p <\infty$. The conclusion is therefore provided by Proposition~\ref{prop-besov-sufficient}.
\hfill $\Box$

\vspace{.2cm}

\noindent{\bf Proof} of Proposition~\ref{prop:dirac2d}. (i) By assumption, there is some $M>0$ such that $\nu_h((-\epsilon,\epsilon))\leq C \epsilon^\gamma$ for all $0 \leq \epsilon \leq M$.  For $\gamma\geq 2$ the result follows directly from Proposition~\ref{prop-besov-sufficient} since the resolvent is Sobolev-differentiable $h^{-1} \in W^1_1(\Mm)$  by the proof of Proposition~\ref{prop:pseudogap_sufficient}. For $\gamma < 2$, however, one cannot expect any (integer-)Sobolev norm of $(h-\imath \epsilon)^{-1}$ to stay finite as $\epsilon \to 0$. This makes it more challenging to establish the  estimate.

\vspace{.1cm}

Choose a function $\chi_0 \in C^\infty_0(\RM)$ supported on  $(-2,-\frac{1}{4})\cup (\frac{1}{4},2)$, which is equal to $1$ on $(-1,-\frac{1}{2})\cup (\frac{1}{2},1)$ and such that the scaled functions
$$
\chi_{k}(\lambda)
\;=\; 
\chi_0(2^{-k-1} M\, \lambda)
\;,
\qquad
k\in\ZM\,,\;\;k\leq 0\,,
$$ 
satisfy $\sum_{k=-\infty}^0 \chi_{k}(\lambda)=1$ for all $0 < \abs{\lambda}\leq M$. Such a dyadic decomposition always exists (see Section~\ref{sec-BesovDef}). Now set
$$
\chi(h)
\;=\;
\sum_{k=-\infty}^0 \chi_{k}(h)
\;,
\qquad
\chi^c(h)
\;=\;
\one\,-\, \sum_{k=-\infty}^0  \chi_{k}(h)
\;.
$$
Then $\chi^c$ can be taken to be a smooth function since $0$ is not an eigenvalue of $h$ and
$$
\one
\;=\; 
\chi(h) + \chi^c(h)
\;.
$$
In the following let us further abbreviate $\chi_k = \chi_k(h)$. By assumption on the density of states combined with functional calculus there are constants $C_1$ and $C_2$ such that uniformly in $-\infty < k \leq 0$ 
\begin{equation}
	\label{eq:dirac_scaling}
	\norm{\chi_{k}\,\frac{1}{h}}_1 
	\;\leq\; 
	C_1 \,2^{(\gamma-1) k}\;, 
	\qquad \norm{\chi_{k} \,\frac{1}{h}}_\infty 
	\;\leq\; 
	C_2\, 2^{-k}
	\;.
\end{equation}
Observe that
\begin{equation}
	\label{eq:cutoff_scaling}
	\norm{\nabla_i \chi_k} \leq 2^{-k} C_3 \norm{\nabla_i h}
\end{equation}
for some constant $C_3$ depending only on the function $\chi_0$ since $\chi_k = \chi_0(2^{-k} h)$ can be written with the Dynkin-Helffer-Sj\"ostrand formula and differentiated under the integral sign
$$
\nabla_i \chi_k
\;=\; 
-\frac{1}{2 \pi}\int_D (\partial_{\overline{z}}\widetilde{\chi}_0)(z) \,\frac{1}{2^{-k} h-z}\, (2^{-k} \nabla_i h) \,\frac{1}{2^{-k} h-z} \, \difd{z}
$$
for $D\subset \CM$ compact and $\widetilde{\chi}_0$ a suitable quasi-analytic extension of $\chi_0$, satisfying at least $\abs{(\partial_{\overline{z}}\widetilde{\chi}_0)(z)}\leq C \abs{\Im m(z)}^2$. 

\vspace{.1cm}

The goal is to bound
\begin{equation}
	\label{eq:torusdecomp}
	\norm{f(h)}_{B^{s}_{1,1}} 
	\;\leq \;
	\norm{\chi(h) \,f(h)}_{B^{s}_{1,1}} 
	\;+\; 
	\norm{\chi^c(h) \, f(h)}_{B^{s}_{1,1}}
	\;,
\end{equation}
for functions $f$ of $h$ which are smooth everywhere except possibly at $0$. As $\chi^c \, f$ is a smooth function the Besov-norms of $\chi^c(h) f(h)$ are all finite by Lemma~\ref{lem-SmoothHSmoothF}. Let us hence focus on the singular part $\chi(h) f(h)$. From Lemma~\ref{lem:diff_besov} and for $W_\ScaleInd (t) = \varphi(2^{-\ScaleInd } |t|)$ as in \eqref{eq-WkChoice}, one has that 
$$
\norm{\widehat{W}_j*a}_1 
\;\leq\; 
2^{-j} C \sum_{i=1}^d \norm{\nabla_i a}_1 
\;,
$$
uniformly for $j>0$. Let us use the notation $\norm{a}_{\dot{W}_1^1}=\sum_{i=1}^d \norm{\nabla_i a}_1$. Then for arbitrary $0 < \delta < 1$ 
\begin{align}
&
\!\!\!\!\!
\norm{\chi(h) \,f(h)}_{{B}^{s}_{1,1}} 
\nonumber
\\
&
\;\leq\; 
\norm{\chi(h) \,f(h)}_1 \;+\; \sum_{j=1}^\infty \sum_{k=-\infty}^0 2^{js} \norm{\widehat{W}_j*(\chi_k \,f(h))}_{1} \nonumber \\
&
\;=\; 
\norm{\chi(h) \,f(h)}_1 \;+\; \sum_{j=1}^\infty \sum_{k=-\infty}^0 2^{js} \norm{\widehat{W}_j*(\chi_k \,f(h))}_{1}^\delta \norm{\widehat{W}_j*(\chi_k \,f(h))}_{1}^{1-\delta} \nonumber \\
&
\;\leq\; 
\norm{\chi(h) \,f(h)}_1 \;+\; C_\delta \sum_{j=1}^\infty \sum_{k=-\infty}^0 2^{j(s-\delta)}  \norm{\chi_k \,f(h)}_{\dot{W}_1^1}^\delta \norm{\chi_k \,f(h)}_{1}^{1-\delta}
\;. 
\label{eq:besovnormdecomp}
\end{align}
Note that the shifted cutoff functions $\tilde{\chi}_{k}=\chi_{k-1}$ satisfy $\chi_{k}=\chi_{k}\,\tilde{\chi}_{k}$. One can differentiate $f(h)=h^{-1}$ using the Leibniz identity $\nabla_i \frac{1}{h} = -\frac{1}{h+\imath \epsilon} (\nabla_i h) \frac{1}{h+\imath \epsilon}$ which extends to the limit $\epsilon=0$ as long as there is a smooth cut-off excluding the singularity. Hence
\begin{align*}
	\norm{\nabla_i \chi_{k} \, \frac{1}{h}}_{1} 
	&\;=\; \norm{\nabla_i \chi_{k} \, \frac{1}{h} \tilde{\chi}_{k}}_{1} \\
	&
	\;\leq\; 
	\norm{(\nabla_i \chi_{k}) \,\frac{1}{h}\tilde{\chi}_{k} + \chi_{k} \,\frac{1}{h}(\nabla_i \tilde{\chi}_{k}) \;-\;\chi_{k} \,\frac{1}{h}(\nabla_i h)\frac{1}{h}\tilde{\chi}_{k}}_1 \nonumber \\
	&
	\;\leq\; 
	\norm{\nabla_i\chi_{k}}_\infty\,\norm{\tilde{\chi}_{k} \,\frac{1}{h}}_1 \;+\; \norm{\nabla_i\tilde{\chi}_{k}}_\infty\,\norm{\chi_{k} \,\frac{1}{h}}_1 \;+\;\norm{{\chi}_{k} \,\frac{1}{h} (\nabla_i h)}_\infty\, \norm{\tilde{\chi}_{k} \frac{1}{h}}_1 \nonumber \\
	&
	\;\leq\; 
	\left(4\,C_3\, 2^{-k}  \norm{\nabla_i h}_\infty \;+\; C_2 \,2^{-k} \norm{\nabla_i h}\right) C_1 \,2^{(\gamma-1)k}
	\;,
\end{align*}
where the scaling relations \eqref{eq:dirac_scaling} and \eqref{eq:cutoff_scaling} were used. Thus there is a constant $C$ such that 
$$
\norm{\chi_{k} \, \frac{1}{h}}_{\dot{W}^1_1} 
\;=\; 
\sum_{i=1}^d \norm{\nabla_i \chi_{k} \, \frac{1}{h}}_{1} 
\;\leq\; 
C_4\, 2^{(\gamma-2)k}
\;.
$$
Substituting into \eqref{eq:besovnormdecomp} with $f(h)=h^{-1}$ one gets
\begin{align*}
	\norm{\chi\,\frac{1}{h}}_{\dot{B}^{s}_{1,1}}
	&\;\leq\; \norm{\chi(h) \,\frac{1}{h}}_1 \;+\; \tilde{C}_\delta \sum_{j=1}^\infty \sum_{k=-\infty}^0 2^{j(s-\delta)}  2^{(\gamma-2)k\delta} 2^{(\gamma-1)k(1-\delta)} \\
	&\;=\; \norm{\chi(h) \,\frac{1}{h}}_1 \;+ \;\tilde{C}_\delta \sum_{j=1}^\infty \sum_{k=-\infty}^0 2^{j(s-\delta)}  2^{(\gamma-1-\delta)k}
\end{align*}
and the sum is finite whenever $\gamma-1-\delta > 0$ and $\delta > s$, thus the l.h.s. is finite if $\gamma-1>s$ which completes the proof of part (i) of the proposition.

\vspace{.1cm}

For part (ii) one needs to obtain estimates for $f_\epsilon(h)=\frac{1}{h}-\frac{1}{h+\imath \epsilon}$. By the resolvent identity and the proof of Lemma~\ref{lem-SmoothHSmoothF}  one first checks that $\norm{\chi^c(h) f_\epsilon}_{B^{s}_{1,1}} \leq C \abs{\epsilon}$ for small enough $\epsilon$ since $\chi^c f_\epsilon$ converges uniformly to $0$ with all its derivatives. Moreover, again using the resolvent identity combined with \eqref{eq:dirac_scaling} one obtains bounds that are uniform in $k$ and $\epsilon$:
\begin{align*}
\norm{\chi_{k}\, \frac{1}{h+\imath \epsilon}}_1  
&\;\leq\; 
\norm{\chi_{k}\, \frac{1}{h}}_1 
\;+\; 
\norm{\chi_{k}\, \frac{1}{h}\, \epsilon\, \frac{1}{h+\imath \epsilon}}_1 
\\
&\;\leq\; 
C_1\, 2^{(\gamma-1)k}\left(1+\abs{\epsilon} \, \norm{\frac{1}{h+\imath \epsilon}}_\infty\right) 
\\
&\; \leq\; 
2\, C_1\, 2^{(\gamma-1)k}
\;,
\end{align*}
since $\norm{\frac{1}{h+\imath \epsilon}}_\infty \leq \frac{1}{\abs{\epsilon}}$. Likewise
$$
\norm{\chi_{k} \,\frac{1}{h+\imath \epsilon}}_\infty 
\;\leq \;
2 \,C_2 \,2^{-k}
\;.
$$
For any $0 < \theta < 1$ the H\"older inequality and log-convexity of the $p$-normn (see Appendix~\ref{app-Interpol}) imply
\begin{align*}
	\norm{\chi_{k}\left(\frac{1}{h}-\frac{1}{h+\imath \epsilon}\right)}_1 
	&\;=\; 
	\norm{\chi_{k}\left(\frac{1}{h}\epsilon\frac{1}{h+\imath \epsilon}\right)}_1 
	\\
	&
	\;\leq\;
	\abs{\epsilon}\,\norm{\chi_{k}\frac{1}{h}}_p \norm{\chi_{k}\frac{1}{h+\imath \epsilon}}_q 
	\\ 
	&
	\;\leq\; 
	\abs{\epsilon}\, \norm{\chi_{k}\frac{1}{h}}_1^{1-\theta}\, \norm{\chi_{k}\frac{1}{h}}_\infty^{\theta} \, \norm{\chi_{k}\frac{1}{h+\imath \epsilon}}^{\theta}_1\,
	\norm{\chi_{k}\frac{1}{h+\imath \epsilon}}^{1-\theta}_\infty\\
	&
	\;=\; 
	\abs{\epsilon}(C_1 2^{(\gamma-1)k})^{1-\theta}\, (C_2 2^{-k})^\theta\, (2 C_1 2^{(\gamma-1)k})^{\theta} \abs{\epsilon}^{\theta-1}
	\\
	&
	\;\leq\; 
	C_5 \,2^{(\gamma-1-\theta)k} \abs{\epsilon}^\theta
\end{align*}
for any $1=\frac{1}{p}+\frac{1}{q}$ with $\frac{1}{p}=1-\theta$. Assuming $\theta < \gamma -1$ one can sum over $k$ to obtain $\norm{\chi(h) \left(\frac{1}{h}-\frac{1}{h+\imath \epsilon}\right)}_1 \leq C_6 \abs{\epsilon}^\theta$ which provides a $L^1$-norm estimate sufficient for the inhomogeneous part of the Besov norm. To bound the homogeneous part, let us again estimate the derivatives 
\begin{align*}
	&\left\lVert \nabla_i \,\chi_{k}\left(\frac{1}{h}\,-\,\frac{1}{h+\imath \epsilon}\right)
	\right\rVert_1 
\\
&\;\leq\; 
	\norm{(\nabla_i\chi_{k})\left(\frac{1}{h}\,-\,\frac{1}{h+\imath \epsilon}\right)\tilde{\chi}_{k}}_1 \;+\; \norm{ \chi_{k}\left(\frac{1}{h}-\frac{1}{h+\imath \epsilon}\right)\nabla_i \tilde{\chi}_{k}}_1 \\
	&
	\;\;\;\;\;\;\; + \;\norm{\chi_{k}\left(\frac{1}{h}-\frac{1}{h+\imath \epsilon}\right)(\nabla_i h) \frac{1}{h}\tilde{\chi}_{k}}_1 
	\;+ \;\norm{\chi_{k}\,\frac{1}{h+\imath \epsilon}(\nabla_i h) \left(\frac{1}{h}-\frac{1}{h+\imath \epsilon}\right)\tilde{\chi}_{k}}_1 
	\\
	&
	\;\leq\; \norm{\tilde{\chi}_{k}\left(\frac{1}{h}\,-\,\frac{1}{h+\imath \epsilon}\right)}_1 
	\big(3 C_3 2^{-k} \norm{\nabla_i h}_\infty \;+\; \norm{\nabla_i h}_\infty\, 3\, C_2 2^{-k}\big)
\end{align*}
and hence for any $0< \theta <1$ there is a constant such that
$$
\norm{\chi_{k}\left(\frac{1}{h}-\frac{1}{h+\imath \epsilon}\right)}_{\dot{W}^1_1} \;\leq\; C_7\, 2^{(\gamma-2-\theta)k} \,\abs{\epsilon}^\theta
\;.
$$
Substituting into \eqref{eq:besovnormdecomp} gives
\begin{align*}
&
\!\!\!\!\!\!\!\!\!
\norm{\chi(h)\,f_\epsilon(h)}_{\hat{B}^{s}_{1,1}}
\\
&\leq\;  C_6 \epsilon^{\theta} + 
	\abs{\epsilon}^\theta C_\delta \sum_{j=1}^\infty \sum_{k=-\infty}^0 2^{j(s-\delta)} (C_7 2^{(\gamma-2-\theta)k})^\delta (C_6 2^{(\gamma-1-\theta)k})^{1-\delta}\\
&\leq\;  C_6 \epsilon^{\theta} + 
	\abs{\epsilon}^\theta \tilde{C}_\delta \sum_{j=1}^\infty \sum_{k=-\infty}^0 2^{j(s-\delta)} 2^{(\gamma-1-\delta-\theta)k}
\end{align*}
and if one first picks any $\delta < s < \gamma-1$ and then $0 < \theta < \gamma-1-\delta$ the sum is finite and thus the l.h.s. satisfies the desired inequality.
\hfill $\Box$

\vspace{.2cm}

To apply Proposition~\ref{prop:dos_splitting2} one therefore requires merely finite-hopping range and a pseudogap of order $\gamma > \frac{3}{2}$. In particular, this applies to non-disordered Dirac-semimetals in $d=2$ but also, for example, nodal line semimetals in $d=3$ with linear dispersion, for which the Fermi surface $S \subset \bbT^3$ is now a sufficiently regular $1$-dimensional submanifold and the dispersion relation
$$
\abs{h(k)}
\;\geq \;
c\; \mathrm{dist}(k, S)
$$
holds in a neighborhood of $S$. Hence it appears that the majority of periodic Hamiltonians with pseudogaps are covered by the combination Proposition~\ref{prop:dos_splitting} or Proposition~\ref{prop:dos_splitting2}. Also note that despite the restriction to vanishing disorder in the bulk, the half-space Hamiltonian is allowed to have fairly general disorder at the boundary which is subject only to the conditions of Proposition~\ref{prop:dos_splitting2}.

\vspace{.2cm}

As already explained above, one should not expect the existence of pseudogaps in a disordered regime, unless the disorder respects a chiral symmetry. In this situation, it is helpful to use fractional moments to establish bounds for the $L^p$-quasinorms for some $0 < p < 1$:

\begin{lemma}
\label{lemma:dos_splitting_tech}
Let $h=h^* \in M_N(\Mm)$ and assume that the Aizenman-Molchanov-bound \eqref{eq:exploc} is satisfied in an interval $\Delta$ containing $0$ for some exponent $0<s<1$. For any $0 < p < s$ there are constants $c_1,c_2,c_3$ such that for $P_I = \chi(\hat{X}_\Halfspaceaction\in I)$ and $I \subset \bbR$ a compact interval one has
\begin{align}
\label{eq:am_resolvent}
\norm{P_{I} \,\frac{1}{h-z_1}}_{p} 
\;&\leq\; 
c_1\, \norm{P_{I}}_{p},\\
\label{eq:am_resolvent2}
\norm{P_{I} \left(\frac{1}{h-z_1}-\frac{1}{h-z_2}\right)}_{p} 
\;&\leq\; 
c_2 \abs{z_1-z_2} \, \abs{\Im m (z_1)}^{-\frac{p}{s}}\, \norm{P_{I}}_{p}\\
\label{eq:am_resolvent3}
\norm{P_{I} \frac{1}{(h-z_1)^2}}_{p} 
\;&\leq\; 
c_3 \abs{\Im m (z_1)}^{-\frac{p}{s}}\, \norm{P_{I}}_{p}
\end{align}
uniformly in $z_1,z_2$ with $\mathrm{dist}(z_i, \sigma(h)\setminus \Delta) > \delta$ for some $\delta > 0$.
\end{lemma}

\noindent{\bf Proof.}
Let us note that if \eqref{eq:exploc} holds for some $0 < s < 1$ then it also holds for all fractional exponents $0 < p \leq s$ with adapted constants the reason being  the $L^p$-inclusion for probability spaces resulting from $\mathbb{E}\abs{f}^{p} \leq (\mathbb{E}\abs{f}^s)^{\frac{p}{s}}$.

\vspace{.1cm}
	
Let us write $\frac{1}{h-z} = \sum_{x\in \bbZ^d} \phi_{x,z} u^x$ as a Fourier series with coefficient functions $\phi_{x,z}\in M_N(C(\Omega))\subset M_N(\bbT^d_{\BB,\Omega})$ satisfying $\abs{\phi_{x,z}(\omega)}=\abs{\langle 0| \pi_\omega(\frac{1}{h-z})| x\rangle}$. Since $\phi_{x,z}$ is $\Halfspaceaction$-invariant, it commutes with $\hat{X}_\Halfspaceaction$ and thus $P_{I}$ (and any function of $D_\Halfspaceaction$) such that

\begin{align*}
\norm{P_{I}\phi_{x,z} u^x}^p_{p} 
	\;&=\; \hat{\Tt}_\Halfspaceaction(\left(P_{I}\abs{\phi_{x,z}}^2 P_{I}\right)^{\frac{p}{2}}) 
\\
&
\;=\;  \hat{\Tt}_\Halfspaceaction\left(P_{I}\abs{\phi_{x,z}}^p P_{I}\right) 
\\
&
\;= \;\norm{P_{I}}_1 \Tt(\abs{\phi_{x,z}}^p) \\
	&
	\;=\; 
	\norm{P_{I}}_1 \int_{\Omega} \norm{\langle 0| (\pi_\omega(h)-z)^{-1}| x\rangle}_p^p \,\mathbb{P}(\difd \omega)\\
	&
	\leq \;\norm{P_{I}}_1  \int_{\Omega} N\,\norm{\langle 0| (\pi_\omega(h)-z)^{-1}| x\rangle}^p \,\mathbb{P}(\difd \omega) \\
	&
	\leq\; \norm{P_{I}}_1  N\, A_p e^{-\mu_p \abs{x}}.
\end{align*}
Hence the first inequality follows from
\begin{align*}
\norm{P_{I} \frac{1}{h-z}}^p_{p}\; &\leq\; \sum_{x\in \bbZ^d} \norm{P_{I}\phi_{x,z} u^x}^p_{p} 
\;\leq\; \norm{P_{I}}_p^p \sum_{x\in \bbZ^d} N\, A_p e^{-\mu_p \abs{x}} 
\;<\; 
\infty
\;.
\end{align*}
The remaining estimates are based on the expansion 
\begin{align*}
P_{I} \left(\frac{1}{h-z_1}\, \frac{1}{h-z_2}\right)
&\;=\; \sum_{x,y\in \bbZ^d} P_{I}\,\phi_{x,z_1} u^x \phi_{y,z_2} u^y 
\\
&\;=\; \sum_{x,y\in \bbZ^d} P_{I}\,\phi_{x,z_1} u^x P_{(I-\Halfspaceunitvector\cdot x)} \, \phi_{y,z_2} u^y
\end{align*}
where the projections $P_{(I-\Halfspaceunitvector\cdot x)}$ could be inserted since the $\Halfspaceaction$-spectrum satisfies $\sigma_{\Halfspaceaction}(u^x)= \{ \Halfspaceunitvector\cdot x\}$. Using the H\"older inequality and $\log$-convexity of the $L^p$-(quasi-)norms one has
$$
\norm{ab}_{p} 
\;\leq\; 
\norm{a}_{\frac{ps}{s-p}}\, \norm{b}_{s}
\;\leq\; 
\norm{a}_{p}^{1-\frac{p}{s}} \, \norm{a}_\infty^{\frac{p}{s}} \, \norm{b}_{s}
\;,
$$ 
and hence one obtains
\begin{align*}
&
\!\!\!\!
\norm{P_{I} \left(\frac{1}{h-z_1}\,\frac{1}{h-z_2}\right)}^{p}_{p} \;
\\
&
=\; 
\Big\|
\sum_{x,y\in \bbZ^d} P_{I}\,\phi_{x,z_1} u^x P_{(I-\Halfspaceunitvector\cdot x)} \, \phi_{y,z_2} u^y
\Big\|^{p}_{p}\\
&\leq  \;\sum_{x,y\in \bbZ^d} \norm{P_{I} \,\phi_{x,z_1} u^x P_{(I-\Halfspaceunitvector\cdot x)} \, \phi_{y,z_2} u^y}^{p}_{p} \\
&
\leq \;\sum_{x,y\in \bbZ^d} \norm{P_{I}\,\phi_{x,z_1} u^x }^{p\frac{p}{s}}_{\infty} \norm{P_{I}\,\phi_{x,z_1} u^x }^{p(1-\frac{p}{s})}_{p}\norm{P_{(I-\Halfspaceunitvector\cdot x)} \, \phi_{y,z_2} u^y}^{p}_{s} 
\\
&
\leq  \;\norm{P_I}_{1}\abs{\Im m (z_1)}^{-p\frac{p}{s}} \sum_{x,y\in \bbZ^d} \left(N A_{p}e^{-\mu_{p}\abs{x}} \right)^{1-\frac{p}{s}} \left(N A_{s}e^{-\mu_{s}\abs{y}} \right)^{\frac{p}{s}}\, 
\\
& < \;C  \norm{P_I}_{1}\abs{\Im m (z_1)}^{-p\frac{p}{s}}
\;,
\end{align*}
which implies \eqref{eq:am_resolvent3} and \eqref{eq:am_resolvent2} follows from this and the resolvent identity.
\hfill $\Box$

\begin{proposition}
\label{prop:dos_splitting3}
Let $h \in M_N(\Mm )$ and $\hat{h} = 
P h P \,+\, \tilde{k} \in M_N(P \Nn_{\Halfspaceaction} P)$ satisfy the {\rm CH} and the finite hopping range conditions. Assume that the Aizenman-Molchanov-bound \eqref{eq:exploc} is satisfied in an interval $\Delta$ containing $0$ for some exponent $0<s<1$. Then $$\sgn(\hat{h})\,-\,P\, \sgn(h)\, P 
\;\in\; L^p(P\Nn_{\Halfspaceaction} P)\;, 
\qquad \forall \;p\in (0,\infty] 
\;.
$$ 
Thus condition {\rm (ii)} of {\rm Proposition~\ref{prop:flat_band_suff}} holds.
\end{proposition}

\noindent {\bf Proof.} 
The proof is a modification of that of Proposition~\ref{prop:dos_splitting} to work with the quasi-Banach norms of $L^p(\Nn)$ for any $0 < p < s$, such that one can apply the resolvent estimates from Lemma~\ref{lemma:dos_splitting_tech}. To estimate the contour integrals arising from Lemma~\ref{lemma:contours} it is therefore necessary to use the theory of integration in quasi-Banach spaces outlined in Appendix~\ref{app-p<1}. The main difference to the normed case is that the usual triangle inequality for integrals fails, such that one must instead use a series expansion with term-wise estimates. 

\vspace{.1cm}

Let us again  write the difference $\Delta_\epsilon = \sgn_\epsilon(\hat{h}) - P\, \sgn_\epsilon(h)P$ as 
\begin{equation}
\label{eq-cRep2}
\Delta_\epsilon  
\;=\;
\frac{1}{2 \pi \imath}\sum_{\sigma\in\{-,+\}}
\int_{\calC^\sigma_\epsilon}  \frac{P}{ \hat{h}-zP}\, V\, P_{[-m,m]}\,\frac{1}{h-z}\, P\;\difd{z} 
\;.
\end{equation}
with $V=P h(\one-P)  - k$. Then apply \eqref{eq:am_resolvent} from Lemma~\ref{lemma:dos_splitting_tech} to bound
$$
\Big\| P_{[-m,m]}\, \frac{1}{h-z}  \Big\|_p
\;\leq\; 
c_1 \norm{ P_{[-m,m]}}_p
\;,
$$
uniformly in $z$ with $\mathrm{dist}(z, \sigma(h)\setminus \Delta)>\delta>0$. Choosing $\delta$ small enough the bound holds for all $z$ in a neighborhood of $\bigcup_{\epsilon>0} \Cc^\pm_\epsilon$. This implies that the integrand in \eqref{eq-cRep2} is analytic in the quasi-Banach space $L^p(\Mm)$ at any point $z \in \calC^\pm_\epsilon$ and hence the integral exists as a convergent Riemann integral in the $L^p$-quasinorm. For $z_L= \imath\, 2^{-L}$ set $g_L(z) = \frac{1}{h-z} - \frac{1}{h-z_L}$ and note that 
\begin{equation*}
\begin{split}
 \Delta_\epsilon 
 &
 \;=\; 
 \frac{1}{2 \pi \imath} \sum_{\sigma\in\{-,+\}} \int_{\calC^\sigma_\epsilon}   \,\frac{P}{\hat{h}-zP} \,V \,P_{[-m,m]} \,\left(g_L(z)\,+\,\frac{1}{h-z_L}\right) P \difd{z}\\
 &
\;=\; 
\,\sgn_\epsilon(\hat{h}) \,V\, P_{[-m,m]}\,\frac{1}{h-z_L}\,P 
\;+\; 
\frac{1 }{2 \pi \imath}\sum_{\sigma\in\{-,+\}}\int_{\calC^\sigma_\epsilon}  \,\frac{P}{ \hat{h}-zP} \,V\, P_{[-m,m]}\,g_L(z) \,P\;\difd{z}
\;,
\end{split}
\end{equation*}
where the $L^p$-Riemann integral can be evaluated since the $z$-dependent part converges in $L^\infty$-norm. The first term is uniformly bounded in $\epsilon$ and $L$ w.r.t. the $L^p(\Nn_{\Halfspaceaction})$-quasi-norm (in fact, it is even Cauchy due to \eqref{eq:am_resolvent2}) and the same will now be shown for the second term. Again, most of the straight line parts can be bounded trivially in $\epsilon$ and $L$ simply since the Riemann integrals exist due to analyticity of the resolvent. The only critical parts are the four line segments that approach the spectrum of $\hat{h}$ at $z=0$. Each of them is bounded in the same way, so let us only consider
$$
\int_\epsilon^1\, F_L(z)\; \frac{\difd{z} }{2 \pi} 
\qquad
\mbox{ with }
\;\;
F_L(z)\;=\;
\frac{P}{\imath P z - \hat{h}} \,V\, P_{[-m,m]}\, \left(\frac{1}{h-\imath z} -\frac{1}{h - \imath z_L} \right)\, P
\;.
$$
From this point on, we will choose $\epsilon=2^{-L}$ and bound the integrals uniformly for $L\to \infty$. The integrals are analyzed separately on the dyadic intervals $I_j = (z_{j+1},z_{j})$, $0 \leq j < L$ with $z_j = 2^{-j}$. On $I_j$ one has the convergent series expansion
$$
F_L(z)
\;=\; 
\sum_{n=0}^\infty x^{(j)}_{L,n} \,f^{(j)}_{n}(z)
\;,
$$
with $f_{n}^{(j)}(z) = -(\imath)^n(z-z_j)^n$ and
\begin{align*}
x^{(j)}_{L,n} 
\,&=\,
\sum_{k=0}^n\frac{P}{(\hat{h}-\imath Pz_j)^{k+1}}\, V\, P_{[-m,m]}\left(\frac{1}{(h-\imath z_j)^{n-k+1}} \right) P 
\\
&
\;\;\;\;\;\,-\, 
\frac{P}{(\hat{h}-\imath Pz_j)^{n+1}} \,V\, P_{[-m,m]}\, \frac{1}{h-\imath z_L}\,P\\
&=\,
\sum_{k=0}^{n-1}\frac{P}{(\hat{h}-\imath Pz_j)^{k+1}}\, V\, P_{[-m,m]}\left(\frac{1}{(h-\imath z_j)^{n-k+1}} \right) P \\
&\;\;\;\;\;\;+\, 
\frac{P}{(\hat{h}-\imath Pz_j )^{n+1}} \,V\, P_{[-m,m]}\, \left(\frac{1}{h-\imath z_j}\,-\frac{1}{h-\imath z_L}\right)\,P
\end{align*}
In the second line each term of the sum contains the factor $P_{[-m,m]}\frac{1}{(h-\imath z_j)^{2}}$ that can be estimated with \eqref{eq:am_resolvent3} and in the third line one can apply \eqref{eq:am_resolvent2}. Bounding all other resolvents in each term with the standard resolvent estimate  $\norm{\frac{1}{h-\imath z}}_\infty \leq \frac{1}{\abs{z}}$ one obtains uniformly in $n,j,L$
\begin{align*}
\norm{x^{(j)}_{L,n}}_p^p 
&
\;\leq\; 
\left(c_1 \sum_{k=0}^{n-1} \abs{z_j}^{-np-p\frac{p}{s}}\, + c_2 \abs{z_j}^{-(n+1)p-p\frac{p}{s}}\, \abs{z_j-z_L}^{p} \right) \norm{V} \, \norm{P_{[-m,m]}}^p_p\\
&
\;\leq\; 
(c_3 n + c_4) 2^{jp(n+\frac{p}{s})}
\;,
\end{align*}
where $\abs{z_j-z_L}\leq \abs{z_j}+\abs{z_L} \leq 2 \abs{z_j}$ since $j < L$. One also has the trivial estimate
$$
\norm{f_{n}^{(j)}}_\infty 
\;=\; 
\sup_{z \in [z_{j+1},z_j]} \abs{f_n^{(j)}(z)} 
\;\leq\; 
2^{-(j+1)n}
\;.
$$ 
and applying the triangle inequality to the termwise integral \eqref{eq:quasi_int} thus gives
\begin{align*}
\Big\|\int_{z_{j+1}}^{z_j} F_L(z) \,\difd{z}\Big\|_p^p 
&\;\leq\; 
\sum_{n=0}^\infty (c_3 n+c_4) 2^{jp(n+\frac{p}{s})} 2^{-p(j+1)n} \, \abs{z_j - z_{j-1}}^p \\
&
\;= \; \sum_{n=0}^\infty (c_3 n+c_4) 2^{jp(n+\frac{p}{s})} 2^{-p(j+1)n} \, 2^{-(j+1)p}\\
&
\;= \; \sum_{n=0}^\infty (c_3 n+c_4) 2^{-pn+ pj(\frac{p}{s}-1) - p}\\
&= c_5 2^{jp(\frac{p}{s}-1)}
\;,
\end{align*}
such that $p<s$ implies that taking the sum over $j$ gives a uniform upper bound for the original integral.
\vspace{.1cm}
Hence the sequence $\Delta_{2^{-L}}$ is uniformly bounded in both the $L^p$-quasi-norm and also in operator norm since $\norm{\Delta_\epsilon} \leq 2$. This also implies boundedness w.r.t. each of the $L^q$-(quasi-)norms for $p \leq q \leq \infty$ and thus the strong limit $\sgn(h)-\sgn(\hat{h})$ is also in $L^q(\Nn_\Halfspaceaction)$ for all $p \leq q \leq \infty$ by Lemma \ref{lemma:convergence}(v). This is enough to complete the proof, though with slightly more effort one can, moreover, show that $\Delta_\epsilon$ actually converges to its strong limit in the $L^p(\Nn_{\Halfspaceaction})$-quasinorm.
\hfill $\Box$

\vspace{.2cm}

\noindent {\bf Proof} of Theorem~\ref{theo-SurfaceIntro}.
For sake of completeness, let us here collect the above results to show that each of the conditions of Theorem~\ref{theo-SurfaceIntro} separately implies that the two conditions (Besov regularity and boundary compactness) of Proposition~\ref{prop:flat_band_suff} hold, so that \eqref{eq-BBCIntro} indeed follows. For (i), namely that $h$ is gapped,  Besov and Sobolev and regularity are obvious and boundary compactness follows from Proposition~\ref{prop-hInvertible} (which is essentially already in Proposition~\ref{prop-index_map_bb_sm}). Case (ii), namely the case of a mobility gap, follows by combining Proposition~\ref{prop-MBG} and Proposition~\ref{prop:dos_splitting3}. For the case (iii), Proposition~\ref{prop:pseudogap_sufficient} and Proposition~\ref{prop:dos_splitting} apply if the pseudogap is of order $\gamma>2$, otherwise for $\frac{3}{2} < \gamma \leq 2$ one applies Proposition~\ref{prop:dos_splitting2} whose conditions are satisfied by Proposition~\ref{prop:dirac2d}. Note also that the pseudogap is sufficient for Proposition~\ref{prop:pseudogap_sufficient} to show the required Sobolev regularity of the Fermi unitary. 
\hfill $\Box$

\section{Application to graphene}
\label{sec-Graphene}

\begin{figure}

\begin{center}

\begin{tikzpicture}[
    scale=1.2,
    line cap=round
]

\path (-0.86602540378cm,-2.5cm) node[circle,draw,fill=white,inner sep=.15cm](3){};
\path (-2*0.86602540378cm,-3cm) node[circle,fill=black,inner sep=.15cm](4){};
\path (-2*0.86602540378cm,-4cm) node[circle,draw,fill=white,inner sep=.15cm](5){};
\path (-3*0.86602540378cm,-4.5cm) node[circle,fill=black,inner sep=.15cm](6){};
\path (-3*0.86602540378cm,-5.5cm) node[circle,draw,fill=white,inner sep=.15cm](7){};

\path (2*0.86602540378cm+0,-1cm) node[circle,draw,fill=white,inner sep=.15cm](8){};
\path (2*0.86602540378cm+-0.86602540378cm,-1.5cm) node[circle,fill=black,inner sep=.15cm](9){};
\path (2*0.86602540378cm+-0.86602540378cm,-2.5cm) node[circle,draw,fill=white,inner sep=.15cm](10){};
\path (2*0.86602540378cm+-2*0.86602540378cm,-3cm) node[circle,fill=black,inner sep=.15cm](11){};
\path (2*0.86602540378cm+-2*0.86602540378cm,-4cm) node[circle,draw,fill=white,inner sep=.15cm](12){};
\path (2*0.86602540378cm+-3*0.86602540378cm,-4.5cm) node[circle,fill=black,inner sep=.15cm](13){};
\path (2*0.86602540378cm+-3*0.86602540378cm,-5.5cm) node[circle,draw,fill=white,inner sep=.15cm](14){};

\path (4*0.86602540378cm+0,-1cm) node[circle,draw,fill=white,inner sep=.15cm](15){};
\path (4*0.86602540378cm+-0.86602540378cm,-1.5cm) node[circle,fill=black,inner sep=.15cm](16){};
\path (4*0.86602540378cm+-0.86602540378cm,-2.5cm) node[circle,draw,fill=white,inner sep=.15cm](17){};
\path (4*0.86602540378cm+-2*0.86602540378cm,-3cm) node[circle,fill=black,inner sep=.15cm](18){};
\path (4*0.86602540378cm+-2*0.86602540378cm,-4cm) node[circle,draw,fill=white,inner sep=.15cm](19){};
\path (4*0.86602540378cm+-3*0.86602540378cm,-4.5cm) node[circle,fill=black,inner sep=.15cm,label=left:$A$](20){};
\path (4*0.86602540378cm+-3*0.86602540378cm,-5.5cm) node[circle,draw,fill=white,inner sep=.15cm](21){};

\path (6*0.86602540378cm+-0.86602540378cm,-1.5cm) node[circle,fill=black,inner sep=.15cm](22){};
\path (6*0.86602540378cm+-0.86602540378cm,-2.5cm) node[circle,draw,fill=white,inner sep=.15cm](23){};
\path (6*0.86602540378cm+-2*0.86602540378cm,-3cm) node[circle,fill=black,inner sep=.15cm](24){};
\path (6*0.86602540378cm+-2*0.86602540378cm,-4cm) node[circle,draw,fill=white,inner sep=.15cm](25){};
\path (6*0.86602540378cm+-3*0.86602540378cm,-4.5cm) node[circle,fill=black,inner sep=.15cm](26){};
\path (6*0.86602540378cm+-3*0.86602540378cm,-5.5cm) node[circle,draw,fill=white,inner sep=.15cm](27){};

\path (7*0.86602540378cm+-0.86602540378cm,-1cm) node[circle,draw,fill=white,inner sep=.15cm](28){};
\path (8*0.86602540378cm+-0.86602540378cm,-1.5cm) node[circle,fill=black,inner sep=.15cm](29){};
\path (8*0.86602540378cm+-0.86602540378cm,-2.5cm) node[circle,draw,fill=white,inner sep=.15cm](30){};
\path (8*0.86602540378cm+-2*0.86602540378cm,-3cm) node[circle,fill=black,inner sep=.15cm](31){};
\path (8*0.86602540378cm+-2*0.86602540378cm,-4cm) node[circle,draw,fill=white,inner sep=.15cm](32){};
\path (8*0.86602540378cm+-3*0.86602540378cm,-4.5cm) node[circle,fill=black,inner sep=.15cm](33){};
\path (8*0.86602540378cm+-3*0.86602540378cm,-5.5cm) node[circle,draw,fill=white,inner sep=.15cm](34){};
\path (10*0.86602540378cm+-3*0.86602540378cm,-4.5cm) node[circle,fill=black,inner sep=.15cm](35){};
\path (10*0.86602540378cm+-3*0.86602540378cm,-5.5cm) node[circle,draw,fill=white,inner sep=.15cm](36){};
%


\path (0,-1cm) node[circle,draw,fill=white,inner sep=.15cm](1){};
\path (-0.86602540378cm,-1.5cm) node[circle,fill=black,inner sep=.15cm](2){};
\path (-2*0.86602540378cm,-1cm) node[circle,draw,fill=white,inner sep=.15cm](0){};
\path (-3*0.86602540378cm,-1.5cm) node[circle,fill=black,inner sep=.15cm](-1){};
\path (-3*0.86602540378cm,-2.5cm) node[circle,draw,fill=white,inner sep=.15cm](-2){};

\path [black,ultra thick] (20) edge[->] (18);
\path [black,ultra thick] (20) edge[->] (19);
\path [black,ultra thick] (20) edge[->] (26);

\path (-1*0.86602540378cm+3.1cm,-4.8cm) node {$v_1$}
(-1*0.86602540378cm+2.1cm,-3.4cm) node {$v_2$}
(-1*0.86602540378cm+2.97cm,-3.9cm) node {$B$}
(-1*0.86602540378cm+2.6cm,-4.35cm) node {$w$};

\draw [black,ultra thick,dotted] (3)--(4) (-1)--(0) (5)--(6) (2)--(1) (10)--(11) (12)--(13) (19)--(20) (8)--(9);
\draw [black,ultra thick,dotted]  (25)--(26) (23)--(24) (22)--(28) (30)--(31) (32)--(33) (17)--(18) (15)--(16);

\draw [black] (3)--(4) (4)--(5) (6)--(7);

\draw [black] (9)--(10) (11)--(12) (13)--(14);

\draw [black] (16)--(17) (18)--(19) (20)--(21);

\draw [black] (22)--(23) (24)--(25) (26)--(27);

\draw [black] (28)--(29) (29)--(30) (31)--(32) (33)--(34) (35)--(36);

\draw [black] (23)--(31) (25)--(33) (32)--(35);
\draw [black]  (3)--(11) (5)--(13) (8)--(16) (10)--(18) (12)--(20) (15)--(22) (17)--(24) (19)--(26); 


\draw [black] (-2)--(4) (-2)--(-1) (0)--(2) (2)--(3) (1)--(9);

\end{tikzpicture}

\end{center}

\caption{The hexagonal lattice split into two triangular lattices, the $A$-lattice and the $B$-lattice. The two sites connected by a dashed line form a unit cell.}
\label{fig-grapheneAll}
\end{figure}
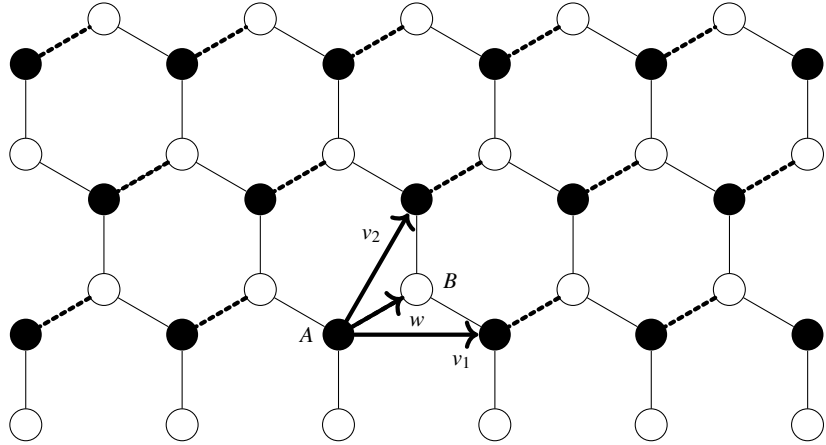

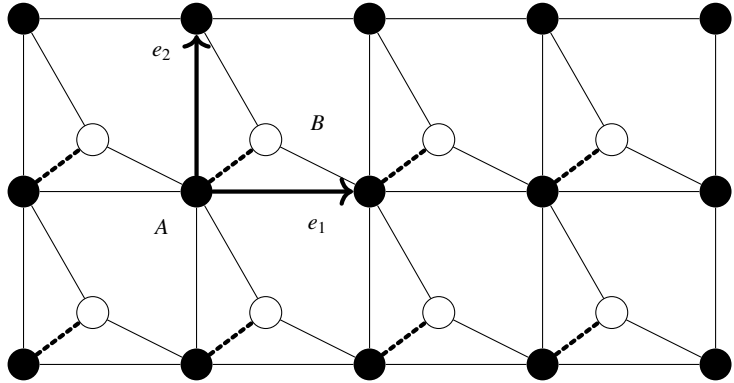
\begin{figure}

\begin{center}

\begin{tikzpicture}[
    scale=2.3,
    line cap=round
]

\path (0*1.0cm,0*1.0cm) node[circle,fill=black,inner sep=.15cm](1){};
\path (1*1.0cm,0*1.0cm) node[circle,fill=black,inner sep=.15cm](2){};
\path (2*1.0cm,0*1.0cm) node[circle,fill=black,inner sep=.15cm](3){};
\path (3*1.0cm,0*1.0cm) node[circle,fill=black,inner sep=.15cm](4){};
\path (4*1.0cm,0*1.0cm) node[circle,fill=black,inner sep=.15cm](5){};

\path (0*1.0cm,-1*1.0cm) node[circle,fill=black,inner sep=.15cm](6){};
\path (1*1.0cm,-1*1.0cm) node[circle,fill=black,inner sep=.15cm](7){};
\path (2*1.0cm,-1*1.0cm) node[circle,fill=black,inner sep=.15cm](8){};
\path (3*1.0cm,-1*1.0cm) node[circle,fill=black,inner sep=.15cm](9){};
\path (4*1.0cm,-1*1.0cm) node[circle,fill=black,inner sep=.15cm](10){};

\path (0*1.0cm,-2*1.0cm) node[circle,fill=black,inner sep=.15cm](11){};
\path (1*1.0cm,-2*1.0cm) node[circle,fill=black,inner sep=.15cm](12){};
\path (2*1.0cm,-2*1.0cm) node[circle,fill=black,inner sep=.15cm](13){};
\path (3*1.0cm,-2*1.0cm) node[circle,fill=black,inner sep=.15cm](14){};
\path (4*1.0cm,-2*1.0cm) node[circle,fill=black,inner sep=.15cm](15){};

\path (0*1.0cm+0.4cm,-1*1.0cm+0.3cm) node[circle,draw,fill=white,inner sep=.15cm](16){};
\path (1*1.0cm+0.4cm,-1*1.0cm+0.3cm) node[circle,draw,fill=white,inner sep=.15cm](17){};
\path (2*1.0cm+0.4cm,-1*1.0cm+0.3cm) node[circle,draw,fill=white,inner sep=.15cm](18){};
\path (3*1.0cm+0.4cm,-1*1.0cm+0.3cm) node[circle,draw,fill=white,inner sep=.15cm](19){};

\path (0*1.0cm+0.4cm,-2*1.0cm+0.3cm) node[circle,draw,fill=white,inner sep=.15cm](21){};
\path (1*1.0cm+0.4cm,-2*1.0cm+0.3cm) node[circle,draw,fill=white,inner sep=.15cm](22){};
\path (2*1.0cm+0.4cm,-2*1.0cm+0.3cm) node[circle,draw,fill=white,inner sep=.15cm](23){};
\path (3*1.0cm+0.4cm,-2*1.0cm+0.3cm) node[circle,draw,fill=white,inner sep=.15cm](24){};

\path [black,ultra thick] (7) edge[->] (2);
\path [black,ultra thick] (7) edge[->] (8);

\path (0.8cm,-0.2cm) node {$e_2$}
(1.7cm,-1.2cm) node {$e_1$}
(1.7cm,-0.6cm) node {$B$}
(0.8cm,-1.2cm) node {$A$};


\draw [black,ultra thick,dotted] (6)--(16) (7)--(17) (8)--(18) (9)--(19);
\draw [black,ultra thick,dotted] (11)--(21) (12)--(22) (13)--(23) (14)--(24);

\draw [black] (1)--(2) (2)--(3) (3)--(4) (4)--(5);
\draw [black] (6)--(7) (7)--(8) (8)--(9) (9)--(10);
\draw [black] (11)--(12) (12)--(13) (13)--(14) (14)--(15);

\draw [black] (1)--(6) (2)--(7) (3)--(8) (4)--(9) (5)--(10);
\draw [black] (6)--(11) (7)--(12) (8)--(13) (9)--(14) (10)--(15);

\draw [black] (1)--(16) (2)--(17) (3)--(18) (4)--(19);
\draw [black] (7)--(16) (8)--(17) (9)--(18) (10)--(19);

\draw [black] (6)--(21) (7)--(22) (8)--(23) (9)--(24);
\draw [black] (12)--(21) (13)--(22) (14)--(23) (15)--(24);

\end{tikzpicture}

\end{center}

\caption{The hexagonal lattice mapped onto a decorated square lattice.}
\label{fig-grapheneAllSquare}
\end{figure}

A prototypical and minimal example of a chiral Hamiltonian with non-trivial weak topological invariants is given by the discrete (graph) Laplacian on a honeycomb lattice, which is often used as a basic model for graphene. The description of this model is illustrated in Figure~\ref{fig-grapheneAll}. The hexagonal lattice can be split into two triangular sublattices, usually denoted as $A$-lattice $\Lambda_A$ and $B$-lattice $\Lambda_B$ respectively. The $B$-lattice is obtained simply by shifting the $A$-lattice  by a fixed vector $w$ to a neighboring site on the $B$-lattice. A minimal choice for a unit cell is then a pair of points $(A,B)$ connected by this vector $w$. In the Hilbert space, combining any two lattice points that differ by $w$ leads to an identification $\ell^2(\Lambda_A\cup\Lambda_B) \simeq \ell^2(\Lambda_A) \otimes \CM^2$, that is one equivalently has a single triangular lattice with two on-site degrees of freedom. 
Moreover, the triangular $A$-lattice can be mapped into a square lattice: choosing two linearly independent vectors $v_1,v_2$ connecting neighboring $A$-lattice sites, any point $x\in\Lambda_A$ is given by $x=n_1v_1+n_2v_2$ where $n_1,n_2\in\ZM$. This establishes a bijection $x\in\Lambda_A\mapsto (n_1,n_2)\in\ZM^2$ which corresponds to the identifications $e_1\cong v_1$ and $e_2\cong v_2$ of the basis vectors with standard basis vectors of $\ZM^2$. This is illustrated in Figure~\ref{fig-grapheneAllSquare}. Thus the Hilbert space is $\ell^2(\Lambda_A\cup\Lambda_B)$ is identified with $\ell^2(\ZM^2)\otimes\CM^2$.  

The graphene Hamiltonian $h$ simply consists of the nearest neighbor hopping terms, so it is the adjacency matrix on the hexagon lattice. Let us write it out on the Hilbert space $\ell^2(\ZM^2)\otimes\CM^2$ where the upper component is the $A$-lattice and the lower one the $B$-lattice. Then a given $A$-site has three nearest neighbors. As they all lie in the $B$-lattice (and vice versa), the Hamiltonian  $h$ is off-diagonal containing three summands in each off-diagonal entry. Recalling that $u_1$ and $u_2$ are the lattice translations on $\ZM^2$ as in Proposition~\ref{prop-CovRep}, the choice of the vectors $v_1$ and $v_2$ leads (according to Figure~\ref{fig-grapheneAll}) to
\begin{equation}
\label{eq:graphene_example}
h \;=\; 
\begin{pmatrix}
0 & 1 + u_1 +  u_2\\
1 + u_1^*+u_2^* & 0
\end{pmatrix}
\;.
\end{equation}
The magnetic field $\BB$ is supposed to vanish here. The off-diagonal nature of $h$ is a chiral symmetry in the sense of Definition~\ref{def-CH} with $J$ as in \eqref{eq-JDef} with $N=2$. Let us note that any next-nearest neighbor hopping term would destroy this chiral symmetry. 


\vspace{.2cm}

Let us now begin with the spectral analysis of $h$. After Fourier transform one now gets a function  $h: \bbT^2 \to M_2(\bbC)$ given by
\begin{equation}
\label{eq:honeycomb}
h(k)\, =\, \begin{pmatrix}
0 & 1 + e^{2\pi\imath k_1} + e^{2\pi\imath k_2}\\
1 + e^{- 2\pi\imath k_1} + e^{- 2\pi\imath k_2} & 0
\end{pmatrix}
\;.
\end{equation}
The spectrum of 
$$
\sigma(h(k))
\;=\; 
\big\{\pm \, \big| \,1 + e^{2\pi\imath k_1} + e^{2\pi\imath k_2}\big|\big\}
$$ 
consists of two bands that touch at the Dirac points $k_{\pm}= (\pm \frac{1}{3},\pm \frac{1}{3})$. The Fourier transform of the Fermi unitary is given by 
$$
u_F(k) 
\;=\; 
\frac{1 + e^{2 \pi\imath  k_1} + e^{2 \pi\imath  k_2}}{\lvert 1 + e^{2 \pi\imath k_1} + e^{2 \pi\imath k_2}\rvert}
\;
$$
and is manifestly not continuous at the Dirac points. This is directly related to the topology of the Hamiltonian since indeed $u_F(k)$ has winding numbers $\pm 1$ around the Dirac points, which could not be possible if $u_F$ were a smooth function. Of course, this is a result of zeros of the off-diagonal entries $a(k)=1 + e^{2\pi\imath k_1} + e^{2\pi\imath  k_2}$ at the Dirac points.

\vspace{.2cm}

Next let us compute the weak Chern numbers $\Ch_{\Tt,\xi}(u_F)$ for this model, namely the weak winding numbers. By the discussion following \eqref{eq:dirac_weyl} the weak Chern numbers are well-defined.  Due to linearity it is sufficient to do this for the standard basis vectors $e_1,e_2$, see \eqref{eq-WindLin}. Clearly, the Hamiltonian is symmetric under the exchange of $e_1$ and $e_2$, hence $\Ch_{\Tt,e_1}(u_F)=\Ch_{\Tt,e_2}(u_F)$. The winding number w.r.t. $e_1$ is
$$
\Ch_{\Tt,e_1}(u_F)
\;=\; 
\Tt(u \nabla_1 u^*) 
\;=\; 
\int_{[0,1]}  \int_{[0,1]}\,u_F(k) \partial_{k_1} u_F^*(k)\;\difd{k_1}\,\difd{k_2}
\;.
$$
The integral for fixed $k_2$ is just the winding number along one slice and the homotopy invariance implies that it must be locally constant in $k_2$, {\it i.e.} the value of the inner integral can only change at the discontinuities of $u_F$ which are given by the two Dirac points. Clearly, these are also the winding numbers of $k_1\in[0,1)\mapsto a(k_1,k_2)=1 + e^{2 \pi\imath  k_1} + e^{2 \pi\imath  k_2}$ which are equal to $1$ as long as $k_2\in(\frac{1}{3},\frac{2}{3})$ and vanish otherwise. As this is a third of the $k_2$-torus, one deduces
\begin{equation}
\label{eq-Graphene1}
\Ch_{\Tt,e_1}(u_F) \;=\; \frac{1}{3}\;,
\end{equation}
implying that also
\begin{equation}
\label{eq-Graphene2}
\Ch_{\Tt,e_2}(u_F) \;=\; \frac{1}{3}\;.
\end{equation}
Let us note that the value $\frac{1}{3}$ can also be interpreted as the distance between the two Dirac points in momentum space. 

\vspace{.2cm}

One can also modify the model slightly by introducing an additional parameter $\lambda$ for the hopping strength within each unit cell:
\begin{equation}
\label{eq:honeycombMod}
h_\lambda(k)\, =\, \begin{pmatrix}
0 & \lambda + e^{2\pi\imath k_1} + e^{2\pi\imath k_2}\\
\lambda + e^{- 2\pi\imath k_1} + e^{- 2\pi\imath k_2} & 0
\end{pmatrix}
\;.
\end{equation}
The Fermi unitary is denoted by $u_{F,\lambda}$. This model has a spectral gap for $\abs{\lambda}>2$ and a linear pseudogap for $\lambda < 2$ with two Dirac points $k^{(1)}$ and $k^{(2)}$. One can still compute the weak Chern numbers $\Ch_{\Tt,e_1}(u_{F,\lambda})$ and $\Ch_{\Tt,e_2}(u_{F,\lambda})$ by the techniques above in terms of the components of $k^{(1)}-k^{(2)}$.  For $\abs{\lambda} > 2$ the Hamiltonian is a topologically trivial gapped insulator with $\Ch_{\Tt,\xi}(u_{F,\lambda})=0$.  For $\abs{\lambda}=2$ the two Dirac points merge to form a so-called semi-Dirac point, which still results in a pseudogap with a square root DOS $\mu_h([-\epsilon,\epsilon])\sim \abs{E}^{1+\frac{1}{2}}$, which is enough to ensure the existence of the weak Chern numbers even though they vanish.  From the above expression in terms of the positions of the Dirac points in momentum space, one sees that the Chern numbers change continuously throughout the perturbation. This also holds more generally (see Proposition~\ref{prop-WeakInvCont}) and demonstrates that the weak Chern numbers in the absence of a bulk gap do not have to be quantized in general. Let us remark that this may change when one introduces additional constraints such as lattice symmetries that the perturbations also have to respect. In the example above, the hexagonal symmetry of the lattice is violated for all $\lambda \neq 1$ and it is clear that if it was enforced then the Dirac points could only be located at certain high-symmetry points in the Brillouin torus, thereby leading to an effective quantization.

\vspace{.2cm}

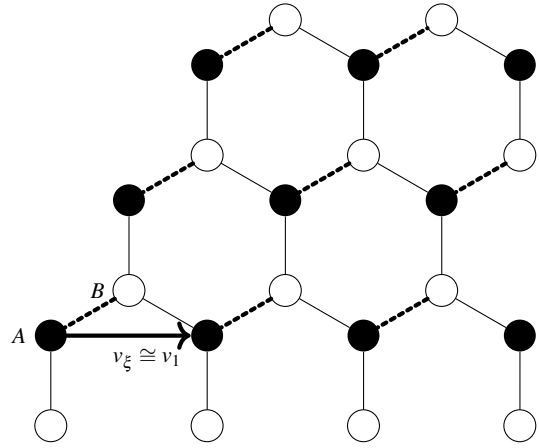
\begin{figure}

\begin{center}


\begin{tikzpicture}[
    scale=1.2,
    line cap=round
]



\path (4*0.86602540378cm+0,-1cm) node[circle,draw,fill=white,inner sep=.15cm](15){};
\path (4*0.86602540378cm+-0.86602540378cm,-1.5cm) node[circle,fill=black,inner sep=.15cm](16){};
\path (4*0.86602540378cm+-0.86602540378cm,-2.5cm) node[circle,draw,fill=white,inner sep=.15cm](17){};
\path (4*0.86602540378cm+-2*0.86602540378cm,-3cm) node[circle,fill=black,inner sep=.15cm](18){};
\path (4*0.86602540378cm+-2*0.86602540378cm,-4cm) node[circle,draw,fill=white,inner sep=.15cm,label=left:$B$](19){};
\path (4*0.86602540378cm+-3*0.86602540378cm,-4.5cm) node[circle,fill=black,inner sep=.15cm,label=left:$A$](20){};
\path (4*0.86602540378cm+-3*0.86602540378cm,-5.5cm) node[circle,draw,fill=white,inner sep=.15cm](21){};

\path (6*0.86602540378cm+-0.86602540378cm,-1.5cm) node[circle,fill=black,inner sep=.15cm](22){};
\path (6*0.86602540378cm+-0.86602540378cm,-2.5cm) node[circle,draw,fill=white,inner sep=.15cm](23){};
\path (6*0.86602540378cm+-2*0.86602540378cm,-3cm) node[circle,fill=black,inner sep=.15cm](24){};
\path (6*0.86602540378cm+-2*0.86602540378cm,-4cm) node[circle,draw,fill=white,inner sep=.15cm](25){};
\path (6*0.86602540378cm+-3*0.86602540378cm,-4.5cm) node[circle,fill=black,inner sep=.15cm](26){};
\path (6*0.86602540378cm+-3*0.86602540378cm,-5.5cm) node[circle,draw,fill=white,inner sep=.15cm](27){};

\path (7*0.86602540378cm+-0.86602540378cm,-1cm) node[circle,draw,fill=white,inner sep=.15cm](28){};
\path (8*0.86602540378cm+-0.86602540378cm,-1.5cm) node[circle,fill=black,inner sep=.15cm](29){};
\path (8*0.86602540378cm+-0.86602540378cm,-2.5cm) node[circle,draw,fill=white,inner sep=.15cm](30){};
\path (8*0.86602540378cm+-2*0.86602540378cm,-3cm) node[circle,fill=black,inner sep=.15cm](31){};
\path (8*0.86602540378cm+-2*0.86602540378cm,-4cm) node[circle,draw,fill=white,inner sep=.15cm](32){};
\path (8*0.86602540378cm+-3*0.86602540378cm,-4.5cm) node[circle,fill=black,inner sep=.15cm](33){};
\path (8*0.86602540378cm+-3*0.86602540378cm,-5.5cm) node[circle,draw,fill=white,inner sep=.15cm](34){};
\path (10*0.86602540378cm+-3*0.86602540378cm,-4.5cm) node[circle,fill=black,inner sep=.15cm](35){};
\path (10*0.86602540378cm+-3*0.86602540378cm,-5.5cm) node[circle,draw,fill=white,inner sep=.15cm](36){};
%



\path [black,ultra thick] (20) edge[->] (26);

\path (-1*0.86602540378cm+2.8cm,-4.77cm) node {$\Halfspaceunitvector \cong v_1$};

\draw [black,ultra thick,dotted] (19)--(20);
\draw [black,ultra thick,dotted]  (25)--(26) (23)--(24) (22)--(28) (30)--(31) (32)--(33) (17)--(18) (15)--(16);



\draw [black] (16)--(17) (18)--(19) (20)--(21);

\draw [black] (22)--(23) (24)--(25) (26)--(27);

\draw [black] (28)--(29) (29)--(30) (31)--(32) (33)--(34) (35)--(36);

\draw [black] (23)--(31) (25)--(33) (32)--(35);
\draw [black]  (15)--(22) (17)--(24) (19)--(26); 



\end{tikzpicture}
\end{center}

\caption{For the choice $\Halfspaceunitvector =e_1$ one obtains a zigzag edge, here on the l.h.s. of the figure. Note that none of the unit cells is broken up.}
\label{fig-grapheneZigzag}
\end{figure}

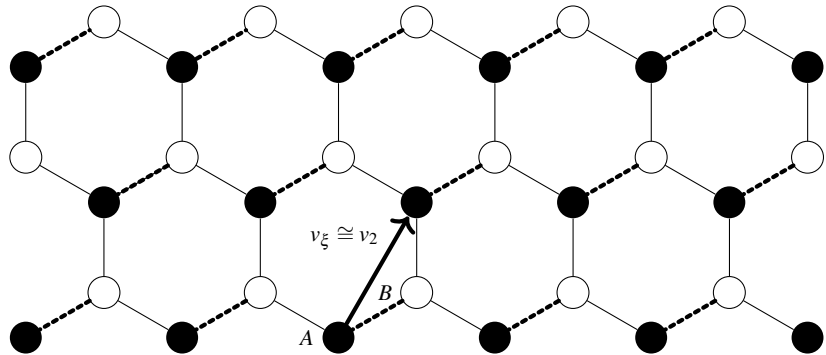
\begin{figure}

\begin{center}

\begin{tikzpicture}[
    scale=1.2,
    line cap=round
]

\path (-0.86602540378cm,-2.5cm) node[circle,draw,fill=white,inner sep=.15cm](3){};
\path (-2*0.86602540378cm,-3cm) node[circle,fill=black,inner sep=.15cm](4){};
\path (-2*0.86602540378cm,-4cm) node[circle,draw,fill=white,inner sep=.15cm](5){};
\path (-3*0.86602540378cm,-4.5cm) node[circle,fill=black,inner sep=.15cm](6){};

\path (2*0.86602540378cm+0,-1cm) node[circle,draw,fill=white,inner sep=.15cm](8){};
\path (2*0.86602540378cm+-0.86602540378cm,-1.5cm) node[circle,fill=black,inner sep=.15cm](9){};
\path (2*0.86602540378cm+-0.86602540378cm,-2.5cm) node[circle,draw,fill=white,inner sep=.15cm](10){};
\path (2*0.86602540378cm+-2*0.86602540378cm,-3cm) node[circle,fill=black,inner sep=.15cm](11){};
\path (2*0.86602540378cm+-2*0.86602540378cm,-4cm) node[circle,draw,fill=white,inner sep=.15cm](12){};
\path (2*0.86602540378cm+-3*0.86602540378cm,-4.5cm) node[circle,fill=black,inner sep=.15cm](13){};

\path (4*0.86602540378cm+0,-1cm) node[circle,draw,fill=white,inner sep=.15cm](15){};
\path (4*0.86602540378cm+-0.86602540378cm,-1.5cm) node[circle,fill=black,inner sep=.15cm](16){};
\path (4*0.86602540378cm+-0.86602540378cm,-2.5cm) node[circle,draw,fill=white,inner sep=.15cm](17){};
\path (4*0.86602540378cm+-2*0.86602540378cm,-3cm) node[circle,fill=black,inner sep=.15cm](18){};
\path (4*0.86602540378cm+-2*0.86602540378cm,-4cm) node[circle,draw,fill=white,inner sep=.15cm,label=left:$B$](19){};
\path (4*0.86602540378cm+-3*0.86602540378cm,-4.5cm) node[circle,fill=black,inner sep=.15cm,label=left:$A$](20){};

\path (6*0.86602540378cm+-0.86602540378cm,-1.5cm) node[circle,fill=black,inner sep=.15cm](22){};
\path (6*0.86602540378cm+-0.86602540378cm,-2.5cm) node[circle,draw,fill=white,inner sep=.15cm](23){};
\path (6*0.86602540378cm+-2*0.86602540378cm,-3cm) node[circle,fill=black,inner sep=.15cm](24){};
\path (6*0.86602540378cm+-2*0.86602540378cm,-4cm) node[circle,draw,fill=white,inner sep=.15cm](25){};
\path (6*0.86602540378cm+-3*0.86602540378cm,-4.5cm) node[circle,fill=black,inner sep=.15cm](26){};

\path (7*0.86602540378cm+-0.86602540378cm,-1cm) node[circle,draw,fill=white,inner sep=.15cm](28){};
\path (8*0.86602540378cm+-0.86602540378cm,-1.5cm) node[circle,fill=black,inner sep=.15cm](29){};
\path (8*0.86602540378cm+-0.86602540378cm,-2.5cm) node[circle,draw,fill=white,inner sep=.15cm](30){};
\path (8*0.86602540378cm+-2*0.86602540378cm,-3cm) node[circle,fill=black,inner sep=.15cm](31){};
\path (8*0.86602540378cm+-2*0.86602540378cm,-4cm) node[circle,draw,fill=white,inner sep=.15cm](32){};
\path (8*0.86602540378cm+-3*0.86602540378cm,-4.5cm) node[circle,fill=black,inner sep=.15cm](33){};
\path (10*0.86602540378cm+-3*0.86602540378cm,-4.5cm) node[circle,fill=black,inner sep=.15cm](35){};
%


\path (0,-1cm) node[circle,draw,fill=white,inner sep=.15cm](1){};
\path (-0.86602540378cm,-1.5cm) node[circle,fill=black,inner sep=.15cm](2){};
\path (-2*0.86602540378cm,-1cm) node[circle,draw,fill=white,inner sep=.15cm](0){};
\path (-3*0.86602540378cm,-1.5cm) node[circle,fill=black,inner sep=.15cm](-1){};
\path (-3*0.86602540378cm,-2.5cm) node[circle,draw,fill=white,inner sep=.15cm](-2){};

\path [black,ultra thick] (20) edge[->] (18);

\path (-1*0.86602540378cm+1.8cm,-3.4cm) node {$\Halfspaceunitvector \cong v_2$};

\draw [black,ultra thick,dotted] (3)--(4) (-1)--(0) (5)--(6) (2)--(1) (10)--(11) (12)--(13) (19)--(20) (8)--(9);
\draw [black,ultra thick,dotted]  (25)--(26) (23)--(24) (22)--(28) (30)--(31) (32)--(33) (17)--(18) (15)--(16);

\draw [black] (3)--(4) (4)--(5);

\draw [black] (9)--(10) (11)--(12);

\draw [black] (16)--(17) (18)--(19);

\draw [black] (22)--(23) (24)--(25);

\draw [black] (28)--(29) (29)--(30) (31)--(32);

\draw [black] (23)--(31) (25)--(33) (32)--(35);
\draw [black]  (3)--(11) (5)--(13) (8)--(16) (10)--(18) (12)--(20) (15)--(22) (17)--(24) (19)--(26); 


\draw [black] (-2)--(4) (-2)--(-1) (0)--(2) (2)--(3) (1)--(9);

\end{tikzpicture}

\end{center}

\caption{For the choice $\Halfspaceunitvector =e_2$ one also obtains a zigzag edge, but now on the bottom of the figure. }
\label{fig-grapheneZigzag2}
\end{figure}

\begin{figure}

\begin{center}

\begin{tikzpicture}[
    scale=1.2,
    line cap=round
]

\path (-0.86602540378cm,-2.5cm) node[circle,draw,fill=white,inner sep=.15cm](3){};
\path (-2*0.86602540378cm,-3cm) node[circle,fill=black,inner sep=.15cm](4){};
\path (-2*0.86602540378cm,-4cm) node[circle,draw,fill=white,inner sep=.15cm](5){};
\path (-3*0.86602540378cm,-4.5cm) node[circle,fill=black,inner sep=.15cm](6){};
\path (-3*0.86602540378cm,-5.5cm) node[circle,draw,fill=white,inner sep=.15cm](7){};

\path (2*0.86602540378cm+0,-1cm) node[circle,draw,fill=white,inner sep=.15cm](8){};
\path (2*0.86602540378cm+-0.86602540378cm,-1.5cm) node[circle,fill=black,inner sep=.15cm](9){};
\path (2*0.86602540378cm+-0.86602540378cm,-2.5cm) node[circle,draw,fill=white,inner sep=.15cm](10){};
\path (2*0.86602540378cm+-2*0.86602540378cm,-3cm) node[circle,fill=black,inner sep=.15cm](11){};
\path (2*0.86602540378cm+-2*0.86602540378cm,-4cm) node[circle,draw,fill=white,inner sep=.15cm](12){};
\path (2*0.86602540378cm+-3*0.86602540378cm,-4.5cm) node[circle,fill=black,inner sep=.15cm](13){};
\path (2*0.86602540378cm+-3*0.86602540378cm,-5.5cm) node[circle,draw,fill=white,inner sep=.15cm](14){};

\path (4*0.86602540378cm+0,-1cm) node[circle,draw,fill=white,inner sep=.15cm](15){};
\path (4*0.86602540378cm+-0.86602540378cm,-1.5cm) node[circle,fill=black,inner sep=.15cm](16){};
\path (4*0.86602540378cm+-0.86602540378cm,-2.5cm) node[circle,draw,fill=white,inner sep=.15cm](17){};
\path (4*0.86602540378cm+-2*0.86602540378cm,-3cm) node[circle,fill=black,inner sep=.15cm](18){};
\path (4*0.86602540378cm+-2*0.86602540378cm,-4cm) node[circle,draw,fill=white,inner sep=.15cm,label=left:$B$](19){};
\path (4*0.86602540378cm+-3*0.86602540378cm,-4.5cm) node[circle,fill=black,inner sep=.15cm,label=left:$A$](20){};

\path (6*0.86602540378cm+-0.86602540378cm,-1.5cm) node[circle,fill=black,inner sep=.15cm](22){};
\path (6*0.86602540378cm+-0.86602540378cm,-2.5cm) node[circle,draw,fill=white,inner sep=.15cm](23){};
\path (6*0.86602540378cm+-2*0.86602540378cm,-3cm) node[circle,fill=black,inner sep=.15cm](24){};

\path (7*0.86602540378cm+-0.86602540378cm,-1cm) node[circle,draw,fill=white,inner sep=.15cm](28){};
\path (8*0.86602540378cm+-0.86602540378cm,-1.5cm) node[circle,fill=black,inner sep=.15cm](29){};
%


\path (0,-1cm) node[circle,draw,fill=white,inner sep=.15cm](1){};
\path (-0.86602540378cm,-1.5cm) node[circle,fill=black,inner sep=.15cm](2){};
\path (-2*0.86602540378cm,-1cm) node[circle,draw,fill=white,inner sep=.15cm](0){};
\path (-3*0.86602540378cm,-1.5cm) node[circle,fill=black,inner sep=.15cm](-1){};
\path (-3*0.86602540378cm,-2.5cm) node[circle,draw,fill=white,inner sep=.15cm](-2){};

\path [black,ultra thick] (20) edge[->] (11);

\path (-1*0.86602540378cm+0.5cm,-3.25cm) node {$\Halfspaceunitvector $};

\draw [black,ultra thick,dotted] (3)--(4) (-1)--(0) (5)--(6) (2)--(1) (10)--(11) (12)--(13) (19)--(20) (8)--(9);
\draw [black,ultra thick,dotted]  (23)--(24) (22)--(28) (17)--(18) (15)--(16);

\draw [black] (3)--(4) (4)--(5) (6)--(7);

\draw [black] (9)--(10) (11)--(12) (13)--(14);

\draw [black] (16)--(17) (18)--(19);

\draw [black] (22)--(23);

\draw [black] (28)--(29);

\draw [black]  (3)--(11) (5)--(13) (8)--(16) (10)--(18) (12)--(20) (15)--(22) (17)--(24); 


\draw [black] (-2)--(4) (-2)--(-1) (0)--(2) (2)--(3) (1)--(9);

\end{tikzpicture}

\end{center}

\caption{The choice $\Halfspaceunitvector =2^{-\frac{1}{2}}(e_2-e_1)$ produces an armchair edge on the lower right of the figure.}
\label{fig-grapheneArmchair}
\end{figure}
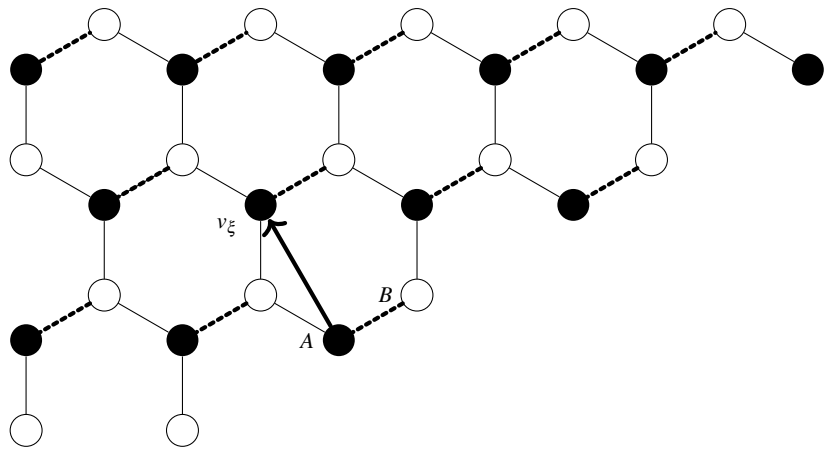

The results of Section~\ref{sec-flat} show that the weak Chern numbers dictate the signed density of surface states for a half-space graphene Hamiltonian. Let us begin by an explicit discussion of the type of edges considered, namely how precisely the half-space Hamiltonian $\hat{h}$ is constructed. According to Definition~\ref{def-HalfSpaceHamConst}, it is of the form $\hat{h}=P h P + \tilde{k}$ with a (in general random) boundary term $\tilde{k}$ being supported on a strip of finite width along the boundary.  Here $P= \chi(\hat{X}_\Halfspaceaction > 0)$ is the half-space projection in direction $\Halfspaceaction$ (which is the vector in the representation on the square lattice).  The physical representations are then given by
\begin{equation}
\label{eq-HhatRep}
\hat{h}_{r,\omega}
\;=\;
\chi(\hat{X}_\Halfspaceaction +r> 0)
\,h_\omega\,
\chi(\hat{X}_\Halfspaceaction +r> 0)
\;+\;
\tilde{k}_{r,\omega}
\;,
\end{equation}
An important point is that $P$ it commutes with $J$, so that it does break up the chosen unit cells. Let us first present the standard choices of $\Halfspaceunitvector $ which lead to the well-studied zigzag and armchair edges. Both of the choices $\Halfspaceunitvector=e_1$ and $\Halfspaceunitvector= e_2$ lead to zigzag edges, as is illustrated in Figures~\ref{fig-grapheneZigzag} and \ref{fig-grapheneZigzag2}. On the other hand, choosing $\Halfspaceunitvector= 2^{-\frac{1}{2}}(e_2-e_1)$ corresponds to an armchair edge, see Figure~\ref{fig-grapheneArmchair}. It is also possible to produce edges with dangling modes with other choices of $\Halfspaceunitvector$, see Figure~\ref{fig-grapheneZigzagDangling} for an example.  We do not attempt to plot examples of other choices of $\Halfspaceunitvector $, but stress once again that the edge cuts considered here never break up the unit cells and that this may produce so-called dangling sites. Another way to produce (possibly randomly distributed) dangling sites is by means of the boundary perturbation $\tilde{k}$.

\vspace{.2cm}

Another comment is that the constructions up to here are dependent on several choices that could also be taken in a different manner. First of all, one can choose different basis vectors $v_1$ and $v_2$ for the $A$-lattice. This changes the Hamiltonian and also the values of the weak Chern numbers. For example, if one chooses $v'_1=v_2$ and $v'_2=v_1+v_2$, then $\Ch_{\Tt,e_1}(u_F) =0$ and $\Ch_{\Tt,e_2}(u_F)=\frac{1}{3}$ and furthermore $\Halfspaceunitvector =e_1$ produces an armchair edge and $\Halfspaceunitvector =e_2$ a zigzag edge (note that this produces, of course, consistent values for the signed density of surface states). Furthermore, one can choose larger unit cells containing more than just $2$ lattice sites.

\vspace{.2cm}

\begin{figure}

\begin{center}

\begin{tikzpicture}[
    scale=1.2,
    line cap=round
]


\path (2*0.86602540378cm+0,-1cm) node[circle,draw,fill=white,inner sep=.15cm](8){};
\path (2*0.86602540378cm+-0.86602540378cm,-1.5cm) node[circle,fill=black,inner sep=.15cm](9){};
\path (2*0.86602540378cm+-0.86602540378cm,-2.5cm) node[circle,draw,fill=white,inner sep=.15cm](10){};
\path (2*0.86602540378cm+-2*0.86602540378cm,-3cm) node[circle,fill=black,inner sep=.15cm](11){};

\path (4*0.86602540378cm+0,-1cm) node[circle,draw,fill=white,inner sep=.15cm](15){};
\path (4*0.86602540378cm+-0.86602540378cm,-1.5cm) node[circle,fill=black,inner sep=.15cm](16){};
\path (4*0.86602540378cm+-0.86602540378cm,-2.5cm) node[circle,draw,fill=white,inner sep=.15cm](17){};
\path (4*0.86602540378cm+-2*0.86602540378cm,-3cm) node[circle,fill=black,inner sep=.15cm](18){};
\path (4*0.86602540378cm+-2*0.86602540378cm,-4cm) node[circle,draw,fill=white,inner sep=.15cm,label=left:$B$](19){};
\path (4*0.86602540378cm+-3*0.86602540378cm,-4.5cm) node[circle,fill=black,inner sep=.15cm,label=left:$A$](20){};

\path (6*0.86602540378cm+-0.86602540378cm,-1.5cm) node[circle,fill=black,inner sep=.15cm](22){};
\path (6*0.86602540378cm+-0.86602540378cm,-2.5cm) node[circle,draw,fill=white,inner sep=.15cm](23){};
\path (6*0.86602540378cm+-2*0.86602540378cm,-3cm) node[circle,fill=black,inner sep=.15cm](24){};
\path (6*0.86602540378cm+-2*0.86602540378cm,-4cm) node[circle,draw,fill=white,inner sep=.15cm](25){};
\path (6*0.86602540378cm+-3*0.86602540378cm,-4.5cm) node[circle,fill=black,inner sep=.15cm](26){};
\path (6*0.86602540378cm+-3*0.86602540378cm,-5.5cm) node[circle,draw,fill=white,inner sep=.15cm](27){};

\path (7*0.86602540378cm+-0.86602540378cm,-1cm) node[circle,draw,fill=white,inner sep=.15cm](28){};
\path (8*0.86602540378cm+-0.86602540378cm,-1.5cm) node[circle,fill=black,inner sep=.15cm](29){};
\path (8*0.86602540378cm+-0.86602540378cm,-2.5cm) node[circle,draw,fill=white,inner sep=.15cm](30){};
\path (8*0.86602540378cm+-2*0.86602540378cm,-3cm) node[circle,fill=black,inner sep=.15cm](31){};
\path (8*0.86602540378cm+-2*0.86602540378cm,-4cm) node[circle,draw,fill=white,inner sep=.15cm](32){};
\path (8*0.86602540378cm+-3*0.86602540378cm,-4.5cm) node[circle,fill=black,inner sep=.15cm](33){};
\path (8*0.86602540378cm+-3*0.86602540378cm,-5.5cm) node[circle,draw,fill=white,inner sep=.15cm](34){};
\path (10*0.86602540378cm+-3*0.86602540378cm,-4.5cm) node[circle,fill=black,inner sep=.15cm](35){};
\path (10*0.86602540378cm+-3*0.86602540378cm,-5.5cm) node[circle,draw,fill=white,inner sep=.15cm](36){};
%


\path (0,-1cm) node[circle,draw,fill=white,inner sep=.15cm](1){};
\path (-0.86602540378cm,-1.5cm) node[circle,fill=black,inner sep=.15cm](2){};

\path [black,ultra thick] (20) edge[->] (24);

\path (-1*0.86602540378cm+5.2cm,-3.2cm) node {$v_\xi\parallel e_1+e_2$};

\draw [black,ultra thick,dotted] (1)--(2)  (10)--(11) (19)--(20) (8)--(9);
\draw [black,ultra thick,dotted]  (25)--(26) (23)--(24) (22)--(28) (30)--(31) (32)--(33) (17)--(18) (15)--(16);


\draw [black] (9)--(1) (9)--(10);

\draw [black] (16)--(17) (18)--(19);

\draw [black] (22)--(23) (24)--(25) (26)--(27);

\draw [black] (28)--(29) (29)--(30) (31)--(32) (33)--(34) (35)--(36);

\draw [black] (23)--(31) (25)--(33) (32)--(35);
\draw [black]  (8)--(16) (10)--(18)  (15)--(22) (17)--(24) (19)--(26); 



\end{tikzpicture}

\end{center}

\caption{The choice $\Halfspaceunitvector =2^{-\frac{1}{2}}(e_1+e_2)$ produces a zigzag edge with dangling $A$-sites to the lower left. For this edge, one has thus has $\hat{\Tt}\big(J \,\Ker(\hat{h})\big)=\frac{\sqrt{2}}{3}$.}
\label{fig-grapheneZigzagDangling}
\end{figure}

Let us finally come to the index-theoretic formula for the signed surface state density of the flat band of edge states in graphene. By Proposition~\ref{prop:dirac2d} the graphene Hamiltonian satisfies all conditions of the BBC. Therefore taking into account that $\Ch_{\Tt,e_1}(u_F) =\Ch_{\Tt,e_2}(u_F)=\frac{1}{3}$ by \eqref{eq-Graphene1} and \eqref{eq-Graphene2},Proposition~\ref{prop:flat_band_suff} implies that  \eqref{eq-GrapheneIntro} holds for the half-space Hamiltonian $\hat{h}$:
\begin{equation}
\label{eq-GrapheneAgain}
\hat{\Tt}\big(J \,\Ker(\hat{h})\big) \;=\;\frac{1}{3}\,(\xi_1 +\xi_2)
\;,
\qquad
\Halfspaceunitvector
\,=\,
\begin{pmatrix}
\xi_1 \\ \xi_2
\end{pmatrix}
\;,
\end{equation}
notably almost every realization $\hat{h}_{r,\omega}$ of the half-space Hamiltonian $\hat{h}$ has a flat band of zero energy edges states with signed density given by the r.h.s. of \eqref{eq-GrapheneAgain}. In particular, for the zigzag edges there has to be a flat band of edge states (as previously found, see \cite{RH,DUM}). The same can be said for any other angle, except for $\xi_1 +\xi_2=0$ which corresponds to the armchair edge. The equality \eqref{eq-GrapheneAgain} is known in the physics literature for rationally dependent normal vectors \cite{DUM}, but here the claim is also shown for irrational angles and its stability under random boundary disorder is established. 

\vspace{.2cm}

Let us briefly comment on higher dimensional cases. One can also write down chiral Hamiltonians which satisfy the conditions of the BBC, for example, by stacking the model above and adding a weak interlayer coupling. 
A more interesting possibility in $d=3$ is given by nodal line semimetals where the spectrum at the Fermi level consists of a loop in momentum space. One may again have non-vanishing winding numbers, which in turn lead to flat bands of surface states whose signed density depends linearly on the components of the surface normal vector \cite{MatsuuraEtAl}.

\newpage

\appendix

\chapter{Appendices}

\section{$(p)$-Banach-spaces and integration}
\label{app-p<1}

This appendix collects some basic results on $(p)$-Banach spaces, $p>0$, and the integration of analytic functions (see {\it e.g.} \cite{Kalton2003} for a review). A quasi-norm on a complex vector space $E$ is a positive functional $\norm{\cdot}: E \to [0,\infty)$ satisfying for $\lambda \in \bbC$, $x,y\in E$ and some constant $K>0$,
$$
\norm{x} \;=\;0 \, \iff \, x \;=\; 0\;,
\qquad
\norm{\lambda x}    \;=\; \abs{\lambda}\, \norm{x}\;,
\qquad
\norm{x+y} \;\leq \;K (\norm{x}+\norm{y})\;.
$$
A quasi-normed space in which every Cauchy-sequence converges is called a quasi-Banach space. If there is some $0 < p < 1$ such that the modified triangle inequality
$$
\norm{x+y}^p 
\;\leq\; 
\norm{x}^p \,+\, \norm{y}^p
\;
$$
is satisfied, then the norm is called a $(p)$-norm. A quasi-Banach-space with a $(p)$-norm is a complete topological spaces in the metric
$$
d(x,y) \;=\; \norm{x-y}^p
\;.
$$
Our main application for $(p)$-Banach-spaces consists of estimating the norms of elements in a non-commutative $L^p$-space that are obtained by holomorphic functional calculus and hence by integration along curves. The Riemann integral for functions taking values in a quasi-normed space is defined exactly as in the normed case, however, there is the caveat that not all continuous function are integrable. Moreover, it is difficult to estimate $\norm{\int_{\mathcal{C}} f(z) \difd{z}}$ since neither of the triangle inequalities give a useful estimate for limits of Riemann sums.

\begin{definition}
\label{def-pAnalytic}
Let $E$ be a $(p)$-Banach-space and $\mathcal{C}\subset \bbC$ a region. A function $f: \mathcal{C} \to E$ has a local expansion in $z_0 \in \mathcal{C}$ if there is an open neighborhood $U(z_0) \subset \mathcal{C}$  such that 
\begin{equation}
\label{eq-panalytic}
f(z) 
\;=\; 
\sum_{n=0}^\infty x_n\, f_n(z)
\end{equation}
for all $z \in U(z_0)$ with analytic functions $f_n : U(z_0) \to \bbC$ and $x_n \in E$ satisfying
$$
\sum_{n=0}^\infty \abs{x_n}^p \, \norm{f_n}^p_\infty \;<\; \infty\;.
$$
The function $f$ is called $(p)$-analytic on $\mathcal{C}$ if it has a local expansion in every point $z_0 \in \mathcal{C}$.
\end{definition}

This definition of analyticity is chosen since there are some subtleties involving the convergence of power series in $(p)$-Banach-spaces. Analytic functions in $(p)$-Banach spaces are still integrable:

\begin{theorem}[\cite{Gramsch1965}]
Let $E$ be a $p$-Banach space, $\Gamma \subset \bbC$ a finite-length curve and $f: \Gamma \to E$ an $(p)$-analytic function on a neighborhood of $\Gamma$. Then $f$ is Riemann integrable. In particular, if $f$ has a local expansion \eqref{eq-panalytic}, then 
\begin{equation} 
\label{eq:quasi_int} 
\int_{\Gamma} f(z) \,\difd{z} 
\;=\;  
\sum_{n=0}^\infty x_n  \int_{\Gamma} f_n(z)\, \difd{z}
\;,
\end{equation}
and
$$
\norm{\int_{\Gamma} f(z) \,\difd{z}}^p
\;\leq \;
\sum_{n=0}^\infty \norm{x_n}^p \norm{f_n}_\infty^p \, \abs{\Gamma}^p
\;.
$$
\end{theorem}
%


\section{Non-commutative $L^p$-spaces}
\label{app-Lp}

Section~\ref{sec:traces} constructed, starting from an $\alpha$-invariant trace $\Tt$ on a $C^*$-dynamical system $(\Aa,G,\alpha)$, a von Neumann algebra $\Mm=L^\infty(\Aa,\Tt)$ with a s.n.f. trace $\Tt$. The following Section~\ref{sec-DualTraces} constructed a von Neumann crossed product algebra $L^\infty(\Aa,\Tt)\rtimes_\alpha G=L^\infty(\Aa\rtimes_\alpha G,\hat{\Tt})$ equipped with a s.n.f.  trace $\hat{\Tt}$. This appendix briefly recalls some of the consequences of this data by outlining the general theory of non-commutative $L^p$-spaces, based on the review \cite{Pisier03}. 

\vspace{.2cm}

Hence let $\Mm$ be a von Neumann algebra with a  s.n.f.  trace $\Tt$. We write $L^\infty(\Mm) = \Mm$ for convenience and assume that $\Mm$ acts on the GNS-Hilbert space $\Hh=\Hh_\Tt$. For an element $a \in \Mm$ define the support projection by
$$
\text{supp}(a) 
\;=\; 
\inf \{e \in \calP(\Mm) \;:\; a e = a\}
\;,
$$
where $\calP(\Mm)$ denotes the set of projections in $\Mm$. Then $S_+$ denotes the set of positive operators $x\in\Mm^+$ with finite support projection, namely $\Tt(\text{supp}(a) )<\infty$. Further let $S$ be the linear span of $S_+$. As $\Tt$ is semi-finite and normal, any projection is the increasing strong limit of $\Tt$-finite projections and as projections generate $\Mm$, the set $S$ is weakly dense in $\Mm$. Note that $a \in S$ implies $\lvert a \rvert^p \in S_+$ and hence one can show that 
$$
\lVert a \rVert_p 
\;=\; \Tt (\lvert a \rvert^p)^{\frac{1}{p}}
$$
defines a norm on $S$ for $1 \leq p < \infty$ and a $p$-quasi-norm for $0 < p < 1$. The non-commutative $L^p$-space is defined as the completion of $S$ in this (quasi-)norm and is denoted $L^p(\Mm)$. The H\"older-inequality still holds in the usual form $\lVert ab \rVert_r \leq \lVert a \rVert_p \lVert b \rVert_q$ where $\frac{1}{r} = \frac{1}{p} + \frac{1}{q}$ with $ 0 < r,p,q \leq \infty$. Hence $a \in L^p(\Mm)$ and $ b \in L^q(\Mm)$ implies $ab \in L^r(\Mm)$. If $\Tt$ is finite, this implies that $L^p(\Mm)\subset L^q(\Mm)$ for all $p>q$. Furthermore, this allows to make sense of products involving factors in $L^p$-spaces and the density of $L^p(\Mm) \cap \Mm$ implies that the cyclicity of the trace extends to $\Tt(ab)=\Tt(ba)$ for all $ a \in L^p(\Mm)$ and $b \in L^q(\Mm)$. Moreover, for $1 \leq p <\infty$ one still has the natural duality $L^p(\Mm) ^* = L^q(\Mm)$. In particular, $L^p(\Mm)$ is reflexive for $1 < p < \infty$ and $L^1(\Mm)$ is the pre-dual $\Mm_*$ of $\Mm$. 

\vspace{.2cm}

An important property of the spaces $L^p(\Mm)$ is that they are realized as (possibly unbounded) operators on $\Hh$. Recall  \cite{Dixmier81}  that a closed operator $a$ defined on a dense subset of $\Hh$ is called affiliated to $\Mm$ if it commutes with any unitary from the commutant of $\Mm$, or equivalently, if the partial isometry $u$ from its polar decomposition $a=u\abs{a}$ and every bounded Borel function of $\abs{a}$ are elements of $\Mm$. Furthermore \cite{FackKosaki}, an operator $a$ is called $\Tt$-measurable if it is affiliated to $\Mm$ and there is some $t\in \bbR$ such that $\Tt(\chi(\abs{a} > t)) < \infty$. For any two $\Tt$-measurable operators $a,b$ one can find dense domains such that their sum and product are well-defined, closable and $\overline{a+b}$, $\overline{ab}$  are $\Tt$-measurable. The set of $\Tt$-measurable operators $\overline{\Mm}$ is therefore a $*$-algebra with these operations. It is made into a topological $*$-algebra with the measure topology, which is the topology generated by the open neighborhoods
$$
N_{\epsilon,\delta} (x) 
\,=\,
\{y \in \overline{\Mm}: \text{ there is a projection }e \in \Mm \text{ with } \, \norm{(x-y)e} < \epsilon, \, \Tt(1-e)<\delta \}
.
$$
The trace $\Tt$ extends to an unbounded positive functional on the positive cone $\overline{\Mm}_+$ by defining 
$$
\Tt(x) 
\;=\; 
\sup_{n\in \bbN}\;\Tt\left( \int_{0}^n \lambda \,\difd{E_\lambda}\right)
\;,
$$
for every $x\in \overline{\Mm}_+$ with spectral resolution $E_\lambda$.
Every element $a\in L^p(\Mm)$ corresponds to an element of $a \in \overline{\Mm}$ such that
$$
\norm{a}_p \;=\; \Tt(\abs{a}^p)^{\frac{1}{p}}\;.
$$ 
All these facts are proved in \cite{FackKosaki} or references therein, just as the following generalizations classical convergence theorems:

\begin{theorem}
\label{theorem-measurable}
Let $a_n \in \overline{\Mm}$ be a sequence converging in the measure topology to $a \in \overline{\Mm}$.
\begin{enumerate}
\item[{\rm (i)}] {\rm (Fatou)} If all $a_n$ are positive, then $\Tt(a) \leq \liminf_{n\to \infty} \Tt(a_n)$.
\item[{\rm (ii)}] {\rm (Monotone Convergence)} If $0 \leq a_n \leq a$, then $\Tt(a) = \lim_{n\to \infty} \Tt(a_n)$.
\item[{\rm (iii)}] {\rm (Dominated Convergence)} Let $0 < p < \infty$. If $S_n \in L^p(\Mm)$ converges to $S \in L^p(\Mm)$ in measure topology and, moreover, $\lim_{n\to \infty} \norm{S_n}_p = \norm{S}_p$, and $\abs{a_n} \leq \abs{S_n}$, then 
$$
\lim_{n\to \infty}\norm{a-a_n}_p \;=\; 0
\;.
$$
\end{enumerate}
\end{theorem}

The measure topology has convenient properties, but in practice it can be difficult to show convergence in measure. In contrast, when working with bounded elements of $\Mm$ the strong and weak operator topology are more convenient, in particular, since continuous functional calculus preserves convergence of sequences. The following lemma relates the strong operator topology with convergence in $L^p(\Mm)$:

\begin{lemma}
\label{lemma:convergence}
Let $(a_n)_{n \in \bbN}$ be a sequence in $\Mm$ converging strongly to $a\in \Mm$ {\rm (}and hence the sequence is uniformly bounded in norm{\rm )}. 

\begin{enumerate}

\item[{\rm (i)}] The sequence converges in the strong operator topology of $\Mm$ on $\Hh$ if and only if it converges in the 
strong operator topology of $\Mm$ represented on $L^2(\Mm)$.

\item[{\rm (ii)}] Let $0 < p < \infty$. If $a_n\in \Mm \cap L^p(\Mm)$  and $(a_n)_{n \in \bbN}$ converges in the $L^p$-(quasi-)norm to some $\tilde{a} \in \Mm \cap L^p(\Mm)$, then $a = \tilde{a}$. 

\item[{\rm (iii)}] Let $1 < p <\infty$. If $a_n \in L^p(\Mm)$ holds for all $n$ and $\limsup_{n\to \infty} \norm{a_n}_p < \infty$, then $(a_n)_{n\in\NM}$ converges in the weak $\sigma(L^p(\Mm),L^q(\Mm))$-topology where $1=\frac{1}{p} + \frac{1}{q}$. 

\item[{\rm (iv)}] One has $\norm{a}_p \leq \liminf_{n\to \infty}\norm{a_n}_p$ for all $1 < p < \infty$. If in addition $\slim_{n\to \infty} a_n^* =a^*$, then the same also holds for $0 < p \leq 1$.

\item[{\rm (v)}] For a sequence $(b_n)_{n\in \bbN}$ in $L^p(\Mm)$, $0 < p < \infty$ that converges in (quasi-)norm, one has $\lim_{n\to \infty} a_n b_n = a b_n$ with convergence in $L^p(\Mm)$.
\end{enumerate}
\end{lemma}

\noindent {\bf Proof.} 
(i) A strongly convergent sequence is bounded and therefore also $\sigma$-strongly convergent, as the strong and $\sigma$-strong topologies coincide on bounded sets. The representation of $\Mm$ on $L^2(\Mm)$ is normal and hence $\sigma$-strongly continuous, {\it i.e.}  $(a_n)_{n\in\NM}$ converges in the $\sigma$-strong topology of $\calB(L^2(\Mm))$ and because it is bounded, also in the strong topology of $\calB(L^2(\Mm))$. Exchanging the roles of $\Hh$ and $L^2(\Mm)$ shows the other direction.

\vspace{.1cm}

(ii) By (i) it is sufficient to show $\slim_{n\to\infty}a_n = \tilde{a}$ for $\Mm$ acting on $L^2(\Mm)$ and as there is a uniform bound on the operator norm, it is enough to show  $a_n b \to \tilde{a} b $ for $b$ in the dense subset $\Mm \cap L^2(\Mm)$. For $p \leq 2$ one has
$$
\lVert (\tilde{a}-a_n)b \rVert^2_2 
\;=\; 
\Tt\big(|(\tilde{a}-a_n)b|^2\big)
\;\leq \;
\Vert \tilde{a}-a_n \rVert_\infty^{2-p} \, \lVert \tilde{a}-a_n \rVert^p_p \, \lVert b \rVert^2_\infty,
\;
$$
which converges to zero as $\norm{a_n}_\infty$ is uniformly bounded and for $p>2$ one can use the H\"older inequality in the form
$$
\lVert (\tilde{a}-a_n)b \rVert_2 
\;\leq\; 
\lVert \tilde{a}-a_n \rVert_p \, \lVert b\rVert_{\frac{2p}{p-2}}
$$
to conclude the same.

\vspace{.1cm}

(iii) It has to be shown that $a\in L^p(\Mm)$ and $\Tt(a_n b) \to \Tt(a b)$ for all $b\in L^q(\Mm)$. As $\norm{a_n}_p$ is uniformly bounded, it is enough to assert convergence for the dense subspace $b \in L^q(\Mm) \cap L^1(\Mm)$. For fixed $b \in L^1(\Mm)$ the map $x\in \Mm \mapsto \Tt(x b)$ is $\sigma$-weakly continuous and thus strongly continuous on bounded sets, which implies 
$$
\lim_{n\to \infty}\Tt(a_n b) 
\;=\; 
\Tt\big((\slim_{n \to \infty}a_n)b\big) 
\;=\; 
\Tt(ab)
\;, 
\qquad 
\forall\; b \in L^q(\Mm) \cap L^1(\Mm)
\;.
$$ 
To show $a\in L^p(\Mm)$, note that the functional 
$$
\psi\;:\; L^1(\Mm)\cap L^q(\Mm) \to \bbC\;, 
\qquad 
b \mapsto \Tt(ab) \,=\, \lim_{n\to \infty} \Tt(a_n b)
\;,
$$ 
is densely defined and bounded with $\abs{\psi(b)}\leq \limsup_{n\to\infty}\norm{a_n}_p \, \norm{b}_q$ and hence extends to a bounded functional on $L^q(\Mm)$. By duality $\phi$ must thus have the form $\phi(b)=\Tt(\tilde{a} b)$ for some $\tilde{a} \in L^p(\Mm)$ and $a=\tilde{a}$ must hold since $\Tt((a-\tilde{a})x)=0$ for all $\, x\in L^1(\Mm)$.

\vspace{.1cm}

(iv) One may assume $\liminf_{n \to \infty} \norm{a_n}_p < \infty$ and, after passing to a subsequence, also $\liminf_{n \to \infty} \norm{a_n}_p  = \lim_{n \to \infty} \norm{a_n}_p$. If $1 < p < \infty$, then (iii) implies that $a_n \to a$ converges weakly to $a\in L^p$ and hence $\norm{a}_p=\lim_{n\to \infty}\norm{a_n}_p$.
For $0 < p \leq 1$ the additional requirement implies $\abs{a}=\slim_{n\to\infty} \abs{a}$ and, since $\abs{\cdot}^{\frac{2}{p}}$ is a continuous function, the sequence $\abs{a_n}^{\frac{p}{2}} \in L^2(\Mm)$ converges strongly to $\abs{a}^{\frac{p}{2}}$ and is uniformly bounded in $L^2$-norm. Hence $\abs{a}^{\frac{p}{2}}\in L^2(\Mm)$ and thus $\abs{a}\in L^p(\Mm)$ with 
$$
\norm{a}_p
\;=\;
\norm{\,\abs{a}\,}_p 
\;\leq \;
\left(\liminf_{n\to \infty} \norm{\abs{a_n}^{\frac{p}{2}}}_2\right)^{\frac{2}{p}} 
\;=\; \liminf_{n\to \infty} \norm{a_n}_p
\;.
$$

(v) Due to $\norm{a_n (b-b_n)}_p \leq \norm{b-b_n}_p \, \sup_{n\in \bbN}\norm{a}$ it is enough to prove this for a constant sequence $b_n = b$. By density one can assume $b \in \Mm \cap L^p(\Mm) \cap L^2(\Mm)$. The statement holds for $p=2$ since this is the definition of strong convergence in the GNS-representation $L^2(\Mm)$. For $p<2$ write $b = b_1 b_2$ with $b_1, b_2 \in \Mm \cap L^2(\Mm)$ then $$\norm{(a_n -a) b}_p \leq \norm{(a_n-a) b_1}_2 \, \norm{ b_2}_{\frac{p}{2-p}}$$
and for $p>2$
$$
\norm{(a_n-a)b}_p 
\;\leq\; 
\norm{(a_n-a)b}^{\frac{2}{p}}_2 \, \norm{(a_n-a)b}^{1-\frac{2}{p}}_\infty
\;,
$$
which implies the claim.
\hfill $\Box$


\section{Complex interpolation theory}
\label{app-Interpol}

This appendix recall some basic notions and results from complex interpolation theory, {\it e.g.} from \cite{Lunardi2018}. Let $E_0$ and $E_1$ be Banach spaces over $\CM$ that are embedded in a  Hausdorff topological vector space such that $E_0 \cap E_1$ and $E_0 + E_1$ are well-defined and Banach spaces with respective norms 
$$
\norm{x}_{E_0 \cap E_1} 
\;=\; 
\max\{\norm{x}_{E_0},\norm{x}_{E_1}\}
$$
and
$$
\norm{x}_{E_0 + E_1} 
\;=\; 
\inf
\big\{\norm{x_0}_{E_0}+\norm{x_1}_{E_1}\;:\;
x=x_0+x_1\,,\;x_0\in E_0\,,\;x_1\in E_1
\big\}
\;.
$$  
Then $(E_0,E_1)$ is called a (complex) interpolation couple and one can generate intermediate spaces using real or complex methods. We will only need the complex method, which is based on the maximum principle in the form of Hadamard's three line theorem. Define $\scrF(E_0,E_1)$ to be the set of functions $f: S \to E_0+E_1$ on the vertical strip
$$
S
\;=\; 
\{z \in \bbC\,: \, 0 \leq \Re e( z) \leq 1\}
\;,
$$ 
that are continuous and bounded on $S$, analytic on the interior $S^\circ$ and satisfy on the boundaries that $t \mapsto f(\imath t) \in C_b(\bbR,E_0)$ and $t \mapsto f(1+\imath t)\in C_b(\bbR,E_1)$. Then $\scrF(E_0,E_1)$ is a Banach space with the norm
$$
\norm{f}_{\scrF(E_0,E_1)}
\;=\; 
\max
\big\{
\sup_{t\in\bbR}\norm{f(\imath t)}_{E_0},\,\sup_{t\in\bbR}\norm{f(1+\imath t)}_{E_1}
\big\}
\;,
$$
and for any $\theta \in (0,1)$ one can construct an intermediate Banach space
$$
(E_0,E_1)_\theta
\;=\; 
\{f(\theta)\,:\, f \in \scrF(E_0,E_1)\} 
\;=\; \scrF(E_0,E_1) / \Nn_\theta
\;,
$$
where $\Nn_\theta = \{f \in \scrF(E_0,E_1), f(\theta)=0\}$ and whose norm is the quotient norm
\begin{equation}
\label{eq-interpolnorm}
\norm{x}_\theta 
\;=\; 
\inf_{f \in \scrF(E_0,E_1)\,,\, f(\theta) = x} \norm{f}_{\scrF(E_0,E_1)}
\;.
\end{equation} 
The three line theorem then gives the log-convexity inequality for the norms
\begin{equation}
\label{eq-interpolnorm2}
\norm{x}_\theta \leq \norm{x}_0^{1-\theta} \norm{x}_1^{\theta}\;, 
\qquad \forall\; x \in E_0 \cap E_1
\;,
\end{equation}
which shows that $E_0 \cap E_1 \subset E_\theta \subset E_0 + E_1$ is indeed a space intermediate between $E_0$ and $E_1$.
Let us give the outcome of these constructions for two examples that turn out to be relevant later on.

\vspace{.2cm}

\noindent {\bf Example 1} ({\it e.g.} \cite{Kur})
Let $\Mm $ be a von Neumann algebra with a s.n.f.  trace $\Tt$. Then $(L^{p_0}(\Mm),L^{p_1}(\Mm))$ is an interpolation couple for $1 \leq p_0 < p_1 \leq \infty$ and
\begin{equation}
\label{eq-InterpolLp}
(L^{p_0}(\Mm),L^{p_1}(\Mm))_\theta 
\;=\; 
L^{p}(\Mm)
\;,
\end{equation}
where $\frac{1}{p} = \frac{1-\theta}{p_0} + \frac{\theta}{p_1}$. In particular, one has the log-convexity inequality  
\begin{equation}
\label{eq-LogConvex}
\|a\|_p\;\leq\; \|a\|_{p_0}^{1-\theta}\|a\|_{p_1}^{\theta}
\end{equation} 
for the $p$-norms.
\hfill $\diamond$

\vspace{.2cm}

\noindent {\bf Example 2} ({\it e.g.} \cite[Section 5.6]{BerghLofstrom76})
For a Banach space $E$ and $s\in \bbR$, $q\in [1,\infty)$ define the sequence spaces 
$$
\ell^q_s(E)
\;=\; 
\big\{x \in E^\bbN\;:\;  \sum_{k\geq 0} 2^{sqk} \norm{x_k}^q_E < \infty
\big\}
$$
and
$$
\ell^\infty_s(E)
\;=\; 
\{x \in E^\bbN
\;:\;  \sup_{k\geq 0} 2^{sk} \norm{x_k}_E < \infty
\big\}
\;,
$$
with norms
$$ 
\norm{x}_{\ell_s^q(E)}
\;=\;
\Big(\sum_{k\geq 0} 2^{skq} \norm{ x_k}^q_E\Big)^{\frac{1}{q}}
\;,
\qquad
\norm{x}_{\ell^\infty_s(E)}
\;=\;
\sup_{k\geq 0} 2^{sk} \norm{ x_k}_E
\;.
$$
If then $(E_0,E_1)$ is an interpolation couple, 
\begin{equation}
\label{eq-InterpollittleLp}
\left(\ell_{s_0}^{q_0}(E_0),\ell_{s_1}^{q_1}(E_1)\right)_\theta
\;=\;
\ell_s^q((E_0,E_1)_\theta)
\;,
\end{equation}
with $s = (1-\theta)s_0 + \theta s_1$ and $\frac{1}{q} = \frac{1-\theta}{q_0}+\frac{\theta}{q_1}$, 
\hfill $\diamond$

\vspace{.2cm}

In Section~\ref{sec-HigherPeller} the following interpolation theorem for analytic families of operators is used. It is a slight generalization of \cite[Theorem 2.7]{Lunardi2018}:

\begin{theorem}
\label{theo-Interpol}
Let $(E_0,E_1)$ and $(F_0,F_1)$ be interpolation couples and $\calD \subset E_0 \cap E_1$ a dense subspace w.r.t. the combined norm. If $z \in S \mapsto T_z \in \Bb(\calD, F_0 + F_1)$ is a family of operators on the strip $S$ such that for any $x\in \calD$ the map $z \mapsto T_z x$ is continuous and bounded on $S$, analytic in $S^\circ$, $T_{\imath t}$ and $T_{1+\imath t}$ take values in $F_0$ and $F_1$ respectively for all $t \in \bbR$ and there are constants $M_0$, $M_1$ such that
\begin{align*}
\sup_{t \in \bbR} \norm{T_{\imath t} x}_{F_0} \;\leq\; M_0 \norm{x}_{E_0}
\;,
\qquad
\sup_{t \in \bbR} \norm{T_{1+\imath t} x}_{F_1} \;\leq\; M_1 \norm{x}_{E_1}
\;,
\end{align*}
for all $x \in \calD$, then for any $\theta \in (0,1)$  one can  extend $T_\theta$ to a bounded operator from $(E_0,E_1)_\theta$ to $(F_0,F_1)_\theta$ satisfying
$$
\norm{T_\theta}_{\Bb((E_0,E_1)_\theta, (F_0,F_1)_\theta)} 
\;\leq \;
M_0^{1-\theta} M_1^\theta
\;.
$$
\end{theorem}

\noindent {\bf Proof.}
It is known \cite{Lunardi2018} that the finite linear span of the functions  $\calF(\bbC,\bbC) \cdot (E_0\cap E_1)$ is dense in $\calF(E_0,E_1)$ and hence the same holds for for $\calV(\calD) = \rm{span} \, \calF(\bbC,\bbC) \cdot \calD$.  By replacing $T_z $ by $ (M_0^{1-z} M_1^z)^{-1}T_z$ one may assume that $M_0=M_1=1$ and that $T_z$ is a contraction. These properties imply that the map $T: \calV(\calD) \to \calF(F_0,F_1)$ defined by
$$
T\big((f(z))_{z\in S}\big)
\;=\;
(T_z f(z))_{z\in S}
$$
is well-defined and bounded. Hence it extends to $\calF(E_0,E_1)$.  Using suitable linear combinations one can show that $\calN_\theta(E_0,E_1)\cap \calV(\calD)$ is dense in $\calN_\theta(E_0,E_1)$  (compare \cite[Remark 2.5]{Lunardi2018}), which implies that $T$ maps $\calN_\theta(E_0,E_1)$ into $\calN_\theta(F_0,F_1)$ and therefore lifts to a contractive map of the equivalence classes $$\tilde{T}_\theta: (E_0,E_1)_\theta \to (F_0,F_1)_\theta$$
that coincides with $T_\theta$ on $\calD$ using the natural identifications. 
\hfill $\Box$

\vspace{.2cm}

Let us finally comment on the interpolation of quasi-Banach spaces. The definition of interpolation couples can be extended to quasi-normed spaces and the real interpolation methods generalize straightforwardly to that situation. The complex method can be formulated for quasi-Banach spaces with some modifications (though there are several slightly different versions, {\it e.g.} \cite{CwikelEtAl86,CobosPeetrePersson98,KaltonMitrea98}), but due to the failure of the maximum principle in arbitrary quasi-Banach spaces the interpolation spaces $(E_0,E_1)_\theta$ obtained are in general only abstract completions of (a quotient of) $E_0 \cap E_1$ and not always interpolation spaces in the actual sense, namely $E_0 \cap E_1 \subset (E_0,E_1)_\theta \subset E_0 + E_1$ does in general not hold with continuous embeddings. However, as an important special case the non-commutative version of the Riesz-Thorin theorem also holds for $0 < p < 1$:

\begin{theorem}[\cite{Xu90}]
\label{theorem:rieszthorin}
Let $\Mm $ be a von Neumann algebra with a s.n.f. trace $\Tt$. Then, for $0 < p_0 < p_1 \leq \infty$,
$(L^{p_0}(\Mm,\Tt),L^{p_1}(\Mm,\Tt))$ is an interpolation couple  and
\begin{equation}
\label{eq-InterpolLp<}
(L^{p_0}(\Mm),L^{p_1}(\Mm))_\theta 
\;=\; 
L^{p}(\Mm)
\;,
\end{equation}
where $\frac{1}{p} = \frac{1-\theta}{p_0} + \frac{\theta}{p_1}$.
In particular, if $T: L^{p_0}(\Mm) \cap L^{p_1}(\Mm)\to L^{p_0}(\Mm) + L^{p_1}(\Mm)$ is a bounded operator w.r.t. the $\norm{\cdot}_{p_j\to p_j}$-quasi-norms, then $T$ extends from $L^{p_0}(\Mm) \cap L^{p_1}(\Mm)$ to a bounded operator
$$
T\;:\; L^{p_\theta}(\Mm)\; \to\; L^{p_\theta}(\Mm)
$$
with operator norm
$$
\norm{T}_{p_\theta \to p_\theta} 
\;\leq \;
\norm{T}_{p_0 \to p_0}^{1-\theta} \, \norm{T}_{p_1 \to p_1}^{\theta}
\;.
$$
\end{theorem}

\section{Compact and Breuer-Fredholm operators}
\label{app-Breuer}

This appendix outlines the theory of Breuer-Fredholm operators w.r.t. a semi-finite von Neumann algebra $(\Mm,\Tt)$. It originated from \cite{Breuer73} and was extended to real-valued indices by \cite{PR}. More recent reviews can be found in \cite{BCP,CPRS2,CPRS3}. A projection $e \in \Mm$ is called finite if it is in $L^1(\Mm)$. Let $\calF_\Tt$ denote the smallest algebraic ideal of $\Mm$ that contains the finite projections. Its norm-closure $\calK_\Tt$ is the ideal of the so-called $\Tt$-compact elements. While $\calK_\Tt$ is always a $C^*$-algebra, it is not necessarily separable. Another useful characterization of $\calK_\Tt$ is given in \cite{Lesch91}, namely for all $1 \leq p < \infty$ and with a closure in the operator norm,
\begin{equation}
\label{eq-TauComp}
\calK_\Tt \;=\; \overline{\Mm \cap L^p(\Mm)}
\;.
\end{equation}
There is a basic result concerning projections in $\calK_\Tt$.

\begin{proposition}[\cite{Kaad2012}]
Any projection $e \in \Kk_\Tt$ is already finite, that is, in $\Ff_\Tt$. The algebra $\Ff_\Tt$ is closed under the holomorphic functional calculus of its closure $\Kk_\Tt$. Hence it is a local $C^*$-algebra and the inclusion into $\Kk_\Tt$ induces an isomorphism  $i_*: K_0(\Ff_\Tt) \to K_0(\Kk_\Tt)$ of $K$-groups.
\end{proposition}

Since $\Tt$ defines a $0$-cycle on $\Ff_\Tt$, one therefore has an induced homomorphism $\Tt_*=\langle \Tt, \cdot\rangle:K_0(\Ff_\Tt) \to \bbC$ that extends to $K_0(\Kk_\Tt) \simeq K_0(\Ff_\Tt)$. The Calkin-algebra is defined as $\Qq_\Tt = \Mm / \Kk_\Tt$ and is also a $C^*$-algebra. An element $T \in \Mm$ is called $\Tt$-Fredholm (or Breuer-Fredholm w.r.t. $\calK_\Tt$) if its image $T + \Kk_\Tt \in \Mm / \Kk_\Tt$ is invertible in the Calkin-algebra.

\vspace{.2cm}

For any $a \in \Mm$ denote by $N_a$ the projection onto the kernel of $a$ and by $R_a$ the projection onto the closure of the range of $a$. Both of these projections are elements of $\Mm$ since they can be written as strong limits and there are the convenient expressions
$$
N_a \;= \;\sup\{e \in \calP(\Mm) \;:\; ae=0\}
\;,
\qquad
R_a\; =\; \inf\{e \in \calP(\Mm) \;:\; ea=a\}
\;.
$$ 
where $\calP(\Mm)$ denotes the set of projections in $\Mm$. There is the following generalization of Atkinson's theorem:

\begin{theorem}[\cite{PR}]
$a \in \Mm$ is $\Tt$-Fredholm if and only if there exists a projection $e \in \Kk_\Tt$ such that
$$
N_a\; \in\; \Kk_\Tt
\;,
\qquad
\Ran(\one-e)\, \subset \,\Ran(a)
\;.
$$
\end{theorem}

The second condition implies $\one-R_a \leq e \in \Kk_\Tt$ and is strictly stronger than this, as $\Tt$-Fredholm elements in general need not have a closed range. As a direct consequence one also has $N_{a^*} = \one - R_a \in \Kk_\Tt$. This allows to associate two indices to a $\Tt$-Fredholm operator $a$, namely the $K_0$-index
$$
\Ind(a) \;=\; [N_a]_0 \,-\, [N_{a^*}]_0 \;\in\; K_0(\Kk_\Tt)
$$
and the numerical index
$$
\Tt\text{-}\Ind(a) 
\;=\; 
\Tt(N_a) \,-\, \Tt(N_{a^*}) 
\;=\; \Tt_*(\Ind(a))
\; \in\; 
\bbR\;.
$$
The $K_0$-index is closely related to the index map $\Ind: K_1(\Qq_\Tt) \to K_0(\Kk_\Tt)$ corresponding to the Calkin-extension $ 0 \to \Kk_\Tt \to \Mm \to \Qq_\Tt \to 0$.  Indeed, consider the polar decomposition $a = u \lvert a \rvert$ with $u \in \Mm$ being the unique partial isometry that satisfies this formula and for which $\one-u^*u= N_a$ and $\one-uu^* = N_{a^*}$. The image $u+\Kk_\Tt$ is a unitary element of $\Qq_\Tt$ and since it lifts to the partial isometry $u$, 
$$
\Ind([u+\Kk_\Tt]_1)
\;=\;
 [\one-u^*u]_0 \,-\, [\one-uu^*]_0 
 \;=\; 
 [N_a]_0 \,-\, [N_{a^*}]_0 
 \;=\; 
 \Ind(a)
 \;.
 $$
The numerical index is not in general an isomorphism in $K$-theory unless $\Mm$ is a factor. It still has the following invariance properties \cite[Corollary 3.8]{CPRS3}:

\begin{proposition}
\label{prop-fredholminv}
\begin{enumerate}
\item[{\rm (i)}] If $a\in \Mm$ is $\Tt$-Fredholm, then there is some $\delta > 0$ such that $a + b$ is also $\Tt$-Fredholm with $\Tt\text{-}\Ind(a+b)=\Tt\text{-}\Ind(a)$ for all $b \in \Mm$ with $\norm{b}<\delta$. Thus the set of $\Tt$-Fredholm operators is open in the norm-topology and the numerical index is constant on each connected component.
\item[{\rm (ii)}] If $a$ is $\Tt$-Fredholm and $k \in \Kk_\Tt$ then $a + k$ is also $\Tt$-Fredholm and $\Tt\text{-}\Ind(a+k)=\Tt\text{-}\Ind(a)$.
\end{enumerate}
\end{proposition}

Let us close this appendix with a useful criterion and formula (see \cite{PS} for a proof, or \cite{Lesch91} in the case $n=1$, $m=1$).

\begin{theorem}[Semifinite Calderon-Fedosov formula]
\label{prop:fedosov}
Let $a \in \Mm$ such that  $(\one - aa^*)^n \in L^1(\Mm)$ and $(\one - a^*a)^n \in L^1(\Mm)$ for some $n \in \bbN \setminus \{0\}$. Then $a$ is $\Tt$-Fredholm and for all $m \geq n$,
\begin{equation}
\label{eq:fedosov}
\Tt\textup{-Ind}(a) 
\;=\; 
\Tt((\one - a^*a)^m) \;-\; \Tt((\one - aa^*)^m)
\;.
\end{equation}
\end{theorem}

\chapter{Acronyms and Notations}

\begin{tabbing} aaaaaaaaaaaaa \= aaaaaaaaaaaaaaaaaaaaaaaaaaaaaaaaaaaaaaaaaaaaaaaaaaaa \= aaaaaaaaa \kill
$\imath$ \> imaginary unit $\sqrt{-1}$ \>
\\[0.8ex]
$| \cdot |$ \> absolute value \>
\\[0.8ex]
SOT \> strong operator topology \> 
\\[0.8ex]
s.n.f. \> semi-finite normal and faithful (trace) \> 
\\[0.8ex]
$\TM\cong[0,1)$ \> torus \>
\\[0.8ex] 
$G=\bbT^{n_0} \oplus \bbR^{n_1}$ \> abelian group with $n=n_0+n_1$ parameters \>
\\[0.8ex]
$\hat{G}=\ZM^{n_0} \oplus \bbR^{n_1}$ \> dual group
\\[0.8ex]
$\scrS(\RM^n)$ \> Schwartz functions \> 
\\[0.8ex]
$E$ \> Banach space \> 
\\[0.8ex]
$x$ \> elements in Banach space $E$ \> 
\\[0.8ex]
$\beta$ \> $G$-action on Banach space $E$ \> 
\\[0.8ex]
$\Ff$ \> Fourier  transform \> Section~\ref{sec-CStar}
\\[0.8ex]
$FA(\hat{G})$ \> Fourier algebra $\calF L^1(G)$\>
\\[0.8ex]
$\chi$ \> characteristic function of a set or event \> 
\\[0.8ex]
${\rm sgn}$ \> sign function $\sgn(x)=\chi(x>0)-\chi(x<0)$ \> 
\\[0.8ex]
$\chi_s$ \> smooth characteristic function of a set from $C_{0,*}(\RM)$ \> Equation~\eqref{eq-SmoothCutOff}
\\[0.8ex]
$\imath$ \> imaginary unit $\sqrt{-1}$  \> 
\\[0.8ex]
$\gamma_i$ \> Clifford generators \> Equation~\eqref{eq:CliffGen}
\\[0.8ex]
$\Aa$ \> $C^*$-algebra $\Aa$ \> 
\\[0.8ex]
$a$ \> element of  $\Aa$ \> 
\\[0.8ex]
$\Aa^\sim$ \> unitization of an algebra $\Aa$ \> 
\\[0.8ex]
$M_N(\Aa)$ \> $N\times N$ matrices with entries in $\Aa$ \> 
\\[0.8ex]
$\diag(A,B)$ \> block diagonal matrix built from $A$ and $B$  \> 
\\[0.8ex]
$\mbox{\rm Tr}$ \> standard trace over Hilbert spaces \> 
\\[0.8ex]
$\Bb(\Hh)$ \> bounded operators on Hilbert space $\Hh$ \> 
\\[0.8ex]
$\calK(\Hh)$ \> compact operators on Hilbert space $\Hh$ \> 
\\[0.8ex]
$\one$ \> identity operator  \> 
\\[0.8ex]
$(\Aa,G,\alpha)$ \> $C^*$-dynamical system \>  Section~\ref{sec-CStar}
\\[0.8ex]
$\Aa \rtimes_\alpha G$ \> $C^*$-crossed product \> Section~\ref{sec-CStar}
\\[0.8ex]
$(\pi,U)$ \> left regular representation \> Equation~\eqref{eq-RegRepr}
\\[0.8ex]
$D$ \> generator in left regular representation $(\pi,U)$ \>
Equation~\eqref{eq-GenDefReg}
\\[0.8ex]
$(\Mm,G,\alpha)$ \> $W^*$-dynamical system \> Section~\ref{sec-WStar}
\\[0.8ex]
$\Nn=\Mm\rtimes_\alpha G$ \> $W^*$-crossed product \> Section~\ref{sec-WStar}
\\[0.8ex]
$\one_N$ \> identity in $M_N(\CM)$ or $M_N(\Aa)$ \>  
\\[0.8ex]
$\Tt$ \> s.n.f. trace \> Section~\ref{sec:traces}
\\[0.8ex]
$(\pi_\Tt,V)$ \> GNS representation (covariant) of $(\Mm,G,\alpha)$ on $L^2(\Mm)$
\> Section~\ref{sec:traces}
\\[0.8ex]
$X$ \> generator of $V$ in GNS representation $(\pi_\Tt,V)$ \> Equation~\eqref{eq-GenDefGNS}
\\[0.8ex]
$\calK_\Tt$ \> $\Tt$-compact operators \> Equation~\eqref{eq-TauComp}
\\[0.8ex]
$\Qq_\Tt$ \> Calkin algebra w.r.t. $\Tt$ \> Appendix~\ref{app-Breuer}
\\[0.8ex]
$\Tt$-$\Ind$ \>  Breuer-Fredholm index w.r.t. $\Tt$ \> Appendix~\ref{app-Breuer}
\\[0.8ex]
$\sigma_\alpha(x)$ \> Arveson spectrum of $x$ w.r.t. action $\alpha$ \> Definition~\ref{def-AversonSpec}
\\[0.8ex]
$\hat{\Tt}_\alpha$ \> dual trace \> Section~\ref{sec-DualTraces}
\\[0.8ex]
$\hat{\alpha}$ \> dual action of dual group $\hat{G}$ on crossed product  \> Section~\ref{sec-Duality}\\
[0.8ex]
$i_T$ \> Takai duality isomorphism  \> Section~\ref{sec-Duality}
\\[0.8ex]
$\nabla_v$ \> directional derivation of action  \> Section~\ref{sec-DiffElements}
\\[0.8ex]
$C^m(\Aa, \alpha) $ \> $m$ times differentiable elements of $\Aa$ \> Section~\ref{sec-DiffElements}
\\[0.8ex]
$\Aa_{\Tt,\alpha}$ \> Fr{\a'e}chet algebra of smooth elements \>
Section~\ref{sec-DiffElements}
\\[0.8ex]
$L^p(\Mm)$ \> non-commutative $L^p$-space w.r.t. $\Tt$ \> Appendix~\ref{app-Lp}
\\[0.8ex]
$W^m_{p}(\Mm)$ \> non-commutative Sobolev spaces \> Equation~\eqref{eq-SobolevNorm}
\\[0.8ex]
$\Mmc$ \> space of integrable operators with compact spectrum \> Equation~\eqref{eq-Mmc}
\\[0.8ex]
$(W_\ScaleInd )_{\ScaleInd \in\bbN}$ \> dyadic decomposition \> Equation~\eqref{eq-WkChoice}
\\[0.8ex]
$B^s_{p,q} (\Mm)$ \> scale of Besov spaces w.r.t. a $G$-action \> Equation~\eqref{eq-LpBesovDef}
\\[0.8ex]
$B_n(\Mm)$ \>  short notation for Besov space $B^{\frac{n}{n+1}}_{n+1,n+1}(\Mm)$ \> Equation~\eqref{eq-BesovAbbrev}
\\[0.8ex]
$P_I$ \> spectral projections of $D$ on $I \subset \bbR^n$  \> Equation~\eqref{eqref:spec_generator}
\\[0.8ex]
$H_a$ \> Hankel operator with symbol $a$  \> Definition~\ref{def-HankelToep}
\\[0.8ex]
$\hat{H}_a$ \> two-sided Hankel operator with symbol $a$  \> Equation~\eqref{eq:Hankel_commutator}
\\[0.8ex]
$\Ch_{\Tt,\alpha}$ \> Chern cocycle for $G$-action $\alpha$ \>  Definition~\ref{def-ChernCocycle}
\\[0.8ex]
$\Halfspaceaction$ \> $\RM$-action on $\Mm$  \> Chapter~\ref{sec-DualityToep}
\\[0.8ex]
$\Ch_{\Tt,\theta}$ \> Chern cocycle for $\RM^n$-action $\theta$ \> Definition~\ref{def-ChernCocycleFrechet}
\\[0.8ex]
$\DD$ \> Dirac operator associated to action \> Equation~\eqref{eq:dirac_def}
\\[0.8ex]
$\PP$ \> positive spectral (Hardy) projection of $\DD$ \> Definition~\ref{def-HankelToep}
\\[0.8ex]
$\Toep$ \> smooth Toeplitz extension  \> Definition~\ref{def-ToepExt}
\\[0.8ex]
$\partial^\xi_i$ \> Connes-Thom isomorphism \> Section~\ref{sec-ConnectSmoothToep}
\\[0.8ex]
$s_i$ \> suspension maps in $K$-theory \> Section~\ref{sec-ConnectSmoothToep}
\\[0.8ex]
\>  \>
\\[0.8ex]
\>  \>

\end{tabbing}

\vspace{.3cm}

\newpage

\noindent {\bf Acronyms and specific meaning in the applications of Chapter~\ref{sec-Applications}}

\vspace{.3cm}

\begin{tabbing} aaaaaaaaaaaaa \= aaaaaaaaaaaaaaaaaaaaaaaaaaaaaaaaaaaaaaaaaaaaaaaaaaaa \= aaaaaaaaa \kill
BGH \> bulk gap hypothesis \> Definition~\ref{def-BGH}
\\[0.8ex]
CH \> chiral symmetry hypothesis \> Definition~\ref{def-CH}
\\[0.8ex]
MGR \> mobility gap regime \> Definition~\ref{def-MGR}
\\[0.8ex]
DOS \> density of states \> Definition~\ref{def-DOS}
\\[0.8ex]
BBC \> bulk-boundary correspondence \> 
\\[0.8ex]
$(\Omega,\ZM^d,\PM)$ \> disorder configuration probability space with $\ZM^d$-action \>  Section~\ref{sec-AlgSetUp}
\\[0.8ex]
$E_F$ \> Fermi energy  \> 
\\[0.8ex]
$p_F$ \> Fermi projection $\chi(h\leq E_F)$  \> 
\\[0.8ex]
$\sgn_\epsilon$ \> approximate sign function \> Lemma~\ref{lemma:contours}
\\[0.8ex]
$u_F$ \> Fermi unitary operator for chiral system \>  
\\[0.8ex]
$J$ \> chiral symmetry operator \> Equation~\eqref{eq-JDef}
\\[0.8ex]
$\Aa=\bbT^d_{\BB,\Omega}$ \> disordered non-commutative torus, $C^*$-algebra  \> Section~\ref{sec-AlgSetUp}
\\[0.8ex]
$\pi_\omega$ \> physical representations of bulk on lattice Hilbert space  \> Equation~\eqref{eq-GNSfiber}
\\[0.8ex]
$\Tt$ \>  trace per unit volume \> Equation~\eqref{eq-TV}
\\[0.8ex]
$\Mm$ \> von Neumann algebra $L^\infty(\bbT^d_{\BB,\Omega},\Tt)$ of bulk observables  \> Section~\ref{sec-AlgSetUp}
\\[0.8ex]
$\Ee$  \>  $C^*$-algebra $\bbT^d_{\BB,\Omega} \rtimes_\Halfspaceaction G$ of edge operators \> Section~\ref{sec-HalfSpace}
\\[0.8ex]
$\hat{\Aa}$ \> $C^*$-algebra  ${\rm T}(\bbT^d_{\BB,\Omega}, G, \Halfspaceaction)$ of half-space operators \> Section~\ref{sec-HalfSpace}
\\[0.8ex]
$\Nn_{\Halfspaceaction}=\Mm  \rtimes_\Halfspaceaction G$ \> von Neumann algebra of half-space observables \> Section~\ref{sec-HalfSpace}
\\[0.8ex]
$X_\Halfspaceaction\;=\;\Halfspaceunitvector \cdot X $ \> position operator off the surface \> Section~\ref{sec-HalfSpace}
\\[0.8ex]
$\Halfspaceaction$ \> $\RM$-action induced by $v_\Halfspaceaction$ perpendicular on half-space \> Equation~\eqref{eq-ActionHalfSpace}
\\[0.8ex]
$\Halfspaceaction$ \> $\RM$-action shifting perpendicular to surface \> Equation~\eqref{eq-ActionHalfSpace}
\\[0.8ex]
$\Aa \rtimes_\Halfspaceaction G$ \> $C^*$-algebra of edge operators \> Section~\ref{sec-HalfSpace}
\\[0.8ex]
$P$ \> half-space projection $P=\chi(X_\xi>0)$  \>
\\[0.8ex]
$\Pp$ \> soft half-space restriction  \> Section~\ref{sec-BoundaryCurrents}
\\[0.8ex]
$h$ \> Hamiltonian from $\Aa$  \>
\\[0.8ex]
$\hat{h}$ \> half-space Hamiltonian  \> Proposition~\ref{prop-index_map_bb} and Eq.~\eqref{eq-HamHalfSpace}
\\[0.8ex]
$\nu_h$ \> density of states measure \> Definition~\ref{def-DOS}
\\[0.8ex]
$\theta$ \> $\RM^n$-action specifying (weak) invariants \> Equation~\eqref{eq-ThetaAct}
\\[0.8ex]
$\Ch_{\Tt,\theta}(p_F)$ \>  even bulk Chern numbers \> Equations~\eqref{eq-ChernNumberEven} 
\\[0.8ex]
$\Ch_{\Tt,\theta}(u_F)$ \> odd bulk Chern numbers \> Equation~\eqref{eq-ChernNumberOdd}
\\[0.8ex]
$\hat{\Tt}_\Halfspaceaction$ \> averaged trace per surface area \> Equation~\eqref{eq-ThatForm}
\\[0.8ex]
$\hat{\Tt}_{\omega,r}$ \> trace per surface area \> Propositions~\ref{prop-rationaltrace} and \ref{prop-irrationaltrace}
\end{tabbing}



\end{document}